\RequirePackage[l2tabu, orthodox]{nag}

\documentclass[11pt,phd,a4paper,twoside]{ucl_thesis}

\usepackage[T1]{fontenc}
\usepackage[utf8]{inputenc}
\usepackage{lmodern}

\usepackage{blindtext}
\blindmathtrue

\usepackage{emptypage}

\usepackage{graphicx}
\graphicspath{{./Figures/}}

\usepackage{float}

\usepackage{amsmath}

\usepackage{gensymb}
\usepackage{textcomp}

\usepackage{setspace}
\usepackage{multirow}

\usepackage{bibentry} 

\usepackage[format=hang,font=small,labelfont=bf]{caption}

\usepackage{etoolbox}

\usepackage{textgreek}
\usepackage{mathtools}

\usepackage{amssymb} 
\usepackage{amsthm}
\usepackage{bm}
\theoremstyle{plain}
\newtheoremstyle{slanted}
  {}
  {}
  {\slshape}
  {}
  {\bfseries}
  {}
  { }
  {}
\theoremstyle{slanted}
\newtheorem{thm}{Theorem}
\newtheorem{lem}{Lemma}

\theoremstyle{definition}
\newtheorem{defn}{Definition}

\newtheorem{exmp}{Example}

\theoremstyle{remark}

\usepackage{xcolor}%
\usepackage{booktabs}
\usepackage{pdflscape}
\usepackage[font={small}]{caption}
\usepackage{dcolumn}
\usepackage{stackengine}
\usepackage{setspace}
\usepackage{algpseudocode,algorithm,algorithmicx}
\floatstyle{plain}
\newfloat{myalgo}{tbhp}{mya}
{\begin{myalgo}[#1]
\centering
\begin{minipage}{#2}
\begin{algorithm}[H]\setstretch{1.15}}%
{\end{algorithm}
\end{minipage}
\vspace{0.5cm}
\end{myalgo}}
\usepackage{physics}
\renewcommand{\braket}[1]{\ensuremath\langle #1\rangle}
\usepackage{complexity}
\usepackage{enumerate}
\usepackage{enumitem}
\usepackage{hhline}
\usepackage{array}
\newcolumntype{P}[1]{>{\centering\arraybackslash}p{#1}}

\usepackage[textsize=footnotesize]{todonotes}

\usepackage{soul}

\usepackage{tocloft}

\usepackage{bibentry}
\makeatletter\let\saved@bibitem\@bibitem\makeatother
\usepackage[pdftex,hidelinks]{hyperref}
\makeatletter\let\@bibitem\saved@bibitem\makeatother
\makeatletter
    \hypersetup{
        pdfsubject={Thesis Subject},
        pdfkeywords={Thesis Keywords},
        pdfauthor={Author},
        pdftitle={Title},
        colorlinks=true,       
	    linkcolor=blue!40!black,          
	    citecolor=blue!40!black,
	    filecolor=blue!40!black,
	    urlcolor=blue!40!black,
	    runcolor=blue!40!black,
    }
\makeatother
    
\usepackage[caption=false]{subfig} 
\renewcommand{\trace}[1]{\ensuremath\operatorname{Tr}\left( #1 \right)}

\renewcommand{\p}{\ensuremath\partial}

\usepackage{imakeidx}
\makeindex

\usepackage{nomencl}
\makenomenclature 

\usepackage{tabularx,ragged2e,booktabs,caption}

\usepackage{dsfont}

\usepackage{siunitx}

\usepackage[noadjust]{cite}

\usepackage{tabularx,ragged2e,booktabs}

  \makeatletter
    \def\@pnumwidth{2em}
  \makeatother
  
\begin{document}






\setstretch{1.3}


\title{Quantum metrology with optomechanical systems in the nonlinear regime}
\author{Sofia \textsc{Qvarfort}}
\department{Department of Physics and Astronomy}
\supervisor{Prof. Alessio \textsc{Serafini}}
\date{\today}

\maketitle
\makedeclaration

\begin{abstract} 
This thesis focuses on the mathematical description and application of nonlinear cavity optomechanical systems. 
The first part is concerned with solving the dynamics of the standard nonlinear optomechanical Hamiltonian with an additional time-dependent mechanical displacement and single-mode squeezing term. The solution is based on identifying a minimal and finite Lie algebra that generates the time-evolution of the system, which reduces the problem to considering a finite set of coupled ordinary differential equations of real functions. We derive analytic solutions when possible, and otherwise consider perturbation theory for specific cases. 
The second part of this thesis applies the solutions of the extended optomechanical Hamiltonian to the study of non-Gaussianity. We compute the non-Gaussian character of an optomechanical state as a function of the parameters in the Hamiltonian, and investigate the interplay between the non-Gaussianity, the strength of the nonlinear coupling and the strength of a single-mode mechanical squeezing term. While we find that the strength and form of the nonlinear coupling strongly impacts the non-Gaussianity, its relationship with the squeezing term is highly complex.  
The third part of this thesis concerns the use of nonlinear optomechanical systems as quantum sensors. We derive a general expression of the quantum Fisher information given the extended optomechanical Hamiltonian, and demonstrate its applicability through three concrete examples: estimating the strength of a nonlinear light--matter coupling, the strength of a time-modulated mechanical displacement, and the strength of a single-mode mechanical squeezing parameter, all of which are modulated at resonance. In the last Chapter of the thesis, we consider the estimation of a constant gravitational acceleration with an optomechanical system. We derive the fundamental limits to gravity sensing and prove that homodyne detection saturates the quantum Fisher information bound. Our results suggest that optomechanical systems could, in principle, be used as powerful quantum sensors.

\end{abstract}

\begin{impactstatement}
The use of sensors to probe the natural world constitutes an integral part of the scientific method. Recent scientific interest has focused intensely on sensors operating in the quantum regime since quantum properties (such as coherence and entanglement) have been shown to vastly enhance the sensitivities that can be achieved by quantum systems compared with classical systems.

One family of systems with significant sensing potential is cavity optomechanical systems. They consist of light resonating inside a cavity interacting with a small mechanical element. Their sensitivity to extremely small displacements make them an ideal candidate for force sensing and accelerometry. As a result, optomechanical systems are emerging as one of the most promising force sensing quantum technologies, and they have been identified as one of the key quantum sensing platforms in the European Quantum Flagship Initiative. The first optomechanical and opto-electrical prototype sensors are envisioned to be built within the next 10 years. 

The results presented in this thesis aid the theoretical development and understanding of optomechanical systems operating in the nonlinear regime. They expand on a large body of research concerning systems in the linear regime, which approximates the full nonlinear dynamics but fails to reproduce certain useful properties of nonlinear systems, such as the ability to generate non-Gaussian states. The main result in this thesis is the analytic solution (up to two differential equations) of a nonlinear optomechanical system with time-dependent displacement and squeezing terms. Subsequent investigations into non-Gaussianity and quantum metrology shed light on the  complexity of nonlinear optomechanical states and provides a set of tools that can be used to compute the expected fundamental sensitivity of the optomechanical system. A specific application of these results in this thesis concerns the measurement of constant gravitational acceleration, also known as gravimetry, where it is shown that sensitivities of up to $10^{-15}\,\rm{ms}^{-2}$ could in principle be achieved. 

Within academia, the results presented in this thesis allow for a better and deeper understanding of the evolution and nature of optomechanical systems. They could potentially be used to explain effects seen in the laboratory due to the nonlinear dynamics, which is currently not possible with methods for linear systems. In terms of sensing, the methods developed in this thesis allows for the modelling of arbitrary, time-dependent mechanical displacements and single-mode squeezing effects, which cover a substantial number of physical processes that can act on the optomechanical system. When placed on satellites, optomechanical sensors can be used to measure gravity anomalies of the Earth or other planets. These gravimetric surveys can aid astrophysical research and even climate crisis research by, for example, tracking large bodies of meltwater through their gravitational signature. From the perspective of fundamental physics, the use of quantum optomechanical sensors opens up the possibility of measuring extremely weak gravitational effects. Probing the nature of gravity and searching for deviations from the predictions of general relativity could yield insights into physics beyond the Standard Model, and help solve the outstanding problem of developing a unified theory of quantum gravity. The results in this thesis are likely to aid the effort of developing nonlinear optomechanical sensors that can be used for this purpose and test deviations beyond the nano-meter scale, which is the current bound. 

Outside academia, quantum-enhanced optomechanical sensors could, in the long term, be used to build highly sensitive  earthquake detection systems, ground water sensors and sensors for geological and mineralogical surveys. The results presented here imply that further study into building commercial optomechanical sensors is well-motivated. 

\end{impactstatement}

\begin{acknowledgements}
First of all, my deepest thanks go to my supervisor Alessio Serafini. His insight, guidance and support made the time as a PhD student an absolute pleasure. I also thank him for the freedom he gave me in following and developing my own research ideas. This allowed me to grow into a creative and independent researcher. I also thank his family for their wonderful hospitality on our visit to Italy. 

I also thank Sougato Bose, who, as my secondary supervisor, set me on the path of studying optomechanics and quantum metrology. I especially thank him for his continued support throughout my PhD. 

This thesis was much improved by the thoughtful comments provided by my two examiners Ivette Fuentes and Andrew Fisher. I thank them for their careful reading of this work as well as their encouraging feedback. 

My thanks also goes to my co-authors and collaborators, Peter Barker, David Edward Bruschi, Daniel Braun, Dennis R\"{a}tzel Fabienne Schneiter, and Andr\'{e} Xuereb. I thank them for their guidance and teaching, especially David and Dennis, from whom I learned a lot. My PhD and research would not be what it is today without their insight, support and hard work. 

My time at University College London was fantastic in so many ways thanks to the other members of my PhD cohort. I thank Paul Brookes, Nathana\"{e}l Bullier, Padraic Calpin, Danial Dervovic, Gavin Dold, Alex Morgan, James Morley, Simon Shaal, Michael Vassmer, and David Wise for their support and  friendship. I also thank them for the many great memories from the trips and events that we attended together. 

There are many people who made this PhD project special and fantastic. I would like to thank, in no particular order, Peter Barker, Tania Monteiro, Agnese Abrusci, Alfred Harwood, Tom Rivlin, Abbie Bray, Lia Li, Matt Flinders, TK Le, Ryan Marshman, Tamara Kohler, Andy Maxwell, Jacob Lang, Uther Shackerley-Bennet, and Marko Toros. Special thanks go to Paul, Alex, and Alfred, who shared my office and with whom I enjoyed many great discussions and conversations. 

Throughout my PhD,  I enjoyed many inspiring discussions with the following researchers: Daniel Goldwater, Doug Plato, Marco Genoni, Stefano Liberati, Magdalena Zych, Igor Pikovski, \v{C}aslav Brukner, Niels Linnemann, Juani Bermejo-Vega, Robin Lorenz, Johannes Kleiner, Alessio Belenchia, Flavio del Santo, Anja Metelmann, Gavin Morley, James Millen, and Michael Vanner. Special thanks go to Daniel for reading parts of my thesis and to Anja for inviting me to Berlin and introducing me to everyone there. 

I had the privilege to visit the University of Vienna for two-and-a-half months in 2018. Many people made the stay very special. I thank Markus Aspelmeyer for the opportunity to come visit, and Sahar Sahebdivan, Flavio del Santo and Chiara Cardelli, Fabienne Schneiter, and Dennis R\"{a}tzel for fantastic discussions and some lovely evenings of classical music in the Musikverein. I especially thank David Edward Bruschi and his partner Leila for welcoming me to their home, especially over the Easter holidays.

While I completed my PhD far away from home, my family was always there for me and cheered me on. I thank my sister Anna, my mother Theresa, my father \r{A}ke and his wife Lovisa for their love and support. I hope that I have made them proud. I also thank my dear friend Elena Un for the many good meals and good times we enjoyed in London and elsewhere. 

Finally, I thank my husband Stephen for his love and support, and especially for marrying me in the middle of my PhD. I cannot express in words the joy and happiness the last few years have brought, and I am delighted to be sharing this journey with him. 

\end{acknowledgements}

\newpage

\begin{publications}
This thesis is based on the following publications. 

\begin{itemize}
\item[\cite{qvarfort2018gravimetry}] Sofia Qvarfort, Alessio Serafini, Peter F Barker, and Sougato Bose. \href{
https://doi.org/10.1038/s41467-018-06037-z}{Gravimetry through non-linear optomechanics}. \textit{Nature Communications}, \textbf{9}, 3690 (2018).
\item[\cite{qvarfort2019enhanced}]  Sofia Qvarfort, Alessio Serafini, Andr\'{e} Xuereb, Dennis R\"{a}tzel, and David  Edward Bruschi. \href{https://doi.org/10.1088/1367-2630/ab1b9e}{Enhanced continuous generation of non-Gaussianity through optomechanical modulation}. \textit{New Journal of Physics} (2019).
\item[\cite{qvarfort2019time}] Sofia Qvarfort, Alessio Serafini, Andr\'{e} Xuereb, Daniel Braun, Dennis R\"{a}tzel, and David Edward  Bruschi. \href{https://doi.org/10.1088/1751-8121/ab64d5}{Time-evolution of nonlinear optomechanical systems: Interplay of arbitrary mechanical squeezing and non-Gaussianity}. \textit{Journal of Physics A: Mathematical and Theoretical}, (2019).
\item[\cite{schneiter2019optimal}] Fabienne Schneiter, Sofia Qvarfort, Alessio Serafini, Andr\'{e} Xuereb, Dennis R\"{a}tzel, and David Edward Bruschi. \href{https://journals.aps.org/pra/abstract/10.1103/PhysRevA.101.033834}{Optimal estimation with quantum optomechanical systems in the nonlinear regime.} \textit{Physical Review A}, 101.3 (2020): 033834. 
\end{itemize}

The following publications were not included in this thesis. 
\begin{itemize}
\item[\cite{qvarfort2018mesoscopic}] Sofia Qvarfort, Sougato Bose, Alessio Serafini. \href{https://arxiv.org/abs/1812.09776}{Mesoscopic entanglement from central potential interactions}. \textit{arXiv preprint arXiv:1812.09776}, (2019)
\end{itemize}

\end{publications}

\setcounter{tocdepth}{2} 

\newpage

\tableofcontents

\cleardoublepage

\listoffigures
\raggedbottom

\cleardoublepage
\cleardoublepage
\listoftables
\cleardoublepage


\renewcommand{\nomname}{List of Symbols}
\renewcommand{\nompreamble}{The next list describes several symbols that will be later used within the body of the document.
If not using \texttt{latexmk}, will need to run code \texttt{makeindex Main.nlo -s nomencl.ist -o Main.nls}.
In any case, compile twice 
}

\nomenclature{$\hbar$}{Reduced Planck constant}
\nomenclature{$\omega_{\rm{c}}$}{Optical oscillation frequency}
\nomenclature{$\omega_{\rm{m}}$}{Mechanical oscillation frequency}
\nomenclature{$\mathcal{G}(t)$}{Time-dependent light--matter coupling}
\nomenclature{$\mathcal{D}_1(t)$}{Time-dependent linear mechanical displacement coupling}
\nomenclature{$\mathcal{D}_2(t)$}{Time-dependent single-mode mechanical squeezing term}
\nomenclature{$\hat H$}{Hamiltonian}
\nomenclature{$\hat a, \hat a^\dag$}{Annihilation and creation operators for a single optical mode}
\nomenclature{$\hat b$, $\hat b^\dag$}{Annihilation and creation operators for a single mechanical phonon mode}
\nomenclature{$\alpha$}{Bogoliubov coefficient}
\nomenclature{$\beta$}{Bogoliubov coefficient}
\nomenclature{$\delta$}{Measure of non-Gaussianity}
\nomenclature{$c$}{Speed of light in a vacuum inertial frame}
\nomenclature{$g_0$}{Optomechanical coupling amplitude}
\nomenclature{$d_1$}{Linear displacement amplitude}
\nomenclature{$d_2$}{Single-mode mechanical squeezing amplitude}
\nomenclature{$\omega_{g}$}{Optomechanical coupling oscillation frequency}
\nomenclature{$\omega_{d_1}$}{Linear mechanical displacement oscillation frequency}
\nomenclature{$\omega_{d_2}$}{Single-mode mechanical squeezing oscillation frequency}
\nomenclature{$\mathcal{I}_Q$}{Quantum Fisher information}
\nomenclature{$V_{\rm{c}}$}{Cavity mode volume}
\nomenclature{$\epsilon_0$}{Permittivity of free space}
\nomenclature{$\epsilon$}{Relative permittivity}
\printnomenclature


\part{Theory of nonlinear optomechanical systems}

\chapter{Introduction}
\label{chap:introduction}

The main focus of this thesis is the theory of optomechanical systems operating in the nonlinear regime. As part of this investigation, we develop the theoretical description of optomechanical systems, and apply this to study the non-Gaussianity of optomechanical states, and the use of optomechanical systems for quantum metrology tasks. In this first Chapter, we aim to provide a comprehensive introduction and background to the research results that we later present. 

This introductory chapter is divided into three parts. We begin by introducing the field of theoretical nonlinear optomechanics, where we also define the notions of nonlinear as opposed to linear dynamics. We then proceed with the topic of non-Gaussianity in quantum theory, for which we define Gaussian states and provide a brief introduction to the covariance matrix formalism. The discussion centres on a measure of non-Gaussianity based on the relative entropy of a Gaussian and non-Gaussian state. Finally, we introduce the field of quantum metrology and define the key quantity of interest: the quantum Fisher information (QFI) and the Cram\'{e}r--Rao bound. Some of the insights in this Chapter, especially in Section~\ref{chap:introduction:non:Gaussianity:general:behaviour} were contributed by David Edward Bruschi. 
 
Before we proceed with the technical content, we provide a layperson's summary of the results in this thesis. This summary is meant to be a non-technical introduction intended for an audience with little or no knowledge of physics or mathematics.

\section{Layperson's summary}
In physics, it is common to study the evolution of a physical system in order to predicts its behaviour. By \textit{system}, we mean an isolated physical object which can be described independently of its environment, such as a single atom trapped in the laboratory. We predict the behaviour of the systems by modelling them mathematically, either by hand or by using a computer to solve the equations numerically. An intuitive example is throwing a ball and calculating the endpoint of its trajectory: If we know the starting velocity and the direction of the ball, we can predict exactly where it will land. 

\begin{figure}[h!]
\centering
  \includegraphics[width=.7\linewidth, trim = 14mm 0mm 10mm 12mm]{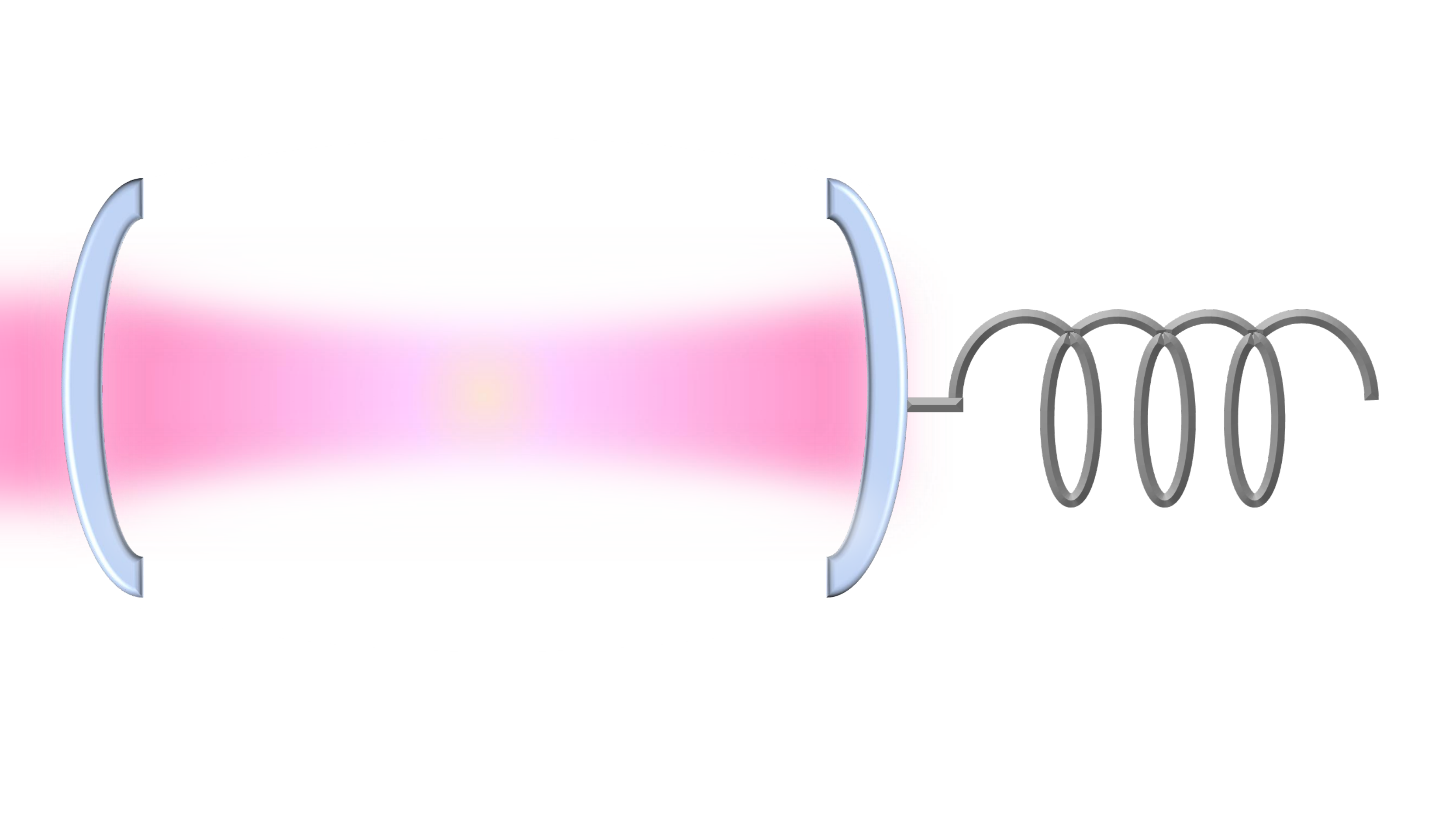}%
\caption[A Fabry-P\'{e}rot moving-end mirror system]{An optomechanical system consisting of a moving end mirror.  }
\label{chap:introduction:fig:mirror}
\end{figure}

\begin{figure}[h!]
\centering
{\includegraphics[width=.5\linewidth, trim = -10mm 0mm 0mm -10mm]{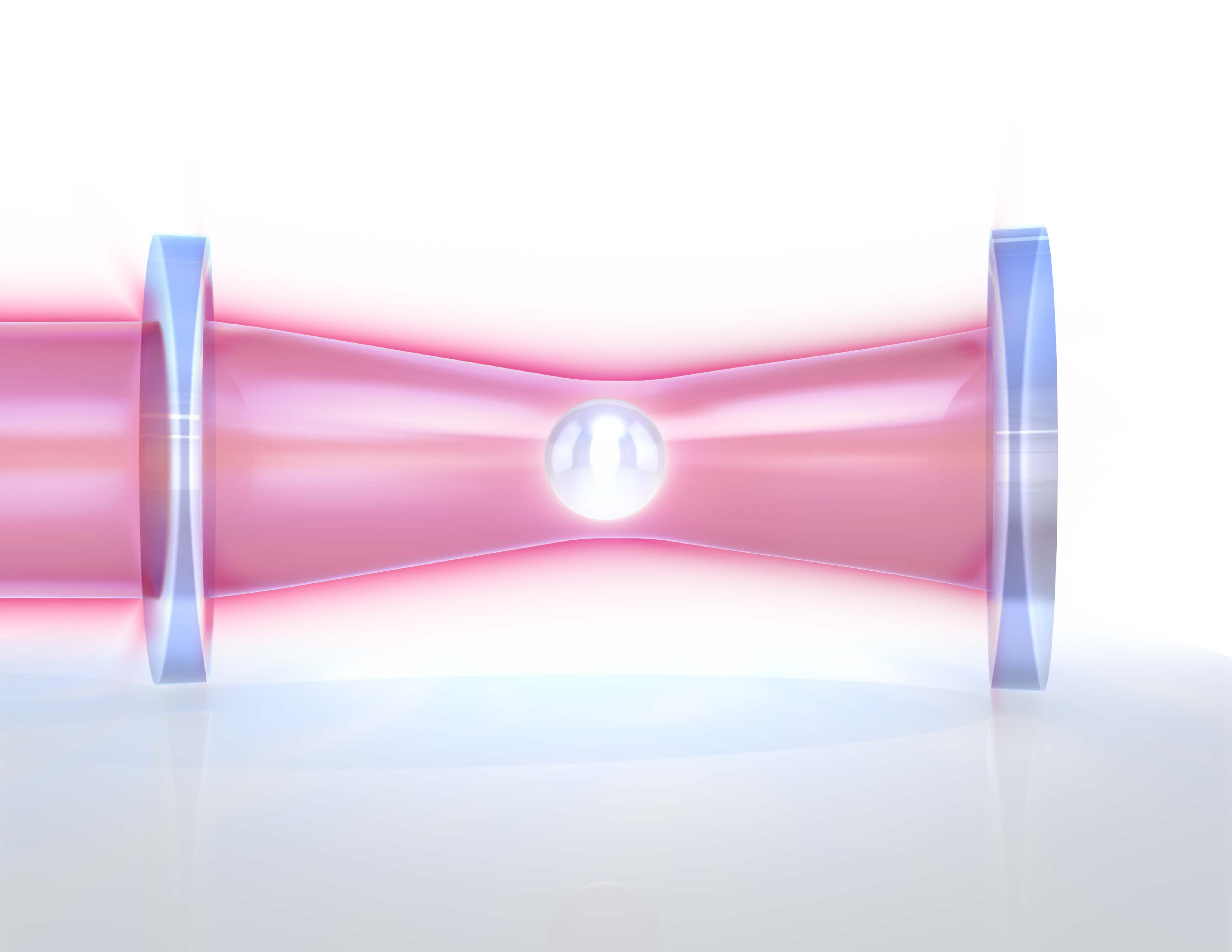}}\hfill
\caption[A levitated nanosphere in a cavity]{A levitated nanosphere in a cavity. Image credits: Mark Mazaitis.}\label{chap:introduction:fig:mirror:cavity}
\end{figure}

The same can be done for more complicated systems, such as quantum optomechanical systems. 
In this thesis, we study so-called \textit{cavity optomechanical systems}. They are systems where light (`opto') has been trapped in a cavity that consists of two mirrors. The light bounces back and forth between the mirrors, and can stay in the cavity for a fairly long time until imperfections in the mirrors causes it to be absorbed or reflected out of the cavity. The light interacts with a small \textit{mechanical element} that is placed in the cavity. As the mechanical element, we can take one of the mirrors of the cavity, as in Figure~\ref{chap:introduction:fig:mirror}. It is attached to a small spring, thereby allowing it to move slightly around a fixed point. The mirror must be extremely small, often no larger than the width of a hair. Alternatively, the mechanical element can be a small sphere made of diamond or silica that is suspended with other lasers in the middle of the cavity, as in Figure~\ref{chap:introduction:fig:mirror:cavity}. The bead must be suspended to prevent it from falling due to gravity. In comparison with the mirror, the bead is much smaller, often around one nanometer in diameter. This is 10,000 smaller than the width of a human hair, and it can often not be seen with the naked eye.

There are many more systems that fall under the category of 'optomechanics'. For example, the bead can be replaced with a cloud of trapped atoms, which collectively behave just like a single mechanical element. The different systems all have specific advantages and disadvantages, and many of them are being studied in laboratories around the world. Fortunately, they all behave in very similar ways, which means that we can study them all with the same tools. We can describe the photons and the movement of the mechanical element mathematically,  which makes it possible to predict how they evolve and interact.

The reason for the cavity is so that the light can interact with the mechanical element many times. What does this interaction look like? The light is made up of particles called photons, and while photons have no mass, they do carry a very small amount of energy. When the photons reach the mirror or the levitated bead, they bounce off it and give it a small push. Just like hitting tennis balls against a door would make it slowly swing open, the impacts of the photons cause the mirror or the bead to move slightly. This, in turn, changes the length of the cavity, and this dictates which \textit{wavelengths} of light we can trap inside the cavity. The result is an  interesting self-interacting interplay between the light and the position of the mechanical element: The light pushes the mechanics; this changes the type of light we can have in the cavity. This, in turn, pushes differently on the mechanical element,  this changes the light that we can have in the cavity\ldots and so on. In summary, the behaviour of the system is determined by the behaviour of the system, which might sound complicated, but it is a prime example of what we refer to as \textit{nonlinear} dynamics. 

Now, things becomes even more interesting when the mechanical element is a quantum system. When it is cooled down to extremely low temperatures, the mechanical element can no longer be described with the laws of physics that describe things we see in our everyday life. Instead, we must use the laws of \textit{quantum mechanics}, which are distinctly different to the \textit{classical} laws that we are used to. Quantum mechanics  allows for fascinating phenomena such as \textit{superpositions}. The most famous example of a quantum superposition is Schr\"{o}dinger's cat, which is a thought experiment, where a hypothetical cat is said to be both dead and alive at the same time. Another example of a superposition is a \textit{spatial superposition}, where an atom is in a state that, when measured, \textit{collapses} (according to some interpretations) into one of two or more distinct spatial positions. Another interesting quantum property is \textit{entanglement}. Two quantum systems, such as two particles, can become entangled after interacting, and they remain entangled even when separated by large distances. 
The effect of entanglement is that the particles exhibit correlations that are stronger than those allowed by classical physics. If the first particle is measured, the second one ends up in a state that is perfectly correlated with the first one. For optomechanical systems in the quantum regime, the mechanical element can be placed in highly non-classical states, such as spatial superpositions, and it can even become entangled with the light. 

We are motivated to study optomechanical systems for several reasons, but there are two main points: Firstly,  optomechanical systems could potentially be used to test fundamental physics. One major unsolved question in theoretical physics is that of quantum gravity. Objects that behave quantum-mechanically are so small that we cannot measure their gravitational field, and we therefore have no data on how gravity behaves on the scales of atoms or electrons. By making the quantum objects larger and larger, so that we can eventually measure the gravitational field of, say, the levitated beads mentioned inside the cavity, we hope to one day gather enough information to learn a little bit more about how gravity affects these systems. Secondly, the mechanical element in an optomechanical system is extremely sensitive to small displacements, from being pushed by a force or experiencing an acceleration. As a result, optomechanical systems could potentially become powerful quantum-enhanced sensors that can help us measure very small forces. The fact that they are quantum helps too. Certain quantum-mechanical properties, such as superpositions and entanglement, can be utilised to enhance the sensitivity of the system, and small systems are generally easier to control in the laboratory. Some potential applications of optomechanical sensors include earthquake detection systems, accelerometers, and gravity sensors for astrophysical and geological research. 

This thesis is concerned with the mathematical description of optomechanical systems and their application as sensors. To describe the optomechanical system in an accurately manner is quite difficult, since the equations are hard to solve, both by hand and by using computers. As a result, it has not been possible to model the system and its behaviour in the laboratory in full generality. Therefore, one of the key goals of this thesis is to provide a more complete mathematical description of optomechanical systems. Additional goals include studying specific ways in which optomechanical system changes with time, and deriving the best-possible sensitivity of optomechanical sensors. 

What follows is a brief summary of the results in the thesis, chapter by chapter. Our main result can be found in Chapter~\ref{chap:decoupling}, where we provide a mathematical description of both standard and more complex optomechanical systems. While the basic evolution of systems with constant interactions were previously known, we are able to solve the dynamics for systems with time-dependent effects, which is generally much harder to do. We then proceed in Chapters~\ref{chap:non:Gaussianity:coupling} and~\ref{chap:non:Gaussianity:squeezing} to compute the \textit{non-Gaussianity} of the system; a property of the quantum state that could potentially be beneficial for sensing. Our results include the fact that there are many different ways to create non-Gaussian states with optomechanical systems, and that a strong enhancement can be achieved by modulating the light--matter interaction in time, which was not previously known. In Chapter~\ref{chap:metrology}, we instead focus on using the optomechanical system as a quantum sensor. By looking at how the system reacts to external effects, such as being slightly displaced, we can calculate how well the system cam measure external effects, and we provide a way to compute the fundamental bound to the sensitivity. In general, we find that optomechanical systems could, in principle, be used as very powerful sensors. To demonstrate this, in Chapter~\ref{chap:gravimetry} we consider the effects of gravity, and we compute the sensitivity by which the gravitational acceleration can be measured with an optomechanical system. While it was known that optomechanical sensors could be used to measure acceleration, this is the first time that the fundamental bounds have been computed given the quantum dynamics of the system. A technical summary of the thesis can be found in Chapter~\ref{chap:conclusions}. 

The results in this thesis show that optomechanical systems could potentially be used as extremely powerful quantum sensors, but more work is needed to determine exactly how good they can be in a realistic setting.  One question this thesis does not answer is what happens when noise is included, which every real experimental system suffers from. 

We hope that the results described in this thesis will be used to further describe and understand optomechanical systems and their applications as quantum-enhanced sensors. Should you, the reader, have any questions relating to these results, the thesis author is more than happy to answer them via email. Please contact the author using the following email address: \href{sofiaqvarfort@gmail.com}{sofiaqvarfort@gmail.com}.

\section{Theory of nonlinear optomechanical systems} \label{chap:introduction:theory:of:nonlinear:optomechanical:systems}

This thesis is concerned with the analytical treatment of optomechanical systems operating in the nonlinear regime. Our main goal is to solve the dynamics of optomechanical systems and apply the solutions to study the generation of non-Gaussian states and the use of optomechanical systems for quantum metrology tasks. In order to explore these applications, we must first introduce the fundamental mathematical and physical tools needed to model these systems. 

In this Section, we  provide a basic introduction to the theory of nonlinear quantum optomechanical systems. We start by briefly describing optomechanical systems and motivate their study by providing a brief review of their role in tests of fundamental physics and quantum technology applications. Next, we define the optomechanical Hamiltonian and discuss results currently present in the literature. 
Finally, to demonstrate the link between the linear and nonlinear regimes, we outline the linearisation process as applied to optomechanical systems. 

\subsection{Background and motivation}


An optomechanical system consists of light that interacts with a mechanical element. By measuring and controlling the light, the state of the mechanics can be indirectly accessed and manipulated, which allows for unprecedented precision and insight into the quantum nature of macroscopic objects. Cavity optomechanics refers to systems where a cavity is added in order to enhance the light--matter interaction~\cite{aspelmeyer2014cavity}. When the mechanical element is constrained to move along a single axis and allowed to act as one of the mirrors in the cavity, its position determines the resonant frequency of the cavity mode. Since the photons carry momentum, the radiation pressure slightly displaces on the mechanical element further, which changes the cavity length and the optical frequency. That in turn changes the radiation pressure, which is the hallmark of a nonlinear interaction between the optical modes and the mechanical displacement. This nonlinear interaction can be used to create highly non-Gaussian states, a topic we elaborate on in Chapters~\ref{chap:non:Gaussianity:coupling} and~\ref{chap:non:Gaussianity:squeezing}, and the sensitivity of the mechanical element to extremely small displacements means that optomechanical systems show great potential as quantum sensors. We discuss this in detail in Chapter~\ref{chap:metrology}.

Optomechanical systems are, to date, the most massive systems that can be controlled in the lab while retaining their quantum properties. See Section~\ref{chap:introduction:section:examples} for further discussions of specific experimental platforms and the masses that can be achieved. 
The large mass provides an opportunity for testing a number of fundamental physics proposals and for building powerful quantum sensor's.  We here provide a brief overview of some major areas of interest where the study of optomechanical systems can be applied to great effect. 

The measurement problem is perhaps one of the most elusive outstanding problems in quantum theory~\cite{landau2013quantum}. In short, the measurement problem states that there are currently no confirmed dynamical mechanisms by which we can describe how quantum states collapse into a specific eigenstate. One potential solution to the problem involve so-called collapse theories~\cite{bassi2013models}, which modify the Schrödinger equation to include a dynamical collapse mechanism. Since many collapse theories predict an increased collapse rate with the number of constituent interacting quantum systems, with the aim to explain the emergence of the quantum--classical transition, they can  be extensively tested with optomechanical systems~\cite{bahrami2014proposal, goldwater2016testing, carlesso2018non}. Prominent collapse theories include the Continuous Spontaneous Localisation Model (CSL), which stipulates a continuous stochastic evolution in the Hilbert space~\cite{pearle1989combining,ghirardi1990markov} and the Diosi--Penrose model, which suggests that superposition of gravitational fields induces collapse~\cite{diosi1989models, penrose1996gravity}. A number of tests of collapse theories with optomechanical systems have already been performed~\cite{vinante2016upper,helou2017lisa},  however while the bounds on some of the collapse parameters have been narrowed, no conclusive evidence that support the validity of the models exists. At the time of writing, additional tests are planned for the near future, including one notable project, the MAQRO project, which aims to test spatial superpositions of macroscopic objects by sending a satellite with the possibility to release spheres into a trap to one of the Lagrange points between the Earth and the moon ~\cite{kaltenbaek2016macroscopic}. The unique environment allows for longer free-fall times in microgravity that are inaccessible in ground-based experiments, which enables tests of the quantum-classical transition and alternative theories in hitherto unreachable parameter regimes. 

Furthermore, the interplay between quantum theory and the low-energy limit of gravity could potentially be probed with optomechanical systems. Recent proposals suggest that the quantum nature of gravity can be determined by searching for gravitationally induced entanglement between massive systems~\cite{bose2017spin, marletto2017gravitationally}. That is, if gravity is a quantum force, it should be possible to detect entanglement due to the gravitational interaction. By considering spatial superpositions of massive quantum systems and the resulting Newtonian interaction, it can be shown that the final state is entangled. While the time-scales for detecting gravitational entanglement are reasonable when considering the largest levitated system possible, it is extremely difficult to generate the superpositions needed, as the two spatial eigenstates must be separated by large distances. Subsequent investigations by the thesis author, Sougato Bose and Alessio Serafini indicate that gravitational entanglement could similarly be generated using only Gaussian resources, where the two systems interact when adjacently trapped, which foregoes the need for creating large spatial superpositions~\cite{qvarfort2018mesoscopic}. Detecting gravitational entanglement does however become significantly more difficult in this setting, since all other effects in the trapped system exceed the gravitational interaction. A number of publications have since explored similar settings~\cite{nguyen2019entanglement, krisnanda2019observable}. 
Successfully detecting gravitational entanglement has implications for quantum gravity theories such as string theory~\cite{dienes1997string} and quantum loop gravity~\cite{rovelli2008loop}, since the low-energy limit of these theories reduces to an effective field theory, which should generate gravitational entanglement~\cite{carney2018massive}. If gravitational entanglement is not found in future experiments, all quantum gravity theories that reduce to an effective field theory at low energies will subsequently be proven invalid. 

Beyond fundamental physics, optomechanical systems show remarkable promise as quantum sensors and are being pursued with the purpose of developing a major quantum technology. This is also one of the key topics of this thesis. Substantial experimental challenges remain, including fully controlling the quantum properties of the system and operating in the nonlinear regime. Possible applications includes sensing of extremely weak forces~\cite{mason2019continuous} and accelerometry~\cite{Cervantes2014} have already been demonstrated experimentally~\cite{Cervantes2014}. 

\subsection{The optomechanical Hamiltonian}


An optomechanical system consist of light that interacts with a mechanical element. Sometimes, a cavity is added, which the light resonates in, which in turn enhances the light--matter interaction. The light resonates in the cavity and can be used to measure and manipulate the mechanical element. The photons carry momentum, which means that the radiation pressure displaces the mechanical element. The displacement, in turn, changes the resonant frequency of the cavity. 

In this Section, we discuss the fundamental building blocks of the theory of quantum optomechanical systems that operate in the nonlinear regime. Several extensive works have been written on this topic, including reviews~\cite{kippenberg2008cavity, aspelmeyer2014cavity},  reviews on levitated optomechanics~\cite{yin2013optomechanics, millen2019optomechanics}, and a book by Bowen and Milburn~\cite{bowen2015quantum}. The discussion presented in this thesis covers only a small percentage of the literature, and we therefore encourage the reader to seek out additional information in the included references when needed. 

In quantum mechanics, systems can often be approximated to oscillate around a stable equilibrium point, which allows us to model them a quantum harmonic oscillator. The unitary dynamics of a system is typically encoded in a Hamiltonian $\hat H$, which represents the energy of a system. The Hamiltonian of a single-mode quantum harmonic oscillator is defined as
\begin{equation} \label{eq:introduction:QHO:Hamiltonian}
\hat H_{\rm{QHO}} = \frac{1}{2} m \omega^2 \hat x^2 + \frac{\hat p^2}{2m}, 
\end{equation}
where $m$ is the effective mass of the system and $\omega$ is the oscillation frequency. Furthermore, $\hat x$ and $\hat p$ are the position and momentum operators acting on the system, which satisfy the canonical commutator relation $[\hat x, \hat p] = i \hbar$. 

Sometimes, it is easier to consider single excitations in the system through the equivalent framework of annihilation and creation operators $\hat a$ and $\hat a^\dag$. These obey the commutator relation $[\hat a, \hat a^\dag] \equiv \hat a \hat a^\dag - \hat a^\dag \hat a = 1$ and act on a single Fock state as $\hat a\ket{n} = \sqrt{n} \ket{n-1}$, where $n$ is an integer value that describes the number of excitations in the system. We identify the relationship between the position and momentum operators $\hat x$ and $\hat p$ and the annihilation and creation operators as 
\begin{align} 
&\hat x = \sqrt{\frac{\hbar}{2m \omega}} \left( \hat a ^\dag + \hat a \right)\, , &&\mbox{and} && \hat p =i \sqrt{\frac{\hbar m \omega}{2}}\left( \hat a ^\dag - \hat a \right).
\end{align}
The Hamiltonian of the quantum harmonic oscillation $\hat H_{\rm{QHO}}$ in Eq.~\eqref{eq:introduction:QHO:Hamiltonian} can therefore be written as
\begin{equation}
\hat H_{\rm{QHO}} = \hbar \omega \left( \hat a^\dag \hat a+ \frac{1}{2} \right) .
\end{equation}
In quantum optics, the boundary conditions imposed by the resonator allow for the quantisation of the electromagnetic field and a subsequent description of the optical modes in terms of annihilation and creation operators. In this thesis, we denote the optical field operators by $\hat a$ and $\hat a^\dag$.

The mechanical modes, on the other hand, arise through interactions within the mechanical element, which gives rise to bosonic quasi-particles that collectively behave just like free particles that interact with the optical field. We denote these by $\hat b$ and $\hat b^\dag$. 

Key to the description of optomechanical systems is the light--matter interaction, which induces a change in the cavity length. The precise nature of the light--matter interaction differs from system to system, but can be described through use of fairly general mathematical framework. This change can be treated formally by describing the interaction between the optics and the mechanics. 

A Hamiltonian description of an optomechanical system must include the free evolution of the optical mode, the mechanical phonon mode and the interaction between them. In Appendix~\ref{app:optomechanical:Hamiltonian}, we provide two derivations of the optomechanical Hamiltonian in Eq.~\eqref{chap:introduction:eq:standard:nonlinear:Hamiltonian} from first principle for a moving-end mirror in Section~\ref{app:optomechanical:mirror}, and for a levitated nanosphere in Section~\ref{app:optomechanical:hamiltonian:nanosphere}. The derivations are based on those in Refs~\cite{law1995interaction}, and~\cite{romero2011optically}, respectively. 

The standard optomechanical Hamiltonian reads 
\begin{equation} \label{chap:introduction:eq:standard:nonlinear:Hamiltonian}
\hat H = \hbar \, \omega_{\rm{c}} \, \hat a^\dag \hat a + \hbar \, \omega_{\rm{m}} \, \hat b^\dag \hat b - \hbar \, \mathcal{G}(t) \, \hat a^\dag \hat a \, \left( \hat b^\dag + \hat b \right) \, , 
\end{equation}
where $\hat a, \hat a^\dag$ are the annihilation and creation operators of the optical mode which oscillates with a frequency $\omega_{\rm{c}}$, and $\hat b, \hat b^\dag$ are the annihilation and creation operators for the phonons in the mechanical element, which oscillate with a frequency $\omega_{\rm{m}}$. The nonlinear light--matter interaction term takes the form $\hat a^\dag\hat a\bigl( \hat b^\dag + \hat b \bigr)$, which can be interpreted as the number of photons ($\hat a ^\dag \hat a$) coupling to the amplitude of the mechanical displacement $ \hat x_{\rm{m}} \sim \bigl( \hat b^\dag + \hat b \bigr)$, where $\hat x_{\rm{m}}$ corresponds to the position operator acting on the mechanics. The coupling is nonlinear, in that the equations of motion for $\hat a$ and $\hat b$ cannot be written in terms of linear contributions.  The interaction is weighted by the function $\mathcal{G}(t)$, which takes different forms for different systems. For many systems, $\mathcal{G}(t)$ is assumed to be constant $\mathcal{G}(t) \equiv g_0$, but there are systems, such as hybrid Paul traps~\cite{Millen2015iontrap}, where the coupling is generally time-dependent. In this thesis, we examine the effects of both a constant coupling and a modulated coupling. 

It is generally beneficial to work with a dimensionless quantities, which is what we do in for the remainder of this thesis. We therefore rescale $\hat H$ by the mechanical frequency $\omega_{\rm{m}}$ and define a dimensionless time as $\tau = \omega_{\rm{m}} \, t$. At $\tau = 2\pi$, the mechanical element has completed one oscillation. The optical frequency becomes $\Omega_{\rm{c}}= \omega_{\rm{c}}/\omega_{\rm{m}}$ and we denote the coupling by $\tilde{\mathcal{G}}(\tau) = \mathcal{G}(t)/\omega_{\rm{m}}$. The Hamiltonian becomes
\begin{equation} \label{chap:introduction:eq:standard:nonlinear:Hamiltonian:dimensionless}
 \hat{\tilde{H}} = \hat H/(\omega_{\rm{m}} \hbar) = \Omega_{\rm{c}} \, \hat a^\dag \hat a + \hat b^\dag \hat b - \tilde{\mathcal{G}}(\tau) \, \hat a^\dag\hat a \, \left( \hat b^\dag + \hat b \right) \, . 
\end{equation} 
We  generally denote dimensionless quantities by the addition of a tilde. The action of this Hamiltonian on some state $\ket{\Psi}$ can be determined by determining the action of the time-evolution operator $\hat U(\tau)$, which is defined by
\begin{equation}\label{chap:introduction:eq:general:time:evolution:operator}
\hat{U}(t)=\overset{\leftarrow}{\mathcal{T}}\,\exp\left[-\frac{i}{\hbar}\int_0^{t} dt'\,\hat{H}(t')\right],
\end{equation}
where $\overset{\leftarrow}{\mathcal{T}}$ denotes time-ordering. 

The main goal of this thesis is to solve the dynamics such that $\hat U(t)$ in Eq.~\eqref{chap:introduction:eq:general:time:evolution:operator} can be written in a simple, analytic form which can be straight-forwardly used to compute various quantities of interest. We do so in Chapter~\ref{chap:decoupling}, where we solve the dynamics for an extended version of the Hamiltonian in Eq.~\eqref{chap:introduction:eq:standard:nonlinear:Hamiltonian}.  We proceed in this Chapter to consider additional properties of optomechanical systems. 

\subsection{Initial states of the optics and mechanics} \label{chap:introduction:initial:states}

To examine the evolution of optomechanical systems, we start with an initial state $\ket{\Psi(t = 0)}$ that we evolve under $\hat U(t)$ in Eq.~\eqref{chap:introduction:eq:general:time:evolution:operator}. The initial state of the system must be as accurate as possible if the theoretical description is to match the experiment. 

Laser light is naturally coherent, and so the most natural initial state for the cavity mode is a coherent state $\ket{\mu_{\rm{c}}}$, which is defined as the eigenstate of the annihilation operator: $\hat a \ket{\mu_{\rm{c}}} = \mu_{\rm{c}} \ket{\mu_{\rm{c}}}$, where $\mu_{\rm{c}} \in \mathbb{C}$ is the eigenvalue of the state $\ket{\mu_{\rm{c}}}$. A coherent state admits an expansion in the Fock basis according to
\begin{equation} \label{chap:introduction:eq:coherent:state}
\ket{\mu_{\rm{c}}} = e^{- |\mu_{\rm{c}}|^2/2} \sum_{n = 0}^\infty \frac{\mu_{\rm{c}}^n}{\sqrt{n!}} \ket{n} \, .
\end{equation}
Light can also be generated as a single Fock state $\ket{n}$, or superposition of Fock states $\ket{\Psi} = \frac{1}{\sqrt{2}} \left( \ket{0} + \ket{n}\right)$, however it is extremely difficult to generate Fock states with high $n$ as cavity states, so we only briefly consider this configuration in Chapter~\ref{chap:gravimetry}. 

The mechanical mode, on the other hand, is most often found in a thermal state. This is an initially mixed state which follows from assuming a non-zero temperature. In the Fock basis, this state is given by 
\begin{equation} \label{chap:introduction:eq:thermal:state:one}
\hat \rho_{\rm{th}} = \sum_{n = 0}^\infty \frac{\tanh^{2n}r_T}{\cosh^2 r_T} \ketbra{n} \, , 
\end{equation}
where $\tanh r_T = \exp( - \hbar \omega_{\rm{m}}/(2 k_{\rm{B}} T) )$, for which $k_{\rm{B}}$ is Boltzmann's constant and $T$ is the temperature. There is an additional, equivalent description of thermal states in the coherent state basis. It is sometimes more beneficial to work in this basis, so we write
\begin{equation} \label{chap:introduction:eq:thermal:state:two}
\hat \rho_{\mathrm{th}} = \frac{1}{\bar{n} \pi} \int \mathrm{d}^2 \mu_{\rm{m}} \, e^{- |\mu_{\rm{m}}|^2/\bar{n}} \ket{\mu_{\rm{m}}}\bra{\mu_{\rm{m}}} \, .
\end{equation} 
where $\mu_{\rm{m}}$ is the coherent state parameter and $\bar{n}$ is the average number of phonons in the system~\cite{vanner2011pulsed}.

In summary, we will consider three different states in this thesis: 
\begin{enumerate}
\item  A coherent state of the optics and mechanics:
\begin{equation} \label{chap:introduction:eq:initial:state:coherent:coherent}
\ket{\Psi} = \ket{\mu_{\rm{c}}} \otimes \ket{\mu_{\rm{m}}} \, .
\end{equation}
\item A coherent state of the optics and a thermal state of the mechanics in the coherent state basis:
\begin{align}\label{chap:introduction:eq:initial:state:coherent:thermal:2}
\hat \rho(t = 0) &= \ketbra{\mu_{\rm{c}}} \otimes \sum_{n = 0}^\infty \frac{\tanh^{2n}r_T}{\cosh^2 r_T} \ketbra{n} \, , 
\end{align}
\item A Fock state of the optics and a coherent state of the mechanics:
\begin{equation}
\ket{\Psi} = \frac{1}{\sqrt{2}} \left( \ket{0} + \ket{n} \right) \otimes \ket{\mu_{\rm{m}}} \, . 
\end{equation}
\end{enumerate}
It should be noted that while the thermal states describe some initial noise, they do not account for thermal noise affecting the system throughout the evolution, which we discuss next. 

Injecting the initial coherent optical state in the cavity brings additional challenges, which we do not discuss here, mainly because the schemes we consider do not easily allow us to consider an open cavity. 

\subsection{Solutions in the literature}

The first solutions of the standard optomechanical Hamiltonian in Eq.~\eqref{chap:introduction:eq:standard:nonlinear:Hamiltonian} with a constant coupling $\mathcal{G}(t) \equiv g_0$ and an initial analysis of the resulting states were provided by Bose \textit{et al}.~\cite{bose1997preparation}, and Mancini \textit{et al}.~\cite{ mancini1997ponderomotive}. They showed that
the time-evolution operator $\hat U(t)$ can  be written as 
\begin{align} \label{chap:introduction:eq:time:evolution:operator:Bose}
\hat U(\tau) =& \exp[- i \,  \Omega_{\rm{c}} \hat a^\dag \hat a  ] \, \exp[ i \, \tilde{g}_0 ^2 \, (\hat a ^\dag \hat a )^2 (\tau - \sin(\tau)) ] \, \nonumber \\
&\times\exp [ \tilde{g}_0 \, \hat a^\dag \hat a \, (\eta \,  \hat b^\dag - \eta^* \,  \hat b )] \, \exp[ - i \, \hat b^\dag \hat b \, \tau] \, , 
\end{align}
where we recall that $\Omega_{\rm{c}} = \omega_{\rm{c}}/\omega_{\rm{m}}$, $\tau = \omega_{\rm{m}} \, \tau$, $\tilde{g}_0 = g_0/\omega_{\rm{m}}$ and where we defined $\eta = 1 - e^{- i \tau}$. We have changed the notation used in Ref~\cite{bose1997preparation} to coincide with that used in the rest of this thesis.

By applying the operator in Eq.~\eqref{chap:introduction:eq:time:evolution:operator:Bose} to an initially coherent state as that in Eq.~\eqref{chap:introduction:eq:initial:state:coherent:coherent}: $\ket{\Psi(0)} = \ket{\mu_{\rm{c}}} \otimes \ket{\mu_{\rm{m}}}$, it has been shown that the states evolves into\footnote{We note that the second phase term in Eq.~\eqref{chap:introduction:constant:plain:optomechanical:state} is missing in Ref~\cite{bose1997preparation}}
\begin{equation} \label{chap:introduction:constant:plain:optomechanical:state}
\ket{\Psi(\tau)} = e^{- |\mu_{\rm{c}}|^2 /2} \sum_{n = 0}^\infty \frac{\mu_{\rm{c}}^n}{\sqrt{n!}} e^{i \, \tilde{g}_0^2 \, n^2 \, ( \tau - \sin(\tau))} e^{ \tilde{g}_0 \, n \, \left( \eta  \, \mu_{\rm{m}} - \eta^* \,  \mu_{\rm{m}}^* \right)} \ket{n} \otimes \ket{\phi_n(\tau)} \, ,
\end{equation}
where $\phi_n(\tau) = \mu_{\rm{m}} \, e^{- i \, \tau} + \tilde{g}_0 \,  n  \, ( 1 - e^{- i \, \tau})$ is a coherent state of the mechanics. We remark on some interesting properties of the state:
\begin{itemize}
\item The system is periodic, meaning that at $\tau = 2\pi$, the mechanics returns to its initial state. 
\item At $\tau = 2\pi$, the optical state is completely disentangled from the mechanics. This feature becomes important in Chapter~\ref{chap:gravimetry}, where we consider sensing of constant gravitational acceleration. 
\item At $\tau = \pi$, the state is maximally entangled, and depending on the value of $g_0$, the optical state can be prepared in a number of highly non-classical cat states (see~\cite{bose1997preparation}). 
\end{itemize}
It is sometimes useful to consider the optical and mechanical subsystems separately. We have that the optical subsystem becomes
\begin{align}
\hat \rho_{\rm{c}} =& \,  e^{- |\mu_{\rm{c}}|^2} \sum_{n = 0, \, n' = 0}^\infty  \frac{\mu_{\rm{c}}^n \mu_{\rm{c}}^{* n'} }{\sqrt{n! n'!}} \,  e^{i \,  \left( \tilde{g}_0^2 \, (n^2 - n^{\prime 2} ) \right) ( \tau - \sin(\tau) )} \, e^{\tilde{g}_0 \, \left( n - n' \right)\left( \eta \, \beta - \eta^* \, \beta^*\right)/2} \, \nonumber \\
&\quad\quad\quad\quad\quad\quad\quad\quad\quad\quad\times e^{- |\phi_n|^2 /2 - |\phi_{n'}|^2/2 + \phi^*_{n'} \phi_n } \, \ket{n} \bra{n'} \, .
\end{align}
The solution in Eq.~\eqref{chap:introduction:eq:time:evolution:operator:Bose} is valid for a constant light--matter coupling $\mathcal{G}(t) \equiv g_0$. The goal of this thesis is to extend these results to a wider range of optomechanical systems.

\subsection{Examples of optomechanical systems}  \label{chap:introduction:section:examples}

To date, a number of different optomechanical systems have been experimentally refined to the point that they can be controlled in the laboratory. A non-exhaustive list of realised experimental platforms include moving-end mirrors, also known as Fabry-P\'{e}rot cavities~\cite{favero2009optomechanics, arcizet2006radiation}, levitated silica spheres~\cite{barker2010cavity, yin2013large, Millen2015iontrap} or diamond beads~\cite{neukirch2015multi}, which can also sometimes have implanted nitrogen vacancy (NV) centres that allow for greater experimental control~\cite{neukirch2015multi}. Additional setups include whispering gallery modes, where the light resonates inside a sphere with a high enough refractive index for internal reflection~\cite{schliesser2010cavity, barker2010doppler}, clamped membrane optomechanics~\cite{tsaturyan2017ultracoherent}, membrane-in-the-middle configurations~\cite{jayich2008dispersive}, optomechanical crystals~\cite{eichenfield2009picogram, safavi2014two}. and Brillouin optomechanics~\cite{van2016unifying}. Yet another approach replaces the rigid mechanical element with an ensemble of cold atoms, which collectively act as the mechanical element. They can either be cold atoms~\cite{purdy2010tunable} or a Bose-Einstein condensate~\cite{brennecke2008cavity}. 

The nonlinear light--matter interaction is different for each experimental platform. In this Section, we look closer at three specific systems: Fabry-P\'{e}rot cavities, levitated spheres, and atomic ensembles. We present the specific coupling constant for each case and discuss experimental parameters of the systems. What follows is by no means a comprehensive review of the full properties of the systems. 

We start with an optomechanical  moving-end mirror, for which we derived the Hamiltonian in Appendix~\ref{app:optomechanical:Hamiltonian}. The mirror is mounted on a spring; a setup referred to as a Fabry-P\'{e}rot cavity. The light--matter coupling is given by 
\begin{eqnarray} \label{chap:introduction:eq:coupling:Fabry:Perot}
g_{\mathrm{FP}} = \frac{\omega_{\mathrm{c}}}{L} \sqrt{\frac{\hbar }{2  m \omega_{\rm{m}}}} \, ,  
\end{eqnarray}
where $L$ is the length of the cavity and $m$ is the mass of the mirror. Mirrors used in this way can generally be made rather heavy, with masses as large as $10^{-7}$ \si{kg}~\cite{arcizet2006radiation}. The mechanical frequencies that can typically be achieved with these systems are around $10^3$ \si{Hz}, however most systems operate at higher frequencies, such as around $10^5$ \si{Hz}~\cite{groblacher2009observation}. 

A levitated nano- or micro-crystal (e.g. a diamond or silicon bead), on the other hand, has a coupling given by~\cite{chan2011laser, chang2010cavity}
\begin{eqnarray}\label{chap:introduction:eq:coupling:levitated}
g_{\mathrm{Lev}} =  \frac{P}{4V_{\mathrm{c}}\epsilon_0}  \sqrt{\frac{\hbar }{2m \omega_{\rm{m}}}} k_{\mathrm{c}} \omega_{\mathrm{c}} \, , 
\end{eqnarray} 
where $\epsilon_0$  is the permittivity of free space,  $V_{\mathrm{c}}$ is the cavity mode volume,  and $k_{\mathrm{c}}$ is the wavevector of the laser, given by $2\pi/\lambda$, where $\lambda$ is the laser wavelength. $P= 3V\epsilon_0(\epsilon - 1)/(\epsilon + 2)$ is the polarisability of the levitated object of volume $V$ and $\epsilon $ is the relative electric permittivity.  Levitated systems have recently demonstrated exceptionally long coherence times, of orders $10^5$ \si{s}~\cite{pontin2019ultra}, and they can achieve extremely low mechanical oscillation frequencies of $\omega_{\rm{m}} \sim 100 $ \si{Hz}~\cite{bullier2019characterisation}. The mass of the levitated system is considerably lower than that of the mirror at a typical $10^{-14}$ \si{kg}~\cite{pino2018chip}, which for diamond with a density of $\rho = 3,539$ \si{kg.m^{-3}} yields a radius of $R = 0.87 \times 10^{-6}$ \si{m}. 

The last coupling we consider in this thesis arises for cold atoms trapped in a cavity. Here, the collective motion of the ensemble acts as the massive oscillator. For this system, the  coupling constant is given by~\cite{brennecke2008cavity, munstermann1999dynamics}
\begin{equation} \label{chap:introduction:eq:coupling:Atomic}
g_{\mathrm{Atom}} = \frac{\sqrt{N} g_a^2 k_1}{\Delta_{\mathrm{c}a}} \sqrt{\frac{\hbar}{2 M \omega_{\rm{m}}}},
\end{equation}
where $N$ is the number of atoms in the ensemble, $g_a$ is the single-atom cavity QED coupling rate, $M = N m $ is the collective mass of all the trapped atoms with individual mass $m$, $k_l $ is the wavevector of the laser and  $\Delta_{\mathrm{ca}} = \omega_p - \omega_{\mathrm{c}}$ with pumping frequency $\omega_p$. It has been demonstrated that at least $10^5$ atoms can be trapped in this way~\cite{brennecke2008cavity}. 

We will return to these expressions when we consider the limits to gravity sensing in Chapter~\ref{chap:gravimetry}.

\subsection{Open system dynamics}\label{chap:introduction:open:system:dynamics}

In the laboratory, quantum systems experience decoherence as a result of interactions with the environment. The result of the interaction is degradation of the quantum properties of the system, which can severely impact the sensitivity of the system. 

There are a number of noise sources that affect optomechanical systems. These include optical scattering off or within the mechanical element (if transparent),  shot noise (which arises from random fluctuations due to the quantised nature of the photon), gas collisions from an imperfect vacuum, and stray electric and magnetic fields. For non-levitated setups, such as Fabry--P\'{e}rot moving-end and clamped membranes, mechanical vibrations and thermal noise due to insufficient cooling are often the dominant noise source. 

For our purposes, and given the tools available to us for simulating noise in the nonlinear regime, we here focus on systems in which the bulk temperature of the mechanical oscillator is low (such that we can ignore blackbody radiation), the gas pressure is low (such that we can ignore gas collisions), and in which the laser field intensity is low enough that shot noise is negligible compared with scattering. 
In this regime, there are two major remaining sources of noise in an optomechanical system: photons leaking from the cavity, or photons being absorbed by imperfections in the mirrors and damping of the oscillator motion, both of which manifests as phonon dissipation. 

These effects can be modelled with the Lindblad master equation, which can be derived from considering Markovian dynamics (which means that the environment 'forgets' information about the system). The general idea of the derivation involves identifying suitable environmental operators that encode the evolution of the system at small time-scales with the Markovian constraint, see e.g.~\cite{pearle2012simple}. 

The Lindblad master equation is given by~\cite{gardiner2004quantum}
\begin{equation} \label{chap:introduction:eq:Lindblad}
\dot{\hat{\rho}}(t) = - \frac{i}{\hbar} [\hat H, \hat \rho(t)] + \hat L \, \hat \rho(t) \, \hat L^\dag - \frac{1}{2}  \{\hat L^\dag \hat L,\hat  \rho(t) \} \, , 
\end{equation}
where $\dot{\hat{\rho}}(t)$ is the time-derivative of the quantum state, $\hat H$ is the Hamiltonian of interest and  $\hat L$ is the Lindblad operator, which is arbitrary, meaning that it does not have to be Hermitian nor unitary. Finally, $\{\hat L^\dag \hat L,\hat  \rho(t) \} \equiv \hat L^\dag \hat L\hat  \rho(t) + \hat  \rho(t)\hat L^\dag \hat L$ denotes the anti-commutator. 

In optomechanical systems, photon leakage from the  cavity can be modelled by the Lindblad operator $\hat L_{\rm{c}} = \sqrt{\kappa_{\rm{c}}} \, \hat a$, and phonon dissipation by the Lindblad operator $\hat L_{\rm{m}} = \sqrt{\kappa_{\rm{m}}} \, \hat b$. There are currently no known solutions to the Lindblad equation in Eq.~\eqref{chap:introduction:eq:Lindblad} for photon decay with $\hat L \equiv \hat L_{\rm{c}}$. However, approximate solutions have been found for $\kappa_{\rm{c}} \ll 1$~\cite{mancini1997ponderomotive}. The full open system state evolution must be solved numerically, which we do for certain investigations in this thesis. When required, we use the Python library \textit{QuTiP} where open system dynamics is simulated with a 12th order Runge--Kutta method~\cite{johansson2013qutip}. We discuss the challenges with simulating open optomechanical system in the nonlinear regime below in Section~\ref{chap:introduction:sec:numerical:challenges}. 

Damping of the mechanics through phonon damping is another matter, since it has been shown that the Lindblad equation can be analytically solved for phonon dissipation~\cite{bose1997preparation}. The phonon decoherence manifests as a gradual damping of the oscillator motion, which moves the state towards a mixture in the coherent state basis~\cite{zurek1981pointer, paz1993reduction, anglin1996decoherence, zurek1993coherent}. 

Given an initially coherent state $\hat{\rho}(t = 0) = \ketbra{\mu_{\rm{c}}} \otimes \ketbra{\mu_{\rm{m}}}$, and a constant optomechanical coupling $\tilde{g}_0$,  the Lindblad equation  in Eq.~\eqref{chap:introduction:eq:Lindblad} can be solved for $\hat L \equiv \hat L_{\rm{m}}$, such that  the final state becomes~\cite{bose1997preparation}
\begin{align} \label{chap:introduction:traced:out:optics}
\hat\rho(\tau)  =& \,  e^{- |\mu_{\rm{c}}|^2 } \sum_{n = 0, m= 0}^\infty \frac{\mu_{\rm{c}}^n \mu_{\rm{c}}^{*m}}{\sqrt{n! m!} } e^{i \,  \tilde{g}_0^2 \,  (n^2 - m^2 ) (\tau - \sin(\tau))} e^{- D(n,m,\tilde{\kappa}_{\rm{m}}, \tau)} \,\nonumber \\
&\quad\quad\quad\quad\quad\quad\quad\quad\quad\quad\times \ket{n}\bra{m} \otimes \ketbra{\phi_n(\tilde{\kappa}_{\rm{m}}, \tau)} \, ,
\end{align}
where $\tilde{\kappa}_{\rm{m}} = \kappa_{\rm{m}}/\omega_{\rm{m}}$, and the coherent states are now given by 
\begin{equation}
\phi_n (\bar{\kappa}_{\rm{m}}, \tau) = \frac{ i \, \tilde{g}_0 \, n}{i + \tilde{\kappa}_{\rm{m}}/2 } \left( 1 - e^{- ( i + \tilde{\kappa}_{\rm{m}}/2) \tau} \right) \, ,
\end{equation}
and where 
\begin{align}
D(n,m,\tilde{\kappa}_{\rm{m}}, \tau) &= \frac{\tilde{g}_0^2 \,  (n-m)^2 \,  \tilde{\kappa}_{\rm{m}}}{2 ( 1 + \tilde{\kappa}_{\rm{m}}/4)} \biggl[ \tau + \frac{1 - e^{- \tilde{\kappa}_{\rm{m}} \tau}}{\tilde{\kappa}_{\rm{m}}} \nonumber \\
&\quad\quad\quad\quad\quad\quad\quad\quad- \left( \frac{e^{(i - \tilde{\kappa}_{\rm{m}}/2)\tau} - 1}{ i - \tilde{\kappa}_{\rm{m}}/2} - \frac{e^{- ( i + \tilde{\kappa}_{\rm{m}}/2)\tau} - 1}{i + \tilde{\kappa}_{\rm{m}}} \right) \biggr] \, .  
\end{align} 
We consider open dynamics in Chapter~\ref{chap:gravimetry} when we compute the non-Gaussianity of an optomechanical state, and in Chapter~\ref{chap:non:Gaussianity:coupling}, where we consider the measurement of constant gravitational acceleration. In both cases, we will solve the open system dynamics numerically.

\subsection{Numerical challenges} \label{chap:introduction:sec:numerical:challenges}

When numerically simulating a quantum system, the state is generally represented as a finite-dimensional matrix which approximates the infinite-dimensional representation. If the state information is spread across the entire Hilbert space  $\mathcal{H}$,  any truncation will necessarily remove some information about the state. It manifests as a type of `decoherence', since  certain values are negatively affected, however the state remains pure as this is often a normalisation conditions imposed by the solvers. 

\begin{figure*}[t!]
\subfloat[
\label{chap:introduction:good:quadrature}]{%
  \includegraphics[width=.4\linewidth, trim = 0mm 0mm 0mm 0mm]{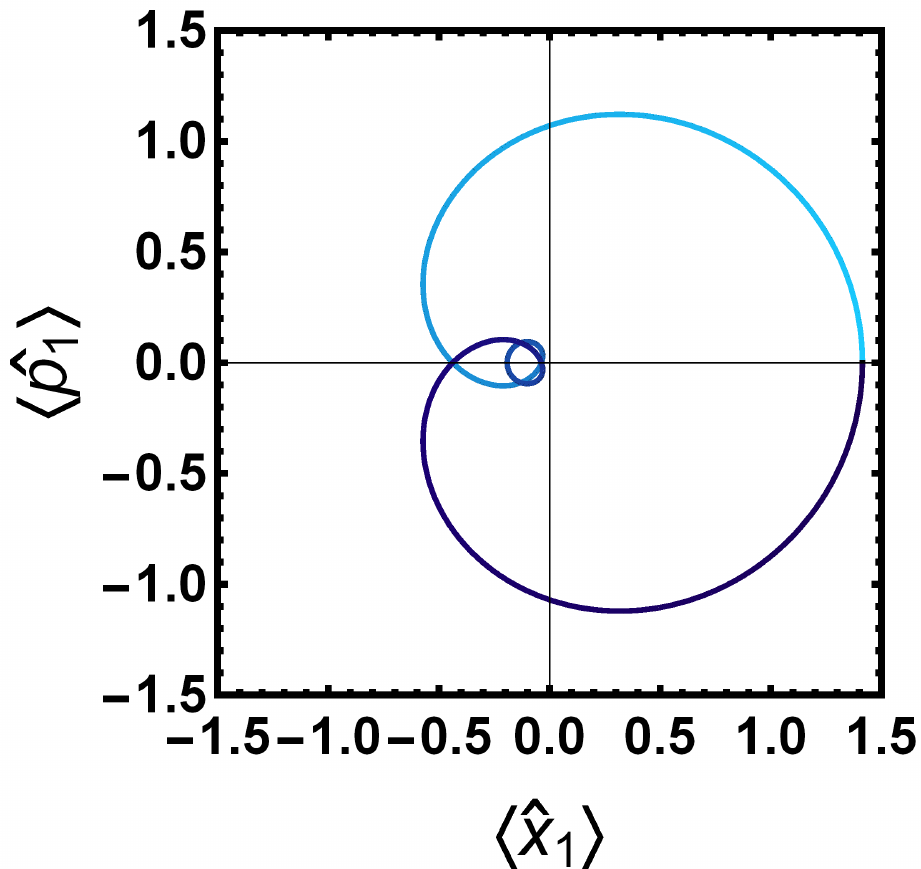}%
}\hfill
\subfloat[
\label{chap:introduction:bad:quadrature}]{%
  \includegraphics[width=.4\linewidth, trim = 0mm 0mm 0mm 0mm]{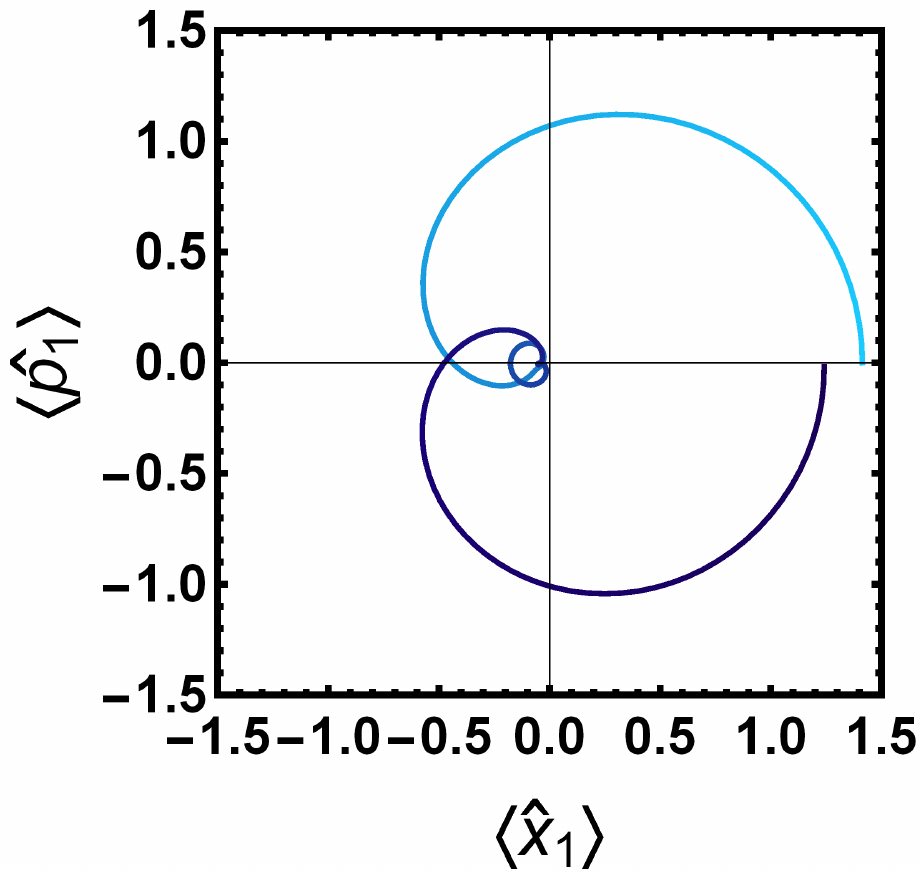}%
}\hfill
\caption[Numerically obtained optical quadrature of an optomechanical systems]{The optical quadrature of an optomechanical system evolving with the standard Hamiltonian in Eq.~\eqref{chap:introduction:eq:standard:nonlinear:Hamiltonian}. The begins in a light-blue colour and gradually becomes darker as time $\tau$ increases. The parameters are $\tilde{g}_0 = \mu_{\rm{c}} = \mu_{\rm{m}} = 1$. \textbf{(a)} A plot of the analytic optical quadrature in Eq.~\eqref{app:exp:values:eq:optical:quadratures}. \textbf{(b)} A plot of the same quadrature but this time simulated with a Hilbert space that is too small. The system `decoheres' from the insufficient computational resources and fails to return to its original state, as expected from the state in Eq.~\eqref{chap:introduction:constant:plain:optomechanical:state}.}
\label{chap:introduction:bad:quadratures}
\end{figure*}

Nonlinear systems are notoriously difficult to simulate because the nonlinearity causes information to quickly spread across all sectors in the Hilbert space. For example, we can compute the quadratures of the state in Eq.~\eqref{chap:introduction:traced:out:optics}, which are the expectation values of the $\hat x_{\rm{c}}$ and $\hat p_{\rm{c}}$ with respect to the evolving state. The analytic expressions for the optical and mechanical quadratures can be found in Eqs.~\eqref{app:exp:values:eq:optical:quadratures} and~\eqref{app:exp:values:eq:mechanical:quadratures} in Appendix~\ref{app:exp:values}. We recall that when $\tilde{g}_0^2$ is an integer, the system returns to its original state. We use this fact to compare the analytic quadrature of the system with that from a numerical simulation where the Hilbert space is too small. The results can be found in Figure~\ref{chap:introduction:bad:quadratures}, where the state performs one single trajectory in phase space and returns to its starting position. The line starts as light-blue at $\tau = 0$, then gradually becomes darker until it is completely black at $\tau = 2\pi$. Both plots use parameters  $\mu_{\rm{c}} = \mu_{\rm{m}} = \tilde{g}_0 = 1$. In Figure~\ref{chap:introduction:good:quadrature}, we plot the analytic quadrature, where the state can clearly be seen to return to its original state at $\tau = 2\pi$. 
However, in Figure~\ref{chap:introduction:bad:quadrature}, the Hilbert space dimension for one of the modes has been set to $N = 10$, which is not large enough for analytic evolution. As a result, the phase-space trajectory `decays' and the system fails to return to its original state. The apparent loss of unitarity cannot be detected by tracking the purity of the state, since the numerical methods will ensure that the state remains normalised.  Thus the `unitarity' of the quadratures serves to determine when the Hilbert space dimension is large enough to properly simulate the state.

As $\tilde{\mathcal{G}}$ increases, we require a larger Hilbert space dimension to simulate the state. If both the optical and mechanical subsystems have a Hilbert space $\mathcal{H}$ with dimension $\dim{\mathcal{H}} \in N$, the tensor product of the bipartite state scales with $N^2$. Furthermore, if we consider open system dynamics, where the full density matrix must be simulated, the state scales with $N^4$. Simulating large spaces quickly becomes numerically infeasible, as the memory required for these simulations quickly grows. A Hilbert space of dimension $N = 100$ will require $10^8$ complex entries. 
This unfavourable scaling sometimes be mitigated by using Monte Carlo methods, which compute statistic over a number of simulated state collapses to model the density matrix. While the computational burden is reduced, it is not removed altogether. 

In this thesis, we will simulate the open system dynamics of optomechanical systems with the Python library  \textit{QuTiP}~\cite{johansson2013qutip}, which uses a 12th-order Runge--Kutta method to solve the Lindblad equation in Eq.~\eqref{chap:introduction:open:system:dynamics} for both photon and phonon decoherence. The numerical method is limited to narrow parameter regimes and often cannot be used to simulate systems with experimentally accurate parameters. Therefore, our results should be seen as indicative only.

\subsection{Linearised optomechanics} \label{chap:introduction:linearisation}

To date, most experiments operate in the linear regime and can be accurately simulated without considering the fully nonlinear nature of the light--matter interaction. Together with the decrease in mathematical complexity that follows the linearisation procedure and the lack of numerical challenges to simulating noise discussed in the previous section, linearised dynamics is the preferred method for modelling optomechanical systems. 

We here present the linearisation procedure which takes a nonlinear optomechanical system to a linear one. The basis for the linearisation procedure is considering a strong steady-state laser drive which produces a strong coherent state in the cavity. The strong coherent state in the input laser beam is almost classical and very strong, which means that the smaller quantum corrections can be neglected. 

Starting from the nonlinear Hamiltonian in~\eqref{chap:introduction:eq:standard:nonlinear:Hamiltonian}, we perturb the optical field mode operators $\hat a$ around a strong coherent drive with a real amplitude $|\alpha|$ as $ \hat a= \hat a' - \alpha(t)$, where $\alpha(t)$ can be understood as multiplied by the identity operator. The same can be done for the phonon operator as $\hat b =\hat b' - \beta(t) $. The addition of a complex scalar to an operator does not alter the canonical commutator relation $[\hat x, \hat p] = i \hbar$. For the linearisation procedure to be valid, we require that $|\braket{\hat a }|\ll |\alpha(t)|$ and $|\braket{\hat b}| \ll |\beta(t)| $. 
We then insert this expression into the optomechanical Hamiltonian in Eq.~\eqref{chap:introduction:eq:standard:nonlinear:Hamiltonian} to find
\begin{equation} \label{chap:introduction:Hamiltonian:linearised}
\hat{H}_{\rm{Lin}} = \omega(l) \hat{a}^{\dag} \hat{a}+ \omega_{m} \hat{b}^{\dag}\hat{b}
- \mathcal{G}(\tau) \,  (\hat{b}+\hat{b}^{\dag})|\alpha(t)|^2 + \mathcal{G}(\tau) \,  (\hat{b}+\hat{b}^{\dag})(\alpha(t)\hat{a}^{\prime \dag} +\alpha^*(t) \hat{a}')  \, .
\end{equation}
where $\omega(l)$ is the cavity frequency as a function of the cavity length $l$ and where we have discarded terms with $\hat a^2$, since they are extremely small,  according to our initial assumptions. The addition of the strong coherent drive implies that the cavity is open, which means that light can both enter and exit the cavity. To include this in our considerations, we model additional decoherence effects from photons leaking from the cavity. We do so by considering the Langevin equations~\cite{lemons1997paul}, which describe the stochastic evolution of the annihilation and creation operators in the Heisenberg picture. We find that they are given by
\begin{align} \label{chap:introduction:Langevin}
\dot{\hat{a}}'(t) &\approx -\frac{\kappa_{\rm{c}}}{2} \hat{a}' -i  \, \omega(l) \,  \hat{a}' - i  \, \mathcal{G}(\tau) \,  \alpha(t)   \, (\hat{b}+\hat{b}^{\dag})  + \sqrt{\kappa_{\rm{c}}} \, \hat{a}_{in}(t) \nonumber \; ,\\
\dot{\hat{b}}'(t) &\approx -\left(\frac{\kappa_m}{2}+i\omega_m\right) \hat{b}' - i  \, \mathcal{G}(\tau)\,  |\alpha(t)|^2 
+ i  \, \mathcal{G}(\tau)\,  [\alpha(t) \hat{a}^{\dag}+\alpha^{*}(t)\hat{a}]  + \sqrt{\kappa_m} \, \hat{b}_{in}(t) \; .
\end{align}
where $\kappa_{\rm{c}}$ is the optical decoherence rate, $\kappa_{\rm{m}}$ is the mechanical decoherence rate, and $\hat a_{\rm{in}}(t)$ and $\hat b_{\rm{in}}(t)$ are external modes that can be considered further in the so-called input-output-formalism~\cite{gardiner1985input, walls2007quantum, serafini2017quantum}. From the equations in Eq.~\eqref{chap:introduction:Langevin}, a number of properties of the system can be studied, including the average length of the cavity, and sideband driving. 

The linearised Hamiltonian in Eq.~\eqref{chap:introduction:Hamiltonian:linearised} is quadratic in its arguments, which means that an input Gaussian state will remain Gaussian throughout the evolution. We discuss this in detail in Section~\ref{chap:introduction:sec:non:Gaussianity}, where we also review the covariance matrix formalism.

 %
 \section{Non-Gaussianity in quantum theory} \label{chap:introduction:sec:non:Gaussianity}

In this thesis, we are specifically interested in optomechanical systems operating in the nonlinear regime. One consequence of nonlinear dynamics is that input Gaussian states can be transformed into non-Gaussian states. To study the nonlinear character of the system, we can therefore study its ability to generate non-Gaussian states. Key to this investigation is the definition of a measure of non-Gaussianity. 

In this Section, we first introduce the notion of Gaussian and non-Gaussian states in quantum theory and discuss why non-Gaussian states in particular are important. We then show that the non-Gaussianity of a state can be accurately quantified with the help of a relative entropy measure. As part of this review, we also introduce the basis of the covariance matrix formalism and we show how it can be used to compute the non-Gaussianity of a state. 

\subsection{Definition of Gaussian and non-Gaussian states}

The notion of Gaussianity is key to this section, we begin by providing a definition of Gaussian and non-Gaussian states. 

We start with the formal definition of a Gaussian state. 
\begin{defn}[Gaussian states]
Any Gaussian state $\hat \rho_{\rm{G}}$ can be written as~\cite{serafini2017quantum}
\begin{equation} \label{chap:introduction:eq:definition:thermal:state}
\hat \rho_{\rm{G}}= \frac{e^{- \beta \hat H}}{\trace{e^{- \beta \hat H}}} \, ,
\end{equation}
where $\beta\in \mathbb{R}^+$ is the inverse temperature and $\hat H$ is a quadratic Hamiltonian. This is generally a mixed state, however the limiting case, 
\begin{equation}
\hat \rho_{\rm{G}} = \lim_{\beta \rightarrow \infty} \frac{e^{- \beta  \hat H}}{\trace{e^{- \beta \hat H}}} \, ,
\end{equation}
which corresponds to a state at zero temperature, is a pure state. 

A Gaussian state can also be defined in terms of its Wigner function~\cite{wigner1997quantum}. The Wigner function is a phase-space representation of a quantum state commonly used in quantum optics.  The Wigner function $W(x,p)$ of a single-mode Gaussian state reads:
\begin{equation}
W(x,p) = \frac{2}{\pi} \int_{\mathbb{R}} \mathrm{d} x' \, e^{2 \, i \, p \, x'} \bra{x + x'} \hat \rho \ket{x - x'} \, 
\end{equation}
where the integration occurs over the full real line, $x$, $x'$ and $p$ are real parameters and $\ket{x}$ is an improper quadrature eigenstate, which is defined as the eigenstate of the operator $\hat x$. The Wigner function captures certain notions  of non-classicality; when it is negative, the state is non-classical~\cite{kenfack2004negativity}. All Gaussian states can be associated with a positive Wigner functions, which means that they can always be described by a natural realistic hidden variable description. As a result, Gaussian states cannot be used to violate a Bell inequality if restricted to Gaussian measurements. However, through the use of non-Gaussian measurements, such as dichotomic variables that partition the Hilbert space in an appropriate manner, two-mode squeezed entangled states (which are Gaussian) can indeed violate a Bell inequality. Thus, Gaussian states can display strongly non-local behaviour~\cite{serafini2017quantum}. 
\end{defn}
Any quantum state can be expressed completely in terms of its moments. By moments, we refer to the statistical moments of the state $\psi(x)$, which are defined by 
\begin{equation} \label{chap:introduction:eq:statistical:moments}
E_n = \int^\infty_{- \infty}\mathrm{d}x  \, \psi^*(x) \,  x^n  \, \psi(x)  \, .
\end{equation}
The first moment is the expectation value $E_1 = \braket{x}$ of the state $\psi(x)$, the second moment is the variance $E_2 = \braket{x^2}$ and the third moment is the skewness $E_3 = \braket{x^3}$. We have used wavefunction notation in Eq.~\eqref{chap:introduction:eq:statistical:moments}, but this is equivalent to any other formulation of quantum theory, including the use of a bosonic field theory. 

For full tomography of a quantum state, and using quantum optics notation, we must compute the statistical moments of the position operator $\hat x$ and the momentum operator $\hat p$. For a general, non-Gaussian continuous variable state, this amounts to compute an infinite number of moments. 

A Gaussian state, however, is uniquely determined by its first and second moments. This means that any higher moments can be computed with the use of the first and second moments. Consequently, it suffices to track the first and second moments of a Gaussian state to completely capture the full information of the state, including features such as coherence and entanglement. Examples of Gaussian states include coherent states (defined in Eq.~\eqref{chap:introduction:eq:coherent:state}, the thermal state defined in Eq.~\eqref{chap:introduction:eq:definition:thermal:state}, and all single and two-mode squeezed coherent states, which are defined as 
\begin{equation}\label{chap:introduction:definition:squeezed:state}
\ket{\mu, z} = \hat D_{ \alpha} \hat S_{z} \ket{0} \,,  
\end{equation}
where $\alpha, z \in \mathbb{C}$ and where
\begin{align}
&\hat D_{\alpha} = e^{\alpha \hat a - \alpha^* \hat a} \, , &&\mbox{and } && \hat S_z = e^{\frac{1}{2} (z^* \hat a^2 - z \hat a^{\dag 2} )} \, ,
\end{align}
The physical intuition behind squeezing can be understood by considering the standard deviation of $\hat x$ and $\hat p$, which must always satisfy the uncertainty principle $\Delta x \Delta p \geq \hbar/2$. By increasing the standard deviation $\Delta x$, it is possible to decrease the momentum deviation $\Delta p$ to the degree that measurements of momentum are extremely precise, while ensuring that the uncertainty principle is not violated. All states presented here can be generalised to multi-mode states, where the squeezing occurs between two different modes. 

\begin{figure}[ht!]
\centering
  \includegraphics[width=.5\linewidth, trim = 0mm 0mm 0mm 0mm]{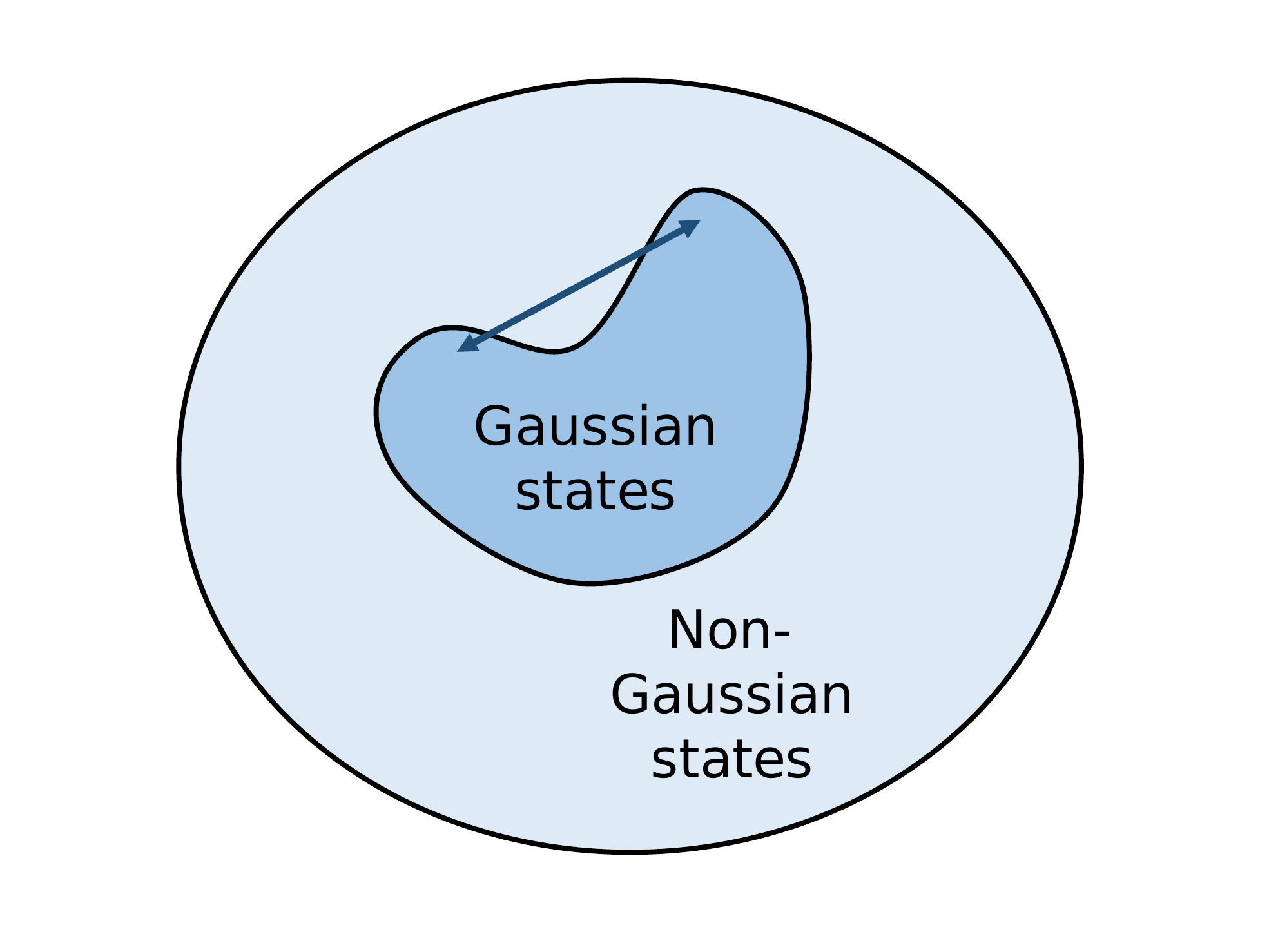}%
\caption[The convex set of Gaussian states]{The set of Gaussian states contained within the full set of non-Gaussian states.  The set is concave, since the addition of two Gaussian states can be a non-Gaussian state. }
\label{chap:introduction:set:Gaussian:states}
\end{figure}

The set of Gaussian states forms a  concave subset of all states, as per Figure~\ref{chap:introduction:set:Gaussian:states}. This fact can be readily understood by considering the superposition of the Wigner functions of two Gaussian states. We call them $W_1$ and $W_2$, and their linear combination reads $W_3 = p \, W_1 + ( 1 - p) \, W_2$, where $p$ is a real parameter that satisfies $0 \geq p \geq1$. Since the addition of two Gaussian probability distributions does not generally form another Gaussian distribution, the resulting state is non-Gaussian. Hence the  set of Gaussian states is concave. A concrete example is a superposition of two coherent states $\ket{\alpha_1} $ and $\ket{\alpha_2}$ into  $\ket{\psi} = \left( \ket{\alpha_1} + \ket{\alpha_2}\right)/\sqrt{2}$, also known as a cat-state. 

As we will see in Section~\ref{chap:introduction:sec:covariance:matrix:formalism}, the fact that Gaussian states can be uniquely described by their first and second moments alone is the basis of the covariance matrix formalism. It is a powerful tool in quantum information theory, which transfers the task of describing states in an infinite-dimensional Hilbert space into the treatment of finite-dimensional matrices.

\subsection{The power of non-Gaussian states}

Understanding nonlinear, interacting physical systems is paramount across many areas in physics. Specifically, ``nonlinear'' (or ``anharmonic'') dynamical systems include all those whose Hamiltonian cannot be expressed as a second-order polynomial in the quadrature operators. Crucially, these systems allow us to generate non-Gaussian states, which is not possible given only quadratic couplings. 

Recently, the intrinsic value of nonlinear systems, as opposed to the aforementioned limitations that linear systems face, has been made clearer and more rigorous. It has been shown that nonlinearities in the form of non-Gaussian states constitute an important resource for quantum teleportation protocols~\cite{dell2010teleportation}, universal quantum computation~\cite{lloyd1999quantum, menicucci2006universal}, quantum error correction~\cite{niset2009no}, and entanglement distillation~\cite{eisert2002conditions, fiuravsek2002gaussian, giedke2002characterization}. This view of non-Gaussianity as a resource for information-processing tasks has inspired recent work on developing a resource theory based on non-Gaussianity~\cite{zhuang2018resource, takagi2018convex, albarelli2018resource}. In addition, it has been found that non-Gaussianity provides a certain degree of robustness in the presence of noise~\cite{sabapathy2011robustness, nha2010linear}. 

In the context of quantum information and computation, there has been a drive towards the realisation of anharmonic Hamiltonians as well as more general methods and control schemes capable of generating and stabilising non-Gaussian states~\cite{silberhorn2001generation, lvovsky2001quantum,  heersink2003polarization, ourjoumtsev2006quantum, ourjoumtsev2007generation, parigi2007probing}. 
On the one hand, this is motivated by the fact that, in order to obtain effective qubits from the truncation of infinite dimensional systems, one needs unevenly spaced energy levels, such that only the transition between the two selected energy levels may be targeted and driven. In turn, this requires a sufficiently anharmonic Hamiltonian. On the other hand, it has always been clear that protocols entirely restricted to Gaussian preparations, manipulations and read-outs, through quadratic Hamiltonians and general-dyne detection, are classically simulatable, as their Wigner functions 
may be mimicked by classical probability distributions~\cite{mari2012positive}.\footnote{We should note here that, since uncertainties in quantum Gaussian systems are fundamentally bounded by the Heisenberg uncertainty relation, which in principle does not hold for classical systems, Gaussian operations are in fact sufficient to run some protocols requiring genuine quantum features, such as continuous variable quantum key distribution.} 

There are indications that non-Gaussian states can be used to enhance sensing schemes. A number of results indicate that non-Gaussian states constitute an important resource for sensing. Schr\"{o}dinger cat states~\cite{mancini1997ponderomotive, bose1997preparation}, compass states~\cite{zurek2001sub, toscano2006sub} and hypercube states~\cite{howard2018hypercube} -- which are all non-Gaussian states -- have all been found to have applications in sensing. 

In optomechanical systems~\cite{aspelmeyer2014cavity}, where electromagnetic radiation is coherently coupled to the motion of a mechanical oscillator, the light--matter interaction induced by radiation pressure is inherently nonlinear~\cite{mancini1997ponderomotive, bose1997preparation, ludwig2008optomechanical}. The nonlinear coupling enables the creation of optical cat states in the form of superpositions of coherent states~\cite{mancini1997ponderomotive, bose1997preparation}. These cat states can also be transferred to the mechanics~\cite{palomaki2013coherent}, which opens up the possibility of using massive superpositions for testing fundamental phenomena such as collapse theories~\cite{goldwater2016testing} and, potentially, signatures of gravitational effects on quantum systems at low energies~\cite{bose2017spin,marletto2017gravitationally}. 
This combination of sensing with nonlinear states and fundamental applications makes it imperative to explore the nonlinear properties of the optomechanical systems. See Section~\ref{chap:introduction:section:examples} for a discussion of different optomechanical platforms. 

While several experiments have demonstrated genuine nonlinear behaviour (see for example~\cite{sankey2010strong, doolin2014nonlinear,brawley2016nonlinear,  leijssen2017nonlinear}), most experimental settings can however be fully modelled with linear dynamics~\cite{kippenberg2007cavity, leijssen2017nonlinear}, which we briefly introduced in Section~\ref{chap:introduction:linearisation}. Since the linearised Hamiltonian $\hat H_{\rm{Lin}}$ in Eq.~\eqref{chap:introduction:Hamiltonian:linearised} is quadratic, Gaussian input states remain Gaussian, and therefore linear optomechanical systems cannot generate non-Gaussian states. The study of non-Gaussian states in nonlinear optomechanical systems is a major focus of this thesis, which we thoroughly explore in Chapters~\ref{chap:non:Gaussianity:coupling} and~\ref{chap:non:Gaussianity:squeezing}. 

\subsection{Introduction to the covariance matrix formalism} \label{chap:introduction:sec:covariance:matrix:formalism}

The covariance matrix formalism is a powerful alternative formulation of quantum theory, which  is entirely sufficient for the treatment of Gaussian states. The formalism subverts the problem of studying continuous variable systems in an infinite Hilbert space by allowing them to be described by finite-dimensional matrices. This significantly simplifies the analytical and numerical methods that we employ to simulate the system, especially with regards to simulating open system dynamics. 

We here provide a brief introduction to the covariance matrix formalism, where we closely follow Ref~\cite{serafini2017quantum}, in terms of both exposition and notation. We set $\hbar = 1$ in this Section for clarity, and all position and momentum operators $\hat x$ and $\hat p$ are dimensionless. However, we note that later in this thesis, we will use a different basis for the first and second moments rather than the position--momentum basis used in Ref~\cite{serafini2017quantum}. 

The canonical commutator-relation $[\hat x, \hat p] = i \hbar$ can be straight-forwardly extended to multiple modes, where the operators commute for different modes, such that $[\hat x_j, \hat p_k] = i \, \delta_{jk}$. The same holds for $[\hat a_j, \hat a_k^\dag] = \delta_{jk}$. 

We now introduce an $2n$-dimensional vector of canonical operators $\hat{\vec{r}} = (\hat x_1, \hat p_1, \hat x_2, \hat p_2, \ldots , \hat x_n, \hat p_n)^{\rm{T}}$, where $n$ is the number of bosonic modes of the system. Furthermore, we introduce the real, canonical and anti-symmetric symplectic form $\boldsymbol{\Omega}$, given by the direct sum of identical $2\times 2$ block matrices:
\begin{align} \label{chap:introduction:eq:symplectic:form:definition}
&\boldsymbol{\Omega} = \bigoplus^n_{j = 1} \boldsymbol{\Omega}_1\, ,&& \mbox{with } && \boldsymbol{\Omega}_1 = \begin{pmatrix} 
0 & 1 \\
-1 &0 
\end{pmatrix} \, .
\end{align}
In full, the symplectic form is a $2n\times 2n$ matrix that encodes the commutator relations of a system with $n$ bosonic modes. The anti-symmetry of the symplectic form implies that $\boldsymbol{\Omega} = - \boldsymbol{\Omega}^{\rm{T}}$, from which it follows that $\boldsymbol{\Omega}^2 = - \mathbb{I}_{2n}$, where $\mathbb{I}_{2n}$ is the $2n \times 2n$ identity matrix.

The canonical commutator relations of the vector $\hat{\vec{r}}$ of operators can then be expressed in the following elegant manner:
\begin{equation}
[\hat{\vec{r}}, \hat{\vec{r}}^{\rm{T}} ] = i \, \boldsymbol{\Omega} \, .
\end{equation}
It should here be noted that we can change the basis of the vector $\hat{\vec{r}}$, such that it is comprised of annihilation and creation operators increase, e.g. $\hat{\vec{q}} = ( \hat a_1, \hat a_1^\dag, \hat a_2, \hat a_2^\dag \ldots \hat a_n, \hat a_n^\dag )^{\rm{T}}$. This form is related to $\hat{\vec{r}}$ through a unitary transformation, which similarly changes the symplectic form in Eq.~\eqref{chap:introduction:eq:symplectic:form:definition}. 

We then consider the fact that any Hamiltonian $\hat H$ that is at most quadratic in its arguments can be written in terms of the vector $\hat{\vec{r}}$ of operators as 
\begin{equation} \label{chap:introduction:Hamiltonian:matrix}
\hat H_Q = \frac{1}{2} \vec{\hat{r}}^{\rm{T}} \, \boldsymbol{H} \, \vec{\hat{r}} + \hat{\vec{r}}^{\rm{T}} \vec{r} \, ,
\end{equation}
where $\boldsymbol{H}$ is the Hamiltonian matrix and $\vec{r}$ is a $2n$-dimensional real vector of first moment coordinates. $H$ is symmetric, since any asymmetric part contribute at most a constant addition to the Hamiltonian, and we also require that $\boldsymbol{H}$ is positive $\boldsymbol{H} >0$ to ensure thermal stability. By positive, we mean that all eigenvalues $\{\lambda_j \}$ of $\boldsymbol{H}$ satisfy $\lambda_j > 0 \, , \, \forall j$. 

We stated above that Gaussian states can be fully characterised by their first and second moments. We define the vector of first moments $\vec{\bar{r}}$ as
\begin{equation}
\vec{\bar{r}} = \trace{ \hat \rho_{\rm{G}} \, \hat{\vec{r}} } \, ,
\end{equation}
where $\hat \rho_{\rm{G}}$ is a Gaussian state. 
In the Heisenberg picture, operators evolve in time  with the Heisenberg equation:
\begin{equation}
\dot{\hat{r}}_j = i [\hat H, \hat r_j] \, .
\end{equation}
When we consider the full vector of operators $\vec{\hat{r}}$, the time evolution can be compactly written as
\begin{equation}
\dot{\hat{\vec{r}}} = \boldsymbol{\Omega}  \, \boldsymbol{H} \,  \hat{\vec{r}}\, .
\end{equation}
The solutions to this equation, for when the Hamiltonian matrix $H$ is time-independent, are
\begin{equation}
e^{\frac{i}{2} \hat{\vec{r}} ^{\rm{T}} \, \boldsymbol{H} \, \hat{\vec{r}}} \, \hat{\vec{r}} \, e^{- \frac{i}{2} \hat{\vec{r}}^{\rm{T}} \, \boldsymbol{H}\, \hat{\vec{r}}} = e^{\boldsymbol{\Omega} \, \boldsymbol{H}} \hat{\vec{r}} \, .
\end{equation}
For ease of notation, we define
\begin{equation}\label{chap:introduction:eq:definition:of:symplectic:operator}
\hat S_H = e^{\frac{i}{2} \hat{\vec{r}}^{\rm{T}} \, \boldsymbol{H} \, \hat{\vec{r}}} \, ,
\end{equation}
which means that the action of a quadratic Hamiltonian on the first moments can be written as
\begin{equation}
\hat S_H \, \hat{\vec{r}} \,\hat  S_H^\dag = \boldsymbol{S}_H \, \hat{\vec{r}} \, ,
\end{equation} 
where $\boldsymbol{S}_H = e^{\boldsymbol{\Omega} \, \boldsymbol{H}}$ is a $2n\times 2n$ matrix which is part of the linear symplectic group, often referred to as $Sp_{2n , \mathbb{R}}$. 
 
For the second moments, we introduce the $4\times4$ two-mode covariance matrix ${\sigma}$, defined as 
\begin{equation}
\boldsymbol{\sigma}(t) = \mathrm{Tr} \left[  \{\hat{\vec{r}}, \hat{\vec{r}}^{\mathrm{T}} \} \, \hat \rho_{\rm{G}}(t) \right] \, , 
\end{equation}  
for the Gaussian state $\hat{\rho}_{\rm{G}}(t)$ and the vector 
of operators $\hat{\vec{r}} = (\hat{x}_1, \, \hat{p}_1,\,  \hat{x}_2, \, \hat{p}_2)^{\mathrm{T}}$, and where the bracket $\{\bullet, \bullet \}$ denotes the symmetrised outer product, in the sense that $\{\hat x, \hat p  \} = \hat x \hat p + (\hat x \hat p)^{\rm{T}}$. In this basis, the covariance matrix is a real symmetric matrix

While the covariance matrix is generally considered in the $(\hat x, \hat p)$-basis, in this thesis, we use the following basis $\hat{\vec{r}}' = (\hat a, \hat b, \hat a^\dag, \hat b^\dag )^{\rm{T}}	$, which means that the covariance matrix elements are given by:
\begin{align} \label{chap:introduction:covariance:matrix:elements:ab:basis}
\sigma_{11} &= \sigma_{33} = 1 + 2 \, \braket{\hat a^\dag \hat a} - 2 \braket{\hat a^\dag } \braket{\hat a }  \, ,\nonumber \\
\sigma_{31} &=2 \braket{\hat{a}^2} - 2 \braket{\hat{a}}^2\, , \nonumber \\
\sigma_{22} &= \sigma_{44} = 1 + 2 \braket{\hat{b}^\dag \hat{b}} - 2\braket{\hat{b}^\dag} \braket{\hat{b}} \, , \nonumber \\
\sigma_{42}  &= 2 \braket{\hat{b}^2} - 2 \braket{\hat{b}}^2 \, , \nonumber \\
\sigma_{21} &= \sigma_{34} = 2 \braket{\hat{a} \hat{b}^\dag} - 2 \braket{\hat{a}} \braket{\hat{b}^\dag} \, ,  \nonumber \\
\sigma_{41} &= \sigma_{32} = 2 \braket{\hat{a} \hat{b}} - 2 \braket{\hat{a} }\braket{\hat{b}} \, .
\end{align}
All other elements follow from the fact that $\sigma = \sigma^\dag$, since the covariance matrix must be Hermitian in this basis. Furthermore, the symplectic form in this basis is defined as
\begin{align} \label{chap:introduction:symplectic:form:ab:basis}
&\boldsymbol{\Omega} = \bigoplus_{ j = 1}^n \boldsymbol{\Omega}_1' &&\mbox{with} && \boldsymbol{\Omega}_1' = i \begin{pmatrix} 1 &0 \\0 & - 1\end{pmatrix} \, .
\end{align}
The covariance matrix evolves in time as 
\begin{equation}
\boldsymbol{\sigma}(t)=\boldsymbol{S}(t)\,\boldsymbol{\sigma}_0\,\boldsymbol{S}^{\dag}(t) \, ,
\end{equation}
where $\sigma_0$ corresponds to the covariance matrix of the initial state at time $t = 0$. 

We can compute the covariance matrix of a few different example initial states. In the $\{\hat x, \hat p\}$--basis, a two-mode coherent state has the simple form $\boldsymbol{\sigma}_{\rm{coh}} = \mathrm{diag}(1,1,1,1)$. Similarly, a single mode squeezed state can be represented with $\boldsymbol{\sigma} = \mathrm{diag}(z, 1/z)$, where one quadrature is increased by $z\geq 1$, and the other is decreased. 

A powerful property of the covariance matrix is the fact that it can be decomposed in terms of its fundamental modes. Williamson's theorem ~\cite{williamson1936algebraic} guarantees that any $2n\times2n$ Hermitian matrix, such as the covariance matrix $\boldsymbol{\sigma}$, can be decomposed as $\boldsymbol{\sigma}=\boldsymbol{S}^{\dag}\,{\nu}_{\oplus}\,\boldsymbol{S}$, where $\boldsymbol{S}$ is an appropriate symplectic matrix. The diagonal matrix ${\nu}_{\oplus}=\textrm{diag}(\nu_1,\dots,\nu_n,\nu_1,\dots,\nu_n)$ is known as the Williamson form of the state and $\nu_n:=\coth(\frac{\hbar\,\omega_n}{2\,k_B\,T})\geq1$ 
(where we have introduced normal frequencies $\omega_n$ and a nominal temperature $T$)
are the symplectic eigenvalues of the state~\cite{williamson1936algebraic}. 

Williamson's form ${\nu}_{\oplus}$ contains information about the local and global mixedness of the state of the system
\cite{adesso2014continuous}, and can also be used to computed the entanglement between different modes. The state is pure if $\nu_n = 1$ for all $n $ and is mixed whenever $\nu_n > 1$. As an example, the thermal state
$\boldsymbol{\sigma}_{th}$ of a $n$-mode bosonic system is simply given by its Williamson form, i.e., $\boldsymbol{\sigma}_{th}={\nu}_{\oplus}$. 

In Section~\ref{chap:introduction:linearisation}, we showed how a nonlinear optomechanical system can be linearised to obtain a Hamiltonian with at most quadratic operators. This Hamiltonian maps input Gaussian states into output Gaussian states, which means that the covariance matrix formalism is sufficient for modelling systems in this regime. It is beyond the scope of this thesis to provide a full review of the literature of linear optomechanics, as this is a rich and thriving research field. The interested reader is directed to the reviews in Refs~\cite{kippenberg2008cavity, marquardt2009optomechanics,  meystre2013short, yin2013optomechanics, aspelmeyer2014cavity, millen2019optomechanics} for further reading. 

\subsection{Measures of deviation from Gaussianity} \label{chap:introduction:sec:measure:non:gaussianity}

To study the influence of nonlinear dynamics on the state, we ask the following question: \textit{can we quantify how much the non-Gaussian state $\hat \rho(\tau)$ deviates from a Gaussian state at time $\tau$}? 

This question stems from the following observation. The dynamics of our system is nonlinear. Therefore, we expect an initial Gaussian state, characterised by a Gaussian Wigner function, to become a non-Gaussian state at later times. In fact, the only way for a Gaussian state remain Gaussian is to evolve through a \textit{linear} transformation, which is induced by a Hamiltonian with at most quadratic terms in the quadrature operators~\cite{serafini2017quantum}.

To answer our question we first need to find a suitable measure of deviation from Gaussianity. In this work we choose to employ a measure for pure states, which we denote $\delta(\tau)$, that is based on the comparison between the entropy of the final state and that of the closest possible Gaussian reference state~\cite{genoni2008quantifying}. A similar measure has been used to compute features of mixed systems~\cite{park2019faithful}. 

Let us detail here the construction of the non-Gaussianity quantifier $\delta(\tau)$ for our nonlinear dynamics. First, our initial state $\hat{\rho}(0)$ evolves into the state $\hat{\rho}(\tau)$ at time $\tau$ through the Schr\"{o}dinger equation. Then, we construct a state $\hat{\rho}_\textrm{G}(\tau)$,  which is the Gaussian state defined by the first and second moments that coincide with those of $\hat{\rho}(\tau)$. In fact, it has been shown that $\hat \rho_{\rm{G}}(\tau)$ is indeed the state is the closest possible Gaussian reference state to $\hat \rho(\tau) $~\cite{marian2013relative}. 

Now, we recall that a Gaussian state is \textit{fully} defined by its first and second moments. Therefore, if two Gaussian states $\hat{\rho}_1$ and $\hat{\rho}_2$ have equal first and second moments  they are the same state~\cite{Adesso:Ragy:2014,serafini2017quantum}. However, the non-Gaussian state we wish to quantify will not be equal to its Gaussian reference state. This implies that we can introduce a measure $\delta(\tau)$ that quantifies how $\hat{\rho}(\tau)$ deviates from $\hat{\rho}_\textrm{G}(\tau)$:
\begin{equation}\label{chap:introduction:measure:of:non:gaussianity}
\delta(\tau):=S(\hat{\rho}_\textrm{G}(\tau))-S(\hat{\rho}(\tau)),
\end{equation}
where $S(\hat{\rho})$ is the von Neumann entropy of a state $\hat{\rho}$, defined by 
\begin{equation} \label{chap:introduction:relative:entropy}
S(\hat{\rho}):=-\Tr(\hat{\rho}\,\ln\hat{\rho})\ 
\end{equation} 
This measure has been shown to capture the intrinsic non-Gaussianity of the system, and it vanishes if and only if $\hat{\rho}(\tau)$ is a Gaussian state~\cite{genoni2008quantifying, marian2013relative}. In other words, if at all times  the measure returns $\delta(\tau) = 0$, we know that our state is fully Gaussian, which also means that the dynamics are fully linear.

Since $\hat{\rho}_\textrm{G}(\tau)$ is a Gaussian state, it is uniquely determined by its first and second moments moments, as discussed in Section~\ref{chap:introduction:sec:covariance:matrix:formalism}. Furthermore, it turns out that the relative entropy in Eq.~\eqref{chap:introduction:relative:entropy} can be computed straight from the covariance matrix $\sigma$ of a Gaussian state~\cite{Adesso:Ragy:2014,serafini2017quantum}. 
This is convenient, as the construction of $\hat{\rho}_{\mathrm{G}}$ involves finding the first of second moments of $\hat{\rho}$ anyway. 

To compute the entropy $S(\hat{\rho}_{\textrm{G}}(\tau))$, we must find the symplectic eigenvalues $\{\nu_j\}$ of $\sigma(\tau)$, where it always holds that $\nu_j(\tau)\geq1$ for all physical states. The symplectic eigenvalues can be computed by finding the eigenvalues of the object $i\,\boldsymbol{\Omega}\,\boldsymbol{\sigma}(\tau)$, where $\boldsymbol{\Omega}$ is the symplectic form. We defined $\boldsymbol{\Omega}$ for the $\{\hat x, \hat p\}$--basis in Eq.~\eqref{chap:introduction:eq:symplectic:form:definition}. However, in this basis, $\boldsymbol{\Omega}$ is given by $\boldsymbol{\Omega}=\mathrm{diag}(-i,-i,i,i)$ in this basis. The von Neumann entropy $S(\boldsymbol{\sigma})$ is then given in this formalism by 
\begin{equation} \label{chap:introduction:von:Neumann:definition}
S(\boldsymbol{\sigma}) = \sum_{j = 1}^n  s_V(\nu_j)\, ,
\end{equation}
 where $s_V$ is the binary entropy defined by 
 \begin{equation}
 s_V(x) := \frac{x + 1}{2} \ln \left( \frac{x+1}{2} \right) - \frac{x - 1}{2} \ln \left( \frac{x - 1}{2} \right) \, .
 \end{equation}
In summary, the state $\hat{\rho}(\tau)$ is non-Gaussian at time $\tau$ if and only if $\delta(\tau)>0$. We will use this measure in Chapter~\ref{chap:non:Gaussianity:coupling} and~\ref{chap:non:Gaussianity:coupling} to determine the non-Gaussian character of an initially Gaussian state that evolves with the optomechanical Hamiltonian.

\subsection{General behaviour of the measure of non-Gaussianity} \label{chap:introduction:non:Gaussianity:general:behaviour}

Let us now infer some characteristics of the measure of non-Gaussianity $\delta(\tau)$, which we can use to infer its general behaviour. The following analysis only holds when the system is pure. As mentioned above, the symplectic eigenvalues $\nu_j$ satisfy $\nu_j\geq1$~\cite{serafini2017quantum}, where $ \nu_j = 1$ for pure states. 
When we compute the covariance matrix of the non-Gaussian state $\hat \rho$, we essentially neglect some information that $\hat \rho$ contains in its higher moments. The reference Gaussian state $\hat \rho_{\rm{G}}$ is therefore, in effect, mixed, since we have discarded information to create it. 

A mixed state will have $\nu_j >1$. If we start with a pure state with $\nu_j  = 1$, nonlinear evolution will cause the Gaussian reference state $\hat \rho_{\rm{G}}$ to have the symplectic eigenvalues $\nu_j \geq 0$. Therefore, we can conveniently write $\nu_j=1+\delta\nu_j$, where $\delta \nu_j\geq0$ captures any deviation from purity early on in the evolution. 
In this case, we define $\nu_j=\nu_{0,j}+\delta\nu_j$ with $\nu_{0,j}>1$. Then, we would have that $\nu_j(0)=\nu_{0,j}$ and, again, linear evolution would imply that $\nu_j(\tau)=\nu_{0,j}$. The preceding statements imply that $\delta\nu_\pm$ are functions of the nonlinear contributions alone. Thus, when the nonlinearity tends to vanish, then $\delta\nu_\pm\rightarrow0$. Among the possible asymptotic regimes we have that $\delta\nu_\pm\rightarrow+\infty$ or that it becomes constant.

These observations are important. To understand their implications we use the expression $\nu_j=1+\delta\nu_j$ to compute the general deviation from Gaussianity as $\delta(\tau)= \sum_j s_V(1+\delta\nu_j)$. Using this form, we see that in the nearly linear (Gaussian) regime with only small contributions from the nonlinear dynamics, we will have $\delta\nu_\pm\ll1$ and therefore
\begin{align}\label{chap:introduction:general:measure:of:non:gaussianity:small}
\delta(\tau)\approx&-\sum_j \frac{\delta\nu_j}{2}\ln\frac{\delta\nu_j}{2} \, .
\end{align}
On the contrary, in the highly nonlinear (non-Gaussian) regime we have $\delta\nu_\pm\gg1$ and therefore 
\begin{equation}
\delta(\tau)\approx \sum_j \ln\frac{\delta\nu_j}{2} \, .
\end{equation}
If the symplectic eigenvalues depend on a large parameter $x\gg1$, then one will in general find that they have the asymptotic form 
\begin{equation}
\nu_j\sim x^{N_j}\sum^{N_j}_{n=0}\nu_j^{(n)}\,x^{-n} \, ,
\end{equation} 
 for some appropriate real coefficients $\nu_j^{(n)}$, where $N_{j}$ constitutes the upper limits of the sum~\cite{ahmadi2014quantum}. 
A careful asymptotic expansion of the measure of nonlinearity in this regime gives
 \begin{equation}\label{large:parameter:prediction:expansion}
 \delta(\tau)\sim  \,\ln x \sum_j N_j \, .
\end{equation}
These general results allow us to anticipate the behaviour of the non-Gaussianity. We return to these results in Chapter~\ref{chap:non:Gaussianity:coupling}.

\section{Quantum metrology and estimation theory} \label{chap:introduction:quantum:metrology}

Quantum metrology is the study of sensing schemes that utilise the unique properties of quantum systems, such as coherence and entanglement to enhance the sensitivity of a system. At the same time, quantum systems impose inherent restrictions on the sensitivity that can be achieved, for example through the Heisenberg uncertainty relations, which stipulates that complementary variables cannot be measured simultaneously to infinite precision. To determine how quantum properties enhance but also restrict the sensitivity of the system we wish to use as a probe, we model its dynamics and focus on a key quantity referred to as the quantum Fisher information (QFI).

In this Section, we review the basics of estimation theory and quantum metrology. We begin by introducing basic concepts in estimation theory before we discuss the quantum advantage to sensing schemes. 
We then proceed to derive the classical Fisher information (CFI), which is an information-measure that can be linked to the variance of a parameter. We then show how the CFI can be generalised for all possible measurements to the QFI, which is the focus of the third part of this thesis. The motivation for studying these quantities becomes clear when we introduce and derive the Cram\'{e}r--Rao bound, which relates the quantum and classical Fisher information to the variance of a parameter. 

This discussion is based on lecture notes provided to the author by Animesh Datta~\cite{datta2016notes}, but the general concepts are discussed in a vast number of sources~\cite{gobel2015quantum, simon2017quantum}. 

\subsection{Introduction to estimation theory} \label{chap:introduction:estimation:theory}

Estimation theory lies at the heart of physics and scientific discovery. It is often invoked to determine when enough data has been accumulated for a physical discovery to be accepted by the scientific community. 
Many major discoveries in physics are officially announced once the data support the discovery by $5 \sigma$, where $\sigma$ is the standard deviation. Here, $5\sigma$ corresponds to there being a 1 in $3.5\times 10^6$ chance that the conclusions are incorrect given the data at hand. For example, the existence of the Higgs boson  was determined to at least $5\sigma $~\cite{aad2012observation}, which means that there is a 1 in $3.5\times10^6$ chance that the data accumulated by the Large Hadron Collider would be of this extreme nature if the Higgs boson did not exist. 

Estimation theory provides a framework for this process by defining the parameter of interest and establishing how its value determined given a set of data. We here review a number of basic concepts, which will later be put to use in the derivation of the CFI and QFI. 

We begin by considering a sample of values $X_1, X_2, \ldots, X_n$ of size $n$ from an iid source, where `iid' stands for `independent and identically-distributed'. Our goal is to determine some parameter $\theta$ of interest from this data. While we could just determine $\theta$ from a single data-point $X_1$, a better strategy would be to use all data points whose average value is given by 
\begin{equation}
\bar{X}_n = \frac{1}{n} \sum_i X_i \, .
\end{equation}
We call $X = \{X_1, X_2, \ldots , X_n\}$ the set of data-points, and $\theta \in \mathcal{O}$ is the parameter space. We then define the probability functions $p(X, \theta)$, such that 
\begin{align}
&p(X, \theta) \geq 	0 \,,  &&\mbox{and} && \int \mathrm{d}X p(X, \theta) = 1 \, . 
\end{align}
We can related $p(X,\theta)$ to the conditional probability $p(X|\theta)$, which describes the probability of measuring $X$ given a specific $\theta$. The conditional probability is related to $p(X,\theta)$ via Bayes' Theorem, which states $p(X, \theta) = p(\theta) \, p(X|\theta)$, where $p(\theta)$ is known as the prior probability.

We then define the estimator $\bar{\theta}$, which  approximates the true value of the parameter $\theta$ given our set of data. As an example, consider the case where the `true' value of some parameter is $\theta = 1$. The estimator could then, for example, yield a value of $\bar{\theta} = 1.001$, which can be deemed `close' to $\theta$,  given some suitable definition of `close'. We define the difference between $\bar{\theta} - \theta $ as the \textit{error} in the estimator.

The bias $b$ of an estimator encodes this error, and we define it as
\begin{equation}
  b = E_1 \left[ \bar{\theta}(X) \right] - \theta \, ,
\end{equation}
where $E_1$ is the expectation value of $\bar{\theta}(X)$ defined by 
\begin{equation} \label{chap:introduction:metrology:definition:of:expectation:value}
E_1\left[ \bar{\theta}(X) \right] = \int \mathrm{d} X \, \bar{\theta}(X) \, p(X, \theta).
\end{equation}
When $b = 0$, $\bar{\theta} $ is referred to as an unbiased estimator. 

Another quantity that is key to this discussion is the score, which we call $V_\theta$. The score is the gradient of the log-likelihood function and therefore indicates the sensitivity of the system to an infinitesimal change to the parameter value. The score is defined as
\begin{equation} \label{chap:introduction:eq:definition:score}
V_\theta = \partial_\theta \ln p(X, \theta) = \frac{1}{p(X,\theta)} \, \p_\theta p(X, \theta) \, .
\end{equation} 
The mean has a few important properties. Firstly, the mean of the score is zero, such that
\begin{equation}
E_1[ V_\theta] = 0 \, .
\end{equation}
We can prove this relation by noting that
\begin{equation}
E_1[V_\theta] = \int \mathrm{d}X \, p(X,\theta) \left[ \frac{1}{p(X ,\theta) } \frac{\partial }{\partial \theta} p(X ,\theta) \right] = \frac{\partial}{\partial \theta} \int \mathrm{d}X \, p(X,\theta) = 0 \, ,
\end{equation}
which follows because the probabilities integrate to unity. 

Secondly, the variance of the score is, in fact, the classical Fisher information (CFI), which is defined as
\begin{align}
I_\theta &= \mathrm{Var}(V_\theta)  \nonumber \\
&= E_1 \left[ \left(  \partial_\theta \ln p(X, \theta) \right)^2 \right] \, \nonumber \\
&= \int \mathrm{d} X \frac{1}{p(X, \theta)  } \left( \frac{\partial p(X, \theta) }{\partial X} \right)^2 \, .
\end{align}
As mentioned above, the CFI is a key quantity that we will expand on in Section~\ref{chap:introduction:sec:ClassicalFisher}. We now have the tools required to consider the CFI and QFI. First, however, we motivate the study of quantum systems as sensors.

\subsection{The quantum advantage to sensing} \label{chap:introduction:quantum:metrology:quantum:advantage}

Quantum systems, due to properties such as coherence and entanglement, show extraordinary promise as quantum sensors. There are several reasons for why quantum systems are able to outperform classical setups.

Firstly, and perhaps most importantly, the classical measurement limit can be exceeded by using quantum systems~\cite{giovannetti2004quantum}. For classical systems, the sensitivity scales as $\Delta \theta \propto 1/\sqrt{N}$ due to the Central Limit Theorem, where $N$ is the number of probes used for the measurement. By probes, we mean the number of measurements performed, or the number of probe systems that the source system is interacting with. These could be the number of photons in a cavity or the atoms in a sensor array.  The scaling is sometimes explicitly written as $\Delta \theta \propto 1/\sqrt{N \, M}$, where $N$ is the number of probes and $M$ is the number of measurements. 
This scaling is also known as the Standard Quantum Limit (SQL), which was first discussed in the literature in the context of measuring the position of a single mass in the search for gravitational waves~\cite{braginskiui1975quantum}. 

Quantum systems, in comparison, can achieve a better scaling compared with $1/\sqrt{N}$. Injecting squeezed vacuum in one of the ports of an interferometer was shown to increase the scaling to $1/N^{3/4} $~\cite{caves1981quantum, barnett2003ultimate}. Furthermore, the addition of entanglement into scheme, which was included by injecting entangled states into the interferometer ports yields a scaling of $1/N$, which is a factor $\sqrt{N}$ improvement~\cite{yurke19862, dowling1998correlated}. 
This limit is often referred to the \textit{Heisenberg limit} of interferometry, and results indicate that this is indeed the true quantum limit~\cite{bollinger1996optimal, ou1997fundamental}. 

The definition of the Heisenberg limit is a topic of debate in the community; its ambiguous definition became apparent when some results seemed to indicate scaling beyond the $1/N$ limit defined above. For example, using a Kerr nonlinear optical system with a photon number operator $\hat N_a^2 = (\hat a^\dag \hat a)^2$ yielded a scaling of $1/N^{3/2}$~\cite{boixo2007generalized}, and it was demonstrated that measurement schemes with parallel probes can achieve a scaling of $N^{-k}$ with $k \in \mathbb{N}$~\cite{boixo2007generalized}. 
Yet another scheme demonstrated a scaling of $2^{-N}$~\cite{roy2008exponentially}. While these schemes supposedly exceed the Heisenberg limit to provide extremely high sensitivities in the large $N$ limit, it was argued in Ref~\cite{zwierz2010general} that the Heisenberg limit should not always be defined in terms of the number of physical probes in the system. Rather, the stronger measure of translation in the system sets the scaling of the system, and the Heisenberg limit should be determined with that in mind. Thus given the Kerr nonlinearity with $\hat N_a^2$, the Heisenberg limit is really $1/N^2$, which means that the scaling reported in Ref~\cite{boixo2007generalized} does not in fact exceed it.  We return to this idea in Chapter~\ref{chap:metrology} in order to interpret one of the main results in this thesis that concerns the QFI for optomechanical systems.

\subsection{Classical Fisher information} \label{chap:introduction:sec:ClassicalFisher}

We already came across the classical Fisher information in Section~\ref{chap:introduction:estimation:theory}, which is defined as the variance of the score $V_\theta$. In this Section, we  formally define the classical Fisher information and discuss its properties. 

Given a system that is influenced by some effect parametrised by $\theta$, we are often not able to directly measure the parameter $\theta$. Instead, we measure another parameter and obtain the measurement outcome $x$, from which we estimate $\theta$. Intuitively, the CFI is a measure of how much information we can infer about $\theta$ given a specific measurement of $X$. 

We proceed with the formal definition of the CFI.

\begin{defn}[Classical Fisher Information] \label{chap:introduction:definition:Fisher:information}
Given a parameter $\theta$ that we wish to estimate and given a measurement with outcome $x$, the Fisher Information $I_\theta$ represents the information we acquire about $\theta$ given the measurement outcome $x$. It is given by 
\begin{equation} \label{chap:introduction:eq;definition:Fisher:information}
I_\theta =  \int \mathrm{d}x \frac{1}{p(x|\theta) } \left( \frac{\p p(x|\theta)}{\p x} \right)^2, 
\end{equation}
where $p(x | \theta)$ is a conditional probability density function resulting from a measurement of $x$. The integrals runs over all measurement outcomes $x$. 
\end{defn}

 Definition~\ref{chap:introduction:definition:Fisher:information} is valid for all systems, both classical and quantum. Since we are interested in quantum metrology, we obtain the probability distribution $p(x|\theta)$ from the measurement of a quantum state $\hat \rho_\theta$, which depends on the quantity $\theta$. We define a POVM, which stands for `positive operator valued measure'~\cite{barnett2009quantum, nielsen2002quantum}, with elements $\hat \Pi_x $ parametrised by $x$. The POVM elements satisfy the properties $\hat \Pi_x \geq 0$ for all $x$, and 
\begin{equation} \label{chap:introduction:metrology:POVM:completeness}
\int \mathrm{d}x \, \Pi_x = \mathds{1} \, .
\end{equation}
 where $\mathds{1}$ is the infinite-dimensional identity matrix. 
 
 The POVM elements relate the state $\hat \rho_\theta$ to the conditional probability distribution as
 \begin{equation}
 p(x| \theta) = \trace{\hat{\rho_\theta} \,\hat \Pi_x} \, . 
 \end{equation}
Let us list some properties of $I_\theta$:
\begin{itemize}
\item Compared with other information measures, such as the Shannon entropy~\cite{shannon1948mathematical} or the mutual information, the CFI is dimensionful. We acquire units because the probability densities $p(x |\theta)$ generally are dimensionful, and because the derivative $\p_\theta$ has units of $\theta^{-1}$. This necessary to relate $I_\theta$ to the variance $\mbox{Var}(\theta)$ through the Cram\'{e}r--Rao inequality (see Section~\ref{chap:introduction:Cramer:Rao}). 
\item $I_\theta$ is a strictly positive quantity. This can be shown by considering a more general form of $I_\theta$ that takes into account many variables $\{\theta_1, \theta_2 , \ldots, \theta_N\}$ and examining the form of the resulting matrices, which can be shown to be positive-semidefinite~\cite{zegers2015fisher}. 
\item $I_\theta$ is linked to the mutual information, but they are fundamentally different quantities. To gain an intuitive understanding, the mutual information $I(X;Y)$ measures the \textit{correlation} between the random variables $X$ and $Y$, whereas $I_\theta$ is concerned with finding the \textit{likelihood} that some estimator $\bar{\theta}$ closely approximates the `true' value of $\theta$. 
\end{itemize}

\subsection{Quantum Fisher information} \label{chap:introduction:sec:QuantumFisher}

While the classical Fisher Information $I_\theta$ quantifies the sensitivity that can be inferred through a single measurement, the quantum Fisher information (QFI), which we call $\mathcal{I}_\theta$, optimises the measurement over all possible POVMs .This means that the QFI provides the fundamental bound to the sensitivity that can be achieved with a single measurement. We define it formally as follows. 

\begin{defn}[Quantum Fisher information] The quantum Fisher Information (QFI) $\mathcal{I}_\theta$ is given by
\begin{equation} \label{chap:introduction:eq:quantum:Fisher:information:definition}
\mathcal{I}_\theta \leq \mathrm{Tr}\left[ \hat \rho_\theta \, \hat L_\theta^2 \right] \, , 
\end{equation}
where $\hat \rho_\theta$ is the quantum state that depends on the parameter $\theta$, and  $\hat L_\theta$ is the \textit{symmetric logarithmic derivative} (SLD), defined by 
\begin{equation} \label{chap:introduction:eq:logarithmic:symmetric:derivative}
\p_\theta \hat \rho_\theta = \frac{\hat L_\theta  \, \hat \rho_\theta + \hat \rho_\theta \, \hat L_\theta}{2} \, .
\end{equation}
The classical Fisher information $I_\theta$ saturates the QFI $\mathcal{I}_\theta$ over all possible POVMs:
\begin{equation}
\mathcal{I}_\theta(\theta) = \max_{\{\Pi_x\}} I_\theta(\theta) \, .
\end{equation}
\end{defn}
We are generally interested in finding the optimal POVM since it could potentially be implemented in the laboratory to optimise the sensitivity of the system. However, identifying the optimal POVM is often a difficult task. The relation in Eq.~\eqref{chap:introduction:eq:logarithmic:symmetric:derivative} is also known as the Lyapunov matrix equation, and it  must be solved for the symmetric logarithmic derivative $\hat L_\theta$. The general solution reads~\cite{paris2009quantum}
\begin{equation} \label{chap:introduction:metrology:Lyapunov:equation}
\hat L_\theta = 2 \int^\infty_0 \mathrm{d} q \, e^{- \hat \rho_\theta \, q} \, \partial_\theta \hat \rho_\theta \, e^{- \hat \rho_\theta \, q} \, ,
\end{equation}
where $q $ is a scalar parameter. Writing $\hat \rho_\theta$ in terms of its eigenbasis as $\hat \rho_\theta = \sum_n p_n \ketbra{\psi_n}$, Eq.~\eqref{chap:introduction:metrology:Lyapunov:equation} can be rewritten as
\begin{equation}
\hat L_\theta = 2 \sum_{n, m} \frac{\bra{\psi_m} \partial_\theta \hat \rho_\theta \ket{\psi_n} }{p_n + p_m} \ket{\psi_m } \bra{\psi_n } \, , 
\end{equation}
where the sum includes only the terms which satisfy $p_n + p_m \neq 0$. 

Our analysis in this thesis is mostly concerned with computing the QFI for closed systems. There are methods for inferring noisy bounds on the Cram\'{e}r--Rao inequality~\cite{alipour2014quantum}, and similar method for many-body systems was proposed in~\cite{beau2017nonlinear}. In general, many investigations resort to numerical methods, where efficient methods can be used for classes of specific systems~\cite{liu2016quantum}

To gain intuition for the origin of the QFI, we  present the proof of Eq.~\eqref{chap:introduction:eq:quantum:Fisher:information:definition}. 
 We begin by considering the CFI in Eq.~\eqref{chap:introduction:eq;definition:Fisher:information}, where our goal is to find an expression for the derivative $\partial_\theta p(x|\theta)$ of the conditional probability distribution $p(x |\theta)$ and relate it to the SLD in Eq.~\eqref{chap:introduction:eq:logarithmic:symmetric:derivative}. We write 
\begin{align}
\p_\theta \, p(x | \theta) &= \mathrm{Tr} \left[\hat \Pi\_x,  \p_\theta  \hat  \rho_\theta \right] \nonumber \\
&= \frac{1}{2}\mathrm{Tr} \left[\hat \Pi_x  \, \hat L_\theta \, \hat \rho_\theta \right] +\frac{1}{2} \mathrm{Tr} \left[\hat  \Pi_x  \, \hat \rho_\theta \, \hat L_\theta \right]  \, ,
\end{align}
where in the second line we have substituted in the expressions for the SLD in Eq.~\eqref{chap:introduction:eq:logarithmic:symmetric:derivative}. We then note that the second term can be written in terms of its conjugate transpose to find
\begin{align} \label{chap:introduction:eq:probability:redefinition}
\p_\theta \, p(x |\theta) &= \frac{1}{2}\mathrm{Tr}\left[\hat \rho_\theta \,\hat  \Pi_x \, \hat L _\theta \right] + \frac{1}{2}\mathrm{Tr} \left[ \left( \hat \rho_\theta \, \hat \Pi_x \, \hat L_\theta \right)^\dagger \right]  \nonumber \\
&= \Re \mathrm{Tr} \left[\hat \rho_\theta \, \hat \Pi_x \, \hat L_\theta \right] \, , 
\end{align}
which follows from the fact that $\hat L_\theta $ is Hermitian:  $ \hat L_\theta = \hat L_\theta^\dagger $. The fact that $\hat L_\theta$ is Hermitian follows from the Hermiticity of $\hat \rho_\theta$ through:
\begin{equation}
\left( \frac{\p \hat \rho_\theta}{\p \theta} \right)^\dagger = \frac{\p \hat \rho_\theta}{\p \theta} \, . 
\end{equation}
We can now put everything together. We substitute the  expression in Eq.~\eqref{chap:introduction:eq:probability:redefinition} into the integral for the QFI to find
\begin{align}
\mathcal{I}_\theta &= \int \mathrm{d}x \frac{\mathrm{Re}\left( \mathrm{Tr} \left[\hat  \rho_\theta \, \hat \Pi_x \, \hat L_\theta \right] \right)^2}{\mathrm{Tr}\left[\hat \rho_\theta \, \hat \Pi_x \right] } \, .
\end{align}
We then use the fact that the real part of any complex number $z$ is not larger than its absolute value:
\begin{align}
\mathrm{Re}(z)^2 \leq |z|^2, 
\end{align}
to write
\begin{align}
\int \mathrm{d}x \frac{\mathrm{Re}\left( \mathrm{Tr} \left[\hat  \rho_\theta \, \hat \Pi_x \, \hat L_\theta \right] \right)^2}{\mathrm{Tr}\left[\hat \rho_\theta \, \hat \Pi_x \right] }\leq \int \mathrm{d}x \left| \frac{\mathrm{Tr}\left[ \hat \rho_\theta \, \hat \Pi_x\, \hat  L_\theta \right]}{\sqrt{\mathrm{Tr}\left[\hat \rho_\theta \, \hat \Pi_x \right] }} \right|^2 \, .
\end{align}
We have brought the denominator $\mathrm{Tr} \left[ \hat \rho_\theta \, \hat \Pi_x \right]$ into the absolute-value sign since it is real by definition. We then rewrite the integral in the following manner to simplify the subsequent analysis:
\begin{align} \label{chap:introduction:metrology:rewritten:QFI:integral}
\mathcal{I}_\theta \leq  \int \mathrm{d}x \left| \mathrm{Tr} \left[ \frac{\sqrt{\hat \rho_\theta} \sqrt{\hat \Pi_x} \cdot \sqrt{\hat \Pi_x} \, \hat L_\theta \sqrt{\hat \rho_\theta}}{\sqrt{\mathrm{Tr}\left[\hat \rho_\theta \,  \hat \Pi_x \right] }} \right] \right|^2 \, ,
\end{align}
where we have use the cyclical property of the trace operation and the fact that a positive-semidefinite matrix has a unique positive-semidefinite square root, which holds because $\hat \rho_\theta $ and $\hat \Pi_x$ are both positive-semidefinite matrices. 

We then define the two following operators from the terms under the integral in Eq.~\eqref{chap:introduction:metrology:rewritten:QFI:integral} as $\hat A$ and $\hat B$:
\begin{align}
\hat A &= \frac{\sqrt{\hat \rho_\theta} \, \sqrt{\hat \Pi_x}}{\sqrt{\trace{\hat\rho_\theta \, \hat \Pi}}} \,, \nonumber \\
\hat B &= \sqrt{\hat \Pi_x} \, \hat L_\theta \sqrt{\hat \rho_\theta} \, .
\end{align}
Given the Cauchy-Schwartz inequality, which states that
\begin{equation} \label{chap:introduction:metrology:Cauchy:Schwartz}
\left| \mathrm{Tr}\left[ \hat A^\dagger \hat B \right] \right|^2 \leq \mathrm{Tr}\left[\hat A^\dagger \hat A \right] \mathrm{Tr} \left[\hat B^\dagger \hat B \right] \, ,
\end{equation}
 it follows that $\mathrm{Tr} \bigl[ \hat A^\dag \hat A\bigr] = 1$, and we are left with 
\begin{align}
\mathcal{I}_\theta &\leq \int \mathrm{d}x \mathrm{Tr} \left[\hat \Pi_x \, \hat L_\theta \hat \rho_\theta \hat L_\theta \right] \, \nonumber \\
&\leq \mathrm{Tr}\left[\hat \rho_\theta \, \hat L_\theta^2 \right] \, ,
\end{align}
which is equivalent to Eq.~\eqref{chap:introduction:eq:quantum:Fisher:information:definition}. 

As we have already seen, computing a value of the QFI generally involves finding an expression for the SLD in Eq.~\eqref{chap:introduction:eq:logarithmic:symmetric:derivative}. This can be difficult, especially if the state considered is mixed. However, the QFI in Eq.~\eqref{chap:introduction:eq:quantum:Fisher:information:definition} can be cast in a more convenient form, which simplifies the QFI even for mixed states.
 
Given a unitary channel $\hat U_\theta$, the QFI can be cast in more compact form if the eigenstates $\ket{\lambda_n}$ and eigenvalues $\lambda_n$ of the initial state $\hat \rho_\theta = \sum_n \lambda_n \ketbra{\lambda_n}$ are known. The QFI can then be written as~\cite{pang2014quantum, jing2014quantum}, 
\begin{align}\label{chap:introduction:eq:QFI:definition:of:QFI:mixed:states}
\mathcal{I}_\theta
=& \;4\sum_n \lambda_n\left(\bra{\lambda_n}\mathcal{\hat H}_\theta^2\ket{\lambda_n} - \bra{\lambda_n}\mathcal{\hat H}_\theta\ket{\lambda_n}^2 \right)\nonumber\\
&\;-8\sum_{n\neq m}
\frac{\lambda_n \lambda_m}{\lambda_n+\lambda_m}
\left| \bra{\lambda_n}\mathcal{\hat H}_\theta \ket{\lambda_m}\right|^2\;,
\end{align}
where $\lambda_n$ is the eigenvalue of the eigenstate $\ket{\lambda_n}$, and where the Hermitian operator $\mathcal{\hat H}_\theta$ is defined by $\mathcal{\hat H}_\theta=-i\hat U^\dagger_\theta \partial_\theta{\hat U}_\theta$. We derive this expression in Section~\ref{app:QFI:derivation:of:QFI:mixed:states} in Appendix~\ref{app:QFI}. 

The QFI can now be fully determined by computing the expectation value of $\hat{\mathcal{H}}_\theta$ and $\hat{\mathcal{H}}_\theta^2$, which requires differentiating the evolution operator $\hat U_\theta$ with respect to the parameter of interest $\theta$. The expression in Eq.~\eqref{chap:introduction:eq:QFI:definition:of:QFI:mixed:states} is the key figure of merit that we focus on in Chapter~\ref{chap:metrology}. 

Finally, the QFI has an even simpler expression when the state is pure. For a pure density matrix, we have that $\hat \rho_\theta^2 = \hat \rho_\theta$. This implies that
\begin{equation}
\p_\theta \hat  \rho_\theta = \p_\theta (\hat \rho_\theta)^2= (\p_\theta \hat \rho_\theta)\hat \rho_\theta + \hat \rho_\theta ( \p_\theta \hat \rho_\theta) \, .
\end{equation}
Then, we compare this expression with the SLD in Eq.~\eqref{chap:introduction:eq:logarithmic:symmetric:derivative}. 
We write the density matrix in terms of the pure state $\ket{\Psi_\theta}$, such that $\hat \rho_\theta = \ket{\Psi_\theta}\bra{\Psi_\theta}$. This allows us to write
\begin{align}
\hat L_\theta &= 2\, \p_\theta \hat \rho_\theta = 2 \, \left[ \ket{\Psi_\theta} \bra{\p_\theta \Psi_\theta} + \ket{\p_\theta \Psi_\theta}\bra{\Psi_\theta} \right] \, . 
\end{align}
Inserting this expression into Eq.~\eqref{chap:introduction:eq:quantum:Fisher:information:definition}, we arrive at a simplified expression for the QFI $\mathcal{I}_\theta$ for pure states:
\begin{equation}
\mathcal{I}_\theta = 4 \left( \langle \p_\theta \Psi_\theta|\p_\theta \Psi_\theta \rangle - |\langle \Psi_\theta|\p_\theta \Psi_\theta \rangle|^2 \right). 
\end{equation}
This quantity is  easier to estimate since it does not depend on the symmetric logarithmic derivative $\hat L_\theta$.

\subsection{The Cram\'{e}r--Rao bound} \label{chap:introduction:Cramer:Rao}
As mentioned in the sections above, the Cram\'{e}r--Rao inequality links the CFI and QFI to a real measurement uncertainty in the form of the variance of a parameter $\theta$, which we denote $\mathrm{Var}(\theta)$. We here define the bound and prove its validity. 

\begin{defn}[Cram\'{e}r--Rao bound] 
The Cram\'{e}r--Rao inequality states that  ~\cite{cramer1946contribution, wolfowitz1947efficiency, rao1992information} 
\begin{equation} \label{chap:introduction:eq:Cramer:Rao}
\mbox{Var}(\bar{\theta}) \geq \frac{1}{\mathcal{I}_\theta} \, , 
\end{equation}
where $\bar{\theta}$ is an unbiased estimator (which we defined in Section~\ref{chap:introduction:estimation:theory}) for $\theta$ and $\mathcal{I}_\theta$ is the QFI. The QFI can therefore be used to bound the variance of the parameter $\theta $ from below. 

As discussed in previous sections, the CFI provides the bound for a specific measurement, while the QFI provides the fundamental and optimal bound. These two cases are sometimes referred to the classical Cram\'{e}r--Rao bound (CCRB) and the quantum Cram\'{e}r--Rao bound (QCRB). 
\end{defn}

Let us now provide a proof of the Cram\'{e}r--Rao bound in Eq. \eqref{chap:introduction:eq:Cramer:Rao}. We consider the mean $E_1$, which we defined in Eq. \eqref{chap:introduction:metrology:definition:of:expectation:value}, and the score $V_\theta$, which we defined in Eq.~\eqref{chap:introduction:eq:definition:score}. We now seek to derive a relation between the CFI and   $\mathrm{Var}(\bar{\theta})$. By using the Cauchy-Schwartz inequality in Eq.~\eqref{chap:introduction:metrology:Cauchy:Schwartz}, we find the following relation:
\begin{equation}\label{chap:introduction:metrology:weird:relation}
\left( E_1\left[ \bigl( V_\theta- E_1[V_\theta] \bigr) \bigl( \bar{\theta}  - E_1[\bar{\theta}] \bigr) \right] \right)^2 \leq E_1 \left[ V_\theta - E_1[V_\theta] \right]^2 E_1\bigl[ \bar{\theta} - E_1[\bar{\theta}] \bigr]^2,  
\end{equation}
where $V_\theta$ is the score and $\bar{\theta}$ is an unbiased estimator. We must now simplify both sides of Eq. \eqref{chap:introduction:metrology:weird:relation} to find the correct relation. 

We begin by recalling that the score satisfies the following condition
\begin{equation} \label{chap:introduction:eq:mean:of:score:zero}
E_1 \left[V_\theta\right] = 0 \, ,
\end{equation}
and from the fact that $\bar{\theta}$ is an unbiased estimator, it follows $E_1[\bar{\theta}] = \theta$. The left-hand side of  Eq~\eqref{chap:introduction:metrology:weird:relation} can therefore be simplified to 
\begin{equation} \label{chap:introduction:metrology:simplified:relation:bound}
E_1\left[ \bigl( V_\theta- E_1[V_\theta] \bigr) \bigl( \bar{\theta}  - E_1[\bar{\theta}] \bigr) \right] = E_1 \left[ (\bar{\theta} - \theta)\, V_\theta \right] \, .
\end{equation}
By using the linearity of $E_1$, the expression in Eq.~\eqref{chap:introduction:metrology:simplified:relation:bound} becomes
\begin{equation}
E_1 \left[ ( \bar{\theta} - \theta) \, V_\theta  \right] = E_1\left[  \bar{\theta} \, V_\theta \right] - \theta \, E_1\left[V_\theta\right] = E_1\left[ V_\theta \, \bar{\theta} \right] \, , 
\end{equation}
where in the second equality we have used the result from Eq.~\eqref{chap:introduction:eq:mean:of:score:zero}. Next, we examine the right-hand side of Eq.~\eqref{chap:introduction:metrology:simplified:relation:bound}. We find
\begin{equation}
 E_1 \left[V_\theta^2\right]  E_1 \left[ \bar{\theta} - \theta \right]  = \mathcal{I}_ \theta \,  \mbox{Var}(\bar{\theta}) \, ,
\end{equation}
If we substitute this into the inequality in Eq.~\eqref{chap:introduction:metrology:weird:relation}, we find
\begin{equation} \label{chap:introduction:metrology:inequality:almost:done}
 \left( E_1\left[ V_\theta \,  \bar{\theta} \right] \right)^2 \leq \mbox{Var}(\bar{\theta}) \mathcal{I}_\theta  \, .
\end{equation}
We are almost there. It remains to prove that the left-hand side of Eq.~\eqref{chap:introduction:metrology:inequality:almost:done} is equal to unity:  $\left( E_1\left[ V _\theta \, \bar{\theta} \right] \right)^2 = 1$. We use the definition of $E_1$ in Eq.~\eqref{chap:introduction:metrology:definition:of:expectation:value} to write
\begin{align}
E_1\left[ V_\theta \bar{\theta} \right] &= \int_x \mathrm{d}x  \, \frac{1}{p (x|\theta) } \frac{\p p(x|\theta)}{\p \theta} \bar{\theta}(x) p(x| \theta)  = \int_x \mathrm{d} x  \, \frac{\p p(x| \theta)}{\p \theta} \bar{\theta}(x)  \, .
\end{align}
We now pull the derivative out in front of the integral to find
\begin{align}
E_1 \left[ V_\theta \bar{\theta} \right]&= \frac{\p }{\p \theta} \int_x \mathrm{d} x  \, p(x| \theta)  \, \bar{\theta}(x) = \frac{\p \theta}{\p \theta}= 1 \,,  
\end{align}
where we used the fact that $E_1[\bar{\theta}] = \theta$ for an unbiased estimator. We then take the previous expression and move $\mathcal{I}_\theta$ to the other side, giving the correct expression for the Cram'{e}--Rao bound in Eq.~\eqref{chap:introduction:eq:Cramer:Rao}:
\begin{equation}
\mbox{Var}(\bar{\theta}) \geq \frac{1}{\mathcal{I}_\theta}. 
\end{equation}
This concludes our derivation of the Cram\'{e}r--Rao inequality. 
We are now ready to proceed with the main theoretical result in this thesis: the solution of the dynamics for an extended optomechanical Hamiltonian.

\cleardoublepage

\chapter{Solving the dynamics of nonlinear optomechanical systems}
\label{chap:decoupling}
In this Chapter we solve the dynamics of a fully time-dependent optomechanical system with an additional mechanical displacement term and a single-mode mechanical squeezing term. The methods, solutions and notation presented here will serve as a basis for all further Chapters in this thesis.

We begin this Chapter by providing some basic definitions and examples of Lie algebras and related concepts from group theory. We then proceed to review the Decoupling Theorem~\cite{wei1963lie}, 
which shows that by identifying a minimal and finite Lie algebra that generates the time-evolution of the system, it is possible to solve the dynamics by considering a finite set of coupled ordinary differential equations of real functions.   
In the second part of this Chapter, we apply the Decoupling Theorem to the optomechanical system mentioned above. We solve the dynamics up to up two second-order differential equations and derive a number of relations to simplify their solution. 

This Chapter is based on the work in Ref~\cite{qvarfort2019time}. 
The main decoupling procedure in Section~\ref{chap:decoupling:optomechanical:decoupling} was first derived by David Edward Bruschi, and later extended by the author and the other authors of Ref~\cite{qvarfort2019time}. Specifically, Sections~\ref{chap:decoupling:link:J:Bogoliubov} and~\ref{chap:decoupling:interpretation:of:evolution} were contributed by Dennis R\"{a}tzel.

\section{Background}

The Decoupling Theorem relies on a specific property of a Lie algebra called closure. To better understand this property and Lie algebras in general, we begin by presenting a summary of basic group-theoretical concepts. For each concept discussed, we will provide a formal definition followed by an intuitive example. This section is intended for those who are not overly familiar with group theory. Following the introduction, we outline the proof in Ref~\cite{wei1963lie}, complemented with details from the related work~\cite{wilcox1967exponential}.

\subsection{Lie groups}
Before discussing Lie algebras, we first discuss Lie groups since these two concepts are intrinsically connected. We start by recounting the basic definition of a group. 
\begin{defn}[Groups]
A group $(G,*)$ is a set $G$ with group elements $g$ and a binary operation $*$ such that $G\times G \rightarrow G$, which satisfies three conditions:
\begin{enumerate}
\item \textit{Associativity}: For any three elements $x,y,z \in G$, we have $(x*y) *z = x* (y*z)$. 
\item \textit{Identity}: The group must include an identity element $\epsilon \in G$ such that $\forall g \in G$. In other words, multiplying $g$ by the identity element leaves $g$ invariant, such that $\epsilon * g = g * \epsilon = g$. 
\item \textit{Inverse}: For each element in the group, there must be an inverse element. That is, for each $g \in G$ there is some $\bar{g} \in G$ such that $g * \bar{g} = \bar{g} * g = \epsilon$.
\end{enumerate}
\end{defn}
Groups arise in many separate context in physics. The general study of groups can therefore provide powerful tools to treat a number of cases. We proceed with an example of a group. 
\begin{exmp}[Square rotation group $Z_4$] Perhaps one of the simplest groups is the set of discrete rotation that leave a square invariant. The rotation group $Z_4$ contain rotations of $90^\circ$, $180^\circ$, $270^\circ$ and $360^\circ$ degrees. Let us see how the set of rotations satisfies the group axioms. Any combination of the group elements will satisfy the same symmetry; that is, we can rotate the square by first $90^\circ$, then $360^\circ$, and it remains invariant.  The associativity criterion is satisfied by a proper representation of the group elements, such as a matrix. The identity is equivalent to doing nothing, and the inverse can easily be constructed by rotating the square in the reverse direction. 
\end{exmp}

As hinted at in the above examples, groups are also closely connected with symmetries, but this is not a topic we elaborate on here. We are interested in Lie algebras, and therefore proceed to discuss Lie groups. They are \textit{continuous} groups and play an ubiquitous role in physics and mathematics. In quantum mechanics, the set of unitary time-evolution operators form a Lie-group, as we will see below. In this Chapter, we are interested in the relationship between Lie groups and Lie algebras. 
We begin with a formal definition of Lie groups. 

\begin{defn}[Lie groups]
A Lie group is a set $G$ with two structures: 
Firstly, $G$ is a group with the structure discussed in the definition of a group and secondly, $G$ is a smooth and real manifold. Smoothness means that the group multiplication and inverse map are differentiable. The group is described by a set of real parameters that describe the group elements. 
\end{defn}
As opposed to the finite rotation group $Z_4$, the Lie group must admit infinitesimal rotations. One example of a Lie algebra is the SO(3) rotation groups, which describes the rotations of a sphere in three dimensions. Let us look at three examples of Lie groups.

\begin{exmp}[The real line] Perhaps one of the simplest examples of a Lie group is the real line, where the group operation is addition. The real line constitutes a Lie group because the elements are continuous. The inverse elements can be constructed by adding a minus sign to the element, and the identity element is zero. 
\end{exmp}

\begin{exmp}[The special unitary group SU($n$)] \label{chap:decoupling:example:unitary:group}
The special unitary group SU($n$) is a group of $n\times n$ unitary matrices $U$ with determinant 1 (hence the word `special'). The unitarity condition is $U^\dag = U^{-1}$ and $U^{-1} U = UU^{-1} = \mathds{1}_{n}$, where $\mathds{1}$ is the $n\times n$ identity matrix. Associativity is naturally satisfied by matrix multiplication, the $\mathds{I}_n$ identity matrix is the identity element, and the inverse is obtained through complex conjugation. This group is a Lie group because each element $U$ is parameterised by a real and continuous parameter $\theta$, such that, for example, $U_j = e^{- i \theta_j G_j}$, where $G_j$ is a generator (see Section~\ref{chap:decoupling:sec:lie:algebras} below). 
\end{exmp}
\begin{exmp}[The special orthogonal group SO(3)] \label{chap:decoupling:example:SO3}
The rotations of vectors in three spatial dimensions can be described by orthogonal matrices with unit determinant. Orthogonality refers to the following condition $O^{\rm{T}}O = OO^{\rm{T}}= \mathds{I}_n$. Matrix multiplication automatically satisfies associativity, the $3\times3$ identity matrix is the identity element, and the inverse of a oration can be straight-forwardly defined. This group is a Lie group, because the rotations are continuous and parameterised in a similar way to the above. 
\end{exmp}
In fact, most continuous rotation groups are Lie groups. An extended list of examples can be found in Ref~\cite{fulton2013representation}. 

\subsection{Lie algebras} \label{chap:decoupling:sec:lie:algebras}
Now we are finally in a position to properly define a Lie algebra. They are the key object of interest in this Chapter, and the link between Lie groups and Lie algebras underpins the Decoupling Theorem. Given a Lie group, it is always possible to construct a Lie algebra from the group. Sometimes there are advantages to studying the algebra rather than the group itself. 

We begin with the basic definition of a Lie algebra~\cite{gutowski2007symmetry}. 
\begin{defn}[Lie algebra] A Lie algebra is a vector space $\vec{g}$ over some field $F$, together with a binary operation $[\cdot, \cdot] : g \times g \rightarrow g$ (the Lie bracket) which must satisfy the following axioms:
\begin{enumerate}
\item Bilinearity, such that $[ax + by, z]  = a[x,z] + b[y,z]$ and $[z, ax + by] = a[z, x] + b[z, y]$ for all scalars $a, b$ in the field $F$ and all elements $x,y,z$ in $g$. 
\item Alternativity, which means that the Lie bracket is zero for the same element: $[x,x] = 0$. 
\item The Jacoby identity, which states that 
\begin{equation}
\left[ x, ]y,x] \right] + \left[ z, [x,y] \right] + \left[ y, [z, x] \right] = 0
\end{equation}
\end{enumerate}
\end{defn}
It might already be clear that  the commutator bracket $[A, B] = AB - BA$ satisfies these criteria. In fact, the commutator bracket is often used and is a measure of how non-commutative the algebra is.

The above definition might seem a bit abstract, and the connection between Lie groups and Lie algebras can be made clearer. We therefore provide the following, more intuitive explanation. Consider a Lie group $L$ with elements $G(\alpha) \in L$, where $\alpha$ is some real parameter. To determine the action of the element near identity, we slightly perturb $G(\alpha)$ to find
\begin{equation}
G(\alpha) \approx  + i \delta \alpha_j X_j \, ,
\end{equation}
where we have defined 
\begin{equation}
X_j \equiv - i \frac{\partial}{\partial \alpha_j} D(\alpha) \biggl|_j \, .
\end{equation}
While $\delta \alpha_j$ denotes the amount of the `direction' we perturb in, we call $X_j$ a \textit{generator} which determines the direction of, for example, rotations. 

We can perform this infinitesimal perturbation many times to find
\begin{equation}
D(\alpha) = \left( 1 + 	\frac{i \alpha_j X_j}{k} \right)^k \equiv e^{i \alpha_j X_j} \, .
\end{equation}
This is, in fact, the definition of the exponential map. The $X_j$ are generators, which form a Lie algebra. This Lie algebra generates the group together with the real parameters $\alpha_j$. In other words, given a Lie algebra with a set of $n$ elements one can always use the exponentiation map to generate a Lie group. The connection with the Lie algebra  bracket is made precise by the Baker-Campbell-Hausdorff  formula (see Lemma~\ref{chap:decoupling:lemma:BCH} below). 

\begin{exmp}[Algebra of rotations in three dimensions] In example~\ref{chap:decoupling:example:SO3} we showed that the set of orthogonal rotation matrices in three dimensions constitutes a group called SO(3). The Lie algebra associated with SO(3) consists of $3\times3$ skew-symmetric matrices. A common basis is the following:
\begin{align}
&L_x = \begin{pmatrix} 	
0 & 0 & 0 \\
0 & 0 & - 1 \\
0 & 1 & 0 \end{pmatrix} \, , &L_y = \begin{pmatrix}
0 & 0 & 1\\
0 & 0  &0 \\ 
 - 1 & 0 & 0 \end{pmatrix} \, , &&L_z = \begin{pmatrix}
 0  & - 1& 0 \\
 1 & 0 & 0 \\
 0 & 0 & 0 \end{pmatrix} \, .
 \end{align}
When exponentiated, these three matrices form rotation matrices which can be applied to three-dimensional vectors. 
\end{exmp}

Before we proceed, we list some additional properties of Lie algebras:
\begin{itemize}
\item A Lie algebra can be either finite or infinite. We will see later that it becomes straight-forward to solve a finite algebra, while an infinite algebra poses significant challenges. 
\item A Lie algebra of operators also contains what is known as \textit{structure constants}. That is, if two elements $A_i$ and $A_j$ yield a third operator $A_k$ through application of the Lie bracket as $[A_i, A_j] = f_{ijk} A_k$, then $f_{ijk}$ are the structure constants. 
\item A Lie algebra must satisfy a condition called \textit{closure}. This means that the combination of each two elements $A_i$ and $A_j$ via the Lie bracket must yield an operator $A_k$ that is contained by the algebra. For the Decoupling Theorem, this is the most significant property of the Lie algebra that we shall be using. 
\end{itemize}

\section{The Lie algebra decoupling method } \label{chap:decoupling:Lie:algebra:decoupling:method}
In this Section, we present a method developed by Wei and Norman in 1963  that makes use of Lie algebras to solve linear differential equations~\cite{wei1963lie}. The method effectively transforms the problem of solving an operator-valued linear differential equation (such as the Schr\"{o}dinger equation) into solving a coupled system of differential equations of real coefficients. The same methods have been independently developed and applied in a number of additional works, see e.g. Refs~\cite{bose2003vacuum} and~\cite{Bruschi2013Time}.

\subsection{The Decoupling Theorem}
This section closely follows the proof in Ref~\cite{wei1963lie}. We use slightly different notation in order to be consistent with conventions in quantum information and quantum optics (such as denoting operators by hats), and to align more closely with the notation used in the remainder of this thesis. We also specialise to considering Hermitian operators in the Hamiltonian, which ensures that the final scalar functions are real rather than complex. 

Consider the first-order differential equation 
\begin{equation} \label{chap:decoupling:eq:to:solve}
\frac{d \hat U(t)}{dt} = \hat H(t) \, \hat U(t) \, ,
\end{equation}
where $\hat H(t)$ generally is the Hamiltonian and $\hat U(t)$ is a time-evolution operator. 

Often, the Hamiltonian $\hat H(t)$ can be written as a finite sum of constant operators $\hat H_i$ and general time-dependent coefficients $G_j(t)$ as
\begin{equation}
\hat H(t) = \sum_{j = 1}^m G_j(t) \, \hat H_j \, .
\end{equation}
The set $\{ \hat H_i \}$ with $i = 1, 2, \ldots , m$ reproduces the Hamiltonian $\hat H(t)$. It can be extended to a larger set with $n \geq m$ elements by commuting the elements in $\{\hat H_i \}$. The full set of elements $\{\hat H_1 , \hat H_2, \hat H_3 , \cdots , \hat H_n\}$ form a Lie algebra $L$ under commutation of dimension $n$.  That is, the Hamiltonian $\hat H(t)$ generates a Lie algebra, and we can find it by considering all the elements $\hat H_i$, and the Lie bracket is the commutator relation $[\hat H_i, \hat H_j] \equiv \hat H_i \hat H_j - \hat H_j \hat H_i$. The Lie algebra $L$ is constructed from all operators in $\hat H(t)$, plus all the Lie products
\begin{equation}
\left[ \hat H_{\alpha_1} , \left[ \hat H_{\alpha_2} , \left[ \hat H_{\alpha_3}, \cdots \left[ \hat H_{\alpha_{r-1}}, \hat H_{\alpha_r} \right] \cdots \right] \right] \right] \, , 
\end{equation}
where $\alpha_i = 1 $ to $m$, plus all linear combinations of such products.

If, through consecutive commutation, we find a finite number of element, the Lie algebra is finite. This is always true if the $\hat H_j$ are finite-dimensional matrices. However, there are also cases where the Lie algebra is infinite, for which commutation of two or more operators continuously produce new elements that are not already part of the algebra.  When we consider optomechanical systems, the operators are necessarily infinite-dimensional, which is why we also sometimes recover an infinite Lie algebra (see Section~\ref{chap:decoupling:sec:infinite:algebra}). 

We will now show that the existence of such a finite Lie algebra $L$ enables the `decoupling' of the time-evolution operator $\hat U(t)$ into a product of operators, namely
\begin{equation} \label{chap:decoupling:eq:decoupled:U}
\hat U(t) = \hat U_1 (t) \, \hat U_2 (t) \, \cdots \hat U_n(t) \, , 
\end{equation}
where each component operator $\hat U_i(t)$  is an operator satisfying 
\begin{equation}
\frac{d}{dt} \hat U_j(t) = \dot{F}_j \, \hat H_j \, \hat U_j(t) \, .
\end{equation}
and where the functions $F_j$ are the real functions that we wish to determine. 

In quantum theory, the advantage of writing $\hat U(t)$ in the form in Eq.~\eqref{chap:decoupling:eq:decoupled:U} is that when the action of each $\hat U_j(t)$ is known, it becomes straight-forward to apply them to a quantum state and obtain an analytic expression for its evolution. The advantage of such a method over solvers which use finite-dimensional matrices is significant, as the key task shifts from evolving the state to obtaining analytic expressions for the $F_j$-functions. 

The operator $\hat U(t)$, which is the solution to Eq.~\eqref{chap:decoupling:eq:to:solve} lies in the special unitary group (see Example~\ref{chap:decoupling:example:unitary:group}) which is in the subset of all linear operators $GL$. The general linear group $(GL)$ is a set of $n\times n$ invertible matrices where the group operation is matrix multiplication. The specialisation from $GL$ to a specific subset arise because we require the solution $\hat U(t)$ to fulfil  certain conditions, such as unitary $\hat U^\dag (t) \hat U(t) = 1$. 

This subset is called the associative algebra $R$, which is generated by the Hamiltonian $\hat H(t)$ including all products
\begin{equation} \label{chap:decoupling:eq:all:products}
\hat H^{\beta_1}_{\alpha_1} \,\hat H^{\beta_2}_{\alpha_2} \, \hat H^{\beta_3}_{\alpha_3} \cdots \hat H^{\beta_r}_{\alpha_r} \, , 
\end{equation}
where $\alpha_i = 1$ to $m$ and where $\beta_i = 1,2,3 \cdots$. 

Every associative algebra gives rise to a Lie algebra or a commutator algebra $L$.  In turn, every element of $L$ generated by $\hat H(t)$ is also an element of $R$ generated by $\hat H(t)$. The reverse is not true, which we can show through a simple example. The Lie product $\hat H_1 \hat H_2 - \hat H_2 \hat H_1$ is an element of $L$, but $\hat H_1\hat H_2$ in general is not. We therefore identify $L$ as a subset of $R$, which is defined by the structural constants of the Lie algebra, which we obtain through commutation:
\begin{equation}
\left[ \hat H_i, \hat H_j\right] = \sum_{r = 1}^n f_{ijk} \hat H_k \, . 
\end{equation}
We are now at a point where we can concisely state the Decoupling Theorem. 
\begin{thm}[Decoupling Theorem] \label{chap:decoupling:theorem:decoupling}
 Suppose that the linear operator $\hat H(t)$ can be expressed in the form
\begin{equation} \label{chap:decoupling:eq:hamiltonian:sum}
\hat H(t) = \sum_{j = 1}^m G_j(t) \hat H_j \, ,
\end{equation}
where $m$ is a finite integer and where the functions $G_j(t)$ are scalar functions of time $t$ and the $\hat H_j$ are time-independent Hermitian operators which live in a Hilbert space $\mathcal{H}$. The requirement that $\hat H_j$ are Hermitian ensures that the functions $G_j(t)$ are real. The dimension of $\mathcal{H}$ can be either finite or infinite. Let the Lie algebra $L$ generated by $\hat H(t)$  be of finite dimension $n$. Then there exists a neighbourhood of $t = 0$ in which the solution of the equation 
\begin{equation} \label{chapt2:eq:schrodinger:equation:definition}
\frac{d \hat U(t)}{dt} = \hat H(t) \, \hat U(t) \, ,
\end{equation}
with the initial condition $\hat U(0) = 1$ may be expressed in the form
\begin{equation} \label{chap:decoupling:time:evolution:ansatz}
\hat U(t) = \exp[ - i \, F_1(t) \, \hat H_1]\, \exp[ - i \, F_2(t) \, \hat H_2] \cdots \exp[ - i \, F_n (t) \, \hat H_n(t) ] \, ,
\end{equation}
where $\hat H_1, \hat H_2 , \cdots , \hat H_n$ is a basis for $L$ and the set $\{F_j(t)\}$ are scalar functions of time $t$. The functions $F_j(t)$ depend only on the Lie algebra $L$ and the initial functions $G_j(t)$. The same decoupling of an evolution operator can also be performed for the symplectic matrices when considering the evolution under any quadratic Hamiltonian. See Eq. 17 in Ref~\cite{Bruschi2013Time}. 
\end{thm}

\subsection{Proof of the Decoupling Theorem}
Our goal is to prove  Theorem~\ref{chap:decoupling:theorem:decoupling}. The Decoupling Theorem is based on two lemmas: The first is the well-known Baker-Campbell-Hausdorff Lemma and the second one comprises a specific property of the Lie algebra. We begin with Lemma 1, which states:

\begin{lem}[Baker-Campbell-Hausdorff] \label{chap:decoupling:lemma:BCH}\footnote{In Ref~\cite{wei1963lie}, this is referred to as the Baker-Hausdorff lemma.} If two operators $X, Y\in L$, then $e^XYe^{-X} \in L$ and 
\begin{equation} \label{chap:decoupling:eq:BCH:result}
e^X Y e^{-X} = Y + [X,Y] + \frac{1}{2!}[X, [X, Y]]  + \frac{1}{3!} [X, [X, [X, Y]]] + \cdots \, .
\end{equation}
We define the new operator $adX$, where $adX, X \in L$ by the equation
\begin{equation}
(ad X) Y = [X,Y] \, ,
\end{equation}
where $Y \in L$. Then we define powers of this equation as
\begin{equation}
(adX)^2 Y = [X, [X, Y]] \, , 
\end{equation}
and so on. Thus the Baker-Hausdorff formula can be written as 
\begin{equation}
e^X Y e^{-X} = (e^{ad X})Y \, .
\end{equation}
\end{lem}

\begin{proof}[Proof of Lemma~\ref{chap:decoupling:lemma:BCH}]
We begin by defining a function
\begin{equation} \label{chap:decoupling:eq:BCH:series}
F(a) = e^{a X} Y e^{- a X} = \sum^\infty_{n = 0} \frac{1}{n!} C_n a^n  \, ,
\end{equation}
where the $C_n$ are coefficients, which can also be operators but which are independent of $a$. When $a = 1$, the coefficients correspond to the case we are considering. 

Our goal is to derive expressions for these coefficients in the form of a recursion relation. We first note that 
\begin{equation} \label{chap:decoupling:eq:BCH:differential}
\frac{d}{da} F(a) = \left[ X, F(a) \right] \, . 
\end{equation}
Inserting Eq.~\eqref{chap:decoupling:eq:BCH:series}  into Eq.~\eqref{chap:decoupling:eq:BCH:differential} we find
\begin{equation}
\sum_{n = 1}^\infty C_n \frac{1}{( n- 1)! } a^{n - 1} = \sum_{n = 0}^\infty \frac{1}{n!} \left[ X, C_n \right] a^n  \, .
\end{equation}
The sum on the left-hand side can be rewritten by setting $n \rightarrow n + 1$, such that we find the formula 
\begin{equation} \label{chap:decoupling:eq:BCH:coefficients}
C_{n + 1} = \left[ X, C_n\right] a^n \, .
\end{equation}
This way, all coefficients can be generated through repeated commutation with $X$. We also have that $C_0 = Y$, which follows from simply Taylor expanding the exponentials in Eq.~\eqref{chap:decoupling:eq:BCH:series}. The coefficients in Eq.~\eqref{chap:decoupling:eq:BCH:coefficients} can then be used to generate all the coefficients in Eq.~\eqref{chap:decoupling:eq:BCH:result}. 

The last lines in the lemma follow from the definition of $(ad X)$ as can be seen by writing
\begin{align}
e^Y X e^{- Y} &= \left( e^{ad X} \right) Y \nonumber \\
&= \sum_n \frac{1}{n!} (adX)^n Y \nonumber \\
&= Y + [X, Y]  + \frac{1}{2!} \left[ X, [X, Y] \right]  + \frac{1}{3!} \left[ X, \left[ X, [X, Y] \right] \right] + \ldots \, .
\end{align}
This concludes the proof of Lemma~\ref{chap:decoupling:lemma:BCH}. 
\end{proof}

We proceed with the second lemma. 
\begin{lem}[Lie algebra basis]\label{chap:lemma:Lie:algebra}
Let $\hat H_1, \hat H_2 , \cdots , \hat H_n$ be a basis for the Lie algebra $L$. Then it follows that
\begin{align} \label{chap:decoupling:lemma:Lie:algebra:basis}
\left( \prod_{j = 1}^r \exp[ - i \, F_j \hat H_j] \right) \, \hat H_k \, \left( \prod_{j = r}^1 \exp[ i \, F_j \hat H_j ] \right) = - i \,  \sum_{j = 1}^n \xi_{jk} \hat H_j \, , 
\end{align}
where $r = 1, \cdots , n, $ and where each $\xi_{jk} \equiv \xi_{jk} (g_1, \cdots, g_r)$, is a function of all its arguments. 
\end{lem}

\begin{proof}[Proof of Lemma~\ref{chap:lemma:Lie:algebra}] Our goal is to establish that the $\xi_{jk}$ are analytic functions. This follows from repeatedly applying Lemma~\ref{chap:decoupling:lemma:BCH} to Eq.~\eqref{chap:decoupling:lemma:Lie:algebra:basis}. We demonstrate the first few lines of this proof.

Consider $r = 1$, which is the simplest case. We find, by using Eq.~\eqref{chap:decoupling:eq:BCH:result},
\begin{align} \label{chap:decoupling:eq:proof:lemma:2}
\exp[- i \,  F_1 \hat H_1 ] \hat H_k \exp[ i \, F_1 \hat H_1 ] =& \,  \hat H_k   - i \, F_1 [\hat H_1, \hat H_k] + \frac{( - i \, F_1)^2}{2!} [\hat H_1,[\hat H_1,  \hat H_k ]] \nonumber \\
&+ \frac{( - i \, F_1)^3}{3!} [\hat H_1, [\hat H_1, [\hat H_1, \hat H_k ]]] + \ldots  \, . 
\end{align}
Now, since the Lie algebra is closed under commutation, it means that one of the terms reads
\begin{equation}
[\hat H_1,  (\rm{ad} \hat H_1)^n \, \hat H_k] = [\hat H_1, \hat H_1]  = 0 \ ,
\end{equation}
for some integer $n$, which means that Eq.~\eqref{chap:decoupling:eq:proof:lemma:2} contains a finite number of terms with different powers of $F_1$.  Since all $F_1$ are analytic, the resulting functions $\xi_{jk}$ in the right-hand side of Eq.~\eqref{chap:decoupling:lemma:Lie:algebra:basis} are necessarily analytic. The same argument can be made for multiplication of additional terms when $r \neq 1$. This concludes the proof of Lemma~\ref{chap:lemma:Lie:algebra}.

\end{proof}

We are now ready to prove the Decoupling Theorem (Theorem~\ref{chap:decoupling:theorem:decoupling}). The proof makes use of both Lemma~\ref{chap:decoupling:lemma:BCH} and ~\ref{chap:lemma:Lie:algebra}. 

\begin{proof}[Proof of Theorem~\ref{chap:decoupling:theorem:decoupling}]
We first note that we can write 
\begin{equation}
A(t) = \sum_{j = 1}^n G_j(t) \, \hat H_i  \,\quad \quad \mbox{instead of} \quad \quad 
A(t) = \sum^m_{j = 1} G_j(t) \, \hat H_i \, ,
\end{equation}
where we changed the upper limit of the sum from $m$, which is the number of terms in the Hamiltonian $\hat H(t)$, to $n$, which is the dimension of the Lie algebra. This follows because we can always set $G_i(t) \equiv 0$ for any $i \geq m$. Note also that the boundary condition of the differential equation we are attempting to solve in Eq.~\eqref{chapt2:eq:schrodinger:equation:definition} is $\hat U(0 ) = 1$ at $t = 0$. This implies that all $F_j(0) = 0$. 

We make the ansatz that $\hat U(t)$ can be written as a product of operators $\hat U_i(t)$ as shown in Eq.~\eqref{chap:decoupling:time:evolution:ansatz}. When we differentiate the ansatz, we find
\begin{align} \label{chap:decoupling:eq:differentiated:U}
\frac{d \hat U(t)}{dt} =  - i \, \sum_{j = 1}^n \dot{F}_j (t) \left( \prod_{k = 1}^{j - 1} \exp[- i \, F_k\, \hat H_k] \right) \, \hat H_j \, \left( \prod_{k = j}^n \exp[ - i \, F_k \, \hat H_k] \right) \, ,
\end{align}
We then use the fact that $d \hat U(t) /dt = \hat H(t) \, \hat U(t)$ and the expression for $\hat H(t)$ in Eq.~\eqref{chap:decoupling:eq:hamiltonian:sum}, and we multiply by the inverse operator $\hat U^{-1}(t)$ on the right-hand-side to find
\begin{align} \label{chap:decoupling:eq:part:solved}
\sum_{j = 1}^n G_j(t) \, \hat H_j &= - i \,  \sum_{j = 1}^n \dot{F}_j (t) \left( \prod_{k = 1}^{j - 1} \exp[ - i \, F_k \, \hat H_k ] \right) \, \hat H_j \, \left( \prod_{k = j - 1}^1 \exp[ i \, F_k \, \hat H_k ] \right) \nonumber \\
&= - i \,  \sum_{j = 1}^n \dot{F}_j (t) \left( \prod_{k = 1}^{j - 1} \exp[ - i \, F_k \, ad\hat H_k ] \right) \, \hat H_j \, .
\end{align}
By applying Lemma~\ref{chap:lemma:Lie:algebra} to the last line of Eq.~\ref{chap:decoupling:eq:part:solved}, we find
\begin{equation} \label{chap:decoupling:eq:differential:equation:sum}
\sum_{k = 1}^n G_k (t) \, \hat H_k =  - i \, \sum_{j = 1}^n \sum_{k = 1}^n \dot{F}_j (t) \, \xi_{kj} \, \hat H_k \, .
\end{equation}
We note now that the operators $\hat H_k$ are linearly independent. We use this to find linear relations between the $G_k(t)$ functions and the $\dot{F}_j(t)$. They are related by the elements $\xi_{kj}$ of the transformation matrix $\boldsymbol{\xi}$, which are analytic functions of the $F_i$ functions. We define the vector $\vec{G} = (G_1, G_2, \cdots, G_n)^{\rm{T}}$ and the vector $\dot{\vec{F}} = (\dot{F}_1 , \dot{F}_2 \,, \cdots \, \dot{F}_n)^{\rm{T}}$, such that 
\begin{align}
\vec{G} = - i \,  \boldsymbol{\xi}  \, \dot{\vec{F}} \, . 
\end{align}
Explicitly, this reads
\begin{align}
\begin{pmatrix}
G_1 \\ G_2 \\ \vdots \\ G_n 
\end{pmatrix} = - i \,  \begin{pmatrix} 
\xi_{11} & \xi_{12} & \cdots & \xi_{1n} \\
\xi_{21} & \xi_{22} &\cdots & \xi_{2n} \\
\vdots & \vdots & \cdots & \cdots \\
\xi_{n1} & \xi_{n2} & \cdots & \xi_{nn}
\end{pmatrix} 
\begin{pmatrix}
\dot{F}_1 \\
\dot{F}_2 \\
\vdots \\
\dot{F}_n 
\end{pmatrix} \, ,
\end{align}
which forms a coupled set of differential equations. The system can be solved if $\boldsymbol{\xi}$ is invertible. To determine if that is the case, we note that the determinant $\det{\boldsymbol{\xi}}$ is non-zero at $t = 0$, because $\boldsymbol{\xi}(0) = 1$. Thus, the matrix is invertible in some neighbourhood of $t = 0$. Since the matrix is invertible, we can write
\begin{equation}
\frac{d \vec{F}}{dt} = f(\vec{G}, \vec{F}) = i \,  \boldsymbol{\xi}^{-1} \, \vec{G} \,. 
\end{equation}
With the boundary condition $\vec{F}(0) = 0$, we are ensured that there exists a unique solution to the system. However, there is no guarantee that the resulting differential equations can be analytically solved. This concludes the proof of the Decoupling Theorem. 
\end{proof}

\subsection{A recipe for decoupling}
We here provide a summary of the decoupling methods in the form of a simple recipe that can be applied to any problem with a finite Lie algebra. 
\begin{enumerate}
\item Write the Hamiltonian $\hat H(t)$ as 
\begin{equation} \label{chap:decoupling:eq:recipe:Hamiltonian}
\hat H(t) = \sum_j^m G_j(t) \hat H_j \, ,
\end{equation}
and identify the functions $G(t)$ and the Hermitian operators $\hat H_j$. 
\item Identify the Lie algebra $L$ that is generated by the Hamiltonain by commuting the Hamiltonian terms $\hat H_j$ in Eq.~\eqref{chap:decoupling:eq:recipe:Hamiltonian} until they produce a set of $n$ operators that is closed under commutation. For example,  $[\hat H_1 , \hat H_2] \propto \hat H_3$.  Subsequently, $[\hat H_1, \hat H_3] \propto \hat H_4$, and so on, until the algebra is closed. 
\item State the ansatz for the time-evolution operator, 
\begin{equation}
\hat U(t) = \exp[ - i \, F_1(t) \, \hat H_1] \, \exp[ - i \, F_2(t) \, \hat H_2] \cdots \exp[ - i \, F_n (t) \, \hat H_n(t) ] \, ,
\end{equation}
where the $F_j$-coefficients are real, time-dependent functions and the generators $\hat H_j$ are part of the Lie algebra $L$. 
\item Differentiate the ansatz to find
\begin{align} \label{chap:decoupling:eq:recipe:differentiating}
\frac{d}{dt} \hat U(t) =& - i  \, \dot{F}_1 \, \hat H_1 \,  \hat{U}(t)  - i \, \dot{F}_2 \, \hat U_1 \hat H_2  \prod_{j = 2}^n \hat U_j  - i \, F_3 \, \hat U_1 \, \hat U_2 \, \hat H_3 \, \prod_{j = 3}^n \hat U_j  + \ldots \nonumber \\
&- i \, \dot{F}_n \prod_{j = 1}^{n-1} \hat U_j \hat H_n \hat U_n \, .
\end{align}
\item Multiply Eq.~\eqref{chap:decoupling:eq:recipe:differentiating} by $\hat U^{-1}$ on the right-hand side, and set this ansatz equal to the original Hamiltonian $\hat H(t)$:
\begin{equation}
\hat H(t) = \dot{F}_1 \, \hat H_2  + \dot{F}_2 \, \hat U_1 \, \hat H_2 \, \hat U_1^{-1} + \dot{F}_3 \, \hat U_2 \, \hat H_3 \, \hat U_2^{-1} + \ldots \, .
\end{equation}
\item Since the elements $\hat H_j$ of $L$ constitute a basis, use linear independence to construct a set of differential equations, where the solutions for $F_j$ depend on the original Hamiltonian coefficients $G_j(t)$. The equations are given by 
\begin{equation} \label{chap:decoupling:eq:recipe:differential:equation:sum}
\sum_{k = 1}^n G_k (t) =  - i \, \sum_{j = 1}^n \sum_{k = 1}^n \dot{F}_j (t) \, \xi_{kj}  .
\end{equation}
\item Obtain analytic or numerical solutions to Eq.~\eqref{chap:decoupling:eq:recipe:differential:equation:sum} and use them to determine the time-evolution operator $\hat U(t)$. 
\end{enumerate}
In the next section, we demonstrate the  use of this method to solve the dynamics of an optomechanical Hamiltonian with additional terms.

\section{Decoupling an extended nonlinear optomechanical Hamiltonian} \label{chap:decoupling:optomechanical:decoupling}
We now come to the main result of this thesis, which is an exact solution of a nonlinear optomechanical system with additional mechanical terms.  In the field of theoretical optomechanics, these methods have previously enabled the analytic solution to an `mechano-optical' system~\cite{bruschi2018mechano}, where the light--matter coupling term is given by $\left( \hat a^\dag + \hat a \right) \, \hat b^\dag \hat b$ instead of the usual term shown in Eq.~\eqref{chap:introduction:eq:standard:nonlinear:Hamiltonian}.  More generally, similar techniques have been  developed independently and enabled the analytical solution of the dynamics for many different systems, including many-body models with applications to BECs~\cite{fuentes2007family}, as well as models for mode-exchange collisions in two-mode BECs~\cite{barberis2008mode}. The application of Lie algebras has also allowed for the analytical decoupling of two-mode bosonic fields, which were then studied in the context of quantum metrology in a relativistic setting~\cite{Bruschi2013Time}. These are but a small selection of the works where Lie algebra methods have been used to solve the system dynamics. A full literature review of their application is however beyond the scope of this thesis.

\subsection{The extended optomechanical Hamiltonian}
The characteristic light--matter interaction in optomechanical systems is generally taken to be constant. There are, however, specific systems, such as levitated spheres in hybrid-Paul traps, which exhibit a time-dependent coupling~\cite{Millen2015iontrap}. We therefore consider an arbitrary time-dependent light--matter coupling, which we call $\mathcal{G}(t)$. Furthermore, we  include a time-dependent mechanical displacement term $\mathcal{D}_1(t) (\hat b^\dag + \hat b)$, which can be used to describe the effect of linearised forces. Finally, we add a time-dependent mechanical squeezing term $\mathcal{D}_2(t) (\hat b^\dag + \hat b)^2$.  Mechanical squeezing is an effect where one of the phase space quadratures of the mechanical state is narrowed, which could potentially benefit metrology schemes~\cite{wollman2015quantum}. It can be dynamically generated by modulating the mechanical trapping frequency $\omega_{\rm{m}}$~\cite{rashid2016experimental}. See also Appendix~\ref{app:mathieu} for an explicit derivation of this effect. 

The Hamiltonian that is the main object of study in this thesis is given by 
\begin{equation} \label{chap:decoupling:eq:Hamiltonian}
\hat H = \hbar \, \omega_{\rm{c}} \, \hat a^\dag \hat a + \hbar \, \omega_{\rm{m}} \, \hat b^\dag \hat b - \hbar \, \mathcal{G}(t) \, \hat a^\dag \hat a \left( \hat b^\dag + \hat b\right) + \hbar \, \mathcal{D}_1 \, \left( \hat b^\dag + \hat b \right) + \hbar \, \mathcal{D}_2 \, \left( \hat b^\dag + \hat b \right)^2 \, ,
\end{equation}
where $\hat a, \hat a^\dag$ are the annihilation and creation operators of the optical field with oscillation frequency $\omega_{\rm{c}}$, and  $\hat b, \hat b^\dag$ are the annihilation and creation operators of the phonons in the mechanics with oscillation frequency $\omega_{\rm{m}}$. As already mentioned, $\mathcal{G}(t)$ is the light--matter coupling, and $\mathcal{D}_1(t)$ and $\mathcal{D}_2(t)$ are weighting functions multiplying additional mechanical interaction terms. 

In the remainder of this Chapter and this thesis, it will be beneficial to work with dimensionless units. We therefore rescale the laboratory time $t$ by the mechanical frequency, such that $t = \omega_{\rm{m}}\tau$. When $\tau = 2\pi$, the mechanical element has performed one full oscillation. We rescale all other couplings as $\tilde{\mathcal{G}}(\tau) = \mathcal{G}(t)/\omega_{\rm{m}}$. In general, any quantity with a tilde denotes a dimensionless, rescaled quantity. 

The Hamiltonian in Eq.~\eqref{chap:decoupling:eq:Hamiltonian} can therefore be written as 
\begin{align}\label{chap:decoupling:eq:Hamiltonian:rescaled}
\hat{H}/( \hbar \omega_{\rm{m}}) &= \Omega_{\mathrm{c}}\,\hat{a}^\dag\hat{a}+\hat{b}^\dag\hat{b}- \tilde{\mathcal{G}}(\tau)\,\hat a^\dagger\hat a \,\left(\hat b+ \hat b^\dagger\right)+ \tilde{\mathcal{D}}_1(\tau)\,\left(\hat b+ \hat b^\dagger\right)+  \tilde{\mathcal{D}}_2(\tau) \, \left(\hat b+ \hat b^\dagger \right)^2  \, ,
\end{align}
where we have denoted the rescaled optical frequency by $\Omega_{\rm{c}} = \omega_{\rm{c}} / \omega_{\rm{m}}$. We make one more simplification, namely defining the following operators:
\begin{align}\label{chap:decoupling:operator:Lie:algebra}
	 	\hat{N}^2_a &:= (\hat a^\dagger \hat a)^2 \nonumber \\
	\hat{N}_a &:= \hat a^\dagger \hat a &
	\hat{N}_b &:= \hat b^\dagger \hat b \nonumber\\
	\hat{B}_+ &:=  \hat b^\dagger +\hat b &
	\hat{B}_- &:= i\,(\hat b^\dagger -\hat b) &
	 & \nonumber\\
	\hat{B}^{(2)}_+ &:= \hat b^{\dagger2}+\hat b^2 &
	\hat{B}^{(2)}_- &:= i\,(\hat b^{\dagger2}-\hat b^2) \, . &
	 &  
\end{align}
This set of operators will, perhaps not unexpectedly, play a key role in identifying the Lie algebra generated by the Hamiltonian $\hat H(t)$. With these operators, we write $\hat H(t)$ in Eq.~\eqref{chap:decoupling:eq:Hamiltonian:rescaled} in the following compact form
\begin{align}\label{chap:decoupling:eq:Hamiltonian:rescaled:operator:expressions}
\hat {H}/( \hbar \omega_{\rm{m}}) &= \Omega_{\mathrm{c}}\,\hat N_a + \hat N_b- \tilde{\mathcal{G}}(\tau)\,\hat N_a \, \hat B_++ \tilde{\mathcal{D}}_1(\tau)\,\hat B_+ +  \tilde{\mathcal{D}}_2(\tau) \, \hat B_+^2  \, .
\end{align}

\subsection{Time-evolution}

The time evolution of a system with time dependent Hamiltonian $\hat{H}(t)$ is 
\begin{equation}\label{general:time:evolution:operator}
\hat{U}(t)=\overset{\leftarrow}{\mathcal{T}}\,\exp\left[-\frac{i}{\hbar}\int_0^{t} \mathrm{d}t'\,\hat{H}(t')\right],
\end{equation}
where $\overset{\leftarrow}{\mathcal{T}}$ is the time-ordering operator. This expression simplifies dramatically when the Hamiltonian $\hat{H}$ is time independent, in which case one  obtains $\hat U(t)=\exp[-\frac{i}{\hbar}\,\hat{H}\,t]$ as a solution to the time-dependent Schr\"{o}dinger equation. However, we are interested in Hamiltonians with time dependent parameters, as it allows us to consider a number of effects, including resonances. We proceed to find the solution of the decoupled dynamics.

\subsection{Identifying the Lie algebra}

We begin by identifying the Lie algebra that will serve as a basis for decoupling the time-evolution operator  $\hat U(t)$. The operators in the Hamiltonian $\hat H(t)$ in Eq.~\eqref{chap:decoupling:eq:Hamiltonian:rescaled} are $\hat N_a$, $\hat N_b$, $\hat N_a \, \hat B_+$ and $\hat B_+^{(2)}$. We add them to the algebra $L$. 

We must now generate the remaining terms in $L$ through commutation and prove closure of $L$. 
We first note that all terms commute with $\hat N_a$. However, the mechanical terms have non-trivial commutation relations. We start by commuting the phonon number operator $\hat N_b$ with the nonlinear term:
\begin{equation}
\left[ \hat N_b , \hat N_a \, \hat B_+ \right] = \hat N_a \, \hat B_- \,.
\end{equation}
We identify the new term $ \hat N_a \, \hat B_-$ and add it to $L$. This term does however not commute with the original nonlinear term $\hat N_a \, \hat B_+$. When we commute them, we find 
\begin{equation}
\left[ \hat N_a \, \hat B_+,\hat N_a \, \hat B_- \right] = 2 \, i \, \hat N_a^2 \, .
\end{equation}
This operator $\hat N_a^2 = (\hat a^\dag \hat a)^2$ is not a linear combination of any of the other operators, so we add it to $L$. This term specifically encodes the nonlinear nature of the optomechanical system. 

Next, $\hat H(t)$ in Eq.~\eqref{chap:decoupling:eq:Hamiltonian:rescaled:operator:expressions} contains a mechanical displacement term with $ \hat B_+$. We commute it with $\hat N_b$ to find 
\begin{equation}
\left[ \hat N_b,  \hat B_+\right] = i \, \hat B_-.
\end{equation}
We add $\hat B_-$  to $L$, and note that since we already know the outcome of $[\hat N_a \hat B_+, \hat N_a \hat B_-]$, the commutator $[\hat N_a \hat B_+ , \hat B_-]$ yields terms already in $L$. 

We note that the operators $\hat N_a,\hat N_a^2, \hat B_+,\hat B_-,\hat N_a\,\hat B_+,\hat N_a\,\hat B_-$ form a closed subalgebra of $L$. This is a key observation which will simplify the decoupling. The Hamiltonian in Eq.~\eqref{chap:decoupling:eq:Hamiltonian:rescaled:operator:expressions} does however contain an additional squeezing term with $\hat B_+^2 = \hat b^{\dag 2} + \hat b^2 + 2 \hat N_b + 1$. $\hat N_b$ is already in the algebra, so we add $\hat B^{(2)}_+ = \hat b^{\dag 2} + \hat b^2$ to $L$. To ensure the algebra is closed, we commute 
\begin{equation}
[\hat N_b, \hat B_+^{(2)} ] = 2 \, \hat b^{\dag 2} - 2 \, \hat b^2 \, ,
\end{equation}
which prompts us to add $\hat B_-^{(2)} = i \, \left( \hat b^{\dag 2} - \hat b^2\right)$ to $L$. We also note that $\hat N_b$, $\hat B_+^{(2)}$ and $\hat B_-^{(2)}$ form another subalgebra that is closed under commutation. It is also possible to confirm that $\hat B^{(2)}_+ $ and $\hat B_-^{(2)}$ generate only terms already contained in $L$ when commuted with any other term in $L$. 

With this finding, we have identified the Lie algebra that generates the evolution of $\hat H(t)$. It is given by 
\begin{align}\label{chap:decoupling:eq:Lie:algebra}
	 	\hat{N}^2_a &:= (\hat a^\dagger \hat a)^2 \nonumber \\
	\hat{N}_a &:= \hat a^\dagger \hat a &
	\hat{N}_b &:= \hat b^\dagger \hat b \nonumber\\
	\hat{B}_+ &:=  \hat b^\dagger +\hat b &
	\hat{B}_- &:= i\,(\hat b^\dagger -\hat b) &
	 & \nonumber\\
	\hat{B}^{(2)}_+ &:= \hat b^{\dagger2}+\hat b^2 &
	\hat{B}^{(2)}_- &:= i\,(\hat b^{\dagger2}-\hat b^2) &
	 &  \nonumber\\
	\hat{N}_a\,\hat{B}_+ &:= \hat{N}_a\,(\hat b^{\dagger}+\hat b) &
	\hat{N}_a\,\hat{B}_- &:= \hat{N}_a\,i\,(\hat b^{\dagger}-\hat b) \, ,&
	 & 
\end{align}
where the two subalgebras are
\begin{align}\label{chap:decoupling:eq:Lie:algebra:sub:1}
	 	\hat{N}^2_a &:= (\hat a^\dagger \hat a)^2 \nonumber \\
	\hat{N}_a &:= \hat a^\dagger \hat a &
	\hat{N}_b &:= \hat b^\dagger \hat b \nonumber\\
	\hat{B}_+ &:=  \hat b^\dagger +\hat b &
	\hat{B}_- &:= i\,(\hat b^\dagger -\hat b) &
	 & \nonumber\\
	\hat{N}_a\,\hat{B}_+ &:= \hat{N}_a\,(\hat b^{\dagger}+\hat b) &
	\hat{N}_a\,\hat{B}_- &:= \hat{N}_a\,i\,(\hat b^{\dagger}-\hat b) \, ,&
	 & 
\end{align}
and
\begin{align}\label{chap:decoupling:eq:Lie:algebra:sub:2}
	\hat{N}_b &:= \hat b^\dagger \hat b \nonumber\\
	\hat{B}^{(2)}_+ &:= \hat b^{\dagger2}+\hat b^2 &
	\hat{B}^{(2)}_- &:= i\,(\hat b^{\dagger2}-\hat b^2) \, . &
	 &  
\end{align}
We are now ready to apply the Decoupling Theorem.

\subsection{Separation into subalgebras}

While we could proceed with brute force and decouple the full Hamiltonian in Eq.~\eqref{chap:decoupling:eq:Hamiltonian:rescaled}, the many terms in $L$ would give rise to ten coupled differential equations. In an attempt to simplify the problem at hand, we proceed to divide the Hamiltonian into two parts which each correspond to the terms in the subalgebra.  

We separate the Hamiltonian into a non-quadratic part that is generated by the subalgebra in Eq.~\eqref{chap:decoupling:eq:Lie:algebra:sub:1} and a quadratic part that is generated by the other subalgebra in Eq.~\eqref{chap:decoupling:eq:Lie:algebra:sub:2}. It is straight-forward to model the action of the quadratic part on the nonlinear part with the covariance matrix formalism (see in Section~\ref{chap:introduction:sec:covariance:matrix:formalism}). This allows us to separate the Hamiltonian into two distinct parts, in a manner similar to going into the interaction picture, which is often done in quantum field theory. We then apply the Decoupling Theorem separately to the nonlinear and the quadratic part, which shows that the coupled differential equations can be solved separately. 

We begin by rewriting the Hamiltonian $\hat H(t)$ as 
\begin{align}\label{chap:decoupling:eq:main:Hamiltonian}
	\hat {H}=& \Omega_{\mathrm{c}}\,\hat{a}^\dag\hat{a}+ \tilde{\mathcal{D}}_1(\tau)\,\left(\hat b+ \hat b^\dagger\right)- \tilde{\mathcal{G}}(\tau)\,\hat a^\dagger\hat a \,\left(\hat b+ \hat b^\dagger\right)+  2\,\left(\frac{1}{2}+\tilde{\mathcal{D}}_2(\tau)\right)\,\hat{b}^\dag\hat{b}\nonumber \\
	&\quad +\tilde{\mathcal{D}}_2(\tau)\,\left(\hat b^2+ \hat b^{\dagger2} \right).
\end{align}
Then, the time evolution operator Eq.~\eqref{general:time:evolution:operator} also has the alternative form
\begin{align}\label{formal:solution:to:the:decoupling:easier}
\hat{\tilde{U}}(\tau):=e^{-i\,\Omega_\mathrm{c} \hat a^\dagger \hat a\,\tau}\,\hat{\tilde{U}}_{\mathrm{sq}}\,\overset{\leftarrow}{\mathcal{T}}\,\exp\left[-\frac{i}{\hbar}\,\int_0^\tau\,\mathrm{d}\tau'\,\hat{U}_{\mathrm{sq}}^\dag(\tau')\,\hat {H}_1(\tau')\,\hat{U}_{\mathrm{sq}}(\tau')\right],
\end{align}
where we have introduced
\begin{subequations}
\begin{align}
\hat {H}_{\mathrm{sq}}(\tau):=&2\,\left(\frac{1}{2}+\tilde{\mathcal{D}}_2(\tau)\right)\,\hat{b}^\dag\hat{b}+\tilde{\mathcal{D}}_2(\tau)\,\left(\hat b^2+ \hat b^{\dagger2} \right)\, , \label{chap:decoupling:eq:Hamiltonian:squeezing:subsystem}\\
\hat {H}_1(\tau):=&\tilde{\mathcal{D}}_1(\tau)\,\left(\hat b+ \hat b^\dagger\right)- \tilde{\mathcal{G}}(\tau)\,\hat a^\dagger\hat a \,\left(\hat b+ \hat b^\dagger\right)\, , \label{chap:decoupling:free:and:coupling:Hamiltonian}\\
\hat{\tilde{U}}_{\mathrm{sq}}(\tau):=&\overset{\leftarrow}{\mathcal{T}}\,\exp\left[-i\,\int_0^\tau\,\mathrm{d}\tau'\,\hat {H}_{\mathrm{sq}}(\tau')\right].
\end{align}
\end{subequations}
Transition to the expression Eq.~\eqref{formal:solution:to:the:decoupling:easier} is similar to moving from the Heisenberg (or Schr\"odinger) picture to the interaction picture. We now define $\mathbb{X}:=(\hat b, \hat b^\dag)^{\mathrm{T}}$. From standard symplectic (i.e., Bogoliubov) theory we know that 
\begin{align} \label{eq:Bogoliubov:transform}
\mathbb{X}'
=
\hat{\tilde{U}}_{\mathrm{sq}}^\dag
\,\mathbb{X}\,
\hat{\tilde{U}}_{\mathrm{sq}}
=
\begin{pmatrix}
\hat{\tilde{U}}_{\mathrm{sq}}^\dag\,\hat b\,\hat{\tilde{U}}_{\mathrm{sq}} \\
\hat{\tilde{U}}_{\mathrm{sq}}^\dag\,\hat b^\dag\,\hat{\tilde{U}}_{\mathrm{sq}}
\end{pmatrix}
=
\boldsymbol{S}_{\mathrm{sq}}(\tau)\,
\mathbb{X},
\end{align}
where the $2\times2$ symplectic matrix
$\boldsymbol{S}_{\mathrm{sq}}(\tau)$ is the symplectic representation
of $\hat {H}_{\rm{sq}}(\tau)$ 
and satisfies
$\boldsymbol{S}_{\mathrm{sq}}^\dag(\tau)\,\boldsymbol{\Omega}\,\boldsymbol{S}_{\mathrm{sq}}(\tau)=\boldsymbol{\Omega}$. Here
$\boldsymbol{\Omega}=\text{diag}(-i,i)$ is the symplectic form. 

The matrix $\boldsymbol{S}_{\mathrm{sq}}(\tau)$ therefore has the
expression
$\boldsymbol{S}_{\mathrm{sq}}(\tau)=\overset{\leftarrow}{\mathcal{T}}\,\exp[\boldsymbol{\Omega}\,\int_0^\tau\,d\tau'\,\tilde{\boldsymbol{H}}_{\rm{sq}}(\tau')]$. Here
one has that 
\begin{align}
\hat {H}_{\rm{sq}}
=\frac{1}{2}
\mathbb{X}^\dag
\tilde{\boldsymbol{H}}_{\rm{sq}}
\mathbb{X},
\,\,\,\,\,\,\,\,\,\,\,\,\,\,\,\,\,\,\text{with}\,\,\,\,\,\,
\tilde{\boldsymbol{H}}_{\rm{sq}}
=
\begin{pmatrix}
1+2\,\tilde{\mathcal{D}}_2(\tau) & 2\,\tilde{\mathcal{D}}_2(\tau)\\
2\,\tilde{\mathcal{D}}_2(\tau) & 1+2\,\tilde{\mathcal{D}}_2(\tau) 
\end{pmatrix}.
\end{align}
Therefore, we have that 
\begin{align}\label{main:general:expression:time:dependent:squeezing}
\boldsymbol{S}_{\mathrm{sq}}(\tau)=\overset{\leftarrow}{\mathcal{T}}\,\exp\left[-i\,\begin{pmatrix}
1 & 0\\
0 & -1 
\end{pmatrix}\,\int_0^\tau\,\mathrm{d}\tau'\,\begin{pmatrix}
1+2\,\tilde{\mathcal{D}}_2(\tau') & 2\,\tilde{\mathcal{D}}_2(\tau')\\
2\,\tilde{\mathcal{D}}_2(\tau') & 1+2\,\tilde{\mathcal{D}}_2(\tau') 
\end{pmatrix}\right].
\end{align}
It is easy to show that the time independent orthogonal matrix $\boldsymbol{M}_{\text{ort}}$ with expression
\begin{align}
\boldsymbol{M}_{\text{ort}}
=\frac{1}{\sqrt{2}}
\begin{pmatrix}
1 & 1\\
-1 & 1 
\end{pmatrix}\, , 
\end{align}
puts the Hamiltonian matrix $\tilde{\boldsymbol{H}}_s$ in diagonal form through
 \begin{align}
\begin{pmatrix}
1+2\,\tilde{\mathcal{D}}_2(\tau) & 2\,\tilde{\mathcal{D}}_2(\tau)\\
2\,\tilde{\mathcal{D}}_2(\tau) & 1+2\,\tilde{\mathcal{D}}_2(\tau) 
\end{pmatrix}
=
\frac{1}{2}
\begin{pmatrix}
1 & -1\\
1 & 1 
\end{pmatrix}
\begin{pmatrix}
1+4\,\tilde{\mathcal{D}}_2(\tau) & 0\\
0 &1 
\end{pmatrix}
\begin{pmatrix}
1 & 1\\
-1 & 1 
\end{pmatrix}
.
\end{align}
This allows us to manipulate Eq.~\eqref{main:general:expression:time:dependent:squeezing} and find
\begin{align}\label{main:general:expression:time:dependent:squeezing:manipulated}
\boldsymbol{S}_{\mathrm{sq}}(\tau)=
\boldsymbol{M}_{\text{ort}}^{T}\,\overset{\leftarrow}{\mathcal{T}}\,\exp\left[i\,\int_0^\tau\,\mathrm{d}\tau'\,\begin{pmatrix}
0 & 1 \\
1+4\,\tilde{\mathcal{D}}_2(\tau') & 0 
\end{pmatrix}\right]
\,\boldsymbol{M}_{\text{ort}}.
\end{align}
This means that we have 
\begin{align}
\mathbb{X}'=
\boldsymbol{M}_{\text{ort}}^{T}\,\overset{\leftarrow}{\mathcal{T}}\,\exp\left[i\,\int_0^\tau\,\mathrm{d}\tau'\,\begin{pmatrix}
0 & 1 \\
1+4\,\tilde{\mathcal{D}}_2(\tau') & 0 
\end{pmatrix}\right]
\,\boldsymbol{M}_{\text{ort}}\,
\mathbb{X}.
\end{align}
We introduce the new vector $\mathbb{Y}:=\boldsymbol{M}_{\text{ort}}\,\mathbb{X}$, which is just a rotation of the operators. Then we have 
\begin{align}\label{matrix:time:ordered:exponential}
\mathbb{Y}'=
\overset{\leftarrow}{\mathcal{T}}\,\exp\left[i\,\int_0^\tau\,\mathrm{d}\tau'\,\begin{pmatrix}
0 & 1 \\
1+4\,\tilde{\mathcal{D}}_2(\tau') & 0 
\end{pmatrix}\right]
\,\mathbb{Y}.
\end{align}
We are now in a position where we can solve the time-evolution of the quadratic system and use the results in the decoupling of the nonlinear part.

\subsection{Solving the matrix time-ordered exponential}
Here we seek a formal expression for Eq.~\eqref{matrix:time:ordered:exponential}.
We start by noticing that, if we wrote down the time ordered exponential we would be able to write
\begin{align}\label{initial:useful:conditions}
\overset{\leftarrow}{\mathcal{T}}\,\exp\left[i\,\int_0^\tau\,\mathrm{d}\tau'\,\begin{pmatrix}
0 & 1 \\
1+4\,\tilde{\mathcal{D}}_2(\tau') & 0 
\end{pmatrix}\right]=\boldsymbol{P}+i\,\int_0^\tau \mathrm{d}\tau'\,\boldsymbol{K}\,\boldsymbol{P},
\end{align}
in terms of the matrix $\boldsymbol{K}$ defined as
\begin{align}\label{matrix:time:ordered:exponential:explicit}
\boldsymbol{K}:=
\begin{pmatrix}
0 & 1 \\
1+4\,\tilde{\mathcal{D}}_2(\tau') & 0 
\end{pmatrix},
\end{align}
and the matrix $\boldsymbol{P}$, which we will determine and which is
diagonal. This follows from the fact that the matrix on the left-hand
side of Eq.~\eqref{initial:useful:conditions} is diagonal when squared,
and therefore any even powers in the expansion will be diagonal. We
use the fact that 
\begin{align}
\frac{d}{d\tau}\overset{\leftarrow}{\mathcal{T}}\,\exp\left[i\,\int_0^\tau\,\mathrm{d}\tau'\,\boldsymbol{K}(\tau')\right]
=
i\,\boldsymbol{K}\,
\overset{\leftarrow}{\mathcal{T}}\,\exp\left[i\,\int_0^\tau\,\mathrm{d}\tau'\,\boldsymbol{K}(\tau')\right]\, , 
\end{align}
to find the equation
\begin{align}
-\boldsymbol{K}\,\int_0^\tau \mathrm{d}\tau'\,\boldsymbol{K}\,\boldsymbol{P}=\dot{\boldsymbol{P}}.
\end{align}
Since $\boldsymbol{K}$ is invertible, we manipulate this equation and obtain, after some algebra,
\begin{align}\label{differential:equation:for:matrix}
\ddot{\boldsymbol{P}}- 4\,\frac{\dot{\tilde{\mathcal{D}}}_2(\tau)}{1 \,+4\,\tilde{\mathcal{D}}_2(\tau)}
\begin{pmatrix}
0 &  0\\
0 & 1 
\end{pmatrix}
\dot{\boldsymbol{P}}
+(1+4\,\tilde{\mathcal{D}}_2(\tau))\,\boldsymbol{P}=0.
\end{align}
We can now solve the four differential equations contained in Eq.~\eqref{differential:equation:for:matrix} which read
\begin{align}\label{chap:decoupling:differential:equation:written:down}
\ddot{P}_{11}+(1+4\,\tilde{\mathcal{D}}_2(\tau))\,P_{11}= \, &0 \, ,\nonumber\\
P_{12}=P_{21}=& \, 0\, ,\nonumber\\
\ddot{P}_{22}-4\frac{\dot{\tilde{\mathcal{D}}}_2(\tau)}{1+4\,\tilde{\mathcal{D}}_2(\tau)}\,\dot{P}_{22}+(1+4\,\tilde{\mathcal{D}}_2(\tau))\,P_{22}= & \, 0 \, .
\end{align}
The differential equations in Eq.~\eqref{chap:decoupling:differential:equation:written:down} must be supplemented by initial conditions. We note that since the left hand side of Eq.~\eqref{initial:useful:conditions} is the identity matrix for $\tau=0$ we have that $\boldsymbol{P}(0)=\mathds{1}$ which implies $P_{11}(0)=P_{22}(0)=1$. In addition, taking the time derivative of both sides of Eq.~\eqref{initial:useful:conditions} and setting $\tau=0$ implies $\dot{P}_{11}(0)=\dot{P}_{22}(0)=0$.

By introducing the integral $I_{P_{22}} = \int^\tau_0 \mathrm{d} \tau' P_{22} (\tau')$, one can rewrite the second equation as 
\begin{equation} \label{chap:decoupling:eq:IP22}
\ddot{I}_{P_{22}}  + ( 1 + 4 \, \tilde{\mathcal{D}}_2 (\tau) ) I_{P_{22}} = 0 \, ,
\end{equation}
so that it becomes the same as that for $P_{11}$. The boundary conditions are now $I_{P_{22}}(0) = 0$ and $\dot{I}_{P_{22}} = 1$. We were not able to find a general solution to the differential equation for $P_{11}$ Eq.~\eqref{chap:decoupling:differential:equation:written:down} and the equation for $I_{P_{22}}$ Eq.~\eqref{chap:decoupling:eq:IP22}, but they can be integrated numerically when an explicit form of $\tilde{\mathcal{D}}_2(\tau)$ is given. For specific choices of $\tilde{\mathcal{D}}_2(\tau)$, which we discuss in the main text of this work, the equations become the well-studied Mathieu equation. We derive perturbative solutions for the Mathieu equation in Section~\ref{app:mathieu:perturbative:solutions} in Appendix~\ref{app:mathieu}. 

This has allowed us to find
\begin{align}
\overset{\leftarrow}{\mathcal{T}}\,\exp\left[i\,\int_0^\tau\,\mathrm{d}\tau'\,\begin{pmatrix}
0 & 1 \\
1+4\,\tilde{\mathcal{D}}_2(\tau') & 0 
\end{pmatrix}\right]=&\boldsymbol{P}+i\,\int_0^\tau \mathrm{d}\tau'\,\boldsymbol{K}\,\boldsymbol{P} \nonumber \\
=& 
\begin{pmatrix}
P_{11} &  i\,\int_0^\tau\,\mathrm{d}\tau'\,P_{22}\\
i\,\int_0^\tau\,\mathrm{d}\tau'\,(1+4\,\tilde{\mathcal{D}}_2(\tau'))\,P_{11} & P_{22} 
\end{pmatrix}.
\end{align}
In turn, this allows us to get
\begin{align} 
\boldsymbol{S}_{\mathrm{sq}}(\tau) &=\begin{pmatrix}
\alpha(\tau) & \beta(\tau) \\
\beta^*(\tau) & \alpha^*(\tau) 
\end{pmatrix}
=\boldsymbol{M}_{\text{ort}}^{T}\,
\begin{pmatrix}
P_{11} & i\,\int_0^\tau\,\mathrm{d}\tau'\,P_{22}\\
i\,\int_0^\tau\,\mathrm{d}\tau'\,(1+4\,\tilde{\mathcal{D}}_2(\tau'))\,P_{11} & P_{22} 
\end{pmatrix}\,
\boldsymbol{M}_{\text{ort}},
\end{align}
where we have introduced the Bogoliubov matrix $\boldsymbol{S}_{\mathrm{sq}}(\tau)$ with coefficients $\alpha(\tau)$ and $\beta(\tau)$.
After a little algebra, we write the Bogoliubov coefficients as 
\begin{align} \label{chap:decoupling:eq:bogoliubov:coeffs:expression}
\alpha(\tau)=&\frac{1}{2}\,\left[P_{11}+P_{22}-i\,\int_0^\tau\,\mathrm{d}\tau'\,P_{22}-i\,\int_0^\tau\,\mathrm{d}\tau'\,(1+4\,\tilde{\mathcal{D}}_2(\tau'))\,P_{11}\right] \, , \nonumber\\
\beta(\tau) =&\frac{1}{2}\,\left[P_{11}-P_{22}+i\,\int_0^\tau\,\mathrm{d}\tau'\,P_{22}-i\,\int_0^\tau\,\mathrm{d}\tau'\,(1+4\,\tilde{\mathcal{D}}_2(\tau'))\,P_{11}\right] \, .
\end{align}
These equations can also be rewritten in terms of $I_{P_{22}}$ to find
\begin{align} \label{app:eq:bogoliubov:expressions}
\alpha(\tau) =& \frac{1}{2}\,\left[ P_{11} - i I_{P_{22}} + i\frac{d}{d \tau } ( P_{11}  - i I_{P_{22}} )\right] \, ,\nonumber\\
\beta(\tau) =& \frac{1}{2}\,\left[ P_{11} + i I_{P_{22}} + i\frac{d}{d\tau} ( P_{11}  + i I_{P_{22}} )\right].
\end{align}
This means that we have
\begin{align} \label{eq:Bogoliubov:transform:solved}
\mathbb{X}'=
\hat{\tilde{U}}_{\mathrm{sq}}^\dag
\,\mathbb{X}\,
\hat{\tilde{U}}_{\mathrm{sq}}
=
\begin{pmatrix}
\alpha(\tau) & \beta(\tau) \\
\beta^*(\tau) & \alpha^*(\tau) 
\end{pmatrix}
\,\mathbb{X} \, .
\end{align}
The Bogoliubov (symplectic) identities $|\alpha(t)|^2-|\beta(t)|^2=1$ read
\begin{align}
P_{11}\,P_{22}+\left(\int_0^\tau\,\mathrm{d}\tau'\,P_{22}\right)\,\left(\int_0^\tau\,\mathrm{d}\tau'\,(1+4\,\tilde{\mathcal{D}}_2(\tau'))\,P_{11}\right)=1 \, .
\end{align}
We can now go back to the time evolution operator Eq. 
\eqref{formal:solution:to:the:decoupling:easier} which we reprint here
\begin{align}
\hat{\tilde{U}}(\tau)=e^{-i\,\Omega_\textrm{c} \hat N_a\,\tau}\,\hat{\tilde{U}}_{\mathrm{sq}}(\tau)\,\overset{\leftarrow}{\mathcal{T}}\,\exp\left[-\frac{i}{\hbar}\,\int_0^\tau\,\mathrm{d}\tau'\,\hat{\tilde{U}}_{\mathrm{sq}}^\dag(\tau')\,\hat {H}_1(\tau')\,\hat{\tilde{U}}_{\mathrm{sq}}(\tau')\right] \, .
\end{align}
Our work above allows to obtain
\begin{align}
\hat{\tilde{U}}(\tau)=e^{-i\,\Omega_\textrm{c} \hat N_a\,\tau}\,\hat{\tilde{U}}_{\mathrm{sq}}(\tau)\,\overset{\leftarrow}{\mathcal{T}}\,&\exp\biggl[-i\,\int_0^\tau\,\mathrm{d}\tau'\,\bigl(\tilde{\mathcal{D}}_1(\tau') \bigl(\xi(\tau')\,\hat b+ \xi^*(\tau')\,\hat b{}^\dagger\bigr) \nonumber \\
&\quad\quad\quad\quad\quad\quad-\tilde{\mathcal{G}}(\tau') \,  \hat N_a\,\bigl(\xi(\tau')\,\hat b+ \xi^*(\tau')\,\hat b{}^\dagger\bigr)\bigr)\biggr] \, ,
\end{align}
which can be conveniently rewritten as 
{\small\begin{align}\label{almost:final:time:evolution:operator}
\hat{\tilde{U}}(t)&=e^{-i\,\Omega_\textrm{c}  \hat N_a\,\tau}\,\hat{\tilde{U}}_{\mathrm{sq}}\, \nonumber \\
&\times\overset{\leftarrow}{\mathcal{T}}\,\exp\biggl[-i\,\int_0^\tau\,\mathrm{d}\tau'\,\biggl(\tilde{\mathcal{D}}_1(\tau') \Re\xi(\tau')\,\hat B_+-i \, \tilde{\mathcal{D}}_1(\tau')\,\Im\xi(\tau')\,\hat B_- \nonumber \\
&\quad\quad\quad\quad\quad\quad\quad\quad\quad\quad\quad-\tilde{\mathcal{G}}(\tau') \,\Re\xi(\tau')\,\hat \hat N_a\,\hat B_++i \, \tilde{\mathcal{G}}(\tau') \,\Im\xi(\tau')\,\hat N_a\,\hat B_-\biggr)\biggr] \, .
\end{align}}
Here we have introduced 
\begin{align} \label{chap:decoupling:eq:definition:of:xi}
\xi(\tau):=&\alpha(\tau)+\beta^*(\tau)=P_{11}-i\,\int_0^\tau\,\mathrm{d}\tau'\,P_{22} \, .
\end{align}
for conveniency of presentation. We also find that 
\begin{align} \label{app:eq:xi:alpha:beta:relation}
\alpha(\tau) &= \frac{1}{2}(\xi(\tau) + i \dot{\xi}(\tau)) &\mbox{and} &&\beta(\tau) = \frac{1}{2}(\xi^*(\tau) + i \dot{\xi}^*(\tau)) \, .
\end{align}
This quantity encodes the evolution induced by the quadratic part. 

\subsection{Decoupling the nonlinear Hamiltonian}
Now that we have described the action of the quadratic Hamiltonian on the nonlinear Hamiltonian, we proceed with decoupling the nonlinear Hamiltonian.  We recognise that the functions $G_j$ identified in Eq.~\eqref{chap:decoupling:eq:hamiltonian:sum} consist of the coefficients $\tilde{\mathcal{G}}(\tau), $ $\tilde{\mathcal{D}}_1(\tau)$ and $\tilde{\mathcal{D}}_2(\tau)$. 

The remaining part of the operator Eq. 
\eqref{almost:final:time:evolution:operator}  reads 
\begin{align}
\overset{\leftarrow}{\mathcal{T}}\,&\exp\biggl[-i\,\int_0^\tau\,d\tau'\,\biggl(\tilde{\mathcal{D}}_1(\tau') \Re\xi(\tau')\,\hat B_+-\tilde{\mathcal{D}}_1(\tau')\,\Im\xi(\tau')\,\hat B_- \nonumber \\
&\quad\quad\quad\quad\quad\quad\quad\quad\quad\quad-\tilde{\mathcal{G}}(\tau') \,\Re\xi(\tau')\,\hat N_a\,\hat B_++\tilde{\mathcal{G}}(\tau') \,\Im\xi(\tau')\,\hat N_a\,\hat B_-\biggr)\biggr] \, .
\end{align}
We now make the ansatz
\begin{align}\label{decoupling:ansatz}
\overset{\leftarrow}{\mathcal{T}}\,&\exp\biggl[-i\,\int_0^\tau\,\mathrm{d}\tau'\,\biggl(\tilde{\mathcal{D}}_1(\tau') \Re\xi(\tau')\,\hat B_+-\tilde{\mathcal{D}}_1(\tau')\,\Im\xi(\tau')\,\hat B_- \nonumber \\
&\quad\quad\quad\quad\quad\quad\quad\quad\quad\quad-\tilde{\mathcal{G}}(\tau') \,\Re\xi(\tau')\, \hat N_a\,\hat B_++\tilde{\mathcal{G}}(\tau') \,\Im\xi(\tau')\,\hat N_a\,\hat B_-\biggr)\biggr]\nonumber\\
=&e^{-i\,F_{\hat{N}_a}\,\hat{N}_a}\,e^{-i\,F_{\hat{N}^2_a}\,\hat{N}^2_a}\,e^{-i\,F_{\hat{B}_+}\,\hat{B}_+}\,e^{-i\,F_{\hat{N}_a\,\hat{B}_+}\,\hat{N}_a\,\hat{B}_+}\,e^{-i\,F_{\hat{B}_-}\,\hat{B}_-}\,e^{-i\,F_{\hat{N}_a\,\hat{B}_-}\,\hat{N}_a\,\hat{B}_-} \, .
\end{align}
Taking the time derivative on both sides and find
\begin{align}\label{main:docupling:Hamiltonian:technique:long}
&\tilde{\mathcal{D}}_1(\tau) \Re\xi(\tau)\,\hat B_+-\tilde{\mathcal{D}}_1(\tau')\,\Im\xi(\tau)\,\hat B_--\tilde{\mathcal{G}}(\tau) \,\Re\xi(\tau)\,\hat N_a\,\hat B_++\tilde{\mathcal{G}}(\tau) \,\Im\xi(\tau)\,\hat N_a\,\hat B_-\nonumber\\
=\, &\dot F_{\hat{N}_a}\,\hat{N}_a
+\dot F_{\hat{N}^2_a}\,\hat{N}^2_a
+ \dot F_{\hat{B}_+}\,\hat{B}_+
+ \dot F_{\hat{N}_a\,\hat{B}_+}\,\hat{N}_a\,\hat{B}_+\nonumber\\
&+ \dot F_{\hat{B}_-}\,e^{-i\,F_{\hat{B}_+}\,\hat{B}_+}\,e^{-i\,F_{\hat{N}_a\,\hat{B}_+}\,\hat{N}_a\,\hat{B}_+}\,
\hat{B}_-
\,e^{i\,F_{\hat{N}_a\,\hat{B}_+}\,\hat{N}_a\,\hat{B}_+}\,e^{i\,F_{\hat{B}_+}\,\hat{B}_+}\nonumber\\
&+ \dot F_{\hat{N}_a\,\hat{B}_-}\,e^{-i\,F_{\hat{B}_+}\,\hat{B}_+}\,e^{-i\,F_{\hat{N}_a\,\hat{B}_+}\,\hat{N}_a\,\hat{B}_+}\,
\hat{N}_a\,\hat{B}_-
\,e^{i\,F_{\hat{N}_a\,\hat{B}_+}\,\hat{N}_a\,\hat{B}_+}\,e^{i\,F_{\hat{B}_+}\,\hat{B}_+}\nonumber\\
=\, &\dot F_{\hat{N}_a}\,\hat{N}_a
+\dot F_{\hat{N}^2_a}\,\hat{N}^2_a
+ \dot F_{\hat{B}_+}\,\hat{B}_+
+ \dot F_{\hat{N}_a\,\hat{B}_+}\,\hat{N}_a\,\hat{B}_+\nonumber\\
&+ (\dot F_{\hat{B}_-}+\dot F_{\hat{N}_a\,\hat{B}_-}\,\hat{N}_a)\,e^{-i\,F_{\hat{B}_+}\,\hat{B}_+}\,e^{-i\,F_{\hat{N}_a\,\hat{B}_+}\,\hat{N}_a\,\hat{B}_+}\,
\hat{B}_-
\,e^{i\,F_{\hat{N}_a\,\hat{B}_+}\,\hat{N}_a\,\hat{B}_+}\,e^{i\,F_{\hat{B}_+}\,\hat{B}_+}\nonumber\\
=\, &\dot F_{\hat{N}_a}\,\hat{N}_a
+\dot F_{\hat{N}^2_a}\,\hat{N}^2_a+ \dot F_{\hat{B}_+}\,\hat{B}_+
+ \dot F_{\hat{N}_a\,\hat{B}_+}\,\hat{N}_a\,\hat{B}_+\nonumber \\
&+ (\dot F_{\hat{B}_-} +\dot F_{\hat{N}_a\,\hat{B}_-}\,\hat{N}_a)\,(\hat{B}_-+2\, F_{\hat{B}_+}+2\, F_{\hat{N}_a\,\hat{B}_+}\,\hat{N}_a)\, .
\end{align}
Therefore our main differential equations can be obtained by equating the coefficient of the different, linearly independent operators of the Lie algebra, in accordance with  Eq.~\eqref{chap:decoupling:eq:differential:equation:sum}. We find
\begin{align}\label{main:docupling:Hamiltonian:technique:simplified}
&\tilde{\mathcal{D}}_1(\tau) \Re\xi(\tau)\,\hat B_+-\tilde{\mathcal{D}}_1(\tau')\,\Im\xi(\tau)\,\hat B_--\tilde{\mathcal{G}}(\tau) \,\Re\xi(\tau)\,\hat N_a\,\hat B_++\tilde{\mathcal{G}}(\tau) \,\Im\xi(\tau)\,\hat N_a\,\hat B_-\nonumber\\
=\, &\dot F_{\hat{N}_a}\,\hat{N}_a
+\dot F_{\hat{N}^2_a}\,\hat{N}^2_a
+ \dot F_{\hat{B}_+}\,\hat{B}_+
+ \dot F_{\hat{N}_a\,\hat{B}_+}\,\hat{N}_a\,\hat{B}_+\nonumber \\
&+ (\dot F_{\hat{B}_-}+\dot F_{\hat{N}_a\,\hat{B}_-}\,\hat{N}_a)\,(\hat{B}_-+2\, F_{\hat{B}_+}+2\, F_{\hat{N}_a\,\hat{B}_+}\,\hat{N}_a) \, .
\end{align}
The solutions with operators proportional to $\hat B_\pm$ can be independently solved. However, the solutions for  $F_{\hat N_a}$, $F_{\hat N_a^2} $, and $F_{\hat N_a \, \hat B_\pm}$ are less straight forward. We find the following four coupled differential equations:
\begin{align}\label{eq:diff:eqs:explicit}
\dot{F}_{\hat N_a}  &= - 2  \, \dot{F}_{\hat B_-} \, F_{\hat N_a \, \hat B_+} - 2 \, F_{\hat B_+} \, \dot{F}_{\hat N_a \, \hat B_-} \, , \nonumber \\
\dot{F}_{\hat N_a^2} &= - 2 \, \dot{F}_{\hat N_a \, \hat B_-} \, F_{\hat N_a \, \hat B_+} \, ,  \nonumber \\
\dot{F}_{\hat N_a \, \hat B_+} &=- \tilde{\mathcal{G}} ( \tau) \, \Re \xi(\tau) \, , \nonumber \\
\dot{F}_{\hat N_a \, \hat B_-} &= \tilde{\mathcal{G}}  \Im \xi(\tau) \, .
\end{align}
By first solving the equations for $\dot{F}_{\hat N_a \, \hat B_\pm}$ and $\dot{F}_{\hat B_\pm}$, it is then possible to insert the solutions into the expressions for $\dot{F}_{\hat N_a} $ and $\dot{F}_{\hat N_a^2}$. We find the following key expression for this work
\begin{align}\label{chap:decoupling:eq:sub:algebra:decoupling:solution}
F_{\hat{N}_a}(\tau)=& -2 \,\int_0^\tau\,\mathrm{d}\tau'\,\tilde{\mathcal{D}}_1(\tau')\,\Im\xi(\tau')\int_0^{\tau'}\mathrm{d}\tau''\,\tilde{\mathcal{G}}(\tau'')\,\Re\xi(\tau'')\, ,  \nonumber \\
&-2 \, \int^\tau_0\,\mathrm{d}\tau' \,\tilde{\mathcal{G}}(\tau')\, \Im \xi(\tau') \, \int^{\tau'}_0 \,\mathrm{d}\tau''\, \tilde{\mathcal{D}}_1(\tau'') \, \Re \xi(\tau'') \, ,  \nonumber\\
F_{\hat{N}^2_a}(\tau)= &  \, 2 \,\int_0^\tau\,\mathrm{d}\tau'\,\tilde{\mathcal{G}}(\tau')\,\Im\xi(\tau')\int_0^{\tau'}\mathrm{d}\tau''\,\tilde{\mathcal{G}}(\tau'')\,\Re\xi(\tau'')\, , \nonumber\\
F_{\hat{B}_+}(\tau)=& \int_0^\tau\,\mathrm{d}\tau'\,\tilde{\mathcal{D}}_1(\tau')\,\Re\xi(\tau')\, , \nonumber\\
F_{\hat{B}_-}(\tau)=&- \int_0^\tau\,\mathrm{d}\tau'\, \tilde{\mathcal{D}}_1(\tau')\,\Im\xi(\tau')\, , \nonumber\\
F_{\hat{N}_a\,\hat{B}_+}(\tau)=&- \int_0^\tau\,\mathrm{d}\tau'\,\tilde{\mathcal{G}}(\tau')\,\Re\xi(\tau')\, , \nonumber\\
F_{\hat{N}_a\,\hat{B}_-}(\tau)=&\int_0^\tau\,\mathrm{d}\tau'\,\tilde{\mathcal{G}}(\tau')\,\Im\xi(\tau') \, .
\end{align}
This result concludes the decoupling part of our work. The expressions Eq.~\eqref{chap:decoupling:eq:sub:algebra:decoupling:solution}, together with the decoupling form Eq.~\eqref{decoupling:ansatz}, can be used in the  expression for $\hat{U}(\tau)$ in Eq.~\eqref{formal:solution:to:the:decoupling:easier} to obtain an explicit (up to a formal solution for $\xi(t)$) time-evolved expression for the observables of the system. 
Let us once more write down the full expression of the time-evolution operator
\begin{align} \label{chap:decoupling:eq:final:evolution:operator}
\hat U(\tau)=& \, e^{-i\,\Omega_\mathrm{c} \hat N_a\,\tau}\,\hat{U}_{\mathrm{sq}}(\tau)\,e^{-i\,F_{\hat{N}_a}\,\hat{N}_a}\,e^{-i\,F_{\hat{N}^2_a}\,\hat{N}^2_a}\,e^{-i\,F_{\hat{B}_+}\,\hat{B}_+}\,e^{-i\,F_{\hat{N}_a\,\hat{B}_+}\,\hat{N}_a\,\hat{B}_+}\,\nonumber \\
&\times e^{-i\,F_{\hat{B}_-}\,\hat{B}_-}\,e^{-i\,F_{\hat{N}_a\,\hat{B}_-}\,\hat{N}_a\,\hat{B}_-} \, , 
\end{align}
to be complemented with the functions listed in Eq.~\eqref{chap:decoupling:eq:sub:algebra:decoupling:solution}. 

The form of $\hat U(\tau)$ in Eq.~\eqref{chap:decoupling:eq:final:evolution:operator} constitutes one of the major results in this thesis, and we use this expression in Chapters~\ref{chap:non:Gaussianity:coupling} and~\ref{chap:non:Gaussianity:squeezing} to compute the non-Gaussianity of an optomechanical state. 

\subsection{Decoupling the quadratic mechanical subsystem}
The mechanical subsystem operator $\hat{\tilde{U}}_{\rm{sq}}$ has thus far been left out since we can easily model the evolution of the first and second moments by using the Bogoliubov coefficients that it generates. This suffices for our treatment of the non-Gaussianity of an optomechanical state in Chapters~\ref{chap:non:Gaussianity:coupling} and~\ref{chap:non:Gaussianity:squeezing}. 

For other applications, however, such as the sensing schemes we consider in Chapter~\ref{chap:metrology}, it is beneficial to determine the action of this operator on a state. Specifically for the purpose of metrology, we will later see how measurement of parameters in the squeezing function $\tilde{\mathcal{D}}_2(\tau)$ requires computing the derivatives of $\hat{\tilde{U}}_{\rm{sq}}$. 

We therefore proceed to apply the Decoupling Theorem to $\hat{\tilde{U}}_{\rm{sq}}$ directly. The operator $\tilde{\hat{U}}_{\rm{sq}}$ is given by 
\begin{align}\label{chap:decoupling:eq:Usq:to:be:decoupled}
\hat{\tilde{U}}_{\rm{sq}} &= \overleftarrow{T} \exp\biggl[ - i \int^\tau_0 \mathrm{d}\tau'\left(\,(1 \,+2 \, \tilde{\mathcal{D}}_2(\tau'))\hat{N}_b+ \tilde{\mathcal{D}}_2(\tau')\hat{B}^{(2)}_+\right) \biggr] \, .
\end{align}
We wish to find an analytic expression composed of operators that we can treat individually. We make the following ansatz:
\begin{align} \label{chap:decoupling:eq:Usq:decoupled}
\hat{\tilde U}_{\rm{sq}}= \exp[-i  \, J_b  \, \hat{N}_b]\,\exp[-i \,  J_+ \hat B_+^{(2)}]\,\exp[ - i  \, J_- \, \hat B_- ^{(2)} ] \,, 
\end{align}
where $J_b$ and $J_\pm$ are real, time-dependent functions and the operators are the generators which are part of the Lie algebra. 

We then  differentiate the ansatz with respect to time $\tau$ to obtain 
\begin{align}  \label{app:subsystem:decoupling:differentiated}
\dot{\hat{\tilde{U}}}_{\rm{sq}} \, \hat{\tilde{U}}_{\rm{sq}}^\dag =&  - i \,  \dot{J}_\theta  - i  \, \dot{J}
_+ e^{ - i \, J_b  \, \hat{N}_b } \, \hat B_+^{(2)} \, e^{ i  \, J_b \, \hat{N}_b}  \nonumber \\
&- i \, \dot{J}_- \, e^{ - i \,  J_b \, \hat{N}_b } \,e^{ - i J_+ \hat B_+^{(2)} } \hat B_-^{(2)} \, e^{i J_+ \, \hat B_+^{(2)}} \, e^{i  \, J_b \, \hat{N}_b}\, .
\end{align}
By using the commutator relations in Eq.~\eqref{app:commutators:eq:commutators:1} and Eq.~\eqref{app:commutators:eq:commutators:2} in Appendix~\ref{app:commutators}, Eq.~\eqref{app:subsystem:decoupling:differentiated} can be written purely as terms proportional to the operators $\hat N_b$, $\hat B_+^{(2)}$ and $\hat B_-^{(2)}$: 
\begin{align}
\dot{\hat{\tilde{U}}}_{\rm{sq}} \, \hat{\tilde{U}}_{\rm{sq}}^\dag =& \,   - i  \, \dot{J}_\theta \, \hat{N}_b - i \dot{J}_+  \left(   \cos(2 J_b) \hat B_+^{(2)} - \sin (2 J_b) \hat B_-^{(2)} \right)  \nonumber \\
&- i \dot{J}_- \, \left[ \cosh(4\,J_+)\,\left( \cos(2 J_b)  \, \hat B_-^{(2)} + \sin(2 J_b) \hat B_+^{(2)} \right) + 2\,\sinh(4\,J_+)\,\hat{N}_b-4\,J_+\right]\, .
\end{align}
Now we set this equal to the expression under the integral Eq.~\eqref{chap:decoupling:eq:Usq:to:be:decoupled},
\begin{align}
\,(1 \,+2 \, \tilde{\mathcal{D}}_2(\tau))\hat{N}_b+ \tilde{\mathcal{D}}_2(\tau)\hat{B}^{(2)}_+ =& \dot{J}_\theta \, \hat N_b +  \dot{J}_+  \left(   \cos(2 J_b) \hat B_+^{(2)} -  \sin (2 J_b) \hat B_-^{(2)} \right)  \nonumber \\
&+  \dot{J}_- \, \biggl[ \cosh(4\,J_+)\,\left( \cos(2 J_b)  \, \hat B_-^{(2)} +  \sin(2 J_b) \hat B_+^{(2)} \right) \nonumber \\
&\quad\quad\quad\quad+ 2\,\sinh(4\,J_+)\,\hat N_b-4\,J_+ \biggr] \, .
\end{align}
We then use the linear independence of the operators in order to write down the following differential equations
\begin{align}
(1 \,+2 \, \tilde{\mathcal{D}}_2(\tau)) &=  \dot{J}_b  + 2\,  \, \dot{J}_- \, \sinh(4\,J_+) \, , \nonumber \\
\tilde{\mathcal{D}}_2 ( \tau) &= \dot{J}_+\cos(2 J_b) + \dot{J}_- \, \cosh(4\,J_+)\,\sin(2 J_b) \, ,  \nonumber \\
0 &= - \dot{J}_+ \sin(2 J_b) + \dot{J}_- \, \cosh(4 J_+) \cos( 2 J_b)  \, , 
\end{align}
which can be simplified into the following first-order coupled differential equations: 
\begin{align} \label{chap:decoupling:eq:diff:equations:Js}
\dot{J}_b &=  1 + 2\, \tilde{\mathcal{D}}_2(\tau) \,  \left( 1 - \sin(2 J_b) \tanh(4 J_+) \right)\,,  \nonumber \\
\dot{J}_+ &=    \tilde{\mathcal{D}}_2(\tau) \,  \cos(2 J_b) \, , \nonumber \\
\dot{J}_-  &= \tilde{\mathcal{D}}_2(\tau) \,  \frac{\sin(2 J_b)}{\cosh(4 J_+)} \, .
\end{align}
These equations do not in general allow for analytic solutions. In the main text, we proceed with estimations of parameters in $\tilde{D}_2(\tau)$ by evaluating these equations numerically.  

With these coefficients, we can write the time-evolution operator $\hat U(t)$ in Eq.~\eqref{chap:decoupling:eq:final:evolution:operator} in the following fully decoupled form:
\begin{align} \label{chap:decoupling:eq:final:evolution:operator:J:coefficients}
\hat U(\tau)=& \, e^{-i\,\Omega_\mathrm{c} \hat N_a\,\tau}\, e^{ - i \, J_b \, \hat N_b} \, e^{- i \, J_+ \, \hat B_+^{(2)}} \, e^{- i \, J_- \, \hat B_-^{(2)}} 
\,e^{-i\,F_{\hat{N}_a}\,\hat{N}_a}\,e^{-i\,F_{\hat{N}^2_a}\,\hat{N}^2_a}\,\nonumber \\
&\times e^{-i\,F_{\hat{B}_+}\,\hat{B}_+}\,e^{-i\,F_{\hat{N}_a\,\hat{B}_+}\,\hat{N}_a\,\hat{B}_+}\,e^{-i\,F_{\hat{B}_-}\,\hat{B}_-}\,e^{-i\,F_{\hat{N}_a\,\hat{B}_-}\,\hat{N}_a\,\hat{B}_-} , .
\end{align}
We have a few more relations to explore, which are useful to future explorations. 

%
\subsection{Link between the $J$-coefficients the Bogoliubov coefficients}\label{chap:decoupling:link:J:Bogoliubov}
The $J$-coefficients in Eq.~\eqref{chap:decoupling:eq:diff:equations:Js} are related to the Bogoliubov coefficients in Eq.~\eqref{chap:decoupling:eq:bogoliubov:coeffs:expression}, which describe the action of the squeezing subsystem operator $\hat{\tilde{U}}_{\rm{sq}}(\tau)$. It is beneficial to know the relationship between them, so that the solution to the differential equations  in Eq.~\eqref{chap:decoupling:differential:equation:written:down} and~\ref{chap:decoupling:eq:IP22}, which can be used to compute the Bogoliubov coefficients, can then also be used to compute the $J$-coefficients. 

To determine the relationship, we start by examining $\hat{\tilde{U}}_{\rm{sq}}$. Its action on a state should amount to applying the transformation with the Bogoliubov coefficients. In its decoupled form in Eq.~\eqref{chap:decoupling:eq:Usq:decoupled}, we  it contains a rotation $e^{- i \, J_b\, \hat N_b}$ and two squeezing operations $e^{- i \, J_+ \, \hat B_+^{(2)}}$ and $e^{ - i \, \hat B_-^{(2)}}$. If we know the symplectic form of this operator in terms of $J_b$ and $J_\pm$, we can relate them to $\alpha(\tau)$ and $\beta(\tau)$. We omit the dependence of $\tau$ in the following paragraphs for clarity of notation. 

We begin by noting that both of the squeezing terms above can be written in terms of the general squeezing operator notation 
\begin{equation}
\hat U_s = e^{- \frac{r}{2} \left( e^{i \, \theta} \, \hat b^{\dag 2} + e^{- i \, \theta} \, \hat b^2 \right)} \, ,
\end{equation}
where $r$ is a squeezing amplitude and $\theta$ is the phase of the squeezing. The symplectic representation  of $\hat U_s$ (see Section~\ref{chap:introduction:sec:non:Gaussianity} in Chapter~\ref{chap:introduction}) in the $\hat{\vec{r}} = (\hat b , \hat b^\dag )^{\rm{T}}$ basis reads 
\begin{equation}
\boldsymbol{\Omega} = \left(\begin{array}{cc} -i & 0 \\ 0 & i \end{array} \right) \,.
\end{equation}
This leads to the symplectic form of the squeezing operation
\begin{equation}
\boldsymbol{S}_\mathrm{s}(r,\theta)  = \left(\begin{array}{cc} \cosh(r) & - e^{i \, \theta} \sinh(r) \\  -  e^{-i \, \theta} \sinh(r) & \cosh(r) \end{array} \right)\,.
\end{equation}
Therefore, we can write two successive squeezing operations as 
\begin{eqnarray}\label{eq:doublesqueeze:first}
\boldsymbol{S}_\mathrm{s}(r_1,\theta_1) \boldsymbol{S}_\mathrm{s}(r_2,\theta_2)  & = \begin{pmatrix} S_{11} & S_{12} \\ S_{21} & S_{22} \end{pmatrix} \, , 
\end{eqnarray}
where the matrix elements are given by 
\begin{align} \label{chap:decoupling:eq:squeezing:elements:first}
S_{11} &=\cosh(r_1) \, \cosh(r_2) + e^{i \, (\theta_1 - \theta_2)} \sinh(r_1) \, \sinh(r_2) \, , \nonumber \\
S_{12} &= S_{21}^*=   -\left(e^{i \, \theta_1} \sinh(r_1) \, \cosh(r_2) + e^{i \, \theta_2} \cosh(r_1)\, \sinh(r_2) \right)  \, , \nonumber \\
S_{22} &= \cosh(r_1) \, \cosh(r_2) + e^{-i \, (\theta_1 - \theta_2)} \sinh(r_1)\sinh(r_2) \, ,
\end{align}
Now, the representation of the rotation in this basis is 
\begin{equation}
	\hat{U}_R = e^{-\frac{i \, a}{2}(\hat{b}^\dag\hat{b} + \hat{b}\hat{b}^\dag)} \, ,
\end{equation}
which corresponds to the symplectic matrix
\begin{equation}
\boldsymbol{S}_{R}(a) = \left(\begin{array}{cc} e^{-i \, a} & 0 \\ 0 & e^{i \, a} \end{array} \right)\,.
\end{equation}
A consecutive application of a squeezing and a rotation gives
\begin{equation}\label{eq:rotsqueeze}
\boldsymbol{S}_{R}(a) \boldsymbol{S}_\mathrm{s}(r_3,\theta_3)  = \left(\begin{array}{cc} e^{-i \, a}\cosh(r_3) & - e^{i \, (\theta_3-a)} \sinh(r_3) \\  - e^{-i \, (\theta_3-a)} \sinh(r_3) & e^{i \, a}\cosh(r_3) \end{array} \right)\,.
\end{equation}
Identification of the elements in  Eq.~\eqref{eq:doublesqueeze:first} and Eq.~\eqref{eq:rotsqueeze} leads to
\begin{align}
	\cosh(r_3)= & |\cosh(r_1) \, \cosh(r_2) + e^{i \, (\theta_1 - \theta_2)} \sinh(r_1) \, \sinh(r_2)| \, ,\nonumber \\
	\sinh(r_3)= & |\cosh(r_1) \, \sinh(r_2) + e^{i \, (\theta_1 - \theta_2)} \sinh(r_1) \, \cosh(r_2)|\,.
\end{align}
Furthermore, 
\begin{equation}
	e^{i\theta_3} = \frac{\cosh(r_3)}{\sinh(r_3)}  \, \frac{  e^{i \, \theta_1} \sinh(r_1) \, \cosh(r_2) + e^{i \, \theta_2} \cosh(r_1) \, \sinh(r_2) }  { \cosh(r_1) \, \cosh(r_2) + e^{i \, (\theta_1 - \theta_2)} \sinh(r_1) \, \sinh(r_2) } \, ,
\end{equation}
and, dividing $S_{11}$ by $S_{22}$,
\begin{equation}
	e^{-2 \, i \, a} = \frac{ \cosh(r_1) \, \cosh(r_2) + e^{i \, (\theta_1 - \theta_2)} \sinh(r_1) \, \sinh(r_2) }{ \cosh(r_1) \, \cosh(r_2) + e^{-i \, (\theta_1 - \theta_2)} \sinh(r_1) \, \sinh(r_2)} \, .
\end{equation}
Defining $t_j = \tanh(r_j)  \, e^{i \, \theta_j}$, we find
\begin{equation} \label{app:eq:identities:for:tj}
	t_3 = \tanh(r_3)  \, e^{i \, \theta_3} = \frac{  t_1 + t_2 }  { 1 + t_1  \, t_2^*} \, ,\quad \mathrm{and} \quad e^{-2 \, i \, a} = \frac{ 1 + t_1 \, t_2^*}{ 1 + t_1^* \, t_2} \, ,
\end{equation}
and the composition law for squeezing operators
\begin{equation} \label{chap:decoupling:eq:squeezing:composition:law}
\boldsymbol{S}(z_1)\boldsymbol{S}(z_2) = e^{\frac{1}{4}\ln\left(\frac{1 + t_1 \, t_2^*}{1 + t_1^* \,t_2}\right) (\hat{b}^\dag \hat{b} + \hat{b}\hat{b}^\dag)} \boldsymbol{S}(z_3)\,.
\end{equation}
where we recall that $z_j = r_j \, e^{i \, \theta_j }$. 

We are now in a position to derive the relation between the functions $J_b$, $J_+$ and $J_-$ and the $P_{11}$ and $I_{P_{22}}$ functions. We recall that 
\begin{equation}
	\boldsymbol{S}_{s} \hat{\mathbb{X}} = \left(\begin{array}{cc} \alpha(\tau) & \beta(\tau) \\ \beta^*(\tau) & \alpha^*(\tau) \end{array}\right)\hat{\mathbb{X}} \, ,
\end{equation}
and ask for it to be equivalent to
\begin{eqnarray}\label{eq:doublesqueeze}
\boldsymbol{S}_{R}(J_b) \boldsymbol{S}_\mathrm{s}(2J_+,\pi/2) \boldsymbol{S}_\mathrm{s}(2J_-,\pi)  & = \begin{pmatrix} S_{11} & S_{12} \\ S_{21} & S_{22} \end{pmatrix} \, , 
\end{eqnarray}
where we find  analogously to our result in Eq.~\eqref{chap:decoupling:eq:squeezing:elements:first} that the matrix elements are given by 
 \begin{align} \label{app:eq:squeezing:elements}
\alpha = S_{11} &=e^{-i \, J_b}\left(\cosh(2J_+) \, \cosh(2J_-) - i \, \sinh(2J_+)\, \sinh(2J_-)\right) \, , \nonumber \\
\beta = S_{12} &=  -e^{-i \, J_b}\left( i \, \sinh(2J_+) \, \cosh(2J_-) -  \cosh(2J_+)\, \sinh(2J_-) \right)  \, .
\end{align}
A particular set of solutions to these equations is given as 
\begin{align}\label{chap:decoupling:eq:squeezing:relation}
	J_+ = & \frac{1}{4}\rm{arcosh}(|\alpha^2 - \beta^2|) \, ,\nonumber \\
	J_- = & \frac{1}{4} \rm{arcosh}\left(\frac{(2 \, |\alpha|^2 -1)}{|\alpha^2 - \beta^2|}\right) \, ,\nonumber \\
	J_b = & -\frac{1}{2} \rm{Arg}\left(\frac{\alpha^2 - \beta^2}{|\alpha^2 - \beta^2|}\right) \, .
\end{align}
We arrived at the expressions for $J_b$ from the fact that  
\begin{equation}
e^{- 2 \,  i \,  J_b } = \left(\frac{\alpha^2 - \beta^2}{|\alpha^2 - \beta^2|}\right) \, .
\end{equation}
Taking the logarithm of a complex number gives  $\ln z = \ln |z| + i \rm{Arg} (z)$, where $\rm{Arg}$ is defined as $\rm{Arg}( x + i y) = \rm{arctan}(y/x)$, and where $z \in \mathbb{C}$. In this case, $|e^{ - 2 i J_b}| = 1$, which means that we arrive at the expression in Eq.~\eqref{chap:decoupling:eq:squeezing:relation} above. 

It is then straight-forward to relate the $J$-coefficients to $P_{11}$ and $I_{P_{22}}$ by using the expressions in Eq.~\eqref{app:eq:bogoliubov:expressions}. 

%
\subsection{Interpretation of the evolution operator}\label{chap:decoupling:interpretation:of:evolution}
Using  the general composition law for two squeezing operators in Eq.~\eqref{chap:decoupling:eq:squeezing:composition:law}, we can write $\hat{\tilde{U}}_{\rm{sq}}$ in Eq.~\eqref{chap:decoupling:eq:Usq:to:be:decoupled} as 
\begin{equation} \label{eq:compact:squeezing}
	\hat{\tilde U}_{\rm{sq}} \dot{=} e^{-i ( J_b +  \varphi_J ) \hat{N}_b }\, \hat{S}_b(\rm{arctanh}( |\zeta_J| )e^{i\arg(\zeta_J)}) \,,  
\end{equation}
where $\dot{=}$ indicates equivalence up to a global phase, and where
\begin{align} \label{eq:squeezing:new:coeffs}
	\varphi_J = & \,  \arctan(\tanh(2J_+)\tanh(2J_-)) \, , \nonumber \\
	\zeta_J = & \, \frac{i\tanh(2J_+) - \tanh(2J_-)}{1-i\tanh(2J_+)\tanh(2J_-)}\, .
\end{align}
With the commutation law for displacement and squeezing,
we obtain 
 \begin{align}\label{chap:decoupling:eq:intuitive:form}
\hat U(\tau)=& 
\,e^{-i(\Omega_{\rm{c}} \,\tau + \mathcal{\hat{F}}_{\hat N_a})\hat{N}_a - i \mathcal{\hat{F}}_+ \mathcal{\hat{F}}_- }\, e^{-i ( J_b + \varphi_J)\hat{N}_b } \hat{D}_b(\hat\gamma)  \hat{S}_b\left(\rm{arctanh}(|\zeta_J|) e^{i\arg(\zeta_J)}\right) \,,
\end{align}
where 
\begin{align}
	\hat\gamma = &  \, \frac{(\mathcal{\hat{F}}_- - i\mathcal{\hat{F}}_+)}{\sqrt{1-|\zeta_J|^2}} - e^{i\arg(\zeta_J)} \frac{(\mathcal{\hat{F}}_- + i\mathcal{\hat{F}}_+)|\zeta_J|}{\sqrt{1-|\zeta_J|^2}} \,.
\end{align}
The  time-evolution operator $\hat U(\tau)$ in~Eq.\eqref{chap:decoupling:eq:intuitive:form} can now be interpreted as follows: The mechanics experiences a photon-number dependent displacement through $\hat D_b(\hat{\mathcal{F}}-  i \, \hat{\mathcal{F}})$, followed by two squeezing operations $\hat S_b(2 \, i \, J_+)$ and $\hat S_b(- 2 \, J_-)$, and a rotation $e^{ -  i \, J_b \, \hat N_b}$. The cavity field is rotated through $e^{ - i (\Omega_{\rm{c}} + F_{\hat N_a}) \hat N_a}$ and then strongly translated by a nonlinear Kerr self-interaction term: $e^{- i \, F_{\hat N_a^2} \hat N_a^2}$. The full time evolution operator can be reordered and interpreted as subsequent photon number dependent squeezing, displacement and rotation. 

\subsection{State evolution}

We present here the full state derived under the evolution with $\hat{U}(\tau)$ for two initially coherent states. The full state is
\begin{equation}
\ket{\Psi(t = 0)} = \ket{\mu_{\rm{c}}} \otimes \ket{\mu_{\rm{m}}} \, ,
\end{equation}
where $\ket{\mu_{\rm{c}}}$ and $\ket{\mu_{\rm{m}}}$ are coherent states of the optics and mechanics, respectively, such that $\hat a \ket{\mu_{\rm{c}}} = \mu_{\rm{c}} \ket{\mu_{\rm{c}}}$ and $\hat b\ket{\mu_{\rm{m}}} = \mu_{\rm{m}} \ket{\mu_{\rm{m}}}$.

The full state evolved under the Hamiltonian in Eq.~\eqref{chap:decoupling:eq:Hamiltonian:rescaled} is then given by 
\begin{align}\label{chap:decoupling:eq:non:linear:state:evolution}
\ket{\Psi(\tau)} =& \,    e^{ - i  \,\left( F_{\hat B_+}\,  F_{\hat B_-}+\Im\left(K  \, \mu_{\rm{m}} \right)\right) } \,e^{- \frac{1}{2}|\mu_{\rm{c}}|^2}\,\sum_n \biggl[ \frac{\mu_{\rm{c}}^n}{\sqrt{n!}} \,   e^{- i \,\left(F_{\hat{N}^2_a}+  F_{\hat N_a \, \hat B_+} \, F_{\hat N_a \, \hat B_-}  \right)\, n^2}\nonumber \\
&e^{ - i   \, \left( \Omega_\textrm{c}\,\tau+F_{\hat{N}_a}+F_{\hat N_a \, \hat B_+} \, F_{\hat B_-} + F_{\hat N_a \, \hat B_-} \, F_{\hat B_+}+\Im\left(K_{\hat N_a}\, \mu_{\rm{m}} \right)\right)\,  n} \, 
\ket{n} \otimes  \ket{\phi_n, \tilde{\mathcal{D}}_2(\tau)} \biggr] \, ,
\end{align}
where we have defined $K:=F_{\hat{B}_-} + i \, F_{\hat{B}_+}$ and
$K_{\hat{N}_a}:=F_{\hat{N}_a \, \hat{B}_- }+ i \, F_{\hat{N}_a \, \hat{B}_+}$, where $|\phi_n(\tau), \tilde{\mathcal{D}}_2(\tau) \rangle$ is a mechanical coherent squeezed state where 
$|\phi_n(\tau), \tilde{\mathcal{D}}_2(\tau) \rangle = \hat{U}_{\mathrm{sq}}(\tau) |\phi_n(\tau)\rangle $,  and where $|\phi_n(\tau)\rangle$ is a coherent state with $\phi_n(\tau):= K^*+ n \, K_{\hat N_a}^*+\mu_{\mathrm{m}}$. Note that, in the above, we have kept the dependence on $\tau$ implicit 
but, in general, all exponentials oscillate in time as determined by the Hamiltonian coefficients. We also note that the state Eq.~\eqref{chap:decoupling:eq:non:linear:state:evolution} contains all terms that have been considered in the literature before, including the contributions from a constant nonlinear light--matter term~\cite{bose1997preparation}. The main addition here is $\hat{\tilde{U}}_{\mathrm{sq}}$, which appears only if $\tilde{\mathcal{D}}_2 \neq 0$, and the fact that we can consider arbitrary time-dependent terms in the Hamiltonian in Eq.~\eqref{chap:decoupling:eq:Hamiltonian}. We intend now to explore how the addition of the mechanical squeezing affects the non-Gaussianity of the state.

We note here that the expression of Eq.~\eqref{chap:decoupling:eq:non:linear:state:evolution} allows us to  compute the reduced state of the mechanics $\hat{\rho}_{\text{Mech}}(\tau)$ at any time $\tau$, which reads
\begin{align}\label{non:linear:reduced:mechanical:state:evolution}
\hat{\rho}_{\text{Mech}}(\tau) &=   e^{- |\mu_{\rm{c}}|^2}\,\sum_n \frac{|\mu_{\rm{c}}|^{2\,n}}{n!} 
\ket{\phi_n, \tilde{\mathcal{D}}_2(\tau)} \bra{\phi_n, \tilde{\mathcal{D}}_2(\tau)}.
\end{align}
We hereby conclude the decoupling of the Hamiltonian in Eq.~\eqref{chap:decoupling:eq:Hamiltonian}. Before ending this Chapter, however, we briefly discuss the addition of terms to the Hamiltonian that require the resulting Lie algebra to have an infinite number of elements.

\subsection{Extensions to infinite Lie algebras} \label{chap:decoupling:sec:infinite:algebra}
While the Hamiltonian in Eq.~\eqref{chap:decoupling:eq:Hamiltonian} is generated by a finite Lie algebra, there are terms which render the algebra infinite. One such term is a linear optical driving term $\left( \hat a^\dag + \hat a \right)$, which can be used to model systems where a laser beam continuously drives the cavity. This short example demonstrates the limitations of the decoupling method. 
When we attempt to commute this term with the nonlinear term, we find 
\begin{equation}
\left[ \hat a^\dag + \hat a , \hat a ^\dag \hat a \, \left( \hat b^\dag + \hat b \right) \right] = \left( \hat a - \hat a^\dag \right) \left(\hat b^\dag + \hat b \right) \, .
\end{equation}
The result is a new term in the algebra, so we must commute it with the nonlinear term one more time to ensure closure. We find
\begin{equation}
\left[ \left(\hat a - \hat a^\dag \right) \left( \hat b^\dag + \hat b \right) , \hat a^\dag \hat a \left( \hat b^\dag + \hat b \right) \right] =  \left( \hat a ^\dag + \hat a \right) \left( \hat b^\dag + \hat b \right)^2 \, .
\end{equation}
This term is yet again new, so must be commuted with the previous two terms. Doing so will yield additional powers of $\left(\hat b^\dag + \hat b \right)$, until an infinite number of terms have been generated. The resulting Lie algebra is therefore infinite. While there might be ways to solve the ensuring differential equations, doing so is considerably harder. There are cases, however, where the algebra takes on a particularly simple form such that analytic solutions can be obtained~\cite{bruschi2019time}.

\part{Non-Gaussian character of optomechanical states}

\chapter{Interplay of non-Gaussianity and the nonlinear coupling}
\label{chap:non:Gaussianity:coupling}

In this Chapter, we examine the non-Gaussian character of optomechanical systems that evolve with the standard optomechanical Hamiltonian. We are especially interested in how the form of the nonlinear light--matter coupling affects the non-Gaussianity of the state. For this purpose, we use the decoupled solution presented in Chapter~\ref{chap:decoupling} and apply measure of non-Gaussianity that we introduced in Section~\ref{chap:introduction:sec:measure:non:gaussianity} in Chapter~\ref{chap:introduction}. 

This Chapter is based on Ref~\cite{qvarfort2019enhanced}. The code used to simulate the open system dynamics can be found in the following \href{https://github.com/sqvarfort/QM-Nonlinearities}{GitHub Repository}. We thank Fabienne Schneiter, Daniel Braun, Nathana\"{e}l Bullier, Antonio Pontin, Peter F. Barker, Ryuji Takagi, Francesco Albarelli, Marco G. Genoni, James Bateman and the reviewers for helpful comments and discussions that improved the work in this Chapter. 

It should be noted that the published version of Ref~\cite{qvarfort2019enhanced} contains a mistake; the numerical plots showing the non-Gaussianity of open system dynamics are incorrect, since the entropy of the fully non-Gaussian state has not been subtracted from the entropy of the Gaussian reference state. The plots therefore overestimate the amount of non-Gaussianity retained by the state. This mistake has been rectified in this thesis, and as a result, the plots in Figure~\ref{chap:non:Gaussianity:coupling:fig:measure:decoherence:constant:coupling} and~\ref{chap:non:Gaussianity:coupling:fig:measure:decoherence:resonance} are slightly different compared with those in Ref~\cite{qvarfort2019enhanced}. Furthermore, the analysis for the thermal state as presented in Ref~\cite{qvarfort2019enhanced} is not valid, so we omit it here. Finally, we have here slightly changed the notation compared with the published manuscript so that it coincides with the notation used in the rest of this thesis.

\section{Introduction}\label{intro}
In optomechanical systems~\cite{aspelmeyer2014cavity}, the light--matter interaction induced by radiation pressure is inherently nonlinear, which allows for a number of interesting applications. In particular, the nonlinear coupling enables the creation of optical cat states in the form of superpositions of coherent states~\cite{mancini1997ponderomotive, bose1997preparation}. These cat states can also be transferred to the mechanics~\cite{palomaki2013coherent}, which opens up the possibility of using massive superpositions for testing fundamental phenomena such as collapse theories~\cite{goldwater2016testing} and, potentially, signatures of gravitational effects on quantum systems at low energies~\cite{bose2017spin,marletto2017gravitationally}. In addition to cat states, other non-Gaussian states such as compass states~\cite{zurek2001sub, toscano2006sub} and hypercube states~\cite{howard2018hypercube},  and also all been found to have excellent sensing capabilities.

A number of different optomechanical systems have been experimentally implemented, as discussed in Section~\ref{chap:introduction:section:examples} in Chapter~\ref{chap:introduction}. 
However, it is generally difficult to access the fully nonlinear regime. As a result, significant effort has been devoted to the question of how the non-linearity can be further enhanced. Most approaches focus on the few-excitation regime, where increasing the inherent light--matter coupling allows for detection of the non-linearity. This enhancement can be achieved, for example, by using a large-amplitude, strongly detuned mechanical parametric drive~\cite{lemonde2016enhanced}, or by modulating the spring constant~\cite{yin2017nonlinear}. Similar work has shown that the inclusion of a mechanical quartic anharmonic term can be nearly optimally detected with homodyne and heterodyne detection schemes, which are standard measurements implemented in the laboratory~\cite{latmiral2016probing}. 

A natural question that arises considering the approaches above  is: \textit{Are there additional methods by which the amount of non-Gaussianity in an optomechanical system can be further increased}? One such proposal was put forward in~\cite{liao2014modulated} where it was suggested that the nonlinearity in electromechanical systems could be enhanced by several orders of magnitude by modulating the light--matter coupling. This is achieved by driving the system close to mechanical resonance and takes a simple form in the rotating-wave approximation. Here, we seek to fully quantify the non-Gaussianity of the exact, nonlinear optomechanical state for both ideal and open systems. More precisely, given an initial Gaussian state evolving under the standard optomechanical Hamiltonian, we quantify how non-Gaussian the state becomes as a function of time and the parameters of the Hamiltonian in question. To do so, we make use of recently developed analytical techniques to study the time-evolution of time-dependent systems~\cite{bruschi2018mechano}, and employ a specific measure of non-Gaussianity based on the relative entropy of the state~\cite{genoni2008quantifying}.

This Chapter is organised as follows. In Section~\ref{chap:non:Gaussianity:coupling:system} we present the Hamiltonian of interest and review the solution of the dynamics. In Section~\ref{chap:non:Gaussianity:coupling:general:results} we derive some generic results based on the measure which apply in different regimes. We then proceed to examine the behaviour of the non-Gaussianity in optomechanical systems for two cases: a constant light--matter coupling in Section~\ref{chap:non:Guassianity:coupling:sec:constant:coupling}; and a time-dependent coupling in Section~\ref{chap:non:Gaussianity:coupling:sec:time:dependent:coupling}, where we also show that driving the coupling results in continuously generated non-Gaussianity. Both preceding sections also include an analysis of the open system dynamics. Finally, in Section~\ref{chap:non:Gaussianity:coupling:sec:discussion} we discuss our results and propose various methods by which the modulation of the optomechanical coupling can be achieved. Section~\ref{chap:non:Gaussianity:coupling:sec:conclusions} concludes this Chapter.

\section{System and dynamics}\label{chap:non:Gaussianity:coupling:system}
The work in this Chapter is based on part of the solutions of the dynamics of the extended Hamiltonian in Eq.~\eqref{chap:decoupling:eq:Hamiltonian}, which was solved in Chapter~\ref{chap:decoupling}. We review the key aspects of the solution here. 

\subsection{Hamiltonian}\label{chap:non:Gaussianity:}
To model the optomechanical system, we consider two bosonic modes corresponding to an electromagnetic mode and a mechanical oscillator. The radiation pressure induces a nonlinear interaction between the light and mechanics, and the whole systems is modelled by the  optomechanical Hamiltonian:
\begin{align}\label{chap:non:Gaussianity:coupling:eq:standard:Hamiltonian}
	\hat {H} &= \hbar\,\omega_\mathrm{c}\,\hat a^\dagger \hat a + \hbar\,\omega_\mathrm{m} \,\hat b^\dagger \hat b - \hbar\,\mathcal{G}(t)\,\hat a^\dagger\hat a \bigl(\hat b+ \hat b^\dagger\bigr),
\end{align}
where $\omega_\mathrm{c}$ and $\omega_\mathrm{m}$ are the frequencies of the cavity mode and the mechanical mode respectively, and $\mathcal{G}(t)$ drives the, potentially time-dependent, nonlinear light--matter coupling. The light--matter coupling strength $\mathcal{G}(t)$ takes on different functional forms for different optomechanical systems. 

As before, to simplify our notation and expressions, we rescale the laboratory time $t$ by the frequency $\omega_\mathrm{m}$, therefore introducing the dimensionless time $\tau:=\omega_\mathrm{m}\,t$, the dimensionless frequency $\Omega_{\rm{c}}:=\omega_\mathrm{c}/\omega_\mathrm{m}$, and the dimensionless coupling $\tilde{\mathcal{G}}(\tau):=\mathcal{G}(t\omega_\mathrm{m})/\omega_\mathrm{m}$. This renormalisation effectively is equivalent to the use of time $\tau$ and the Hamiltonian
 \begin{align}\label{chap:non:Gaussianity:coupling:eq:standard:Hamiltonian:dimensionless}
	\hat{\tilde{H}} &= \hat H/(\hbar \, \omega_{\rm{m}}) = \Omega_{\rm{c}} \,  \hat a^\dagger \hat a + \hat b^\dagger \hat b -\tilde{\mathcal{G}}(\tau) \hat a^\dagger\hat a \bigl(\hat b+ \hat b^\dagger\bigr).
\end{align}
To determine the action of this Hamiltonian on initial states, we now proceed to consider the resulting dynamics. 

\subsection{Solution of the dynamics}\label{chap:non:Gaussianity:coupling:solution}
We seek an expression for the time evolution operator $\hat{U}(\tau)$ for a system evolving with Eq.~\eqref{chap:non:Gaussianity:coupling:eq:standard:Hamiltonian}. The unitary time-evolution operator reads 
\begin{equation}\label{time:evolution:operator}
\hat{U}(\tau):=\overset{\leftarrow}{\mathcal{T}}\,\exp\biggl[-i \, \int_0^{\tau} d\tau'\,\hat{H}(\tau')\biggr],
\end{equation}
where $\overset{\leftarrow}{\mathcal{T}}$ is the time-ordering operator. This expression simplifies dramatically when the Hamiltonian $\hat{H}$ is independent of time, in which case one simply has $\hat{U}(\tau)=\exp[-i\,\hat{H}\,\tau ]$. As we will here consider time-dependent light--matter couplings $\tilde{g}(\tau)$, we instead seek to solve the full dynamics of the time-dependent Hamiltonian. 

Given the solution in Chapter~\ref{chap:decoupling}, the time-evolution operator $\hat U(\tau)$ in Eq.~\eqref{chap:decoupling:eq:final:evolution:operator:J:coefficients} encodes the dynamics of not only the standard optomechanical Hamiltonian in Eq.~\eqref{chap:non:Gaussianity:coupling:eq:standard:Hamiltonian}, but also the effects of additional terms, such as $\tilde{\mathcal{D}}_1(t) \, \left( \hat b^\dag + \hat b \right)$ and $\tilde{\mathcal{D}}_2 \, \left( \hat b^{\dag 2} + \hat b^2\right)$. If we now set $\tilde{\mathcal{D}}_1(\tau) = \tilde{\mathcal{D}}_2 = 0$, we can recover the solution to the standard optomechanical Hamiltonian from the expression in Eq.~\eqref{chap:non:Gaussianity:coupling:eq:standard:Hamiltonian}. 

We find, for the evolution with the standard optomechanical Hamiltonian in Eq.~\eqref{chap:non:Gaussianity:coupling:eq:standard:Hamiltonian}, the $\hat U(\tau)$ can be written in the following form:
\begin{align}\label{chap:decoupling:eq:explicit:time:evolution:operator}
\hat U(t)=e^{-i\,\hat{N}_b\,\tau}\,e^{-i\,F_{\hat{N}^2_a}\,\hat{N}^2_a}\,e^{-i\,F_{\hat{N}_a\,\hat{B}_+}\,\hat{N}_a\,\hat{B}_+} e^{-i\,F_{\hat{N}_a\,\hat{B}_-}\,\hat{N}_a\,\hat{B}_-}, 
\end{align} 
where we have transformed into a frame rotating with $\Omega_{\mathrm{c}} \, \hat{N}_a$ in order to neglect the phase-term $e^{- i \, \Omega_{\rm{c}} \,  \tau}$, and where the operators, given by, 
\begin{align}\label{basis:operator:Lie:algebra}
	\hat{N}_a &:= \hat a^\dagger \hat a &
	\hat{N}_b &:= \hat b^\dagger \hat b 
	 & \hat{N}^2_a &:= (\hat a^\dagger \hat a)^2\nonumber\\
	\hat{N}_a\,\hat{B}_+ &:= \hat{N}_a\,(\hat b^{\dagger}+\hat b) &
	\hat{N}_a\,\hat{B}_- &:= \hat{N}_a\,i\,(\hat b^{\dagger}-\hat b), &
	 & 
\end{align}
form a closed Lie algebra under commutation,  and where the coefficients that determine the evolution in Eq.~\eqref{chap:decoupling:eq:explicit:time:evolution:operator} are given by 
\begin{align}\label{sub:algebra:decoupling:solution:text}
F_{\hat{N}^2_a}&= 2  \,\int_0^\tau\,\mathrm{d}\tau'\,\tilde{\mathcal{G}}(\tau')\,\sin(\tau')\int_0^{\tau'}\mathrm{d}\tau''\,\tilde{\mathcal{G}}(\tau'')\,\cos(\tau''),\nonumber\\
F_{\hat{N}_a\,\hat{B}_+}&=- \int_0^\tau\,\mathrm{d}\tau'\,\tilde{\mathcal{G}}(\tau')\,\cos(\tau'),\ \text{and}\nonumber\\
F_{\hat{N}_a\,\hat{B}_-}&= -\int_0^\tau\,\mathrm{d}\tau'\,\tilde{\mathcal{G}}(\tau')\,\sin(\tau').
\end{align}
Given an explicit form of $\tilde{\mathcal{G}}(\tau)$, it is then possible to write down a full solution for $\hat{U}(\tau)$. Since $\tilde{\mathcal{D}}_1 = \tilde{\mathcal{D}}_2 = 0$ in this section, it follows that the remaining coefficients $F_{\hat N_a}$, $F_{\hat B_+}$, $F_{\hat B_-}$, $J_b$, $J_+$, and $J_-$ are all zero.

\subsection{Initial state of the system} \label{sec:initial:state}
In this Chapter, we examine the non-Gaussianity of the evolved state given a coherent state of the optics and mechanics. By starting in an initial Gaussian state, we ensure that any non-Gaussianity revealed by our investigations is due to the nonlinear coupling in Eq.~\eqref{chap:non:Gaussianity:coupling:eq:standard:Hamiltonian}. Indeed, the only way an initially Gaussian state may at any time be non-Gaussian is for the corresponding Hamiltonian to induce some nonlinear evolution~\cite{Adesso:Ragy:2014}. We do however note that while the measure of non-Gaussianity that we shall make use of has a clear and operational notion of the measure for pure states, it is harder to make statements about the non-Gaussianity of states that are mixed. See Section~\ref{chap:non:Gaussianity:coupling:sec:discussion} for a discussion of the properties of the relative entropy measure. 

The initial states that we consider  in this thesis were introduced in Section~\ref{chap:introduction:initial:states} in Chapter~\ref{chap:introduction}, but we recount their basic properties here. 
We consider the case when both the optical and the mechanical modes are in a coherent state, which we denote $\ket{\mu_\mathrm{c}}$ and $\ket{\mu_\mathrm{m}}$ respectively. These states satisfy the relations $\hat{a}\ket{\mu_\mathrm{c}}=\mu_\mathrm{c}\ket{\mu_\mathrm{c}}$ and $\hat{b}\ket{\mu_\mathrm{m}}=\mu_\mathrm{m}\ket{\mu_\mathrm{m}}$. For the optical field, this is a readily available resource, since coherent states model laser light quite well. The mechanical element in optomechanical systems is most often found in a thermal state or, assuming perfect preliminary cooling, in its ground state, with $\mu_\mathrm{m} = 0$.
The initial state $\ket{\Psi(0)}$ of the compound system will therefore be
\begin{equation}\label{initial:state:two}
\ket{\Psi(0)} = \ket{\mu_{\mathrm{c}}} \otimes \ket{\mu_{\mathrm{m}}}. 
\end{equation} 
The evolved state can be computed through
\begin{equation} \label{chap:non:Gaussianity:coupling:evolved:state}
\hat \rho(\tau) = \hat U(\tau) \, \hat \rho_0 \, \hat U^\dag(\tau) \, ,
\end{equation}
where $\hat U(\tau)$ is the time-evolution operator in Eq.~\eqref{chap:decoupling:eq:explicit:time:evolution:operator}. 
We are now ready to consider the covariance matrix elements. 

\subsection{Covariance matrix elements}
In Chapter~\ref{chap:introduction}, we introduced the notion of non-Gaussianity, as well as a measure of non-Gaussianity $\delta(\tau)$, which is defined as the relative entropy between the fully non-Gaussian state  $\hat \rho$ in Eq.~\eqref{chap:non:Gaussianity:coupling:evolved:state} and a Gaussian reference state $\hat \rho_{\rm{G}}$. The measure is given by
\begin{equation}
\delta(\tau) = S(\hat \rho_{\rm{G}}) - S(\hat \rho) \, ,
\end{equation}
where $S(\hat \rho)$ is the von Neumann entropy of the state $\hat \rho$. 

We must now determine the Gaussian reference state $\hat \rho_{\rm{G}}$. We do so by computing the covariance matrix elements of $\hat \rho$. The full calculation can be found in Appendix~\ref{app:exp:values}, where we consider the expectation values of the modified optomechanical Hamiltonian in Eq.~\eqref{chap:decoupling:eq:Hamiltonian}. Since we here consider the simpler case with  $\tilde{\mathcal{D}}_1(\tau) = \tilde{\mathcal{D}}_2(\tau) = 0$, we find $\Gamma(\tau) = 0$, where $\Gamma(\tau)$ is defined in Eq.~\eqref{app:exp:values:definition:of:Gamma:Delta}. 
Before we proceed, we define the following two parameters:
\begin{align} \label{eq:combined:coefficients}
\theta(\tau) &= 2 \, \left( F_{\hat{N}_a^2} + F_{\hat{N}_a \, \hat{B}_+} F_{\hat{N}_a \hat{B}_-} \right) \, , \nonumber \\
K_{\hat N_a} &= F_{\hat{N}_a \, \hat{B}_-} + i F_{\hat{N}_a \, \hat{B}_+} \, . 
\end{align}
For the initial coherent state $\ket{\Psi(\tau = 0)} = \ket{\mu_{\mathrm{c}}} \otimes \ket{\mu_{\mathrm{m}}}$ in Eq.~\eqref{initial:state:two} and ignoring the global phases $e^{- i \, \Omega_{\rm{c}} \,\tau}$, by transforming into a frame rotating with $\Omega_{\rm{c}} \hat{a}^\dag \hat{a}$, we obtain
\begin{align} \label{app:expectation:values:coherent}
\braket{\hat a (\tau) } &:= e^{- i \frac{1}{2} \theta(\tau) } \, e^{|\mu_{\mathrm{c}}|^2 (e^{- i \, \theta(\tau)}-1)} \, e^{-\frac{1}{2}|K_{\hat N_a}|^2} \, e^{K_{\hat N_a}^* \mu_{\mathrm{m}}^* - K_{\hat N_a} \mu_{\mathrm{m}}} \, \mu_{\mathrm{c}}\, , \nonumber \\
\braket{\hat b(\tau) } &:= e^{- i  \, \tau} \mu_{\mathrm{m}}  +  e^{- i \, \tau} \, K_{\hat N_a}^* \, |\mu_{\mathrm{c}}|^2 \, , \nonumber \\
\braket{\hat{a}^2(\tau)} &:= e^{- 2 \, i \, \theta(\tau)} \,   e^{|\mu_{\mathrm{c}}|^2 (e^{- 2\, i \,  \theta(\tau)} - 1)} \,  e^{ - 2 |K_{\hat N_a}|^2    } \, e^{2 \,  ( K_{\hat N_a}^* \mu_{\mathrm{m}}^* - K_{\hat N_a} \mu_{\mathrm{m}})} \,  \mu_{\mathrm{c}}^2\, ,\nonumber \\
 \braket{\hat b^{2}(\tau)} &:= e^{-2 \, i \, \tau} \left(  \mu_{\mathrm{m}} + K_{\hat N_a}^*\, |\mu_{\mathrm{c}}|^2 \right)^2  + e^{- 2 \, i \,  \tau } K_{\hat N_a}^{*2} \, |\mu_{\mathrm{c}}|^2 \, ,\nonumber \\
 \braket{\hat a^\dag(\tau) \hat a(\tau) } &:= |\mu_{\mathrm{c}}|^2\, , \nonumber \\
 \braket{\hat b ^\dag(\tau) \hat b(\tau)}  &:=  \left|  \, \mu_{\mathrm{m}}  + K_{\hat N_a}^*\, |\mu_{\mathrm{c}}|^2 \right|^2 + |K_{\hat N_a}|^2 \, |\mu_{\mathrm{c}}|^2\, ,\nonumber \\
\braket{\hat a(\tau) \hat b(\tau) } &:= e^{- i \,  \frac{1}{2}\theta(\tau)}  \,e^{ |\mu_{\mathrm{c}}|^2 ( e^{- i \,  \theta(\tau)}  - 1)}\, e^{ -\frac{1}{2}  |K_{\hat N_a}|^2   } \, e^{K_{\hat N_a}^* \,  \mu_{\mathrm{m}}^* - K_{\hat N_a} \, \mu_{\mathrm{m}}} \, \nonumber \\
&\quad\quad\times  \mu_{\mathrm{c}} \, e^{- i \,  \tau}\,  \left[\mu_{\mathrm{m}}  + \left( |\mu_{\mathrm{c}}|^2 e^{- i \, \theta(\tau)} + 1\right)K_{\hat N_a}^* \right] \, , \nonumber \\
\braket{\hat a (\tau) \, \hat b^\dag(\tau) }: &=e^{- \frac{1}{2}i \, \theta(\tau)}  \, e^{ |\mu_{\mathrm{c}}|^2 ( e^{- i \, \theta(\tau)}  - 1)}  \,e^{- \frac{1}{2}   |K_{\hat N_a}|^2 }\, e^{K_{\hat N_a}^* \,  \mu_{\mathrm{m}}^* - K_{\hat N_a} \,  \mu_{\mathrm{m}}} \, \nonumber \\
&\quad\quad\times \mu_{\mathrm{c}} \,  e^{i \,  \tau} \,  \left[  \mu_{\mathrm{m}}^* + |\mu_{\mathrm{c}}|^2 e^{- i \, \theta(\tau)}  K_{\hat N_a} \right] \, .
\end{align}
Similarly, the covariance matrix elements have been computed explicitly in Eq.~\eqref{app:exp:values:CM:elements} in Appendix~\ref{app:exp:values}. Given that $\Gamma(\tau) = 0$, they read 
\begin{align}\label{chap:non:Gaussianity:coupling:full:elements:covaraince:matrix}
\sigma_{11} = \sigma_{33} &= 1 + 2|\mu_{\mathrm{c}}|^2 \left( 1 - e^{-4 \, |\mu_{\mathrm{c}}|^2 \sin^2{\theta(\tau)/2}} \, e^{- |K_{\hat N_a}|^2} \, e^{K_{\hat N_a}^* \,  \mu_{\mathrm{m}}^* - K_{\hat N_a} \,  \mu_{\mathrm{m}}} \right)  \,, \nonumber \\
\sigma_{31} &=  2 \, \mu_{\mathrm{c}}^2 \, e^{- i  \, \theta(\tau)} \, e^{- |K_{\hat N_a}|^2} \, \biggl(  e^{- i  \, \theta(\tau)} \, e^{|\mu_{\mathrm{c}}|^2  \, ( e^{-  2 \, i  \, \theta(\tau)} - 1)} \, e^{- \,|K_{\hat N_a}|^2} \, e^{2 \, (K_{\hat N_a}^* \,  \mu_{\mathrm{m}}^* - K_{\hat N_a} \,  \mu_{\mathrm{m}})} \nonumber \\
&\quad\quad\quad\quad\quad\quad\quad\quad\quad\quad\quad- e^{2 \, |\mu_{\mathrm{c}}|^2 ( e^{- i  \, \theta(\tau)} - 1)} \,  e^{2 \, (K_{\hat N_a}^* \,  \mu_{\mathrm{m}}^* -K_{\hat N_a} \,  \mu_{\mathrm{m}})}\biggr)\,, \nonumber \\
\sigma_{22} =  \sigma_{44} &= 1  + 2 \, |\mu_{\mathrm{c}}|^2 \, |K_{\hat N_a}|^2\,,  \nonumber \\
\sigma_{42}  &= 2 \, e^{- 2 \, i \,\tau} \,  |\mu_{\mathrm{c}}|^2 \, K_{\hat N_a}^{*2} \,, \nonumber \\
\sigma_{21} = \sigma_{34} &=  2  \, K_{\hat N_a}\,\mu_{\mathrm{c}}\, |\mu_{\mathrm{c}}|^2 \, (e^{- i\,  \theta(\tau)} - 1) \, e^{- i \frac{1}{2} \theta(t)} \,  e^{i\,\tau} \, e^{|\mu_{\mathrm{c}}|^2 ( e^{- i  \, \theta(\tau)} - 1)} e^{- \frac{1}{2}|K_{\hat N_a}|^2} \, \nonumber \\
&\times e^{K_{\hat N_a}^* \,  \mu_{\mathrm{m}}^* - K_{\hat N_a} \,  \mu_{\mathrm{m}}} \,,\,  \nonumber \\
\sigma_{41}= \sigma_{32}  &= 2\,  K_{\hat N_a}^*\,\mu_{\mathrm{c}} \,  \left(|\mu_{\mathrm{c}}|^2 (e^{- i \,  \theta(\tau)} -1) + 1  \right) \,  e^{- \frac{i}{2} \theta(\tau)}  \, e^{- i\, \tau} \,e^{ |\mu_{\mathrm{c}}|^2 ( e^{- i \,  \theta(\tau)}  - 1)} \, \nonumber \\
&\times e^{- \frac{1}{2}|K_{\hat N_a}|^2 }\,e^{K_{\hat N_a}^*  \, \mu_{\mathrm{m}} ^* - K_{\hat N_a} \,  \mu_{\mathrm{m}}}  \,.
\end{align}

\section{General results} \label{chap:non:Gaussianity:coupling:general:results}
Given the covariance matrix elements in Eq.~\eqref{chap:non:Gaussianity:coupling:full:elements:covaraince:matrix}, we are now ready to compute the full measure of non-Gaussianity $\delta(\tau)$. First, however, we consider some general results with regard to the behaviour of the measure. We mainly consider the two symplectic eigenvalues $\nu_+$ and $\nu_-$ of the covariance matrix $\sigma$, which can be computed by taking the eigenvalues of the object $i \, \Omega \, \sigma$, where $\Omega$ is the symplectic form defined in Eq.~\eqref{chap:introduction:symplectic:form:ab:basis} in this particular basis. 

\subsection{Asymptotic behaviour for a small optical coherent state parameter}
We begin by looking at the case where $|\mu_{\mathrm{c}}|^2\ll1$ and where the mechanics is initially in a coherent state. Here, one can take the covariance matrix elements in Eq.~\eqref{chap:non:Gaussianity:coupling:full:elements:covaraince:matrix} and, after some algebra, show that the perturbative expansion of the symplectic eigenvalues gives
\begin{align} 
\nu_+\sim&1+\left(1-|K_{\hat N_a}|^2\,e^{-|K_{\hat N_a}|^2}\right)\,|\mu_{\mathrm{c}}|^2 \, ,\nonumber\\
\nu_-\sim&1+\left(1-e^{-|K_{\hat N_a}|^2}\right)\,|K_{\hat N_a}|^2\,|\mu_{\mathrm{c}}|^2 \, .
\end{align}
This implies that the behaviour of $\delta(\tau)$ for small $|\mu_{\mathrm{c}}|$ scales as 
\begin{equation} \label{eq:delta:small}
\delta(\tau) \sim -\left(1+\left(1-2\,e^{-|K_{\hat N_a}|^2}\right)\,|K_{\hat N_a}|^2\right)\,|\mu_{\mathrm{c}}|^2\,\ln|\mu_{\mathrm{c}}|,
\end{equation}
in perfect agreement with Eq.~ \eqref{chap:introduction:general:measure:of:non:gaussianity:small}, where we noted that a perturbative expansion of the measure should behave in this way. This approximation suggests that $\delta(\tau)$ scales with $\sim |K_{\hat N_a}|^2 |\mu_{\mathrm{c}}|^2 \ln |\mu_{\mathrm{c}}|$ to leading order. 

These expressions do not hold if the mechanical element is initially mixed. However, we will find in the next section that initial phonon occupation only marginally affects the non-Gaussianity.

\subsection{Asymptotic behaviour for a large optical coherent state parameter} \label{sec:asymptotic:large:coherent:state}

We now investigate the case where $|\mu_\mathrm{c}|\gg1$. Our goal is to derive an analytic expression for the non-Gaussianity that can be used to analyse the overall features of $\delta(\tau)$. Before making any quantitative evaluation, we recall that the measure will have the form in Eq.~\eqref{large:parameter:prediction:expansion}, where now $x\equiv|\mu_\mathrm{c}|^2$. Let us proceed to demonstrate this result analytically for this specific case. 

For large $\mu_{\mathrm{c}}$ and for the mechanics in the ground-state $\mu_{\mathrm{m}} = 0$, it is clear that whenever $\theta(\tau) \neq 2\pi n $ for integer $n$, the matrix elements $\sigma_{31}$, $\sigma_{21}$ and $\sigma_{41}$ in Eq.~\eqref{chap:non:Gaussianity:coupling:full:elements:covaraince:matrix} vanish, due to the exponentials containing the  factor $|\mu_{\mathrm{c}}|^2$. Therefore, far (enough) from the times where $\theta(\tau) = 2\pi n$  we are left with the following covariance matrix elements
\begin{align}\label{covariance:matrix:elements:large:input:photons:far:from:critical:times}
\sigma_{11} &\sim \sigma_{33} = 1 + 2 \,|\mu_{\mathrm{c}}|^2 \left( 1 - e^{-4 \, |\mu_{\mathrm{c}}|^2 \sin^2{\theta(\tau)/2}} \, e^{- |K_{\hat N_a}|^2}  \right) \, , \nonumber \\
\sigma_{22} &\sim \sigma_{44} = 2 \, |\mu_{\mathrm{c}}|^2 |K_{\hat N_a}|^2 + 1\, ,\nonumber \\
\sigma_{42} &\sim \sigma_{24}^* =2 \, |\mu_{\mathrm{c}}|^2  \, e^{- 2 \, i \, \tau} K_{\hat N_a}^{*2} \, ,
\end{align}
and all other elements  are zero. We have kept the full expression for $\sigma_{11}$ because it reproduces some key elements of $\delta$ which we shall discuss later. Therefore, we do not expect the thermal occupation of the mechanics to significantly affect the non-Gaussianity that can be accessed in this system. Note also that we need to keep the next leading order in each element of Eq.~\eqref{covariance:matrix:elements:large:input:photons:far:from:critical:times}, which is a constant in the case of $\sigma_{11}$ and $\sigma_{22}$. Naively neglecting of this element would give an incorrect result when computing the entropy, as the neglected factor becomes significant in the logarithm~\cite{ahmadi2014quantum}. If the thermal element is in a coherent thermal state, however, the expectation values of $\braket{a}$ changes slightly and $\sigma_{11}$ in Eq.~\eqref{covariance:matrix:elements:large:input:photons:far:from:critical:times} will look different.  However, if we approximate $\sigma_{11}$ as $\sigma_{11} \approx 1  + 2 \, |\mu_{\mathrm{c}}|^2$, which follows from that $\braket{a} \sim 0 $ for very large $\mu_{\mathrm{c}}$, the non-zero covariance matrix elements of the coherent thermal mechanical state are the same as in~Eq. \eqref{covariance:matrix:elements:large:input:photons:far:from:critical:times}. We therefore conclude that an initially thermal coherent mechanical state will also exhibit most of the non-Gaussianity we will examine for coherent mechanical states. 

With this simplified matrix, we are able to find a simple and analytic expression for the symplectic eigenvalues, which reads
\begin{align} \label{eq:approx:symplectic:eigenvalues}
\nu_+ &\sim 1 + 2 \, |\mu_{\mathrm{c}}|^2 \left( 1 - e^{-4 \, |\mu_{\mathrm{c}}|^2 \sin^2{\theta(\tau)/2}} \, e^{- |K_{\hat N_a}|^2}  \right) \, , \nonumber \\
\nu_- &\sim\sqrt{4 \, |\mu_{\mathrm{c}}|^2 \, |K_{\hat N_a}|^2 +1} \, .
\end{align}
We note that both eigenvalues grow with $|\mu_{\mathrm{c}}|$, as expected from our analysis in Section~\ref{chap:introduction:non:Gaussianity:general:behaviour}. The amount of non-Gaussianity for large $\mu_{\mathrm{c}}$ is now given by the following expression
\begin{equation} \label{eq:delta:large}
\delta(\tau) \sim  s_V \left( 1 + 2 \, |\mu_{\mathrm{c}}|^2 \left( 1 - e^{-4 \, |\mu_{\mathrm{c}}|^2 \sin^2{\theta(\tau)/2}} \, e^{- |K_{\hat N_a}|^2}  \right)\right) + s_V(\sqrt{4 \, |\mu_{\mathrm{c}}|^2 \, |K_{\hat N_a}|^2 +1}),
\end{equation}
which scales asymptotically as $\delta(\tau)\sim\tilde{\delta}(\tau) :=  4 \, \ln\,|\mu_{\mathrm{c}}|$, in perfect agreement with Eq.~\eqref{large:parameter:prediction:expansion}.
Note that Eq.~\eqref{eq:delta:large} is also valid for a time-dependent light--matter coupling $\tilde{\mathcal{G}}(\tau)$. In all cases, the nonlinearity grows as $\ln{|\mu_{\mathrm{c}}|}$ to leading order. In Sections~\ref{subsec:large:coherent:state} and~\ref{subsec:large:coherent:state:resonance} we will compare the asymptotic measure $\tilde{\delta}(\tau)$ with the full measure $\delta$ for different cases.

\section{Non-Gaussianity from a constant nonlinear coupling} \label{chap:non:Guassianity:coupling:sec:constant:coupling}
Let us now move on to a quantitative analysis of the evolving non-Gaussianity in different contexts. 
We begin by considering the case where the nonlinear light--matter interaction is constant: $\tilde{\mathcal{G}}(\tau)\equiv\tilde{g}_0$. To a large extent, this is the case for most experimental systems. The coefficients  which determine the time-evolution are those found in Eq.~\eqref{app:coefficients:constant:g0} in Appendix~\ref{app:coefficients}, and we note that the function $K_{\hat N_a}$, defined in Eq.~\eqref{eq:combined:coefficients}, which appears in the covariance matrix elements $\sigma_{nm}$ is now given by $K_{\hat N_a} = \tilde{g}_0 \, ( 1 - e^{- i \, \tau})$. 

\begin{figure}[t!]
\centering
\subfloat[ \label{fig:measure:vs:time:various:mu}]{%
  \includegraphics[width=.5\linewidth, trim = 0mm 0mm 0mm 0mm]{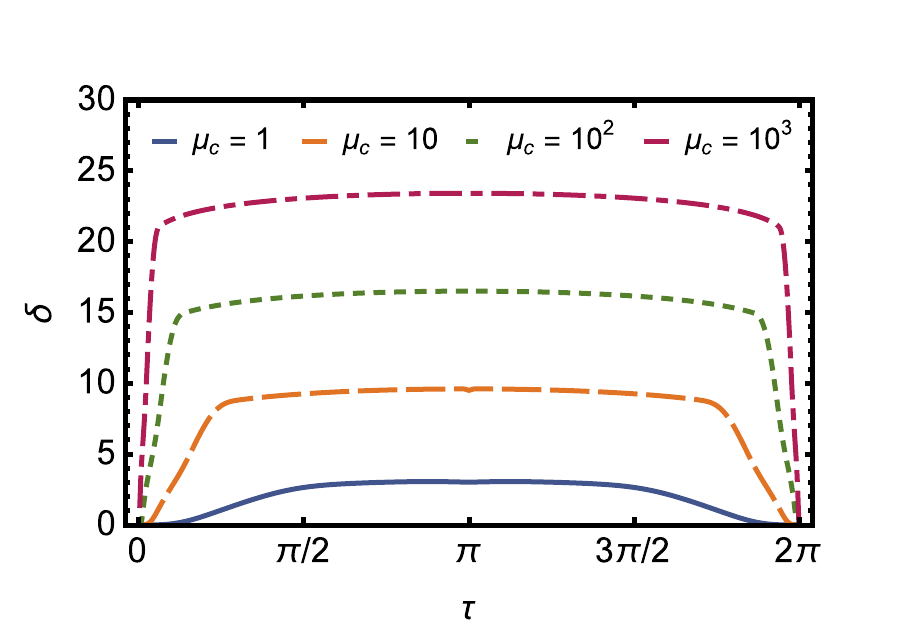}%
}\hfill
\subfloat[ \label{fig:measure:vs:time:various:mu:around:pi}]{%
  \includegraphics[width=.5\linewidth, trim = 0mm 0mm 0mm 0mm]{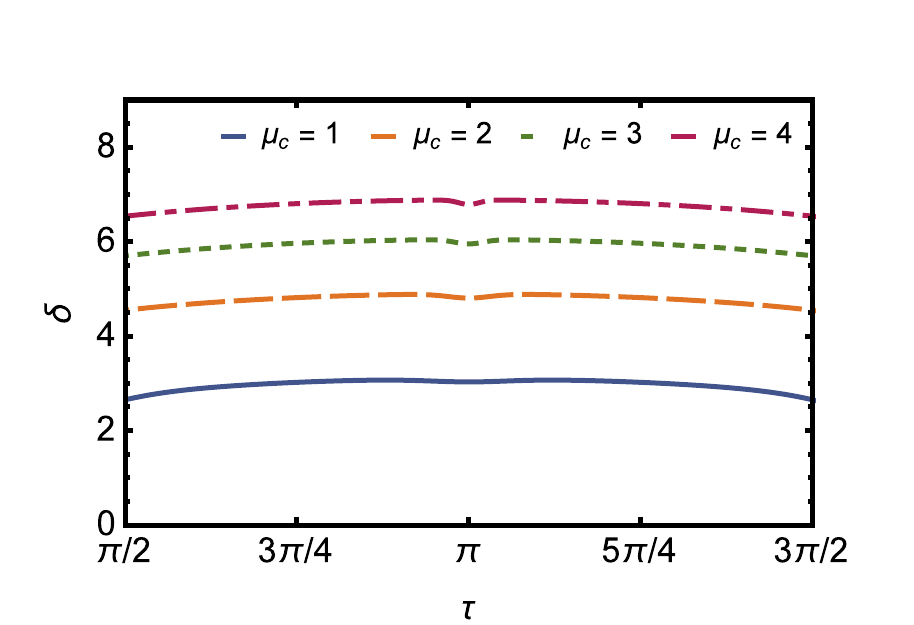}%
}\hfill
\subfloat[
\label{fig:measure:for:small:tau}]{%
  \includegraphics[width=.5\linewidth, trim = 0mm 0mm 0mm 0mm]{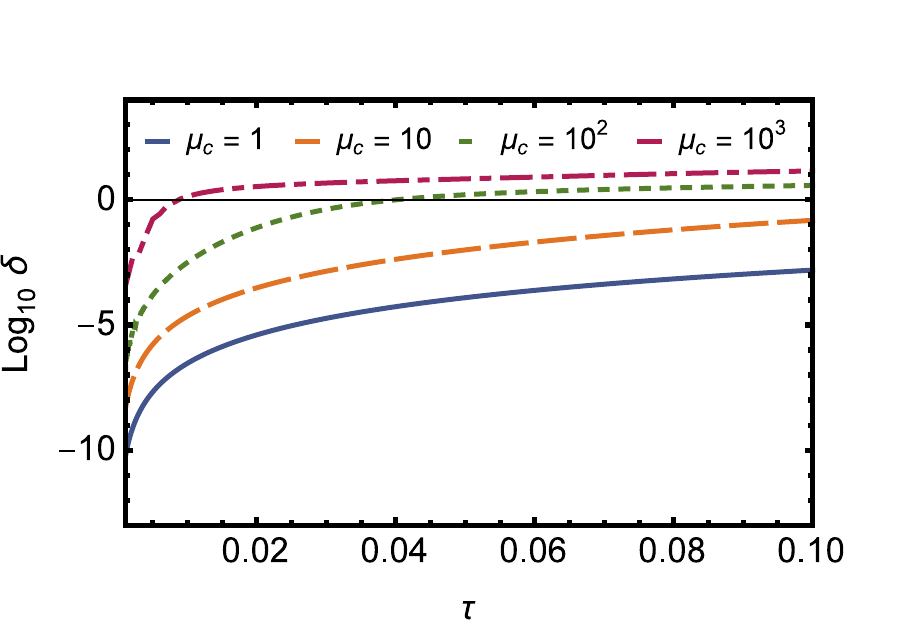}%
}\hfill
\caption[Measure of non-Gaussianity vs. time for a constant nonlinear coupling]{The measure of non-Gaussianity  $\delta(\tau)$ vs. time $\tau$ for systems with constant nonlinear coupling $\tilde{g}_0$. \textbf{(a)} shows a plot of $\delta(\tau)$ as a function of time $\tau$ for different coherent state parameters $\mu_{\mathrm{c}}$. The rescaled coupling is $\tilde{g}_0 = 1$ and the mechanics is in the ground state with  $\mu_{\mathrm{m}} = 0$. \textbf{(b)} shows a plot of $\delta$ vs. time $\tau$ near $\tau = \pi$ for varying  $\mu_{\mathrm{c}}$. The measure displays a local minimum centered around $\tau = \pi$ that becomes sharper with larger $\mu_{\mathrm{c}}$. Here $\tilde{g}_0 = 1$ and $\mu_{\mathrm{m}} = 0$. \textbf{(c)} shows plot of $\log_{10} \delta(\tau)$ at very small times $\tau$ for different coherent state parameters $\mu_{\mathrm{c}}$, $\tilde{g}_0 = 1$ and $\mu_{\mathrm{m}} = 0$. The measure increases exponentially at first before it plateaus towards a constant value, which is the overall behaviour we observe in~(a). }
\label{fig:constant:measure:vs:time}
\end{figure}

\begin{figure}[t!]
\subfloat[
\label{fig:measure:vs:C1}]{%
  \includegraphics[width=.5\linewidth, trim = 0mm 0mm 0mm 0mm]{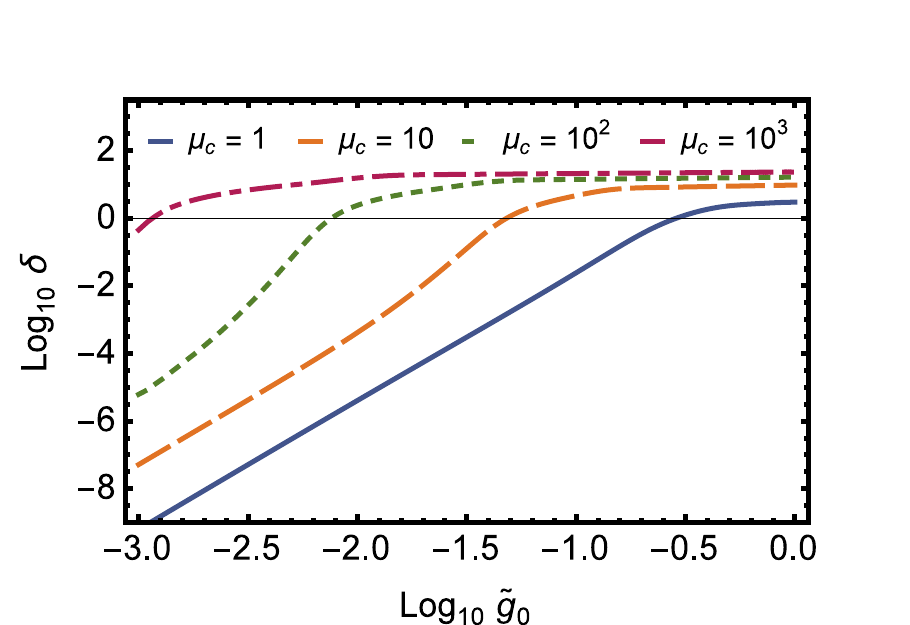}%
} \hfill
\subfloat[
\label{fig:measure:vs:mu}]{%
  \includegraphics[width=.5\linewidth, trim = 0mm 0mm 0mm 0mm]{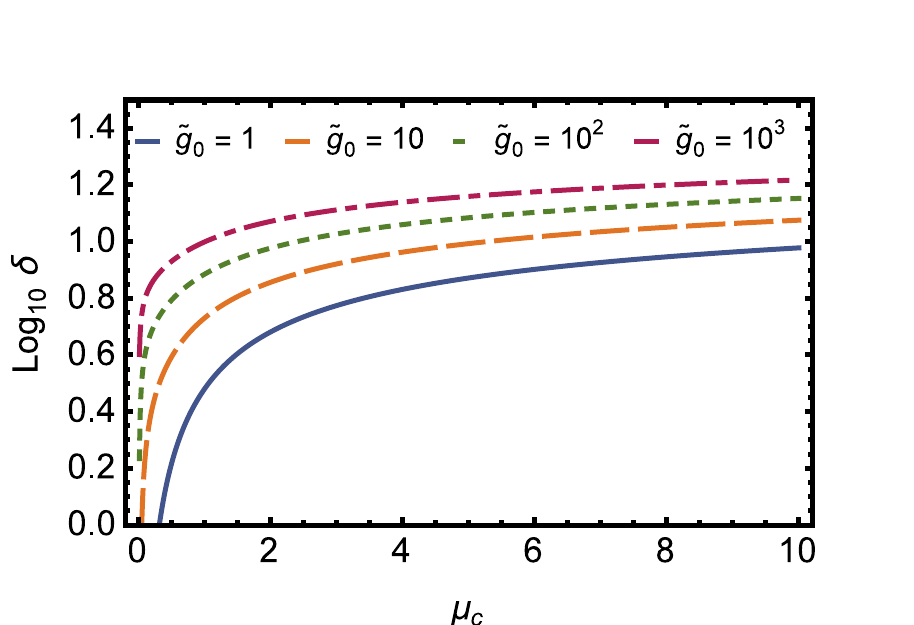}%
} \hfill
\caption[Measure of non-Gaussianity at $\tau = \pi$ for a constant coupling]{The behaviour of the measure of non-Gaussianity  $\delta(\tau)$ at $\tau = \pi$ for systems with constant nonlinear coupling $\tilde{g}_0$ starting in coherent states. \textbf{(a)}  shows a log--log plot of $\delta(\tau)$ vs. the rescaled coupling $\tilde{g}_0$. As $\tilde{g}_0$ increases, the state becomes more and more non-Gaussian, polynomially at first but then it quickly tends towards a constant value. \textbf{(b)} shows a log plot of $\delta(\tau)$ vs. the coherent state parameter $\mu_{\mathrm{c}}$ for different values of $\tilde{g}_0$. $\delta(\tau)$ first increases quickly, then plateaus towards a single value. }
\label{fig:constant:measure:scaling}
\end{figure}

We now proceed to compute the exact measure of non-Gaussianity $\delta(\tau)$ for constant coupling $\tilde{g}_0$ and with the system initially in two coherent states. The exact expression is again too long and cumbersome to be reprinted here, but we plot the results in Figure~\ref{fig:constant:measure:vs:time} and Figure~\ref{fig:constant:measure:scaling}. In Figure~\ref{fig:measure:vs:time:various:mu} we plot the measure of non-Gaussianity $\delta(\tau)$ as a function of time $\tau$ for different values of the coherent state parameter $\mu_{\mathrm{c}}$, over the period $0<\tau<2\,\pi$. The other parameters are set to $\tilde{g}_0 = 1$ and $\mu_{\mathrm{m}} = 0$.  It is known that the full nonlinear dynamics is periodic (or recurrent) with period $2\,\pi$ whenever $\tilde{g}^2_0$ is an integer. This is clearly reflected in our plot.  

At $\tau = 2\pi$, the optics and mechanics are no longer entangled, and while the mechanics returns to its initial state,  the final optical state will depends on the value of $\tilde{g}_0$. For example, when $\tilde{g}_0  = 0.5$, the cavity state becomes a superposition of coherent states at $\tau = 2\pi$, also known as a cat state~\cite{bose1997preparation}. However, if $\tilde{g}_0^2$ is integer, we obtain a phase factor of $e^{2  \pi  i \,  \tilde{g}_0^2} = 1$ in the optical state, and the optics returns to an initial state as well. This is the case in Figure~\ref{fig:constant:measure:vs:time},  where $\delta(2\pi) = 0$. We will make use of the asymptotic measure defined in Section~\ref{sec:asymptotic:large:coherent:state} to analyse this behaviour, see Section~\ref{subsec:large:coherent:state}. Furthermore, while it might seem that the non-Gaussianity peaks at $\tau = \pi$, the measure $\delta$ exhibits a local minimum which grows increasingly narrow with larger $\mu_{\mathrm{c}}$. This is apparent from Figure~\ref{fig:measure:vs:time:various:mu:around:pi} where we have shown a close-up of $\delta$ around $\tau = \pi$ for increasing values of $\mu_{\mathrm{c}}$, and for $\tilde{g}_0 = 1$ and $\mu_{\mathrm{m}} = 0$.  The dip occurs because at $\tau = \pi$, we find that $\theta(\pi) = - 2  \pi \, \tilde{g}^2_0$ and $K_{\hat N_a} = - 2 \,  \tilde{g}_0$. 
Thus, for integer $\tilde{g}_0^2$, we have $\sin^2{\theta(\pi)/2} = 0$ and $\sigma_{11}$ becomes $\sigma_{11} = 1 + 2 \,  |\mu_{\mathrm{c}}|^2 \left(1 - e^{- 4 \, \tilde{g}_0^2} \right)$. The non-zero exponent causes the non-Gaussianity to temporarily decrease, and the same behaviour occurs in the other covariance matrix elements, resulting in the dip.

As already noted, increasing $\mu_{\mathrm{c}}$ yields a logarithmic increase in $\delta(\tau)$, which is evident from the approximation in Eq.~\eqref{eq:delta:large}. 
Figure~\ref{fig:measure:vs:time:various:mu} also implies that for closed dynamics, the nonlinear system  almost immediately becomes maximally non-Gaussian. It then retains approximately the same amount of non-Gaussianity until $\tau = 2\pi$, meaning there is a rapid decrease of non-Gaussianity before the system revives again. 
Such plateaus are justified mathematically by analysing the asymptotic behaviour for large $|\mu_{\mathrm{c}}|$ in the next section. The appearance of these plateaus shows that the maximum amount of non-Gaussianity available during one cycle can be accessed almost immediately without requiring the system to evolve for a long time. As a side remark, we note that the functional form of $\delta(\tau)$ in Figure~\ref{fig:measure:vs:time:various:mu}  closely resembles the linear entropy of the traced-out subsystems as found in Ref~\cite{bose1997preparation}.

To better understand the behaviour of $\delta(\tau)$ for small times $\tau$,  we plot the behaviour of $\log_{10} {\delta(\tau)}$ for $\tau \ll 1$ for different values of $\mu_{\mathrm{c}}$ in Figure~\ref{fig:measure:for:small:tau}. We note that $\delta(\tau)$ increases quickly at first, but soon tends to a near-constant value. This means that $\delta(\tau)$ grows linearly for an interval of small times, which can be seen as the increasing and decreasing parts in Figure~\ref{fig:measure:vs:time:various:mu}. 

Finally, we proceed to examine the scaling behaviour of $\delta(\tau)$ at fixed time $\tau = \pi$. Figure~\ref{fig:measure:vs:C1} shows a $\log_{10}$--$\log_{10}$ plot of the measure $\delta(\tau)$ as a function of the nonlinear coupling $\tilde{g}_0$ for different values of $\mu_{\mathrm{c}}$. As $\tilde{g}_0$ increases, the amount of non-Gaussianity first grows linearly in the logarithm, then plateaus as $\tilde{g}_0$ increases further. The same behaviour occurs for larger $\mu_{\mathrm{c}}$, only more rapidly. This suggests that if we wish to increase the non-Gaussianity substantially, it will become increasingly difficult to do so by increasing $\tilde{g}_0$. As such, focusing on increasing the coupling $\tilde{g}_0$ will only give marginal returns. Similarly,~\ref{fig:measure:vs:mu} shows $\log_{10}{\delta(\tau)}$ as a function of increasing $\mu_{\mathrm{c}}$ for various values of $\tilde{g}_0$.

\subsection{Small coherent state parameters}
For a small amplitude coherent state of the optics, with $|\mu_{\mathrm{c}}|^2 \ll 1$, and with the mechanics in a coherent state, we found in Eq.~\eqref{eq:delta:small} that $\delta(\tau)$ scales with $\sim |K_{\hat N_a}|^2 | \, \mu_{\mathrm{c}}|^2 \,  \ln |\mu_{\mathrm{c}}|$. Given the explicit form of $F$, we see that it scales with $K_{\hat N_a}\propto \tilde{g}_0$. Since $\delta(\tau)$ in this regime is proportional to $|K_{\hat N_a}|^2$, it follows that $\delta(\tau)$ grows quadratically with the light--matter coupling in this regime.

\subsection{Large coherent state parameters} \label{subsec:large:coherent:state}

We derived an asymptotic form of $\delta(\tau)$  in Eq.~\eqref{eq:delta:large} for the case $|\mu_{\mathrm{c}}|\gg 1$, which we called $\tilde{\delta}(\tau)$. As argued before, the behaviour of the measure $\delta(\tau)$ in this regime depends crucially on the distance of $\theta(\tau)$ from the value $2\pi$. In our present case we have that  $\theta(\tau)\sim\tau^3$ for $\tau\ll1$ and $\theta(\tau)\sim-\tilde{g}_0^2\tau$ for $\tau \gg1$. The functions that we decided to ignore (except for $\sigma_{11}$) in the derivation of $\tilde{\delta}(\tau)$ are of the form $f_{|\mu_{\mathrm{c}}|}(\theta(\tau))=(1-\exp[-\beta\,|\mu_{\mathrm{c}}|^2\sin(\theta(\tau)/2)])$ or  $f_{|\mu_{\mathrm{c}}|}(\theta(\tau))=(1-\exp[-\beta'\,|\mu_{\mathrm{c}}|^2\sin^2(\theta(\tau)/2)])$ where $\beta$ and $\beta'$ are irrelevant numerical constant of order $1$. We focus on $f_{|\mu_{\mathrm{c}}|}(\theta(\tau))$ and note that a similar argument applies for the other function as well. Finally, we ignore the transient regime of $\tau\ll1$ and focus on times $\theta(\tau)\sim-\tilde{g}_0^2 \, \tau$.

\begin{figure}[t!]
\subfloat[ \label{fig:approximate:measure:vs:time:various:mu}]{%
  \includegraphics[width=.47\linewidth, trim = 0mm 0mm 0mm 0mm]{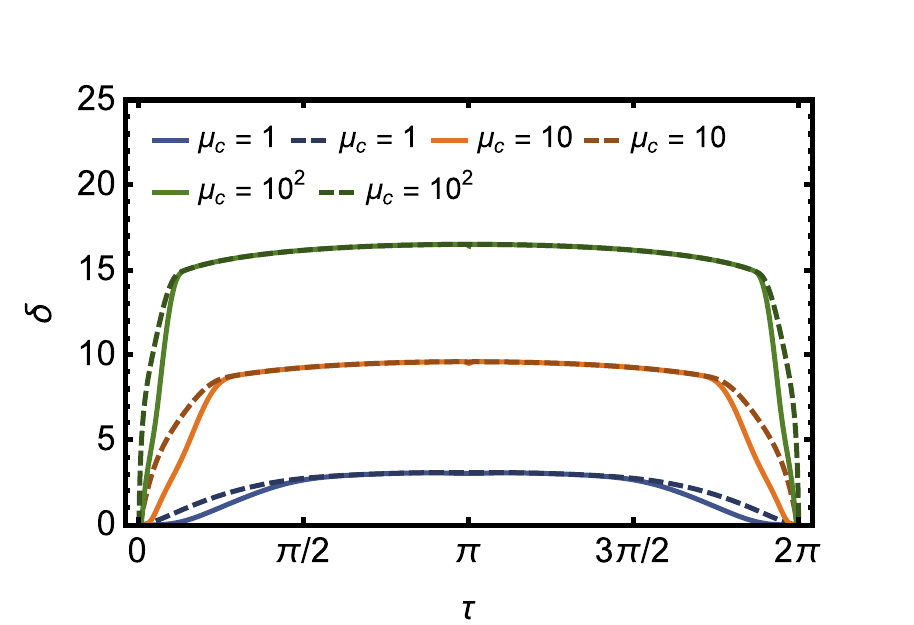}%
}\hfill
\subfloat[
\label{fig:approximate:measure:vs:time:various:g0}]{%
  \includegraphics[width=.47\linewidth, trim = 0mm 0mm 0mm 0mm]{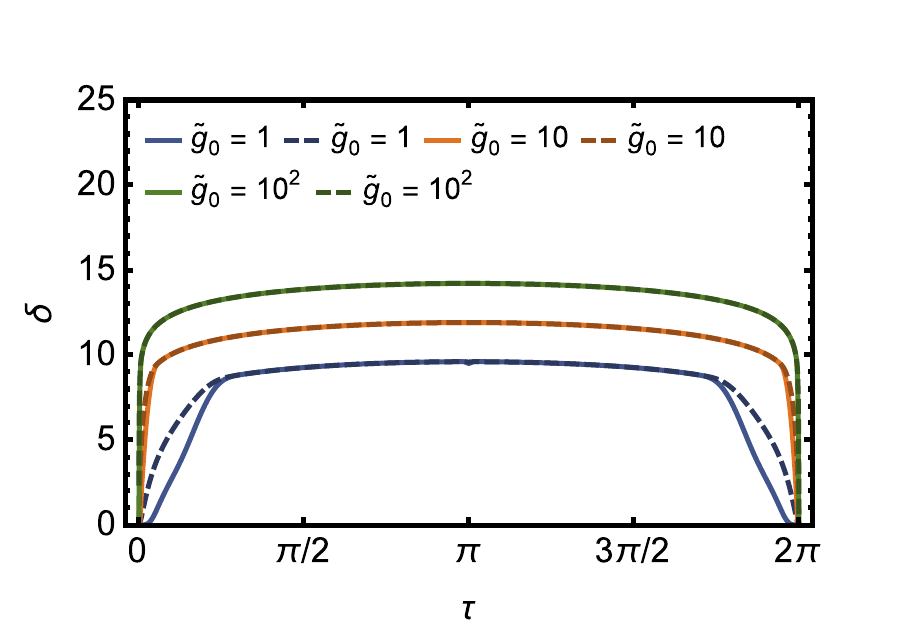}%
}
\caption[Comparison of the full and approximate measure for a constant coupling]{ Comparing the measure of non-Gaussianity $\delta(\tau)$ (solid lines) with the asymptotic form computed in Eq.~\eqref{eq:delta:large} (dashed lines) for coherent states. \textbf{(a)} shows the exact measure (solid line) vs. the approximation for different values of $\mu_{\mathrm{c}}$. As $\mu_{\mathrm{c}}$ increases, the approximation grows increasingly accurate. In this plot, $\tilde{g}_0 = 1$. \textbf{(b)} shows the exact measure (solid line) vs. the approximation for large $\mu_{\mathrm{c}}$ for increasing values of $\tilde{g}_0$ and $\mu_{\mathrm{c}} = 10$. The approximation becomes increasingly accurate as $\tilde{g}_0$ increases, even towards the beginning and end of one oscillation period. }
\label{fig:measure:compared:with:approximation}
\end{figure}

To see how well the asymptotic form $\tilde{\delta}(\tau)$ in Eq.~\eqref{eq:delta:large} approximates the exact measure, we have plotted both the exact form of $\delta(\tau)$ (solid lines) with the asymptotic form (dashed lines) in Figure~\ref{fig:measure:compared:with:approximation}. In Figure~\ref{fig:approximate:measure:vs:time:various:mu}, where we plot $\delta(\tau)$ and $\tilde{\delta}(\tau)$ for different $\mu_{\rm{c}}$, we note that, even for $|\mu_{\mathrm{c}}|\sim 1$, the asymptotic measure $\tilde{\delta}(\tau)$ well approximates the exact value of $\delta(\tau)$. In fact, it becomes even more accurate as the optical coherent state parameter $\mu_{\mathrm{c}}$ increases, which is to be expected given the nature of the approximation. The asymptotic form also becomes  more accurate once we also increase $\tilde{g}_0$, as evident in Figure~\ref{fig:approximate:measure:vs:time:various:g0}. For  $\tilde{g}_0 = 10^2$, the approximation is almost entirely accurate. This occurs because the function $\theta(\tau)$ increases with $\tilde{g}_0$, which further suppresses the off-diagonal covariance matrix elements at the beginning and end of each cycle.

Let us discuss the fact that the measure recurs with $\tau = 2\pi$ for integer $\tilde{g}_0^2$ which we can now address analytically by examining the asymptotic covariance matrix elements in Eq.~\eqref{covariance:matrix:elements:large:input:photons:far:from:critical:times}. We find that $K_{\hat N_a} = 0$ for all $\tau = 2\pi n $ with integer $n$. This means that $\sigma_{42} = 0$ and that $\sigma_{22} = 1$. We also find that $\theta_n(2\pi)  = - 4 \pi \tilde{g}_0^2 $. Thus, if $\tilde{g}_0^2$ is integer, we find that $\sin^2{\theta(\tau)/2} = 0$ and the final covariance matrix element is $\sigma_{11} = 1$. This results in $\mathbf{\sigma} = \mathrm{diag}(1,1,1,1)$ which corresponds to a coherent state, which is fully Gaussian. As a result, the non-Gaussianity vanishes. When $\tilde{g}_0^2$ is not an integer, some non-Gaussianity will be retained, but the fact that $F = 0$ will still result in a reduction at $\tau = 2\pi$. 

\subsection{Non-Gaussianity in open systems with constant coupling} \label{subsec:open:system:constant:coupling}
\begin{figure}[t!]
\subfloat[ \label{fig:open:system:photon:decoherence:constant:coupling}]{%
  \includegraphics[width=.47\linewidth, trim = 0mm 0mm 0mm 0mm]{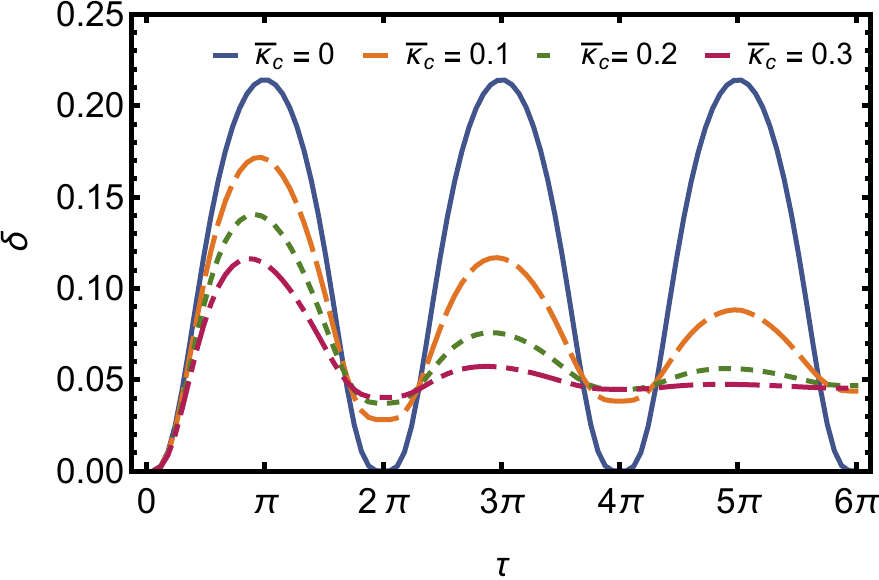}%
}\hfill
\subfloat[
\label{fig:open:system:phonon:decoherence:constant:coupling}]{%
  \includegraphics[width=.47\linewidth, trim = 0mm 0mm 0mm 0mm]{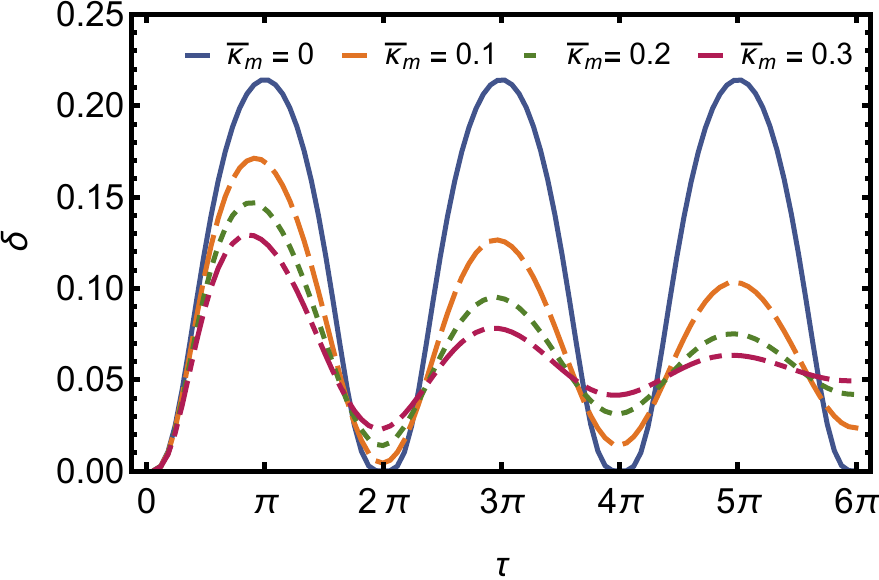}%
}
\caption[Measure of non-Gaussianity of an open system with constant coupling]{Non-Gaussianity of open optomechanical systems with constant light--matter coupling starting in a coherent state. \textbf{(a)} shows the measure of non-Gaussianity $\delta(\tau)$ vs. time $\tau$ for a system with increasing values of photon decoherence $\bar{\kappa}_{\mathrm{c}}$ for $\tilde{g}_0 = 1$, $\mu_{\mathrm{c}} = 0.1$ and $\mu_{\mathrm{m}} = 0$. \textbf{(b)} shows the non-Gaussianity $\delta$ vs. time $\tau$ for a system with increasing values of phonon decoherence $\bar{\kappa}_{\mathrm{m}}$ for $\tilde{g}_0 = 1$, $\mu_{\mathrm{c}} = 0.1$ and $\mu_{\mathrm{m}} = 0$. The non-Gaussianity tends to a steady-state for both cases as $\tau$ increase. A populated mechanical coherent state $\mu_{\mathrm{m}} \neq 0$ does not affect the non-Gaussianity.}
\label{chap:non:Gaussianity:coupling:fig:measure:decoherence:constant:coupling}
\end{figure}

Any realistic system will suffer from decoherence. In Figure~\ref{chap:non:Gaussianity:coupling:fig:measure:decoherence:constant:coupling} we have plotted the non-Gaussianity $\delta$ as a function of time for an optomechanical system with open dynamics. Here, the cavity state and the mechanics are both in initial coherent states in Eq.~\eqref{initial:state:two}. Figure~\ref{fig:open:system:photon:decoherence:constant:coupling} shows the non-Gaussianity for increasing values of the photon decoherence rate $\bar{\kappa}_{\mathrm{c}} = \kappa_{\mathrm{c}}/\omega_{\mathrm{m}}$ with Lindblad operator $\hat L _{\mathrm{c}} = \sqrt{\bar{\kappa}_{\mathrm{c}}} \, \hat a$ and values $\mu_{\mathrm{c}} = 0.1$, $\tilde{g}_0  = 1$ and $\mu_{\mathrm{m}} = 0$. We have chosen a low value of $\mu_{\mathrm{c}} $ to ensure high numerical accuracy of the simulation, as larger values quickly lead to numerical instabilities. We note that the non-Gaussianity $\delta(\tau)$ tends towards a steady value, which is clear from the fact that the higher values of decoherence start to coincide around $\tau = 5\pi$. We also note that around $\tau = 2\pi n$, for integer $n$ the inclusion of noise appears to temporarily increase the non-Gaussianity. This could, however, be due to the fact that the relative entropy measure cannot distinguish between non-Gaussianity induced as a result of genuinely nonlinear dynamics or as a result of classical mixing of the states~\cite{barbieri2010non}. We discuss this further in Section~\ref{chap:non:Gaussianity:coupling:sec:discussion}. Similarly, in Figure~\ref{fig:open:system:phonon:decoherence:constant:coupling} we have plotted the non-Gaussianity $\delta$ for increasing values of phonon decoherence rate $\bar{\kappa}_{\mathrm{m}}$ with Lindblad operator $\hat L _{\mathrm{m}} = \sqrt{\bar{\kappa}_{\mathrm{m}}} \, \hat b$ and the same values as before.

\section{Non-Gaussianity from a time-dependent nonlinear coupling} \label{chap:non:Gaussianity:coupling:sec:time:dependent:coupling}
In all physical systems, such as optomechanical cavities, the confining trap is not ideal. This means that, in general, the coupling $\tilde{g}(\tau)$ is time-dependent as a consequence of, for example, trap instabilities. Time-dependent variations such as phase fluctuations in the laser beam used to trap a levitated bead generally cause the coupling to fluctuate in time. 

In this work, we want to exploit the possibility of controlling the coupling $\tilde{g}(\tau)$ by considering its periodic modulation in time.
In practice, such time-dependent control would be achievable for an optically trapped and levitated dielectric bead that interacts with a cavity field by controlling the optical phase of the trapping laser field. In fact, such phase determines the bead's equilibrium position which, in turn, affects the cavity light-bead coupling through the varying overlap between the bead and the cavity mode function.
Technically, the trapping laser's optical phase may be controlled through an acoustic-optical modulator. 
Alternately, control on a bead's equilibrium position may also be enacted by adopting Paul traps, which work for levitated nanospheres~\cite{millen2015cavity}, and have been used to shuttle ions across large distances, typically for the purpose of quantum information processing~\cite{hensinger2006t, walther2012controlling}. See also Section~\ref{chap:non:Gaussianity:coupling:sec:discussion} for a discussion of additional methods by which the coupling can be modulated. 

\subsection{Modelling the trap modulation}\label{trap:instability:section}
We shall model a time-dependent light--matter coupling $\tilde{\mathcal{G}}(\tau)$ by assuming that the coupling has the simple form 
\begin{align}
\tilde{\mathcal{G}}(\tau) =\tilde{g}_0\,\left( 1+ \epsilon\sin(\Omega_g \,\tau) \right).
\end{align}
Here, $\tilde{g}_0$ is the expected value of the coupling, $\epsilon$ is the amplitude of oscillation and $\Omega_g:=\omega_g/\omega_\textrm{m}$ is the dimensionless frequency that determines how the coupling oscillates in time. We can insert this ansatz in the general expressions in Eq.~\eqref{sub:algebra:decoupling:solution:text} and obtain an explicit form for this case. The full expressions for the coefficients in Eq.~\eqref{sub:algebra:decoupling:solution:text} are again very long and cumbersome, and we do not print them here. They are listed in Eq.~\eqref{app:coefficients:eq:time:dependent:g0} in Appendix~\ref{app:coefficients}

\begin{figure}[t!]
\centering
\subfloat[ \label{fig:time:dependent:plain}]{%
  \includegraphics[width=.45\linewidth, trim = 0mm 0mm 0mm 0mm]{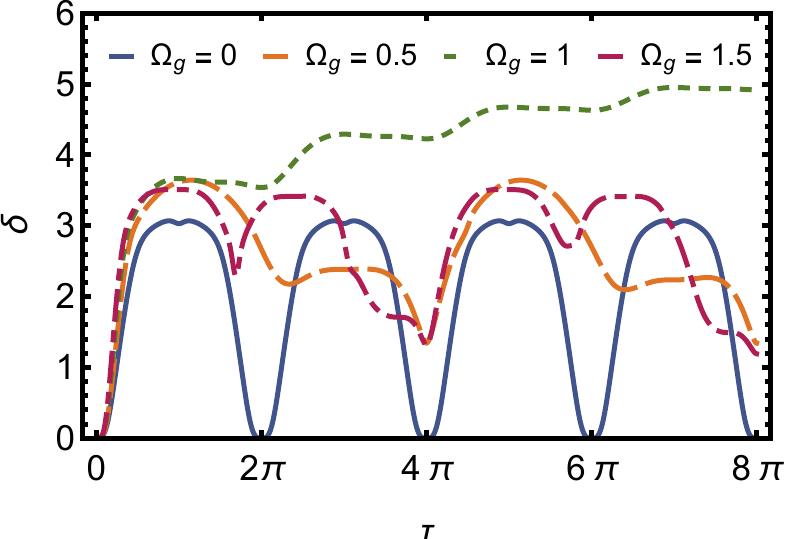}%
}\hfill
\subfloat[
\label{fig:resonance}]{%
  \includegraphics[width=.5\linewidth, trim = 0mm 0mm 0mm 0mm]{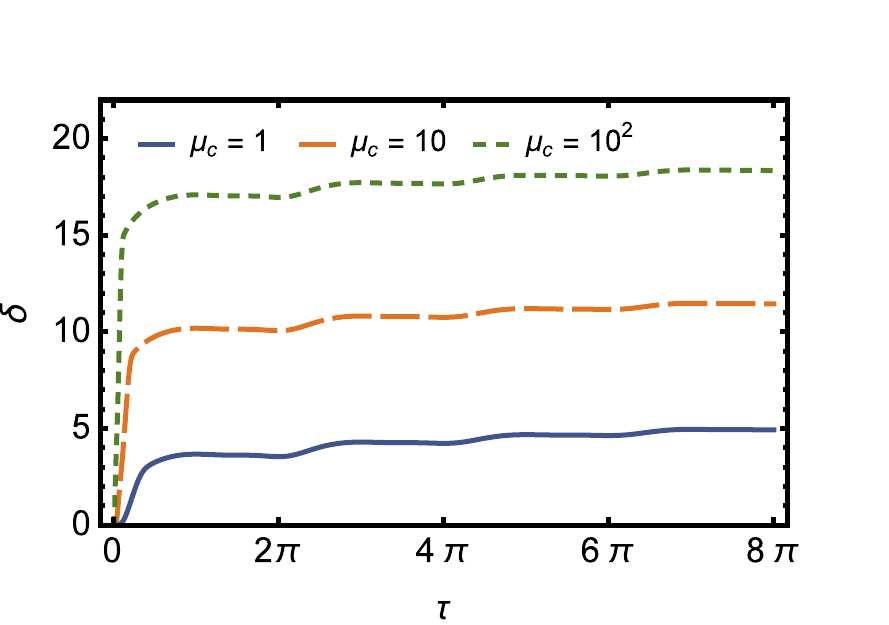}%
}\hfill
\subfloat[
\label{fig:resonance:two}]{%
  \includegraphics[width=.5\linewidth, trim = 0mm 0mm 0mm 0mm]{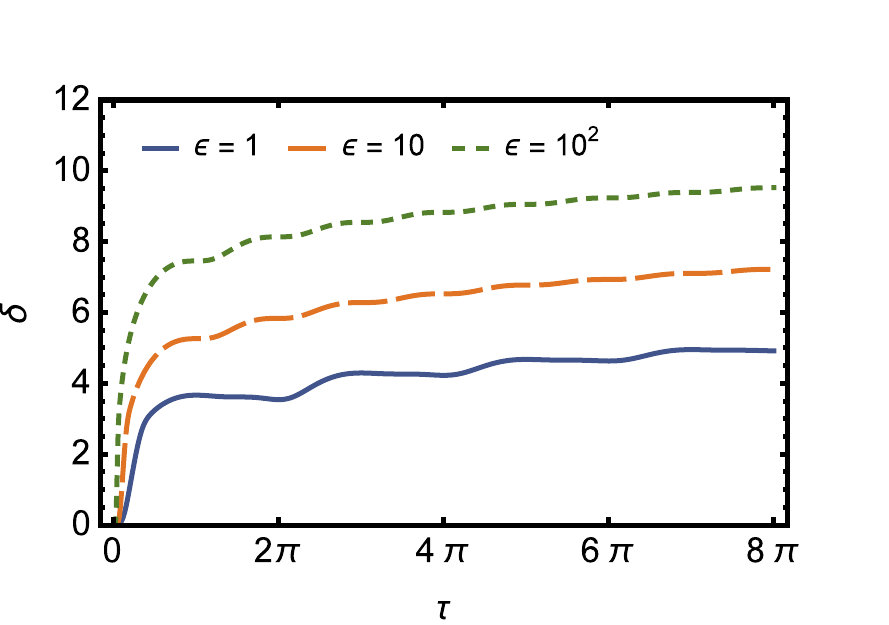}%
}
\caption[Measure of non-Gaussianity as a function of time for a time-dependent coupling]{The measure of non-Gaussianity  $\delta(\tau)$ for systems with time-dependent coupling $\tilde{\mathcal{G}}(\tau) = \tilde{g}_0(1 + \epsilon \sin{(\Omega_g \tau)} )$, where $\epsilon$ is the amplitude and $\Omega_g = \omega_g/\omega_{\mathrm{m}}$ is the modulation frequency. \textbf{(a)} shows the measure $\delta(\tau)$ vs. rescaled time $\tau$ for different values of $\Omega_g$. The case $\Omega_g = 0$ (blue line) corresponds to the time-independent setting. At resonance, with $\Omega_g = 1 $ (green line), the system displays a drastically different behaviour. Other parameters are $\tilde{g}_0 = \epsilon = 1$ and $\mu_{\mathrm{m}} = 0$. \textbf{(b)} shows $\delta(\tau)$ vs. rescaled time $\tau$ at resonance $\Omega_g = 1$ for various values of coherent state parameter $\mu_{\mathrm{c}}$. The system no longer exhibits closed dynamics. Other parameters include $\tilde{g}_0 = \epsilon = 1$ and $\mu_{\mathrm{m}} = 0$. \textbf{(c)} shows $\delta(\tau)$ vs. time $\tau$ for increasing oscillation frequency $\epsilon$ at resonance $\Omega_g = 1$ and with $\mu_{\mathrm{c}}  = 1$. $\delta(\tau)$ increases slowly with $\epsilon$. Again, we have set $\mu_{\mathrm{m}} = 0$. }
\label{fig:time:dependent}
\end{figure}

We can now compute $\delta(\tau)$ for this time-dependent coupling for initial coherent states, and we display the results in Figure~\ref{fig:time:dependent}. In~\ref{fig:time:dependent:plain} we plot $\delta(\tau)$ vs. $\tau$ for different values of the oscillation frequency $\Omega_g$. Note that we here include a larger range of $\tau$ to capture potentially recurring behaviour. In the limit $\Omega_g \rightarrow 0$ we recover the time-independent solution, as expected. Interestingly, when $\Omega_g\neq0$ we see that we can achieve higher values for the nonlinear measure $\delta(\tau)$. This is especially pronounced as $\Omega_g\rightarrow 1$, where the trap oscillation frequency is equal to the mechanical frequency $\omega_{\mathrm{m}}$, for which $\delta(\tau)$ ceases to oscillate periodically, but instead steadily increases. We discuss this case in detail in the following section.

\subsection{Trap modulation on resonance}\label{resonance}
The functions in Eq.~\eqref{app:coefficients:eq:time:dependent:g0} in Appendix~\ref{app:coefficients} contain denominators of the form $\Omega_g-1$. Therefore, among all possible values of $\Omega_g$, we can ask what happens \textit{on resonance}, i.e., when $\Omega_g=1$. Figure~\ref{fig:time:dependent:plain} already provides evidence that the system should behave markedly differently. 

At resonance, where $\Omega_g = 1$, the functions in Eq.~\eqref{app:coefficients:eq:time:dependent:g0} take the relatively simple form
\begin{align}\label{eq:resonance:coefficients}
F_{\hat N_a^2} &=-\frac{1}{16} \tilde{g}_0 \, \bigl[ 16 \, \tau-8 \sin (2\, \tau)+\epsilon \, (32-36 \cos (\tau)+4 \cos (3 \, \tau))\nonumber \\
&\quad\quad\quad\quad\quad\quad\quad +\epsilon ^2 \,   \bigl( 6 \, \tau-4 \sin (2\, \tau)+\sin (2 \, \tau)\,\cos (2 \, \tau) \bigr)\bigr] \, ,\nonumber\\
F_{\hat N_a \, \hat B_+} &= -\tilde{g}_0 \sin (\tau) \left(1+ \frac{\epsilon}{2} \sin (\tau)\right) \, . \nonumber \\
F_{\hat N_a \, \hat B_-} &=\frac{\tilde{g}_0}{4} \epsilon   \,  \left(  \sin (2 \, \tau)-2 \, \tau  \right) -  2 \, \tilde{g}_0 \,  \sin^2\left( \frac{\tau}{2} \right) \, .
\end{align}
We have plotted  the exact measure of non-Gaussianity $\delta(\tau)$ for the resonant case for initially coherent states and for different values of $\mu_{\mathrm{c}}$ in Figure~\ref{fig:time:dependent}. As anticipated, here we no longer have recurrent behaviour. Instead, the non-linearity increases as $\ln{\tau}$. Formally, this growth can continue for arbitrarily large times $\tau$, however, the maximum time $\tau$ that can be achieved in practice is limited by the coherence time of the experiment. Similarly, we plotted $\delta(\tau)$ for various values of $\epsilon $ in Figure~\ref{fig:resonance:two}. We note that $\delta(\tau)$ oscillates increasingly rapidly with larger $\epsilon$ but with decreasing amplitude for increasing $\tau$, as $|K_{\hat N_a}|^2\sim\tilde{g}_0^2\,\epsilon^2\,\tau^2$ becomes the dominant term for $\tau \gg 1$.

As already noted, it is evident from Figure~\ref{fig:time:dependent} that the non-Gaussianity increases continuously. 
The nonlinear coupling in the Hamiltonian is derived by considering the effect of photon pressure on the mechanical element. Given that the overall photon number $\langle \hat a ^\dag \hat a \rangle $ is conserved, the coupling acts as a photon number displacement. If this coupling is time-dependent, this means that the photon pressure displaces with a time-dependence. When this occurs the resonance, this linear displacement grows linearly in time. See also Ref~\cite{liao2014modulated} for further insight once the rotating wave approximation has been applied. 

\subsection{Large coherent state parameters at resonance}\label{subsec:large:coherent:state:resonance}
Using the explicit form of the coefficients Eq.~\eqref{eq:resonance:coefficients}, we note that $|K_{\hat N_a}|^2 = F_{\hat N_a \, \hat B_-}^2 + F_{\hat N_a \, \hat{B}_+}^2$ has the asymptotic behaviour $|K_{\hat N_a}|^2\sim\frac{1}{4}\tilde{g}_0^2\,\epsilon^2\,\tau^2$ for $\tau\gg1$. This implies that $\exp[-|K_{\hat N_a}|^2]\ll1$ for large $\tau$ and therefore we expect, as it happened in Section~\ref{sec:asymptotic:large:coherent:state}, that most covariance matrix elements  will vanish and will not contribute to the asymptotic form of $\delta(\tau)$. This observation allows us to compute the symplectic eigenvalues, which  read $\nu_+=1 + 2|\mu_{\mathrm{c}}|^2 \left( 1 - e^{-4 \, |\mu_{\mathrm{c}}|^2 \sin^2{\theta(\tau)/2}} \, e^{- |K_{\hat N_a}|^2}  \right)$ and $\nu_-=\sqrt{1+4\,|\mu_{\mathrm{c}}|^2\,|K_{\hat N_a}|^2}$, and they 
match the expressions Eq.~\eqref{eq:approx:symplectic:eigenvalues}. We again stress that we have retained the exact expression for $\sigma_{11}$ to capture some crucial features of the non-Gaussianity, such as $\delta(0) = 0$. 

\begin{figure}[t!]
\subfloat[ \label{fig:resonance:comparison:various:mu}]{%
  \includegraphics[width=.5\linewidth, trim = 0mm 0mm 0mm 0mm]{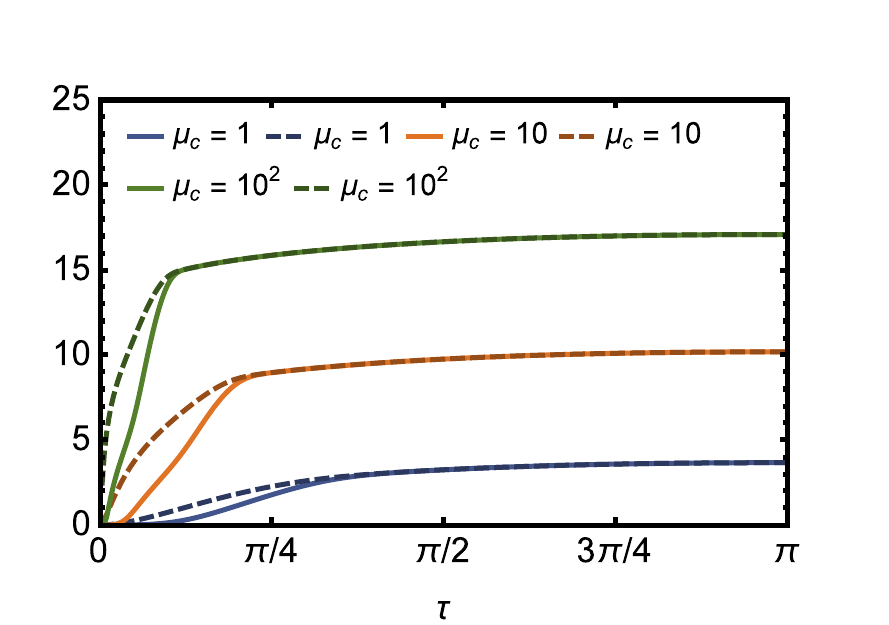}%
}\hfill
\subfloat[
\label{fig:resonance:comparison:various:delta:g}]{%
  \includegraphics[width=.5\linewidth, trim = 0mm 0mm 0mm 0mm]{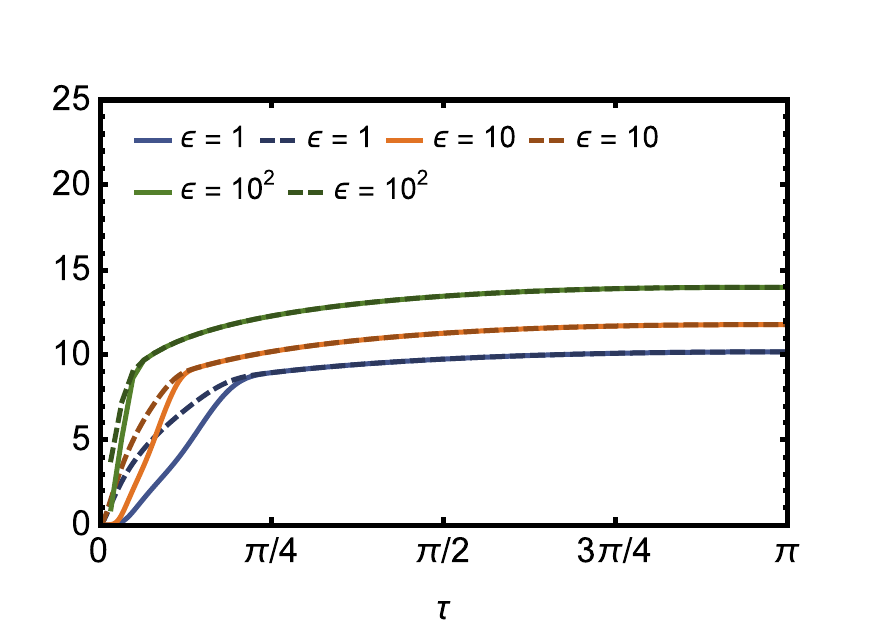}%
}
\caption[Comparison between the full and approximate measure at resonance]{A comparison between the full measure $\delta(\tau)$ (solid line) and the approximate measure (dashed lines) for time-dependent couplings $\tilde{\mathcal{G}}(\tau)$. \textbf{(a)} shows the accuracy of the approximation for different values of $\mu_{\mathrm{c}}$ at $\Omega_g = 0.5$. The approximation becomes very accurate as $\mu_{\mathrm{c}}$ increases. \textbf{(b)}  shows a comparison of the accuracy of $\tilde{\delta}$ for a different values of $\epsilon$ at $\mu_{\mathrm{c}} = 10$. The approximation becomes more accurate as $\epsilon$ increases. }\label{fig:time:dependent:comparison}
\end{figure}

In Figure~\ref{fig:time:dependent:comparison}, we compare the exact measure $\delta(\tau)$ at resonance with the asymptotic form derived  in Eq.~\eqref{eq:delta:large}. The solid lines represent the exact measure $\delta(\tau)$ and the dashed lines represent the asymptotic expression. In Figure~\ref{fig:resonance:comparison:various:mu} we compare them for different values of $\mu_{\mathrm{c}}$. We note that, except for at very small $\tau$, the asymptotic form is entirely accurate and gets even more precise for increasing values of $\mu_{\mathrm{c}}$. This is a consequence, as we noted before, of the exponentials in Eq.~\eqref{chap:non:Gaussianity:coupling:full:elements:covaraince:matrix} that suppress some elements for large $\mu_{\mathrm{c}}$, unless $\theta(\tau)=n\,\pi$. Similarly, in Figure~\ref{fig:resonance:comparison:various:delta:g} we have plotted $\delta(\tau)$ and its asymptotic form for different values of the oscillation amplitude $\epsilon$. Again, the suppression of the exponentials with increasing $\epsilon$ means that larger values of $\epsilon$ yield a more accurate expression.

\subsection{Open system dynamics at resonance}\label{subsec:open:system:resonance}
\begin{figure}[t!]
\subfloat[ \label{fig:open:system:photon:decoherence:resonance}]{%
  \includegraphics[width=.5\linewidth, trim = 0mm 0mm 0mm 0mm]{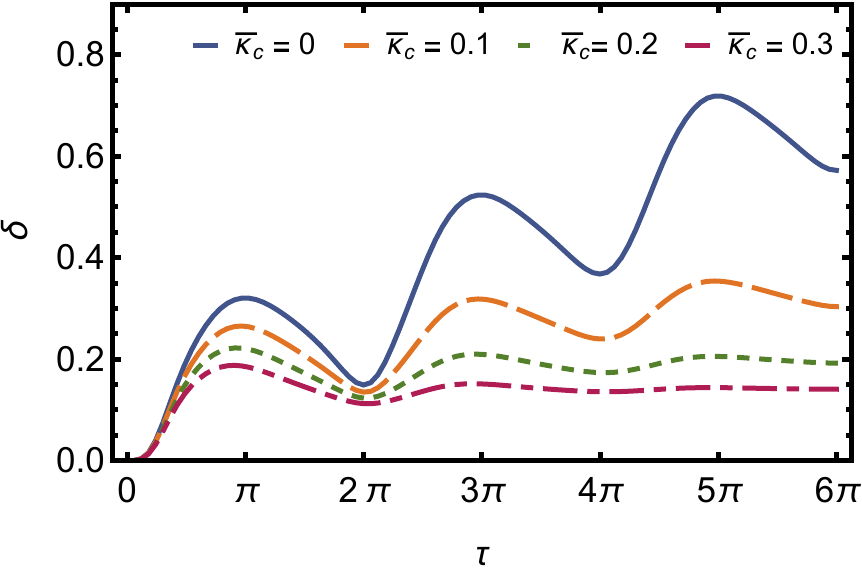}%
}\hfill
\subfloat[
\label{fig:open:system:phonon:decoherence:resonance}]{%
  \includegraphics[width=.5\linewidth, trim = 0mm 0mm 0mm 0mm]{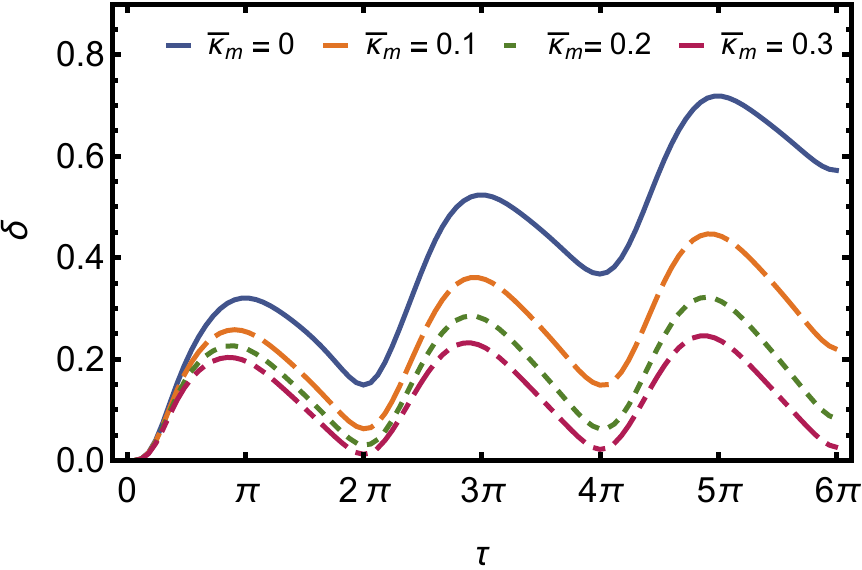}%
}
\caption[Measure of non-Gaussianity for an open optomechanical system at resonance]{Non-Gaussianity for open optomechanical systems at mechanical resonance. \textbf{(a)} shows the non-Gaussianity $\delta $ vs. time $\tau$ for a system with increasing values of photon decoherence $\bar{\kappa}_{\mathrm{c}}$ for $\tilde{g}_0 = 1$, $\mu_{\mathrm{c}} = 0.1$, $\epsilon = 0.5$,  and $\mu_{\mathrm{m}} = 0$. \textbf{(b)} shows the non-Gaussianity $\delta$ vs. time $\tau$ for a system with increasing values of phonon decoherence $\bar{\kappa}_{\mathrm{m}}$ for $\tilde{g}_0 = 1$, $\mu_{\mathrm{c}} = 0.1$, $\epsilon = 0.5$,  and $\mu_{\mathrm{m}} = 0$. The non-Gaussianity tends to a steady-state for both cases as $\tau$ increase, however the amount of non-Gaussianity is about ten times higher compared with when the coupling is constant, as shown in Figure~\ref{chap:non:Gaussianity:coupling:fig:measure:decoherence:constant:coupling}. A populated mechanical coherent state $\mu_{\mathrm{m}} \neq 0$ does not affect the non-Gaussianity. }
\label{chap:non:Gaussianity:coupling:fig:measure:decoherence:resonance}
\end{figure}

If it is possible to continuously increase the non-Gaussianity, the system might have a certain tolerance to noise. That is, there is a level of noise at which the non-Gaussianity essentially reaches a steady-state. In Figure~\ref{chap:non:Gaussianity:coupling:fig:measure:decoherence:resonance} we have plotted the non-Gaussianity $\delta$ as a function of time for different values of photon and phonon decoherence. Figure~\ref{fig:open:system:photon:decoherence:resonance} shows the system at resonance with photons leaking from the cavity with a rate $\bar{\kappa}_{\mathrm{c}} = \kappa_{\mathrm{c}} / \omega_{\mathrm{m}}$ for parameters $\mu_{\mathrm{c}} = 0.1$, $\tilde{g}_0 = 1$, $\epsilon = 0.5$ and $\mu_{\mathrm{m}} = 0$. We note that $\bar{\kappa}_{\mathrm{c}} = 0.3$ yields what is essentially a steady-state of the non-Gaussianity. In Figure~\ref{fig:open:system:phonon:decoherence:resonance} we note the same behaviour but for phonon decoherence with rate $\bar{\kappa}_{\mathrm{m}} = \kappa_{\mathrm{m}} / \omega_{\mathrm{m}}$.

\section{Discussion and practical implementations}\label{chap:non:Gaussianity:coupling:sec:discussion}
We have employed a measure of non-Gaussianity $\delta(\tau)$ in order to quantify the deviation from linearity of an initial Gaussian state induced by the Hamiltonian in Eq.~\eqref{chap:non:Gaussianity:coupling:eq:standard:Hamiltonian}. Our results show that, for a constant light--matter coupling $\tilde{g}_0$, the non-Gaussianity $\delta(\tau)$ scales differently in two contrasting regimes: (i)~For a weak optical input coherent state $\lvert\mu_{\mathrm{c}}\rvert$, the nonlinear character of the state grows as $\tilde{g}_0^2 \, \lvert\mu_{\mathrm{c}}\lvert^2 \, \ln{\lvert\mu_{\mathrm{c}}\lvert}$ if the mechanics is also in a coherent state,  (ii)~conversely, for large $\lvert\mu_{\mathrm{c}}\lvert$, the nonlinear character of the state grows logarithmically with the quantity $\tilde{g}_0 |\mu_{\mathrm{c}}|$, which also holds when the mechanical element is not fully cooled. The same general scaling with $|\mu_{\mathrm{c}}|$ occurs when $\tilde{\mathcal{G}}(\tau)$ is time-dependent.  

Crucially, we also find that the amount of non-Gaussianity can be continuously increased by driving the light--matter coupling at mechanical resonance. This becomes especially useful in the presence of noise. We will now discuss these results in the context of concrete experimental setups, and specifically discuss how the modulated light--matter coupling can be engineered. First, however, we will discuss the measure of non-Gaussianity that we have used in this thesis.

\subsection{Choice of measure}\label{subsec:measure:discussion}
Here, we work with a relative entropy measure of non-Gaussianity, which was first defined in Ref~\cite{genoni2008quantifying}. This measure has previously been extensively used to compute the non-Gaussianity of various states~\cite{tatham2012nonclassical}, as well as in an experimental setting where single photons were gradually added to a coherent state to increase its non-Gaussian character~\cite{barbieri2010non}. Several additional measures for the quantification of non-Gaussianity have been proposed in the literature, linking it to the Hilbert-Schmidt distance~\cite{genoni2007measure} or to quantum correlations~\cite{park2017quantifying}. Specifically, the relative entropy measure was shown to be more general than the Hilbert-Schmidt measure~\cite{genoni2007measure}. Furthermore, a connection has been put forward between the non-Gaussianity of a state and its Wigner function~\cite{genoni2013detecting}, and similarly there appears to be an intrinsic link between the quantum Fisher information and the lowest amount of non-Gaussianity of a state~\cite{yadin2018operational}.

Most crucially, this measure is not upper-bounded. This means that, as opposed to, for example, an entanglement measure where the notion of a maximally entangled state is well-defined, there is no such thing as a maximally non-Gaussian state. This is reflected by our results, where taking $\mu_{\mathrm{c}}$ to infinity yields $\lim_{\mu_{\mathrm{c}} \rightarrow \infty} \delta(\tau) = \infty$. As such, it is only possible to state that one state is more non-Gaussian than another. However, for pure states, there is the relation of the measure to the Hilbert-Schmidt measure. As such, the non-Gaussianity $\delta(\tau)$ of pure states has strong operational implications~\cite{zhuang2018resource}.

For mixed states, the operational meaning is not clear because the measure cannot detect the difference between classical mixtures of Gaussian states, which can be easily prepared by probabilistic sampling of Gaussian states, and inherent non-Gaussianity due to some nonlinear evolution of pure states~\cite{barbieri2010non}. This means that the measure often needs to be used alongside an additional measure of non-classicality, such as the negativity of the Wigner function to indicate when the non-Gaussian states can be used for the quantum information tasks mentioned in Section~\ref{chap:introduction:sec:non:Gaussianity} in Chapter~\ref{chap:introduction}. We know from previous work~\cite{bose1997preparation} that for a constant coupling, the system is maximally entangled at $\tau = \pi$, which satisfies the occurrence of non-classicality in conjunction with the non-Gaussianity. The state is however fully disentangled at $\tau = 2\pi$, and in the case of open system dynamics, this feature of the measure becomes apparent. We note that the non-Gaussianity plotted in Figure~\ref{chap:non:Gaussianity:coupling:fig:measure:decoherence:constant:coupling}  spikes at times $\tau = 2\pi n$ for integer $n$, which is when we usually have no entanglement. This implies that the addition of non-Gaussianity most likely comes from a classical mixture of coherent states that have slightly decohered.

\subsection{Experimental regimes}
There are two relevant experimental regimes for optomechanical systems. They are determined by the magnitude of the light--matter coupling $g$ compared to the other frequencies in the system. In the weak single-photon optomechanical coupling regime, the light--matter coupling $g$ is small compared to the resonant frequency $\omega_{\mathrm{m}}$ and the optical decoherence rate $\kappa_{\mathrm{c}}$. Such experiments usually involve a strong laser drive, which tends to wash out the non-linearity. In the strong single-photon coupling regime, nonlinear effects are in practice small but more significant. Under these conditions, a single photon displaces the mechanical oscillator by more than its zero-point uncertainty and weak optical fields tend to be used~\cite{nunnenkamp2011single}. In summary, most approaches fall into one of two categories: (i) small $g$ and linearised dynamics and (ii) large $g$ and low number of photons. 

Our work suggests that we can further increase the amount of non-Gaussianity by modulating the light--matter coupling. We emphasize that this scheme is applicable in both the weak and strong coupling regimes. This sets it apart from other schemes, which usually focus on enhancing the non--Gaussianity in one of the two categories mentioned above. 

Let us also briefly discuss our results with regard to linearised dynamics. This linearisation of dynamics is fundamentally different to the scenarios considered in this Chapter. When linearising the dynamics, the system is opened and the field operators $\hat a $ are treated as flucutations around a strong optical field as such: $\hat a \rightarrow \hat a = \alpha + \hat a'$, where $\hat a'$ are the fluctuations. See Section~\ref{chap:introduction:linearisation} in Chapter~\ref{chap:introduction} for a demonstration of the linearisation procedure. In this thesis, we have retained the nonlinear dynamics, even when considering open system dynamics. Thus, while we observe that a large coherent state parameter $\mu_{\mathrm{c}}$ increases the non-Gaussianity, we cannot generalise this result to the linearised dynamics.

\subsection{Methods of modulating the light--matter coupling in physical systems}\label{subsec:modulating}
We saw in Section~\ref{chap:non:Gaussianity:coupling:sec:time:dependent:coupling} that the amount of non-Gaussianity in the system increases when the light--matter coupling $\tilde{g}(\tau)$ is modulated. An explanation of this phenomena was provided in~\cite{liao2014modulated}. Consider the force $\vec{F}$ exerted by the photons on the mechanics. For a number of $n$ photons, this force is proportional to $\vec{F} \propto (n + 1/2)$, where $1/2$ comes from the zero-point energy. When the light--matter coupling is constant, this force is constant, and thus we see the periodic evolution. However, when we modulate $\tilde{\mathcal{G}}(\tau)$, the photon-pressure force $\vec{F}$ acts periodically on the mechanics, and is amplified when pushing in tandem with the mechanical resonance. 

While engineering the modulation is challenging, we shall explore several methods that can achieve it. The question is whether the modulation can be performed at mechanical resonance. As a basis for this discussion, we present a derivation of a time-dependent light--matter coupling for levitated nanobeads in Appendix~\ref{app:optomechanical:Hamiltonian}, which is based on the work in~\cite{romero2011optically}. There are several practical ways in which one may envisage to increase the non-linearity by modulating the coupling, depending on the nature of the trap at hand: 
\begin{itemize}
	\item[i)] \textbf{Optically-trapped levitating particles.} The effect that we are looking for can be realised by modulating the phase of the trapping laser beam (which, in turn, can be achieved through an acousto--optical modulator). In our derivation in Section~\ref{app:optomechanical:hamiltonian:nanosphere} in Appendix~\ref{app:optomechanical:Hamiltonian}, this phase is denoted by $\varphi(\tau)$ and it affects the light--matter coupling strength by determining the particle's location with respect to the standing wave of the cavity field. Thus if we let $\varphi(\tau) = \frac{\pi}{2}\left( 1 + \epsilon \sin{\Omega_g \tau} \right)$, with $\Omega_g = \omega_g / \omega_{\mathrm{m}}$, and where $\omega_g$ is the phase modulation frequency, we obtain the expression used in Section~\ref{chap:non:Gaussianity:coupling:sec:time:dependent:coupling}. If, then, the phase frequency is resonant with $\Omega_g =1$, it should be possible to increase the non-Gaussianity even further. 

	\item[ii)] \textbf{Paul traps.} The shuttling of ions has been demonstrated~\cite{walther2012controlling, hensinger2006t} using Paul traps, which are customarily used for ions but which have also recently been used for trapping nanoparticles~\cite{millen2015cavity, fonseca2016nonlinear, aranas2016split}. These works indicate that a modulation of the particle's position, and hence, a modulation of the coupling as per point i), can be obtained in a Paul trap as well. 

	\item[iii)] \textbf{Micromotion in hybrid traps.} Paul traps display three different kinds of particle motion. Firstly, we have \textit{thermal motion}, whereby the particle moves around the trap. Secondly, and most importantly to our scheme, we have \textit{micromotion}, which induces small movements around the potential minimum. Finally, there is \textit{mechanical motion}, which is the harmonic motion in the trap, here denoted by $\omega_{\mathrm{m}}$. Since the micromotion moves the bead around the potential minimum with a frequency $\omega_{\mathrm{d}}$, this already modulates the light--matter coupling, and is, in a way, an equivalent implementation to the ``shaking'' of the trap. If the micromotion can be engineered to occur with a frequency $\omega_{\mathrm{d}}$ equal to $\omega_{\mathrm{m}}$, then one could, instead of averaging it out, adopt the micromotion's variables to increase the non--Gaussianity with the scheme we propose in Section~\ref{resonance}. To date, the micromotion is generally smaller than the mechanical frequency, $\omega_{\mathrm{d}} \leq \omega_{\mathrm{m}}$, but current experimental efforts appear promising. 
\end{itemize}

There are potentially many more ways in which the light--matter coupling could be modulated, including with optomechanically induced transparency~\cite{weis2010optomechanically, karuza2013optomechanically} and by using the Kerr effect to change the refractive index of the oscillator. 

We conclude that the enhancement of the non-linearity predicted by our results can be realised in experiments, given the capabilities mentioned above. There are, of course, many challenges to be overcome. In fact, to take advantage of the rather slow logarithmic scaling with time $\tau$, one must keep the system coherent for longer, which is difficult. However, although our analytical results are restricted to Hamiltonian systems, we note that there is no reason to expect that this enhancement should disappear in a noisy setting.

\subsection{Detecting and measuring non-Gaussianity in optomechanical systems}\label{subsec:detecting}
In practise, how would one proceed to measure the amount of non-Gaussianity in the laboratory? As shown in~\cite{barbieri2010non}, the measure of non-Gaussianity used in this thesis has been measured for the addition of single photons to a coherent state. This requires full state tomography and is thus an expensive process. There are however others ways to proceed. In~\cite{hughes2014quantum} a witness of non-Gaussianity was proposed based on bounding the average photon number in the system from above. While they apply to a single system, they can probably be extended to bipartite systems as well. 

Finally, we here suggest a simple method by which non-Gaussianity can be detected for pure states. We note that the von Neumann entropy $S(\hat{\rho}_{AB})$ of a bipartite state $\hat{\rho}_{AB}$ is bounded by $S(\hat{\rho}_{AB})\geq |S(\hat{\rho}_{A})-S(\hat{\rho}_{B})|$, through the Araki--Lieb inequality~\cite{araki2002entropy} where $\hat{\rho}_{A}$ and $\hat{\rho}_{B}$ are the reduced states of the optical and mechanical subsystems, respectively. Therefore, the measure of non-Gaussianity $\delta(\tau)$ that we defined in Eq.~\eqref{chap:introduction:measure:of:non:gaussianity} is lower-bounded by
\begin{align}\label{bound:on:measure:of:non:gaussianity}
\delta(\tau)\geq|S(\hat{\rho}_{A})-S(\hat{\rho}_{B})|-S(\hat{\rho}(0)),
\end{align}
In this sense, this reduced measure acts as a sufficient (but not necessary) condition for non-Gaussianity. That is, finding that the measure is non-zero does tell us that the state is non-Gaussian, however it does not tell us the full magnitude of the non-Gaussianity. Furthermore, to compute this measure, one would still have to measure the second moments of the optical and mechanical subsystems. This does, however, require fewer measurements than full state tomography on the joint optical and mechanical system. 

\section{Conclusions}\label{chap:non:Gaussianity:coupling:sec:conclusions}

We have quantified the non-Gaussianity of initially Gaussian coherent states evolving under the standard, time-dependent optomechanical Hamiltonian. We used a measure of non-Gaussianity based on the relative entropy of a state and characterised the deviation from Gaussianity of the full system. Our techniques allowed us to derive asymptotic expressions for small and large optical coherent-state amplitudes, see Equation Eq.~\eqref{eq:delta:small} and Eq.~\eqref{eq:delta:large} respectively. We found that for coherent states with amplitude $\lvert\mu_{\mathrm{c}}\rvert\geq 1$, the amount of non-Gaussianity grows logarithmically with the input average number of excitations $|\mu_{\mathrm{c}}|$ and with the light--matter coupling. At resonance, we find that the non-Gaussianity is further enhanced by a logarithmic scaling with the time of interaction. 

An important and promising aspect of our study consists in showing that the amount of non-Gaussianity in the system can be continuously increased by driving a time-modulated optomechanical coupling at mechanical resonance. This allows us to circumvent the usual periodic increase and decrease of non-Gaussianity, and we find that this behaviour effectively yields a non-Gaussian steady-state in the presence of noise. As such, this points to a practically accessible, mechanism to enhance the nonlinear character of optomechanical dynamics at a given light--matter coupling strength. We point out that certain systems, such as hybrid-trap systems, are particularly well-suited for this purpose, as their light--matter interaction is naturally modulated due to the trap characteristics. Finally, we note that our methods can be extended to more complicated Hamiltonians of bosonic modes, and we can include modifications such as squeezing of the mechanical state.

\chapter{Interplay of non-Gaussianity and single-mode mechanical squeezing} \label{chap:non:Gaussianity:squeezing}
\chaptermark{Interplay of non-Gaussianity and single-mode mechanical...}

In this Chapter, we employ the full solution presented in Chapter~\ref{chap:decoupling} to compute the non-Gaussianity of the time-evolved state of the system. Rather than focus on the interplay between the nonlinear light--matter coupling and the non-Gaussianity of the state, as we did in Chapter~\ref{chap:non:Gaussianity:coupling}, we here examine the effect of including a single-mode mechanical squeezing term in the Hamiltonian. 

This Chapter is based on Ref~\cite{qvarfort2019time}. We thank Antonio Pontin, Peter F. Barker, and Robert Delaney for useful comments and discussions that improved the work in this Chapter. The computation of the symplectic eigenvalues shown in Section~\ref{chap:non:Gaussianity:squeezing:symplectic:eigenvalues} was contributed by David Edward Bruschi. 

\section{Introduction}
The mathematical understanding of optomechanical systems operating in the nonlinear quantum regime is a major topic of current interest. While most experiments effectively undergo linear dynamics, governed by quadratic Hamiltonians that emerge following
 a `linearisation' procedure~\cite{aspelmeyer2014cavity,bowen2015quantum, serafini2017quantum}, many experiments now operate in the fully nonlinear regime~\cite{sankey2010strong,leijssen2017nonlinear, fogliano2019cavity} where this procedure fails.  It is therefore highly desirable to provide a complete and analytic characterisation of the fully nonlinear system dynamics. 

As also discussed in Chapter~\ref{chap:non:Gaussianity:coupling}, the inherently nonlinear interaction between the optical field and the mechanical element in an optomechanical system allows for the generation of non-Gaussian states. Starting from a broad class of initial states, including coherent states, the vacuum, and thermal states, this is only possible in the nonlinear regime; quadratic Hamiltonians take input Gaussian states to output Gaussian states. Interestingly, a number of results indicate that non-Gaussian states constitute an important resource for sensing. Schr\"{o}dinger cat states~\cite{mancini1997ponderomotive, bose1997preparation}, compass states~\cite{zurek2001sub, toscano2006sub} and hypercube states~\cite{howard2018hypercube} -- which are all non-Gaussian states -- have all been found to have applications for sensing. 
 More generally, the detection and generation of non-Gaussianity in optomechanical systems has been extensively studied in theoretical proposals~\cite{lemonde2016enhanced,latmiral2016probing, yin2017nonlinear} as well as in experiments~\cite{sankey2010strong, doolin2014nonlinear, leijssen2017nonlinear}. The presence of a nonlinear element is also key to a number of quantum information tasks, such as obtaining a universal gate set for quantum computing~\cite{lloyd1999quantum, menicucci2006universal}, teleportation~\cite{dell2010teleportation}, distillation of entanglement~\cite{fiuravsek2002gaussian, giedke2002characterization}, error correction~\cite{niset2009no}, and non-Gaussianity has been explored as the basis of an operational resource theory~\cite{zhuang2018resource, takagi2018convex, PhysRevA.98.052350}. 
 
Optomechanical systems offer a natural nonlinear coupling which, if strong enough, may lead to substantial non-Gaussianity. It is therefore essential to better understand the dynamics of such systems, with an especial emphasis on
 the interplay between nonlinearities and other Hamiltonian terms in this dynamics. An important question to be answered is thus \textit{how do the different aspects of an optomechanical system affect the non-Gaussianity of the state at a given time?}  Optomechanical systems can exhibit additional, potentially more interesting, effects. An important and non-classical effect that can be included into optomechanical systems is \textit{squeezing} of the optical or mechanical modes. Addition of squeezing can be useful for sensing since it increases the sensitivity in a specific field quadrature. For example, it has been suggested that squeezed light injected into LIGO can be used to enhance the detection of gravitational waves~\cite{aasi2013enhanced}. Similarly, mechanical squeezing 
can aid the amplification and measurement of weak mechanical signals~\cite{clerk2010introduction}. 

In this Chapter, we study the non-Gaussianity of a quantum system of two bosonic modes characterised by the extended optomechanical Hamiltonian studied in Chapter~\ref{chap:decoupling}. We use the full solutions of the decoupled dynamics presented in Section~\ref{chap:decoupling:optomechanical:decoupling} to compute the non-Gaussianity of the system. Interestingly, for time-dependent squeezing modulated at resonance, we find that the dynamics are governed by the well-studied Mathieu equation. We subsequently derive perturbaive solutions and show that these coincide with the physically intuitive rotating-wave approximation for large times. 
Our results indicate that the non-Gaussian character of an initially coherent state \emph{decreases} in general with an increasing squeezing parameter. However, when the squeezing is applied periodically at mechanical resonance, the non-Gaussianity increases approximately linearly with time and the amplitude of the squeezing. The competition between the amount of squeezing and the strength of the nonlinear term is difficult to compute explicitly; instead, we provide asymptotic expressions in terms of upper and lower bounds to the non-Gaussianity in different regimes. A conclusive answer requires further investigation, potentially providing a concise expression where such competition can be easily understood. 

This Chapter is structured as follows. In Section~\ref{chap:non:Gaussianity:squeezing:dynamics}, we introduce the nonlinear Hamiltonian with mechanical squeezing and a short introduction to the methods used to solve the dynamics. The full derivation can be found in Chapter~\ref{chap:decoupling}. Following this, we review the measure of non-Gaussianity and derive expressions for an asymptotic expression and a reduced measure in Section~\ref{chap:non:Gaussianity:squeezing:sec:bounds}. We then specialise to two specific cases and compute the amount of non-Gaussianity for constant squeezing in Section~\ref{chap:non:Gaussianity:squeezing:sec:constant:squeezing}, and for modulated squeezing in Section~\ref{chap:non:Gaussianity:squeezing:sec:modulated:squeezing}. Finally, we conclude with a discussion in Section~\ref{chap:non:Gaussianity:squeezing:discussion} and some final remarks in Section~\ref{chap:non:Gaussianity:squeezing:conclusions}.

\section{System and dynamics}  \label{chap:non:Gaussianity:squeezing:dynamics}
To compute the non-Gaussianity of the state and investigate the influence of the mechanical single-mode term, we must first solve the system dynamics.  We use the full solutions presented in Section~\ref{chap:decoupling:optomechanical:decoupling} in Chapter~\ref{chap:decoupling}. To keep this Chapter self-contained, we review the most important elements of the derivation here. 
\subsection{Hamiltonian}  \label{chap:non:Gaussianity:squeezing:dynamics:Hamiltonian}
While the previous chapter focused on the effects of the nonlinearity on the non-Gaussian nature of an optomechanical state, we here seek to extend the standard optomechanical Hamiltonian to that in Eq.~\eqref{chap:decoupling:eq:Hamiltonian}. We reprint the extended Hamiltonian here for convenience:
\begin{equation} \label{chap:non:Gaussianity:squeezing:eq:Hamiltonian}
\hat H = \hbar \, \omega_{\rm{c}} \, \hat a^\dag \hat a + \hbar \, \omega_{\rm{m}} \, \hat b^\dag \hat b - \hbar \, \mathcal{G}(t) \, \hat a^\dag \hat a \left( \hat b^\dag + \hat b\right) + \hbar \, \mathcal{D}_1 \, \left( \hat b^\dag + \hat b \right) + \hbar \, \mathcal{D}_2 \, \left( \hat b^\dag + \hat b \right)^2 \, ,
\end{equation}
where $\hat a, \hat a^\dag$ are the annihilation and creation operators of the optical field with oscillation frequency $\omega_{\rm{c}}$, and  $\hat b, \hat b^\dag$ are the annihilation and creation operators of the phonons in the mechanics with oscillation frequency $\omega_{\rm{m}}$. $\mathcal{G}(t)$ is the nonlinear light--matter coupling, $\mathcal{D}_1(t)$ is the weighting function for a mechanical displacement term, and $\mathcal{D}_2(t)$ is the weighting function for a single-model mechanical squeezing term. 

In the previous Chapter, we investigated the effects on the non-Gaussianity of a time-dependent nonlinear coupling of the form $\mathcal{G}(t) = g_0 \, \left( 1 + \epsilon \sin(\omega_{\rm{g}} \, t) \right) $, and we found that when the coupling is modulated at mechanical resonance, the non-Gaussianity of the state increases approximately linearly. Here, we are interested in the effect of a single-mode mechanical squeezing term on the non-Gaussianity, and we therefore assume that the nonlinear light--matter coupling remains constant throughout with $\mathcal{G}(t) \equiv g_0$.

As mentioned before, working in dimensionless units is generally beneficial. We rescale the labortory time $t$ to the dimensionless parameter $\tau = \omega_{\rm{m}} \, t$. We also rescale the Hamiltonian in Eq.~\eqref{chap:non:Gaussianity:squeezing:eq:Hamiltonian} by $\hbar\omega_{\rm{m}}$ to find
\begin{equation}
\hat{\tilde{H}} = \Omega_{\rm{c}} \, \hat a^\dag \hat a + \hat b^\dag \hat b - \tilde{\mathcal{G}}(\tau) \, \hat a^\dag \hat a \, \left( \hat b^\dag + \hat b \right) + \tilde{\mathcal{D}}_1 (\tau) \left( \hat b^\dag +  \hat b \right) + \tilde{\mathcal{D}}_2 (\tau) \left( \hat b^\dag + \hat b \right)^2 \, , 
\end{equation}
where $\Omega_{\rm{c}} = \omega_{\rm{c}}/\omega_{\rm{m}}$, $\tilde{\mathcal{G}}(\tau) = \mathcal{G}(\omega_{\rm{m}} \, t)/\omega_{\rm{m}}$, $\tilde{\mathcal{D}}_1(\tau) = \mathcal{D}_1(\omega_{\rm{m}} \, t) /\omega_{\rm{m}}$, and $\tilde{\mathcal{D}}_2(\tau) = \mathcal{D}_2 ( \omega_{\rm{m}}\, t)$. 

\subsection{Solution of the dynamics}  \label{chap:non:Gaussianity:squeezing:solution}
To compute the non-Gaussianity, we must first solve the dynamics of the nonlinear system. We refer the reader to Chapter~\ref{chap:decoupling} for the full details. Here, we provide a brief summary of the main features of the solution. 

Given the Hamiltonian in Eq.~\eqref{chap:non:Gaussianity:squeezing:eq:Hamiltonian}, we showed in Chapter~\ref{chap:decoupling} that the time-evolution operator $\hat U(\tau)$ can be written as 
\begin{align} \label{chap:non:Gaussianity:squeezing:eq:final:evolution:operator}
\hat U(\tau)=& \, e^{-i\,\Omega_\mathrm{c} \hat a^\dagger \hat a\,\tau}\,\hat{\tilde{U}}_{\mathrm{sq}}(\tau)\,e^{-i\,F_{\hat{N}_a}\,\hat{N}_a}\,e^{-i\,F_{\hat{N}^2_a}\,\hat{N}^2_a}\,e^{-i\,F_{\hat{B}_+}\,\hat{B}_+}\,e^{-i\,F_{\hat{N}_a\,\hat{B}_+}\,\hat{N}_a\,\hat{B}_+}\,\nonumber \\
&\times e^{-i\,F_{\hat{B}_-}\,\hat{B}_-}\,e^{-i\,F_{\hat{N}_a\,\hat{B}_-}\,\hat{N}_a\,\hat{B}_-} \, , 
\end{align}
where the evolution operator $\hat{\tilde{U}}_{\rm{sq}} (\tau)$  encodes the quadratic evolution operator of the 
mechanical degree of freedom:
\begin{align}\label{chap:non:Gaussianity:squeezing:decoupling:form:to:be:used}
\hat{\tilde{U}}_{\mathrm{sq}} &= \overleftarrow{T} \exp\biggl[ - i \int^\tau_0 d\,\tau' \, 2\,\left(\frac{1}{2} \,+\tilde{\mathcal{D}}_2(\tau')\right)\,\hat{N}_b+ \tilde{\mathcal{D}}_2(\tau')\,\hat{B}^{(2)}_+ \biggr]\,.
\end{align}
The coefficients in the decoupling above can now be obtained in terms of the following definite integrals:
\begin{align}\label{chap:non:Gaussianity:squeezing:sub:algebra:decoupling:solution}
F_{\hat{N}_a}&= -2\,\int_0^\tau\,\mathrm{d}\tau'\,\tilde{\mathcal{D}}_1(\tau')\,\Im\xi(\tau')\int_0^{\tau'}\mathrm{d}\tau''\,\tilde{\mathcal{G}}(\tau'')\,\Re\xi(\tau'')\, \nonumber \\
&\quad -2 \,\int^\tau_0\,\mathrm{d}\tau' \,\tilde{\mathcal{G}}(\tau')\, \Im \xi \, \int^{\tau'}_0 \,\mathrm{d}\tau''\, \tilde{\mathcal{D}}_1(\tau'') \, \Re \xi(\tau'') \, ,  \nonumber\\
F_{\hat{N}^2_a}&= 2\,\int_0^\tau\,\mathrm{d}\tau'\,\tilde{\mathcal{G}}(\tau')\,\Im\xi(\tau')\int_0^{\tau'}\mathrm{d}\tau''\,\tilde{\mathcal{G}}(\tau'')\,\Re\xi(\tau'') \, , \nonumber\\
F_{\hat{B}_+}&=\int_0^\tau\,\mathrm{d}\tau'\,\tilde{\mathcal{D}}_1(\tau')\,\Re\xi(\tau') \, , \nonumber\\
F_{\hat{B}_-}&=- \int_0^\tau\,\mathrm{d}\tau'\,\tilde{\mathcal{D}}_1(\tau')\,\Im\xi(\tau') \, , \nonumber\\
F_{\hat{N}_a\,\hat{B}_+}&=- \int_0^\tau\,\mathrm{d}\tau'\,\tilde{\mathcal{G}}(\tau')\,\Re\xi(\tau') \, , \nonumber\\
F_{\hat{N}_a\,\hat{B}_-}&=\int_0^\tau\,\mathrm{d}\tau'\,\tilde{\mathcal{G}}(\tau')\,\Im\xi(\tau')\,,
\end{align}
where we have introduced the function
\begin{equation} \label{chap:non:Gaussianity:squeezing:eq:def:xi}
\xi(\tau) = P_{11}(\tau)-i\,\int_0^\tau\,\mathrm{d}\tau'\,P_{22}(\tau) \, ,
\end{equation}
and $P_{11}(\tau)$ and $P_{22}(\tau)$ are defined below.

It can be difficult to obtain an analytical decoupling of Eq. ~\eqref{chap:non:Gaussianity:squeezing:decoupling:form:to:be:used}, but it is straight-forward to obtain an expression for its action on the operators $\hat{b}$ and $\hat{b}^\dag$. First of all, we note that a Bogoliubov transformation of a single mode operator always has  the general expression $\hat{\tilde{U}}_{\mathrm{sq}}^\dag\,\hat{b}\,\hat{\tilde{U}}_{\mathrm{sq}}=\alpha(\tau)\,\hat{b}+\beta(\tau)\,\hat{b}^\dag$, see~\cite{moore2016tuneable}. The challenge is to find an explicit expression for the Bogoliubov coefficients $\alpha(\tau)$ and $\beta(\tau)$, which satisfy the only nontrivial Bogoliubov identity $|\alpha(\tau)|^2-|\beta(\tau)|^2=1$. 
In  Chapter~\ref{chap:decoupling} we show that the Bogoliubov coefficients $\alpha(\tau)$ and $\beta(\tau)$ can be obtained through 
\begin{align} \label{chap:non:Gaussianity:squeezing:eq:bogoliubov:coefficients}
\alpha(\tau)=&\frac{1}{2}\biggl[P_{11}(\tau)+P_{22}(\tau)-i \int_0^\tau \mathrm{d}\tau' P_{22}(\tau')  -i \int_0^\tau \mathrm{d}\tau' (1+4\,\tilde{\mathcal{D}}_2(\tau'))\,P_{11}(\tau')\biggr] \, , \nonumber\\
\beta(\tau) =&\frac{1}{2}\biggl[P_{11}(\tau)-P_{22}(\tau)+i \int_0^\tau \mathrm{d}\tau' P_{22}(\tau')  -i \int_0^\tau \mathrm{d}\tau' (1+4\,\tilde{\mathcal{D}}_2(\tau'))\,P_{11}(\tau')\biggr] \,  ,
\end{align}
whose explicit form can be obtained once an explicit expression of the functions $P_{11}(\tau)$ and $P_{22}(\tau)$ is found. Given the previously defined function $\xi(\tau)$ in Eq.~\eqref{chap:non:Gaussianity:squeezing:eq:def:xi}, we also find  $\alpha(\tau) = \frac{1}{2}(\xi(\tau) + i \dot{\xi}(\tau))$ and $\beta(\tau) = \frac{1}{2}(\xi^*(\tau) + i \dot{\xi}^*(\tau))$, where dotted functions imply differentiation with respect to $\tau$. 

The two functions $P_{11}$ and $P_{22}$ are determined by the two following uncoupled differential equations:
\begin{align}\label{chap:non:Gaussianity:squeezing:differential:equation:written:in:paper}
\ddot{P}_{11}+(1+4\,\tilde{\mathcal{D}}_2(\tau))\,P_{11}&=0 \, ,\nonumber\\
\ddot{P}_{22}-4\frac{\dot{\tilde{\mathcal{D}}}_2(\tau)}{1+4\,\tilde{\mathcal{D}}_2(\tau)}\,\dot{P}_{22}+(1+4\,\tilde{\mathcal{D}}_2(\tau))\,P_{22}&=0 \, ,
\end{align}
where the dot stands for a derivative with respect to $\tau$ and the initial conditions are $P_{11}(0)=P_{22}(0)=1$ and $\dot{P}_{11}(0)=\dot{P}_{22}(0)=0$. Furthermore, the  equation for $P_{22}$ in  Eq. ~\eqref{chap:non:Gaussianity:squeezing:differential:equation:written:in:paper} can be written as
\begin{equation} \label{chap:non:Gaussianity:squeezing:eq:IP22}
\ddot{I}_{P_{22}}  + ( 1 + 4 \, \tilde{\mathcal{D}}_2 (\tau) ) I_{P_{22}} = 0 \, ,
\end{equation}
which now has boundary conditions $I_{P_{22}}(0) = 0$ and $\dot{I}_{P_{22}} = 1$, and where
\begin{equation}
I_{P_{22}} = \int^\tau_0 \mathrm{d}\tau' P_{22} ( \tau') \, .
\end{equation}
The solutions to $P_{11}$ and $P_{22}$ (or $I_{P_{22}}$) can then be used in the expressions Eq.~\eqref{chap:non:Gaussianity:squeezing:sub:algebra:decoupling:solution}, Eq.~\eqref{chap:non:Gaussianity:squeezing:eq:def:xi} and Eq. ~\eqref{chap:non:Gaussianity:squeezing:eq:bogoliubov:coefficients} to find the full dynamics of the state. While the solutions must in general be obtained numerically, we anticipate that there are scenarios, such as constant $\tilde{\mathcal{D}}_2$, where the equations above have analytical solutions.

\subsection{Initial state of the system}  \label{chap:non:Gaussianity:squeezing:dynamics:state}
In this Chapter, we assume that both the optical and mechanical modes are initially in a coherent states, namely $\ket{\mu_{\mathrm{c}}}$ and $\ket{\mu_{\mathrm{m}}}$ respectively, which we defined in Section~\ref{chap:introduction:initial:states} in Chapter~\ref{chap:introduction}. 

For optical fields, this is generally a good assumption. On the other hand, within optomechanical systems the mechanical element is typically found initially in a thermal state. Our choice of initial coherent state can be generalised to that of a thermal state in a straight-forward manner, that is, by integrating over the coherent state parameter with an appropriate kernel 
(as any thermal state may be written as Gaussian average of coherent states, as per its P-representation). 
Restricting ourselves hence to a single coherent state also for the mechanical oscillator, the initial state $|\Psi(0)\rangle$  reads
\begin{equation}\label{chap:non:Gaussianity:squeezing:initial:state:two}
|\Psi(0)\rangle = \ket{\mu_{\mathrm{c}}}\otimes \ket{\mu_{\mathrm{m}}} \, .
\end{equation}
The evolved state is then given by 
\begin{equation} \label{chap:non:Gaussianity:squeezing:evolved:state}
\hat \rho = \hat U(\tau) \ketbra{\mu_{\rm{c}} \mu_{\rm{m}}} \hat U^\dag(\tau) \, .
\end{equation}
We are now ready to compute the covariance matrix elements. 
\subsection{Covariance matrix elements}  \label{chap:non:Gaussianity:squeezing:dynamics:elements}
We are interested in computing the non-Gaussianity of the state, for which we make use of the relative entropy measure introduced in Section~\ref{chap:introduction:sec:non:Gaussianity} in Chapter~\ref{chap:introduction}. 

The decoupled operator $\hat U(\tau)$ in Eq.~\eqref{chap:non:Gaussianity:squeezing:eq:final:evolution:operator} allow us to compute the expectation values of the first and second moments given our initial state in Eq.~\eqref{chap:non:Gaussianity:squeezing:initial:state:two}. The explicit calculations can be found in Appendix~\ref{app:exp:values}. 

We define the following functions for brevity: 
\begin{align} \label{chap:non:Gaussianity:squeezing:definitions:of:quantities}
K_{\hat N_a } :=& F_{\hat N_a \, \hat B_-} + i \, F_{\hat N_a \, \hat B_+} \, , \nonumber \\
\varphi(\tau) :=& \left( F_{\hat{N}_a} + F_{\hat{N}_a^2 } + 2 \,  F_{\hat{N} _a \, \hat{B}_+} \, F_{\hat{B}_-} \right) \nonumber \\ 
\theta(\tau) :=& \,  2 \, \left( F_{\hat{N}_a^2} + F_{\hat{N}a \, \hat{B}_+} \, F_{\hat{N}_a \, \hat{B}_- } \right) \,,\nonumber \\
\Gamma(\tau) :=& \,  (\alpha(\tau) + \beta(\tau)) F_{\hat{B}_-} - i ( \alpha(\tau) - \beta(\tau) ) \, F_{\hat{B}_+} \nonumber \\
\Delta(\tau) :=& (\alpha(\tau) + \beta(\tau)) F_{\hat{N}_a \, \hat{B}_- } - i ( \alpha(\tau) - \beta(\tau)) F_{\hat{N}_a \, \hat{B}_+ }  \, ,\nonumber \\
E_{\hat B _+ \hat B_-} :=& \,   \exp \biggl[ \frac{1}{2} \biggl( - F_{\hat N _a \hat B_-}^2 - F_{\hat N_a \hat B_+}^2 - 2 \,  i \,  F_{\hat N_a \hat B_-} F_{\hat N_a \hat B_+} \, , \nonumber\\
&\quad\quad\quad\quad- 2\, \mu_{\rm{m}} (F_{\hat N_a \hat B_-} + i F_{\hat N _a \hat B_+} ) + 2 \,  \mu_{\rm{m}}^* ( F_{\hat N_a \hat B_-} - i F_{\hat N_a \hat B_+} ) \biggr) \biggr] \, .
\end{align}
The first and second moments are then given by
\begin{align} \label{chap:non:Gaussianity:squeezing:expectation:values}
\braket{\hat a (\tau) } :=\, & e^{- i \varphi } \, e^{|\mu_{\mathrm{c}}|^2  \, (e^{- i \theta}-1)}  \, E_{\hat B_+ \hat B_- } \mu_{\mathrm{c}} \, , \nonumber \\
\braket{\hat b(\tau) } :=\, & \alpha(\tau) \,  \mu_{\mathrm{m}} + \beta(\tau) \,  \mu_{\mathrm{m}}^*  + \Gamma(\tau) + \Delta(\tau)  \, |\mu_{\mathrm{c}}|^2 \, ,\nonumber \\ 
\braket{\hat{a}^2(\tau)} := \, & e^{- 2i \varphi} \, \mu_{\mathrm{c}}^2  \, e^{- i \theta} \,  e^{|\mu_{\mathrm{c}}|^2 \,  (e^{- 2 i \theta} - 1)} \, e^{-|K_{\hat N_a}|^2} \,  E_{\hat B_+ \hat B_-}^2\, , \nonumber  \\
 \braket{\hat b^{2} (\tau)} =&\,  \alpha^2(\tau) \,  \mu_{\mathrm{m}}^2  + \alpha(\tau) \,  \beta(\tau) \,(2 \, |\mu_{\mathrm{m}}|^2 + 1) + \beta^2(\tau) \, \mu_{\mathrm{m}}^{*2}  \nonumber \\
 &+ 2 \, (\alpha(\tau)\,\mu_{\mathrm{m}}+\beta(\tau)\,\mu_{\mathrm{m}}^*) \,  \left[\Gamma(\tau) + \Delta(\tau)\,  |\mu_{\mathrm{c}}|^2 \right] \nonumber \\
 &+ \Gamma^2 (\tau) + 2 \, \Gamma(\tau) \Delta(\tau) \, |\mu_{\mathrm{c}}|^2 + \Delta^2(\tau) \, |\mu_{\mathrm{c}}|^2 \, ( 1 + |\mu_{\mathrm{c}}|^2)\, , \nonumber  \\
\braket{\hat a(\tau) \hat b(\tau) } :=& \, e^{- i \varphi} \,e^{ |\mu_{\mathrm{c}}|^2 \, ( e^{- i \theta}  - 1)} \, \mu_{\mathrm{c}} \, E_{\hat B_+ \hat B_-} \,  \bigl[ \alpha(\tau)\, \mu_{\mathrm{m}} +\beta(\tau)\,(\mu_{\mathrm{m}}^*- K_{\hat N_a}) \nonumber \\
&+ 
\Gamma(\tau) +\left( |\mu_{\mathrm{c}}|^2 \,  e^{- i \theta} + 1\right) \, 
\Delta (\tau)  \bigr]\, ,\nonumber   \\
\braket{\hat a (\tau) \, \hat b^\dag(\tau) }:= \, & e^{- i \varphi} \, e^{ |\mu_{\mathrm{c}}|^2 \, ( e^{- i \theta}  - 1)} \, \mu_{\mathrm{c}} \,E_{\hat B_+ \hat B_-}\,  \bigl[  \alpha^*(\tau)\, (\mu_{\mathrm{m}}^* - K_{\hat N_a}) \nonumber \\
&+\beta^*(\tau)\,\mu_{\mathrm{m}} +
\Gamma^*(\tau)  + \left( |\mu_{\mathrm{c}}|^2 \, e^{- i \theta} + 1\right) \, \Delta^* (\tau)  \bigr]  \, , 
\end{align}
and the number operators are given by 
\begin{align} \label{chap:non:Gaussianity:squeezing:photon:phonon:numbers}
 \braket{\hat a^\dag (\tau)\hat a(\tau) } := \, & |\mu_{\rm{c}}|^2\, , \nonumber \\
 \braket{\hat b ^\dag(\tau) \hat b(\tau)}  := \, &  (|\alpha(\tau)|^2 + |\beta(\tau)|^2) | \, \mu_{\mathrm{m}}|^2 + \alpha^*(\tau) \,  \beta (\tau) \,  (\mu_{\mathrm{m}}^*)^2 + \alpha(\tau) \, \beta^*(\tau) \,  \mu_{\mathrm{m}}^2 \nonumber \\
 &+ (\alpha^*(\tau)\,\mu_{\mathrm{m}}^* +\beta^*(\tau)\,\mu_{\mathrm{m}}) \, \left( \Gamma(\tau) + \Delta(\tau)  \, |\mu_{\mathrm{c}}|^2 \right)\nonumber \\
 &+ (\alpha(\tau)\,\mu_{\mathrm{m}}+\beta(\tau)\,\mu_{\mathrm{m}}^*) \, \left( \Gamma(\tau) + \Delta(\tau) | \, \mu_{\mathrm{c}}|^2 \right)^*\nonumber \\
 &+ (\Gamma^*(\tau) \, \Delta(\tau) + \Gamma(\tau) \,  \Delta^*(\tau)) | \, \mu_{\mathrm{c}}|^2 + |\Delta(\tau)|^2 |\mu_{\mathrm{c}}|^2 \, ( 1 + |\mu_{\mathrm{c}}|^2)\nonumber\\
 &+ |\beta(\tau)|^2  + |\Gamma(\tau)|^2 \, , 
 \end{align}
where we have used slightly different notation compared with the expression in Eq.~\eqref{app:ex:values:summary:of:exp:values} in Appendix~\ref{app:exp:values}. 

Given the above, we can compute an explicit expression the covariance matrix elements. We use the following basis $\hat{\vec{r}} = (\hat a, \hat b, \hat a^\dag, \hat b^\dag)^{\rm{T}}$, for which we find the covariance elements shown in  Eq.~\eqref{chap:introduction:covariance:matrix:elements:ab:basis} in Chapter~\ref{chap:introduction}. The explicit computations can be found in Section~\ref{app:exp:values:derivation:covariance:matrix:elements} in Appendix~\ref{app:exp:values}, and using $K_{\hat N_a}$, they read
\begin{align} \label{chap:non:Gaussianity:squeezing:eq:CM:elements}
\sigma_{11} &=1 + 2\, |\mu_{\mathrm{c}}|^2 \left( 1 - e^{-4 \, |\mu_{\mathrm{c}}|^2 \sin^2{\theta(\tau)/2}} \, |E_{\hat B_+ \hat B_-}|^2 \right)\, , \nonumber \\
\sigma_{31} &=  2\, \mu_{\mathrm{c}}^2 \, e^{- 2 \, i \,  \varphi(\tau)} \left(  e^{- i \,  \theta(t)} \, e^{|\mu_{\mathrm{c}}|^2 ( e^{-  2 \, i \,  \theta(\tau)} - 1)} e^{-|K_{\hat N_a}|^2} \,   - e^{2|\mu_{\mathrm{c}}|^2 ( e^{- i \, \theta(\tau)} - 1)} \right) \, E_{\hat B_+ \hat B_-}^2\, , \nonumber \\
\sigma_{21} &=  2 \,e^{- i \, \varphi(\tau)} \, e^{ |\mu_{\mathrm{c}}|^2 ( e^{- i \, \theta(\tau)}  - 1)}  \,E_{\hat B_+ \hat B_-} \, \mu_{\mathrm{c}}\, \left[\Delta^*(\tau)\left( |\mu_{\mathrm{c}}|^2 (e^{- i \,  \theta(\tau)} - 1) + 1  \right) - \alpha^*(\tau) \, K_{\hat N_a}\right]\, , \nonumber \\
\sigma_{41} &= 2\, e^{- i \,  \varphi(\tau) } \, e^{|\mu_{\mathrm{c}}|^2 (e^{- i \,  \theta(\tau)}-1)} E_{\hat B_+ \hat B_- } \mu_{\mathrm{c}} \, \left[\Delta(\tau) \left( |\mu_{\mathrm{c}}|^2 (e^{- i  \, \theta(\tau)} - 1) + 1\right)- \beta(\tau) K_{\hat N_a}  \right]\, ,\nonumber\\
\sigma_{22} &= 1 + 2\, |\beta(\tau)|^2  + 2\, |\Delta(\tau)|^2 \, |\mu_{\mathrm{c}}|^2\, , \nonumber \\
\sigma_{42}  &=2\,  \alpha(\tau) \, \beta(\tau) +2\,  \Delta^2(\tau) \, |\mu_{\mathrm{c}}|^2 \, .
\end{align}

\subsection{Symplectic eigenvalues of the optical and mechanical subsystems}  \label{chap:non:Gaussianity:squeezing:symplectic:eigenvalues}
For an optomechanical system, the subsystems consist of the traced out optical state $\hat \rho_{\rm{Op}}$ and the traced out mechanical state $\hat \rho_{\rm{Me}}$. To quantify the entropy of the subsystems, we must find the symplectic eigenvalues of the optical and mechanical subsystems, which we call $\nu_{\rm{Op}}$ and $\nu_{\rm{Me}}$ respectively. Lengthy algebra (see Section~\ref{app:exp:values:symplectic:eigenvalues} in Appendix~\ref{app:exp:values}), the use of the Bogoliubov identities $|\alpha|^2=1+|\beta|^2$ and $\alpha\,\beta^*=\alpha^*\,\beta$, and observing that $|E_{\hat B_+ \hat B_-}|^2 = e^{- |K_{\hat N_a}|^2}$   allow us to find
\begin{align}\label{chap:non:Gaussianity:squeezing:sympelctic:eigenvalue:reduced:state}
\nu_{\rm{Op}}^2 =& 1 +4  \, |\mu_{\mathrm{c}}|^2 \left( 1 - e^{-4|\mu_{\mathrm{c}}|^2 \sin^2{\theta/2}} \,e^{- |K_{\hat N_a}|^2} \right) 
+ 4 \, |\mu_{\mathrm{c}}|^4 \biggl( 1 - 2 \,  e^{-4|\mu_{\mathrm{c}}|^2 \sin^2{\theta/2}} e^{- |K_{\hat N_a}|^2 }\nonumber\\
&  -  e^{- 4 |\mu_{\rm{c}}|^2 \sin^2\theta} \, e^{-4 |K_{\hat N_a}|^2 } 2 \, e^{- 3|K_{\hat N_a}|^2} \,  \Re \left\{  e^{i \theta} \, e^{|\mu_{\mathrm{c}}|^2 ( e^{2i \theta} - 1)} \, e^{2 |\mu_{\rm{c}}|^2 (e^{- i \theta} - 1)} \right\}  \biggr) \, , \nonumber \\
\nu_{\mathrm{Me}}^2=&1+4\,|K_{\hat N_a}|^2 |\mu_{\mathrm{c}}|^2 \, ,
\end{align}
where the definitions of   $K_{\hat N_a}$ and $\theta(\tau)$ are given in Eq.~\eqref{chap:non:Gaussianity:squeezing:definitions:of:quantities} above. 

\section{Bounding the measure of non-Gaussianity}  \label{chap:non:Gaussianity:squeezing:sec:bounds}
We introduced a measure of non-Gaussianity in Section~\ref{chap:introduction:sec:non:Gaussianity} in Chapter~\ref{chap:introduction}, which was based on the relative entropy of the fully nonlinear state $\hat \rho$ in Eq.~\eqref{chap:non:Gaussianity:squeezing:evolved:state} and its closest possible Gaussian reference state $\hat \rho_{\rm{G}}$. We construct $\hat \rho_{\rm{G}}$  from $\hat \rho$ through the covariance matrix elements computed in Eq.~\eqref{chap:non:Gaussianity:squeezing:eq:CM:elements}. The measure of non-Gaussianity is given by 
\begin{equation}
\delta(\tau) = S(\hat \rho_{\rm{G}}) - S(\hat \rho) \, .
\end{equation}
We can compute $\delta(\tau)$ with the covariance matrix elements  $\boldsymbol{\sigma}$ and noting that the entropy $S$ can be computed through Eq.~\eqref{chap:introduction:von:Neumann:definition}.  Computing the entropy requires knowing the symplectic eigenvalues, which we computed in Eq.~\eqref{chap:non:Gaussianity:squeezing:sympelctic:eigenvalue:reduced:state}. 

The exact expression for $\delta(\tau)$ is long and cumbersome due to the complex expressions of the covariance matrix elements Eq.~\eqref{chap:non:Gaussianity:squeezing:eq:CM:elements}. We will therefore provide bounds to the measure that can be expressed as simple analytic functions. Since the  full measure $\delta(\tau)$ is an entropy, it can be bounded from above and below by the means of  the Araki--Lieb inequality~\cite{araki2002entropy}, which reads 
\begin{equation} \label{chap:non:Gaussianity:squeezing:eq:araki:lieb}
|S(\hat{\rho}_{A})-S(\hat{\rho}_{B})|\leq S(\hat{\rho}_{AB})\leq S(\hat{\rho}_{A})+S(\hat{\rho}_{B}) \, ,
\end{equation}
 where $\hat \rho_{AB}$ is the full bipartite state and $\hat \rho_A$ and $\hat \rho_B$ are the traced out subsystems. This inequality allows us bound the behaviour of the full measure $\delta(\tau)$ in terms of the subsystem entropies. We therefore proceed to define the lower and upper bounds as 
\begin{align}
&\delta_{\rm{min}}(\tau) :=|S(\hat{\rho}_{A})-S(\hat{\rho}_{B})|\quad\quad \mbox{ and} &&\delta_{\rm{max}}(\tau) := S(\hat{\rho}_{A})+S(\hat{\rho}_{B}) \, .
\end{align} 
We then note that the optical symplectic eigenvalue Eq.~\eqref{chap:non:Gaussianity:squeezing:sympelctic:eigenvalue:reduced:state} is bounded by
\begin{align}
 \nu_{\rm{Op}} < \sqrt{1 + 4|\mu_{\rm{c}}|^2 + 4 |\mu_{\rm{c}}|^4 } \, , 
\end{align}
The bound on $\nu_{\rm{Op}}$ can be inferred by noting that $K_{\hat  N_a}$ is generally given by an oscillating function multiplied by the strength of the optomechanical coupling $\tilde{g}_0$. For specific $\tau$ which ensures that $|K_{\hat N_a}|^2 \neq 0$,  and then considering $\tilde{g}_0 \gg 1$, the exponentials in $\nu_{\rm{Op}}$ in Eq.~\eqref{chap:non:Gaussianity:squeezing:sympelctic:eigenvalue:reduced:state} are suppressed, which means we are left with $\nu_{\rm{Op}} \sim \sqrt{1+4\,|\mu_{\mathrm{c}}|^2+4|\mu_{\mathrm{c}}|^4} $.

 When $S(\hat \rho_{\rm{Op}}) \gg S(\hat \rho_{\rm{Me}})$ or $S(\hat \rho_{\rm{Op}}) \ll S(\hat  \rho _{\rm{Me}})$, the bipartite entropy of the Gaussian reference state $S(\hat \rho_{\rm{G}})$ is approximately equal to one of the subsystem entropies. To determine when this is the case, we consider the maximum values of $\nu_{\rm{Op}} $ and $\nu_{\rm{Me}}$.
In general, when  $|\mu_{\rm{c}}|^2 \gg1$, and when $|K_{\hat N_a}|^2 \gg1$, which requires $\tilde{g}_0 \gg1$ and specific values of $\tau$,  the eigenvalues $\nu_{\mathrm{Op}}$ and $\nu_{\mathrm{Me}}$ tend to their maximum values  $\nu_{\rm{Op}, \rm{max}}$ and $\nu_{\rm{Me},\rm{max}}$, which are  
\begin{align}\label{chap:non:Gaussianity:squeezing:sympelctic:eigenvalue:reduced:state:maximum:value}
\nu_{\mathrm{Op,max}}\sim&1+2\,|\mu_{\mathrm{c}}|^2 \, ,\nonumber\\
\nu_{\mathrm{Me,max}}\sim&2\, |K_{\hat N_a}| \,|\mu_{\mathrm{c}}| \, .
\end{align}
We note that there are three distinct scenarios which arise from the comparison of the coherent state parameter $|\mu_{\rm{c}}|^2$ and the function $K_{\hat N_a}$:
\begin{itemize}
\item[i)]  First, we assume that $1\ll |K_{\hat N_a} |\ll2|\mu_{\mathrm{c}}|$, which implies $\delta(\tau) \sim S(\hat \rho_{\rm{Op}}) = s_V(\nu_{\rm{Op}})$. 
Thus, the non-Gaussianity is well-approximated by 
	\begin{align} \label{chap:non:Gaussianity:squeezing:eq:many:photons}
	S(\hat \rho_{\mathrm{G}}(\tau))\sim s_V(1+2\,|\mu_{\mathrm{c}}|^2) \, .
	\end{align} 
	\item[ii)] Secondly, we assume that $1 \ll 2|\mu_{\rm{c}}|\ll |K_{\hat N_a}| $, which implies that $\delta(\tau) \sim S(\nu_{\rm{Me}}) = s_V(\nu_{\rm{Me}})$. Thus we find that
	\begin{align} \label{chap:non:Gaussianity:squeezing:eq:nG:large:KNa}
	S(\hat \rho_{\mathrm{G}}(\tau))\sim s_V(2\,|K_{\hat N_a}|\,|\mu_{\mathrm{c}}|) \, .
		\end{align}
	\item[iii)] Finally, when $ |K_{\hat N_a}|\sim2|\mu_{\mathrm{c}}|$ and $|\mu_{\mathrm{c}}|\gg1$, we have $S(\hat{\rho}_{A})\sim S(\hat{\rho}_{B})$. In this case, the Araki--Lieb bound is not very informative since the left-hand-side is zero and must evaluate the non-Gaussianity exactly. 
\end{itemize}
Note that the first two cases might occur only for short periods of time $\tau$, because of the oscillating part of $K_{\hat N_a}$.
Furthermore, we note that the squeezing parameter $\tilde{\mathcal{D}}_2(\tau)$ affects the peak value of the non-Gaussianity because it enters into $|K_{\hat N_a}|$ through the $F$-coefficients Eq.~\eqref{chap:non:Gaussianity:squeezing:sub:algebra:decoupling:solution}. The dependence is non-trivial, but we will consider the analytic case for constant squeezing below. However, in general, when  $|\mu_{\rm{c}}|\gg |K_{\hat N_a}|$, we see from Eq.~\eqref{chap:non:Gaussianity:squeezing:eq:many:photons} that the non-Gaussianity is independent of $\tilde{\mathcal{D}}_2(\tau)$ and can be accurately modelled by the standard optomechanical Hamiltonian without mechanical squeezing. 

Let us now consider two specific cases where the squeezing term is either constant or modulated.

\section{Constant squeezing parameter}\label{chap:non:Gaussianity:squeezing:sec:constant:squeezing}
Here we assume that the rescaled squeezing parameter is constant with $\tilde{\mathcal{D}}_2 = \tilde{d}_2$. This case is equivalent to when the mechanical oscillation frequency $\omega_{\mathrm{m}}$ is shifted by a constant amount and where the initial state is a squeezed coherent state, see Section~\ref{app:mathieu:time:varying:trapping:frequency} in Appendix~\ref{app:mathieu}. We begin by deriving analytic expressions for the coefficients in Eq.~\eqref{chap:non:Gaussianity:squeezing:sub:algebra:decoupling:solution} given this choice of parameters. 

\subsection{Decoupled dynamics for a constant squeezing}\label{chap:non:Gaussianity:squeezing:decoupled:dynamics}

\begin{figure*}[t!]
\subfloat[ \label{chap:non:Gaussianity:squeezing:fig:quad:mum0:d20}]{%
  \includegraphics[width=.25\linewidth, trim = 00mm 0mm 0mm 0mm]{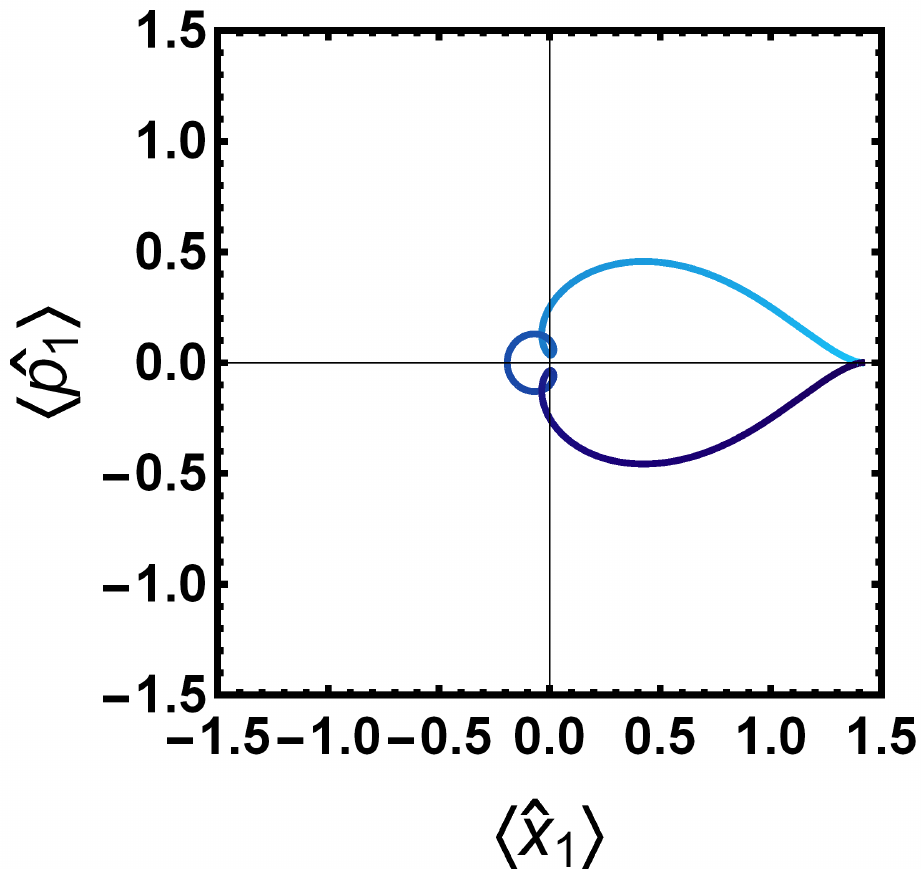}%
}\hfill
\subfloat[ \label{chap:non:Gaussianity:squeezing:fig:quad:mum0:d201}]{%
  \includegraphics[width=.25\linewidth, trim = 00mm 0mm 0mm 0mm]{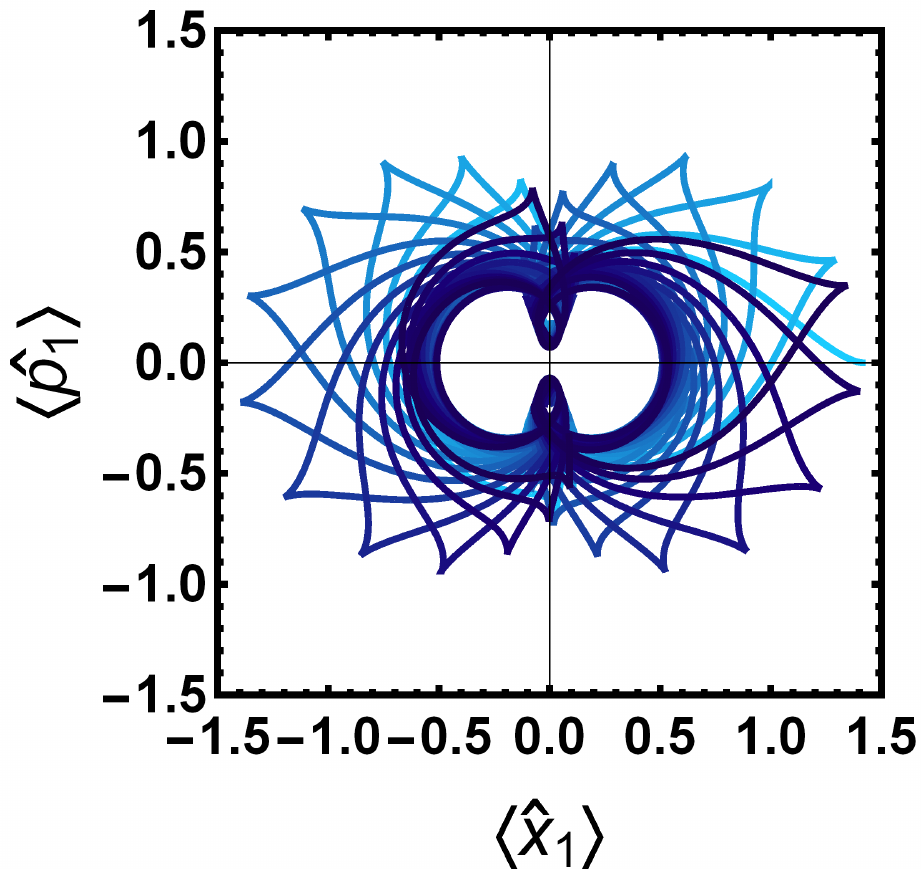}%
}\hfill
\subfloat[
\label{chap:non:Gaussianity:squeezing:fig:quad:mum0:d205}]{%
  \includegraphics[width=.25\linewidth, trim = 00mm 0mm 0mm 0mm]{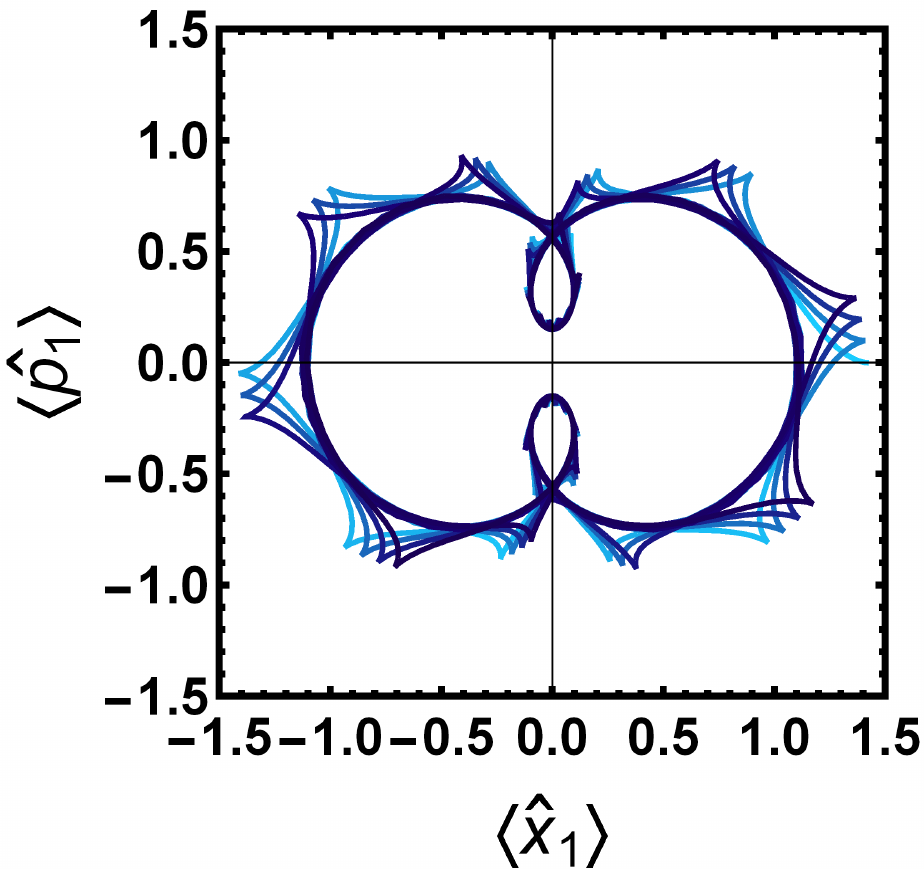}%
}\hfill
\subfloat[
\label{chap:non:Gaussianity:squeezing:fig:quad:mum0:d21}]{%
  \includegraphics[width=.25\linewidth, trim = 00mm 0mm 0mm 0mm]{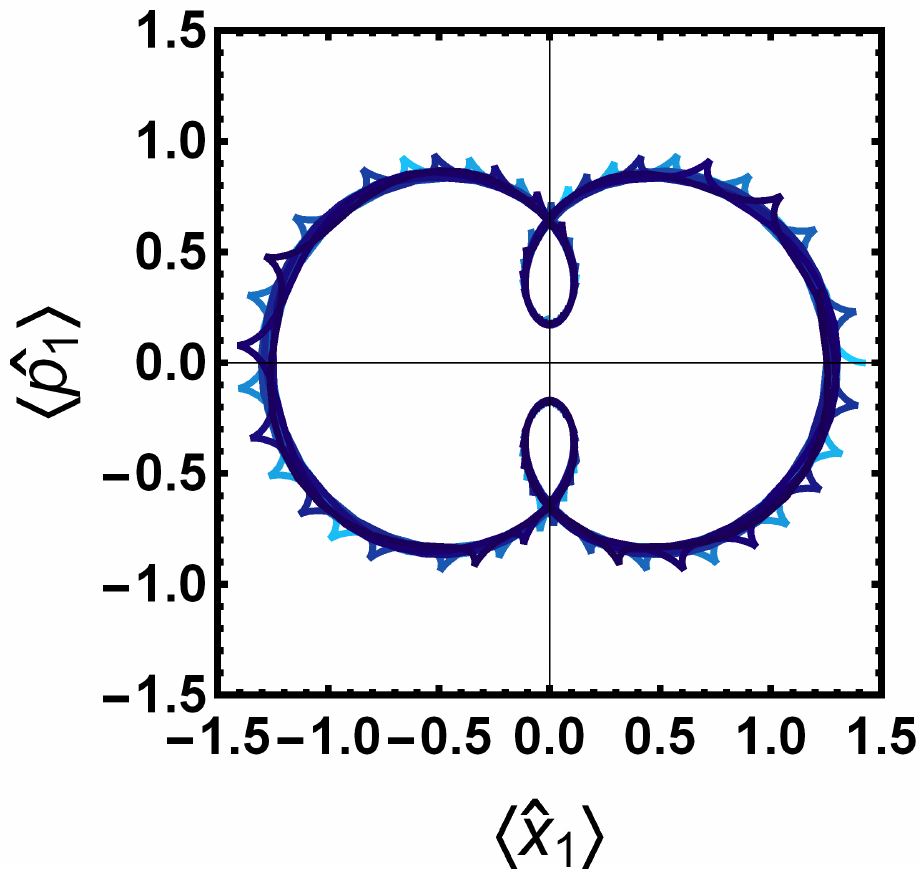}%
}\hfill
\subfloat[
\label{chap:non:Gaussianity:squeezing:fig:quad:mum1:d20}]{%
  \includegraphics[width=.25\linewidth, trim = 00mm 0mm 0mm 0mm]{QuadraturePlotd20Tau1-eps-converted-to}%
}\hfill
\subfloat[
\label{chap:non:Gaussianity:squeezing:fig:quad:mum1:d201}]{%
  \includegraphics[width=.25\linewidth, trim = 00mm 0mm 0mm 0mm]{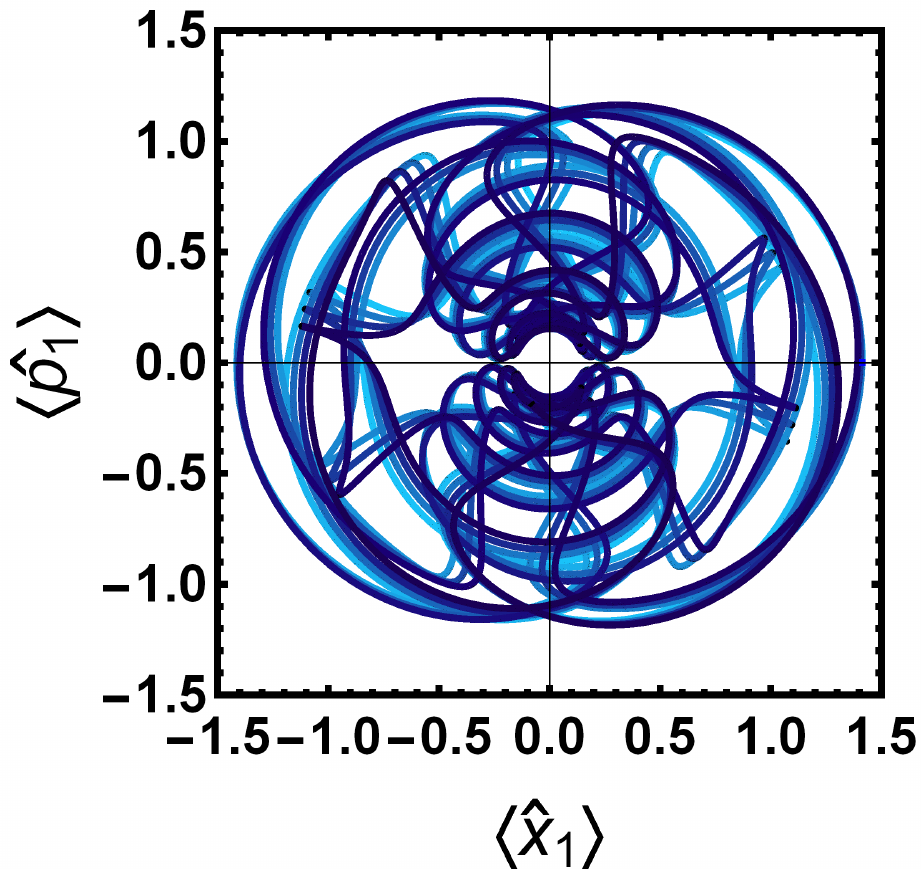}%
}\hfill
\subfloat[
\label{chap:non:Gaussianity:squeezing:fig:quad:mum1:d205}]{%
  \includegraphics[width=.25\linewidth, trim = 00mm 0mm 0mm 0mm]{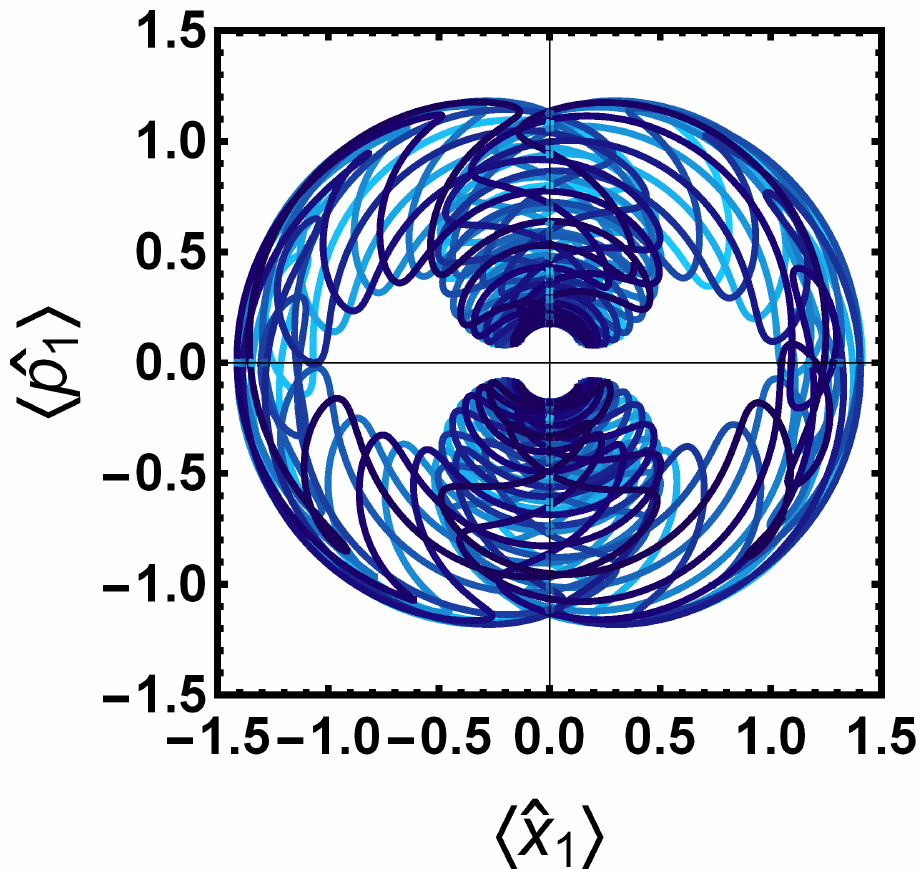}%
}\hfill
\subfloat[
\label{chap:non:Gaussianity:squeezing:fig:quad:mum1:d21}]{%
  \includegraphics[width=.25\linewidth, trim = 00mm 0mm 0mm 0mm]{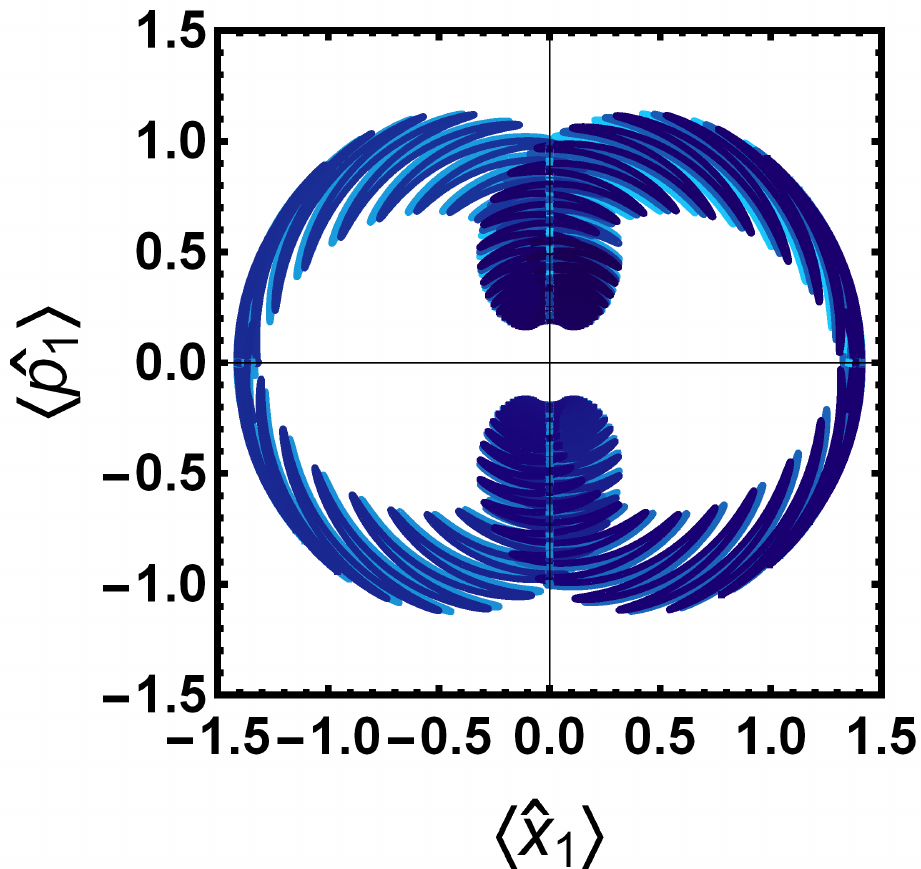}%
}\hfill
\caption[Optical quadratures of an optomechanical system with constant mechanical squeezing parameter]{Optical quadratures of an optomechanical system with constant mechanical squeezing parameter. Both rows show the plots of  $\braket{\hat{x}_{\rm{c}}} = \braket{\hat a^\dag + \hat a}/\sqrt{2}$ vs.\ $\braket{\hat{p}_{\rm{c}}} = i \braket{ \hat a^\dag - \hat a }/\sqrt{2}$. The line starts as light blue at $\tau = 0$ and gradually becomes darker as $\tau$ increase. Plot \textbf{(a)} and \textbf{(e)} show the quadratures for the time range $\tau \in (0, 2\pi)$ and all others have $\tau \in (0, 100\pi)$.  The first row shows the quadratures for values $\mu_{\mathrm{c}} = 1$, $\mu_{\mathrm{m}} = 0$, $\tilde{g}_0 = 1$ and \textbf{(a)} $\tilde{d}_2 = 0$, \textbf{(b)} $\tilde{d}_2 = 0.1$, \textbf{(c)} $\tilde{d}_2 = 0.5$ and \textbf{(d)} $\tilde{d}_2 = 1$. The second row shows the quadratures for values $\mu_{\mathrm{c}} = 1$, $\mu_{\mathrm{m}} = 1$, $\tilde{g}_0 = 1$ and \textbf{(e)} $\tilde{d}_2 = 0$, \textbf{(f)} $\tilde{d}_2 = 0.5$, \textbf{(g)} $\tilde{d}_2 = 1$ and \textbf{(h)} $\tilde{d}_2 = 5$. The increased initial excitation of the mechanical oscillator leads to increased complexity in the quadrature trajectories. A limiting behaviour for large $\tilde{d}_2$ does however appear in which the state is confined to an increasingly narrow trajectory in phase space. Finally, we note that the spikes in \textbf{(b)}, \textbf{(c)}, and \textbf{(d)} appear less pronounced compared with their actual appearance due to restrictions in image resolution. }
\label{chap:non:Gaussianity:squeezing:fig:constant:squeezing:quadratures}
\end{figure*}

Using the methods discussed in Chapter~\ref{chap:decoupling}, we begin by solving the differential equations Eq.~\eqref{chap:non:Gaussianity:squeezing:differential:equation:written:in:paper}. We find the solutions $P_{11} =P_{22}= \cos{ \zeta \tau }$, where we have defined $\zeta = \sqrt{1 + 4 \, \tilde{d}_2}$. This in turn yields the following Bogoliubov coefficients (defined in Eq.~\eqref{chap:non:Gaussianity:squeezing:eq:bogoliubov:coefficients}):
\begin{align}
\alpha(\tau) &= \frac{1}{2} \left( 2 \cos{\zeta \tau} - \frac{i}{\zeta} \left( 1 + \zeta^2  \right)\sin{\zeta \tau} \right),  \nonumber \\
\beta(\tau) &= - 2\,i\,\frac{\tilde{d}_2 }{\zeta} \sin{\zeta \tau}.
\end{align}
Furthermore, we find $\xi(\tau) = \cos{\zeta \tau} - \frac{i}{\zeta} \sin{\zeta \tau}$, which in turn can be integrated to obtain the coefficients Eq.~\eqref{chap:non:Gaussianity:squeezing:sub:algebra:decoupling:solution}, which now read
\begin{align}\label{chap:non:Gaussianity:squeezing:f:functions:constant:squeezing}
&\quad\quad\quad\quad\quad F_{\hat N_a^2} = -\frac{\tilde{g}_0^2}{\zeta^2} \left( 1 - \textrm{sinc}(2\,\zeta\,\tau) \right)\,\tau, \nonumber \\
&F_{\hat N _a \, \hat B_+} = - \frac{\tilde{g}_0}{\zeta} \sin{\zeta \tau}, \quad\quad\quad F_{\hat N _a \, \hat B_-} = \frac{ \tilde{g}_0}{\zeta^2}( \cos{\zeta \tau}-1) \, ,
\end{align}
where $\rm{sinc} (x) = \sin(x) /x$. Since $\tilde{\mathcal{D}}_1 = 0$, all other coefficients are zero. The functions Eq.~\eqref{chap:non:Gaussianity:squeezing:f:functions:constant:squeezing} now fully determine the time evolution through Eq.~\eqref{chap:non:Gaussianity:squeezing:eq:final:evolution:operator}.

\subsection{Quadratures}\label{subsub:time:independent:quadratures}
To gain intuition about the  evolution of the system, we include plots of the optical quadratures. These can be found in Figure~\ref{chap:non:Gaussianity:squeezing:fig:constant:squeezing:quadratures}.  The quadratures are the expectation values of $\hat{x}_{\rm{c}} = (\hat{a}^\dag + \hat{a})/\sqrt{2}$ and $\hat{p}_{\rm{c}} = i (\hat{a}^\dag - \hat{a})/\sqrt{2}$ and would correspond to classical trajectories in phase space. The full expression for the quadratures can be found in Eq.~\eqref{app:exp:values:eq:optical:quadratures} in Appendix~\ref{app:exp:values}.  

In Figures~\ref{chap:non:Gaussianity:squeezing:fig:quad:mum0:d20},~\ref{chap:non:Gaussianity:squeezing:fig:quad:mum0:d201},~\ref{chap:non:Gaussianity:squeezing:fig:quad:mum0:d205} and~\ref{chap:non:Gaussianity:squeezing:fig:quad:mum0:d21}, we have plotted the quadratures for $\mu_{\mathrm{c}} = 1$, $\mu_{\mathrm{m}} = 0$, $\tilde{g}_0 = 1$ and increasing values of $\tilde{d}_2$. Similarly in Figures~\ref{chap:non:Gaussianity:squeezing:fig:quad:mum1:d20},~\ref{chap:non:Gaussianity:squeezing:fig:quad:mum1:d201},~\ref{chap:non:Gaussianity:squeezing:fig:quad:mum1:d205} and~\ref{chap:non:Gaussianity:squeezing:fig:quad:mum1:d21}, we have plotted the quadratures for $\mu_{\mathrm{c}} = 1$, $\mu_{\mathrm{m}} = 1$, $\tilde{g}_0 = 1$ and again increasing values of $\tilde{d}_2$. To show the directionality of the evolution, the colour of the curve starts off as light blue for $\tau = 0$ and becomes increasingly darker as $\tau $ increases. We observe that the addition of mechanical squeezing causes the system to trace out highly complex trajectories, compared with the case when $\tilde{d}_2 = 0$.

\subsection{Measure of non-Gaussianity}\label{measure}
\begin{figure*}[t!]
\subfloat[ \label{chap:non:Gaussianity:squeezing:fig:constant:full:measure:muc1:pi2}]{%
  \includegraphics[width=0.3\linewidth, trim = 0mm 0mm 0mm 0mm]{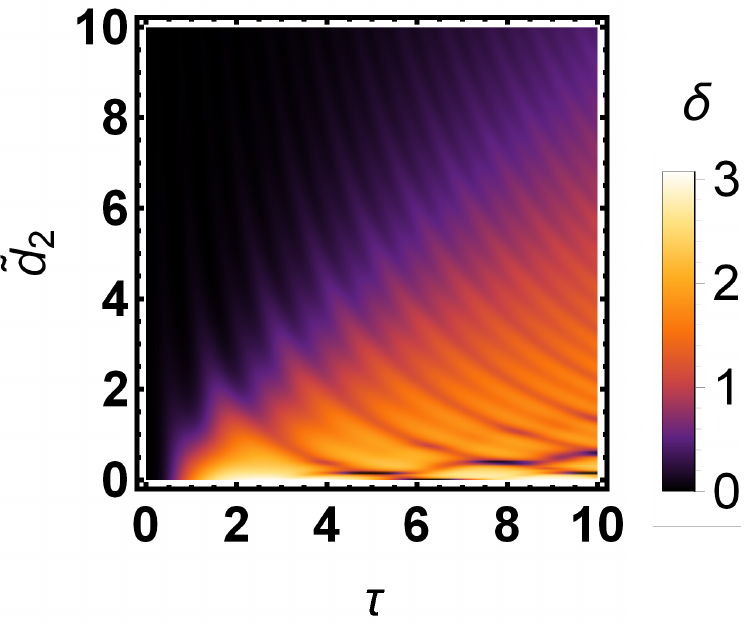}%
}\hfill
\subfloat[ \label{chap:non:Gaussianity:squeezing:fig:constant:reduced:measure:muc1:pi2}]{
  \includegraphics[width=0.315\linewidth, trim = 0mm 0mm 0mm 0mm]{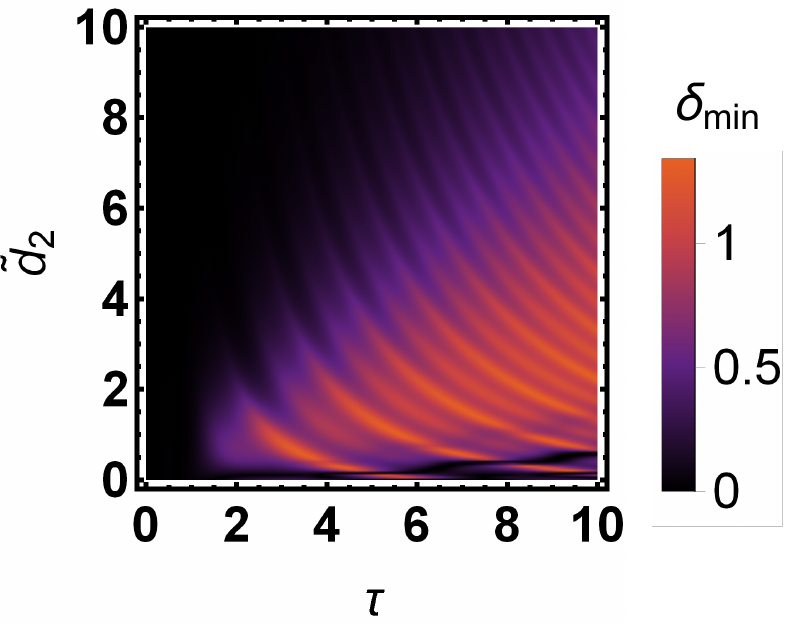}%
  }\hfill
  \subfloat[ \label{chap:non:Gaussianity:squeezing:fig:constant:approximate:measure:muc1:pi2}]{
  \includegraphics[width=0.3\linewidth, trim = 0mm 0mm 0mm 0mm]{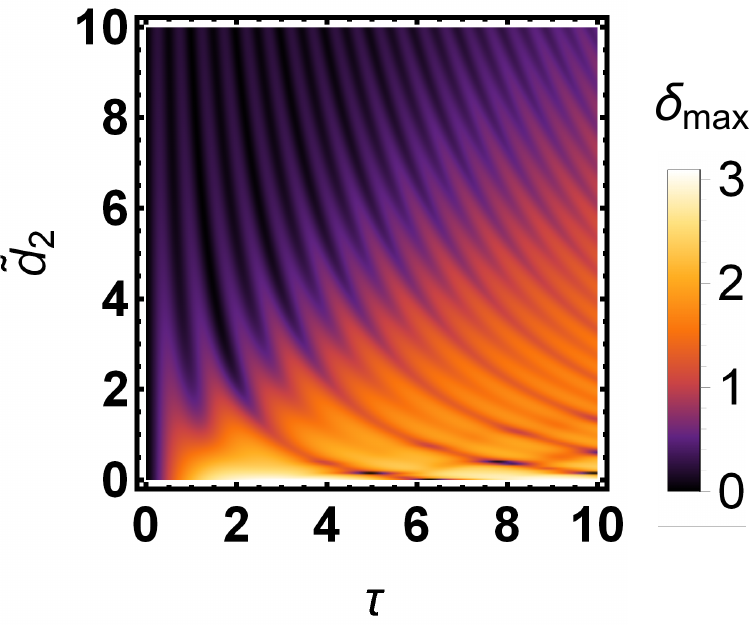}%
}\hfill
\subfloat[ \label{chap:non:Gaussianity:squeezing:fig:constant:full:measure:muc1:pi}]{%
  \includegraphics[width=0.3\linewidth, trim = 0mm 0mm 0mm 0mm]{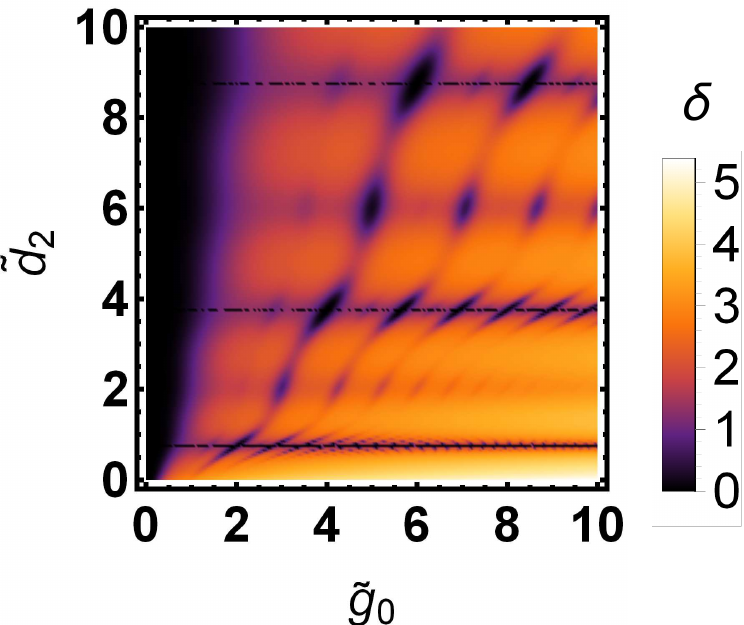}%
}\hfill
\subfloat[ \label{chap:non:Gaussianity:squeezing:fig:constant:reduced:measure:muc1:pi}]{%
  \includegraphics[width=0.3\linewidth, trim = 0mm 0mm 0mm 0mm]{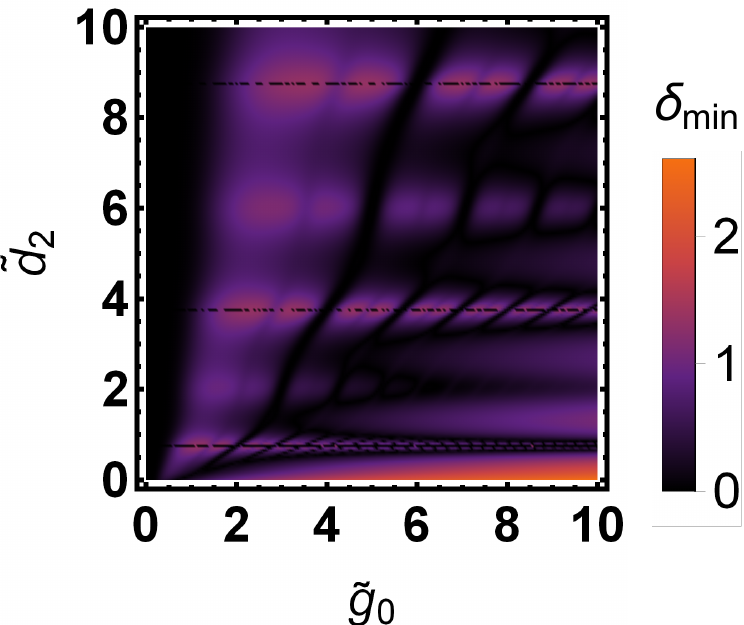}%
}\hfill
\subfloat[ \label{chap:non:Gaussianity:squeezing:fig:constant:approximate:measure:muc1:pi}]{%
  \includegraphics[width=0.3\linewidth, trim = 0mm 0mm 0mm 0mm]{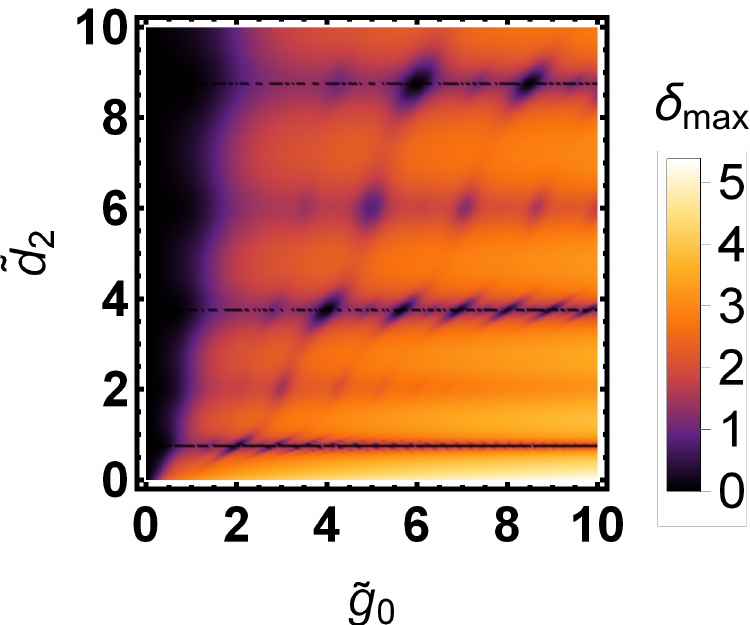}%
}\hfill
\caption[Non-Gaussianity of an optomechanical state with mechanical squeezing]{Non-Gaussianity of an optomechanical state with mechanical squeezing. In each row, the colours have been rescaled to correspond to the same values in the plot. The first row shows the non-Gaussianity as a function of time $\tau$ and the squeezing $\tilde{d}_2$ for $\mu_{\rm{c}} = \tilde{g}_0 = 1$ and $\mu_{\rm{m}} = 0$. \textbf{(a)} shows the full measure $\delta(\tau)$, \textbf{(b)} shows the lower bound $\delta_{\rm{min}}(\tau)$, and \textbf{(c)} shows the upper bound $\delta_{\rm{min}}(\tau)$. The non-Gaussianity generally oscillates in time and does slowly increase for increasing time $\tau$. Furthermore, the upper bound $\delta_{\rm{max}}(\tau)$ approximates the full measure well for these parameters. 
The second row shows the non-Gaussianity $\delta(\tau)$ as a function of the nonlinear coupling $\tilde{g}_0$ and the squeezing parameter $\tilde{d}_2$ for $\mu_{\rm{c}} = 1$ and $\mu_{\rm{m}} = 0$ at time $\tau = \pi$. \textbf{(d)} shows the full measure $\delta(\pi)$, \textbf{(e)} shows the lower bound $\delta_{\rm{min}}(\pi)$ and \textbf{(f)} shows the upper bound $\delta_{\rm{max}}(\pi)$. The non-Gaussianity increases with $\tilde{g}_0$ but decreases with $\tilde{d}_2$. } 
\label{chap:non:Gaussianity:squeezing:fig:constant:squeezing:measure}
\end{figure*}

We now proceed to compute the non-Gaussianity $\delta(\tau)$, defined in Eq.~\eqref{chap:introduction:measure:of:non:gaussianity}, of the state evolving at constant squeezing parameter.  A fully analytic expression for $\delta(\tau)$ exists but is too cumbersome to include here. Instead, we plot the measure of non-Gaussianity in Figure~\ref{chap:non:Gaussianity:squeezing:fig:constant:squeezing:measure}.  In the first row of Figure~\ref{chap:non:Gaussianity:squeezing:fig:constant:squeezing:measure}, we present a comparison between the full measure $\delta(\tau)$ (Figure~\ref{chap:non:Gaussianity:squeezing:fig:constant:full:measure:muc1:pi}) and the lower and upper bounds $\delta_{\rm{min}}(\tau)$ and $\delta_{\rm{max}}(\tau)$ provided by the Araki-Lieb inequality in Figures~\ref{chap:non:Gaussianity:squeezing:fig:constant:reduced:measure:muc1:pi} and ~\ref{chap:non:Gaussianity:squeezing:fig:constant:approximate:measure:muc1:pi}.

We find that the non-Gaussianity increases for large light--matter coupling $\tilde{g}_0$ and large coherent state parameter $\mu_{\mathrm{c}}$, which we already noted in Chapter~\ref{chap:non:Gaussianity:coupling}. However, the most striking feature here is that the larger $\tilde{d}_2$ is, the less non-Gaussian the system becomes. To understand why this is the case, we look at the dependence on $\tilde{d}_2$ in the function $|K_{\hat N_a}|$, since this determines the behaviour of the non-Gaussianity. With the expression Eq.~\eqref{chap:non:Gaussianity:squeezing:f:functions:constant:squeezing} we find 
\begin{equation} \label{chap:non:Gaussianity:squeezing:KNa:constant}
|K_{\hat N_a}|^2 =  \frac{\tilde{g}_0^2}{\zeta^4} \left[ \left( \zeta^2 + 1 \right) \sin^2(\zeta \, \tau) + \cos( 2\, \zeta\, \tau) - 2 \, \cos(\zeta \, \tau) + 1 \right] \, .
\end{equation}
For large $\tilde{d}_2$, and therefore large $\zeta$, the first term inside the brackets dominates and for $\zeta \tau \neq n \pi$ with integer $n$, we are left with $|K_{\hat N_a}|\sim \tilde{g}_0 \, \sin^2( \zeta \, \tau) / \zeta^2$. In general, we find $\lim_{\tilde{d}_2 \rightarrow \infty} |K_{\hat N_a}|^2 = 0$. The consequences for the non-Gaussianity are difficult to predict given the complexity of the expressions, but we note that the mechanical symplectic eigenvalue $\nu_{\rm{Me}}$ decreases, while the optical symplectic eigenvalue $\nu_{\rm{Op}}$ increases. 

Furthermore, the quantity $\theta(\tau) = 2 \, \left( F_{\hat N_a^2} + F_{\hat N_a \, \hat B_+} \, F_{\hat N_a \, \hat B_-} \right)$ is given by 
\begin{equation}
\theta(\tau)  = 2 \, \frac{\tilde{g}_0^2}{\zeta^3}\left( \sin(\zeta \, \tau)  - \zeta \, \tau \right) \, .
\end{equation} 
We find that $\lim_{\tilde{d}_2 \rightarrow \infty} \theta(\tau) = 0 $.  We then look at the symplectic eigenvalues Eq.~\eqref{chap:non:Gaussianity:squeezing:sympelctic:eigenvalue:reduced:state} in this limit. We find that $\nu_{\rm{Me}}\rightarrow 1$, and $\nu_{\rm{Op}} \rightarrow  1 $, which means that both the upper and the lower bounds of the non-Gaussianity tend to zero, and hence $\delta(\tau) \rightarrow 0$ as $\tilde{d}_2$ increases. Hence we conclude that increasing the amount of squeezing in the system reduces the overall non-Gaussianity.

\section{Modulated squeezing parameter}\label{chap:non:Gaussianity:squeezing:sec:modulated:squeezing}
In this Section, we consider a modulated squeezing term. The dimensionless squeezing $\tilde{\mathcal{D}}_2(\tau) = \mathcal{D}_2(t)/\omega_{\rm{m}}$ is time-dependent and of the form 
\begin{equation}  \label{chap:non:Gaussianity:squeezing:eq:modulated:squeezing}
\tilde{\mathcal{D}}_2(\tau) = \,\tilde{d}_2\,  \cos(\Omega_0\, \tau) \, , 
\end{equation}
where $\tilde{d}_2 = d_2/\omega_{\mathrm{m}}$ is the amplitude of the squeezing and $\Omega_0$ denotes the time-scale of squeezing.\footnote{Our rescaled quantities require us to use $\tilde{d}_2 = d_2/\omega_{\mathrm{m}}$ and we define $\Omega_0 = \omega_0/\omega_{\mathrm{m}}$.} 

The differential equations in Eq. ~\eqref{chap:non:Gaussianity:squeezing:differential:equation:written:in:paper}  are not generally analytically solvable for arbitrary choices of $\tilde{\mathcal{D}}_2(\tau)$. However, for the choice of $\tilde{\mathcal{D}}_2(\tau)$ in Eq. ~\eqref{chap:non:Gaussianity:squeezing:eq:modulated:squeezing}, both equations have a known form. Consider the differential equation for $P_{11}$, which we reprint here for convenience, 
\begin{equation} \label{eq:reprint:P11}
\ddot{P}_{11} + \left( 1 + 4 \, \tilde{d}_2 \cos(\Omega_0 \, \tau) \right) P_{11} = 0 \, .
\end{equation}
Equation Eq.~\eqref{eq:reprint:P11} is that of a parametric oscillator, which is used elsewhere in physics to describe, for example, a driven pendulum. The equation for the integral of $P_{22}$ in Eq.~\eqref{chap:non:Gaussianity:squeezing:eq:IP22} takes the same form. 

The equation Eq.~\eqref{eq:reprint:P11} is known as the Mathieu equation. In its most general form, and using conventional notation,   it reads:
\begin{equation} \label{eq:Mathieu}
\frac{d^2 y}{dx^2} + \left[ a - 2  \, q \, \cos(2 \, x) \right] \, y= 0 \, , 
\end{equation}
where $a, q,$ and $x$ are real parameters. The general solutions to this equation will be linear combinations of functions known as the Mathieu cosine $C(a,q,x)$ and Mathieu sine $S(a, q, x)$, the exact form of which will be determined by the boundary conditions for $y$. 

To find which values the $a$, $q$ and $x$ parameters correspond to, we note that the cosine-term in $\tilde{\mathcal{D}}_2(\tau)$ has the argument $\Omega_0 \, \tau$, which means that we must rescale time $\tau $ as $\tau' = \Omega_0 \tau /2$. Inserting our expression for $\tilde{\mathcal{D}}_2(\tau)$ and using the chain-rule to change variables from $\tau$ to $\tau'$, we rewrite the equation for $P_{11}$ as
\begin{equation}
\frac{\Omega_0^2}{4} \frac{d^2 P_{11}}{d \tau^{\prime 2}} + \left( 1 + 4 \, \tilde{d}_2  \cos( 2\, \tau' )  \right) P_{11} = 0 \, ,
\end{equation}
where we identify the variables $a = 4 /\Omega_0^2$, and $q = -8 \, \tilde{d}_2 / \Omega_0^2$.  The boundary conditions $P_{11}(0) = 1$ and $\dot{P}_{11}(0) = 0$ will yield the Mathieu cosine $C(a,q,x)$, and for $I_{P_{22}}$ as the solution, and the boundary conditions $I_{P_{22}}(0) = 0$ and $\dot{I}_{P_{22}}(0) = 1$ will yield the Mathieu sine $S(a,q,x)$ as the solution. For our specific choice of $\mathcal{D}_2(\tau)$ in Eq.~\eqref{chap:non:Gaussianity:squeezing:eq:modulated:squeezing}, the system is resonant at $\Omega_0 = 2$ (see Appendix~\ref{app:mathieu}), which means that $a =1$ and $q = -2\tilde{d}_2$.

\subsection{Approximate solutions at resonance }
The Mathieu equations are notoriously difficult to evaluate numerically. Instead, we use a two-scale method to derive perturbative solutions to $P_{11}$ and $I_{P_{22}}$. The perturbative solutions are valid for  $\tilde{d}_2 \ll 1$ and make use of  specific resonance conditions to ensure that the solutions do not diverge. See Section~\ref{app:mathieu:perturbative:solutions} in Appendix~\ref{app:mathieu} for the full derivation, where we also show that these approximate solutions correspond exactly to the more physically intuitive rotating-wave approximation (RWA) when $\tau \gg1$. For smaller values of $\tau$, the approximate solutions are still valid, but they cannot be interpreted as equivalent to the RWA. 

The squeezing term is resonant when $\Omega_0 = 2$. We then find that the approximate solutions for $P_{11}$ and $I_{P_{22}}$ (the integral of $P_{22}$) are given by, respectively, 
\begin{align}
P_{11} &= \cos (\tau) \,  \cosh (\tilde{d}_2 \, \tau) - \sin (\tau) \,  \sinh ( \tilde{d}_2 \,  \tau) \, \nonumber , \\
I_{P_{22}} &= - \frac{1}{ 1- \tilde{d}_2}\left(\cos (\tau) \,  \sinh (\tilde{d}_2 \, \tau ) - \sin (\tau) \,  \cosh (\tilde{d}_2 \, \tau) \right)  \, .
\end{align}
We then compute $\xi(\tau)$ in Eq.~\eqref{chap:non:Gaussianity:squeezing:eq:def:xi}. We then assume that $\tilde{d}_2\tau\ll 1$ to find 
\begin{align} \label{eq:approx:xi}
\xi(\tau) \approx & \,  e^{- i \tau}\left( 1 + \frac{ \tilde{d}_2^2 \tau^2}{2}\right) + i \, e^{i \, \tau} \, \tilde{d}_2 \, \tau \, ,
\end{align} 
where in the last line we have expanded the hyperbolic functions to second order. By using the relations between $\xi(\tau)$ and the Bogoliubov coefficients in Eq.~\eqref{chap:non:Gaussianity:squeezing:eq:bogoliubov:coefficients}, we find that the Bogoliubov condition is approximately satisfied as:
\begin{equation}
|\alpha(\tau)|^2 - |\beta(\tau)|^2 \approx 1  + \mathcal{O}[(\tilde{d}_2 \tau)^4] \, .
\end{equation}
With this expression, we can now compute the non-zero $F$-coefficients in Eq. ~\eqref{chap:non:Gaussianity:squeezing:sub:algebra:decoupling:solution}:
\begin{align} \label{eq:F:coefficients:modulated:squeezing}
F_{\hat N_a^2} = \, & \tilde{g}_0^2  \tau\, \left( 1 - \tilde{d}_2 \right)  (\mathrm{sinc} (2  \, \tau) - 1 ) + \frac{1}{2} \tilde{g}_0^2 \,  \tilde{d}_2^2 \left(\left(2\,\tau^2 - 3\right) \sin (2 \, \tau) + 2 \, \tau + 4\,\tau \,\cos(2\,\tau)\right)\, ,\nonumber \\
F_{\hat N_a \, \hat B_+} =& - \tilde{g}_0 \, \sin (\tau) - \tilde{g}_0 \, \tilde{d}_2 \, (\tau \,\cos(\tau) - \sin(\tau))  - \frac{1}{2}\,\tilde{g}_0 \, \tilde{d}_2^2 \left[\left(\tau^2 - 2\right) \sin (\tau) + 2\,\tau \,\cos(\tau)\right] ,\nonumber \\
F_{\hat N_a \, \hat B_-} =& - 2\,\tilde{g}_0 \,\sin^2(\tau/2)  + \tilde{g}_0 \, \tilde{d}_2 \, ( \tau\, \sin(\tau) - 2\,\sin^2(\tau/2)  ) \nonumber \\
& + \frac{1}{2} \tilde{g}_0 \,  \tilde{d}_2^2 \left[\left(\tau^2 - 2\right) \cos (\tau) - 2\,\tau\,\sin(\tau) + 2\right]\, ,
\end{align}
where we have discarded terms with $\tilde{d}_2^3$.  With these expressions, we are ready to compute the non-Gaussianity at resonance.

\begin{figure*}[t!]
\subfloat[ \label{chap:non:Gaussianity:squeezing:fig:modulated:full:measure:muc1}]{%
  \includegraphics[width=0.3\linewidth, trim = 0mm 0mm 0mm 0mm]{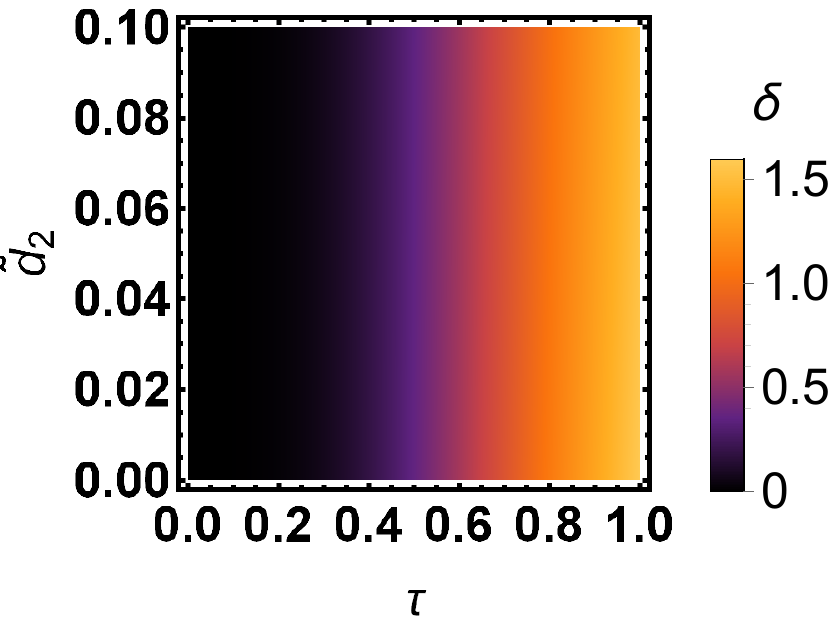}%
}\hfill
\subfloat[ \label{chap:non:Gaussianity:squeezing:fig:modulated:reduced:measure:muc1}]{%
  \includegraphics[width=0.32\linewidth, trim = -2mm 0mm 0mm 0mm]{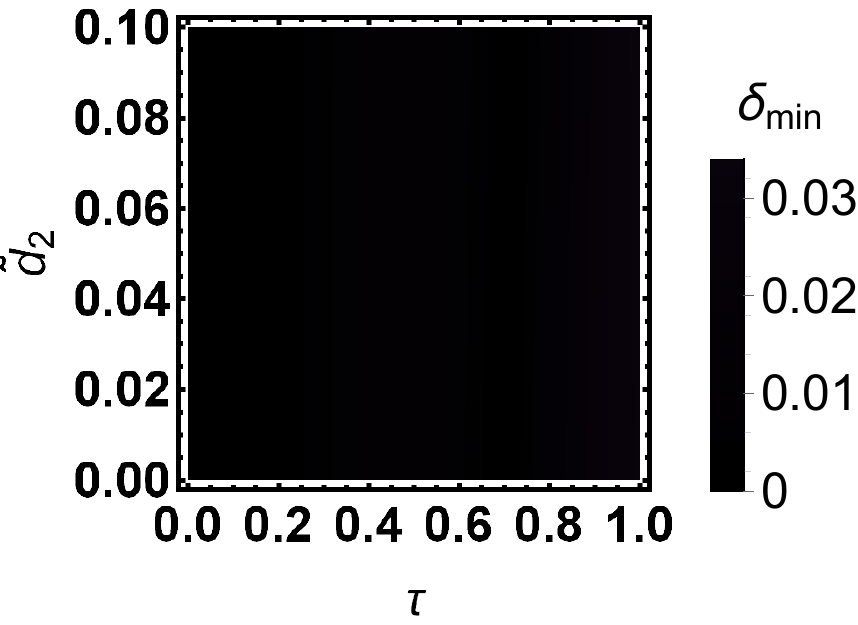}%
}\hfill
\subfloat[ \label{chap:non:Gaussianity:squeezing:fig:modulated:upper:measure:muc1}]{%
  \includegraphics[width=0.3\linewidth, trim = 0mm 0mm 0mm 0mm]{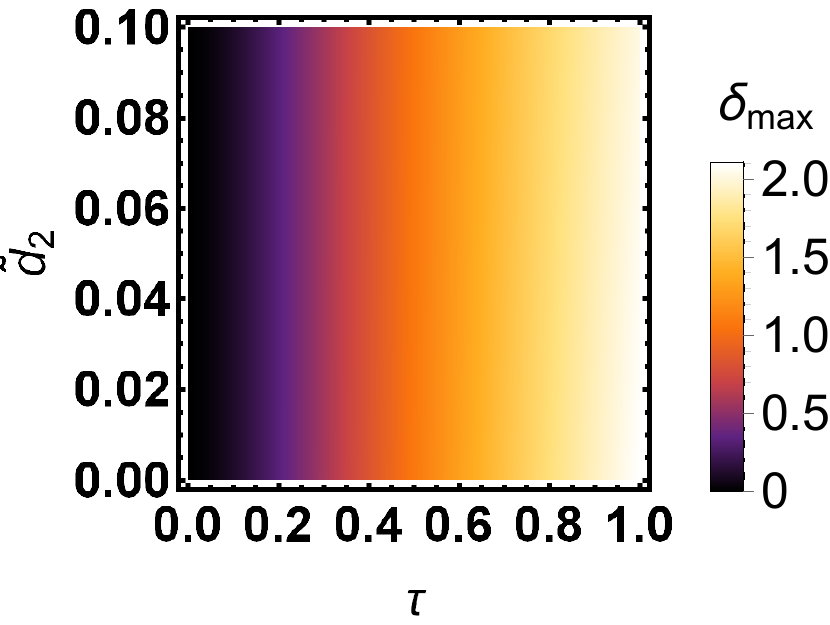}%
}\hfill
\subfloat[ \label{chap:non:Gaussianity:squeezing:fig:modulated:full:measure:muc1:pi}]{%
  \includegraphics[width=0.3\linewidth, trim = 0mm 0mm 0mm 0mm]{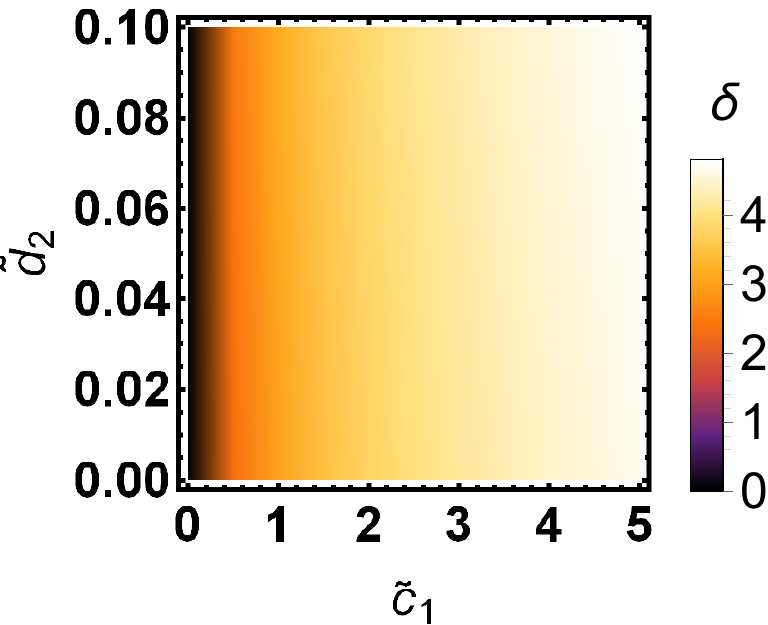}%
}\hfill
\subfloat[ \label{chap:non:Gaussianity:squeezing:fig:modulated:reduced:measure:muc1:pi}]{
  \includegraphics[width=0.311\linewidth, trim = 0mm 0mm 0mm 0mm]{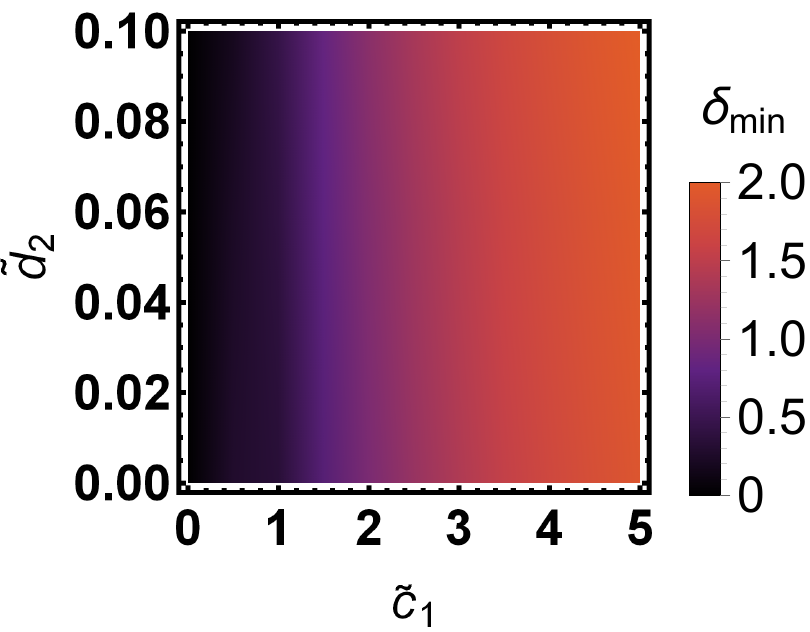}%
  }\hfill
  \subfloat[ \label{chap:non:Gaussianity:squeezing:fig:modulated:upper:measure:muc1:pi}]{
  \includegraphics[width=0.3\linewidth, trim = 0mm 0mm 0mm 0mm]{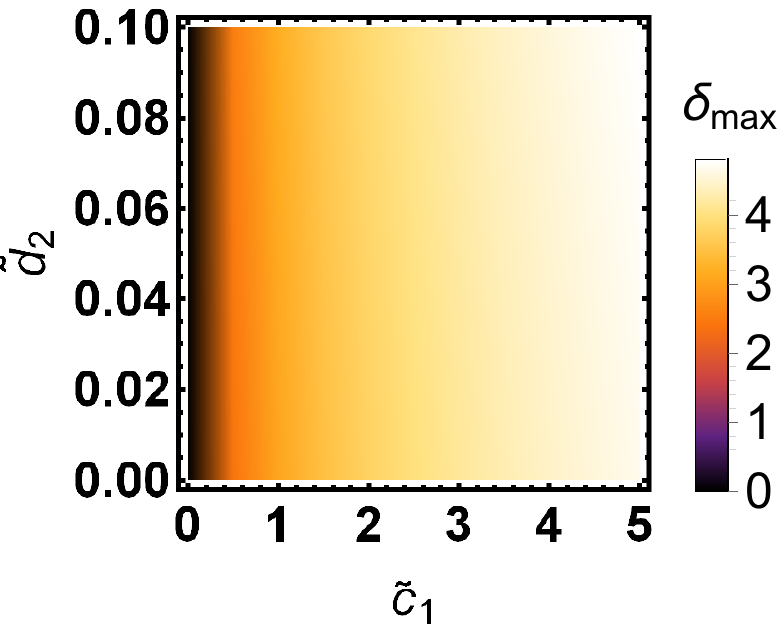}%
}\hfill
\caption[Non-Gaussianity of an optomechanical state with squeezing modulated at mechanical resonance]{Non-Gaussianity of an optomechanical state with squeezing modulated at mechanical resonance. In each row, the colours have been rescaled by the maximum value of $\delta_{\rm{max}}$. The coherent state parameters are $\mu_{\rm{c}} = 1$ and $\mu_{\rm{m}} = 0$ for all plots. 
The first row shows the non-Gaussianity $\delta(\tau)$ as a function of time $\tau$ and squeezing strength $\tilde{d}_2$: \textbf{(a)}  shows the full measure $\delta(\tau)$, \textbf{(b)} shows the lower bound $\delta_{\rm{min}}(\tau)$ and \textbf{(c)} shows the upper bound $\delta_{\rm{max}}(\tau)$. 
The second row shows the non-Gaussianity $\delta(\tau)$ at $\tau = \pi$ as a function of the nonlinear coupling $\tilde{g}_0$ and the squeezing $\tilde{d}_2$: \textbf{(d)} shows the full measure $\delta(\pi)$, \textbf{(e)} shows the lower bound $\delta_{\rm{min}}(\pi)$ and \textbf{(f)} shows the upper bound $\delta_{\rm{max}}(\pi)$. 
The non-Gaussianity increases with $\tilde{g}_0$ and $\tau$. While it is difficult to see for the ranges plotted here, $\delta(\tau)$ also increases very slightly with $\tilde{d}_2$. 
}
\label{chap:non:Gaussianity:squeezing:fig:modulated:squeezing:measure}
\end{figure*}

\begin{figure*}[t!]
\subfloat[ \label{fig:modulated:measure:time}]{%
  \includegraphics[width=0.5\linewidth, trim = 0mm -2mm 0mm 0mm]{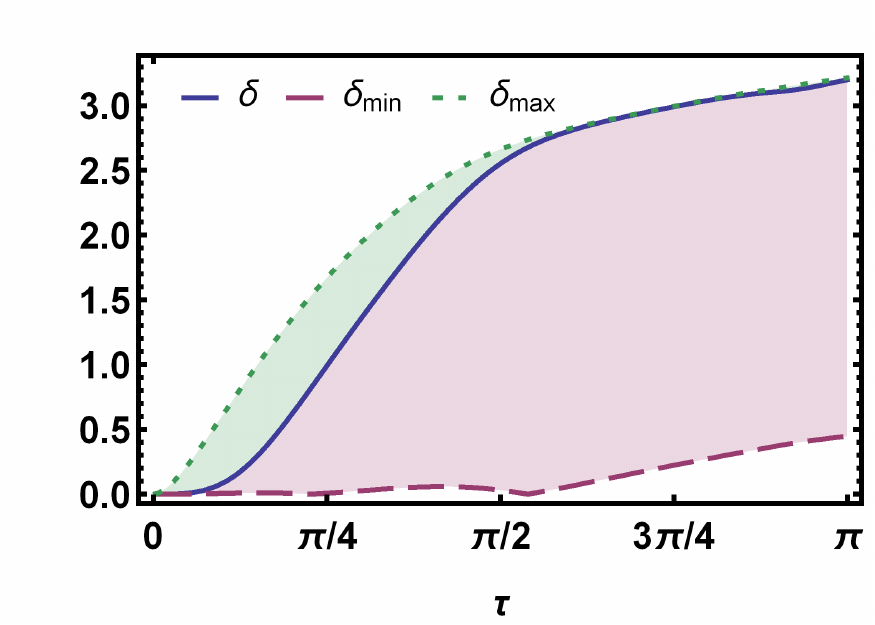}%
}\hfill
\subfloat[ \label{fig:modulated:measure:c1}]{%
  \includegraphics[width=0.49\linewidth, trim = 0mm 0mm 0mm 0mm]{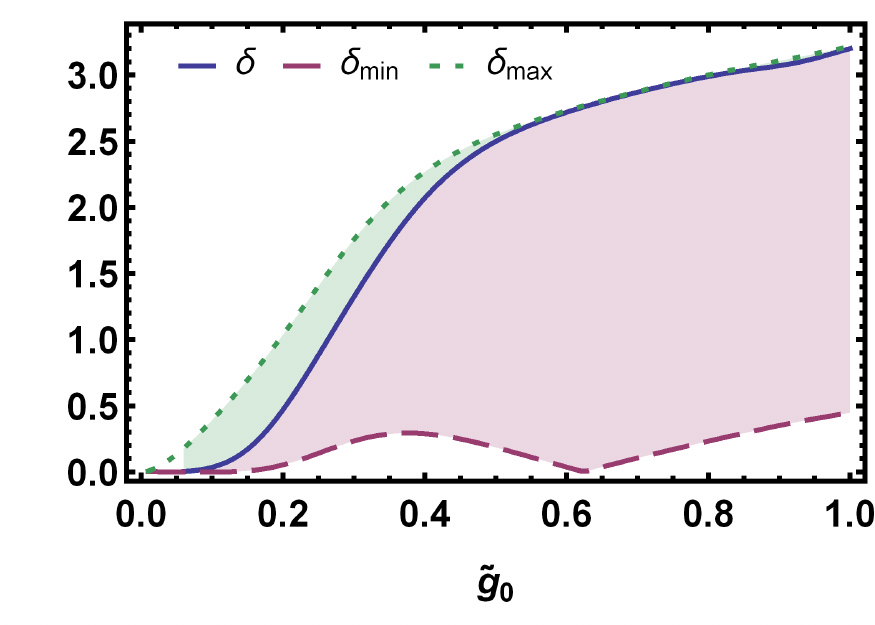}%
}\hfill
\caption[Bounds of the Non-Gaussianity of an optomechanical state with squeezing modulated at mechanical resonance]{Bounds of the Non-Gaussianity of an optomechanical state with squeezing modulated at mechanical resonance. The plots show the non-Gaussianity $\delta(\tau)$ together with its lower bound $\delta_{\rm{min}}(\tau)$ and upper bound $\delta_{\rm{max}}(\tau)$. \textbf{(a)} shows the non-Gaussianity as a function of time $\tau$ at $\tilde{d}_2 = 0.1$, $\mu_{\rm{c}} = 1$, and  $\mu_{\rm{m}} = 0$. \textbf{(b)} shows the non-Gaussianity as a function of $\tilde{g}_0$ at $\tau = \pi$, $\tilde{d}_2 = 0.1$, $\mu_{\rm{c}} = 1$, and $\mu_{\rm{m}} = 0$. The upper bound $\delta_{\rm{max}}(\tau)$ approximates $\delta(\tau)$ increasingly well as $\tilde{g}_0$ and $\tau$ increase. 
}
\label{chap:non:Gaussianity:squeezing:fig:time:dependent:squeezing}
\end{figure*}

\subsection{Measure of non-Gaussianity at resonance}\label{chap:non:Gaussianity:squeezing:subsec:modulated:measure}
We first compute the full measure of non-Gaussianity $\delta(\tau)$ and plot the results in  Figures~\ref{chap:non:Gaussianity:squeezing:fig:modulated:squeezing:measure} and~\ref{chap:non:Gaussianity:squeezing:fig:time:dependent:squeezing}.

In Figure~\ref{chap:non:Gaussianity:squeezing:fig:modulated:squeezing:measure}, we plot the full measure of non-Gaussianity as a function of $\tau $, $\tilde{g}_0$ and $\tilde{d}_2$. The first row shows the  full measure $\delta(\tau)$ in Figure~\ref{chap:non:Gaussianity:squeezing:fig:modulated:full:measure:muc1}, the lower bound $\delta_{\rm{min}}(\tau)$ in Figure~\ref{chap:non:Gaussianity:squeezing:fig:modulated:reduced:measure:muc1}, and the upper bound $\delta_{\rm{max}}(\tau)$ in Figure~\ref{chap:non:Gaussianity:squeezing:fig:modulated:upper:measure:muc1} as a function of time $\tau$ and the squeezing strength $\tilde{d}_2$. Similarly, the second row shows the full measure $\delta(\tau)$ in Figure~\ref{chap:non:Gaussianity:squeezing:fig:modulated:full:measure:muc1:pi}, the lower bound $\delta_{\rm{min}}(\tau)$ in Figure~\ref{chap:non:Gaussianity:squeezing:fig:modulated:reduced:measure:muc1:pi}, and the upper bound $\delta_{\rm{max}}(\tau)$ in Figure~\ref{chap:non:Gaussianity:squeezing:fig:modulated:upper:measure:muc1:pi} as a function of the coupling $\tilde{g}_0$ and the squeezing strength $\tilde{d}_2$. The colours in each row have been rescaled to the largest value displayed in the plots. While it might seem that the non-Gaussianity does not increase in $\tilde{d}_2$, there is a very slight increase. The most crucial feature of these plots is that the non-Gaussianity is no longer oscillating, or indeed decreasing with $\tilde{d}_2$. 

To better demonstrate the behaviour of the measure, we plot the non-Gaussianity as a function of $\tau$ and $\tilde{g}_0$ for fixed values of $\tilde{d}_2$. Figure~\ref{fig:modulated:measure:time} shows the full measure $\delta(\tau)$, the lower bound $\delta_{\rm{min}}(\tau)$ and the upper bound $\delta_{\rm{max}}(\tau)$ as a function of $\tau$ for the parameter  $\tilde{c}_1 = 1$, $\mu_{\rm{c}} = 1$, $\tilde{d}_2 = 0.1$, and $\mu_{\rm{m}} = 0$ as a function of time $\tau$ and the squeezing $\tilde{d}_2$. The second plot in Figure~\ref{fig:modulated:measure:c1} also shows the full measure $\delta(\tau)$, the lower bound $\delta_{\rm{min}}(\tau)$ and the upper bound $\delta_{\rm{max}}(\tau)$ as a function of $\tilde{c}_1$ at $\tau = \pi$, $\tilde{d}_2 = 0.1$, $\mu_{\rm{c}} = 1$, and $\mu_{\rm{m}} = 0$. We find that the non-Gaussianity increases with  $\tilde{c}_1$, as expected.  

 In Figure~\ref{chap:non:Gaussianity:squeezing:fig:time:dependent:squeezing}, we considered $\tilde{d}_2 = 0.1$; a value consistent with the validity of the approximate solutions to the Mathieu equation. For this value, the  non-Gaussianity is found to increase very slightly with $\tilde{d}_2$. To demonstrate this, we consider the regime where $1 \ll 2 |\mu_{\rm{c}}| \ll |K_{\hat N_a}|$, which occurs when $2|\mu_{\rm{c}}| \ll \tilde{g}_0$ for specific values of $\tau$. In this regime, the non-Gaussianity was approximately given by $s_V \left( 2 \, |K_{\hat N_a}| |\mu_{\rm{c}}|  \right)$ in Eq. ~\eqref{chap:non:Gaussianity:squeezing:eq:nG:large:KNa}. 
Given the functions in Eq. ~\eqref{eq:F:coefficients:modulated:squeezing}, we  find that 
\begin{align} \label{eq:modulated:K}
|K_{\hat N_a}|^2 =&  \, 4 \, \tilde{g}_0^2 \,  \sin ^2(\tau/2) - 2\tilde{g}_0^2  \,  \tilde{d}_2 \,  \left( \tau \,  (\sin (\tau) - \sin (2\,\tau)) + (\cos (\tau) - \cos(2\,\tau)) - 2 \sin ^2(\tau/2) \right)\,  \nonumber \\
& + \tilde{g}_0^2 \,  \tilde{d}_2^2 \left(\tau^2 - 2\left(2 -  \tau^2\right) \sin ^2(\tau/2)\right) ,
\end{align}
where  we have again removed terms proportional to $\tilde{d}_2^3$ and $\tilde{d}_2^4$. 
The behaviour of $|K_{\hat N_a}|^2$ is markedly different compared with the constant case in Eq.~\eqref{chap:non:Gaussianity:squeezing:KNa:constant}. Firstly, while $|K_{\hat N_a}|^2$ still oscillates, it also increases with $\tau$ and $\tilde{d}_2$. If we consider the leading term with $\tau^2$, we find that the non-Gaussianity scales with $\delta(\tau) \sim \ln(\tau\, \tilde{d}_2\, \tilde{g}_0)$, which confirms that in this specific regime, the non-Gaussianity increases logarithmically with $\tau$, $\tilde{d}_2$, and $\tilde{g}_0$. We conclude that squeezing is not necessarily detrimental to the non-Gaussianity if the squeezing is modulated at resonance, although more work needs to be done to ascertain the full interplay between the two effects. 
\section{Discussion}\label{chap:non:Gaussianity:squeezing:discussion}

Before  presenting our conclusions, we discuss the applicability and scope of the techniques we developed. 
We also comment on the effect of squeezing on the non-Gaussian character of the system.

\subsection{Advantages over direct numerical simulations}\label{subsec:discussion:numerics}
With our techniques, we have shown that it is possible to solve the dynamics of a nonlinear optomechanical system even when the mechanical squeezing is time-dependent, although those solutions rely on perturbative techniques. To emphasise this point, we wish to compare our approach, which relies on numerically solving the differential equations in Eq. ~\eqref{chap:non:Gaussianity:squeezing:differential:equation:written:in:paper}, with a general numerical method using a standard higher-order Runge-Kutta solver to evolve a state in a truncated Hilbert space, e.g. using the Python library \textit{QuTiP}~\cite{johansson2013qutip}. 

As also discussed in Section~\ref{chap:introduction:sec:numerical:challenges} in Chapter~\ref{chap:introduction}, the dynamics is solved with a Runge-Kutta method, the continuous variable (pure) states are represented as finite-dimensional vectors in a truncated Hilbert space. When the system is nonlinear, information about the state is quickly distributed across large sectors of the Hilbert space. If the computational Hilbert space is too small, numerical inaccuracies  quickly enter into the evolution. It follows that the dimension of the Hilbert space must be large enough to prevent this, which requires significant amounts of computer memory. It is also very difficult to consider parameters of the magnitude $\tilde{g}_0 = 10$ and $\tilde{d}_2 = 10$, as done in this Chapter, since these cause the system to evolve very rapidly and, consequently, require the evolution of the system to be calculated using smaller and smaller time intervals. 

The methods developed here excel at treating systems numerically for large parameters $\tilde{g}_0, \tilde{d}_2, \mu_{\mathrm{c}}$ and $\mu_{\mathrm{m}}$. However, we note that it becomes increasingly difficult to numerically evaluate the dynamics at longer times $\tau$ when the system is numerically solved for arbitrary functions $\tilde{\mathcal{D}}_2(\tau)$. The difficulty is primarily caused by the double integral that determines the coefficient $F_{\hat N_a^2}$ in Eq. ~\eqref{chap:non:Gaussianity:squeezing:sub:algebra:decoupling:solution}, which must  be evaluated numerically. For each value of $\tau$, the integral will be evaluated from $0$ to the final $\tau'$, and then from $0$ to $\tau$. As a result, the integrals take an increasingly long time to evaluate for large $\tau$. We therefore conclude that the key strength in our method lies in evaluating the state of the system at early times $\tau \in (0, 2\pi)$ for large parameters $\mu_{\mathrm{c}}, \tilde{g}_0$, and $\tilde{d}_2 $. We also emphasise that, the computation using our methods is numerically exact, which a naive computation using \textit{QuTiP} or a similar library is not. 

To conclude, our methods allow for the evaluation of the state of the system with large parameters, e.g.  $\tilde{g}_0 = 100$ and $\tilde{d}_2 = 10$, which would be nearly impossible to perform with \textit{QuTiP}  or a similar library unless one had access to significantly more computational resources.

\subsection{Competing behaviours of nonlinearity and squeezing}\label{subsec:discussion:squeezing}
We concluded from Figure~\ref{chap:non:Gaussianity:squeezing:fig:constant:squeezing:measure} that the addition of a constant squeezing term has a detrimental effect on the non-Gaussianity of the system. We also noted that including the constant squeezing is equivalent to changing the mechanical trapping frequency $\omega_{\rm{m}}$ to a specific value and an initially squeezed coherent state (see Section~\ref{app:mathieu:time:varying:trapping:frequency} in Appendix~\ref{app:mathieu}). With this interpretation, our results also show that an initially squeezed states exhibit less non-Gaussianity compared with a coherent states.  The reason for this overall behaviour can be found by simple
  inspection of the total Hamiltonian. If a strong squeezing term is
  included in the Hamiltonian in Eq.~\eqref{chap:non:Gaussianity:squeezing:eq:Hamiltonian},
  it dominates over the interaction term, leading to a decrease in the non-Gaussianity. However, such a process is not fully monotonic, since an increase of the squeezing parameter does not always  decrease the non-Gaussianity. This is, however, reasonable, as it  cannot be expected that only the relative weight of the two parts of  the Hamiltonian matter; the precise dynamics is much more complex, and the non-Gaussianity depends on the entire state, which is driven by the full Hamiltonian.
  
The finding that the non-Gaussianity increases with both time $\tau$ and $\tilde{d}_2$ when modulated at mechanical resonance is interesting and warrants further investigation. We leave this to future work.

\section{Conclusions}\label{chap:non:Gaussianity:squeezing:conclusions}
In this Chapter, we solved the time-evolution of a nonlinear optomechanical system with a time-dependent mechanical displacement term and a time-dependent mechanical single-mode squeezing term. We found analytic expressions for all first and second moments of the quadratures of the nonlinear system and used them to compute the amount of non-Gaussianity of the state. We considered both constant and modulated squeezing parameter, and found that a squeezing parameter  modulated at resonance results in the Mathieu equations, for which we provide approximate solutions equivalent to the rotating-wave approximation.  

In general, we find that the relationship between the squeezing and non-Gaussianity is highly nontrivial. The inclusion of a mechanical squeezing term in the Hamiltonian, which is equivalent to starting with a coherent squeezed state evolving with the standard optomechanical Hamiltonian with a shifted mechanical frequency, decreases the overall non-Gaussianity of the state. If the squeezing term is modulated at mechanical resonance, however, we found that the non-Gaussianity increases with both time and the squeezing parameter in specific regimes. These results hold interesting implications for quantum control of nonlinear optomechanical systems.

Our results also suggest that the combination of non-Gaussian resources and mechanical squeezing may not necessarily be beneficial if the application relies specifically on the non-Gaussian character of the state. However, more work is needed to conclude if this has a significant effect on, for example, applications to sensing. More work is also necessary to properly study the instabilities of the full solutions to the Mathieu equations and how they affect the dynamics. The effect of squeezing the optical rather than mechanical mode is another question we defer to future work. 

The decoupling methods demonstrated here constitute an important step towards fully characterising nonlinear systems with mechanical squeezing and can be used both to model experimental systems and to test numerical methods. The methods used here can also be extended to more complicated quadratic Hamiltonians of bosonic modes, such as Dicke-like models~\cite{emary2003chaos}, which would allow for applications in other areas of physics to be developed.

\part{Quantum metrology with nonlinear optomechanical systems}

\chapter{Optimal estimation with quantum optomechanical systems in the nonlinear regime}
\label{chap:metrology}
\chaptermark{Optimal estimation with quantum optomechanical systems in...}

In this Chapter, we investigate the sensing capabilities of a nonlinear optomechanical systems. We derive a general expression of the quantum Fisher information for the extended optomechanical Hamiltonian in Eq.~\eqref{chap:decoupling:eq:Hamiltonian} and consider three specific examples: (i) estimation of a nonlinear light--matter coupling that is sinusoidally modulated at mechanical resonance, (ii) a mechanical displacement term that is sinusoidally modulated at mechanical resonance, and (iii) a single-mode mechanical squeezing term modulated at parametric resonance. We derive the quantum Fisher information for each case and consider asymptotic cases for simplified expressions. Finally, we input some example values to compute the Fisher information and the variance. 

This Chapter is based on material in Ref~\cite{schneiter2019optimal}. The derivation of the quantum Fisher information was first performed by Fabienne Schneiter and then jointly extended by the thesis author and Dennis R\"{a}tzel. We thank Julien Fra\"isse, Doug Plato, Antonio Pontin, Nathana\"{e}l Bullier, and Peter Barker for useful comments and discussions. 

\section{Introduction}

A key task within the study of quantum metrology entails  investigating the sensing capabilities that can be achieved with different quantum systems. 
Quantum sensing now features prominently in the planning and building of larger-scale experimental efforts, such as the inclusion of squeezed light in Advanced LIGO~\cite{aasi2013enhanced} and space-based tests of microgravity~\cite{van2010bose}. Additional  candidates for quantum sensors include atomic and molecular interferometers for accelerometry and rotation measurements~\cite{lenef1997rotation}. Similarly, Bose-Einstein condensates have been proposed as platforms for testing fundamental physics~\cite{bruschi2014testing, howl2019exploring} and precision measurements of external potentials~\cite{ratzel2019testing}. Quantum advantages in sensing are also furthering the emergence of quantum precision technologies ~\cite{wiseman2009quantum}, which include atomic clocks~\cite{giovannetti2011advances} and extremely precise magnetic field sensors~\cite{liu2019nanoscale, fiderer2018quantum}. 

Optomechanical systems~\cite{marquardt2009optomechanics}, which consist of a mechanical element interacting with light, have previously been shown to be ideal candidates for a number of sensing applications~\cite{arcizet2006high}. In terms of force sensing, microspheres optically trapped in a lattice have been considered~\cite{ranjit2016zeptonewton,hempston2017force}, as well as mesoscopic interferometry for the purpose of gravitational wave detection~\cite{marshman2018mesoscopic}. 
The addition of a cavity to the optomechanical system introduces an inherently nonlinear cubic  interaction between the electromagnetic field and the mechanical element~\cite{aspelmeyer2014cavity}. Optimal estimation schemes for the nonlinear coupling itself have been considered~\cite{bernad2018optimal}. In general, the estimation of anharmonicities present in the system has been a topic of great interest~\cite{rivas2010precision,latmiral2016probing} as well as the enhancement of parameter estimation granted by Kerr nonlinearities~\cite{genoni2009enhancement,rossi2016enhanced}. Additional efforts have focused on parametric driving of the cavity frequency, which manifests itself as a single-mode mechanical squeezing term in the Hamiltonian~\cite{farace2012enhancing}.

To date, due to challenges in solving the dynamical evolution for time-dependent nonlinear systems, most approaches to the full nonlinear case have been restricted to the estimation of static effects. As a result, the proposals considered so far are of limited interest for experimentalists, since static effects are generally difficult to isolate from a random noise floor. Furthermore, if feasible, time-dependent signals also allow for the exploitation of resonances, which can be used to increase the signal-to-noise ratio.

In this Chapter we address this problem by computing the ultimate bounds on the estimation of parameters encoded in an optomechanical Hamiltonian with a time-dependent coupling term, a time-dependent mechanical displacement term and a time-dependent single-mode mechanical squeezing term. 
We use the methods developed in Chapter~\ref{chap:decoupling} to provide a general treatment of the metrological capabilities of an optomechanical system evolving with the Hamiltonian in Eq.\eqref{chap:decoupling:eq:Hamiltonian}. While we focus on the description of optomechanical systems in this thesis, the dynamics we consider can be implemented in different setups such as micro-and nano-cantilevers, membranes levitated nano-spheres and optomechanical resonators~\cite{bowen2015quantum,aspelmeyer2014cavity}. 

This Chapter is organised as follows. We first present the optomechanical Hamiltonian of interest and its analytical solution in Section~\ref{chap:metrology:system:and:dynamics}. We then proceed to define the quantum Fisher information (QFI) in Section~\ref{chap:metrology:sec:QFI:derivation}, which is the key figure of merit we consider. In the same Section, we derive the main result in this Chapter: a general expression for the QFI given the dynamics at hand. Subsequently, in order to demonstrate the applicability of our results, we present three examples of interest in Section~\ref{chap:metrology:three:examples}: (i) Estimation of the strength of a time-dependent optomechanical coupling (Section~\ref{chap:metrology:sec:example:1}), (ii) estimation of the strength of a time-dependent linear displacement term (Section~\ref{chap:metrology:sec:example:2}), and (iii) estimation of the strength of a time-dependent single-mode mechanical squeezing term (Section~\ref{chap:metrology:sec:example:3}). These results are made more concrete in Section~\ref{chap:metrology:sec:applications}, where we compute the QFI given some example experimental parameters. The Chapter is concluded by a discussion of our results in Section~\ref{chap:metrology:sec:discussion}, and some final remarks can be found in Section~\ref{chap:metrology:sec:conclusions}.

\section{System and dynamics} \label{chap:metrology:system:and:dynamics}
In this Chapter, we aim to derive the QFI for an optomechanical system that evolves under the extended optomechanical Hamiltonian in Eq.~\eqref{chap:decoupling:eq:Hamiltonian}. We utilise the fully decoupled  solution derived in Eq.~\eqref{chap:decoupling:eq:final:evolution:operator:J:coefficients} in Chapter~\ref{chap:decoupling}, where the additional terms we considered can be interpreted as external or internal forces that affect the quantum system. 

In this Section, we recap some of the core elements from Chapter~\ref{chap:decoupling} that lie at the foundation of the estimation scheme we consider. 

\subsection{Hamiltonian}

The extended optomechanical Hamiltonian $\hat H$ that we consider in this Chapter is given by
\begin{equation} \label{chap:metrology:eq:Hamiltonian}
\hat H = \hbar \, \omega_{\rm{c}} \, \hat a^\dag \hat a + \hbar \, \omega_{\rm{m}} \, \hat b^\dag \hat b - \hbar \, \mathcal{G}(t) \, \hat a^\dag \hat a \left( \hat b^\dag + \hat b\right) + \hbar \, \mathcal{D}_1(t) \, \left( \hat b^\dag + \hat b \right) + \hbar \, \mathcal{D}_2(t) \, \left( \hat b^\dag + \hat b \right)^2 \, , 
\end{equation}
 where $\hat a, \hat a^\dag$ are the annihilation and creation operators of the optical field with oscillation frequency $\omega_{\rm{c}}$, and  $\hat b, \hat b^\dag$ are the annihilation and creation operators of the phonons in the mechanics with oscillation frequency $\omega_{\rm{m}}$. Furthermore, $\mathcal{G}(t)$ is the light--matter coupling, and $\mathcal{D}_1(t)$ and $\mathcal{D}_2(t)$ are weighting functions that multiply a mechanical displacement term $\left( \hat b^\dag + \hat b\right)$ and a single-mode mechanical squeezing term $\left( \hat b^\dag + \hat b \right)^2$, respectively.

We now ask the following question: \textit{With which precision are we able to estimate the various parameters of $\hat H$?}. We are especially interested in estimating the functions $\mathcal{G}(t)$, $\mathcal{D}_1(t)$ and $\mathcal{D}_2(t)$, or parameters contained in these functions since they can  represent a number of internal and external effects that modify the dynamics of the system. 

Starting with $\mathcal{G}(t)$, it is important to know exactly how the laser light interacts with the mechanical element in any experimental setup. This information is contained in the  nonlinear light--matter coupling weighting function $\mathcal{G}(t)$. To optimally calibrate the system, it is crucial to know how strong the coupling is, or whether it is time-dependent. 
Secondly, linear displacements of the form $\mathcal{D}_1 (t) \, \left( \hat b^\dag + \hat b\right)$ are of major interest in estimation schemes, since any linearised force or potential can be described in this manner. For example, the time-dependent weighting function $\mathcal{D}_1(t)$ can correspond to a time-varying force or acceleration, such as that generated by an electromagnetic field that interacts with an nitrogen-vacancy (NV) centre implanted in the levitated optomechanical sphere.  Estimating any parameters in the function $\mathcal{D}_1(t)$ corresponds to measuring an external effect by using the optomechanical system as a probe. 
Finally, the single-mode mechanical squeezing term $\mathcal{D}_2(t) \, \left( \hat b^\dag + \hat b\right)^2$ corresponds to a time-modulated mechanical trapping frequency, which we show explicitly in Appendix~\ref{app:mathieu}. Whether external or internal effects cause the trapping frequency to oscillate, it is possible to consider an estimation scheme of the amplitude of the change or the time-dependence by estimating parameters contained in $\mathcal{D}_2(t)$. 

Before we proceed, we again consider dimensionless units by dividing the Hamiltonian in Eq.~\eqref{chap:metrology:eq:Hamiltonian} by $\hbar \omega_{\rm{m}}$, such that time becomes $\tau = \omega_{\rm{m}} \, t$. The rescaled Hamiltonian becomes
\begin{equation} 
\hat{\tilde{H}} = \Omega_{\rm{c}} \hat a^\dag \hat a + \hat b^\dag \hat b - \tilde{\mathcal{G}}(\tau) \, \hat a^\dag \hat a \, \left( \hat b^\dag + \hat b \right) + \tilde{\mathcal{D}}_1(\tau) \, \left( \hat b^\dag + \hat b \right) + \tilde{\mathcal{D}}_2(\tau) \, \left( \hat b^\dag + \hat b\right)^2 \, , 
\end{equation}
where $\Omega_{\rm{c}} = \omega_{\rm{c}}/\omega_{\rm{m}}$, $\tilde{\mathcal{G}}(\tau) = \mathcal{G}(\omega_{\rm{m}} t) /\omega_{\rm{m}}$, $\tilde{\mathcal{D}}_1(\tau) = \mathcal{D}_1 (\omega_{\rm{m}} t) /\omega_{\rm{m}}$, and $\tilde{\mathcal{D}}_2(\tau) = \mathcal{D}_2(\omega_{\rm{m}}t)$.

\subsection{Solution of the dynamics}
To derive an expression for the QFI, we must first solve the dynamics generated by the Hamiltonian in Eq.~\eqref{chap:metrology:eq:Hamiltonian}. Only then can we infer how time-dependent effects in particular act on the system and find the fundamental sensitivity of the system. 

From the solution in Chapter~\ref{chap:decoupling}, we know that the decoupled time-evolution operator $\hat U(\tau)$ is given by 
\begin{align} \label{chap:metrology:eq:time:evolution:operator}
\hat U_\theta(\tau)=& \, e^{-i\,\Omega_\mathrm{c} \hat N_a\,\tau}\, e^{ - i \, J_b \, \hat N_b} \, e^{- i \, J_+ \, \hat B_+^{(2)}} \, e^{- i \, J_- \, \hat B_-^{(2)}} 
\,e^{-i\,F_{\hat{N}_a}\,\hat{N}_a}\,e^{-i\,F_{\hat{N}^2_a}\,\hat{N}^2_a}\,\nonumber \\
&\times e^{-i\,F_{\hat{B}_+}\,\hat{B}_+}\,e^{-i\,F_{\hat{N}_a\,\hat{B}_+}\,\hat{N}_a\,\hat{B}_+}\,e^{-i\,F_{\hat{B}_-}\,\hat{B}_-}\,e^{-i\,F_{\hat{N}_a\,\hat{B}_-}\,\hat{N}_a\,\hat{B}_-} \, , 
\end{align}
where we have used the following notation for the operators:
\begin{align}\label{chap:metrology:eq:Lie:algebra}
	 	\hat{N}^2_a &:= (\hat a^\dagger \hat a)^2 \nonumber \\
	\hat{N}_a &:= \hat a^\dagger \hat a &
	\hat{N}_b &:= \hat b^\dagger \hat b \nonumber\\
	\hat{B}_+ &:=  \hat b^\dagger +\hat b &
	\hat{B}_- &:= i\,(\hat b^\dagger -\hat b) &
	 & \nonumber\\
	\hat{B}^{(2)}_+ &:= \hat b^{\dagger2}+\hat b^2 &
	\hat{B}^{(2)}_- &:= i\,(\hat b^{\dagger2}-\hat b^2) &
	 &  \nonumber\\
	\hat{N}_a\,\hat{B}_+ &:= \hat{N}_a\,(\hat b^{\dagger}+\hat b) &
	\hat{N}_a\,\hat{B}_- &:= \hat{N}_a\,i\,(\hat b^{\dagger}-\hat b) \, ,&
	 & 
\end{align}
and where the $F$-coefficients are given by Eq.~\eqref{chap:decoupling:eq:sub:algebra:decoupling:solution}. The $J$-coefficients can either be computed by solving the coupled differential equations for $J_b$ and $J_\pm$ directly in Eq.~\eqref{chap:decoupling:eq:diff:equations:Js}, or indirectly by first solving the differential equations for $P_{11}$ and $P_{22}$ in Eq.~\eqref{chap:decoupling:differential:equation:written:down}, and then relating them to $J_b$ and $J_\pm$ through the expressions in  Eq.~\eqref{chap:decoupling:eq:squeezing:relation}, which in turn rely on the expressions for the Bogoliubov coefficients in Eq.~\eqref{chap:decoupling:eq:bogoliubov:coeffs:expression}. 

The operator in Eq.~\eqref{chap:metrology:eq:time:evolution:operator} can be further simplified by defining the operators $\mathcal{\hat{F}}_\pm :=  F_{\hat B_\pm} + F_{\hat N_a  \, \hat B_\pm} \,  \hat{N}_a$ and $\mathcal{\hat{F}}_{\hat N_a} :=  F_{{\hat N}_a} + F_{{\hat N}^2_a} \,\hat{N}_a$. As $\hat{N}_a$ commutes with all operators in Eq.~\eqref{chap:metrology:eq:time:evolution:operator}, $\hat{N_a}$ and  $\hat{N_a}^2$ can be treated as c-number-valued functions in all manipulations of the exponentials in $\hat{U}(\tau)$. In particular, exponential terms containing only $\hat{N_a}$ and and $\hat{N_a}^2$ and the identity can be freely combined and shifted in $\hat{U}(\tau)$. Then, by using the definition of the Weyl displacement operator in Eq.~\eqref{app:commutators:weyl:operator:combinations} in Appendix~\ref{app:commutators}, we can rewrite the time evolution operator as
 \begin{align}\label{chap:metrology:eq:time:evolution:operator:compact}
\hat U(\tau):=& \,\hat {\tilde{U}}_{\rm{sq}}\, e^{-i(\Omega_{\rm{c}} \,\tau + \mathcal{\hat{F}}_{\hat N_a})\hat{N}_a - i \mathcal{\hat{F}}_+ \mathcal{\hat{F}}_- }\, \hat{D}_b(\mathcal{\hat{F}}_- - i\mathcal{\hat{F}}_+) \, ,
\end{align}
where we used the standard formula for the composition of two displacement operators and defined the operator
\begin{align} \label{chap:metrology:eq:time:evolution:sq:compact}
\hat{\tilde U}_{\rm{sq}} = &  \,  e^{-i \, J_b \hat{N}_b}\,\hat{S}_b(2 \, i \, J_+)\,\hat{S}_b(-2 \, J_-) \, , 
\end{align}
using the definition of the squeezing operator $\hat{S}_b(z):=\exp[\frac{1}{2}(-z \hat{b}^{\dagger 2} + z^* \hat{b}^2)]$, where $z$ is a complex squeezing parameter. 

We note that we have added the subscript $\theta$ to $\hat U(\tau)$ to emphasise the fact that the evolution operator depends on the parameter $\theta$ that we wish to estimate. We also note that while in the previous two chapters (Chapters~\ref{chap:non:Gaussianity:coupling} and~\ref{chap:non:Gaussianity:squeezing}), we used the decoupled version of $\hat U(\tau)$ in Eq.~\eqref{chap:decoupling:eq:final:evolution:operator}, it is more beneficial to use the fully decoupled operator in Eq.~\eqref{chap:metrology:eq:time:evolution:operator} here. This stems from the fact that we must explicitly consider the estimation of parameters in the subsystem evolution operator $\hat U_{\rm{sq}}$. 

\subsection{Initial state of the system}
In this Chapter, we consider the evolution of a initially coherent state in the optics and a thermal state of the mechanics, which we introduced in Section~\ref{chap:introduction:initial:states} in Chapter~\ref{chap:introduction}. We reprint it here for convenience: 
\begin{align}\label{chap:metrology:eq:initial:state:coherent:thermal}
\hat \rho(\tau = 0) &= \ketbra{\mu_{\rm{c}}} \otimes \sum_{n = 0}^\infty \frac{\tanh^{2n}r_T}{\cosh^2 r_T} \ketbra{n} \, , 
\end{align}
The time-evolved state, given by 
\begin{equation}
\hat \rho_\theta = \hat U_\theta (\tau) \, \hat \rho(0) \, \hat U^\dag_\theta(\tau) \, .
\end{equation}
Instead of treating the evolved state, we focus on the evolution operator. 

\section{General quantum Fisher information for optomechanical systems} \label{chap:metrology:sec:QFI:derivation}
As discussed in Section~\ref{chap:introduction:quantum:metrology} in Chapter~\ref{chap:introduction}, quantum metrology provides the tools to compute ultimate bounds on precision measurements of parameters contained in a quantum channel~\cite{paris2009quantum}. The general scheme requires an input state  $\hat \rho(0)$, a channel  $\hat{U}_\theta$ which depends on a classical parameter $\theta$ that will be estimated, and a set of measurements on the final state $\hat \rho(\theta):=\hat{U}_\theta\,\hat \rho(0)\,\hat{U}_\theta^\dag$. 

The quantum Fisher information (QFI), which we denote  $\mathcal{I}_\theta$, allow for the computation of ultimate sensing bounds imposed by the laws of physics~\cite{helstrom1976quantum,holevo2011probabilistic}. The QFI is a dimensionful information measure whose inverse provides a lower bound to the variance $\rm{Var}(\theta)$ of a parameter $\theta$ through the quantum Cram\'er--Rao bound (QCRB) $\mathrm{Var}(\theta)\geq(\mathcal{I}_\theta)^{-1}$~\cite{braunstein1994statistical,braunstein1996generalized,paris2009quantum}. The QCRB is optimized over all possible positive measurements~\cite{peres2006quantum} and all possible unbiased estimator functions. Its importance arises from the fact that it can be saturated in the limit of a large number of measurements. The QCRB hence constitutes an important benchmark for the ultimate sensitivity that can be achieved (at least in principle when all technical noise problems are solved), and only the fundamental uncertainties due to the quantum state itself remain. 

For unitary channels, and given an initial state  $\hat \rho(0) = \sum_n \lambda_n \ket{\lambda_n}\bra{\lambda_n}$, the quantum Fisher information can in general be written in the form~\cite{pang2014quantum,jing2014quantum}
\begin{align}\label{chap:metrology:definition:of:QFI}
\mathcal{I}_\theta
=& \,  4 \, \sum_n \lambda_n\left(\bra{\lambda_n}\mathcal{\hat H}_\theta^2\ket{\lambda_n} - \bra{\lambda_n}\mathcal{\hat H}_\theta\ket{\lambda_n}^2 \right) \nonumber \\
&- 8\sum_{n\neq m}\frac{\lambda_n \lambda_m}{\lambda_n+\lambda_m}\left| \bra{\lambda_n}\mathcal{\hat H}_\theta \ket{\lambda_m}\right|^2\;,
\end{align}
where the operator $\mathcal{\hat H}_\theta$ is defined as $\mathcal{\hat H}_\theta = - i \hat U^\dag_\theta \partial_\theta \hat U_\theta $. We provide an explicit derivation of the expression Eq.~\eqref{chap:metrology:definition:of:QFI} in Section~\ref{app:QFI:derivation:of:QFI:mixed:states} in Appendix~\ref{app:QFI}. 

To find an expression for the QFI, we must compute the derivative $\partial_\theta \hat U_\theta$. Since $\hat U_\theta$ contains non-commuting terms, this must be done with care, as additional terms are generated when the operators are multiplied by congruence. In the following Sections, we first derive a manageable form of $\hat{\mathcal{H}}_\theta$, and we then proceed to apply this expression to derive a compact form of Eq.~\eqref{chap:metrology:definition:of:QFI} for our choice of initial state. 

\subsection{Consequences of the closed Lie algebra}
For metrology purposes, the channel $\hat{U}_\theta$ is the time evolution operator in Eq.~\eqref{chap:metrology:eq:time:evolution:operator:compact}, and the parameter $\theta$ to be estimated is chosen depending on the specific case of interest. 
Since the decoupled $\hat U_\theta$ in Eq.~\eqref{chap:metrology:eq:time:evolution:operator} relies on the closed Lie algebra identified in Chapter~\ref{chap:decoupling}, it follows that the operator $\hat{\mathcal{H}}_\theta$ defined through Eq.~\eqref{chap:metrology:definition:of:QFI} has the same operator structure. This will become clear in the next Section. 

We can therefore write $\hat{\mathcal{H}}_\theta$ in the following form 
\begin{align} \label{chap:metrology:eq:mathcalH:with:coefficients}
 \mathcal{\hat H}_\theta =& \, A\,\hat N_a^2 +B\,\hat N_a +  C_+\,\hat B_+ + C_{\hat N_a,+} \hat N_a \, \hat B_+ + C_-\,\hat B_- + C_{\hat N_a,-} \, \hat N_a \, \hat B_- + E\,\hat N_b \nonumber \\
 &+  F\,\hat B_+^{(2)} + G\,\hat B_-^{(2)} +K \, ,  
\end{align}
where the operators are identified in Eq.~\eqref{chap:metrology:eq:Lie:algebra}, and $A, \, B, \, C_\pm, \, C_{\hat N_a \, \pm} , \, E, \, F, \, G$, and $K$ are real coefficients which follow from the differentiation of $\hat{U}_\theta$. 
The QFI in Eq.~\eqref{chap:metrology:definition:of:QFI} can now be computed by taking the expectation values of the operator-valued terms in Eq.~\eqref{chap:metrology:eq:mathcalH:with:coefficients} and its square with respect to the initial state $\hat{\rho}(0)$ in Eq.~\eqref{chap:metrology:eq:initial:state:coherent:thermal}. 

Before we proceed, we note that for this choice of state, the eigenvectors $\ket{\lambda_n}$ and eigenvalues $\lambda_n$ in Eq.~\eqref{chap:metrology:definition:of:QFI} are given by $\ket{\lambda_n} = \ket{\mu_\textrm{c}}\otimes \ket{n}$ $\lambda_n=\tanh^{2n}(r_T)/\cosh^2(r_T)$ for this case.

\subsection{Derivation of the coefficients} \label{chap:metrology:sec:derivation:coefficients}
Let us proceed to determine the coefficients in Eq.~\eqref{chap:metrology:eq:mathcalH:with:coefficients}. To do so, we first  differentiate the time-evolution operator $\hat U_\theta$ with respect to the parameter $\theta$. The operator $\hat U_\theta$ can be decomposed into the form
$\hat U_\theta=\hat U_a\hat{\tilde U}_{\mathrm{sq}}\hat U_{\hat B_+} \hat U_{\hat B_-}$, where $\hat{\tilde{U}}_{\rm{sq}}$ is defined in Eq.~\eqref{chap:metrology:eq:time:evolution:sq:compact}, and where we have introduced
\begin{align}
 \hat U_{\hat N_a}
&= e^{-i \, \left(\Omega_{\rm{c}} \, \tau +\mathcal{\hat F}_{\hat N_a}\right){\hat N}_a}\;,\nonumber \\
\hat U_{\hat B_+}
&= e^{-i \, \mathcal{\hat F}_+ \, \hat B_+}\;,\nonumber \\
\hat U_{\hat B_-}
&= e^{ - i  \, \mathcal{\hat F}_- \, \hat B_-}  \, ,
\end{align}
and where we recall that $\mathcal{\hat F}_{\hat N_a} = F_{\hat N_a}+F_{\hat N_a^2}\hat N_a$, 
$\mathcal{\hat F}_+ = F_{\hat B_+} + i \, F_{\hat N_a \, \hat B_+} \, \hat N_a$ and $\mathcal{\hat F}_- = F_{\hat B_-} + i \, F_{\hat N_a \,  \hat B_-}\hat{N}_a$. 
Then, we can write $\mathcal{\hat H}_\theta$ as
\begin{align}\label{chap:metrology:Hcal:as:derivatives}
\mathcal{\hat H}_\theta
=& -i \, \biggl( 
\hat U_{ \hat N_a}^\dagger \partial_\theta {\hat U}_{\hat N_a} 
+ \hat U_{\hat B_-}^\dagger \hat U_{\hat B_+}^\dagger \hat{\tilde{U}}_{\mathrm{sq}}^\dagger \partial_\theta{ \hat {\tilde{U}}}_{\mathrm{sq}} \hat U_{\hat B_+}\hat U_{\hat B_-} \nonumber \\
&\quad\quad\quad+\hat U_{\hat B_-}^\dagger \hat U_{\hat B_+}^\dagger \partial_\theta{\hat U}_{\hat B_+} \hat U_{\hat B_-}+ \hat U_{\hat B_-}^\dagger  \partial_\theta \hat U_{\hat B_-}
\biggr).
\end{align}
In order to proceed we need to compute the derivative $\partial_\theta\hat{\tilde U}_{\mathrm{sq}}$, we use the fully decoupled form of $\hat{\tilde U}_{\mathrm{sq}}$, which is
\begin{equation}
 \hat{\tilde U}_{\mathrm{sq}}=  \exp[-i \,  J_b \,  \hat{N}_b]\,\exp[-i \,  J_+ \, \hat B_+^{(2)}]\,\exp[-i \, J_- \, \hat B_-^{(2)}]\;.
\end{equation}
We derive the exact form of $J_b$ and $J_\pm$ in Eq.~\eqref{chap:decoupling:eq:diff:equations:Js} in Chapter~\ref{chap:decoupling} as a solution to a coupled set of differential equations.  If we now assume all three coefficients $J_b, J_+$ and $J_-$ depend on the estimation parameter $\theta$, we differentiate $\hat{\tilde{U}}_{\mathrm{sq}}$  to find 
\begin{align} \label{eq:app:dff:of:Us}
\partial_\theta \hat {\tilde{U}}_{\mathrm{sq}} =& \,   -   i \,  \partial_\theta J_b  \, \hat N_b \, e^{-i \,  J_b   \, \hat N_b}\,e^{-i  \, J_+ \, \hat B^{(2)}_+ }\,e^{- i \, J_- \, \hat B^{(2)}_-} \nonumber \\
&- i  \, \partial_\theta J_+ \,  e^{-i \, J_b \, \hat N_b}\,   \hat B^{(2)}_+ \,  e^{-i \, J_+\, \hat B^{(2)}_+}\,e^{- i \, J_- \, \hat B^{(2)}_-}\nonumber   \\
& - i \, \partial_\theta J_-  \, e^{-i  \, J_b \, \hat N_b}\,e^{- i \,  J_+ \, \hat B^{(2)}_+}\, \hat B^{(2)} _- \, e^{ - i \, J_-\hat B^{(2)}_-} \;.
\end{align}
For simplicity, we then define
\begin{equation}
\hat U^\dagger_{\hat B_-}\hat U^\dagger_{\hat B_+} \hat{\tilde{U}}_{\mathrm{sq}}^\dagger\partial_\theta\hat{\tilde{U}}_{\mathrm{sq}}\hat U_{\hat B_+}\hat U_{\hat B_-} = \hat C_1+\hat C_2+\hat C_3  \, ,
\end{equation} 
where the operator-valued functions are given by 
\begin{align}
\hat C_1=
& - i \partial_\theta J_b \, \hat U_{\hat B_-}^\dag \hat U_{\hat B_+}^\dag  \, e^{i \, J_- \, \hat B^{(2)}_-}\,e^{i  \, J_+ \, \hat B^{(2)}_+ } \,  \hat N_b \, e^{-i  \, J_+ \, \hat B^{(2)}_+ }\,e^{- i \, J_- \, \hat B^{(2)}_-} \hat U_{\hat B_+} \hat U_{\hat B_-}  \nonumber \\
= &  - i \partial_\theta J_b \biggl[ \cosh(4 J_+) \cosh(4 J_-) \biggl(  \hat N_b + \hat B_+\,\hat{ \mathcal{F}}_- +\hat{\mathcal{F}}_-^2\, -\hat B_-\,\hat{\mathcal{F}}_++\hat{\mathcal{F}}_+^2\,  \biggr) \nonumber \\
& \quad\quad\quad\quad+ \frac{1}{2} \cosh(4 J_+) \sinh(4J_-) \biggl( \hat B_+^{(2)} +2\,\hat B_+ \,\hat{\mathcal{F}}_-+2\,\hat{\mathcal{F}}_-^2  \, +2\,\hat B_-\,\hat{\mathcal{F}}_+-2\,\hat{\mathcal{F}}_+^2\,\biggr) \nonumber \\
&\quad\quad\quad\quad +  \cosh(4 J_+) \sinh^2(2 J_-) + \sinh^2(2 J_p)  \nonumber \\
&\quad\quad\quad\quad- \frac{1}{2} \sinh(4 J_+) \biggl( \hat B_-^{(2)} + 2\,\hat B_-\, \hat{\mathcal{F}}_- -2\,\hat B_+\hat{\mathcal{F}}_ +  - 4\,\hat{\mathcal{F}}_-\,\hat{\mathcal{F}}_+ \biggr)  \biggr] \, \nonumber ,
\\
\hat C_2=
 &- i  \partial_\theta J_+ \hat U_{\hat B_-}^\dagger \hat U_{\hat B_+}^\dagger e^{iJ_p\hat B_-^{(2)}}\hat B_+^{(2)}e^{-iJ_-\hat B_-^{(2)}}\hat U_{\hat B_+} \hat U_{\hat B_-} \nonumber   \\
= & - i \partial_\theta J_+  \biggl[ \cosh(4J_- ) \biggl(   \hat B_+^{(2)} +2\,\hat B_+ \,\hat{\mathcal{F}}_-+2\,\hat{\mathcal{F}}_-^2 \,  +2\,\hat B_-\,\hat{\mathcal{F}}_+ -2\,\hat{\mathcal{F}}_+^2\, \biggr) \nonumber \\
&\quad\quad\quad\quad + 2 \sinh(4 J_-) \biggl(  \hat N_b + \hat B_+\, \hat{\mathcal{F}}_-  + \hat{\mathcal{F}}_-^2\, - \hat B_-\,\hat{\mathcal{F}}_++\hat{\mathcal{F}}_+^2\,  \biggr) + \sinh(4 J_-) \biggr] \,, \nonumber \\
\hat C_3=
&- i\,  \partial_\theta J_- \hat U_{\hat B_-}^\dag \hat U_{\hat B_+}^\dag \, \hat B_-^{(2)} \hat U_{\hat B_+} \hat U_{\hat B_-} \nonumber \\
=& - i \, \partial_\theta J_-  \, \biggl[  \hat B_-^{(2)}  + 2\, \hat B_-\, \hat{\mathcal{F}}_- -2\, \hat B_+\hat{\mathcal{F}}_+  - 4\,\hat{\mathcal{F}}_-\,\hat{\mathcal{F}}_+  \biggr]\, .
\end{align}
In computing these expressions, we have used the congruence relations in Eqs.~\eqref{app:commutators:eq:commutators:0}--\eqref{app:commutators:eq:commutators:12} in Appendix~\ref{app:commutators}. 

For the remaining terms in $\hat{\mathcal{H}}$, we obtain
\begin{align}
 \hat U_{ \hat N_a}^\dagger \partial_\theta {\hat U}_{ \hat N_a}  &=  - i\left( \tau \partial_\theta \Omega_{\rm{c}} +  \partial_\theta  \hat{\mathcal{F}}_{N_a}\right)  \hat N_a \;,  \nonumber \\
 \hat U_{\hat B_-}^\dagger  \partial_\theta \hat{ U}_{\hat B_-} &=  - i \partial_\theta \hat{\mathcal{F}}_- \hat B_- \;, \nonumber \\
\hat U_{\hat B_-}^\dagger \hat U_{\hat B_+}^\dagger \partial_\theta{\hat U}_{\hat B_+} \hat U_{\hat B_-} 
&=  - i \partial_\theta \hat{\mathcal{F}}_{+}  \left( \hat B_+ + 2\,\hat{\mathcal{F}}_-\right) \;.
\end{align}
With these expressions, we can then write $\hat{\mathcal{H}}_\theta$ in terms of a clearer operator structure:
\begin{equation}\label{chap:metrology:eq:Hcal:structure}
	\mathcal{\hat H}_\theta \,= \, \mathcal{\hat H}_{\hat N_a} + \sum_{s\in\{+,-\}} \hat{\mathcal{H}}_s \hat{B}_s + E\hat{N}_b + F\hat{B}_+^{(2)} + G \hat{B}_-^{(2)}\,,
\end{equation}
which needs to be supplied by the following definitions:
\begin{align} \label{chap:metrology:eq:mathcal:H:compact}
	\mathcal{\hat H}_{\hat N_a} := &  -\tau \hat{N} _a \,  \partial_\theta\Omega_{\rm{c}}  - \hat{N}_a \,  \partial_\theta F_{\hat N_a} - \hat{N}_a^2 \, \partial_\theta F_{\hat N_a^2}  \nonumber \\
	& - 2 \mathcal{\hat{F}}_- \partial_\theta\mathcal{\hat{F}}_+ + \partial_\theta J_b/2 + E/2   \nonumber \\
	& + 2 \mathcal{\hat{F}}_+ \mathcal{\hat{F}}_- R_{\partial_\theta,0} + \sum_{s\in\{+,-\}} s \, e^{-s 4J_-} \mathcal{\hat{F}}_s^2 R_{\partial_\theta,s} \;,  \nonumber  \\
	\hat{\mathcal{H}}_\pm := & -\partial_\theta \mathcal{\hat{F}}_\pm \pm \mathcal{\hat{F}}_\pm R_{\partial_\theta,0}  - e^{ \pm 4J_-} \mathcal{\hat{F}}_\mp R_{\partial_\theta,\mp}\;,\nonumber   \\
	E := & \, -\left(e^{4J_-} R_{\partial_\theta,-} - e^{-4J_-} R_{\partial_\theta,+} \right)/2 \nonumber \;,\\	
	F := & \,  -\left( e^{4J_-} R_{\partial_\theta,-} + e^{-4J_-} R_{\partial_\theta,+}\right)/4 \nonumber\;, \\
	G := & - R_{\partial_\theta,0}/2  \, , 
\end{align}
where we defined the functions
\begin{align}
	R_{\partial_\theta,0} := & \,   2\partial_\theta J_- - \sinh(4J_+)\partial_\theta J_b \nonumber \,,\\
	R_{\partial_\theta,\pm} := &  \, 2\partial_\theta J_+ \mp  \cosh(4J_+)\partial_\theta J_b\,.
\end{align}
By then comparing the obtained expression for $\mathcal{\hat H}_\theta$ in Eq.~\eqref{chap:metrology:eq:mathcal:H:compact} with the expression in Eq.~\eqref{chap:metrology:Hcal:as:derivatives}, with all the terms we derived, we find the following coefficients for the QFI:
\begin{align}\label{chap:metrology:eq:QFI:coefficients}
A=& 
-\partial_\theta F_{\hat N_a^2}-2 F_{\hat N_a \, \hat B_-}\partial_\theta F_{\hat N_a\, \hat B_+}  + 2 F_{\hat N_a\, \hat B_-}F_{\hat N_a \, \hat B_+}R_{\partial_\theta, 0} \nonumber \\
&+ \sum_{s\in\{+,-\} } s \,e^{-s4J_-} F_{\hat N_a \, \hat B_s}^2 \, R_{\partial_\theta, s} 
\;,  \nonumber\\
B=& 
 -\tau\partial_\theta\Omega_{\rm{c}} - \partial_\theta F_{\hat N_a}-2 \, F_{\hat B_-}\partial_\theta F_{\hat N_a \, \hat B_+}-2 \, F_{\hat N_a\, \hat B_-}\partial_\theta F_{\hat B_+}
\nonumber\\
&+ 2 \left(F_{\hat B_+}F_{\hat N_a\, \hat B_-}+F_{\hat B_-}F_{\hat N_a \, \hat B_+}\right)R_{\partial_\theta, 0}  
+\sum_{s\in\{+,-\} } 2s e^{-s 4J_-}F_{\hat B_s}F_{\hat N_a \, \hat B_s} \, R_{\partial_\theta, s}
\; ,\nonumber\\
 C_\pm = & \, -\partial_\theta F_{\hat B_\pm } \pm  \,  F_{\hat B_\pm } R_{\partial_\theta, 0}
- e^{\pm 4J_-}  \, F_{\hat B_\mp} \, R_{\partial_\theta, \mp}
\;,\nonumber\\
C_{\hat N_a,\pm}=& \, - \partial_\theta  \, F_{\hat N_a \, \hat B_\pm} \pm  \,  F_{\hat N_a \, \hat B_\pm} \, R_{\partial_\theta, 0}
-e^{\pm 4J_-} \, F_{\hat N_a \, \hat B_\mp } \, R_{\partial_\theta, \mp}
\;,\nonumber\\
E = & \,  -\left(e^{4J_-} R_{\partial_\theta,-} - e^{-4J_-} R_{\partial_\theta,+} \right)/2 \;,\nonumber \\	
F = & \, -\left( e^{4J_-} R_{\partial_\theta,-} + e^{-4J_-} R_{\partial_\theta,+}\right)/4 \;,\nonumber \\
G = & \, - R_{\partial_\theta,0}/2   \;,\nonumber \\
K = & \, - 2   F_{\hat B_-} \,  \partial_\theta F_{\hat B_+} + 2 F_{\hat B_-} F_{\hat B_+} R_{\partial_\theta, 0}
 \nonumber\\
&+  \sum_{s\in\{+,-\}} s\, e^{-s4J_-} F_{\hat B_s}^2  \, R_{\partial_\theta, s} \, + \, \partial_\theta J_b /2  + E/2 \, . 
\end{align}
The coefficients $E$ and $K$ will cancel out in the expression for $\mathcal{I}_\theta$, but we include them here for completeness. 

The QFI for an optomechanical system evolving with the extended optomechanical Hamiltonian in Eq.~\eqref{chap:metrology:eq:Hamiltonian} can now be computed for a range of different initial states. The coefficient in Eq.~\eqref{chap:metrology:eq:QFI:coefficients} remains the same, since they are determined by the dynamics rather than the initial state. 

It is clear from the expressions above, that the expressions simplify dramatically when the parameter $\theta$ to estimate is not contained in the coefficients $J_\pm$ and $J_b$, such that $\partial_\theta J_b = \partial_\theta J_\pm = 0$. For that case, we have $E=F=G=0$. When we consider specific examples in Section~\ref{chap:metrology:three:examples}, this observation becomes useful. 

\subsection{General QFI for initially thermal states}
We proceed to compute the expectation values of $\hat{\mathcal{H}}_\theta$ and $\hat{\mathcal{H}}_\theta^2$ with respect to the initial state in Eq.~\eqref{chap:metrology:eq:initial:state:coherent:thermal}. The full derivation can be found in Section~\ref{app:QFI:coherent:thermal:state} in Appendix~\ref{app:QFI}. We keep the notation with the coefficients in Eq.~\eqref{chap:metrology:eq:QFI:coefficients} to keep the expression compact. 

The final result, which is one of the main results in this thesis, reads
\begin{align} \label{chap:metrology:eq:main:result:QFI}
\mathcal{I}_\theta =& 4  \, \biggl[ \left(4|\mu_\textrm{c}|^6+6|\mu_\textrm{c}|^4+|\mu_\textrm{c}|^2\right)A^2
+  \left(4|\mu_\textrm{c}|^4+2|\mu_\textrm{c}|^2\right)AB  +|\mu_\textrm{c}|^2 B^2  \nonumber \\
&\quad+\cosh(2 \, r_T) \sum_{s\in\{+,-\}} C^2_{\hat N_a,s}  |\mu_c|^2 +\frac{1}{\cosh(2 \, r_T)} \sum_{s\in\{+,-\}} (C_s + C_{\hat N_a,s} |\mu_c|^2)^2 \nonumber \\
&\quad  +  4\frac{\cosh ^2(2 r_T)}{\cosh ^2(2 r_T)+1} \left( F^2 + G^2 \right)   \biggr] \, .
\end{align}
This expression simplifies depending on in which function the parameter $\theta$ contained and where it enters into the Hamiltonian in Eq.~\eqref{chap:metrology:eq:Hamiltonian}. As such, there are a number of different estimation schemes that we can consider given this result. In this Chapter, we explore three concrete examples, which have been chosen to demonstrate the fact that these methods allow us to model time-dependent terms in the Hamiltonian. Then, in Chapter~\ref{chap:gravimetry}, we consider estimation of a constant gravitational acceleration. 

Before we proceed, let us briefly comment on the form of the QFI in Eq.~\eqref{chap:metrology:eq:main:result:QFI}. The full explicit expression in Eq.~\eqref{chap:metrology:eq:main:result:QFI} is not particularly revealing, since the coefficients can take different forms depending on the estimation parameter of interest. We note in general that the system scales strongly with the parameter $|\mu_{\rm{c}}|$, in particular with  a term $16|\mu_{\rm{c}}|^6 A^2$. This term arises from the fact that $\hat{\mathcal{H}}_\theta$ contains the term $\hat N_a^2$, which when squared yields the expectation value in Eq.~\eqref{app:commutators:Na4:coherent:state} containing terms of order $|\mu_{\rm{c}}|^8$ and $|\mu_{\rm{c}}|^6$. The eight-order terms cancel, while the leading behaviour of $|\mu_{\rm{c}}|^6$ is retained.  

We also note that the term multiplying the first sum in Eq.~\eqref{chap:metrology:eq:main:result:QFI} scales with the temperature parameter $r_T$, which implies that for certain cases, the QFI will increase with the temperature $r_T$ of the initial state.  The dependence on temperature may be understood by the fact that larger temperatures imply population of the higher phonon states, which will store more information about the light--matter interaction in the overall state.

\section{Estimation of three different cases} \label{chap:metrology:three:examples}
In this Section, we demonstrate the applicability of the main result in Eq.~\eqref{chap:metrology:eq:main:result:QFI} by considering  three concrete scenarios: (i) estimation of the strength of a time-dependent optomechanical coupling, (ii) estimation of the strength of a time-dependent linear displacement, and (iii) estimation of a time-dependent mechanical squeezing term.

\subsection{Example (i): Estimating the strength of an oscillating optomechanical coupling $\tilde{\mathcal{G}}(\tau)$} \label{chap:metrology:sec:example:1}

Characterising the nonlinear coupling in optomechanical systems is a key task when calibrating an experimental system. The case of a constant coupling $\tilde{\mathcal{G}}(\tau) \equiv \tilde{g}_0$ has already been thoroughly considered~\cite{bernad2018optimal}. As an example application of our methods we therefore compute the QFI for estimating the strength $\tilde{g}_0$ of an \textit{oscillating} optomechanical coupling $\tilde{\mathcal{G}}(\tau)$. We assume that it has the functional form 
\begin{equation} \label{chap:metrology:eq:oscillating:light:matter:coupling}
\tilde{\mathcal{G}} (\tau) := \tilde{g}_0 \left( 1 + \epsilon \sin{(\Omega_g \tau)} \right)\;,
\end{equation}
where $\tilde{g}_0 = g_0/\omega_{\rm{m}}$ is the strength of the coupling, $\epsilon$ is the oscillation amplitude and  $\Omega_g = \omega_g/\omega_{\rm{m}}$. We additionally assume that $\tilde{\mathcal{D}}_1=\tilde{\mathcal{D}}_2=0$. 

A nonlinear coupling of this form appears for levitated systems in hybrid Paul traps, where the time-dependent modulation is caused by micromotion of the sphere~\cite{millen2015cavity,fonseca2016nonlinear,aranas2016split}. Using the form of the coupling in Eq.~\eqref{chap:metrology:eq:oscillating:light:matter:coupling} of the coupling we can compute the $F$-coefficients in Eq.~\eqref{chap:decoupling:eq:sub:algebra:decoupling:solution} explicitly. We note that for the choice $\tilde{\mathcal{D}}_1(\tau) =0$, we have $F_{\hat N_a} = F_{\hat B_\pm} = 0$. The remaining nonzero coefficients $F_{\hat N_a^2}$ and $F_{\hat B_\pm \, \hat N_a}$ take on rather lengthy expressions and can therefore be found in Eq.~\eqref{app:coefficients:eq:time:dependent:g0} in Appendix~\ref{app:coefficients}. Furthermore, when $\tilde{\mathcal{D}}_2 = 0$, it follows that $J_b =\tau$ and $J_\pm = 0$. 

The remaining \textit{non-zero} coefficients in Eq.~\eqref{chap:metrology:eq:main:result:QFI} are given by
\begin{align}\label{chap:metrology:eq:g0:nonzero:coefficients}
A &= -\partial_\theta F_{\hat N_a^2}-2 F_{\hat N_a \, \hat B_-}\partial_\theta F_{\hat N_a\, \hat B_+}\;,\\\nonumber
 C_{\hat N_a,\pm} &= -\,\partial_\theta F_{\hat N_a \, \hat B_\pm} \;,
\end{align}
and $C_+ = C_- = F = G = 0$. The QFI for estimating the coupling strength $\tilde{g}_0$ thus becomes
\begin{align} \label{chap:metrology:eq:QFI:g0}
\mathcal{I}_{\tilde{g}_0}= & \, 4 \, |\mu_c|^2\,  \biggl(  
\left( 4 \, |\mu_{\rm{c}}|^4 + 6  \, |\mu_{\rm{c}}|^2 +1 \right) A  +  \cosh(2 r_T)  \, \left(1+ \frac{|\mu_{\rm{c}}|^2}{\cosh^2{(2 r_T)}} \right)  \sum_{s\in\{+,-\}} C_{\hat N_a,s}^2 \biggr) \, . 
\end{align}
We observe that the QFI increases for increasing temperatures $r_T$, which is due to the higher occupied phonon states. The remaining coefficients $A$ and $C_{\hat N_a, \pm}$ in Eq.~\eqref{chap:metrology:eq:QFI:g0} need to be complemented with the appropriate expressions for $F_{\hat N_a^2}$ and $F_{\hat B_\pm \, \hat N_a}$ in Eq.~\eqref{app:coefficients:eq:time:dependent:g0}. 
The explicit expression for the QFI in Eq.~\eqref{chap:metrology:eq:QFI:g0} is long and cumbersome, so we display it in Eq.~\eqref{app:QFI:eq:g0:general:Omega} in Appendix~\ref{app:QFI}. 

\begin{figure*}[t!]
\centering
\subfloat[ \label{chap:metrology:fig:QFI:g0:Omegag:sweep}]{%
  \includegraphics[width=0.48\linewidth, trim = 0mm 1mm 0mm 0mm]{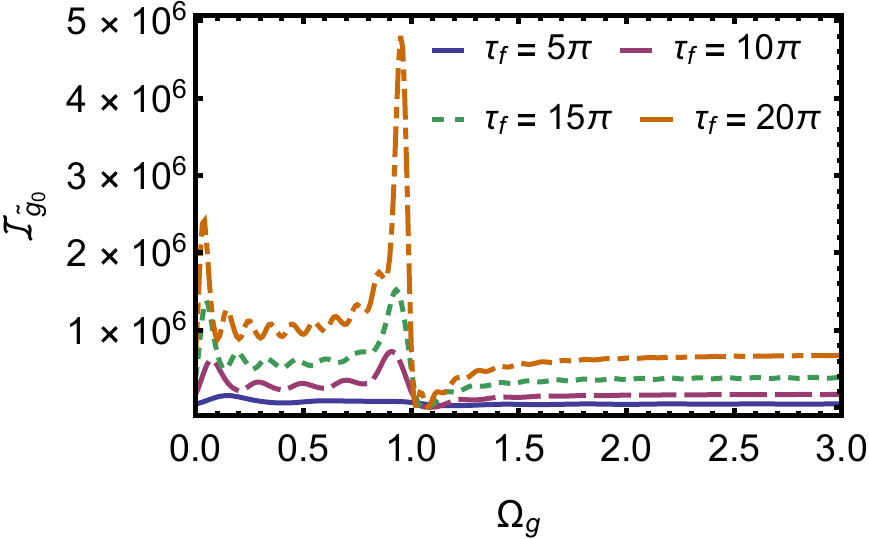}%
}\hfill
\subfloat[ \label{chap:metrology:fig:QFI:g0:various:Omegag}]{%
  \includegraphics[width=0.5\linewidth, trim = 0mm 0mm 0mm 0mm]{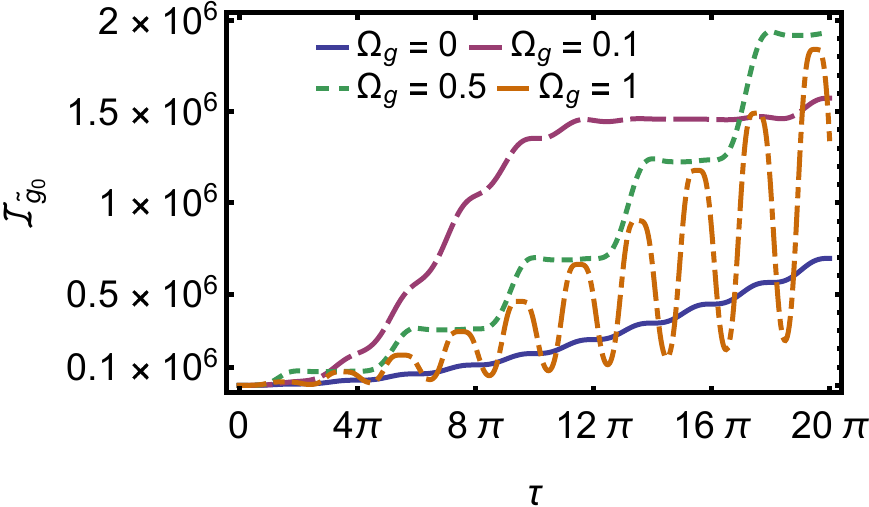}%
}\hfill
\subfloat[ \label{chap:metrology:fig:QFI:g0:various:epsilon}]{%
  \includegraphics[width=0.5\linewidth, trim = 0mm 0mm 0mm 0mm]{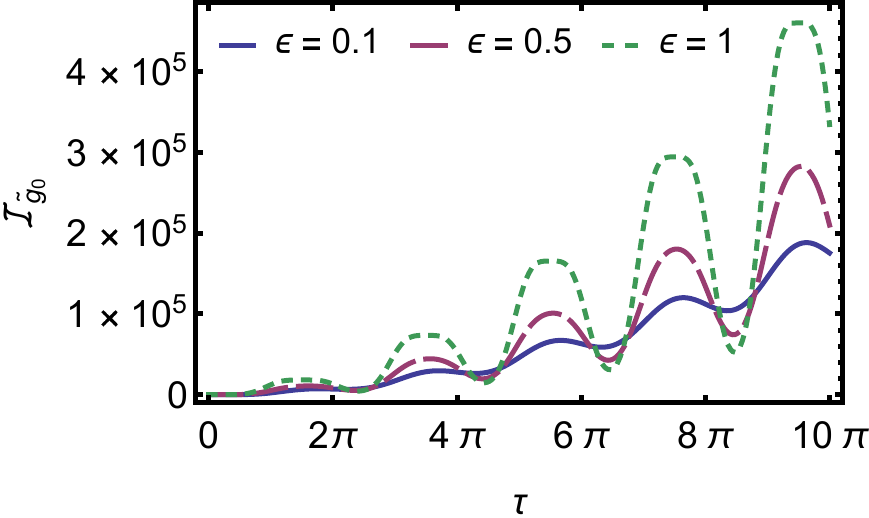}%
}\hfill
\subfloat[ \label{chap:metrology:fig:QFI:g0:various:rT}]{%
  \includegraphics[width=0.5\linewidth, trim = 0mm 0mm 0mm 0mm]{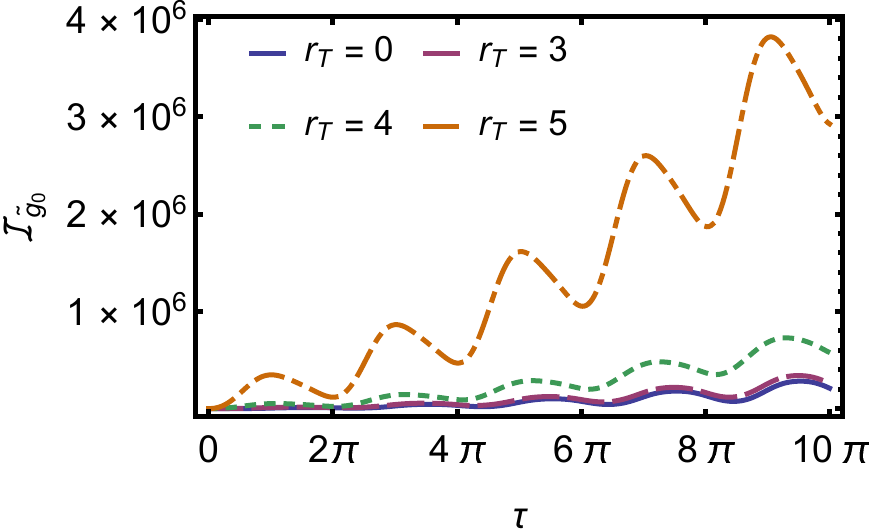}%
}\hfill
\caption[QFI for a measurement of the nonlinear coupling amplitude]{QFI for measurements of $\tilde{g}_0$. Parameters for all plots are $\tilde{g}_0 = 1$ and $\mu_{\rm{c}} = 1$. \textbf{(a)} shows $\mathcal{I}_{\tilde{g}_0}$ for a sweep of the oscillation frequency $\Omega_g$ at different final times $\tau_f$. The peak occurs for different $\Omega_g$ for earlier and later times. \textbf{(b)} shows ${\mathcal{I}}_{\tilde{g}_0} $ as a function of time $\tau$ for different $\Omega_g$. The resonant behaviour clearly differs from the others and slowly diverges. \textbf{(c)} shows $\mathcal{I}_{\tilde{g}_0}$ as a function of time $\tau$ for different oscillation amplitudes $\epsilon$. A larger amplitude corresponds to a stronger signal, which increases the QFI.  \textbf{(d)} shows ${\mathcal{I}}_{\tilde{g}_0} $ as a function of time $\tau$ for different thermal state parameters $r_T$. A higher temperature means the QFI increases. }
\label{chap:metrology:fig:QFI:g0}
\end{figure*}

To explore some properties of $\mathcal{I}_{\tilde{g}_0}$, we plot four different scenarios of the full QFI  as a function of $\Omega_{g}$ in Figure~\ref{chap:metrology:fig:QFI:g0}. In Figure~\ref{chap:metrology:fig:QFI:g0:Omegag:sweep}, we plot a frequency sweep of $\mathcal{I}_{\tilde{g}_0}$ as a function of $\Omega_g$, where the different lines correspond to different final times $\tau_f$. The other values are set to $\tilde{g}_0 = |\mu_{\rm{c}}| = 1$, and $r_T = 0$. We note that the QFI peaks  at the resonance frequency $\Omega_g = 1$, but only at later times $\tau \gg 1$.  At earlier time,  the peak occurs for values of $\Omega_g \leq 1$, which suggests the system needs to settle into the resonant behaviour before the higher values of the QFI can be exploited. 
In Figure~\ref{chap:metrology:fig:QFI:g0:various:Omegag}, we plot $\mathcal{I}_{\tilde{g}_0}$ as a function of time $\tau$ for different frequencies $\Omega_g$. Here we can see the effects of the resonant frequency, in that the QFI strongly oscillates, for at later times, these oscillations appear as the dominant term. 
In Figure~\ref{chap:metrology:fig:QFI:g0:various:epsilon}, we plot $\mathcal{I}_{\tilde{g}_0}$ as a function of time $\tau$ for different oscillation amplitudes $\epsilon$. As expected, the QFI increases with $\epsilon$, since this strengthens the signal. Finally, in Figure~\ref{chap:metrology:fig:QFI:g0:various:rT} we plot $\mathcal{I}_{\tilde{g}_0} $ as a function of time $\tau$ for different temperature parameters $r_T$. The QFI increases with temperature, which we already noted in  Eq.~\eqref{chap:metrology:eq:main:result:QFI}. This might seem counter-intuitive, but it is due to the increased occupation of the phonon states, which as a result hold more information about the coupling strength $\tilde{g}_0$. 

When the coupling modulation occurs at mechanical resonance with $\Omega_g \rightarrow 1$, the QFI takes on a more compact form. We present the full expression in Eq.~\eqref{app:QFI:eq:g0:general:Omega:resonance} in Appendix~\ref{app:QFI}.  We can simplify it even further by noting that, at large-time scales $\tau \gg 1$, the first term of Eq.~\eqref{app:QFI:eq:g0:general:Omega:resonance} dominates. Furthermore, when the mechanics in the vacuum state with $r_T = 0$,  and when the optomechanical coupling is much greater than the oscillation amplitude $\tilde{g}_0 \gg \epsilon$, the expression simplifies significantly to
\begin{align}\label{chap:metrology:eq:QFI:g0:asymptotic}
\mathcal{I}_{\tilde{g}_0}^{(\rm{res})} \sim 16  \, \tilde{g}_0^2 \,  \tau^2  \, | \mu_c|^2 \left(4 |\mu_c|^4+6  \, | \mu_c| ^2+1\right) \, . 
\end{align}
As expected, when $|\mu_{\rm{c}}|^2 $ is zero -- no initial cavity mode excitations -- or $\tilde{g}_0$ is zero -- no coupling -- the QFI vanishes. The same can be seen from the full expression Eq.~\eqref{app:QFI:eq:g0:general:Omega}. 

\subsection{Example (ii): Estimating a parameter in the linear displacement $\tilde{\mathcal{D}}_1(\tau)$} \label{chap:metrology:sec:example:2}

\begin{figure*}[t!]
\centering
\subfloat[ \label{chap:metrology:fig:QFI:d1:Omegad1:sweep}]{%
  \includegraphics[width=0.5\linewidth, trim = 0mm 0mm 0mm 0mm]{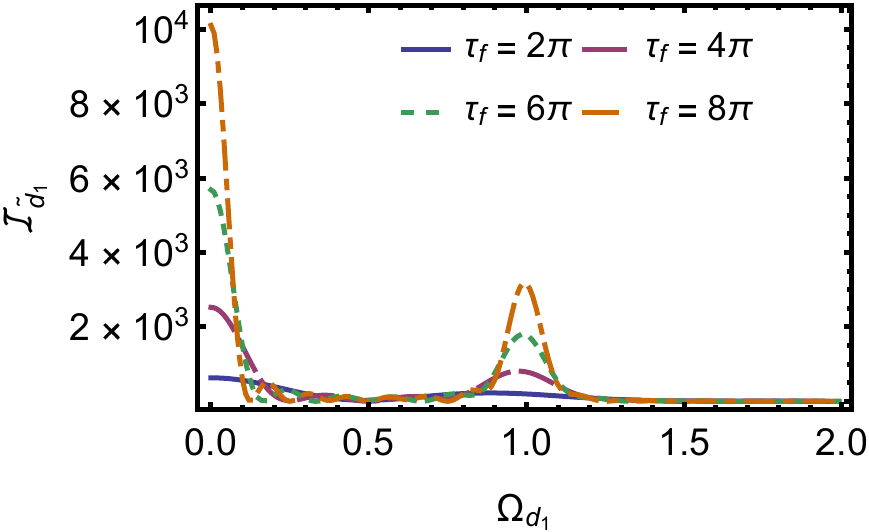}%
}\hfill
\subfloat[ \label{chap:metrology:fig:QFI:d1:various:Omegad1}]{%
  \includegraphics[width=0.48\linewidth, trim = 0mm 0mm 0mm 0mm]{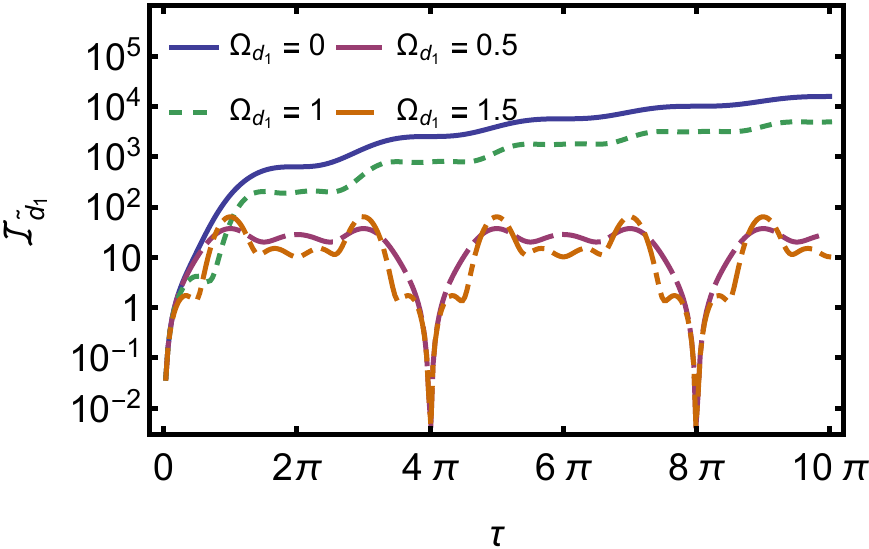}%
}\hfill
\subfloat[ \label{chap:metrology:fig:QFI:d1:various:rT}]{%
  \includegraphics[width=0.5\linewidth, trim = 0mm 0mm 0mm 0mm]{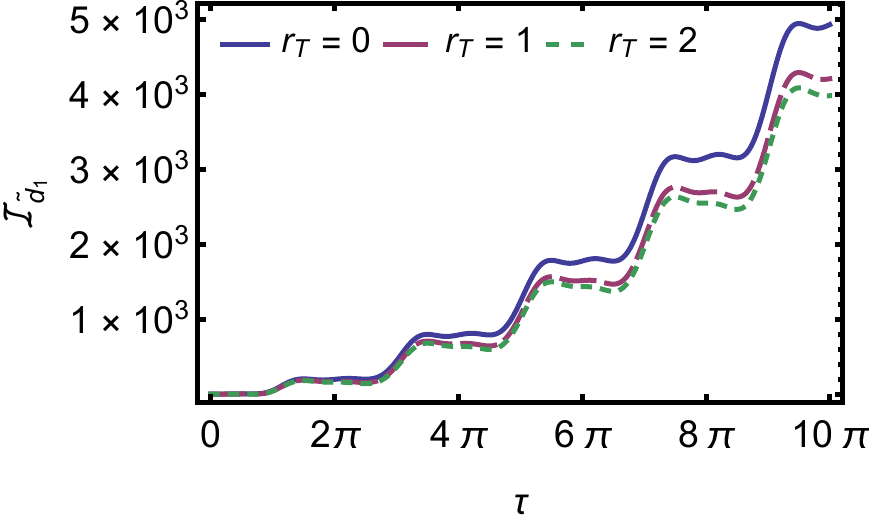}%
}\hfill
\caption[QFI for a measurement of a mechanical displacement amplitude]{QFI for measurements of $\tilde{d}_1$. The parameters are $\tilde{g}_0 = 1$, $\mu_{\rm{c}} = 1$. \textbf{(a)} shows $\mathcal{I}_{\tilde{d}_1}$ for a sweep of the oscillation frequency $\Omega_{d_1}$. Resonance occurs at $\Omega_{d_1} = 1$, but the constant case $\Omega_{d_1} = 0$ performs best overall. \textbf{(b)} shows a linear-log-plot of $\mathcal{I}_{\tilde{d}_1} $ as a function of time $\tau$. The QFI oscillates for arbitrary $\Omega_{d_1}$ but increases continuously at resonance. \textbf{(c)} shows $\mathcal{I}_{\tilde{d}_1}$ as a function of time $\tau$ for increasing thermal state parameters $r_T$. Values of $r_T$ larger than displayed here do not severely affect the QFI for $r_T > 2$. }
\label{chap:metrology:fig:QFI:d1}
\end{figure*}

The case of estimating a constant displacement $\tilde{\mathcal{D}}_1$  corresponds to, for example, estimating a constant gravitational acceleration, which we explore in Chapter~\ref{chap:gravimetry}. Here, we therefore focus on a time-dependent driving $\tilde{\mathcal{D}}_1(\tau)$, which is generally experimentally easier to detect compared with a static signal. 

We consider a periodic modulation of the mechanical driving term $\tilde{\mathcal{D}}_1(\tau)$ of the form 
\begin{equation} \label{chap:metrology:eq:d1:time:dependent}
\tilde{\mathcal{D}}_1 (\tau) = \tilde{d}_1 \, \cos(\Omega_{d_1} \tau) \;,
\end{equation}
where $\tilde{d}_1$ is the dimensionless driving strength and $\Omega_{d_1} = \omega_{d_1}/\omega_{\rm{m}}$ is the oscillation frequency of the driving. A coupling of this form can, for example,  be reproduced in levitating setups by applying an AC field to the system~\cite{hempston2017force}. 

We are interested in estimating the driving strength $\tilde{d}_1$ of the time-dependent coupling.  As opposed to in the last section,  here we assume that the light--matter coupling is constant with $\tilde{\mathcal{G}}(\tau) \equiv \tilde{g}_0$, and we also assume that $\tilde{\mathcal{D}}_2=0$. This implies that   $\partial_\theta F_{\hat N_a^2} = \partial_\theta F_{\hat N_a \, \hat B_\pm}=0$.  Furthermore, since $\tilde{\mathcal{D}}_2=0$, it follows that $ J_b = \tau$,  $  J_\pm = 0$,  and $\xi(\tau) = e^{- i \, \tau}$. As a result, the following coefficients are zero: $A =C_{\hat N_a,+} = C_{\hat N_a,-}  = F = G = 0$ and the only \textit{non-zero} coefficients that appear in the main expression Eq.~\eqref{chap:metrology:eq:main:result:QFI} of the QFI are 
\begin{align}\label{chap:metrology:eq:d1:nonzero:coefficients}
 B&=  - \partial_\theta F_{\hat N_a}-2
F_{\hat N_a \, \hat B_-}\partial_\theta F_{\hat B_+}
 \;,\\\nonumber 
C_\pm &= -\partial_\theta F_{\hat B_\pm}\;.
\end{align}
This implies that the QFI for the estimation of $\tilde d_1$ reduces to the expression
\begin{align}\label{chap:metrology:QFI:estimate:d1}
\mathcal{I}_{\tilde{d}_1} = 4\,B^2\,|\mu_\textrm{c}|^2 +  \frac{4}{\cosh(2r_T)} \left(  C_+^2 + C_-^2\right)\;.
\end{align}
We note in Eq.~\eqref{chap:metrology:QFI:estimate:d1} that the term $4 \, B^2 \, |\mu_{\rm{c}}|^2$ specifically encodes the nonlinearity; that is, when  $\tilde{g}_0 = 0$ it follows that $B = 0$. 

We solve the integrals in Eq.~\eqref{chap:decoupling:eq:sub:algebra:decoupling:solution}  for the general time-dependent function in Eq.~\eqref{chap:metrology:eq:d1:time:dependent}. They are listed in Eq.~\eqref{app:coefficients:F:coefficients:time:dependent:d1} in Appendix~\ref{app:coefficients}. With these expressions, the general expression for $\mathcal{I}_{\tilde{d}_1}$ for all frequencies $\Omega_{d_1}$  is again quite long, so we show it in Eq.~\eqref{app:QFI:eq:formula:oscillating:d1} in Appendix~\eqref{app:QFI}. 

To understand the properties of the QFI, we plot $\mathcal{I}_{\tilde{d}_1}$ in Figure~\ref{chap:metrology:fig:QFI:d1}.  
In Figure~\ref{chap:metrology:fig:QFI:d1:Omegad1:sweep}, we plot $\mathcal{I}_{\tilde{d}_1}$ as a function of the oscillation frequency $\Omega_{d_1}$.  The QFI is clearly maximised at resonance, where $\Omega_{d_1} = 1$, and when the displacement $\tilde{\mathcal{D}}_1(\tau) \equiv \tilde{d}_1$ is constant. However, the QFI is also maximised for a constant displacement with $\Omega_{\tilde{d}_1} = 0$. In Figure~\ref{chap:metrology:fig:QFI:d1:various:Omegad1}, we plot $\mathcal{I}_{\tilde{d}_1}$ as a function of time $\tau$ in a log plot for different frequencies $\Omega_{\tilde{d}_1}$. The constant case $\Omega_{\tilde{d}_1} = 0$ and the resonant case $\Omega_{\tilde{d}_1} = 1$ show a markedly different behaviour compared with any other frequencies listed. In Figure~\ref{chap:metrology:fig:QFI:d1:various:rT}, we plot $\mathcal{I}_{\tilde{d}_1}$ as a function of time $\tau$ for different temperature parameters $r_T$. This time, the QFI decreases with increasing temperature, but since the first term in Eq.~\eqref{chap:metrology:QFI:estimate:d1} is generally dominant, the QFI is only marginally impacted as $r_T \rightarrow \infty$. This implies that the system does not have to be cooled to the ground state to achieve a good sensitivity. 

We can consider some special cases analytically. 
For a constant linear displacement, the $F$-coefficients in Eq.~\eqref{chap:decoupling:eq:sub:algebra:decoupling:solution} can be evaluated analytically. They are listed in Eq.~\eqref{app:coefficients:F:coefficients:constant:d1} in Appendix~\eqref{app:coefficients},  and the QFI reduces to the simpler expression:
\begin{align}\label{chap:metrology:eq:QFI:d1:general}
\mathcal{I}_{\tilde{d}_1}^{\rm{(const)}} =& 16 \,  \biggl(\tilde{g}_0^2 \,  \left| \mu_c \right| ^2 (\tau-\sin (\tau))^2 +\frac{\sin ^2\left(\tau/2\right)}{\cosh(2 \, r_T)} \biggr) \, .
\end{align}
The first contribution in this expression originates from the cavity field and its interaction with the mechanics, while the second contribution originates from the mechanics only, which includes the dependence on the temperature $r_T$. We note that the QFI scales with $\tau^2$, which implies that significant sensitivities could be achieved if the system can be kept resonant for longer.

The second special case we consider is mechanical resonance with  $\Omega_{d_1} = 1$. Here, the expression in Eq.~\eqref{chap:metrology:QFI:estimate:d1} simplifies to 
\begin{align} \label{chap:metrology:eq:QFI:d1:resonance}
\mathcal{I}_{\tilde{d}_1}^{(\rm{res})} =&  4 \, \tilde{g}_0^2 \, |\mu_{\rm{c}}|^2 \left[\tau+\sin(\tau)\left(\cos(\tau)-2\right)\right]^2    + \frac{\tau^2+2\,\tau\,\sin(\tau)\cos(\tau)+\sin^2(\tau)}{\cosh{(2 r_T)}}. 
\end{align}
Notice that $\bigl[\tau+\sin(\tau)\left(\cos(\tau)-2\right)\bigr]^2=\bigl[1+\text{sinc}(\tau)\left(\cos(\tau)-2\right)\bigr]^2\,\tau^2$ and $\tau^2+2\,\tau\,\sin(\tau)\cos(\tau)+\sin^2(\tau)=(1+2\,\text{sinc}(\tau)\cos(\tau)+\text{sinc}^2(\tau))\,\tau^2$, where $\text{sinc}(x):=\frac{\sin x}{x}$ and $\text{sinc}(x)\rightarrow0$ for $x\rightarrow0$. This highlights the appearance of terms proportional to $\tau^2$ in Eq.~\eqref{chap:metrology:eq:QFI:d1:resonance}. Therefore, these terms do not oscillate for $\tau \gg 1$ but grow polynomially, that is, the resonant QFI scales as $\mathcal{I}_{\tilde{d}_1}^{(\rm{res})} \sim 4 \, \tilde{g}_0^2 \, |\mu_{\rm{c}}|^2  \, \tau^2 $,  while the QFI for a constant coupling scales as $\mathcal{I}_{\tilde{d}_1}^{(\rm{const})} \sim 16  \, \tilde{g}_0^2 \,  |\mu_{\rm{c}}|^2 \,  \tau^2$. All together, this implies that $\mathcal{I}^{(\rm{const})}_{d_1} \approx  4  \, \mathcal{I}_{d_1}^{(\rm{res})}$ for $\tau \gg1$.

At higher temperatures, $r_T$ decreases the QFI for both a constant coupling, as we saw in Eq.~\eqref{chap:metrology:eq:QFI:d1:general} and a resonant coupling, as we saw in Eq.~\eqref{chap:metrology:eq:QFI:d1:resonance} displacements. However, the effect differs between the two cases in the $\tau \gg 1$ limit. For $\mathcal{I}_{\tilde{d}_1}^{(\rm{const})}$, the temperature-dependent term oscillates with $\tau$, and therefore is completely negligible as $\tau \gg 1$. For $\mathcal{I}_{\tilde{d}_1}^{(\rm{res})}$, on the other hand, the temperature-dependent term also scales with $\tau$. Hence there is resonant buildup of the information contained in the temperature-dependent term, which leads to a advantage for the resonant case when $r_T$ is small. The difference is however small if $\tilde{g}_0 \gg1$ and $|\mu_{\rm{c}}|^2 \gg 1$, for which the first terms in both Eq.~\eqref{chap:metrology:eq:QFI:d1:general} and Eq.~\eqref{chap:metrology:eq:QFI:d1:resonance} dominate.

\subsection{Example (iii): Estimating a parameter in the mechanical squeezing $\tilde{\mathcal{D}}_2(\tau)$} \label{chap:metrology:sec:example:3}

\begin{figure*}[t!]
\subfloat[ \label{chap:metrology:fig:QFI:d2:Ires:and:Iconst}]{%
  \includegraphics[width=0.45\linewidth, trim = 0mm 0mm 0mm 0mm]{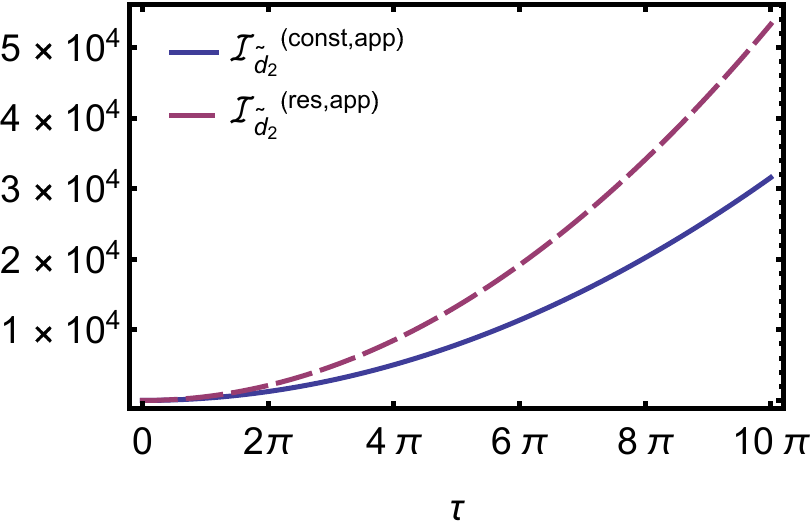}%
} \hspace{1cm}
\subfloat[ \label{chap:metrology:fig:QFI:d2:temperature}]{%
  \includegraphics[width=0.5\linewidth, trim = 0mm 0mm 0mm 0mm]{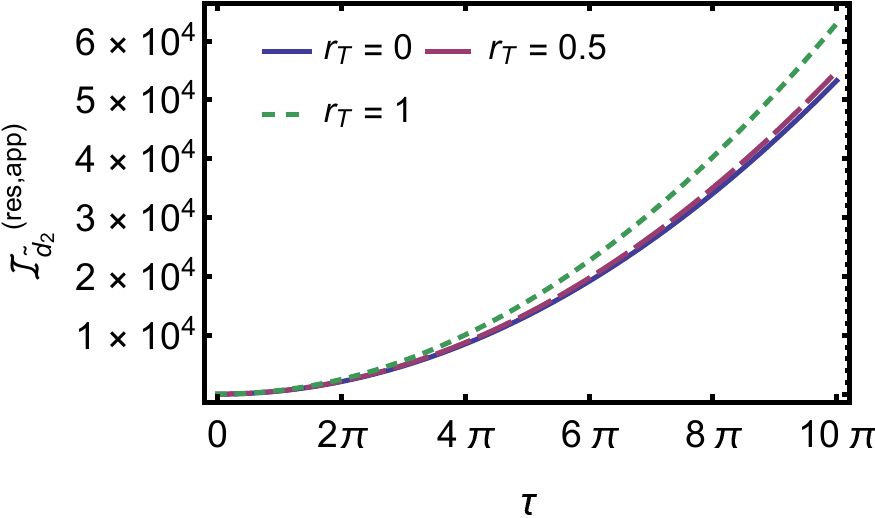}%
}\hfill
\caption[QFI for estimation of a single-mode mechanical squeezing amplitude]{The quantum Fisher information for measurements of $\tilde{\mathcal{D}}_2$ . The parameter are $ \tilde{d}_2 = 0.1$, such that the approximations used to derive the dynamics are valid,  and $\tilde{g}_0 =\mu_{\rm{c}} = 1$. \textbf{(a)} shows a comparison between the constant QFI $\mathcal{I}_{\tilde{d}_2}^{(\rm{const, app})}$ and the resonant QFI $\mathcal{I}_{\tilde{d}_2}^{(\rm{res, app})}$. The scaling for the constant QFI is slightly better compared with when the squeezing is constant.  \textbf{(b)} shows $\mathcal{I}_{\tilde{d}_2}^{(\rm{res, app})}$ as a function of time $\tau$ for different values of $r_T$. The QFI increases with higher temperatures.  }\label{chap:metrology:fig:QFI:d2}
\end{figure*}
\addtocontents{lof}{\protect\newpage}

 In this section, we consider a mechanical squeezing term $\tilde{\mathcal{D}}_2(\tau)$ of the form
\begin{equation} \label{chap:metrology:eq:modulated:squeezing}
\tilde{\mathcal{D}}_2(\tau) =  \tilde{d}_2  \,  \cos( \Omega_{d_2} \, \tau) , 
\end{equation}
where $\tilde{d}_2$ is the oscillation amplitude and $\Omega_{d_2}$ is the frequency of the modulation.  A modulation of this form can arise from an external time-dependent shift of the mechanical frequency $\omega_{\rm{m}}$, achievable in the laboratory~\cite{farace2012enhancing}, as we explicitly demonstrate in Section~\ref{app:mathieu:time:varying:trapping:frequency} in Appendix~\ref{app:mathieu}. Our goal is to estimate the squeezing strength $\tilde{d}_2$ for both constant and modulated squeezing strengths.

A non-zero squeezing term affects the full dynamics of the system by changing the function $\xi(\tau)$, which is defined in Eq.~\eqref{chap:decoupling:eq:definition:of:xi}. This function in turn enters into the $F$-coefficients in Eq.~\eqref{app:coefficients:F:coefficients:constant:d1}. The squeezing parameter is   also contained  in the $J$-coefficients, which we can compute by using the relation in Eq.~\eqref{chap:decoupling:eq:squeezing:relation}. 
 When the squeezing term is constant, that is, $\Omega_{d_2} = 0$, the equations are analytically solvable, as we showed in Section~\ref{chap:non:Gaussianity:squeezing:sec:constant:squeezing} in Chapter~\ref{chap:non:Gaussianity:squeezing}. 

For a time-dependent coupling of the form in Eq.~\eqref{chap:metrology:eq:modulated:squeezing}, the mechanical subsystem equations in Eq.~\eqref{chap:decoupling:differential:equation:written:down} take the form of the Mathieu equation. The Mathieu equation is notoriously difficult to solve numerically, and only has analytic solutions for specific cases. 
However, we show in Section~\ref{app:mathieu:perturbative:solutions} in Appendix~\ref{app:mathieu} that perturbative solutions of the form in Eq.~\eqref{app:mathieu:eq:RWA:solutions} can be obtained at parametric resonance $\Omega_{\tilde{d}_2} = 2$ when $\tilde{d}_2 \ll 1$, i.e., the squeezing strength is small. These solutions correspond to the rotating-wave approximation for times $\tau \gg1$, a fact we prove in Section~\ref{app:mathieu:rotating:wave:approximation} in the same Appendix. 

We first note that  $B = C_\pm = 0$  whenever $\tilde{\mathcal{D}}_1(\tau) = 0$, which means that the general QFI expression in Eq.~\eqref{chap:metrology:eq:main:result:QFI} reduces to
\begin{align}\label{chap:gravimetry:eq:QFI:d2:general}
\mathcal{I}_{\tilde{d}_2} =& 4\, \biggl[  \left( 4\, |\mu_c|^6 + 6\, |\mu_c|^4 + |\mu_c|^2 \right) \, A^2 \nonumber\\
&\quad\quad+ |\mu_c|^2\cosh(2 \, r_T)\left( 1 + \frac{|\mu_c|^2}{\cosh^2(2 \, r_T)} \right) \sum_{s\in\{+,-\}} C^2_{\hat N_a,s}  \nonumber \\
&\quad\quad+  4\frac{\cosh ^2(2 r_T)}{\cosh ^2(2 r_T)+1} \left( F^2 + G^2 \right)\biggr] \, .
\end{align}
We now consider two special cases: Estimating a constant $\tilde{d}_2$ and estimating the strength of a resonant coupling with $\Omega_{d_2} = 2$. For both cases, we assume that $\tilde{d}_2 \ll 1$, and $\tilde{d}_2\tau \sim 1$, which allows us to approximate the form of the coefficients.  

For a constant coupling with $\tilde{\mathcal{D}}_2 ( \tau) \equiv \tilde{d}_2$. We obtain the $F$-coefficients shown in Eq.~\eqref{chap:coefficients:eq:constant:d2} and the  $J$-coefficients in Eq.~\eqref{app:coefficients:eq:constant:d2:Js:coeffs}.  Then, the QFI is given by  
\begin{align} \label{chap:metrology:eq:app:QFI:d2:constant:app}
\mathcal{I}_{\tilde{d}_2}^{(\rm{const,app})} = 16 \,  \tilde{g}_0^2 \,  \tau^2  \, |\mu_{\rm{c}}|^2 \frac{|\mu_{\rm{c}}|^2+\cosh^2 (2 \, r_T)}{\cosh (2  \, r_T)} ,
\end{align}
where the superscript `app' refers to the fact that our solutions to the dynamics are approximate.

With the approximate solutions in Eq.~\eqref{app:mathieu:eq:RWA:solutions}, we compute the $F$-coefficients shown in Eq.~\eqref{app:coefficients:constant:d2} in  Appendix~\ref{app:coefficients}, as well as the $J$-coefficients in Eq.~\eqref{app:coefficients:eq:constant:d2:Js:coeffs}. We use the resulting expressions to compute the QFI  for detection of the squeezing strength $\tilde{d}_2$. The QFI becomes:
\begin{align} \label{chap:metrology:eq:app:QFI:d2:resonance:app}
\mathcal{I}_{\tilde{d}_2}^{(\rm{res,app})} =&   \,4 \, \tau^2 \Bigg(  \tilde{g}_0^4 \,  (4\,|\mu_c|^6 + 6\,|\mu_c|^4 + |\mu_c|^2)   \nonumber \\
&\quad\quad\quad\quad\quad\quad\quad + \tilde{g}_0^2\, |\mu_c|^2 \, \frac{|\mu_c|^2 + \cosh^2(2r_T)}{\cosh(2r_T)} + \frac{\cosh^2(2r_T)}{\cosh^2(2r_T)+1} \Bigg)\, . 
\end{align}
We note that for the resonant case,  $\mathcal{I}_{\tilde{d}_2}^{(\rm{res,app})}$ scales quadratically with $\tau $ and displays a strong dependence on $\mu_{\rm{c}}$ through the term $|\mu_{\rm{c}}|^6$, while for the constant case, $\mathcal{I}^{(\rm{const,app})}_{\tilde{d}_2}$ only scales with $|\mu_{\rm{c}}|^4$. The QFI for the resonant case also scales with $\tilde{g}_0^4$, which indicates that the strength of the nonlinearity is particularly key to sensing of resonantly modulated squeezing.  Just like in Example (i) in Section~\ref{chap:metrology:sec:example:1}, we find that the very last term in Eq.~\eqref{chap:metrology:eq:app:QFI:d2:resonance:app} tends to 1 as $r_T \rightarrow \infty$,  the second-to-last term diverges exponentially as $r_T$ increases, which indicates that a higher temperature $r_T$ contributes positively to the  QFI.  

We plot $\mathcal{I}_{\tilde{d}_2}^{(\rm{const, app})}$ and $\mathcal{I}_{\tilde{d}_2}^{(\rm{res, app})}$ in Figure~\ref{chap:metrology:fig:QFI:d2}. In Figure~\ref{chap:metrology:fig:QFI:d2:Ires:and:Iconst}, we plot both expressions as a function of time $\tau$ for the parameters $\tilde{d}_2 = 0.1$, and $\tilde{g}_0 = \mu_{\rm{c}} = 0$. We keep the parameters small to make sure the approximate solutions are valid. We note that $\mathcal{I}_{\tilde{d}_2}^{(\rm{res, app})}$ increases faster than $\mathcal{I}_{\tilde{d}_2}^{(\rm{const, app})}$, which is to be expected given the expressions in~Eqs. \eqref{chap:metrology:eq:app:QFI:d2:constant:app} and~\eqref{chap:metrology:eq:app:QFI:d2:resonance:app}.  In Figure~\ref{chap:metrology:fig:QFI:d2:temperature}, we plot $\mathcal{I}_{\tilde{d}_2}^{(\rm{res, app})}$ for different temperature parameters $r_T$. As expected from the expression in Eq.~\eqref{chap:metrology:eq:app:QFI:d2:resonance:app}, the QFI increases with $r_T$. 

In the limit $|\mu_{\rm{c}}| \gg 1$, and at zero temperature $r_T = 0$, we find that $\mathcal{I}_{\tilde{d}_2}^{(\rm{const, app})} \sim 16 \, \tilde{g}_0^2 \, \tau^2 \, |\mu_{\rm{c}}|^4$ and $\mathcal{I}_{\tilde{d}_2}^{(\rm{res, app})} = 16 \, \tilde{g}_0^4  \, \tau^2  \,|\mu_{\rm{c}}|^6$, which implies that $\mathcal{I}_{\tilde{d}_2}^{(\rm{res, app})} \sim \tilde{g}_0^2 \, |\mu_{\rm{c}}|^2 \, \mathcal{I}_{\tilde{d}_2}^{(\rm{const, app})}$. Therefore, the resonant sensing scheme benefits especially from strong light--matter couplings.

\section{Applications}\label{chap:metrology:sec:applications}
We  derived a general expression for the QFI  for an optomechanical system operating in the nonlinear regime and discussed three specific examples of parameter estimation scenarios in order to demonstrate how our results can be applied. Our expression can be used to infer the fundamental sensitivity for estimation of any parameters that enter into the Hamiltonian in Eq.~\eqref{chap:metrology:eq:Hamiltonian}.

To further demonstrate the applicability of these methods, we consider some example input values for the following three cases: 
Estimating the coupling $\tilde{g}_0$ at resonance, given the exact expression in Eq.~\eqref{app:QFI:eq:g0:general:Omega:resonance} in Appendix~\ref{app:QFI}, estimating the linear displacement  $\tilde{d}_1$ given in Eq.~\eqref{chap:metrology:eq:QFI:d1:resonance}, and estimating the squeezing parameter $\tilde{d}_2$ given in Eq.~\eqref{chap:metrology:eq:app:QFI:d2:resonance:app}. When we compute the QFI for $\tilde{g}_0$, we set $\tilde{\mathcal{D}}_1 (\tau) = \tilde{\mathcal{D}}_2(\tau) = 0$, and when we compute the QFI for $\tilde{d}_1$ and $\tilde{d}_2$, we keep the optomechanical coupling constant $\tilde{\mathcal{G}}(\tau) \equiv \tilde{g}_0$. In addition, for the estimation of $\tilde{d}_1$ and $\tilde{d}_2$, we set the other coefficient to zero respectively, such that $\tilde{\mathcal{D}}_2(\tau) = 0$ when estimating $\tilde{d}_1$, and $\tilde{\mathcal{D}}_1 = 0$ for estimation of $\tilde{d}_2$. 

The parameters used for all cases include the coupling strength $\tilde{g}_0 = 10^2$, which can be readily achieved with levitated systems~\cite{millen2019optomechanics}, a coherent state parameter of $|\mu_{\rm{c}}|^2= 10^6$, a temperature of $200$ nano-Kelvin, which together with an oscillation frequency $\omega_{\rm{m}} = 10^2$ Hz implies a temperature parameter $r_T = 3.48$. We consider a single measurement performed at the final time $\tau_f= 2\pi$. The results can be found in Table~\ref{chap:metrology:tab:Values}, where dimensions can be restored where required by the multiplication the appropriate number of factors of $\omega_{\rm{m}}$.

\begin{table}[h!]
\caption[Single-shot QFI for three different estimation schemes]{Single-shot QFI for estimating the optomechanical coupling strength $\tilde{g}_0$, a linear mechanical displacement strength $\tilde{d}_1$, and a mechanical squeezing strength $\tilde{d}_2$ (all on resonance). In each scheme we set the other couplings to zero or, in the case of the coupling $\tilde{g}_0$, to a constant. Estimation of $\tilde{g}_0$ and $\tilde{d}_2$ correspond to an internal characterisation of the system, while estimation of $\tilde{d}_1$ yields the sensitivity of the optomechanical system to an external force.}
\begin{tabular}{l c c} \toprule
 \textbf{Parameter} & \textbf{Symbol}  &  \textbf{Value}  \\  \midrule
Time of measurement & $\tau_f = \omega_{\mathrm{m}} \, t $ & $2\pi$ \\ 
Coherent state parameter & $|\mu_{\rm{c}}|^2$ & $10^6$ \\
Mechanical oscillation frequency & $\omega_{\rm{m}}$ & $10^2$ Hz \\
Thermal state temperature & $T$ & $200$ nK \\
Thermal state parameter & $r_T$ & 3.48 \\ \midrule
\multicolumn{3}{c}{\textbf{Estimation of $\tilde{g}_0$}}\\ \midrule
Optomechanical coupling & $ \tilde{g}_0 = g_0/\omega_{\rm{m}}$ & $10^2$   \\
Coupling oscillation & $\epsilon $ & $0.5$   \\ 
\textbf{QFI for estimation of $\tilde{g}_0$} & $\mathcal{I}^{(\rm{res})}_{\tilde{g}_0} $ & $3.02\times 10^{25}$ \\ \midrule
\multicolumn{3}{c}{\textbf{Estimation of $\tilde{d}_1$}}\\ \midrule
Optomechanical coupling & $ \tilde{g}_0 = g_0/\omega_{\rm{m}}$ & $10^2$   \\
Linear displacement & $\tilde{d}_1 = d_1/\omega_{\rm{m}}$ & 1 \\
\textbf{QFI for estimation of $\tilde{d}_1$}& $\mathcal{I}^{(\rm{res})}_{\tilde{d}_1} $ & $1.58 \times 10^{12}$\\ \midrule
\multicolumn{3}{c}{\textbf{Estimation of $\tilde{d}_2$}}\\ \midrule
Optomechanical coupling & $ \tilde{g}_0 = g_0/\omega_{\rm{m}}$ & $10^2$   \\
Squeezing parameter & $\tilde{d}_2 = d_2/\omega_{\rm{m}} $ & 0.1 \\
Coherent state parameter & $\mu_{\rm{c}}$ & $10^3$  \\
\textbf{QFI for estimation of $\tilde{d}_2$} & $\mathcal{I}^{(\rm{res, app	})}_{\tilde{d}_2} $ & $6.32\times 10^{28}$\\ \bottomrule
\end{tabular} \label{chap:metrology:tab:Values}
\end{table}

We now proceed to discuss all three cases in detail.

\begin{itemize}
	\item[i)] \textbf{Estimation of the amplitude $\tilde{g}_0$}. The constant case has  already been thoroughly explored~\cite{bernad2018optimal}. We therefore focus on a time-dependent coupling at mechanical resonance. We set the oscillation amplitude to $\epsilon = 0.5$, and we find the QFI to be $\mathcal{I}_{\tilde{g}_0}^{(\rm{Res})} = 3.02 \times 10^{25}$. This implies a single-shot sensitivity of $\Delta \tilde{g}_0 = 1/(\mathcal{I}_{\tilde{g}_0}^{(\rm{Res})})^{-\frac{1}{2}}  = 1.82\times 10^{-14}$, and a relative sensitivity of $\Delta \tilde{g}_0/\tilde{g}_0 = 1.82 \times 10^{-16}$.  

	\item[ii)] \textbf{Estimation of $\tilde{d}_1$}. The constant case is considered in Chapter~\ref{chap:gravimetry}. For the resonant case, we find $\mathcal{I}_{\tilde{g}_0}^{(\mathrm{res})} = 3.02 \times 10^{25}$, which implies a single-shot sensitivity of $\Delta \tilde{d}_1 = 7.96\times 10^{-7}$. Since $\tilde{d}_1  = 1$ in our example, the relative sensitivity $\Delta \tilde{d}_1/\tilde{d}_1$ takes the same value.  This example can be made more concrete in the context of force sensing. We consider detection of a linear force, which physically corresponds to the system being trapped in a linear but oscillating potential, which causes the mechanical element to become displaced.  Let $\tilde{\mathcal{D}}_1(\tau) = a(\tau) \sqrt{ m/(2 \, \hbar \, \omega_{\rm{m}}^3)}$, where $m$ is the mass of the system, and $a(\tau) = a_0 \, \cos(\Omega_{a} \, \tau)$ is a time-dependent acceleration. We then have that $\tilde{d}_1 = a_0 \, \sqrt{ m/(2 \, \hbar \, \omega_{\rm{m}}^3)}$, in analogy with Example (ii) in Section~\ref{chap:metrology:sec:example:2}. Since we now are interested in estimating $a_0$ rather than $\tilde{d}_1$, we note that $\partial_{a_0} = \partial_{a_0} \tilde{d}_1 \partial_{\tilde{d}_1}$, and hence the (dimensionful) QFI becomes $\mathcal{I}_{a_0}^{(\rm{Res})} = ( \partial_{a_0} \, \tilde{d}_1)^2 \, \mathcal{I}_{\tilde{d}_1}^{(\rm{Res})}$. We consider a levitated setup with a mass $m = 10^{-14}$ kg with an oscillation frequency of $\omega_{\rm{m}} = 10^2$ Hz. Given these values together with the parameters $\tilde{g}_0 = 10^2$, $|\mu_{\mathrm{c}}|^2 = 10^6$, and $T = 200$ nK, which implies $r_T =3.48$, we find $\mathcal{I}_{a_0}^{(\rm{Res})}  = 7.48 \times 10^{25}$ ms$^{-2}$, which leads to the sensitivity $\Delta a_0 = 1.16\times 10^{-15}$ ms$^{-2}$.  This, in turn, allows us to measure forces of $F = m a_0$ of magnitude $\Delta F = m \, \Delta a_0 = 1.16\times 10^{-27}$ N. 

	\item[iii)] \textbf{Estimation of both a constant and parametric modulation of the cavity frequency $\delta \omega_{\rm{m}}$}.  This cases are both of interest since, to our knowledge, they have not been considered before. This setting is analogous to Example (iii) in Section~\ref{chap:metrology:sec:example:3}, we have $\tilde{d}_2 = \delta \omega_{\rm{m}}$. When the squeezing is constant with $\tilde{\mathcal{D}}_2(\tau) \equiv \tilde{d}_2$, and with $\tilde{g}_0 = 10^2$, we find $\mathcal{I}_{\tilde{d}_2}^{(\mathrm{const})} = 1.53\times 10^{16}$, which implies a sensitivity of $\Delta \tilde{d}_2 = 8.08\times 10^{-9}$. For the resonant time-dependent case, our expressions are only valid for small values of $\delta \omega_{\rm{m}} \sim 0.1$. We find $\mathcal{I}_{\delta \omega_{\rm{m}}}^{( \rm{res, approx})} = 6.32 \times 10^{28}$, which implies the sensitivity $\Delta (\delta \omega_{\rm{m}}) = 3.98 \times 10^{-15}$ and a relative sensitivity $\Delta (\delta \omega_{\rm{m}})/\delta \omega_{\rm{m}} = 3.98 \times 10^{-14}$. We conclude that estimation of a time-modulated trapping frequency could potentially be successfully implemented for these systems.
\end{itemize}

\section{Discussion}\label{chap:metrology:sec:discussion}
Our results show that, by analytically solving the system dynamics of an optomechanical system with additional dynamics terms, it is possible to obtain the bounds on the precision of the measurements of relevant experimental parameters contained in the Hamiltonian of the system. We here discuss the results and elaborate on a number of properties of the QFI. 

\subsection{The cavity as a resource}
The addition of the cavity to a mechanical element results in the characteristic cubic optomechanical light--matter interaction $- \tilde{g}_0 \, \hat{a}^\dag\hat{a}(\hat{b}^\dag+\hat{b})$. From our results, we note some distinct advantages of the inclusion of the cavity, in particular for estimation of the linear displacement $\tilde{d}_1$, where the QFI is given in Eq.~\eqref{chap:metrology:QFI:estimate:d1}. When either the optical state is the vacuum state (defined by $|\mu_{\rm{c}}| = 0$), or when the optical coupling is zero (that is, $\tilde{g}_0 = 0$), the contributions from the $B$-coefficient vanish, while the coefficients $C_\pm$ remains non-zero. This situation corresponds to estimating the displacement of a single mechanical element without the cavity.  We note that, in this setting, the enhancement from $|\mu_{\rm{c}}|^2$ is lost, which means that the QFI is reduced overall. As such, we can consider the cavity and the coherent state parameter $|\mu_{\rm{c}}|^2$ as resources which aid the sensing scheme. However, more work is needed to establish the exact role of the nonlinear light--matter coupling in sensing schemes.

\subsection{The Heisenberg limit}
The Heisenberg limit is often associated with a scaling of the sensitivity of a system as $N^{-1}$ (as opposed to $N^{-1/2}$ for classical systems), where $N$ is the number of physical probes in the system.  However, it should be kept in mind that this result is derived under rather specific conditions~\cite{braun2018quantum}: $N$ distinguishable, non-interacting subsystems, finite-dimensional Hilbert spaces, and parameter encoding via a unitary evolution with a parameter-dependent Hamiltonian~\cite{giovannetti2004quantum,giovannetti2006quantum}.   By coincidence, the $1/N$ (respectively $1/\sqrt{N}$) scaling is also the scaling of the sensitivity with the average number of photons with which the phase-shift in a Mach--Zehnder interferometer can be measured. This scaling occurs when a NOON-state (respectively a coherent state) is used, even though the photons are indistinguishable bosons with infinite-dimensional Hilbert space, and the photon number is in both cases only defined on average.  This result follows immediately from the general expression of the pure state QFI in terms of the variance of the generator $\hat G$ that generates the unitary  transformation $\hat U_\alpha$, which encodes the parameter $\alpha$ according to $\hat U_\alpha=e^{i \alpha \hat G}$, together with the phase shift Hamiltonian $\hat H=\alpha \hat a^\dagger \hat a$. It is, however, also well known that the scaling with $N$ can be faster than $1/N$ for the estimation of an interaction parameter~\cite{luis2004nonlinear,braun2011heisenberg}, and this advantage can extend in certain parameter regimes to the estimation of other parameters of an interacting system~\cite{fraisse2015coherent} if one has access to the full system.  

In light of the $1/N$ scaling that is often associated with the Heisenberg limit, our main result in Eq.~\eqref{chap:metrology:eq:main:result:QFI} appears to indicate scaling beyond the Heisenberg limit due the term $|\mu_{\rm{c}}|^6$, which can be written in terms of the initial average number $N_{\rm{ph}}$ of photons as $|\mu_{\rm{c}}|^6=N_{\rm{ph}}^3$. A similar scaling has been predicted for the phase sensitivity of nonlinear optical systems~\cite{boixo2007generalized}. The $N^3$ term corresponds to a sensitivity that scales $\propto N_{\rm{ph}}^{-3/2}$, i.e.~decays faster than the ``Heisenberg limit'' $1/N$. The origin of the $|\mu_{\rm{c}}|^6$  term is the $(\hat a^\dagger \hat a)^2$ term in $\hat{\mathcal{H}}_{N_a}$ (see Eq.~\eqref{chap:metrology:eq:mathcalH:with:coefficients}).  If one restricts the maximum amount of energy available, its contribution to the QFI is maximized  when the light and mechanics form the  aforementioned NOON state~\cite{braun2018quantum}, but $N_{\rm{ph}}$ is replaced by $N_{\rm{ph}}^2$,  i.e.~the true Heisenberg-limit in the sense of the smallest possible uncertainty is  now a $1/N_{\rm{ph}}^2$ scaling of the sensitivity, whereas the coherent state gives the $1/N_{\rm{ph}}^{3/2}$ found above. 

Given that a NOON state is extremely difficult to prepare, especially for highly excited Fock states, the scaling obtained for the coherent state is quite favorable, given this consideration.   
Since the corresponding parameter $F_{\hat N_a}^2$  depends not only on the coupling constant $\tilde{\mathcal{G}}_1$ but also on the  squeezing parameter $\tilde d_2$ relevant for force sensing, we have here the remarkable situation that the nonlinear interaction between the two oscillators not only allows enhanced sensitivity for estimating the interaction (i.e.,~faster than $1/N_{\rm{ph}}$ scaling of the sensitivity, but which cannot be compared to the non-interacting case, as the parameter  $\tilde{g}_0$ does not exist there), but also enables enhanced sensitivity of a parameter of the original non-interacting system! This is a fundamental insight that was possible only through the exact decoupling scheme used here, and it should be highly useful for metrology.  In principle one could  envisage other systems leading to even higher powers of $N_{\rm{ph}}$, if the Lie-algebra of generators in  $\hat H$ closed after more iterations. We note, however, that the sensitivity to linear displacements  with this system scales as $1/N_{\rm{ph}}^{1/2}$,  i.e.~up to a change of prefactor the same sensitivity as for measuring a phase shift with a coherent state.  However, it should be kept in mind that it is the excitation of the optical cavity that determines the sensitivity with which the shift of the mechanical oscillator is measured, and which can be much larger than the initial thermal excitation of the mechanical oscillator.

\subsection{Resonance}
Here we discuss the implications of driving the system at mechanical resonance. The resonance behavior differs for all three examples considered in Section~\ref{chap:metrology:three:examples}, which implies a rich and complicated structure of the QFI.  We here provide a brief discussion of some of the main features observed in this work.   For estimation of $\tilde{g}_0$, it can be seen in   Figure~\ref{chap:metrology:fig:QFI:g0}, where we plotted a frequency sweep of the QFI at various times $\tau_f$, that the onset of the increase of QFI is due to the accumulation of the resonant behavior. In fact, Figure~\ref{chap:metrology:fig:QFI:g0} demonstrates  that driving on resonance only provides a significant advantage as $\tau \gg 1$. 
  
For estimations of a linear drive $\tilde{d}_1$, we found that a constant coupling performs better than a time-dependent one. This observation is most likely due to our choice to let the weighting function $\tilde{\mathcal{D}}(\tau) = \tilde{d}_1 \cos(\Omega_{d_1}  \tau)$ oscillate around zero rather than a fixed displacement. Finally, for estimation of $\tilde{d}_2$, our results are only valid  close to parametric resonance, which occurs when $\Omega_{d_2} = 2$.  In all cases considered here, general, we demonstrated that resonances play an important, but not always beneficial, part in enhancing the sensitivity of a system.

\subsection{Time-dependence}
In all three examples we considered, the QFI was found to increase essentially quadratically with dimenionless time $\tau$ to leading order  at resonance. 
 Optomechanical systems are among the most massive quantum systems that can be controlled  in the laboratory to date, and while impressively narrow linewidths have recently been demonstrated experimentally with levitated nano-particles~\cite{pontin2019ultra}, achieving long quantum coherence times is still a challenging task.  In the pioneering experiments reported in~\cite{o2010quantum} the fitted $T_2$ dephasing time of a nano-mechanical oscillator with resonance frequency of 6\,GHz was about 20\,ns, corresponding to a maximally achievable $\tau\simeq 754$. 
Given a finite available measurement time limited by the decoherence time, our results show that the precise timing of the measurements and the choice of frequency ratios is crucial for optimizing the overall sensitivity per square root of Hertz.  It is a major benefit of our method that the precise time-dependence  of the QFI can be obtained in such a non-linear and possibly driven or parametrically modulated optomechanical system. 
 
\section{Conclusions} \label{chap:metrology:sec:conclusions}

In this Chapter we have derived a general expression for the quantum Fisher information for a  nonlinear optomechanical system with a time-dependent light--matter coupling term, a time-dependent linear mechanical displacement term, and a time-dependent single-mode mechanical squeezing term in the Hamiltonian.  The expression we derived can be used to compute the optimal estimation bounds for any parameter which enters into any of the terms in the Hamiltonian. Most importantly, our methods include the treatment of arbitrary time-dependent effects, which offer significant advantages for experimental schemes since time-varying signals can be more easily distinguished from a random noise floor. 

To demonstrate the applicability of the expression and our methods, we computed the QFI for three specific examples: (i) Estimating the strength of an oscillating optomechanical coupling, (ii) estimating the amplitude of an oscillating linear displacement term, and, (iii) estimating the amplitude of a  resonant time-dependent mechanical squeezing term.  We derived exact and asymptotic expressions for the QFI in the first two cases, as well as an approximate expressions based on perturbative solutions for a squeezing term modulated at resonance. 

Our results include a number of interesting phenomena. Firstly, in all three cases considered, we find that resonances (where the oscillation frequency of the measured effect matches the mechanical oscillation frequency of the system) cause the QFI to   increase continuously and polynomially as a function of time. This suggests that once optomechanical systems reach long coherence times, resonances could be utilised to greatly increase the sensitivity of the system. Secondly, we find that the temperature of the mechanical thermal state is not always detrimental to the system performance, and might even sometimes aid estimation of the parameter in question. More work is needed to establish how this effect can be harnessed and how it might potentially limit the coherence time of the system. 

Finally, our methods can be applied to the detection of a number of internal and external effects that act on the optomechanical systems, as long as they take the form of the coefficients in the Hamiltonian we consider.  A number of questions remain unanswered, including understanding how the different effects in the Hamiltonian interact to enhance or decrease the sensitivity of the system, and which measurements actually saturate the QFI. We leave these questions for future work.


\chapter{Gravimetry through nonlinear optomechanics} \label{chap:gravimetry}
In this Chapter, we apply the tools developed throughout this thesis to determine the sensitivities that can be achieved with optomechanical systems when estimating a constant gravitational acceleration. We start by considering the system dynamics for the addition of a constant gravitational displacement term, then proceed to derive the quantum Fisher information for three different initial states. Since the quantum Fisher information provides an upper bound on the sensitivity, but otherwise does not inform on the measurement that saturates this bound, we consider homodyne measurements and show that these are in fact the optimal measurements at specific times. The Chapter is concluded by a discussion and comparison between the results obtained here and other theoretical and experimental schemes. 

This Chapter is based on Ref~\cite{qvarfort2018gravimetry} and its Supplemental Material. The code used to simulate the open system dynamics can be found in the following \href{https://github.com/sqvarfort/Coherent-states-Fisher-information}{GitHub Repository}. However, many of the results in Ref~\cite{qvarfort2018gravimetry} were generalised through the subsequent work presented in Chapter~\ref{chap:decoupling}. As a result, the exposition in this Chapter has been revised and extended to draw on the results and notation presented in the preceding sections of this thesis. In general, the results presented in this Chapter constitute one of the most straight-forward applications of our methods. They further demonstrate how the results in this thesis can be extended to a number of estimation schemes for optomechanical systems in the nonlinear regime. 

We would like to thank Doug Plato, Abolfazl Bayat, Nathana\"{e}l Bullier, Victor Montenegro, Dennis Schlippert and Stephen Stopyra for useful comments and discussions in relation to this work. It should also be ntoed that we have corrected a typo in Ref \cite{qvarfort2018gravimetry} in Eq. \eqref{chap:gravimetry:eq:traced:out:cavity:state}, where an extra phase was included in error.

\section{Introduction}
The practise of measuring the gravitational acceleration $g$ -- also known as gravimetry -- has led to important advances in both fundamental science and industry. For example, local gravity variations due to mass redistribution driven by climate changes have been mapped with the GRACE satellite~\cite{ray2006tide,  crowley2006land, chen2006satellite}, and more recently, the Juno spacecraft mission reported the measurement of the gravity harmonics of Jupiter~\cite{iess2018measurement}. Furthermore, precise measurements of $g$ can test for small deviations from Newtonian gravity on extremely small scales, which may provide indications of a deeper theory of quantum gravity~\cite{biswas2012towards}. In industry, precision accelerometry is extensively used in inertial navigation technologies and for conducting geological surveys. 

While classical systems have and are still being utilised to perform accurate measurements of $g$, quantum systems offer several useful advantages, including reduced noise levels, a compact setup and most importantly an increased measurement sensitivity achieved through the power of coherence and interferometry. Over the past decade, a variety of quantum systems have been explored to this aim, in both theory and practice. The largest research effort to date has focused on atom interferometry ~\cite{peters2001high,mcguirk2002sensitive,bidel2013compact,hu2013demonstration}, for which the highest achieved sensitivity currently stands at $\Delta g = 4.3\times 10^{-9}$ ms$^{-2}$~\cite{hu2013demonstration}. A similar investigation has been carried out for both on-chip and fountain Bose-Einstein condensate (BEC) interferometry with best sensitivity $\Delta g  = 7.8 \times 10^{-10}$ \si{ms^{-2}}~\cite{abend2016atom}. Finally, a proposal for using magnetically levitated spheres which predicts sensitivities of $2.2 \times 10^{-9}$  \si{ms^{-2}Hz^{-1/2}} has been put forward in Ref~\cite{johnsson2016macroscopic}. For comparison, the current commercial standard is set by the LaCoste FG5-X gravimeter which can achieve a measurement sensitivity of $1.5\times 10^{-9}$ \si{ ms^{-2}Hz^{-1/2}}~\cite{LaCoste2016}. More generally, the broader topic of using quantum systems to probe relativistic phenomena is currently being pursued with great interest (see for example~\cite{dimopoulos2008general,bruschi2014quantum, howl2016gravity, seveso2016can, seveso2017quantum,  asenbaum2017phase, tan2017relativistic, joshi2017space}). 

A key advantage to quantum systems are their interferometric properties. The following question arises: \textit{How can these interferometric properties be enhanced in order to improve the measurement sensitivity?} One possibility is to place a quantum system in the form of a mechanical oscillator in an optical cavity, a research area known as quantum optomechanics ~\cite{aspelmeyer2014cavity}. See Figure~\ref{chap:introduction:fig:mirror:cavity} for an illustration of a nanodiamond trapped in an optical cavity as an example of a class of optomechanical systems. The addition of the cavity allows for a strong coherent coupling between light and oscillator which, as we shall see, cancels out any initial thermal noise and fundamentally improves the measurement sensitivity of the device. 

Within classical optomechanics, the idea of gravimetry and accelerometry by optically detecting the mechanical oscillator has been experimentally realised by Cervantes \textit{et al}.~\cite{Cervantes2014}. Other avenues, such as the detection of high frequency gravitational waves through the driving of resonant mechanical elements was proposed also in Ref~\cite{Arvanitaki2013}. In the related field of electromechanics, Schr\"{o}dinger cat states and a Kerr nonlinearity have recently been found to be useful for the same applications~\cite{jacobs2016quantum}. However, the ensuing fundamental limits on the measurement sensitivity of gravimetry in the quantum regime of optomechanics using its trilinear radiation pressure interaction is yet to be investigated. 

Here we undertake this task and obtain some striking results: Firstly, it is possible, in principle, to surpass the sensitivity $\Delta g $ that has been obtained in atom interferometers and other implementations to date. Secondly, due to the periodic decoupling of light and mechanics, the mechanical element does not require initial cooling to the ground state to improve the fundamental sensitivity of the gravimeter and, finally, the best possible sensitivity is achieved by a simple homodyne measurement of the cavity field, while only a low photon number in the cavity is required. That is, no measurement on the mechanical oscillator is required. Unlike the case of atomic interferometers, in optomechanics the interaction of light and matter is continuous, and we will see that our Hamiltonian cyclically entangles and disentangles the light and mechanics, leading to their decoupling. It follows that the experimental challenge will be to maintain the quantum coherence of the field and mechanics over the duration of each run of the experiment, which we set as one oscillation period of the mechanical element.  This requirement, on which the plausibility of the scheme hinges, will be discussed in some detail. 

This Chapter is structured as follows. In Section~\ref{chap:gravimetry:system:and:dynamics}, we consider the system Hamiltonian, including the gravitational acceleration term, and solve the resulting dynamics. We also discuss a number of properties of the state, including the phase space quadratures, the linear entropy, and when the light and mechanics disentangle. In Section~\ref{chap:gravimetry:estimation:of:g}, we consider the task of estimating the gravitational acceleration and the tools required. This task is then carried out in Section~\ref{chap:gravimetry:sec:QFI}, where we calculate the quantum Fisher information (QFI) of the system. To also consider real-world applications and measurements in the laboratory, we computer the classical Fisher information in Section~\ref{chap:gravimetry:sec:CFI} for a homodyne measurement and show that it is optimal at certain times. Following that, in Section~\ref{chap:gravimetry:open:system:estimation} we present a short investigation into open-system dynamics, and in Section~\ref{chap:gravimetry:computing:sensitivity}, we insert state-of-the-art parameters into the expressions we derive to predict the sensitivity with which an optomechanical system can measure the gravitational acceleration. The Chapter is concluded by a discussion in  Sections~\ref{chap:gravimetry:discussion} and some concluding remarks in Section~\ref{chap:gravimetry:conclusions}. 

\section{System and dynamics} \label{chap:gravimetry:system:and:dynamics}

In this Section, we outline the dynamics and state evolution of an optomechanical system with a constant gravitational acceleration. We also aim to provide intuition for the state evolution by computing the linear entropy and quadratures of the state. A full introduction to the theory of optomechanical systems can be found in Chapter~\ref{chap:introduction}. 

\subsection{Hamiltonian with gravitational acceleration}
Let us begin by considering a general optomechanical system consisting of a mechanical oscillator coupled to a light--field in the cavity. The non-gravitational Hamiltonian that describes the dynamics of an optomechanical system is given by~\cite{mancini1997ponderomotive,bose1997preparation}:
\begin{equation} \label{eq:Hamiltonian}
\hat{H} = \hbar \omega_{\mathrm{c}} \, \hat a^\dagger \hat a + \hbar \omega_m \, \hat b^\dagger \hat b - \hbar \mathcal{G}(t) \hat a^\dagger \hat a \, (\hat b^\dagger + \hat b), 
\end{equation}
where $\hat a, \hat a^\dagger$ are the annihilation and creation operators for the cavity field with frequency $\omega_{\mathrm{c}}$, $\hat b, \hat b^\dagger $ are the annihilation and creation operators for the mechanical oscillator with frequency $\omega_m$, and $\mathcal{G}(t) $  is a coupling constant that determines the interaction strength between the photon number $\hat a^\dagger \hat a $ and the position $\hat{x}_{\mathrm{m}} \propto (\hat b^\dagger + \hat b)$ of the oscillator. In this Chapter, we assume that $\mathcal{G}(t) $ is constant, such that $\mathcal{G}(\tau) \equiv g_0$, where $g_0$ is the strength of the interaction.

In order to introduce a coupling to a gravitational potential in the Hamiltonian, we add a term of the form $ m g \hat{x}_{\mathrm{m}}  \cos{\theta}$. Here, $m$ is the mass of the mechanical oscillator, $g$ is the gravitational acceleration, $\hat{x}_{\mathrm{m}} = \sqrt{\hbar/2m\omega_m}\, (\hat b^\dagger + \hat b)$ is the position operator acting on the mechanical oscillator, and $\theta$ is an angle from the vertical axis that we include in order to describe inclined systems, similar to~\cite{Scala2013a}.  Note that while the mass $m$ appears as a coupling in the Hamiltonian, we later see that measurements of $g$ are mass-independent, which is what we expect from the equivalence of inertial and gravitational mass.  
With the addition of Newtonian gravity, the  Hamiltonian of the system becomes
\begin{align} \label{chap:gravimetry:eq:Hamiltonian:with:gravity}
\hat{H}_{\mathrm{G}} = \,  &\hbar \omega_{\mathrm{c}} \,  \hat a^\dagger \hat a + \hbar \omega_m \, \hat b^\dagger \hat b - \hbar \, g_0\,   \hat a^\dagger \hat a \, ( \hat b^\dagger + \hat b)+ \cos{\theta} \, g  \sqrt{\frac{\hbar m }{2\omega_m}} \, (\hat b^\dagger +\hat b) . 
\end{align}
This Hamiltonian is analogous to the decoupled Hamiltonian in Eq.~\eqref{chap:decoupling:eq:Hamiltonian} in Chapter~\ref{chap:decoupling} with a constant mechanical displacement weighted by the function $\mathcal{D}_1(t) := d_1 = \cos{\theta} \, g  \sqrt{\frac{\hbar m }{2\omega_m}}$, but where the single mode mechanical squeezing term is set to zero with $\mathcal{D}_2(t) = 0$.

It is beneficial to work in dimensionless units, especially when considering the quantum Fisher information. We therefore rescale $\hat{H}_{\rm{G}}$ by $\hbar \omega_{\rm{m}}$ to find:
\begin{align}  \label{chap:gravimetry:eq:rescaled:Hamiltonian}
\hat{\tilde{H}}_{\mathrm{G}} = \Omega_{\rm{c}}\,  \hat a^\dagger \hat a +  \hat b^\dagger \hat b -  \tilde{g}_0 \hat a^\dagger \hat a \, ( \hat b^\dagger + \hat b)+ \tilde{d}_1 \, (\hat b^\dagger +\hat b) \, , 
\end{align}
where $\Omega_{\rm{c}} = \omega_{\rm{c}}/\omega_{\rm{m}}$, and $\tilde{g}_0(\tau) = g_0 /\omega_{\rm{m}}$. Since $d_1$ is not multiplied by $\hbar$, we divide it to obtain 
\begin{equation}
\tilde{d}_1 := \cos{\theta} \, g  \sqrt{\frac{ m }{2 \, \hbar\, \omega_m^3}} \, ,
\end{equation}
where $\tilde{d}_1 = d_1/(\hbar \omega_{\rm{m}})$. 

In what follows, we primarily focus on estimation with respect to $\tilde{d}_1$, and later specialise to estimation of $g$. This treatment is possible due to application of the chain rule, see Section~\ref{chap:gravimetry:sec:QFI}. 

\subsection{Solution of the dynamics}
Since the Hamiltonian $\hat H_{\rm{G}}$ is analogous to that considered in Chapter~\ref{chap:decoupling}, we can use the tools developed there to solve the time-evolution of the system. 
The full solution is given in Eq.~\eqref{chap:decoupling:eq:final:evolution:operator:J:coefficients},   but since here $\mathcal{D}_2(\tau) = 0$, it follows that $J_b = \tau$ and $J_\pm = 0$. We therefore recover the following evolution operator for a constant linear displacement $\tilde{d}_1$:
\begin{align} \label{chap:gravimetry:eq:time:evolution:operator}
\hat U(\tau)=& \, e^{-i\,\Omega_\mathrm{c} \hat N_a\,\tau} \, e^{- i \, \tau\, \hat N_b} \, 
\,e^{-i\,F_{\hat{N}_a}\,\hat{N}_a}\,e^{-i\,F_{\hat{N}^2_a}\,\hat{N}^2_a}\,e^{-i\,F_{\hat{B}_+}\,\hat{B}_+}\,e^{-i\,F_{\hat{N}_a\,\hat{B}_+}\,\hat{N}_a\,\hat{B}_+}\,\nonumber \\
&\times e^{-i\,F_{\hat{B}_-}\,\hat{B}_-}\,e^{-i\,F_{\hat{N}_a\,\hat{B}_-}\,\hat{N}_a\,\hat{B}_-}  \, ,
\end{align}
 where the notation of the operators reads
\begin{align}\label{chap:gravimetry:eq:Lie:algebra}
	 	\hat{N}^2_a &:= (\hat a^\dagger \hat a)^2 \nonumber \\
	\hat{N}_a &:= \hat a^\dagger \hat a &
	\hat{N}_b &:= \hat b^\dagger \hat b \nonumber\\
	\hat{B}_+ &:=  \hat b^\dagger +\hat b &
	\hat{B}_- &:= i\,(\hat b^\dagger -\hat b) &
	 & \nonumber\\
	\hat{N}_a\,\hat{B}_+ &:= \hat{N}_a\,(\hat b^{\dagger}+\hat b) &
	\hat{N}_a\,\hat{B}_- &:= \hat{N}_a\,i\,(\hat b^{\dagger}-\hat b) \, ,&
	 & 
\end{align}
and where the $F$-coefficients are obtained by solving the integrals in Eq.~\eqref{chap:decoupling:eq:sub:algebra:decoupling:solution}. 
Given the constant displacement function $ \tilde{d}_1$, and a constant optomechanical coupling $\tilde{g}_0$, we are able to solve the integrals for the $F$-coefficients in Eq.~\eqref{chap:decoupling:eq:sub:algebra:decoupling:solution} to find
\begin{align} \label{chap:gravimetry:F:coefficients}
F_{\hat N_a} &= \, \tilde{g}_0 \, \tilde{d}_1 \left( \tau - \cos(\tau)\, \sin(\tau) \right) \, ,\nonumber \\
F_{\hat N_a^2} &= \frac{1}{2} \tilde{g}_0 ^2 \left( \sin(2 \, \tau) - 2 \, \tau \right) \, ,\nonumber \\
F_{\hat B_+} &= \tilde{d}_1 \, \sin(\tau) \, ,\nonumber \\
F_{\hat B_-} &= \tilde{d}_1 \, \left( 1- \cos(\tau)\right) \, ,\nonumber \\
F_{\hat N_a\, \hat B_+} &= - \tilde{g}_0 \, \sin(\tau) \, ,\nonumber \\
F_{\hat N_a, \hat B_-} &= - \tilde{g}_0 \, \left( 1 - \cos(\tau) \right) \, .
\end{align}
With these coefficients, the evolution operator becomes
\begin{align} \label{chap:gravimetry:eq:time:evolution:operator:explicit}
\hat U(\tau)=& \, e^{-i\,\Omega_\mathrm{c} \, \tau \, \hat N_a} \, e^{- i \, \tau\, \hat N_b} \, 
\,e^{-i\ \tilde{g}_0 \, \tilde{d}_1 \left( \tau - \cos(\tau)\, \sin(\tau) \right)\,\hat{N}_a}\,e^{-\frac{1}{2}  \, i\, \tilde{g}_0 ^2 \left( \sin(2 \, \tau) - 2 \, \tau \right)\,\hat{N}^2_a}\,\nonumber \\
&\times e^{-i\, \left( \tilde{d}_1 - \tilde{g}_0 \, \hat N_a \right) \, \sin(\tau) \,\hat{B}_+} \, e^{-i\,  \left( \tilde{d}_1 - \tilde{g}_0 \, \hat N_a \right) \left( 1- \cos(\tau)\right)\,\hat{B}_-} \, ,
\end{align}
where we have combined the last four exponentials into two pairs.

In Ref~\cite{qvarfort2018gravimetry}, we provided an alternative analytic decoupling of the operator, which is based on a decoupling provided in Ref~\cite{bose1997preparation}. Since the interaction term can be written as $\bigl( \tilde{g}_0 \, \hat a ^\dag \hat a - \tilde{d}_1\bigr) \bigl(\hat b^\dag + \hat b\bigr)$, where the scalar $\tilde{d}_1$ is added to the photon number operator $\hat a^\dag \hat a$, the same method applies, and the evolution operator can be written in the following form:
\begin{eqnarray} \label{chap:gravimetry:U:alternative:derivation}
\hat U(\tau) &=e^{- i \Omega_{\rm{c}} \hat a^\dagger \hat a\,  t} \, e^{ i ( \bar{g}_0 \,   \hat a^\dagger \hat a -  \tilde{d}_1)^2 (\tau - \sin{\tau}) }  e^{(\tilde{g}_0 \hat a^\dagger \hat a -\tilde{d}_1( \eta \hat b^\dagger - \eta^* \hat b )} e^{- i \hat b^\dagger \hat b \, \tau}, 
\end{eqnarray}
where $\eta = 1 - e^{-it}$. It can be shown that the $\hat U(\tau)$ in Eq.~\eqref{chap:gravimetry:U:alternative:derivation} corresponds to that in Eq.~\eqref{chap:gravimetry:eq:time:evolution:operator} by some rearrangement of the terms.

\subsection{Open system dynamics} \label{chap:gravimetry:open:system:dynamics}
The previous solution is valid for closed system dynamics only. However, as discussed in Section~\ref{chap:introduction:open:system:dynamics} in Chapter~\ref{chap:introduction}, there are a large variety of decoherence effects for optomechanical systems, such as decoherence due to photons leaking from the cavity, or phonons gradually being lost from the mechanical element. We briefly consider decoherence in this Chapter in the context of performing homodyne measurements on the state in Section~\ref{chap:gravimetry:open:system:estimation}. 

In this Chapter, we make the assumption that the phonon decoherence is negligible over one oscillation period of the oscillator. Instead, we consider photon decoherence and its effects on the system, as this is deemed a more severe effect compared with phonon decoherence. The effect of photons leaking from a cavity on a state $\hat \rho(\tau)$ can be modeled using a Lindblad master equation of the form
\begin{equation} \label{chap:gravimetry:eq:Lindblad}
\frac{\partial\hat \rho(\tau)}{\partial \tau} = - \frac{i}{\hbar} [\hat H,\hat \rho(\tau)] +  \hat L_{\rm{c}} \hat \rho(\tau) \hat L^\dagger_{\rm{c}} - \frac{1}{2} \left\{ \hat \rho(\tau),  \hat L^\dagger_{\rm{c}} \hat L_{\rm{c}} \right\},  
\end{equation}
where $\hat H$ is the Hamiltonian (in our case, we consider the rescaled $\hat{\tilde{H}}_{\rm{G}}$ in Eq.~\eqref{chap:gravimetry:eq:rescaled:Hamiltonian}),
 $\hat L_{\rm{c}} = \sqrt{\tilde{\kappa}}_{\rm{c}} \, \hat  a$ are Lindblad operators,  and $\tilde{\kappa}_{\rm{c}} = \kappa_{\rm{c}}/\omega_{\rm{m}}$ is the decoherence rate $\kappa_{\rm{c}}$ with respect to the rescaled time $\tau$. Finally, $\{\bullet, \bullet\}$ denotes the anti-commutator. 
 
 This equation cannot easily be solved analytically for the system we consider,  since the operator $\hat a$ does not commute with the Hamiltonian $\hat{H}_{\mathrm{G}}$ in Eq.~\eqref{chap:gravimetry:eq:Hamiltonian:with:gravity}. Some solutions have been found for specific cases, for example when assuming that the photon leakage occurs only during the injection of the state into the cavity. The decoherence can then be modeled as a series of beamsplitters~\cite{montenegro2015entanglement}. We will not consider these modifications here, but instead solve the Lindblad master equation numerically and compute the Fisher information $\mathcal{I}_{\tilde{d}_1}(\tau)$ for the resulting mixed state. 
\subsection{Initial and evolved states} \label{chap:gravimetry:initial:and:evolved:states}

In this Chapter, we consider three different initial states: coherent states of the optics and mechanics, a coherent state of the optics and a thermal state of the mechanics,  and a Fock state of the otpics and coherent state for the mechanics.  The motivation for using these states is discussed in Section~\ref{chap:introduction:initial:states} in Chapter~\ref{chap:introduction}.
We here consider three different coherent states:
\begin{itemize}
\item[(i)] \textbf{A coherent state of the optics and mechanics}. Laser light is naturally coherent, while the mechanics can be assumed to be in a coherent state. This is the most important state in this Chapter, and most of our analysis focuses on it. 
The initial state $\ket{\Psi(\tau = 0)}$ is given by 
\begin{equation} \label{chap:gravimetry:eq:coherent:coherent}
\ket{\Psi(0)} = \ket{\mu_{\rm{c}}} \otimes \ket{\mu_{\rm{m}}} \,,
\end{equation}
where $\mu_{\rm{c}}$ is the coherent state parameter of the optics, and $\mu_{\rm{m}}$ is the coherent state of the mechanics.

This state evolves under $\hat U(\tau)$ in Eq.~\eqref{chap:gravimetry:eq:time:evolution:operator:explicit} into 
\begin{align} \label{chap:gravimetry:eq:fully:evolved:state}
\ket{\Psi(\tau) } = & \, e^{- |\mu_{\rm{c}}|^2/2} \sum_{n = 0}^\infty \biggl[ \frac{\mu_{\rm{c}}^n}{\sqrt{n!}} e^{i  (\tilde{g}_0^2 \,  n^2 - 2\,  \tilde{g}_0 \, \tilde{d}_1 \, n ) \, ( \tau - \sin\tau)} \,  \nonumber \\
&\quad\quad\quad\quad\quad\quad\times e^{( \tilde{g}_0 \, n - \tilde{d}_1 )\left( \eta \mu_{\rm{m}} - \eta^* \mu_{\rm{m}}^*\right)/2}\ket{n} \otimes \ket{\phi_n(\tau)}\biggr],
\end{align}
where we have defined $\eta = 1 - e^{- i \, \tau}$, and $\ket{\phi_n(\tau)}$ are coherent states of the oscillator given by 
\begin{equation} \label{chap:gravimetry:eq:mechanical:coherent:state:parameter}
\ket{\phi_n(\tau)} = \ket{e^{-i \, \tau} \mu_{\rm{m}} + (\tilde{g}_0 \,  n - \tilde{d}_1 )(1 - e^{-i \, \tau})} \, .
\end{equation}  
In the derivation of this state, we have adopted a rotating frame for the cavity field, thus ignoring the free evolution  induced by the term $\exp[- i  \, \tau \,  \Omega_{\rm{c}}]$.

\item[(ii)] \textbf{A coherent state of the optics and a thermal state of the mechanics}. If the mechanical element is not cooled to the ground state, it will be in a thermal state, which can be thought of as a mixture of coherent state integrated over some appropriate kernel~\cite{barnett2002methods}. Our main purpose for considering this state in this Chapter is to show that the mechanical element does not have to be cooled to the ground state in order to maximise the sensitivity of the system. 

We considered this state in Chapter~\ref{chap:metrology} as well, and it is given by 
\begin{equation} \label{chap:gravimetry:initial:state:coherent:thermal}
\hat \rho_{\rm{th}}(0) = \ketbra{\mu_{\rm{c}}} \otimes \sum_{m=0}^\infty \frac{\tanh^{2m} r_T}{\cosh^2 r_T } \ketbra{m} \,,
\end{equation}
where $r_T$ is a temperature parameter given by $\tanh r_T = \exp[ - \hbar \, \omega_{\rm{m}}/(k_{\rm{B}} T)]$, where $k_{\rm{B}}$ is Boltzmann's constant and $T$ is a temperature. 

Applying the time-evolution operator $\hat U(\tau)$ in Eq.~\eqref{chap:gravimetry:eq:time:evolution:operator:explicit} gives the state
\begin{align} \label{chap:gravimetry:evolved:coherent:thermal:state}
\hat \rho_{\rm{th}}(\tau)=  & \,  \sum_{n,n'} \frac{\mu_{\rm{c}}^n (\mu_{\rm{c}}^*)^{n'}}{\sqrt{n! n'!}} e^{i [\tilde{g}_0^2 (n^2 - n^{\prime 2}) - 2 \tilde{g}_0 \, \tilde{d}_1(n-n')] ( \tau - \sin\tau ) } \ket{n}\bra{n'}  \nonumber \\
&\quad\quad \times e^{- 2 \, i \, \left( \tilde{g}_0^2 ( n^2 - (n^\prime)^2 ) - 2 \, \tilde{d}_1 \, \tilde{g}_0 ( n- n^\prime) \right)  \sin \tau \, \sin^2(\tau/2)} \nonumber \\
&\times  \sum_{m=0}^\infty \frac{\tanh^{2m} r_T}{\cosh^2 r_T }   \hat D(\varphi_n(\tau)) \ket{m}\bra{m} \hat  D^\dag(\varphi_n(\tau)) \, , 
\end{align}
where $\hat D(\varphi_n(\tau))$ is a Weyl displacement operator with the argument $\varphi_n(\tau) =- \, (\tilde{g}_0 \,  n -  \tilde{d}_1) \eta^* $. While we could apply the remaining displacement operators to an expression in terms of so-called displaced number states~\cite{cahill1969ordered}, we leave it general since it is easier to demonstrate the state properties at $\tau = 2\pi$ in this form. 

\item \textbf{A Fock state of the optics and a coherent state of the mechanics}. A Fock state $\ket{n}$ is an eigenstate of the number operator $\hat a^\dag \hat a$, such that $\hat a^\dag \hat a \ket{n} = n \, \ket{n}$. We are interested in the evolution of an initial superposition of two Fock states, namely $\frac{1}{\sqrt{2}} \left( \ket{0} + \ket{n} \right)$. With this Fock state and the mechanics in a coherent state $\ket{\mu_{\rm{m}}}$, the full state is given by 
\begin{equation} \label{chap:gravimetry:initial:state:Fock:coherent}
\ket{\Psi(0)}_{\rm{Fock}} = \frac{1}{\sqrt{2}} \left( \ket{0} + \ket{n} \right) \otimes \ket{\mu_{\rm{m}}} \, .
\end{equation}
It is generally difficult to generate superpositions of Fock states, although superpositions of $n = 10$ have been successfully demonstrated~\cite{ourjoumtsev2007generation}. As a result, we are primarily interested in demonstrating scaling properties of the quantum Fisher information with this state, which we do in Section~\ref{chap:gravimetry:sec:QFI:initial:Fock:states}. 

This state evolves under $\hat U(\tau)$ in Eq.~\eqref{chap:gravimetry:eq:time:evolution:operator:explicit} as 
\begin{align} \label{chap:gravimetry:eq:Fock:state:evolved}
\ket{\Psi(\tau)}_{\rm{Fock}} &= \frac{1}{\sqrt{2}}\biggl[ \ket{0} \ket{\mu_{\rm{m}} \,  e^{-i\, \tau} - \tilde{d}_1 \, \eta} \nonumber \\
&+ e^{ i (\tilde{g}_0^2 \,  n^2 - 2 \tilde{g}_0 \, \tilde{d}_1 \, n) \tau } \, e^{ \tilde{g}_0 \,  n \,  (\eta \,  \mu_{\rm{m}} - \eta^* \,  \mu_{\rm{m}}^*)/2}
\ket{n} \ket{\mu_{\rm{m}} e^{-i\, \tau} - ( \tilde{g}_0 n - \tilde{d}_1) \eta}\biggr].
\end{align}

\end{itemize}
Before proceeding with estimation of the gravitational acceleration, we consider the states and their properties at $\tau = 2\pi$.

\subsection{Disentangling of the optics and mechanics} \label{chap:gravimetry:sec:disentangling}

When we examine the states discussed in Section~\ref{chap:gravimetry:initial:and:evolved:states}, we note that the light and mechanics entangle and disentangle periodically. In this Section, we show explicitly that all the initial states we considered in the previous Section disentangle at $\tau =2\pi$. 

We begin with coherent states. Starting with the expression in Eq.~\eqref{chap:gravimetry:eq:fully:evolved:state} for the full evolution, at $\tau = 2\pi $ we find
\begin{equation} \label{chap:gravimetry:evolved:coherent:coherent:state}
\ket{\Psi(2\pi)} = e^{- |\mu_{\rm{c}}|^2/2} \sum_{n = 0}^\infty  \frac{\mu_{\rm{c}}^n}{\sqrt{n!}} e^{2\pi i  (\tilde{g}_0^2 \,  n^2 - 2\,  \tilde{g}_0 \, \tilde{d}_1 \, n ) \, }  \ket{n} \otimes \ket{\mu_{\rm{m}}} \, , 
\end{equation}
which follows since $\eta = 0$ at $\tau = 2\pi$. This implies that, regardless of the values of $\tilde{g}_0$ and $\tilde{d}_1$, the mechanics returns to its original state after one period of the oscillator motion. 

Similarly to  Section~\ref{chap:introduction:initial:states} in Chapter~\ref{chap:introduction}, we here note a few properties of the state in Eq.~\eqref{chap:gravimetry:evolved:coherent:coherent:state}. 
\begin{itemize}
\item The state in Eq.~\eqref{chap:gravimetry:eq:fully:evolved:state} show us that light and mechanics will entangle and disentangle  periodically, with maximum entanglement occurring at $\tau = \pi$. See the investigation into the linear entropy in the following Section. 
\item At $\tau = 2\pi$, the oscillator state $\ket{\phi_n(\tau)}$ returns to the original mechanical coherent state $\ket{\mu_{\rm{m}}}$, regardless of the values of $\tilde{g}_0, \tilde{d}_1$ and $\mu_{\rm{m}}$, and therefore by extension a thermal state also returns to its initial state because it will undergo the same compact evolution. This means that the initial oscillator state does not impact the fundamental sensitivity of this scheme. 
\item Since the cavity state is completely disentangled from the oscillator at $\tau = 2\pi $, all information about $\tilde{d}_1$ is transferred to the phase of the cavity state.  As a result, any measurement scheme only needs to consider the cavity state after one oscillation period, meaning that direct or indirect access to the oscillator state is not required. 
\item The cavity state also returns to its original state $\ket{\mu_{\rm{c}}}$  at $\tau = 2\pi$  if $\tilde{g}_0$ and $\tilde{d}_1$ are integer values. If this is the case, the phases $e^{2 \pi \, i \, \tilde{g}_0^2 \, n^2}$ and $e^{- 4\pi \, i \, \tilde{g}_0 \, \tilde{d}_1 \, n}$ both become unity. 
\end{itemize}

Next, we proceed with the thermal state in Eq.~\eqref{chap:gravimetry:evolved:coherent:thermal:state}.  At $\tau = 2\pi$, the state evolves to 
\begin{align} \label{chap:gravimetry:evolved:coherent:thermal:state:2pi}
\hat \rho_{\rm{th}}(\tau)=  & \,  \sum_{n,n'} \frac{\mu_{\rm{c}}^n (\mu_{\rm{c}}^*)^{n'}}{\sqrt{n! n'!}} e^{2 \pi \, i \,  [\tilde{g}_0^2 (n^2 - n^{\prime 2}) - 2 \tilde{g}_0 \, \tilde{d}_1(n-n')]} \ket{n}\bra{n'}  \sum_{m=0}^\infty \frac{\tanh^{2m} r_T}{\cosh^2 r_T }   \ket{m}\bra{m} \, , 
\end{align}
since again $\eta = 0$ at $\tau = 2\pi$. Thus the mechanics has returned to its original state, is  completely disentangled from the light. 

In fact, the optical state in Eq.~\eqref{chap:gravimetry:evolved:coherent:thermal:state:2pi} is the same as that in Eq.~\eqref{chap:gravimetry:evolved:coherent:coherent:state}. This means that the thermal state of the mechanics does not impact the optical state at $\tau = 2\pi$, which implies that the mechanics does not have to be cooled to its ground state and that we can focus on the cavity state only. The disentangling of the light and mechanics also implies that an analysis of initially coherent states of the light and mechanics suffices to draw general conclusions about the sensitivity of the system at this time. 

Finally, we briefly consider the initial Fock state of the optics and the thermal state of the mechanics in Eq.~\eqref{chap:gravimetry:eq:Fock:state:evolved}. At $\tau = 2\pi$, the state becomes
\begin{align} \label{chap:gravimetry:eq:Fock:state:evolved:2pi}
\ket{\Psi(2\pi)}_{\rm{Fock}} &= \frac{1}{\sqrt{2}}\left(\ket{0}+ e^{2 \pi i (\tilde{g}_0^2 \,  n^2 - 2 \tilde{g}_0 \, \tilde{d}_1 \, n)  }
\ket{n} \right) \otimes \ket{\mu_{\rm{m}}} \, .
\end{align}
As for the other states considered above, the optics and mechanics disentangle completely at $\tau = 2\pi$, which means that we can focus on the cavity state only. 

The fact that the states disentangle implies a significant advantage for an experimental implementation, as measuring the oscillator state is generally difficult. This convenient property arises from the interferometric properties of the oscillator; its quantum nature allows it to acts as an interferometer to ensure that any initial thermal noise is removed from the cavity field, and thereby our scheme does not require cooling of the oscillator to a pure ground state.  Note, however, that decoherence ensuing from damping to the oscillator motion during the state evolution will adversely affect the final measurement sensitivity and cause the oscillator state to grow increasingly mixed. We will not consider this kind of decoherence in this Chapter, and instead assume that the mechanical element remains coherent over one oscillation period. 

\subsection{Quadratures and linear entropy of coherent states}

\begin{figure*}[t!]
\subfloat[ \label{chap:gravimery:fig:op:d1:1}]{%
  \includegraphics[width=.25\linewidth, trim = 00mm 0mm 0mm 0mm]{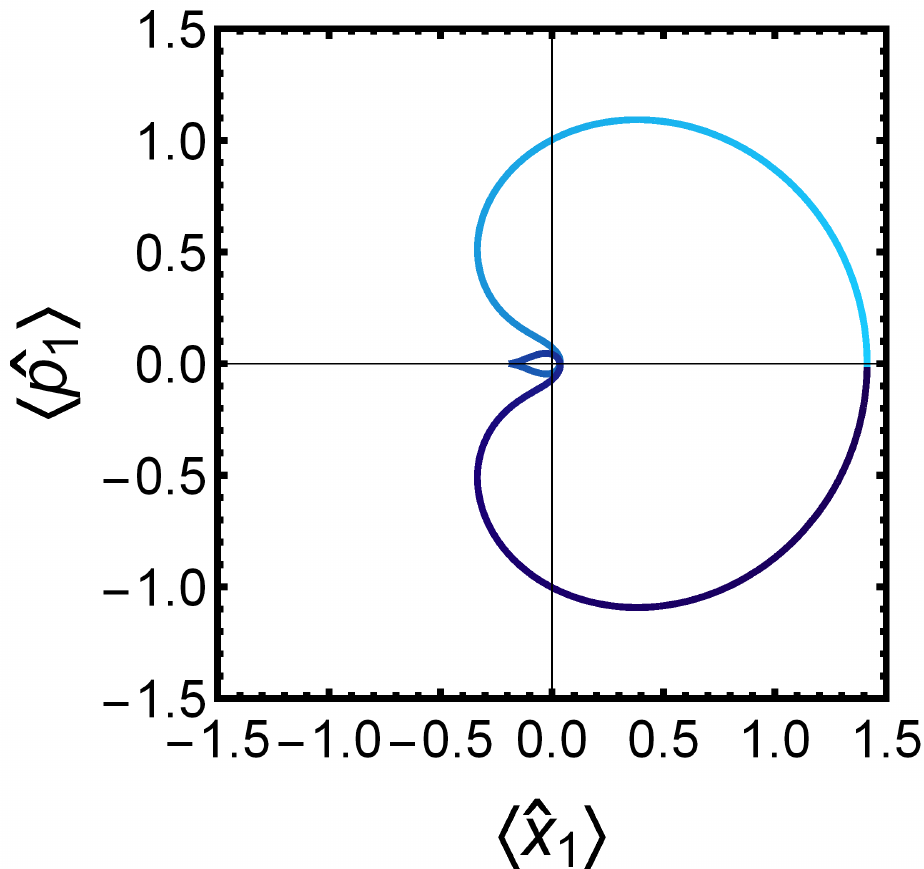}%
}\hfill
\subfloat[ \label{chap:gravimery:fig:op:d1:2}]{%
  \includegraphics[width=.25\linewidth, trim = 00mm 0mm 0mm 0mm]{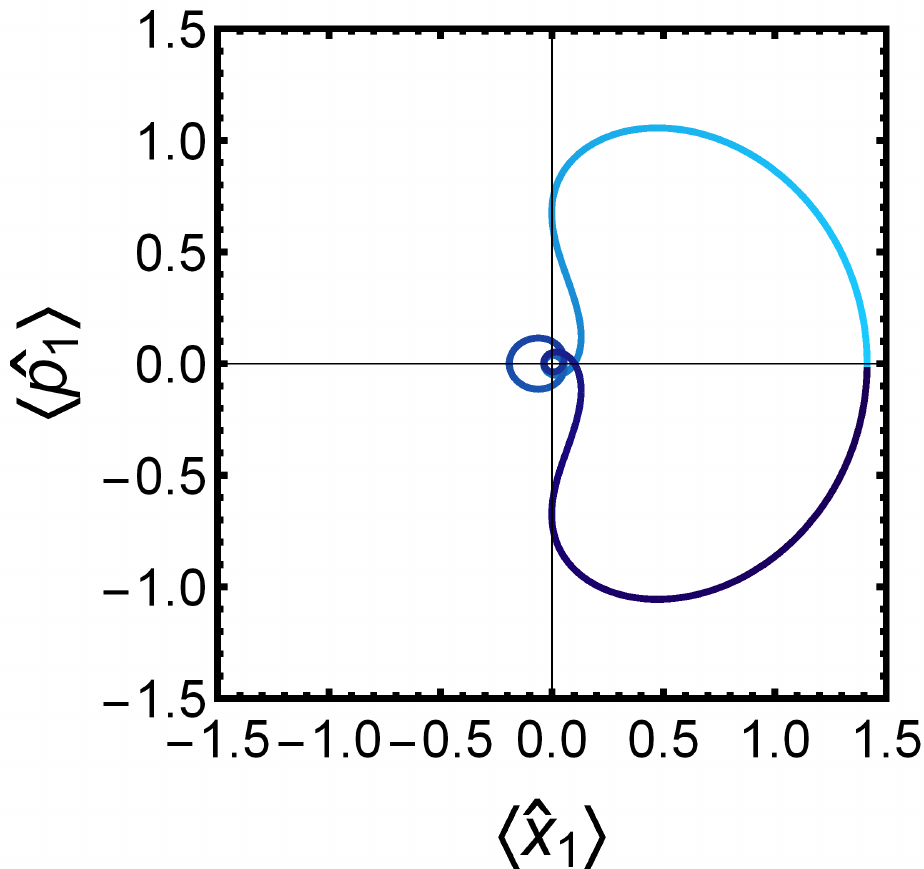}%
}\hfill
\subfloat[
\label{chap:gravimery:fig:op:d1:3}]{%
  \includegraphics[width=.25\linewidth, trim = 00mm 0mm 0mm 0mm]{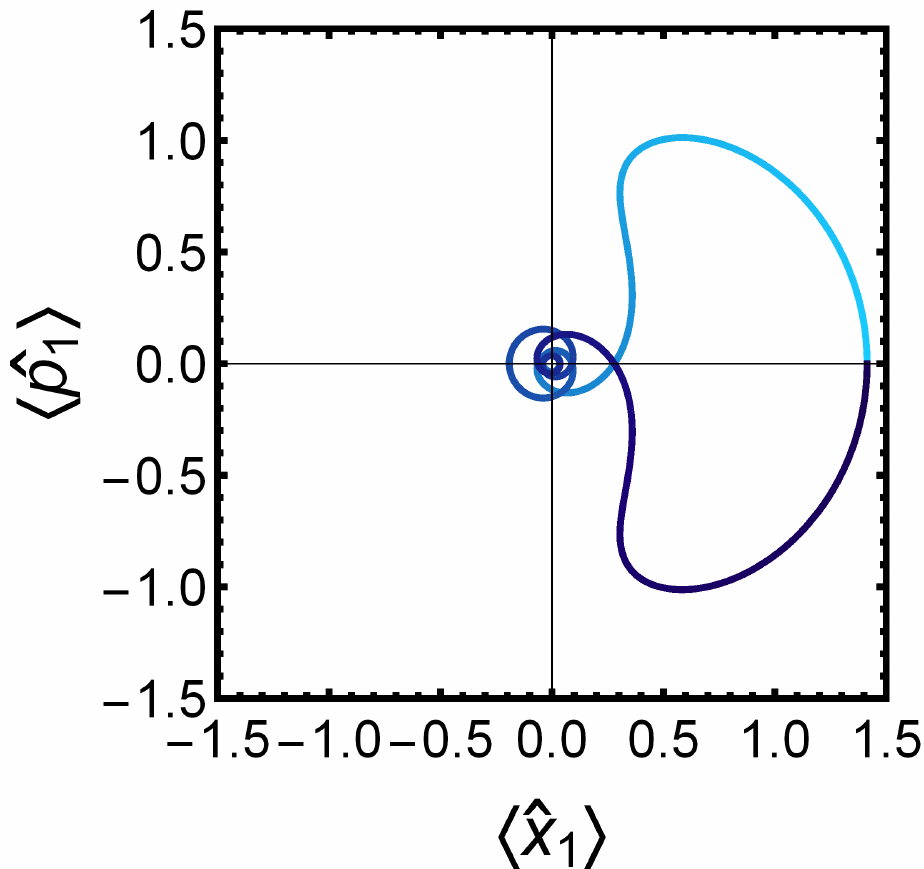}%
}\hfill
\subfloat[
\label{chap:gravimery:fig:op:d1:4}]{%
  \includegraphics[width=.25\linewidth, trim = 00mm 0mm 0mm 0mm]{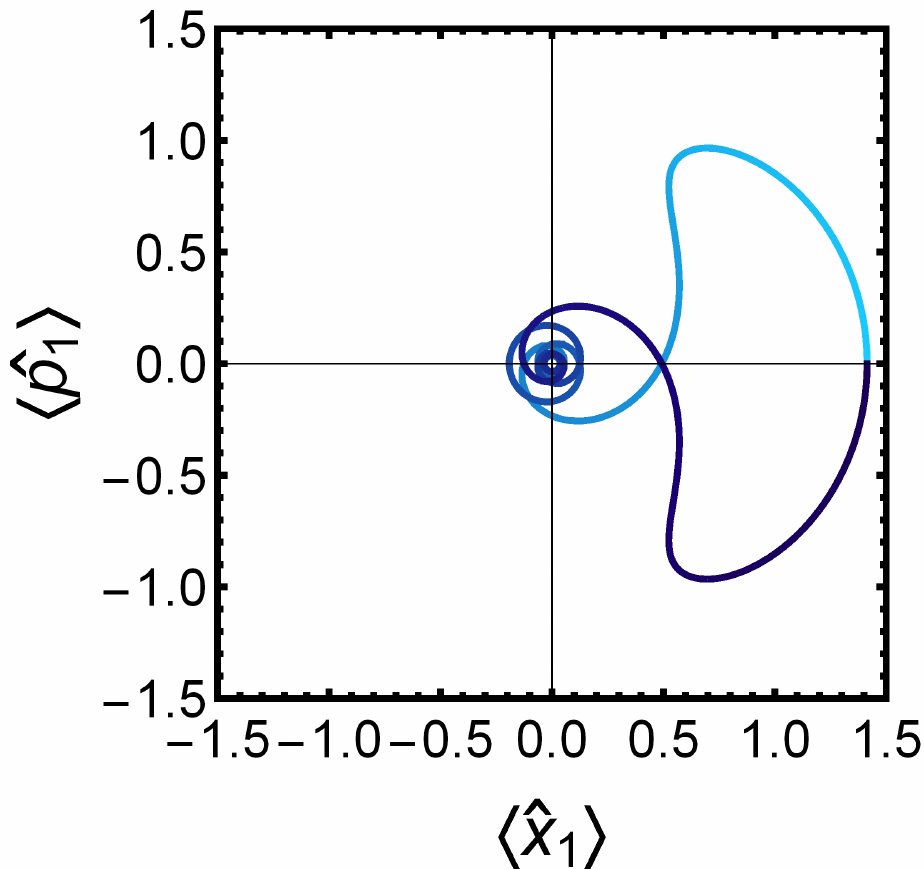}%
}\hfill
\subfloat[
\label{chap:gravimery:fig:mech:d1:1}]{%
  \includegraphics[width=.25\linewidth, trim = 00mm 0mm 0mm 0mm]{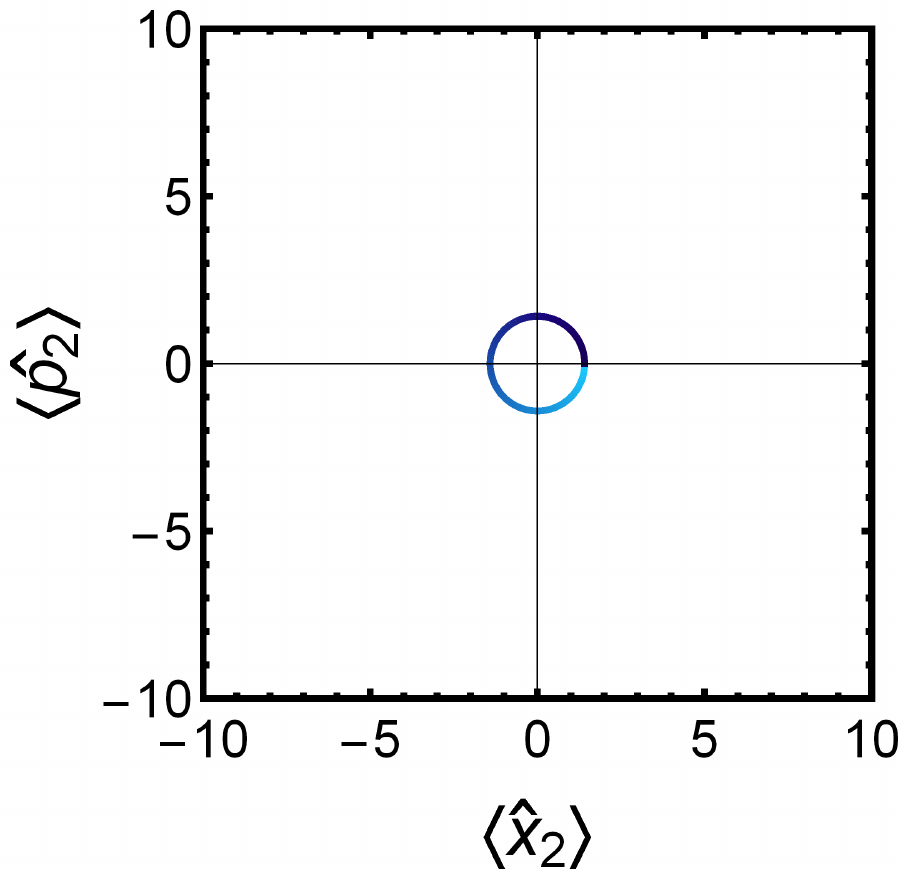}%
}\hfill
\subfloat[
\label{chap:gravimery:fig:mech:d1:2}]{%
  \includegraphics[width=.25\linewidth, trim = 00mm 0mm 0mm 0mm]{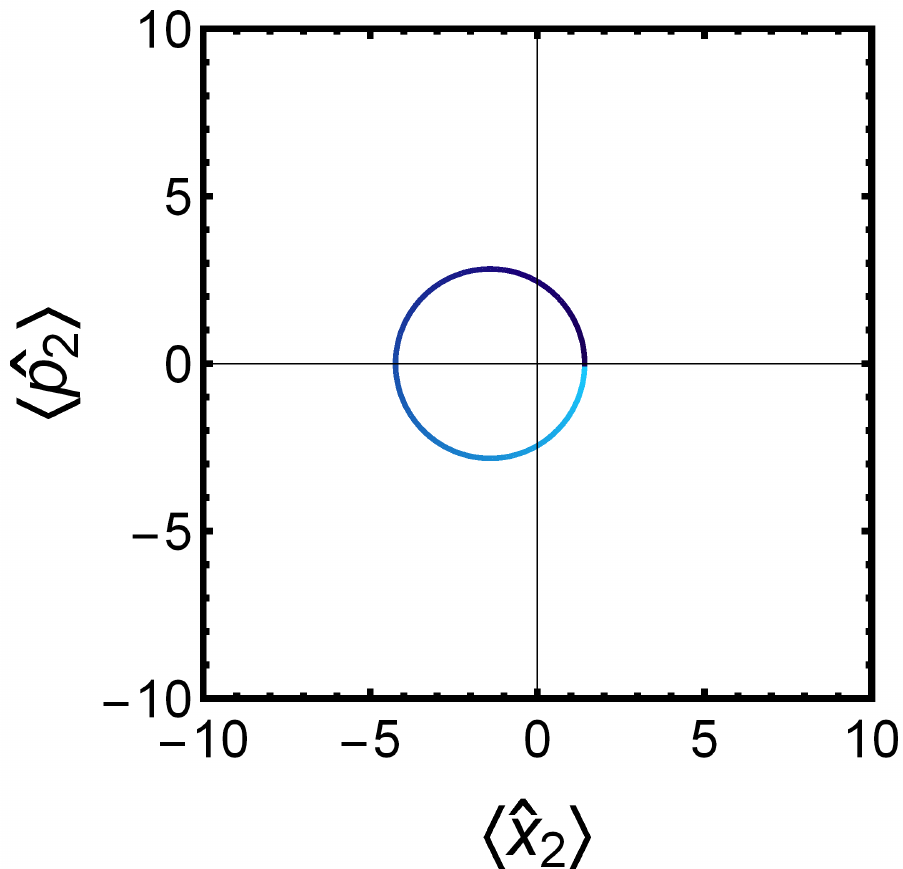}%
}\hfill
\subfloat[
\label{chap:gravimery:fig:mech:d1:3}]{%
  \includegraphics[width=.25\linewidth, trim = 00mm 0mm 0mm 0mm]{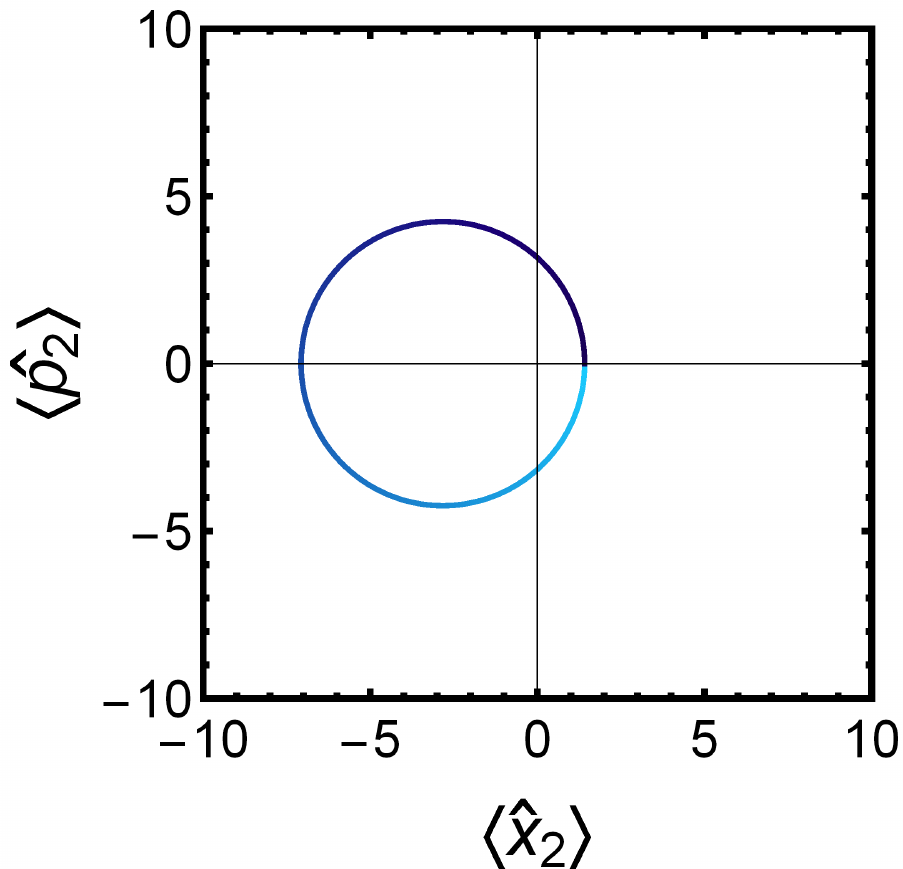}%
}\hfill
\subfloat[
\label{chap:gravimery:fig:mech:d1:4}]{%
  \includegraphics[width=.25\linewidth, trim = 00mm 0mm 0mm 0mm]{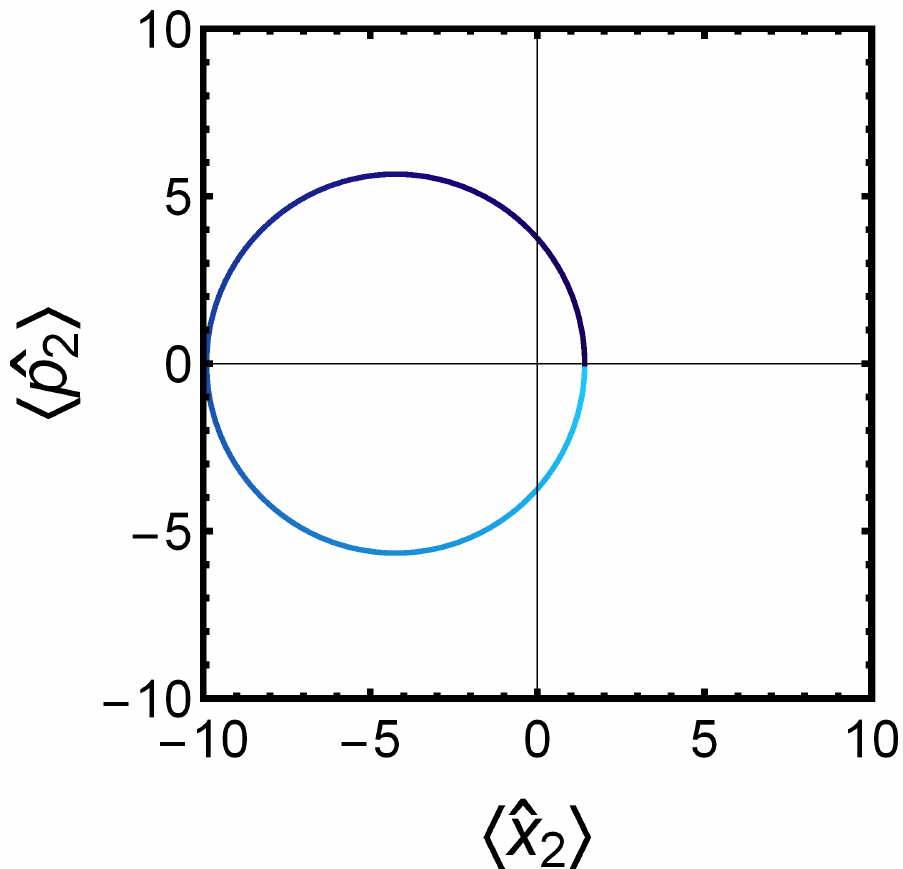}%
}\hfill
\caption[Optical and mechanical quadratures for a mechanical displacement]{Optical and mechanical quadratures for a mechanical displacement. All plots use the parameters $\tilde{g}_0 = \mu_{\rm{c}}= \mu_{\rm{m}} =  1$.  The line starts as light-blue at $\tau = 0$ and ends as dark blue at $\tau = 2\pi$. For $\tilde{d}_1 = 0$, the optical quadratures can be found in Figure~\ref{chap:non:Gaussianity:squeezing:fig:constant:squeezing:quadratures}, and the mechanical quadratures read $\braket{\hat x_2} = \sqrt{2}$ and $\braket{\hat p_2} = 0$ for all values of $\tau$. 
The first row shows the optical quadratures $\braket{\hat x_1}$ and $\braket{\hat p_1}$ for \textbf{(a)} $\tilde{d}_1 = 1$, \textbf{(b)} $\tilde{d}_1 = 2$, \textbf{(c)} $\tilde{d}_1 = 3$, and \textbf{(d)} $\tilde{d}_1 = 4$. The optical quadrature becomes increasingly complex as $\tilde{d}_1$ increases. The second row shows the mechanical quadratures $\braket{\hat x_2} $ and $\braket{\hat p_2} $ for \textbf{(e)} $\tilde{d}_1 = 1$, \textbf{(f)} $\tilde{d}_1 = 2$, \textbf{(g)} $\tilde{d}_1 = 3$, and \textbf{(h)} $\tilde{d}_1 = 4$. Since $\tilde{d}_1 \bigl( \hat b^\dag + \hat b \bigr)$ is a mechanical driving term, the mechanical quadratures become increasingly displaced as $\tilde{d}_1$ increases. }
\label{chap:gravimetry:fig:quadratures}
\end{figure*}

To better understand and picture the state dynamics and evolution, we examine the optical and mechanical quadratures for an initially coherent state. We also consider the linear entropy of the state, which is a simple measure of the entanglement between the light and mechanics. 

Given the evolved state from an initially coherent state in the optics and mechanics in Eq.~\eqref{chap:gravimetry:eq:fully:evolved:state}, the  traced-out cavity state is given by 
\begin{align} \label{chap:gravimetry:eq:traced:out:cavity:state}
\hat \rho_{\mathrm{c}}(\tau) = &e^{-|\mu_{\rm{c}}|^2} \sum_{n, n^\prime}^\infty \biggl[  \frac{\mu_{\rm{c}}^n(\mu_{\rm{c}}^*)^{n^\prime}}{\sqrt{n!n^\prime !}} e^{i \left( \tilde{g}_0^2(n^2 - n^{\prime 2}) - 2\, \tilde{g}_0 \, \tilde{d}_1(n-n') \right)(\tau - \sin\tau)}  \nonumber \\
&\quad\quad\quad\quad\quad\times e^{\tilde{g}_0(n-n^\prime)\left( \eta \, \mu_{\rm{m}} - \eta^* \mu_{\rm{m}}^* \right)/2}\nonumber \\
&\quad\quad\quad\quad\quad\times e^{- |\phi_{n}|^2/2 -  |\phi_{n^\prime}|^2 / 2 + \phi^*_{n^\prime} \phi_n} \ket{n} \bra{n^\prime} \biggr] \, ,
\end{align}
where the coherent state parameter $\phi_n(\tau)$ was given in Eq.~\eqref{chap:gravimetry:eq:mechanical:coherent:state:parameter}, and the traced-out mechanical state is given by
\begin{align}\label{chap:gravimetry:traced:out:mechanics}
\hat{\rho}_{\text{Mech}}(\tau) &=   e^{- |\mu_{\rm{c}}|^2}\,\sum_n \frac{|\mu_{\rm{c}}|^{2\,n}}{n!} 
\ket{\phi_n(\tau)} \bra{\phi_n(\tau)}.
\end{align}
For decoherence-free evolution, the trajectories traced out by the system in phase space are given by $\hat{x}_{\rm{c}} = ( \hat a^\dagger + \hat a)/\sqrt{2}$ and $\hat{p}_{\rm{c}} = i (\hat a^\dagger - \hat a )/\sqrt{2}$, and similarly for $\hat x_{\rm{m}} = (\hat b^\dag + \hat b )/\sqrt{2}$ and $\hat p_{\rm{m}} = i (\hat b^\dag - \hat b )/\sqrt{2}$, The expressions for the optical and mechanical quadratures given two initially coherent states are given explicitly in Eqs. \eqref{app:exp:values:eq:optical:quadratures} and \eqref{app:exp:values:eq:mechanical:quadratures} in Appendix~\ref{app:exp:values}, respectively.  

We plot the result for different values of $\tilde{d}_1$ in Figure~\ref{chap:gravimetry:fig:quadratures}. The optical quadrature for $\tilde{d}_1 = 0$ can be found in Figure~\ref{chap:non:Gaussianity:squeezing:fig:constant:squeezing:quadratures}, and the mechanics yield $\braket{\hat x_2}= \sqrt{2}$ and $\braket{\hat p_2} = 0$. The parameters used for the plot are $\tilde{g}_0 = \mu_{\rm{c}} = \mu_{\rm{m}} = 1$. The subplots in the first row show the optical quadratures for $\tilde{d}_1 = 1$ in Figure~\ref{chap:gravimery:fig:op:d1:1}, $\tilde{d}_1 = 2$ in Figure~\ref{chap:gravimery:fig:op:d1:1}, $\tilde{d}_1 = 3$ in Figure~\ref{chap:gravimery:fig:op:d1:3}, and $\tilde{d}_1 = 4$ in Figure~\ref{chap:gravimery:fig:op:d1:4}. As $\tilde{d}_1$ increases, the quadratures grow increasingly complex. 
The second row of subplots show the mechanical quadrtures for $\tilde{d}_1 = 1$ in Figure~\ref{chap:gravimery:fig:mech:d1:1}, $\tilde{d}_1 = 2$ in Figure~\ref{chap:gravimery:fig:mech:d1:2}, $\tilde{d}_1 = 3$ in Figure~\ref{chap:gravimery:fig:mech:d1:3}, and $\tilde{d}_1 = 4$ in Figure~\ref{chap:gravimery:fig:mech:d1:4}. The quadrature becomes increasingly displaced, which follows from the fact that the term $\tilde{d}_1 \bigl( \hat b^\dag + \hat b \bigr)$ is a mechanical driving term that strongly affects the mechanical state.

\begin{figure*}[t!]
\centering
\subfloat[ \label{chap:gravimetry:subfig:entropy:pure}]{%
  \includegraphics[width=0.45\linewidth, trim = 0mm 0mm 0mm 0mm]{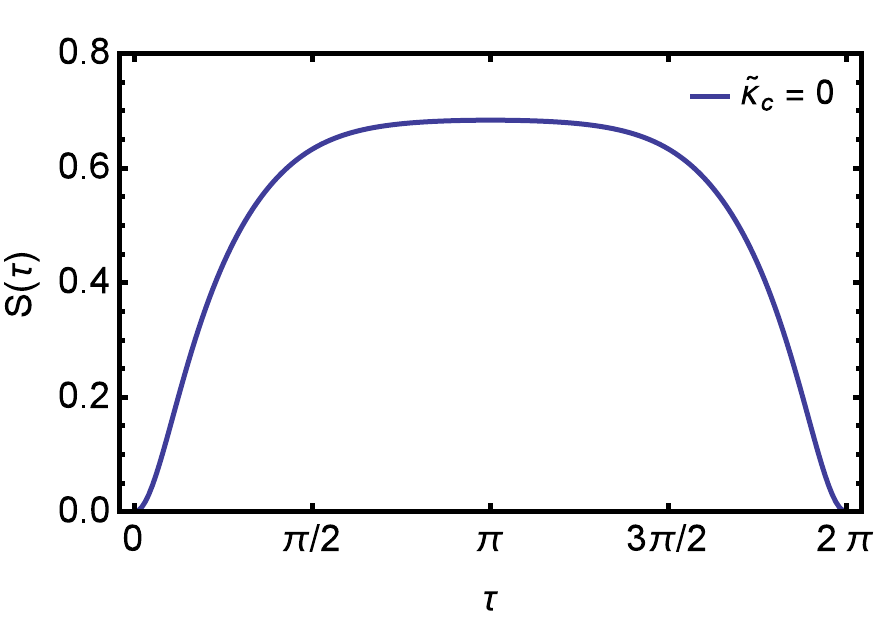}%
}\hfill
\subfloat[ \label{chap:gravimetry:subfig:entropy:mixed}]{%
  \includegraphics[width=0.45\linewidth, trim = 0mm 0mm 0mm 0mm]{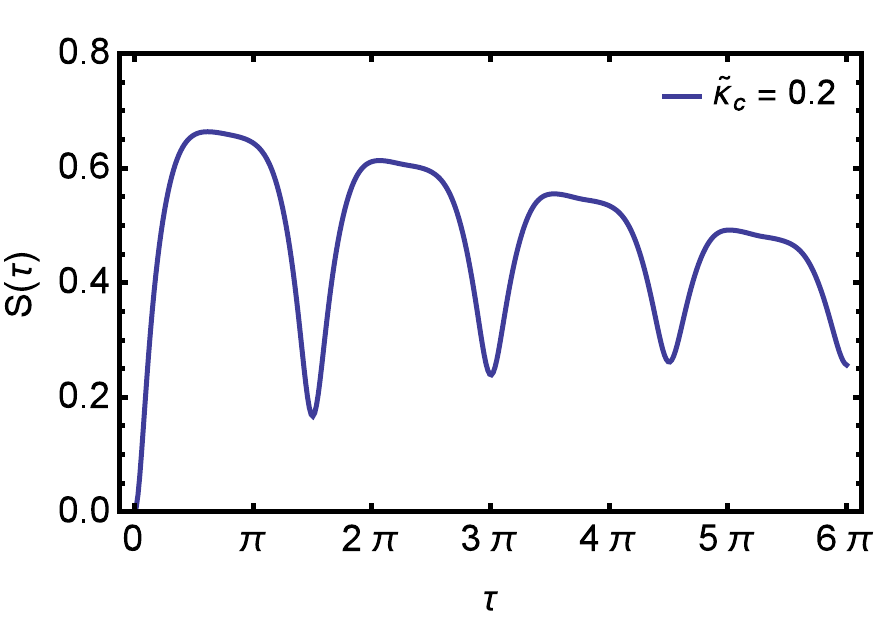}%
}\hfill
\caption[Linear entropy for an optomechanical state given closed and open dynamics]{Linear entropy for an optomechanical state given closed and open dynamics.}
\label{chap:gravimetry:linear:entropy}
\end{figure*}

To better see how the system entangles and disentangles, we compute the linear entropy $S(\tau)$ for the traced out cavity state $\hat \rho_{\mathrm{c}}(\tau)$ in Eq.~\eqref{chap:gravimetry:eq:traced:out:cavity:state}. The linear entropy is defined as 
\begin{equation}
S(\tau) =  1 - \mathrm{Tr}\left[\hat \rho^2_{\mathrm{C}}(\tau)\right] \, . 
\end{equation}
The linear entropy tells us about the entanglement between the cavity and oscillator states. The results can be found  in Figure~\ref{chap:gravimetry:linear:entropy}, where we have plotted $S(\tau)$  for pure state in Figure~\ref{chap:gravimetry:subfig:entropy:pure} and for  states undergoing decoherence with photon dissipation rate in Figure~\ref{chap:gravimetry:subfig:entropy:mixed} for  $\tilde{\kappa}_{\rm{c}} = \kappa_{\rm{c}} / \omega_{\rm{m}}$. To compute the entropy for these dynamics, we solve the Lindblad master equation in Eq.~\eqref{chap:gravimetry:eq:Lindblad} numerically. 

We see that $S(\tau) $ increases until the state is maximally entangled at $\tau = \pi$. While a pure state completely disentangles the light and mechanics at $\tau = 2\pi$ for any values of $\tilde{g}_0$ and $\tilde{d}_1$, a decohering state becomes increasingly mixed and does not return to its original state.

\section{Estimating the gravitational acceleration} \label{chap:gravimetry:estimation:of:g}
We now come to our main results which concern the use of optomechanical systems as gravimeters. The question we wish to answer is: \textit{What is the best fundamental sensitivity $\Delta g$ with which an optomechanical system can measure the gravitational acceleration $g$?} Here, $\Delta g$ denotes the standard deviation of a gravimetric measurement. Our goal is to compute this number. 

We can directly predict $\Delta g$ from the system's dynamics by calculating the quantum Fisher information (QFI) $\mathcal{I}_{g}$ which provides a natural lower bound on the variance $\mathrm{Var}(g)$ of an unknown parameter, in our case $g$. 
This relationship is captured by the Cram\'{e}r-Rao inequality~\cite{cramer1946contribution, wolfowitz1947efficiency, rao1992information} 
\begin{equation} \label{chap:gravimetry:eq:Cramer:Rao:inequality}
\mathrm{Var}( g )\geq \frac{1}{N  \,\mathcal{I}_g} \,.
\end{equation}
where $N$ is the number of measurements performed on the system. Maximising $\mathcal{I}_g$, means that the measurement spread of $g$ is maximised. See Section~\ref{chap:introduction:quantum:metrology} in Chapter~\ref{chap:introduction} for an extended introduction and derivation of the QFI. 

While the QFI provides the fundamental bound on the sensitivity that can be achieved with a system, it does not yield information about the measurement that saturates the bound in Eq.~\eqref{chap:gravimetry:eq:Cramer:Rao:inequality}. In the following two sections, we therefore first compute the QFI for the different initial states discussed in Section~\ref{chap:gravimetry:sec:QFI}. We then consider the classical Fisher information (CFI) for a single measurement of the system in Section~\ref{chap:gravimetry:sec:CFI}.

\section{Quantum Fisher information} \label{chap:gravimetry:sec:QFI}
The quantum Fisher information (QFI), which we denote $\mathcal{I}_{g}$ for estimation of $g$, is computed by optimising over all possible positive-operator valued measures (POVMs) given some initial state $\ket{\Psi(0)}$.   $\mathcal{I}_{g}$ represents the ultimate bound on obtainable information from a system, but it does not reveal which specific measurement is required to achieve it. 

In this Section, we investigate the QFI for the three different initial states that we presented in Section~\ref{chap:gravimetry:initial:and:evolved:states}. 
The QFI for a pure state is given by (see Section~\ref{app:QFI:relation:pure:states} in Appendix~\ref{app:QFI}):
\begin{equation} \label{chap:gravimetry:pure:QFI}
\mathcal{I}_{g} = 4 \left( \braket{\Psi(0)} \hat{\mathcal{H}}_{g}^2 \ket{\Psi(0)} - \braket{\Psi(0) \hat{\mathcal{H}}_g \Psi(0)}^2\right) \, .
\end{equation}
This form of $\hat{\mathcal{H}}_{g}$ is the same as that considered in Example ii in Section~\ref{chap:metrology:sec:example:2} in~\ref{chap:metrology}. It follows from the more general expression for $\hat{\mathcal{H}}_\theta$ in Eq.~\eqref{chap:metrology:eq:mathcalH:with:coefficients}, but since $\mathcal{D}_2 = 0$, we have that the other coefficients are all set to zero: $A = C_{\hat N_a, \pm} = E = F= G = 0$. The expression for 
 $\hat{\mathcal{H}}_{g}$ for estimations of a gravitational acceleration is therefore given by 
\begin{align} \label{chap:gravimetry:eq:mathcalH}
 \mathcal{\hat H}_{g} =  B\,\hat N_a +  C_+\,\hat B_+ + C_-\,\hat B_- \, , 
\end{align}
where the non-zero coefficients $B$ and $C_\pm$ are defined in Eq.~\eqref{chap:metrology:eq:QFI:coefficients}. 

While we are interested in estimating $g$, we defined the dimensionless parameter $\tilde{d}_1$ in the rescaled Hamiltonian in Eq.~\eqref{chap:gravimetry:eq:rescaled:Hamiltonian} and in the evolution operator in Eq.~\eqref{chap:gravimetry:eq:time:evolution:operator:explicit}. It is sometimes easier to first consider estimation with respect to the dimensionless parameter, and then restore dimensions to compute the physical sensitivity. 
For this purpose, we note from the QFI in Eq.~\eqref{chap:gravimetry:pure:QFI} that in applying the chain rule, we are able to easily obtain the sensitivity from $g$ from the dimensionless quantity $\mathcal{I}_{\tilde{d}_1}$.  We find
\begin{equation} \label{chap:gravimetry:QFI:chain:rule}
\mathcal{I}_g = \left( \frac{\partial \tilde{d}_1}{\partial g} \right)^2 \mathcal{I}_{\tilde{d}_1} \, .
\end{equation}
Here, $\mathcal{I}_{\tilde{d}_1}$ is a dimensionless quantity, while the square of the derivative $\partial_g \tilde{d}_1$  provides the correct units of $\mathcal{I}_g$, which are $\si{s^4 m^{-2}}$. In what follows, we focus on deriving an expression for $\mathcal{I}_{\tilde{d}_1}$, and we then specialise to $\mathcal{I}_g$ when needed. 

The first step in computing the QFI is deriving the coefficients $B$ and $C_\pm$ in Eq.~\eqref{chap:gravimetry:eq:mathcalH}. We did so in Eq.~\eqref{chap:metrology:eq:QFI:coefficients} in Chapter~\ref{chap:metrology}, and showed that for estimating a linear displacement, $B$ and $C_\pm$ are given in terms of the $F$-coefficients in Eq.~\eqref{chap:gravimetry:F:coefficients}. For the constant displacement and a constant optomechanical coupling considered in this Chapter, we found the coefficients in Eq.~\eqref{chap:gravimetry:F:coefficients}, the coefficients for the QFI become
\begin{align} \label{chap:gravimetry:QFI:coefficients}
 B &= 2 \, \tilde{g}_0 \left( \sin\tau - \tau \right) \, , \nonumber \\
C_+ &= - \sin \tau \, , \nonumber \\
C_-&= \cos \tau - 1 \, .
\end{align}
The second step in computing the QFI is to consider the expectation value of $\hat{\mathcal{H}}_{\tilde{d}_1}^2$ and $\hat{\mathcal{H}}_{\tilde{d}_1}$ with respect to some initial state. In the following Sections, we consider the three different states introduced in Section~\ref{chap:gravimetry:initial:and:evolved:states}. 

Before we proceed, we note that the QFI of the global system might not seem particularly relevant as the mechanical part of the optomechanical system cannot easily be measured directly. However, we showed in Section~\ref{chap:gravimetry:sec:disentangling} that a number of different initial states disentangle at $\tau = 2\pi$ and the mechanics returned to its original state. As a result, all information about $\tilde{d}_1$ and therefore also $g$ is transferred to the phase of the pure, disentangled cavity state. This is advantageous from an experimental point of view, since the mechanics does not need to be accessed.

\subsection{QFI for initially coherent states}
We begin by considering the initial state in Eq.~\eqref{chap:gravimetry:eq:coherent:coherent}, which is a coherent state of both the optics and mechanics. 
Given this state, it can be shown that (see Section~\ref{app:QFI:coherent:coherent} in Appendix~\ref{app:QFI}), that the QFI for estimating a linear displacement is given by 
\begin{equation}
\mathcal{I}_{\tilde{d}_1}^{(\rm{coh})} = 4 \left( B^2 |\mu_{\rm{c}}|^2 + C_+^2 + C_-^2 \right) \, .
\end{equation}
Using the expression for $B$ and $C_\pm$ for specifically a  constant mechanical displacement in Eq.~\eqref{chap:gravimetry:QFI:coefficients}, we find 
\begin{equation} \label{chap:gravimetry:eq:QFI:coherent:coherent:general}
\mathcal{I}_{\tilde{d}_1} ^{(\rm{coh})} = 16 \, \left( \tilde{g}_0 ^2 \, |\mu_{\rm{c}}|^2  \, \left( \tau - \sin \tau \right)^2 + \frac{1}{2} \left( 1 - \cos \tau \right) \right) \, .
\end{equation}
We are now able to derive a concise expression for the QFI at $\tau = 2\pi$. At this time, the mechanics is fully disentangled from the optical field, and we find
\begin{equation} \label{chap:gravimetry:eq:QFI:coherent:coherent:2pi}
\mathcal{I}_g^{(\rm{coh})}  = \frac{32 \pi^2 \,  \tilde{g}_0^2 \,m \,|\mu_{\rm{c}}|^2 \cos^2{\theta} }{\hbar \omega^3_m}.
\end{equation}
Note that the mass term $m$ in the denominator is canceled by the appearance of $m$ in the coupling constant $\tilde{g}_0$, so that the final accelerometry measurement will be mass-independent. We also note the strong dependence on $\tilde{g}_0$ and $\omega_{\rm{m}}$, and that the expression scales linearly with the number of photons $|\mu_{\rm{c}}|^2$. 

\begin{figure*}[t!]
\centering
\subfloat[ \label{chap:gravimetry:subfig:QFI:coherent:coherent}]{%
  \includegraphics[width=0.49\linewidth, trim = 0mm 0mm 0mm 0mm]{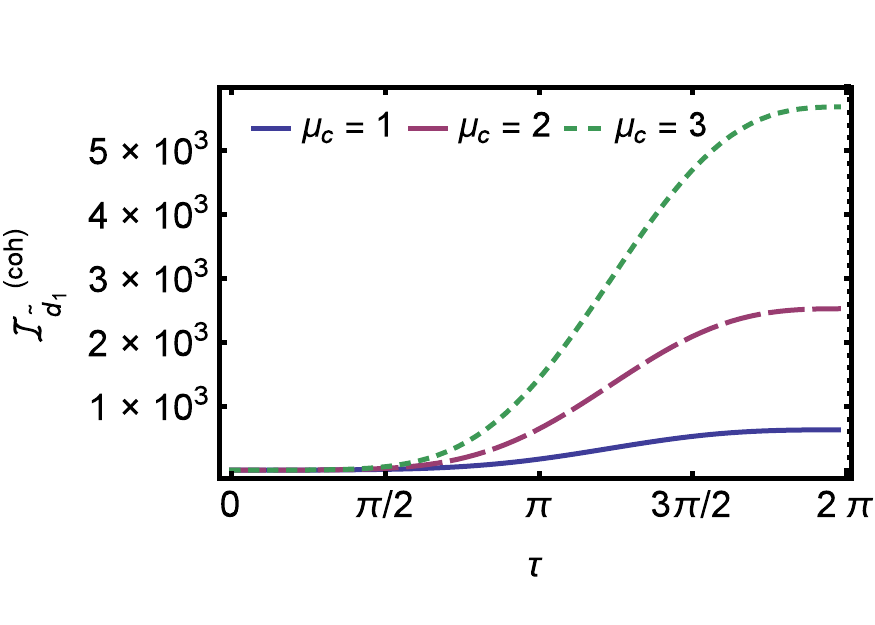}%
}\hfill
\subfloat[ \label{chap:gravimetry:subfig:QFI:coherent:thermal}]{%
  \includegraphics[width=0.5\linewidth, trim = 0mm 1mm 0mm 0mm]{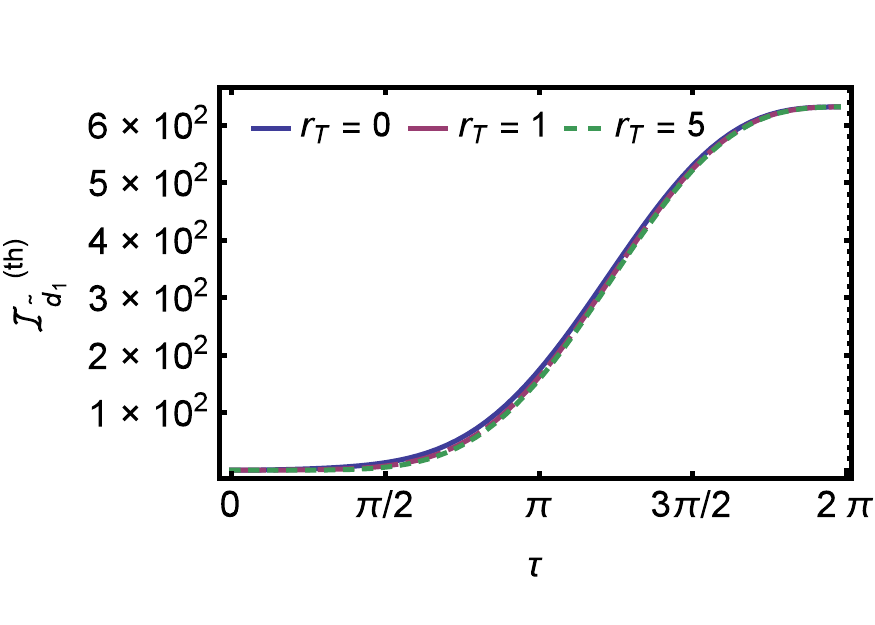}%
}\hfill
\subfloat[ \label{chap:gravimetry:subfig:QFI:Fock:coherent}]{%
  \includegraphics[width=0.5\linewidth, trim = 0mm 0mm 0mm 0mm]{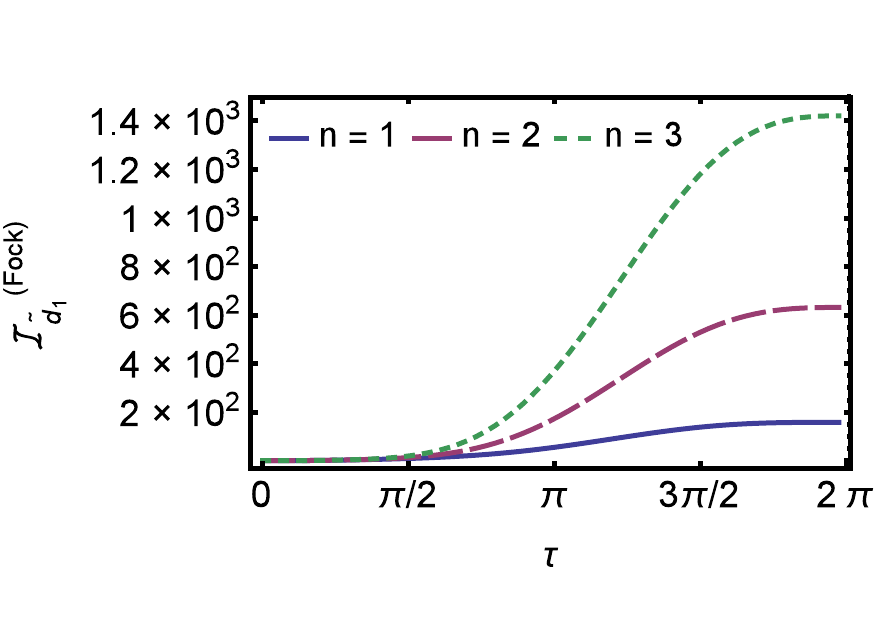}%
}\hfill
\caption[QFI for estimating a constant gravitational acceleration with different initial states]{QFI for estimating a constant gravitational acceleration with different initial states. We set $\tilde{g}_0 = 1$ for all plots. \textbf{(a)} shows the QFI $\mathcal{I}_{\tilde{d}_1}^{(\rm{coh})}$  for two initially coherent states of the optics and mechanics as a function of time for the mechanical coherent state parameter $\mu_{\rm{m}} = 1$. As the value of $\mu_{\rm{c}}$ increases, the QFI increases as well. \textbf{(b)} shows the QFI $\mathcal{I}_{\tilde{d}_1}^{(\rm{th})}$ as a function of time for an initial coherent state in the optics and a thermal state of the mechanics. The optical coherent state parameter is $\mu_{\rm{c}} = 1$, and the QFI is shown for different values of the temperature parameter $r_T$. As $r_T$ increases, the QFI is only somewhat negatively affected. Finally, \textbf{(c)} shows the QFI $\mathcal{I}_{\tilde{d}_1}^{(\rm{Fock})}$ as a function of time for an initial Fock state superposition for the optics and a coherent state of the mechanics with parameter $\mu_{\rm{m}} = 1$. The QFI is shown for different numbers of initial photons $n$, where the scaling is stronger than that for coherent states since $n = |\mu_{\rm{c}}|^2$. }
\label{chap:gravimetry:fig:QFI:different:initial:states}
\end{figure*}

We proceed to plot the QFI for general times $\tau$ in Figure~\ref{chap:gravimetry:fig:QFI:different:initial:states}. We use the following parameters: $|\mu_{\rm{c}}|^2 = 1$ and set $\tilde{g}_0 = \tilde{d}_1 = 1$. The QFI for two coherent states resulting $\mathcal{I}_{\tilde{d}_1}^{(\rm{coh})}$ as a function $\tau $ for different values of $|\mu_{\rm{c}}|^2$ can be found in Figure~\ref{chap:gravimetry:subfig:QFI:coherent:coherent}. 
We note that $\mathcal{I}_{\tilde{d}_1}^{(\rm{coh})} $ reaches its maximum value at $\tau = 2\pi$, which means that $\mathcal{I}_g^{(\rm{coh})}$ in Eq.~\eqref{chap:gravimetry:eq:QFI:coherent:coherent:2pi} returns the largest possible value during one oscillation period for any choice of system.

\subsection{QFI for an initial thermal mechanical states}
Next, we compute the QFI for thermal states, which is a special case of the modulated coupling considered in  Section~\ref{chap:metrology:sec:example:2} in  Chapter~\ref{chap:metrology}. There, we found that the QFI for initially thermal states of the mechanics such as that in Eq.~\eqref{chap:gravimetry:initial:state:coherent:thermal} was given by 
\begin{equation}
\mathcal{I}_{\tilde{d}_1}^{(\rm{th})} = 4 \left( B^2 |\mu_{\rm{c}}|^2 + \frac{1}{\cosh 2 r_T} \left( C_+ ^2 + C_-^2 \right) \right) \, . 
\end{equation}
When inserting the expression for the coefficients form Eq.~\eqref{chap:gravimetry:QFI:coefficients}, we find 
\begin{equation} \label{chap:gravimetry:QFI:coherent:thermal:explicit}
\mathcal{I}_{\tilde{d}_1}^{(\rm{th})}  = 16 \left( \tilde{g}_0^2 \, |\mu_{\rm{c}}|^2 \, \left( \tau - \sin \tau \right)^2  + \frac{1}{\cosh 2 r_T} \sin^2 (\tau/2) \right) \, .
\end{equation}
As already noted in Section~\ref{chap:metrology:sec:example:2} in Chapter~\ref{chap:metrology}, the second term in Eq.~\eqref{chap:gravimetry:QFI:coherent:thermal:explicit} oscillates, while the first term scales with $\tau^2$. This means that a higher temperature decreases the sensitivity of the system. 

We also note that at $\tau = 2\pi$, the second term of $\mathcal{I}_{\tilde{d}_1}^{(\rm{th})}$ is zero, which means that  $\mathcal{I}_{g}^{(\rm{th})} = \mathcal{I}_{g}^{(\rm{coh})}$ at this time. 
 This implies that the initial temperature in the mechanical state does not affect the sensitivity of the system, which in turn means that the system does not have to be cooled to the ground state. This is extremely beneficial from an experimental point of view. 

We plot $\mathcal{I}_{\tilde{d}_1}^{(\rm{th})}$ as a function of $\tau$ for different $r_T$ in Figure~\ref{chap:gravimetry:subfig:QFI:coherent:coherent} for the parameter $\mu_{\rm{c}} = \tilde{g}_0 = 1$. As can be seen, an increasing temperature does not affect the QFI particularly negatively as $\tau$ increases.

\subsection{QFI for initial Fock states} \label{chap:gravimetry:sec:QFI:initial:Fock:states}
The QFI for initially coherent states in the optics and mechanics at $\tau = 2\pi$ in Eq.~\eqref{chap:gravimetry:eq:QFI:coherent:coherent:2pi} scales linearly with the number of photons $|\mu_{\rm{c}}|^2$. As a result, it does not allow us to saturate the Heisenberg limit (see Section~\ref{chap:introduction:quantum:metrology:quantum:advantage} in Chapter~\ref{chap:introduction}). To achieve a quantum speedup in the sensitivity $\Delta g$, one must instead show that the QFI scales with the number of photons squared, that is $|\mu_{\rm{c}}|^4$. This cannot be achieved with initially coherent states. Instead, we investigate the scaling of the QFI given the highly non-classical Fock state in Eq.~\eqref{chap:gravimetry:initial:state:Fock:coherent}. 

The derivation of the QFI for the initial Fock state can be found in Section~\ref{app:QFI:Fock:coherent:states} in Appendix~\ref{app:QFI}. The QFI for estimation of $\tilde{d}_1$ is given by 
\begin{equation}
\mathcal{I}_{\tilde{d}_1}^{(\rm{Fock})} =     n^2 \,  B^2 + 4 \left( C_+^2 + C_-^2 \right) \, , 
\end{equation}
where $n$ is the number of photons. Using the expressions in Eq.~\eqref{chap:gravimetry:QFI:coefficients} for $B$ and $C_\pm$, we find 
\begin{equation} \label{chap:gravimetry:QFI:initial:Fock:state}
\mathcal{I}_{\tilde{d}_1}^{(\rm{Fock})} = 4 \left( \tilde{g}_0 ^2 \, n^2 \, \left( \tau - \sin \tau\right)^2 + \sin^2 (\tau/2) \right) \, . 
\end{equation}
At $\tau = 2\pi$ the expression simplifies. Using the chain rule in Eq.~\eqref{chap:gravimetry:QFI:chain:rule}, we find the following compact expression for the QFI:
\begin{align}
\mathcal{I}_{g}^{(\rm{Fock})}
&= \frac{16   \pi^2  \, m \,  \tilde{g}_0^2  \, n^2}{ \hbar \omega_m^3} \cos^2{\theta} \, , 
\end{align}
where the appearance of $n^2$ indicates that the Standard Quantum Limit has been surpassed. Comparing with our previous results, and setting $n = |\mu_{\rm{c}}|^2$, we find at $\tau = 2\pi$:
\begin{equation}
\mathcal{I}^{(\rm{Fock})}_g = \frac{1}{2} |\mu_{\rm{c}}|^2 \, \mathcal{I}^{(\rm{coh})}_g \, .
\end{equation}
If sufficiently large superpositions of Fock states can be achieved, the scaling of $\mathcal{I}^{(\rm{Fock})}_g $ could potentially be utilised. At the present, however, most cavity state in optomechanical setups are assumed to be coherent states. As a result, the remaining Sections in this Chapter focus on initially coherent states.

\section{Classical Fisher information} \label{chap:gravimetry:sec:CFI}
Let us now consider a specific measurement of $\tilde{d}_1$, which could potentially be performed in the laboratory. In this Section, we focus exclusively on estimation schemes for the initially coherent state in Eq.~\eqref{chap:gravimetry:eq:coherent:coherent}, since we found that the other could to some extent be related to it. 

The classical Fisher information (CFI) $I_{\tilde{d}_1} $ for estimation of the parameter $\tilde{d}_1$ determines the sensitivity of the system given  specific measurement with POVM elements $\{\hat \Pi_x\}$. The CFI $I_{\tilde{d}_1}$ for estimation of $\tilde{d}_1$ is given by the expression
\begin{eqnarray} \label{gravimetry:eq:CFI:definition}
I_{\tilde{d}_1} = \int \mathrm{d} x \, \frac{1}{p(x|\tilde{d}_1) } \left( \frac{\partial p(x|\tilde{d}_1)}{\partial \tilde{d}_1} \right)^2 \, ,
\end{eqnarray}
where $p(x|\tilde{d}_1) = \mathrm{Tr}[\hat \Pi_x \,  \hat \rho_{\tilde{d}_1}]$ is a conditional probability distribution that depends on the values of $\tilde{d}_1$ and the measurement outcome $x$. We also omit the superscript denoting the initial state of the system that we used in the previous section, since we here consider coherent states only.  See Section~\ref{chap:introduction:sec:ClassicalFisher} in Chapter~\ref{chap:introduction} for a discussion and a derivation of the CFI. 

As was the case for the QFI, the chain rule allows us first consider the dimensionless quantity $I_{\tilde{d}_1}$ and relate it to the sensitivity of the system through
\begin{equation}
I_g = \left( \frac{\partial \tilde{d}_1 }{\partial g} \right)^2 I_{\tilde{d}_1} \, ,
\end{equation}
which means that we can focus on $I_{\tilde{d}_1}$
 for general results,  and later specialise to $I_g$. 

\subsection{CFI for homodyne detection} \label{chap:gravimetry:CFI:homodyne:detection}

We now consider a general homodyne measurement on the traced-out cavity state $\rho_{\mathrm{C}}$. For notational convenience, we use a general Hermitian operator 
\begin{equation}
\hat{x}_\lambda =\frac{1}{\sqrt{2}} \left( \hat a \,  e^{- i \lambda} + \hat a^\dagger \, e^{i \lambda } \right) \, , 
\end{equation}
where $\lambda$ denotes a label that rotates between the field quadratures~\cite{barnett2002methods}. 
Any two operators that differ by $\lambda = \pi/2$ form a conjugate pair that satisfies the position-momentum commutator relation. 
In the following, we shall refer to the choices $\lambda=0$ and $\lambda=\pi/2$ as a 'position' and 'momentum' measurement respectively.  

\begin{figure*}[t!]
\centering
\subfloat[ \label{chap:gravimetry:fig:CFI:homodyne:various:g0:position}]{%
  \includegraphics[width=0.49\linewidth, trim = 0mm 0mm 0mm 0mm]{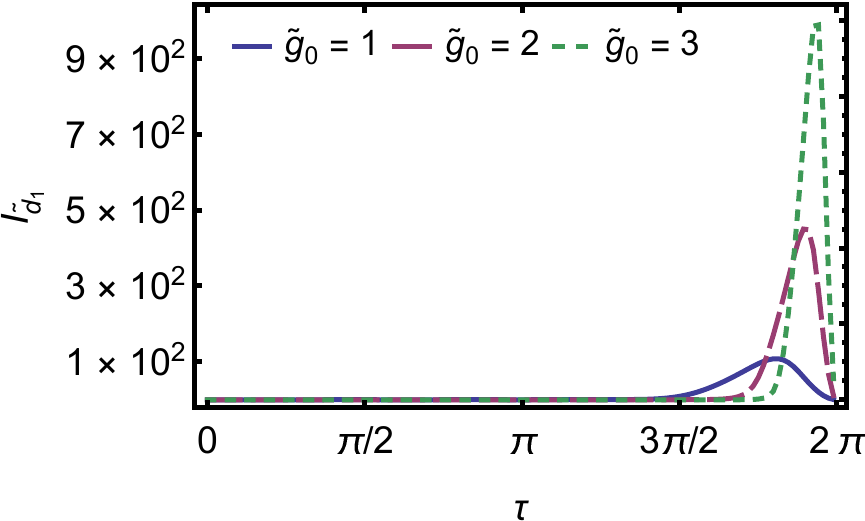}%
}\hfill
\subfloat[ \label{chap:gravimetry:fig:CFI:homodyne:various:g0:momentum}]{%
  \includegraphics[width=0.5\linewidth, trim = 0mm 1mm 0mm 0mm]{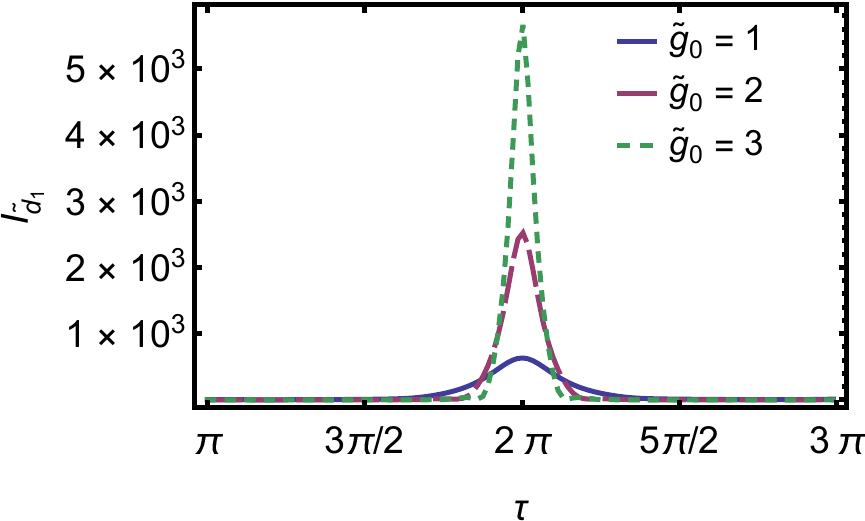}%
}\hfill
\caption[CFI for estimating a constant gravitational acceleration with a homodyne detection scheme for different coupling strengths]{CFI for estimating a constant gravitational acceleration with a homodyne detection scheme for different optomechanical coupling strengths. The parameters are $\tilde{d}_1 = \mu_{\rm{c}} = 1$ and $\mu_{\rm{m}} = 0$. \textbf{(a)} shows $I_{\tilde{d}_1}$ for a position measurement with $\lambda = 0$ as a function of time $\tau$ for different values of $\tilde{g}_0$. The function increase with $\tilde{g}_0$, but the peak grows more narrow and occurs just before $\tau = 2\pi$. \textbf{(b)} shows $I_{\tilde{d}_1}$ for a momentum measurement with $\lambda = \pi/2$ as a function of time $\tau$ for different values of $\tilde{g}_0$. The function peaks at $\tau = 2\pi$ and increases with $\tilde{g}_0$, but again the peak narrows for larger values of $\tilde{g}_0$. }
\label{chap:gravimetry:fig:CFI:homodyne:various:g0}
\end{figure*}

In order to calculate $I_{\tilde{d}_1}$ we must determine the probability distribution $p(x_\lambda| \tilde{d}_1) = \mathrm{Tr}[\ketbra{x_\lambda}  \rho_{\mathrm{c}}(\tilde{d}_1)]$, where $\ketbra{x_\lambda}$ are the eigenstate of $\hat{x}_\lambda$. While the position eigenstates themselves are not proper vectors, we can make use of a standard result from the quantum harmonic oscillator~\cite{barnett2002methods}  
\begin{equation}
\langle n | x_\lambda \rangle = \frac{1}{\pi^{1/4} 2^{n/2} (n!)^{1/2} } \, e^{-x^2_\lambda /2 }  H(x_\lambda)  \, e^{in\lambda} \, ,
\end{equation}
which allows us to write
\begin{align} \label{chap:gravimetry:eq:probability:distribution}
p(x_\lambda| \tilde{d}_1) =  e^{- |\mu_{\rm{c}}|^2} \sum_{n,n'} &\biggl[ \frac{\mu_{\rm{c}}^n(\mu_{\rm{c}}^*)^{n^\prime} }{\sqrt{n!n^\prime!} } e^{i \left( \tilde{g}_0^2(n^2 - n^{\prime2}) - 2  \, \tilde{g}_0  \, \tilde{d}_1(n-n^\prime)\right) (\tau - \sin \tau) } \nonumber \\
&\quad\times  \frac{e^{- x_\lambda^2} }{\pi^{1/2}}  \frac{H_n(x_\lambda ) \,H_{n^\prime}(x_\lambda ) \, e^{- i \lambda  (n-n^\prime)}}{2^{(n+n^\prime)/2} \,\sqrt{n!n^\prime!}}  \nonumber \\
&\quad\times  e^{\left(\tilde{g}_0(n-n^\prime) \right)\left( \eta \, \mu_{\rm{m}} - \eta^* \,  \mu_{\rm{m}}^* \right)/2}\nonumber \\
&\quad\times  e^{- |\phi_n|^2/2 - |\phi_{n^\prime}|^2/2 + \phi_{n^\prime}^* \phi_n } \biggr] \, ,
\end{align}
where $H_n(x)$ are the Hermite polynomials of order $n$. We differentiate $p(\tilde{d}_1 |x)$ with respect to $\tilde{d}_1$ to find
\begin{align} \label{chap:gravimetry:eq:derivative:of:p}
\partial_{\tilde{d}_1} p(x | \tilde{d}_1) =   -e^{- |\mu_{\rm{c}}|^2} \sum_{n,n'} \biggl[ &\left( 2 i  \tilde{g}_0 (n - l )(\tau - \sin \tau) \right) \frac{\mu_{\rm{c}}^n(\mu_{\rm{c}}^*)^{n^\prime} }{{n!n^\prime!} }  \nonumber \\
&\times e^{i \left( \tilde{g}_0^2(n^2 - n^{\prime2}) - 2  \, \tilde{g}_0  \, \tilde{d}_1(n-n^\prime)\right) ( \tau - \sin\tau )} \nonumber \\
&\times  \frac{e^{- x_\lambda^2} }{\pi^{1/2}}  \frac{H_n(x_\lambda ) \,H_{n^\prime}(x_\lambda ) \, e^{- i \lambda  (n-n^\prime)}}{2^{(n+n^\prime)/2} \,}  \nonumber \\
&\times e^{\tilde{g}_0(n-n^\prime) \left( \eta \, \mu_{\rm{m}} - \eta^* \,  \mu_{\rm{m}}^* \right)/2}\nonumber \\
&\times e^{- |\phi_n|^2/2 - |\phi_{n^\prime}|^2/2 + \phi_{n^\prime}^* \phi_n } \biggr] \, .
\end{align}
The derivative follows by noting that the quantity $- |\phi_n|^2/2 - |\phi_{n'}|^2/2 + \phi_{n'}^* \phi_n $ is actually independent of $\tilde{d}_1$. 
Inserting these expressions into Eq.~\eqref{gravimetry:eq:CFI:definition}, the CFI then becomes:
\begin{align} \label{chap:gravimetry:eq:homodyne:CFI}
I_{\tilde{d}_1} =&- 4 \, \tilde{g}_0^2  \left( \tau - \sin \tau \right)^2 \, e^{-|\mu_{\rm{c}}|^2} \int \mathrm{d}x_\lambda \frac{\left[ \sum_{n,n^\prime} (n-n') \, c_{n,n^\prime} \, d_{n,n^\prime}(x_\lambda) \, f_{n,n^\prime}\right]^2}{\sum_{n,n^\prime} c_{n,n^\prime} \, d_{n,n^\prime}(x_\lambda) \, f_{n,n^\prime} },
\end{align}
where
\begin{subequations}
\begin{align}
c_{n,n^\prime} &= \frac{(\mu_{\rm{c}}^*)^{n^\prime} \mu_{\rm{c}}^n }{\sqrt{n!n^\prime!}} \, e^{i \left[\tilde{g}_0^2 (n^2 - n^{\prime2} ) - 2 \, \tilde{g}_0 \, \tilde{d}_1 (n-n^\prime) \right] (\tau - \sin \tau)}, \\
d_{n,n^\prime}(x_\lambda) &= \frac{e^{- x_\lambda^2} }{\pi^{1/2}}  \frac{H_n(x_\lambda )\,  H_{n^\prime}(x_\lambda ) \, e^{- i \lambda  (n-n^\prime)}}{2^{(n+n^\prime)/2} \, \sqrt{n!n^\prime!}}, \\
f_{n,n^\prime} &=   e^{\tilde{g}_0 \, (n-n^\prime) \left( \eta \, \mu_{\rm{m}} - \eta^*  \, \mu_{\rm{m}}^* \right)/2 } \, e^{- |\phi_{n^\prime}|^2/2 - |\phi_n|^2/2 + \phi_{n^\prime}^* \phi_n } \, .
\end{align} 
\end{subequations}

\noindent It is difficult to simplify the CFI in Eq.~\eqref{chap:gravimetry:eq:homodyne:CFI} further. We therefore evaluate the sum and integral in Eq.~\eqref{chap:gravimetry:eq:homodyne:CFI} numerically to understand how the CFI changes with the time $\tau$ and the values of the parameter $\tilde{g}_0$ and $\tilde{d}_1$. There are two measurements that we can perform: $\lambda = 0$, which corresponds to a `position' measurement of the operator $\hat x_1= (\hat a^\dag + \hat a )/\sqrt{2}$, and $\lambda = \pi/2$, which corresponds to a `momentum' measurement of the operator $\hat p_1 = i ( \hat a^\dag - \hat a )/\sqrt{2}$. 

\begin{figure*}[t!]
\centering
\subfloat[ \label{chap:gravimetry:fig:CFI:homodyne:various:d1:position}]{%
  \includegraphics[width=0.45\linewidth, trim = 0mm 0mm 0mm 0mm]{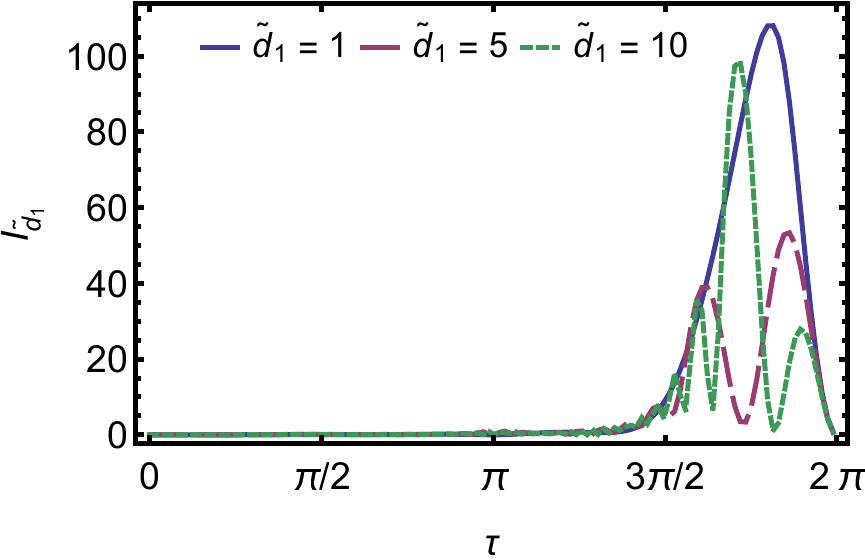}%
}\hfill
\subfloat[ \label{chap:gravimetry:fig:CFI:homodyne:various:d1:momentum}]{%
  \includegraphics[width=0.5\linewidth, trim = 0mm 1mm 0mm 0mm]{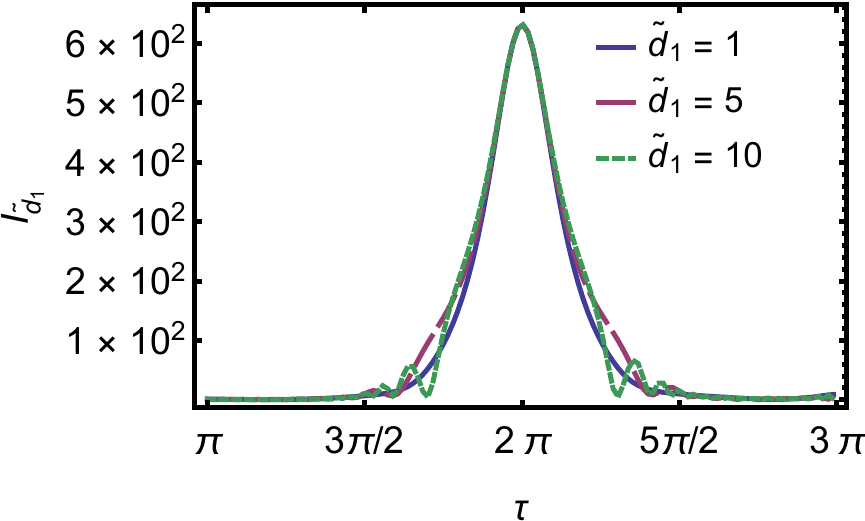}%
}\hfill
\caption[CFI for estimating a constant gravitational acceleration with a homodyne detection scheme with different displacement strengths]{CFI for estimating a constant gravitational acceleration with a homodyne detection scheme with different displacement strengths. The parameters are $\tilde{g}_0 = \mu_{\rm{c}} = 1$ and $\mu_{\rm{m}} = 0$. \textbf{(a)} shows $I_{\tilde{d}_1}$ for a position measurement with $\lambda = 0$ as a function of time $\tau$ for different values of $\tilde{d}_1$. A larger value of $\tilde{d}_1$ actually shifts the peak of the CFI further away from just before $\tau = 2\pi$.  \textbf{(b)} shows $I_{\tilde{d}_1}$ for a momentum measurement with $\lambda = \pi/2$ as a function of time $\tau$ for different values of $\tilde{d}_1$. The CFI displays additional oscillations as $\tilde{d}_1$ increases, but remains unchanged at $\tau = 2\pi$. }
\label{chap:gravimetry:fig:CFI:homodyne:various:d1}
\end{figure*}

The results can be found in Figures~\ref{chap:gravimetry:fig:CFI:homodyne:various:g0} and~\ref{chap:gravimetry:fig:CFI:homodyne:various:d1}. We set $\tilde{d}_1 = \mu_{\rm{c}} = 1$ and $\mu_{\rm{m}} = 0$. In Figure~\ref{chap:gravimetry:fig:CFI:homodyne:various:g0:position},  we  consider measurements of the position quadrature with $\lambda = 0$,  and in Figure~\ref{chap:gravimetry:fig:CFI:homodyne:various:g0:momentum}, we consider measurements of the momentum quadrature with $\lambda = \pi/2$. For position measurements, we find that the CFI peaks just before $\tau = 2\pi$, but for momentum measurements, the CFI peaks around $\tau = 2\pi$, which is also when the QFI in Eq.~\eqref{chap:gravimetry:eq:QFI:coherent:coherent:general} was maximal. From both Figure~\ref{chap:gravimetry:fig:CFI:homodyne:various:g0:position} and~\ref{chap:gravimetry:fig:CFI:homodyne:various:g0:momentum}, we see that increasing $\tilde{g}_0$ increases the magnitude of the CFI at its peak, but they also cause the peak to become more localised. We explore this fact in Section~\ref{chap:gravimetry:time:scale:measurements}. 

Similarly, we plot the CFI for a position and momentum measurement in Figure~\ref{chap:gravimetry:fig:CFI:homodyne:various:d1} for different values of $\tilde{d}_1$. Figure~\ref{chap:gravimetry:fig:CFI:homodyne:various:d1:position} shows the position measurement and Figure~\ref{chap:gravimetry:fig:CFI:homodyne:various:d1:momentum} shows the momentum measurement. While the CFI does not increase with  $\tilde{d}_1$, it does oscillate increasingly quickly for larger values of $\tilde{d}_1$. However, it still always peaks at $\tau = 2\pi$ for a momentum measurement, when the cavity field and mechanics disentangle.

\subsection{Time-scale of measurements} \label{chap:gravimetry:time:scale:measurements}
Let us analyse the expression for the CFI $I_{\tilde{d}_1}$ in Eq.~\eqref{chap:gravimetry:eq:homodyne:CFI}. We note that any terms in the sum with $n = n'$ do not contribute to the Fisher information. The remaining behaviour of $I_{\tilde{d}_1}$ can be inferred from the second exponential in $f_{n,n^\prime}$, which reads $\mathrm{exp}\left[ -|\phi_{n^\prime}|^2/2 - |\phi_n|^2/2 + \phi_{n^\prime}^*\phi_n  \right]$ as this will dominate the entire expression for large values of the coupling $\tilde{g}_0$. 

If we simplify the expression in the exponential, we find that it is equal to
\begin{equation}
-|\phi_{n^\prime}|^2/2 - |\phi_n|^2/2 + \phi_{n^\prime}^*\phi_n  = - \tilde{g}_0^2 (n-n^\prime)^2 (1 - \cos{\tau}) + \frac{1}{2}\tilde{g}(n-n^\prime)  \left[ \mu_{\rm{m}} \eta - \mu_{\rm{m}}^* \eta^* \right] \, . 
\end{equation}
For $n \neq n^\prime$ and large $\tilde{g}_0$, the first term  dominates, and the exponential is small for any $t$ that is not a multiple of $2\pi$. In other words, the Fisher information for a homodyne measurement becomes significant only when light and mechanics are completely disentangled. We saw this already in Figure~\ref{chap:gravimetry:fig:CFI:homodyne:various:g0}, where the CFI becomes increasingly narrow as $\tilde{g}_0$ increases. 

If we examine the expression for the CFI in Eq.~\eqref{chap:gravimetry:eq:homodyne:CFI} and the QFI in Eq.~\eqref{chap:gravimetry:eq:QFI:coherent:coherent:general}, we note that both expressions scale with $\tilde{g}_0^2$. This means systems with a large single-photon coupling are especially well-suited for this task. However, the fact that the peak of the CFI narrows with increasing $\tilde{g}_0$ means that the homodyne measurement must be performed within an increasingly narrow time-window. 

We can estimate the timescale in question by finding the full-width-half-maximum (FWHM) of the peak. To do so, we consider only the dominant first term $- \tilde{g}_0^2 (n-n^\prime)^2 (1 - \cos{\tau})$ for small perturbations in $\tau$ around $\tau = 2\pi$, thus $\cos{(2\pi + \tau')} \approx 1 - \tau^{'2}/2$. That brings the first term into the form $-\tilde{g}_0^2 (n-n^\prime)^2 (\tau')^2/2$, which is now a Gaussian distribution. For a Gaussian function with $\mathrm{exp}\left[ - (x-x_0)^2/(2\sigma^2) \right]$, the full-width-half-maximum (FWHM) is given by $2 \sqrt{2 \ln{2}} \,  \sigma$. In our case, we find $\sigma^2 = [2\tilde{g}_0^2 \, (n-n^\prime)^2]^{-1}$. We already noted that terms with $(n-n')$ do not contribute to the CFI, and any term with $|n-n^\prime|\gg 1$ causes the peak to narrow further. Thus we only consider the terms with $|n-n^\prime| = 1$, leaving us with $\sigma = [2\tilde{g}_0]^{-1}$, and so we conclude that any measurement must be performed roughly on a timescale set by the optomechanical coupling: $(\omega_{\rm{m}} \tilde{g}_0)^{-1} = g_0^{-1}$. 

\subsection{Optimality of homodyne detection}
Let us see if we can simplify the expression for $I_{\tilde{d}_1}$ in Eq.~\eqref{chap:gravimetry:eq:homodyne:CFI} even further and whether it bears any semblance to the QFI. At $\tau = 2\pi$, $\phi_n(2\pi ) = \mu_{\rm{m}}$ and $\eta = 0$, which simplifies the expression. Furthermore, we set $\tilde{g}_0$ and $\tilde{d}_1$ to integer values. Since all phases reduce to unity, the $I_{\tilde{d}_1}$ loses all dependence of $\tilde{d}_1$ at $\tau = 2\pi$. The absence of $\tilde{d}_1$ from $I_{\tilde{d}_1}$  is not a problem for sensing $\tilde{d}_1$ -- it just means that the sensitivity at times $\tau=2\pi$ is independent of the actual value of $\tilde{d}_1$.  Given these assumptions, the coefficients reduce to $c_{n,n'} = (\mu_{\rm{c}}^*)^{n'} \mu_{\rm{c}}^n /\sqrt{n!n^\prime!}$  and $f_{n,n^\prime} = 1$. We are then able to show that  Eq.~\eqref{chap:gravimetry:eq:homodyne:CFI} reduces to the compact expression
\begin{equation} \label{chap:gravimetry:CFI:homodyne:2pi}
I_{\tilde{d}_1} =\frac{8 \,  \pi^2 \tilde{g}_0^2 m}{\hbar \omega_{\rm{m}}^3}  (i \, e^{-i \,  \lambda } \mu_{\rm{c}} - i \, e^{i \, \lambda} \mu_{\rm{c}}^*)^2 \, . 
\end{equation}	
See Section~\ref{app:QFI:homodyne:optimality:proof} in Appendix~\ref{app:QFI} for the full derivation. This expression coincides precisely with the QFI in Eq.~\eqref{chap:gravimetry:eq:QFI:coherent:coherent:2pi} for complementary choices of $\lambda$ and $\mu_{\rm{c}}$. To better see why this is, we rewrite the term in the brackets as $\left[ ( e^{- i \lambda } - e^{i \lambda} )i \Re{\mu_{\rm{c}}} - (e^{- i \lambda } + e^{i \lambda} ) \Im{\mu_{\rm{c}}} \right]^2$.  We now note that when $\lambda = 0 $, only $\Im{\mu_{\rm{c}}}$ contributes to the CFI, whereas at $\lambda = \pi/2$, only $\Re{\mu_{\rm{c}}}$ contributes. For both of these specific choices of $\lambda$, and when matched by $\mu_{\rm{c}}$ being either entirely real or imaginary,  the CFI coincides precisely with the QFI in Eq.~\eqref{chap:gravimetry:eq:QFI:coherent:coherent:2pi} because the term in the brackets reduces to $4\Re{\mu_{\rm{c}}}^2$ or $4 \Im{\mu_{\rm{c}}}^2$, respectively. 

We conclude that the homodyne measurement saturates the QFI bound up to a phase dependence of $\mu_{\rm{c}}$, which can always be accounted for by changing the quadrature of the homodyne measurement.  Note,  however, that  the CFI only saturates the QFI when the light and mechanics have disentangled. The optimality of the homodyne detection for sensing within our scheme is greatly advantageous as it is a routine measurement which is easy to accomplish. It has in fact also been shown to be an optimal measurement~\cite{latmiral2016probing} in other contexts.

\section{Estimation with open dynamics} \label{chap:gravimetry:open:system:estimation}

The calculation above is valid for pure states, but in practice every measurement will suffer various forms of decoherence, as discussed in Section~\ref{chap:gravimetry:open:system:dynamics}. We will here investigate the effects of decoherence on the CFI for a narrow parameter range, as realistic parameters are very difficult to simulate numerically, as discussion in Section~\ref{chap:introduction:sec:numerical:challenges} in Chapter~\ref{chap:introduction}. We shall later use these results as indications of the behaviour of realistic systems.  

Before we proceed to evaluate the CFI, we discuss the numerical methods that we use in this Section.

\subsection{Numerical methods}
To evolve the system, we use the Python library \textit{QuTiP}~\cite{johansson2013qutip} and a 4th order Runge--Kutta--Fehlberg method~\cite{fehlberg1969low} for verification. The probability distribution $p(x| \tilde{d}_1)$ is straight-forward to obtain numerically, since any operator has a finite matrix representation from which we can obtain the eigenstates and use these as our POVM elements. For example, we define the position operator $\hat{x}_{\rm{c}}$ as a finite-dimensional matrix and solve for its eigenstates.  

To obtain $\bar{I}_{\tilde{d}_1}$ we must also compute the derivative $\partial_{\tilde{d}_1} p(x|\tilde{d}_1)$. This can be done in a number of ways. The simplest one is to use a higher order method of the central difference theorem. We obtained good and accurate results with the 4th-order five-point method~\cite{press2007numerical}. For a function $f(x)$ with parameter $x$ and step-size $h$, the first derivative with this method is given by 
\begin{align} \label{chap:gravimetry:eq:numerical:derivative}
f'(x) =\bigl[ &- f(x+ 2h) + 8 f(x+h)  \nonumber \\
&- 8f(x-h) + f(x- 2h) \bigr]/(12h)+ \mathcal{O}(h^4) \, .
\end{align}
As this method requires five data point, it is an expensive computation. However, it was our preferred numerical method since computing the CFI can still be done within reasonable time-scales using the optimised master equation solver provided by the \textit{QuTiP} library. It does however contain two different sources of numerical errors: Errors in the solver and errors that originate from the cut-off in the numerical derivative. 

To verify that the error introduced by the numerical differentiation is not severely affecting the results, we used an additional method which provides an exact result. We reproduce it here for completion. We start by noting that as long as the POVM element $\hat \Pi_x$ does not depend on $\tilde{d}_1$, we can write the derivative as
\begin{equation}
\frac{\partial p(\tilde{d}_1 |x)}{\partial\tilde{d}_1} = \mathrm{Tr} \left[ \frac{\partial \rho_{\tilde{d}_1}}{\partial \tilde{d}_1} \hat  \Pi_x \right], 
\end{equation}
Note that we are differentiating with respect to $\tilde{d}_1$ instead of $g$ and that we have suppressed the dependence of $\tau$ for clarity. This statement also holds for subsystems of $\rho(g)$, which we can see by noting that the derivative distributes over a joint separable system $\rho_{AB}= \rho_A \otimes \rho_B$ as
\begin{equation}
\frac{\partial \hat \rho_{AB}}{\partial \tilde{d}_1} = \frac{\partial \hat \rho_A}{\partial \tilde{d}_1} \otimes \hat \rho_B + \hat \rho_A \otimes \frac{\partial \hat \rho_B}{\partial  \tilde{d}_1}. 
\end{equation} 
Performing a measurement with $\hat \Pi_x$ that only acts on subsystem $A$ then gives
\begin{align}
\mathrm{Tr}_B \left[ \frac{\partial \hat \rho_{AB} }{\partial \tilde{d}_1} \,  \hat \Pi_x \right] &=  \mathrm{Tr}_B\left[ \frac{ \partial \hat \rho_A}{\partial \tilde{d}_1} \, \hat  \Pi_x \otimes \rho_B \right] +   \mathrm{Tr}_B\left[ \hat \rho_A \,  \hat \Pi_x \otimes \frac{\partial \hat \rho_B}{\partial \tilde{d}_1}  \right] . 
\end{align}
The second term reduces to zero because $\mathrm{Tr}[\partial_{\tilde{d}_1} \hat \rho_B] = \partial_{\tilde{d}_1} \mathrm{Tr}[\hat \rho_B] = 0$. While we have shown this for separable states, the same argument can be extended to entangled states by linearity. 

In order to obtain the evolution for this state, we must now solve a modified version of the master equation.  That is, given the Lindblad equation in Eq.~\eqref{chap:gravimetry:eq:Lindblad}, we now differentiate both sides with respect to $\tilde{d}_1$ to obtain
\begin{align} \label{eq:ModMasterEq}
\partial_{\tilde{d}_1} \dot{\rho}_{\tilde{d}_1} = &- \frac{i}{\hbar} \left[ \partial_{\tilde{d}_1} \hat H_{\tilde{d}_1}, \rho_{\tilde{d}_1} \right] - \frac{i}{\hbar} \left[ \hat H_{\tilde{d}_1} , \partial_{\tilde{d}_1} \rho_{\tilde{d}_1} \right]   - \frac{1}{2} \{ \partial_{\tilde{d}_1} \rho_{\tilde{d}_1} , \hat L^\dagger \hat L\}, 
\end{align}
where the Hamiltonian $\hat H_{\tilde{d}_1}$ depends on $\tilde{d}_1$, and where we have used the notation $\partial_{\tilde{d}_1} = \partial / \partial \tilde{d}_1$. A more complicated form is obtained if the Lindblad operators $\hat L$ depend on $\bar{g}$, which here is not the case. In coupled form, we can write 
\begin{align} \label{chap:gravimetry:eq:modified:Lindblad}
&\frac{d}{d\tau} \left(
\begin{matrix}
\hat \rho_{\tilde{d}_1}  \\ \partial_{\tilde{d}_1} \hat \rho_{\tilde{d}_1}
\end{matrix} \right) = 
\left(
\begin{matrix} - \frac{i}{\hbar} \left[ \hat H_{\tilde{d}_1}, \hat \rho_{\tilde{d}_1} \right]  +  \left( \hat L_{\rm{c}} \,  \hat\rho_{\tilde{d}_1} \, \hat L_{\rm{c}}^\dagger - \frac{1}{2} \{ \hat \rho_{\tilde{d}_1}, \hat L_{\rm{c}}^\dagger \hat L_{\rm{c}} \} \right) \\
  - \frac{i}{\hbar} \left( \left[\partial_{\tilde{d}_1} \hat H_{\tilde{d}_1} , \hat \rho_{\tilde{d}_1} \right]  +\left[ \hat H_{\tilde{d}_1} , \partial_{\tilde{d}_1} \, \hat  \rho_{\tilde{d}_1} \right] \right) +  \hat L_{\rm{c}} \partial_{\tilde{d}_1} \hat \rho \hat L_{\rm{c}}^\dagger - \frac{1}{2} \{ \partial_{\tilde{d}_1}, \hat \rho_{\tilde{d}_1} \hat L_{\rm{c}}^\dagger \hat L_{\rm{c}} \} 
\end{matrix}
\right). 
\end{align}
The system can be solved using any standard higher-order method, such as the family of Runge--Kutta ODE solvers. Note that the \textit{QuTiP} Master Equation solver cannot be used as Eq.~\eqref{chap:gravimetry:eq:modified:Lindblad} is not in standard Hamiltonian form. 

Once the time-evolved state $\partial_{\tilde{d}_1} \rho_{\tilde{d}_1} $ has been obtained, we proceed as usual to compute the probability distribution with the set of POVM elements $\{\hat \Pi_x\}$. With this method, we avoid round-off errors that appear in the five-point numerical derivative above. 

\subsection{Homodyne detection for an open system }
\begin{figure*}[t!]
\centering
\subfloat[ \label{chap:gravimetry:fig:CFI:homodyne:open:system:position}]{%
  \includegraphics[width=0.49\linewidth, trim = 0mm 0mm 0mm 0mm]{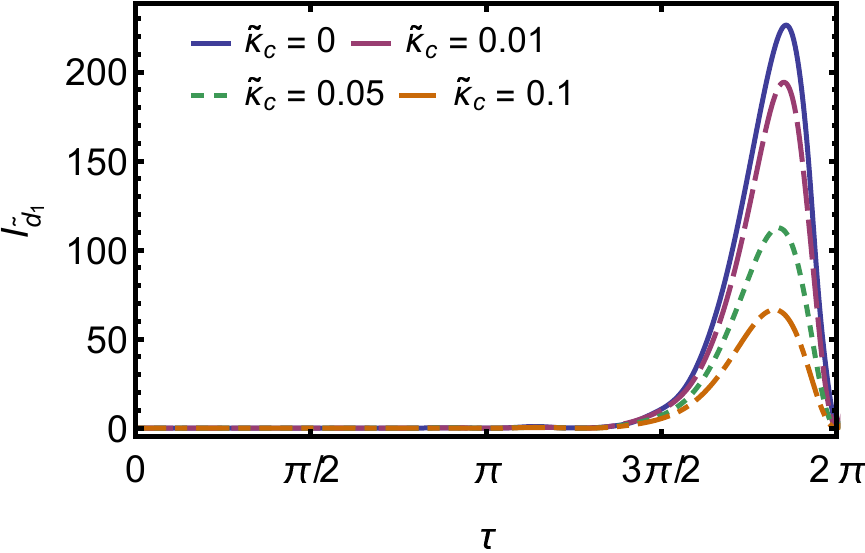}%
}\hfill
\subfloat[ \label{chap:gravimetry:fig:CFI:homodyne:open:system:momentum}]{%
  \includegraphics[width=0.5\linewidth, trim = 0mm 1mm 0mm 0mm]{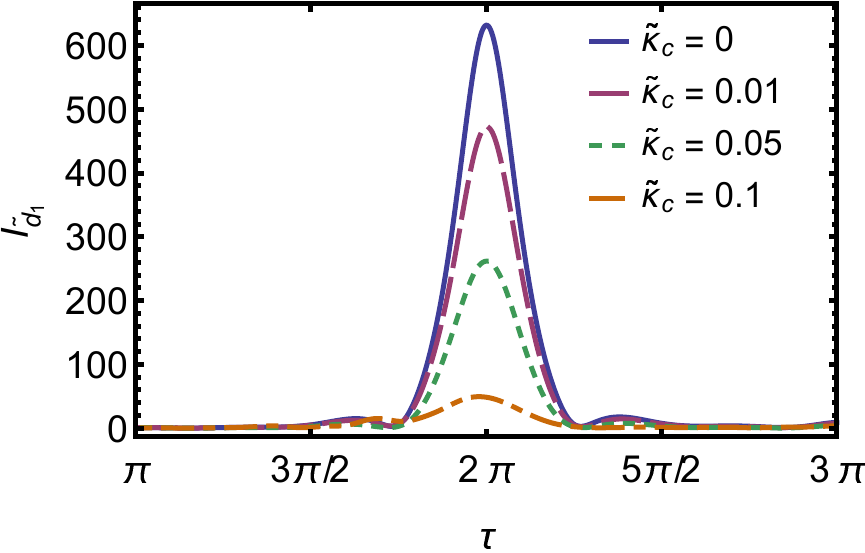}%
}\hfill
\caption[CFI for estimating a constant gravitational acceleration with a homodyne detection scheme for an open system]{CFI for estimating a constant gravitational acceleration with a homodyne detection scheme for an open system.  The parameters are $\tilde{g}_0 = \tilde{d}_1 = \mu_{\rm{c}} = \mu_{\rm{m}} = 1$. \textbf{(a)} shows the CFI $I_{\tilde{d}_1}$ as a function of time $\tau$ for a position measurement with different optical decoherence rates $\tilde{\kappa}_{\rm{c}}$. The CFI peaks just before $\tau = 2\pi$ and decreases as the photons leak from the cavity. \textbf{(b)} shows the CFI $I_{\tilde{d}_1}$ as a function of time $\tau$ for a momentum measurement with different optical decoherence rates $\tilde{\kappa}_{\rm{c}}$. The CFI peaks at $\tau = 2\pi$ and decreases as $\tilde{\kappa}_{\rm{c}}$ increases.  }
\label{chap:gravimetry:fig:CFI:homodyne:open:system}
\end{figure*}

We are now ready to compute the CFI for a homodyne measurement of an open system. In all subsequent numerical evaluations, we will set $\tilde{g}_0 = \tilde{g}_1 = 1$ and $\mu_{\rm{c}} = 1$ (note the choice of $\mu_{\rm{c}} \in \mathbb{R}$, which will optimise the CFI for the choice $\lambda = \pi/2$). We stay with these values, since higher values of the parameters may cause the system to quickly grow numerically unstable due to the inclusion of Kerr nonlinear terms such as $(\hat a^\dagger \hat a)^2$ in the evolution with $\hat U(\tau)$ in  Eq.~\eqref{chap:gravimetry:eq:time:evolution:operator:explicit}. While  $\tilde{g}_0 = 1$ is experimentally achievable with the right choice of parameters, we can justify setting  $\tilde{d}_1 = 1$ by noting that it physically corresponds to a heavily inclined cavity with $\theta \approx \pi/2$. These numerical investigations should only be seen as a indication as to how decoherence will affect $I_{\tilde{d}_1}$, and not as predictions for the sensitivity of a realised device. We later extrapolate from these results to make a prediction for realistic systems. Unfortunately, it is  currently extremely difficult to predict the evolution of a fully realistic system in the nonlinear regime. 

A plot of the CFI $I_{\tilde{d}_1}$ as a function of time $\tau$ can be found in Figure~\ref{chap:gravimetry:fig:CFI:homodyne:open:system} for a position and momentum measurement of an open system with decoherence rate $\tilde{\kappa}_{\rm{c}}$. Figure~\ref{chap:gravimetry:fig:CFI:homodyne:leaking:photons:position} shows the CFI as a function of time $\tau$ when a position measurement is performed on the environment (the leaking photons), and Figure~\ref{chap:gravimetry:fig:CFI:homodyne:leaking:photons:momentum} shows the same scenario for a momentum measurement. We note that larger values of $\bar{\kappa}_{\rm{c}}$ do affect the CFI adversely, but setting $\bar{\kappa}_{\rm{c}} = 0.1$ implies that about 10\% of the pure state CFI is still accessible.

\subsection{Measurements of leaking photons}
\begin{figure*}[t!]
\centering
\subfloat[ \label{chap:gravimetry:fig:CFI:homodyne:leaking:photons:position}]{%
  \includegraphics[width=0.49\linewidth, trim = 0mm 0mm 0mm 0mm]{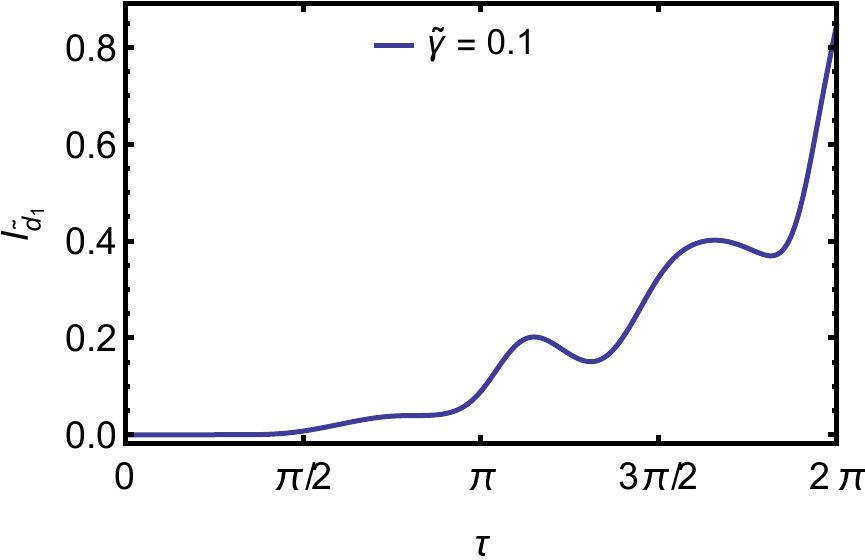}%
}\hfill
\subfloat[ \label{chap:gravimetry:fig:CFI:homodyne:leaking:photons:momentum}]{%
  \includegraphics[width=0.5\linewidth, trim = 0mm 1mm 0mm 0mm]{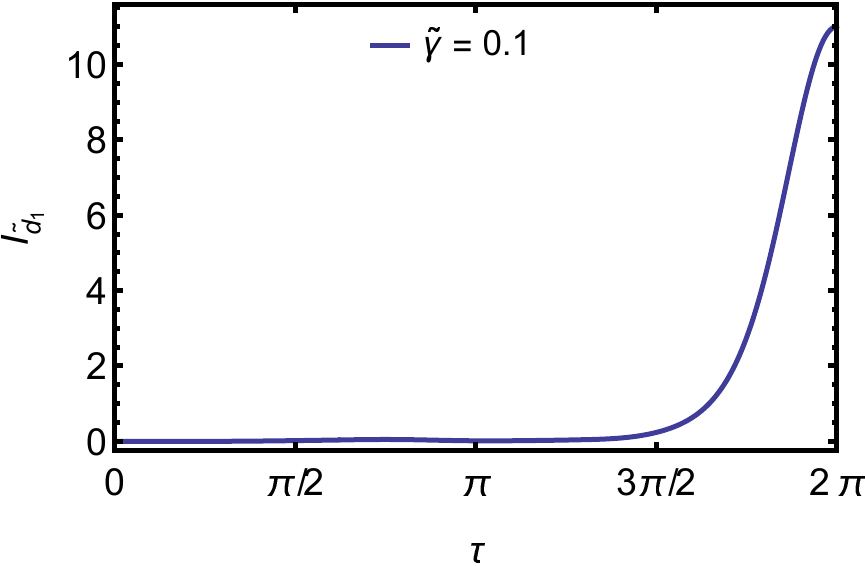}%
}\hfill
\caption[CFI for estimating a constant gravitational acceleration with a homodyne detection scheme for photons leaking from the cavity]{CFI for estimating a constant gravitational acceleration with a homodyne detection scheme for photons leaking from the cavity. The parameters are $\tilde{g}_1 = \tilde{d}_1 = \mu_{\rm{c}} = \mu_{\rm{m}} = 1$. The coupling between the system and the environment has been set to $\tilde{\gamma} = \gamma/\omega_{\rm{m}} = 0.1$. \textbf{(a)} shows the CFI for a position measurement of the environment, and \textbf{(b)} shows the CFI for a momentum measurement of the environment. }
\label{chap:gravimetry:fig:CFI:homodyne:leaking:photons}
\end{figure*}

In practise, a homodyne measurement is performed by monitoring and measuring the photons that continuously leak from the cavity. Aside from the experimental considerations, such a scheme also negates part of the photon dissipation considered above. We briefly estimate the CFI obtained through such a setup by using a simplified model where a pure vacuum state of the environment $\ket{0}$ is added to our original state $\ket{\Psi(\tau)}$ for two initially coherent states given in Eq.~\eqref{chap:gravimetry:eq:coherent:coherent}. This yields the combined initial state 
$\ket{\Psi(\tau)}\otimes \ket{0}$. 

We then add a rotating wave interaction term $\hat H_I$ to the Hamiltonian $\hat{H}_{\mathrm{G}}$ in Eq.~\eqref{chap:gravimetry:eq:Hamiltonian:with:gravity}, of the form
\begin{equation}
\hat{H}_I = \hbar \gamma \left( \hat a^\dagger \hat c + \hat a \hat c^\dagger \right),
\end{equation}
where $\gamma$ is the interaction strength and $\hat c$ and $\hat c^\dagger$ are the creation and annihilation operators of the environment. As before, we rescale the environment coupling with respect to the timescale of mechanical oscillations: $\tilde{\gamma} = \gamma/\omega_{\rm{m}}$. 

The effect of this interaction Hamiltonian is to couple the cavity state to the environment which causes information about $\tilde{d}_1$, and by extension $g$, to slowly leak out from the cavity into $\ket{0}$. This toy model is an approximation, since the unitary dynamics eventually cause information to `leak back' into the system. For short time-scales, however, it is a good approximation. 

We numerically compute the CFI for performing a homodyne measurement on the environment state. As before, we evolve the full state for a single photon $|\mu_{\rm{c}}|^2 = 1$ and with the mechanics in the vacuum state: $\mu_{\rm{m}} = 0$. The other parameters are $ \tilde{g}_0 = \tilde{d}_1 = 1$.  To maximise the CFI, we choose $\mu_{\rm{c}} \in \mathbb{R}$ and $\lambda = \pi/2$. 

The results can be found in Figure~\ref{chap:gravimetry:fig:CFI:homodyne:leaking:photons} for a rescaled coupling strength $\tilde{\gamma} = \gamma/\omega_{\rm{m}} = 0.1$. We keep the coupling small to ensure numerical stability. The position measurement can be found in Figure~\ref{chap:gravimetry:fig:CFI:homodyne:leaking:photons:position}, and the momentum measurement can be found in Figure~\ref{chap:gravimetry:fig:CFI:homodyne:leaking:photons:momentum}. 
As evident from Figure~\ref{chap:gravimetry:fig:CFI:homodyne:leaking:photons}, we suffer a $10^{-2}$ reduction in the information that can be extracted from the system.  Note also that the behaviour of $I_{\tilde{d}_1}$ for this scenario will most likely also resemble a delta function centered around $\tau = 2\pi$ for realistic parameters.

\section{Computing the ideal sensitivity} \label{chap:gravimetry:computing:sensitivity}
In this Section, we calculate the ideal QFI for the three optomechanical systems we discussed in Section~\ref{chap:introduction:section:examples} in Chapter~\ref{chap:introduction} by using the expression for the QFI at $\tau = 2\pi$ in Eq.~\eqref{chap:gravimetry:eq:QFI:coherent:coherent:2pi}. We then discuss the experimental challenges and advantages to an optomechanical gravimeter.  As we here calculate the fundamental sensitivity, which is unlikely to be realised, we will only concern ourselves with order-of-magnitude estimates. These results are meant to showcase the potential of optomechanical systems, and to do so we have chosen state-of-the-art parameters that have been implemented in a variety of systems. For discussions of an experimental implementation including noise, see the Discussion in Section~\ref{chap:gravimetry:discussion}. 

Starting with the Fabry--P\'{e}rot cavity system, we choose a fully vertical cavity with $\theta = 0$ and use the following state-of-the-art experimental parameters: We choose a mass $m = 10^{-6}$  kg, oscillator frequency $\omega_{\rm{m}} = 10^3$ Hz, cavity frequency $\omega_{\mathrm{c}} = 10^{14}$ Hz, cavity length $L= 10^{-5}$ m and a photon number of $|\mu_{\rm{c}}|^2 = 10^6$. For these values, the rescaled coupling constant in Eq.~\eqref{chap:introduction:eq:coupling:Fabry:Perot} becomes $\tilde{g}_{\mathrm{FP}} \approx 2.30$, which gives us a QFI of $\mathcal{I}_g = 1.58 \times 10^{28}$ m$^{-2}$ s$^4$.  This implies a sensitivity of $\Delta g  \approx 7.96 \times 10^{-15}$ ms$^{-2}$. 

Next, we look at a levitated micro-object confined in an ion trap interacting with an optical cavity, as demonstrated very recently in Refs.~\cite{Millen2015iontrap, fonseca2016nonlinear}. Again setting $\theta = 0$ for maximal effect, we use mass $m = 10^{-14}$ kg, oscillator frequency $\omega_{\rm{m}} = 10^2$ Hz, cavity frequency $\omega_{\mathrm{c}} = 10^{14}$ Hz, volume $V = 10^{-18}$ m$^3$, cavity mode volume $V_{\mathrm{c}} = 10^{-14}$ m$^3$, electric permitivity $\epsilon = 5.7$ for nanodiamonds, laser wavelength $\lambda = 1064\times 10^{-9}$ m and a photon number of $|\mu_{\rm{c}}|^2 = 10^6$. From these values we obtain the rescaled coupling from Eq.~\eqref{chap:introduction:eq:coupling:levitated} $\tilde{g}_{\mathrm{ND}} = 1963$  which leads to $\mathcal{I}_g \approx  1.15 \times 10^{29}$ m$^{-2}$s$^4$. This gives us a final sensitivity of $\Delta g  \approx 2.94 \times 10^{-15}$ ms$^{-2}$ for levitated nanospheres. 

Finally, let us also consider cold atoms trapped in a cavity. Based on~\cite{brennecke2008cavity}, we choose the following parameters: A wavelength $\lambda = 780$ nm, implying $\omega_{\mathrm{c}} = 10^{15}$ Hz, a single-atom coupling of $g_0 = 10^7 $ Hz, an atomic oscillation frequency $\omega_{\rm{m}} = 10^2$ Hz, a single-atom mass $m = 10^{-25}$ kg, a detuning of $\Delta_{ca} = 10^{11}$ Hz, and a laser wavevector of $k_l = 10^8$ m$^{-1}$.  With $N =10^5$ atoms trapped in the cavity, we find from Eq.~\eqref{chap:introduction:eq:coupling:Atomic} that the rescaled coupling is given by $\tilde{g}_{BEC} = 2.30 \times 10^6$ and $\mathcal{I}_g \approx 1.58 \times 10^{19}$ m$^{-2}$s$^4$, giving a sensitivity of $\Delta g \approx 2.5 \times 10^{-10}$ ms$^{-2}$. The reason for this disparity seems to be that the polarisability of the collection of cold atoms is not high enough to match the polarisability exhibited by the nanosphere. The number of trapped atoms can hardly match the number of atoms in a single nanosphere. One would either have to increase the number of atoms trapped in the cavity or increase the single-atom coupling strength to increase the Fisher information.

\subsection{Comparison of theoretical and experimental results}
Let us briefly compare the results obtained here with the performance of other quantum systems in the literature. In Table~\ref{chap:gravimetry:tab:experimental:comparison} we have listed a variety of experimentally implemented gravimeter systems with their best achieved sensitivity to date. Table~\ref{chap:gravimetry:tab:theoretical:comparison}, on the other hand, lists the ideal fundamental limits to sensitivities calculated in this work and others. The values for $\Delta g$ and $\Delta g/ \sqrt{\mathrm{Hz}}$ are presented in units of ms$^{-2}$ and ms$^{-2}$ Hz$^{-1/2}$ respectively. The last column in Table~\ref{chap:gravimetry:tab:experimental:comparison} lists the integration time for each experiment, whereas in Table~\ref{chap:gravimetry:tab:theoretical:comparison} the last column lists the experimental cycle time set by the oscillation frequency of the system in question. For atom interferometry, it is suggested in ~\cite{chiow2011102} that sensitivities of $\Delta g \sim 10^{-12}$ ms$^{-2}$ might be achieved, and a study of the fundamental limits has very recently been presented in Ref~\cite{kritsotakis2018optimal}.

\begin{table}[h]
\caption[Comparison between gravimetry sensitivities obtained by various experimental systems]{Comparison between gravimetry sensitivities obtained by various experimental systems. These include the commercial LaCoste FG5-X, atom interferometry, gravimetry through on-chip Bose-Einstein condensate (BEC) and classical optomechanical accelerometry. The second column lists the sensitivity $\Delta g$ in  ms$^{-2}$ and the third column lists the $\sqrt{\mbox{Hz}}$-noise $\Delta g / \sqrt{\mathrm{Hz}}$ in ms$^{-2}$Hz$^{-1/2}$. The last column indicates the integration time needed to achieve each sensitivity. 
\\
*This value was provided to us by the authors of Ref~\cite{Cervantes2014} in private communications.}
\centering
\begin{tabular}{l c c c} \toprule
\multicolumn{4}{c}{\textbf{Experiments }}  \\  \toprule
 \textbf{System} & $\Delta g$  & $\Delta g/\mathrm{\sqrt{Hz}} $&  \textbf{Int. time}\\ 
\midrule 
  LaCoste FG5-X~\cite{LaCoste2016}  & $  1\times 10^{-9}$ & $1.5\times 10^{-7}$ &6.25 hours \\
  Atom intf. ~\cite{hu2013demonstration} & $5 \times 10^{-9}$ & $4.2\times 10^{-8}$ & $100$ s \\
On-chip BEC~\cite{abend2016atom}  & $7.8\times 10^{-10}$  & $ 5.3\times10^{-9}$ & $100$ s  \\
Optomech. accel.~\cite{Cervantes2014} & $3.10\times 10^{-5}$  & $9.81 \times 10^{-7}$   & $10^{-3}$ s$^*$ \\ \bottomrule
\end{tabular} \label{chap:gravimetry:tab:experimental:comparison}
\end{table}

\begin{table}[h]
\caption[Comparison between sensitivities obtained by theoretical predictions for a variety of systems]{Comparison between sensitivities obtained by theoretical predictions for a variety of systems. These include magnetomechanics, a Fabry--P\'{e}rot optomechanical system, a levitated nanosphere optomechanical system and trapped cold atoms. The second column lists the sensitivity $\Delta g$ in  ms$^{-2}$ and the third column lists the $\sqrt{\mbox{Hz}}$-noise $\Delta g / \sqrt{\mathrm{Hz}}$ in ms$^{-2}$Hz$^{-1/2}$. The last column indicates the cycle time or oscillation frequency $\omega_m$ for each system. \\
Values calculated in this work are denoted by *.}
\centering
\begin{tabular}{l c c c}\toprule
\multicolumn{4}{c}{\textbf{Theoretical predictions}}
\\ \toprule
 \textbf{System} & $\Delta g$  & $\Delta g/\mathrm{\sqrt{Hz}} $&  \textbf{Cycle time}\\  \midrule
Magnetomechanics~\cite{johnsson2016macroscopic} & $2.2 \times 10^{-7}$ & $2.2 \times 10^{-9}$ & $10^{-4}$ s \\
Fabry--P\'{e}rot moving-end mirror* & $10 ^{-15}$ & $10^{-16}$ & $10^{-3}$ s\\
Levitated optomechanics*  & $10^{-15}$ & $10^{-16}$ & $10^{-2}$  s\\	
Cold atoms optomechanics* & $10^{-10}$ & $10^{-11}$ & $10^{-2}$ s\\\bottomrule
\end{tabular} \label{chap:gravimetry:tab:theoretical:comparison}
\end{table}

\section{Discussion} \label{chap:gravimetry:discussion}
In this work, we investigated a new scheme for measurements of the gravitational acceleration $g$ using a compact cavity optomechanical system with the usual cubic optomechanical coupling to the cavity field. We derived a fundamental limit to the sensitivity $\Delta g$ by computing the quantum Fisher information and showed that the optimal sensitivity is achieved by a homodyne detection scheme performed on the cavity state at $\tau = 2\pi$. That is, no direct measurement of the mechanical oscillator is required. Using the expression in Eq.~\eqref{chap:gravimetry:eq:QFI:coherent:coherent:2pi} and state-of-the-art experimental parameters, we predict a upper bound on the sensitivity of order $\Delta g \sim 10^{-15}$ ms$^{-2}$ for both a Fabry--P\'{e}rot cavity and a levitated microsphere cavity, and $\Delta g \sim 10^{-10}$ ms$^{-2}$ for trapped cold atoms. These values compare favourably to all other currently available experimental and theoretical gravimetry proposals (see Table~\ref{chap:gravimetry:tab:experimental:comparison} and~\ref{chap:gravimetry:tab:theoretical:comparison}). Furthermore, the quantum nature of the oscillator ensures that any thermal distribution in its initial state does not affect the fundamental sensitivity. However, as our scheme relies on superpositions involving distinct coherent states, we require thermal decoherence during one period of the oscillator motion to be negligible, which we estimate requires a $Q$-factor  of at least $10^6$ for the case of a Fabry--P\'{e}rot cavity (see below). To explore the effects of photons leaking from the cavity, we numerically explored a narrow parameter range with $\tilde{g}_0 = \tilde{d}_1= 1$, which physically corresponds to a nearly horizontally aligned cavity. We found that this form of decoherence does affect the system's performance, but not severely. Finally, we briefly investigated what proportion of $\Delta g$ we retain by performing measurements on the photons that leak from the cavity. Using a simplified noise model, we found a reduction of $10^{-2}$ in the resulting Fisher information. Given these results, we believe that there is significant potential in the use of quantum optomechanical systems for measurements of gravity and acceleration.

\subsection{Experimental challenges}
Let us now address some of the experimental challenges related to this scheme. Due to measurement inefficiencies and additional sources of decoherence not considered here, the final performance of optomechanical systems will naturally be expected to be lower than the values presented in Table~\ref{chap:gravimetry:tab:theoretical:comparison}.  While we have shown that the initial optomechanical state does not need cooling to the ground-state, thermal noise due to external influences during the evolution will gradually decohere the oscillator motion. We estimate that in the case of a Fabry--P\'{e}rot cavity cooled to a temperature of milli-Kelvin, a number of $\hbar \omega_{\rm{m}}/(k_{\mathrm{B}} T_{\mathrm{th}}) = N$ phonons are present in the system at any time. Here, $k_{\mathrm{B}}$ is Boltzmann's constant and $T_{\mathrm{th}}$ is the system's temperature. To retain coherence throughout the evolution, we require that $\kappa_{\rm{m}} N \ll \omega_m $, where $\kappa_{\rm{m}}$ is the phonon dissipation rate. In other words, the timescale of phonon decoherence $\kappa_{\rm{m}}$ must be much less than the characteristic timescale of the system. With $\omega_{\rm{m}} = 1$ kHz, as we assumed for Fabry--P\'{e}rot cavities, we find $N = 10^5$ and $\kappa_{\rm{m}} = 10^{-2}$ Hz. A cavity which achieves such a decoherence rate must have a mechanical $Q$-factor of at least $Q = \omega_{\rm{m}}/\kappa_{\rm{m}} \sim 10^6$ to retain coherence, a regime which is not unprecedented. 

Next, let us discuss which parameters yield the best sensitivities.  Firstly, we note that the QFI in Eq.~\eqref{chap:gravimetry:eq:QFI:coherent:coherent:2pi} ultimately scales with $\omega_{\rm{m}}^{-6}$. In addition to the factor $\omega_{\rm{m}}^{-3}$ in the denominator, we acquire an extra $\omega_{\rm{m}}^{-2}$ from the rescaled coupling constant $\tilde{g}_0 = g_0/\omega_m$. The final factor of $\omega_{\rm{m}}$ comes from the dependence of $\omega_{\rm{m}}$ in $g_0^2$. Given this scaling, we require $\omega_{\rm{m}}$ to be as small as possible. At the same time, we also require the photon dissipation rate $\kappa_{\rm{c}}$ to be low. From our simulations, we saw that we require at least $\tilde{\kappa}_{\rm{c}} =\kappa_{\rm{c}}/\omega_{\rm{m}} = 0.1$. This combination is difficult to achieve as low $\omega_m$ means the cavity must remain coherent over longer timescales. Therefore, the main experimental challenge of this scheme is to reduce $\omega_{\rm{m}}$ and $\kappa_{\rm{c}}$ at the same time. Taking our numerical results as guidance, we essentially require that $\bar{\kappa}_{\rm{c}} = \kappa_{\rm{c}}/\omega_{\rm{m}} \ll 1$, which is exactly the resolved sideband regime~\cite{chan2011laser}.

\subsection{Dissipation rates}

In the above, we used state-of-the-art parameters to calculate the ideal QFI for a variety of systems. However, as we just saw, the photon dissipation rate $\kappa_{\rm{c}}$ must be very low for these sensitivities to be achieved, and this has not yet been experimentally demonstrated for the parameters we used in Table~\ref{chap:gravimetry:tab:theoretical:comparison}. As technology improves we expect that this to be possible in future experiments, but for now, let us estimate the sensitivities that could be achieved today already. One of the best coherence times to date was demonstrated in Ref~\cite{zhao2009vibration}, which achieved a cavity linewidth of $\kappa_{\rm{c}} = 660$ Hz. 
To achieve a rescaled photon rate of $\bar{\kappa}_{\rm{c}} = 0.1$ for this system, we let $\omega_{\rm{m}} = 6600$ Hz and use $L = 9.4$ cm as reported in the paper. We keep $m = 10^{-6}$ kg (since the QFI is ultimately independent of mass) and let $\omega_{\mathrm{c}} = 10^{14}$ Hz as before. Because the oscillation frequency $\omega_{\rm{m}}$ is rather high, we choose to calculate $\mathcal{I}_g$ for the Fabry--P\'{e}rot cavity with a mechanical mirror, as this system performed slightly better for higher $\omega_{\rm{m}}$. The resulting coupling constant is $\tilde{g}_{\mathrm{FP}} = 1.44\times 10^{-5}$, and the Fisher information is $\mathcal{I}_g \approx 2.16 \times 10^{15}$ m$^{-2}$s$^4$. This leads to $\Delta g \approx 2.15 \times 10^{-8}$ ms$^{-2}$. If we now assume that decoherence causes a similar proportion of the Fisher information to dissipate at these parameters compared to the ones chosen in our numerical simulations, we see that we retain about 10\% of the pure-state Fisher information. Using this assumption, we find $\Delta g \approx 6.80\times 10^{-8}$ ms$^{-2}$ and a $\sqrt{\mathrm{Hz}}$-noise of $8.37\times 10^{-10}$ ms$^{-2}$/$\sqrt{\mathrm{Hz}}$. This is directly comparable with the values in Table~\ref{chap:gravimetry:tab:theoretical:comparison}, and so we believe that this scheme could be experimentally realised today, although the experimental challenges are of course substantial. 

Let us briefly discuss ways in which we can decrease $\kappa_{\rm{c}}$ further and how this might affect the Fisher information. A heuristic estimate for $\kappa_{\rm{c}}$ can be given by considering the number of times per second that a  single photon traverses the cavity. Each time the photon is reflected at the mirror, it has a $T = 1 - R$ chance of being transmitted instead of reflected. Here, $T$ is the proportion of transmissions and $R$ is the proportion of the number of reflections. The photon bounces off a mirror $c/L$ times per second, where $c$ is the speed of light. Thus we can take the dissipation rate to be $\kappa_{\rm{c}} = T c/L$, which means that increasing $L$ decrease the photon dissipation rate $\kappa_{\rm{c}}$, as the photon is effectively spending longer inside the cavity. However, increasing $L$ also decreases the single-photon coupling constant, as we saw in the calculation above. This is true for all couplings we quote here, but it is perhaps most clearly seen for the case of the mechanical mirror and a Fabry-Perot cavity, with $g_{\mathrm{FP}}$ given by Eq.~\eqref{chap:introduction:eq:coupling:Fabry:Perot} $g_{\mathrm{FP}} $ scales with $L^{-1}$, and so do the other couplings, through their dependence on the cavity volume $V_{\mathrm{c}}$ or the single-photon coupling $g_0$. We recall that the Fisher information depends on $\tilde{g}_0^2$, which means that it ultimately scales with $L^{-2}$. Thus, changing $L$ by an order of $10$ will decrease the Fisher information by an order of $10^2$. This contributes to the challenges of realising this scheme. However, it is important to note that there are realistic ways of increasing $L$ without changing the single-photon coupling: One such method was explored in Ref~\cite{pontin2018levitated}, where $L$ was increased by adding an optical fibre to the cavity. 

Furthermore, we showed in Section~\ref{chap:gravimetry:sec:CFI} that homodyne detection is optimal, but we also found that such a measurement must be performed within a rather narrow temporal window, of timescale $1/g_0$. Let us here estimate how quickly these measurements have to be performed based on the values we calculated for the coupling constant $g_0$. The nanospheres displayed the highest single-photon coupling with $\tilde{g}_{\mathrm{ND}} \times \omega_{\rm{m}} = 10^5$ Hz for the choice of $\omega_{\rm{m}} = 10^2$ Hz. Thus any homodyne measurement must be performed within $10^{-5}$ s, so we require at most microsecond precision, which is perfectly achievable. In comparison, we calculated $\tilde{g}_{\mathrm{FP}} = 2.30$ for the levitated microsphere, which allows for a very comfortable $\approx 0.19$ s window. 

In spite of these challenges, optomechanical systems come with a number of advantages. They can remain stationary while performing the measurement, in contrast to on-chip BECs or cold-atom fountains which need to be launched, and the short cycle time of optomechanical systems allows for a large number of measurements to be performed very quickly. An additional point which we did not elaborate on above is that the spatial resolution of optomechanical systems will be extremely high since the oscillator is displaced only by a minuscule distance. As a result, it will be possible to determine very fine local variations in $g$, something which is not possible using larger systems. The scheme presented in this work also allows for the creation of macroscopic spatial superpositions, which, as pointed out in Ref~\cite{johnsson2016macroscopic}, is of great interest to testing gravitational collapse models (see for example~\cite{diosi1989models, hu2014gravitational, pfister2016universal}). 

\subsection{Comparison with atom interferometry}

Before we conclude, let us briefly discuss the underlying physical differences between atom interferometry and optomechanical systems for the purpose of gravimetry. A thorough analysis of the fundamental bounds of atom interferometry was since published in Ref~\cite{kritsotakis2018optimal}. 

In atom interferometry, we prepare the atoms in a superposition of a ground state $\ket{g}$ and an excited state $\ket{e}$, such that the full state becomes 
\begin{equation}
\ket{\Psi(0)}_{\rm{atom}} =\frac{1}{\sqrt{2}} (\ket{g} + \ket{e} ) \,.
\end{equation}
Photons are then used to separate the two states by momentum transfer, causing them to take two different paths through a gravitational potential. We then assign a potential gravitational energy $ mg \Delta x$ for the excited state, where $\Delta x$ is the difference in height between the two paths. The phase accumulated by the excited state is then equal to $e^{i mg \Delta  T/\hbar}$, where $m$ is the atomic mass and $T$ is the time of flight. We must now determine $\Delta x$. Ignoring any geometric factors associated with the paths, we assume that the distance roughly depends on the atoms' velocity $v$ and their time of flight $T$. That is, we let  $\Delta x \sim v T$. The total velocity is determined by the momentum transfer from the photons in the laser pulse, and is therefore proportional to the number of photons $n$. The momentum carried by one photon is given by $\hbar k_{\mathrm{c}}$, where $k_{\mathrm{c}}$ is the wavevector of the photon (which we take to be the same as the wavevector of the photons in the cavity). Thus, assuming that each photon transfers all of its momentum to the atom, we find that 
\begin{equation}
\Delta x \sim vT \sim \frac{n\hbar k_{\mathrm{c}} }{m} T \, . 
\end{equation}
If we insert this into the expression for the phase and apply it to the state, we find 
\begin{equation}
\ket{\Psi(T)}_{\rm{atom}} = \frac{1}{\sqrt{2}} \left( \ket{g} + e^{ i ngk_{\mathrm{c}} T^2} \ket{e} \right) \, .
\end{equation}
Calculating the quantum Fisher information for this state is straight-forward. We find that
\begin{align} \label{chap:gravimetry:eq:QFI:atom:interferometry}
\mathcal{I}^{(\rm{atom})}_g&= 4 \left( \langle \partial_g \Psi | \partial_g \Psi \rangle - |\langle \partial_g \Psi | \Psi \rangle |^2 \right) \nonumber \\
&= 4 \left( \frac{n^2 k_{\mathrm{c}}^2 T^4}{2} - \frac{n^2 k_{\mathrm{c}}^2 T^4}{4} \right) \nonumber \\
&= n^2 k_{\mathrm{c}}^2 T^4 \, . 
\end{align}
Since $k_{\mathrm{c}}$ has dimension m$^{-1}$, this expression has the correct units of s$^4$m$^{-2}$. In terms of scalability, we note that this expression surpasses the Heisenberg limit in terms of the number of photons $n$, and that it is highly dependent on the time of flight $T$.

We can now compare this with the optomechanical Fisher information for the Fabry--P\'{e}rot cavity. The explicit Fisher information with $g_{\mathrm{FP}}$ inserted into Eq.~\eqref{chap:gravimetry:eq:QFI:coherent:coherent:2pi} given by 
\begin{equation}
\mathcal{I}_{g}^{ (\mathrm{FP})} = \frac{32\pi^2n  \cos^2{\theta}}{\omega_0^6} \frac{\omega_{\mathrm{c}}^2}{L^2}. 
\end{equation}
where we have replaced $|\mu_{\rm{c}}|^2$ by $n$. To compare the two expressions, we let $\omega_{\rm{m}}^4 \sim 1/T^4$,  $\omega_{\mathrm{c}} =2\pi c /\lambda  $, and $k_{\mathrm{c}} = 2\pi/\lambda$. We set $\theta = 0$ for clarity and then divide them to find that 
\begin{align}
\xi_{\mathrm{FP}} &=  \frac{ 32 \pi^2 n c^2T^4/(\omega_{\rm{m}}^2\lambda^2 L^2) }{ 4\pi^2n^2  T^4 / \lambda^2 } \nonumber \sim \frac{ c^2}{n \omega_{\rm{m}}^2L^2} \, . 
\end{align}
Here, $\xi_{\mathrm{FP}}$ is an enhancement factor. This is due to the cavity confinement, whereby each photon interacts with the oscillator $c/(2L\omega_{\rm{m}})$ times per oscillation cycle, which is also the time period over which the gravimetric phase is accumulated. For the levitated nano-sphere, we find a $\xi_{\mathrm{Lev}} =\xi_{\mathrm{FP}} P^2 /(\epsilon_0 V_{\mathrm{c}}  )^2$, where, again, for a micro-object containing $\sim 10^{13}$ atoms, the polarisability $P$ is much higher than that of a single atom. In practice, however, both of the enhancement factors will be damped by a factor $\sim 1/(\omega_{\rm{m}} T)^4$ with respect to atom interferometry as the time of atomic interferometry $T$ is typically larger than the time $1/\omega_{\rm{m}}$ of our scheme. Thus the sensitivity $\Delta g$ is seen to improve by a factor of $ \sqrt{n} L \omega_{\rm{m}}^3 T^2/c   \sim \sqrt{n} \times 10^{-4}$ in our optomechanical scheme with respect to atomic interferometers. As $n$ increases, the differences level out. However, we saw in Section~\ref{chap:gravimetry:sec:QFI} that different initial states in an optomechanical system, such as the superposition of two Fock states, which also produces a QFI that scales with $n^2$ (see Eq.~\eqref{chap:gravimetry:QFI:initial:Fock:state}).  Strictly speaking, the enhancement is valid for when the cavity field remains coherent for the time $1/\omega_{\rm{m}}$ over which our phase accumulation, i.e., $\kappa_{\rm{c}} \ll \omega_{\rm{m}}$ (the resolved side-band regime). However, our numerical results indicate that even in the presence of finite decoherence, say, $\kappa_{\rm{c}} \sim 0.1 \omega_{\rm{m}}$, the Fisher information is lowered only by a factor of about $10$ compared to the case of loss-less cavities. Finally, we can also compare the treatment presented in this work to a position measurement of a classical oscillator that has been displaced due to gravity. While a classical treatment of the problem returns a preliminary measurement sensitivity similar to what we have derived in this work, it fails to take into account effects such as radiation pressure and the full quantum nature of the cavity field. Most importantly, a classical treatment does not utilise the coherent nature of the oscillator, which as we saw above negates any initial thermal noise in the state, and does not allow for the inclusion of other quantum states, such as squeezed states.

\section{Conclusions} \label{chap:gravimetry:conclusions}
In this Chapter, we investigated a scheme for measurements of the gravitational acceleration $g$ using a compact cavity optomechanical system with the usual nonlinear optomechanical coupling to the cavity field. We derived a fundamental limit to the sensitivity $\Delta g$ by computing the quantum Fisher information and showed that the optimal sensitivity is achieved by a homodyne detection scheme performed on the cavity state only. That is, no direct measurement of the mechanical oscillator is required. 

With our results, and using state-of-the-art experimental parameters, we predict a upper bound on the sensitivity of order $\Delta g \sim 10^{-15}$ ms$^{-2}$ for both a Fabry--P\'{e}rot cavity and a levitated object cavity. This value compares favourably to all other currently available experimental and theoretical gravimetry proposals (see Table~\ref{chap:gravimetry:tab:experimental:comparison} and~\ref{chap:gravimetry:tab:theoretical:comparison}). Furthermore, the quantum nature of the oscillator ensures that any thermal distribution in its initial state does not affect the fundamental sensitivity. However, as our scheme relies on superpositions involving distinct coherent states, we require thermal decoherence \emph{during} one period of the oscillator motion to be negligible, which we estimate requires a $Q$-factor  of at least $10^6$ for the case of a Fabry--P\'{e}rot cavity. 
To explore the effects of photons leaking from the cavity, we numerically explored a narrow parameter range, which physically corresponds to a nearly horizontally aligned cavity. We found that this form of decoherence does affect the system's performance, but not severely.

\addcontentsline{toc}{chapter}{Appendices} 

\chapter{General Conclusions}
\label{chap:conclusions}


In this Chapter, we provide a technical summary of the content of the thesis, discuss the future of this work, and draw general conclusions based on the results we have derived.

\section{Technical summary of the thesis}
In this Section, we provide a chapter-by-chapter summary of the results, including those derived in the appendices. 
The results in Chapter~\ref{chap:introduction}--\ref{chap:gravimetry} are as follow. 

\begin{description}
\item[Chapter~\ref{chap:introduction}] In this Chapter, we introduce a number of concepts that are central to the theory of optomechanical systems. While Chapter~\ref{chap:introduction} does not contain many results, it is intended as an introduction to the theory of optomechanical systems (Section~\ref{chap:introduction:theory:of:nonlinear:optomechanical:systems}), the prevalence of non-Gaussianity in quantum theory (Section~\ref{chap:introduction:sec:non:Gaussianity}), and an introduction to quantum metrology (Section~\ref{chap:introduction:quantum:metrology}), including some derivation of the key figures of merit, such as the quantum Fisher information. 
\item[Chapter~\ref{chap:decoupling}] In this Chapter, we first introduce the Decoupling Theorem in Section~\ref{chap:decoupling:Lie:algebra:decoupling:method}, and then derives the main result in this thesis in Section~\ref{chap:decoupling:optomechanical:decoupling}: a solution of the dynamics of an optomechanical Hamiltonian with an additional mechanical displacement term and a single-mode mechanical squeezing term. The Hamiltonian is shown in Eq.~\eqref{chap:decoupling:eq:Hamiltonian}, and the decoupled evolution operator can be found in two versions: one which retains some quadratic behaviour in Eq.~\eqref{chap:decoupling:eq:final:evolution:operator}, and one which is fully decoupled in Eq.~\eqref{chap:decoupling:eq:final:evolution:operator:J:coefficients}.  The solutions presented here constitute the basis for the remainder of the thesis and hold significant potential for further applications beyond those presented. 
\item[Chapter~\ref{chap:non:Gaussianity:coupling}] This Chapter computes the non-Gaussianity of an initially coherent state evolving with the standard optomechanical Hamiltonian in Eq.~\eqref{chap:non:Gaussianity:coupling:eq:standard:Hamiltonian}.  We compute the first and second moments of the state in Eq.~\eqref{app:expectation:values:coherent} and the covariance matrix elements in Eq.~\eqref{chap:non:Gaussianity:coupling:full:elements:covaraince:matrix}. We find that the non-Gaussianity increases with both the photon number and with the strength of the optomechanical coupling. In Section~\ref{chap:non:Guassianity:coupling:sec:constant:coupling}, we explore a constant coupling, and in Section~\ref{chap:non:Gaussianity:coupling:sec:time:dependent:coupling}, we explore a coupling that is sinusoidally modulated in time. While concise analytical results are difficult to obtain, the key findings are shown in Figures~\ref{fig:constant:measure:vs:time} and~\ref{fig:constant:measure:scaling}. Perhaps the most important finding is that the non-Gaussianity can be continuously increased when the nonlinear coupling is modulated at mechanical resonance, as shown in Figure~\ref{fig:time:dependent}. The results hold even in the presence of decoherence, as shown in Figure~\ref{chap:non:Gaussianity:coupling:fig:measure:decoherence:resonance}. 
\item[Chapter~\ref{chap:non:Gaussianity:squeezing}] In this Chapter, we compute the non-Gaussianity of an initially coherent state evolving with the extended optomechanical Hamiltonian in Eq.~\eqref{chap:non:Gaussianity:squeezing:eq:Hamiltonian}. As before, we derive expressions for the first and second moments of the state in Eqs.~\eqref{chap:non:Gaussianity:squeezing:expectation:values} and~\eqref{chap:non:Gaussianity:squeezing:photon:phonon:numbers}, and the covariance matrix elements in Eq.~\eqref{chap:non:Gaussianity:squeezing:eq:CM:elements}. Here, we are interested in seeing how the non-Gaussianity changes with the strength of the mechanical squeezing term. We explore constant mechanical squeezing in Section~\ref{chap:non:Gaussianity:squeezing:sec:constant:squeezing} and a sinusoidally time-modulated squeezing in Section~\ref{chap:non:Gaussianity:squeezing:sec:modulated:squeezing}.  We generally find that the non-Gaussianity decreases with the squeezing parameter, since the quadratic contribution to the Hamiltonian dominates as the squeezing increases. However, when the coupling is modulated at twice mechanical resonance, there are certain parameter regimes where the non-Gaussianity increases with the squeezing. In general, we uncover a highly complex interplay between the non-Gaussianity and the squeezing. 
\item[Chapter~\ref{chap:metrology}] In this Chapter, we compute a general expression for the quantum Fisher information given the extended optomechanical Hamiltonian in Eq.~\eqref{chap:metrology:eq:Hamiltonian} that is solved in Chapter~\ref{chap:decoupling}. The main result is the expression for the quantum Fisher information shown in Eq.~\eqref{chap:metrology:eq:main:result:QFI} and its coefficients in Eq.~\eqref{chap:metrology:eq:QFI:coefficients}. These results can be used as a starting point for a number of considerations of different sensing schemes. We explore three concrete estimation schemes in Section~\ref{chap:metrology:three:examples}, and find that high sensitivities could be achieved in each of the  cases considered (see Table~\ref{chap:metrology:tab:Values} for a summary). 
\item[Chapter~\ref{chap:gravimetry}] In the final Chapter, we consider an application of the sensing methods derived in Chapter~\ref{chap:metrology}: Estimating a constant gravitational acceleration with a nonlinear optomechanical system. The expression for the quantum Fisher information can be found in Eq.~\eqref{chap:gravimetry:eq:QFI:coherent:coherent:general}, and using state-of-the-art parameters, our results imply that an optomechanical systems could, in principle, achieve a sensitivity of $\Delta g = 10^{-15}$ \si{ms^{-2}}. We consider three different optomechanical setups, the results for which are summarised in Table~\ref{chap:gravimetry:tab:theoretical:comparison}. In addition, we show in Section~\ref{chap:gravimetry:sec:CFI} that a homodyne detection scheme is the optimal measurement for this task. The Chapter is concluded by a discussion in Section~\ref{chap:gravimetry:discussion} of the various parameters required for the system to operate optimally. 
\end{description}

\noindent The results in Appendices~\ref{app:optomechanical:Hamiltonian}--\ref{app:QFI} are as follows:

\begin{description}
\item[Appendix~\ref{app:optomechanical:Hamiltonian}] This Appendix contains two derivations of the optomechanical Hamiltonian and light--matter couplings for the Fabry-P\'{e}rot moving-end mirror system (Section~\ref{app:optomechanical:mirror}) and the levitated nanobead (Section~\ref{app:optomechanical:hamiltonian:nanosphere}), respectively. 
\item[Appendix~\ref{app:coefficients}] This Appendix contains a number of solutions for the $F$-coefficients in Eq.~\eqref{chap:decoupling:eq:sub:algebra:decoupling:solution} and the $J$-coefficients in Eq.~\eqref{chap:decoupling:eq:diff:equations:Js} given different forms of the weighting functions $\mathcal{G}(t)$, $\mathcal{D}_1(t)$ and $\mathcal{D}_2(t)$ in the Hamiltonian in Eq.~\eqref{chap:decoupling:eq:Hamiltonian}. These expressions are used throughout the thesis. 
\item[Appendix~\ref{app:commutators}] This Appendix contains a number of commutators and expectation values that are used primarily in Chapter~\ref{chap:metrology} to compute the quantum Fisher information. A general procedure for deriving expressions for operator multiplication by congruence can be found in Section~\ref{app:commutator:congruence}. 
\item[Appendix~\ref{app:exp:values}] This Appendix contains a number of results for evolution under the extended Hamiltonian in Eq.~\eqref{chap:decoupling:eq:Hamiltonian}. The first and second moments for initially coherent states can be found in Eq.~\eqref{app:ex:values:summary:of:exp:values}, the covariance matrix elements can be found in Eq.~\eqref{app:exp:values:CM:elements}, and the symplectic eigenvalues of the optical and mechanical subsystem can be found in Eq.~\eqref{app:exp:values:sympelctic:eigenvalue:summary}. 
\item[Appendix~\ref{app:mathieu}] This Appendix contains a derivation of a mapping between the optomechanical Hamiltonian with a squeezing term and the standard Hamiltonian in Section~\ref{app:mathieu:time:varying:trapping:frequency}, and a perturbative treatment of the Mathieu equation in Section~\ref{app:mathieu:perturbative:solutions}, which solves the mechanical subsystem dynamics. The perturbative solutions given two different sets of boundary conditions can be found in Eq.~\eqref{app:mathieu:eq:P11:approx} and~\eqref{app:mathieu:eq:IP22:approx}, respectively. 
\item[Appendix~\ref{app:QFI}] This Appendix contains a number of results relating to the discussion of the quantum Fisher information in Chapters~\ref{chap:metrology} and~\ref{chap:gravimetry}. In Section~\ref{app:QFI:derivation:of:QFI:mixed:states}, we derive the mixed state quantum Fisher information expression in Eq.~\eqref{chap:metrology:definition:of:QFI}, and in Sections~\ref{app:QFI:different:initial:states:expressions} and~\ref{app:QFI:expressions:estimation:schemes}, we derive a number of expressions for the quantum Fisher information for different estimation scenarios. The Appendix is concluded by a proof of the optimality of homodyne detection for constant gravitational accelerations in Section~\ref{app:QFI:homodyne:optimality:proof}. 
\end{description}

\section{Discussion}
Before concluding this Chapter and this thesis, we discuss the results obtained and place them in a wider context. 

\subsection{The advantage of analytic solutions}
The main result in this thesis is the solution to the extended optomechanical Hamiltonian presented in Chapter~\ref{chap:decoupling}. There are significant advantages to using analytical solutions over numerical simulations, even if some must be obtained through perturbative methods. We already touched on this topic in Section~\ref{chap:non:Gaussianity:squeezing:discussion} in Chapter~\ref{chap:non:Gaussianity:squeezing}, but we reiterate some of the key points here. 

The difficulties encountered when solving nonlinear systems numerically (which we discussed in Section~\ref{chap:introduction:sec:numerical:challenges} in Chapter~\ref{chap:introduction}), mean that nonlinear optomechanical systems have thus far not been studied in full generality. With the methods developed in this thesis, we have significantly extended the categories of couplings and external effects that can be considered. In particular, being able to solve time-dependent dynamics without having to simulate the full states in a truncated Hilbert space implies a significant step forwards towards modelling real-world effects. 
However, we saw in Chapters~\ref{chap:non:Gaussianity:squeezing} and~\ref{chap:metrology} that not all time-dependent functions can be solved analytically. Generally, we find that when squeezing is included, the complexity of the system dynamics increases dramatically. 

The solutions presented here hold potential for being used as tests for numerical methods that aim to efficiently solve the nonlinear evolution with perturbative methods. One example of such methods is Ehrenfest guided trajectories~\cite{shalashilin2009quantum, ye2012modeling}, which can be used to approximately solve the Schr\"{o}dinger equation. By comparing, for example, the first and second moments of the evolving optomechanical state with the numerically obtained quantities, it is possible to determine the accuracy of the numerical method. 

\subsection{Towards the realisation of nonlinear quantum optomechanical sensors}
The results presented in this thesis mainly concern closed system dynamics. Some numerical results for open systems were presented in Chapters~\ref{chap:non:Gaussianity:coupling} and~\ref{chap:gravimetry}, but they took only extremely limited parameter ranges into account and cannot be easily generalised to cases where the cavity photon number is highly excited.
In order to describe realistic systems, the results presented in this thesis must be complemented by analysis of experimental sources of noise and additional effects. Without this analysis, especially the results in Chapter~\ref{chap:metrology} cannot be put to proper use.  

Another weakness of these methods is the fact that a linear optical driving term, which is a common additions in physical systems, cannot be included as it causes the Lie algebra to become infinite. This is problematic, since realistic systems are open by default (otherwise light cannot be injected into the cavity) and almost always operate in the steady-state. Without an open cavity, it is also not possible to consider effects such as, for example,  cooling the mechanical element to the ground state. For linear systems, the input--output formalism captures these effects perfectly, and while efforts appear to have been made towards explaining some of the aspects mentioned above for nonlinear systems~\cite{shahidani2014steady}, an equivalent formalism has yet to be developed for nonlinear systems.

\subsection{Future work}
The methods and results presented in this thesis apply to a wide range of optomechanical systems and external effects that impact the system dynamics. There are a number of potential research avenues that can be further explored. 

One question includes whether the Lie algebra methods presented in Chapter~\ref{chap:decoupling} can be extended to open system dynamics. If this can be done, it would be possible to describe optomechanical systems to even greater generality, and the results presented in this thesis could be extended to realistic setups. Once open dynamics can be simulated, it would be possible to determine the amount of noise that can be present in a system for a certain effect to be detectable. For example, when measuring extremely weak gravitational effects, it becomes key to predict in advance whether a certain system configuration would be suitable. The noise levels required will influence the choice of system; for example, the absence of physical contact between a levitated nanobead and the cavity prevents the transmission of noise through vibrations or convection heating. 

Another pressing question concerning optomechanical system is the generation and detection of non-classical effects, such as entanglement. With the methods developed in this thesis, it might be possible to investigate the generation of entanglement between the light and the mechanics, or even between two different mechanical elements. Results obtained from such investigations could help shed light on the quantum-to-classical transition, and even help optimise metrology schemes. For example, it might be possible to show that a sensor array of mutually entangled optomechanical systems can achieve sensitivities beyond those attainable by single-system setups. 

Finally, given the previous references to metrology schemes, there are a number of scenarios that can be investigated with the methods developed here. A natural extension to the results presented in Chapter~\ref{chap:gravimetry} include time-dependent gravitational effects, such as the extremely small gravitational signal generated by an oscillating sphere, or even gravitational waves. The interplay between the different effects in the Hamiltonian can also be further investigated, and perhaps employed the enhance estimation schemes. For example, it is possible that a time-dependent light--matter coupling modulated at the same rate as an external time-dependent effect might aid the sensing scheme, or that squeezing the mechanical element enhances its sensitivity to specific effects. 

In general the Lie algebra methods demonstrated here can be used to solve additional examples of nonlinear dynamics. We leave this, and the points discussed above, to future work.

\section{Conclusions}

In this thesis, we studied the mathematical description and application of cavity optomechanical systems evolving in the nonlinear regime. We solved the dynamics of an optomechanical Hamiltonian with an additional time-dependent mechanical displacement and single-mode squeezing term, and used the solutions to derive a number of results. 

To better understand the implications of the nonlinear dynamics, we computed the non-Gaussianity of the optomechanical state using a relative entropy measure. We found that non-Gaussianity increases with the number of photons and the nonlinear coupling, but that the addition of a single-mode mechanical squeezing term generally decreases the non-Gaussianity. These results might provide insights into when the state evolution becomes increasingly complex and deviates from regimes where it is well-approximated by Gaussian states. 

We further studied nonlinear optomechanical systems as quantum sensors and derived a general expression for the quantum Fisher information for estimation of any of the parameters in the extended Hamiltonian. We  found that optomechanical systems generally display high fundamental sensitivities, which means that they could, in principle, be used as powerful quantum sensors. To demonstrate the applicability of our results, we estimated the sensitivity of measuring a constant gravitational acceleration and found that it theoretically could be measured to an accuracy of $10^{-15}$~\si{ms^{-2}}. We also proved that this sensitivity is attainable with standard laboratory measurements. 

We hope that the results presented here will aid the theoretical understanding of nonlinear optomechanical systems and motivate the building of future optomechanical experiments.

\cleardoublepage 

\appendix

\begin{appendix}
\chapter{Derivation of the optomechanical Hamiltonian}
\label{app:optomechanical:Hamiltonian}
In this Appendix, we provide two derivations of the optomechanical Hamiltonian. The first follows the standard derivation of the light--matter interaction term for a moving-end mirror in a Fabry-P\'{e}rot cavity \cite{serafini2017quantum,law1995interaction}, and the second follows the derivation in Ref \cite{romero2011optically} for a levitated nanobead in an optical cavity. The purpose of the first derivation is to build intuition for the physical elements of optomechanical systems and how they interaction, whereas the purpose of the second derivation is to explicitly show how the optomechanical coupling can be modulated as a function of time, which has applications in the generation of non-Gaussianity (see Chapters \ref{chap:non:Gaussianity:coupling} and \ref{chap:non:Gaussianity:squeezing}). 

As part of each derivation, we discuss for which parameter regimes the approximations are valid. 

\section{Hamiltonian of a moving-end mirror}\label{app:optomechanical:mirror}
We begin with the moving-end mirror. There are two ways to outline this derivation. The first takes a rather heuristic approach, and the second recounts the original derivation by Law in Ref \cite{law1995interaction}. 

\subsection{Derivation from heuristic notions}\label{app:optomechanical:mirror:heuristic}

The following derivation is short but captures the central notions involved in modelling the light--matter interaction. We closely follow the derivation in Ref \cite{serafini2017quantum}. 

We start by considering a cavity which is formed by two perfect mirrors, where one of the mirrors is tethered to some stationary frame through a spring or some other mechanical oscillator. The light in the cavity is resonant with the cavity width, and thus any motion of the mirror will induce a change in the resonant frequency of the cavity. We focus on a single mode of the optics and mechanics, for which the Hamiltonian operator of the cavity field is $\hat H_{\rm{c}} = \hbar \hat \omega(\hat x) \, \hat a^\dag \hat a$, 
where $\hat \omega(\hat x)$ is the frequency of the cavity mode. As emphasised by the notation, the frequency is not a scalar function but an operator acting on the mechanical oscillator Hilbert space. It is given by 
\begin{equation}
\hat \omega ( \hat x) = \frac{w}{\hat x} \, ,
\end{equation}
where $w$ is the velocity of the cavity mode, in our case, the speed of light, and $\hat x $ is the position quadrature operator defined by 
\begin{equation}
\hat x = \sqrt{\frac{\hbar}{m \omega_{\rm{m}}} } \frac{\hat b^\dag + \hat b }{\sqrt{2}} \, .
\end{equation}
Here, $m$ is the mass of the moving mirror and $\omega_{\rm{m}}$ is the frequency by which it moves. 

We can now consider pertubations to the motion of the mirror. Let the average length of the cavity be $L$, and let the system perform small motions around this equilibrium. We can therefore write
\begin{equation}
\hat \omega( \hat x) = \frac{w}{L + \hat x } \approx \frac{w}{L} - \frac{w}{L^2 } \hat x^2 \, .
\end{equation}
The light--matter coupling constant is defined by the commutator
\begin{equation}
\left[ \hat \omega( \hat x) , \hat b \right] \approx \frac{w}{L^2} \sqrt{\frac{\hbar}{2 m \omega_{\rm{m}} }} \equiv g_0 \, .
\end{equation}
Inserting these terms into the Hamiltonian for the cavity field $\hat H_{\rm{c}}$ defined above we find
\begin{equation}
\hat H_{\rm{c}} = \hbar \hat \omega(\hat x) \, \hat a^\dag \hat a \approx \hbar \omega_{\rm{c}} \, \hat a^\dag \hat a - g_0 \, \hat a ^\dag \hat a \, \left( \hat b^\dag + \hat b \right) \,, 
\end{equation}
where $\omega_{\rm{c}} = w/L$ is the cavity resonant frequency. Once the free evolution of the phonon mode $\hbar \omega_{\rm{m}} \hat b^\dag \hat b $ has been added, we obtain the optomechanical Hamiltonian in Eq. \eqref{chap:introduction:eq:standard:nonlinear:Hamiltonian}. Additional analysis can be performed to rigorously show how the cavity length $L$ relates to the strength of the light--matter interaction. See Ref \cite{serafini2017quantum} for more details. 

\subsection{Derivation from first principle}\label{app:optomechanical:mirror:first:principle}

Our goal is to derive a Hamiltonian that accurately describes the interaction between the light and the mirror. Compared with the derivation in the previous Section, we  here consider a Lagrangian and Hamiltonian formalism, which allows us to present the quantisation from first principle. This derivation was first proposed in Ref \cite{law1995interaction}, and this Section follows it closely. 

We start by considering a one-dimensional cavity (see Figure \ref{chap:introduction:fig:mirror}) formed by two perfectly reflecting mirrors, where one of the mirrors is fixed at position $x = 0$ and the other moves. We denote the position of the moving mirror by $q(t)$ and its mass by $m$. 
Together, the isolated cavity field and mirror form an energy-conserving system. However, in order to write down the Hamiltonian, we must first define the canonical momentum of the mirror. 

We start by defining the vector potential $A(x, t)$ of the cavity field in the region $0 \leq x \leq q(t)$. Setting $c = 1$, where $c$ is the speed of light, we write the wave-equation as
\begin{equation} \label{app:Hamiltonian:wave:equation}
\frac{\partial^2 A(x, t) }{\partial x^2} = \frac{\partial^2 A(x, t) }{\partial t^2 } \, .
\end{equation}
We then impose the time-dependent boundary conditions
\begin{equation}\label{app:Hamiltonian:boundary:conditions}
A(0, t) = A(q(t), 0) =  0 \, , 
\end{equation}
which ensure that the electric fields are always zero in the rest frame of the mirror's surface. Then, the non-relativistic equation of motion of the mirror is given by 
\begin{equation} \label{app:Hamiltonian:eom}
m \ddot{q} = - \frac{\partial V(q)}{\partial q} + \frac{1}{2} \left( \frac{\partial A(x, t)}{\partial x} \right)^2 \biggr|_{x = q(t)} \, , 
\end{equation}
where the second term corresponds to the radiation pressure force. Eq. \eqref{app:Hamiltonian:eom} can be derived from the radiation pressure force in the rest frame of the mirror. In the co-moving frame, the radiation pressure force is given by $B^{\prime 2}/2$, where $B'$ is the magnetic field on the mirror's surface. 

The system dynamics are completely determined by Eqs. \eqref{app:Hamiltonian:wave:equation}, \eqref{app:Hamiltonian:boundary:conditions}, and \eqref{app:Hamiltonian:eom}. Given our definition of $q(t)$, the mirror's position is strictly positive or zero. We proceed to  define a set of generalised coordiantes $\{Q_k \}$, which correspond to  the mode decomposition of the fields. They are given by 
\begin{equation}
Q_k = \sqrt{\frac{2}{q(t)}} \int^{q(t)}_0 \mathrm{d} x\, A(x, t) \, \sin \left( \frac{kn z}{q(t)} \right) \, ,
\end{equation}
where $k$ is a positive integer that indicates the mode. The mode basis functions are determined by the instantaneous position of the mirror. 

Since the mode functions act as basis functions, they also obey a completeness relation. This allows us to write down an equation of motion for $A(x,t)$ in terms of the boundary conditions in Eq. \eqref{app:Hamiltonian:boundary:conditions}:
\begin{equation} \label{app:Hamiltonian:field:eom}
A(x, t) = \sum_{k = 1}^\infty Q_k (t) \, \sqrt{\frac{2}{q(t)}} \, \sin \left( \frac{k \pi z}{q(t)} \right) \, . 
\end{equation}
Using Eq. \eqref{app:Hamiltonian:field:eom}, and the orthogonality of the mode functions, we can show that Eqs. \eqref{app:Hamiltonian:wave:equation} and \eqref{app:Hamiltonian:eom} are equivalent to the following equations of motion:
\begin{align} \label{app:Hamiltonian:eom:Qk}
\ddot{Q}_k &= - \omega_k^2 Q_k + 2 \frac{\dot{q}}{q} \sum_j g_{kj} \dot{Q}_j + \frac{\ddot{q} q - \dot{q}^2}{q^2} \sum_j g_{kj} Q_j + \frac{\dot{q}^2}{q^2} \sum_{jl} g_{jk} g_{jl} Q_l  \, , 
\end{align}
and
\begin{equation} \label{app:Hamiltonian:eom:q}
m \ddot{q} = - \frac{\partial V(q)}{\partial q} + \frac{1}{q} \sum_{kj} (-1)^{k + j} \, \omega_k  \, \omega_j \, Q_k Q_j \, , 
\end{equation}
where the position dependent frequencies $\omega_k$ are given by 
\begin{equation}
\omega_k (q) = \frac{k\pi}{q} \, ,
\end{equation}
and the dimensionless coefficients $g_{kj}$ are given by 
\begin{equation}
g_{kj} =\begin{cases}
(-1)^{k + j} \frac{2 kj}{j^2 - k^2 } &k\neq j \, ,\\
0 &k = j \, .
\end{cases}
\end{equation} 
The next task is writing the equations of motions in terms of a Lagrangian formulation, and consequently identify the Hamiltonian. We define a Lagrangian $L$, and by examining Eqs. \eqref{app:Hamiltonian:eom:Qk} and \eqref{app:Hamiltonian:eom:q}. We find
\begin{align} 
L(q,\dot{q}, Q_k, \dot{Q}_k ) =& \,  \frac{1}{2} \sum_k \left( \dot{Q}_k^2 - \omega_k^2 (q) Q_k^2 \right) + \frac{1}{2} m \dot{q}^2 - V(q) \nonumber \\
&- \frac{\dot{q}}{q} \sum_ {jk} g_{kj} \dot{Q}_k Q_j + \frac{\dot{q}^2}{2q^2} \sum_{jkl} g_{kj} g_{kl} Q_l Q_j \, .
\end{align}
The Euler--Lagrange quations correspond to the equations in  Eqs. \eqref{app:Hamiltonian:eom:Qk} and \eqref{app:Hamiltonian:eom:q}. The Hamiltonian associated with the Lagrangian then becomes
\begin{equation}
H(P_k , Q_j, p, q) =  p \dot{q} + \sum_k P_k \dot{Q}_k - L(q, \dot{q}, Q_k \dot{Q}_k ) \, ,
\end{equation}
where $P_k$ and $p$ are canonical momenta that are conjugate to $Q_k$ and $q$, respectively. They are given by 
\begin{equation}
P_k = \dot{Q}_k - \frac{\dot{q}}{q} \sum_j g_{kj} Q_j \, ,
\end{equation}
and 
\begin{equation}
p = m \dot{q} - \frac{1}{q} \sum_{jk} g_{kj} P_k Q_j \, .
\end{equation}
We note that the mirror's canonical momentum $p$ is not equal to the kinetic momentum $m \dot{q}$ for nonzero fields. We then identify the explicit expression for the Hamiltonian as
\begin{equation}
H = \frac{1}{2m } \left( p + \frac{1}{q} \sum_{jk} g_{kj} P_k Q_j \right)^2 + V(q) + \frac{1}{2} \sum_k \left( P_k^2 + \omega_k^2 Q_k^2 \right) \, . 
\end{equation}
We can then confirm that $H$ corresponds to the total energy of the system, which is given by 
\begin{equation}
H = H_{\rm{field}} + \frac{1}{2} m \dot{q}^2 + V(q) \, ,
\end{equation}
where $H_{\rm{field}}$ is the cavity field energy defined by 
\begin{equation}
H_{\rm{field}} = \frac{1}{2} \int^{q(t)}_0 \mathrm{d}x \left[ \left( \frac{\partial A(x,t)}{\partial t} \right)^2 + \left( \frac{\partial A(x, t)}{\partial x} \right)^2 \right] \, .
\end{equation}
Once we know the Hamiltonian, we also know the process for quantisation. We promote the variables $p, q, P_k, Q_k$ to operators, which satisfy the commutator relations
\begin{align}
&[\hat q, \hat Q_j ] = [\hat q, \hat P_k ] = [\hat p, \hat Q_j] = [\hat p, \hat P_k] = 0  \, ,\nonumber \\
&[\hat q, \hat p ] = i \hbar \, ,\nonumber \\
&[\hat Q_j, \hat P_k] = i \delta_{jk} \hbar \, .
\end{align}
We can then define the annihilation and creation operators of the cavity field as
\begin{align}
\hat a_k (\hat q) &= \sqrt{\frac{1}{2 \hbar \omega_k(\hat q)}} \left( \hat \omega_k (\hat q) \, \hat Q_k + i \, \hat P_k \right) \, , \nonumber \\
\hat a_k^\dag (\hat q) &= \sqrt{\frac{1}{2 \hbar \omega_k(\hat q) }} \left(\hat \omega_k ( \hat q) \, \hat Q_k - i \, \hat P_k \right) \, , 
\end{align}
where as in the previous Section, we denoted $\hat \omega(\hat q)$ with a hat to emphasize that this is an operator.

The quantum Hamiltonian now reads
\begin{equation} \label{app:Hamiltonian:eq:definition:of:mirror:Hamiltonian}
\hat H = \frac{( \hat p + \hat \Gamma)^2}{2m  } + V(\hat q) + \hbar \sum_k \hat \omega_k (\hat q) \left[ \hat a^\dag _k \hat a_k + \frac{1}{2} \right] \, ,
\end{equation}
where
\begin{equation}
\hat \Gamma \equiv \frac{i \hbar }{2 \hat q} \sum_{kj} g_{kj} \left[ \frac{k}{j} \right]^{1/2} \left[ \hat a_k^\dag \hat a_j^\dag - \hat a _k \hat  a_j + \hat a_k^\dag \hat a _j - \hat a_j^\dag \hat a_k \right] \, .
\end{equation}
where we have used the short notation $\hat a_k = \hat a_k (q)$.  $\hat \Gamma$ can be seen as an effective momentum term which describes the contribution from mixing between the modes. 

The vacuum field energy that appears in Eq. \eqref{app:Hamiltonian:eq:definition:of:mirror:Hamiltonian} is divergent. We can follow the usual procedure to obtained the Casimir energy, which is detailed in Ref \cite{plunien1986casimir}. For a one-dimensional space, we find
\begin{equation}
\hat H = \frac{(\hat p + \hat \Gamma)^2}{2m} + V(q) + \hbar \sum_k \hat \omega_k (\hat q) \hat a^\dag_k \hat a _k - \frac{\hbar \pi}{24 \hat q} \, .
\end{equation}
To obtain the familiar form of the optomechanical Hamiltonian in Eq. \eqref{chap:introduction:eq:standard:nonlinear:Hamiltonian}, we consider a single mode, which means that $\hat \Gamma = 0$. We find
\begin{equation}
\hat H = \frac{\hat p^2}{2m} + V(q) + \hbar \,  \hat \omega_{\rm{c}} (\hat q) \hat a^\dag \hat a  - \frac{\hbar \pi}{24 \hat q} \, , 
\end{equation}
where $\hat \omega_{\rm{c}}$ is the frequency of the single populated mode. If we now assume that the Casimir force is so weak that it is negligible, 
and consider only first order corrections to $\hat \omega_{\rm{c}}$, analogous to the previous Section, becomes
\begin{equation}
\hat \omega_{\rm{c}} = \frac{w}{L + \hat x} \approx \frac{w}{L} - \frac{w}{L^2} \hat x^2 \, ,
\end{equation}
where we used $\hat q = L + \hat x$ for a small displacement $\hat x$ around the equilibrium position $L$ and assumed that $\hat x \ll L$, and where $w$ is a velocity. 

If we then assume that the potential $V(\hat q)$ is harmonic with 
\begin{equation}
V(\hat q) = m \, \omega_{\rm{m}}^2 \, \hat x^2 \, ,
\end{equation} 
we arrive at the familiar expression for the Hamiltonian:
\begin{equation}
\hat H = \hbar \omega_{\rm{c}} \hat a^\dag \hat a + \hbar \omega_{\rm{m}} \hat b^\dag \hat b - g_0 \hat a^\dag \hat a \left( \hat b^\dag + \hat b \right) \, ,
\end{equation}
where we defined 
\begin{equation}
g_0 := \frac{\omega_{\rm{c}}}{L} \sqrt{\frac{\hbar}{2 m \omega_{\rm{m}}}} \, . 
\end{equation}
This concludes our derivation of the optomechanical Hamiltonian for a moving-end mirror system.

\section{Hamiltonian of a levitated nanosphere} \label{app:optomechanical:hamiltonian:nanosphere}

In this Appendix, we show how the time-dependent term used in Section~\ref{chap:non:Gaussianity:coupling:sec:time:dependent:coupling} in Chapter \ref{chap:non:Gaussianity:coupling} can be derived for levitated nanobead systems.  In~\cite{romero2011optically}, a fully general theory of light--matter coupling is presented. We will recount some of the derivation here and show how the cavity volume can be modulated in a manner such that it is useful to our scheme. 

Given a number of assumptions regarding the light--matter interaction (see~\cite{romero2011optically} for a full description) the full Hamiltonian that describes the light--matter interaction for a homogeneous dielectric object is the following: 
\begin{equation}
\hat{H}^{\mathrm{tot}} = \hat{H}^f_m + \hat{H}^f_c + \hat{H}^f_c + \hat{H}^f_{\mathrm{out}} + \hat{H}^f_{\mathrm{free}} + \hat{H}^i_{\mathrm{cav-out}} + \hat{H}^i _{\mathrm{diel}}. 
\end{equation}
The term $\hat{H}^f_m = \hat{p}^2/2M$, where $M$ is the total mass of the system, is the kinetic energy of the centre-of-mass position along the cavity axis. $\hat{H}_c^F = \hbar \omega_{\mathrm{c}}\hat{a}^\dag \hat{a}$ is the energy of the cavity mode. $\hat{H}^f_{\mathrm{out}}$ and $\hat{H}^f_{\mathrm{free}}$ are terms describing an open system, which we shall ignore in this derivation. We likewise ignore $\hat{H}^i_{\mathrm{cav-out}}$ which describes a coupling between the cavity input and the output mode. 

The last term $\hat{H}^i_{\mathrm{diel}}$ describes the light--matter coupling and can be written in the general form
\begin{equation} \label{app:def:dielectric:hamiltonian}
\hat{H}^i_{\mathrm{diel}} = - \frac{1}{2}\int_{V(\vec{r})} \mathrm{d} \vec{x} \, P(\vec{x}) \, \hat{E}(\vec{x}), 
\end{equation}
where $P(\vec{x})$ is the polarization of the levitated objects (which we assume to be a scalar quantity) and $\hat{E}(\vec{x})$ is the total electric field, which can be obtained from solving Maxwell's equations given a set of well-defined boundary conditions. The quantised modes of the electric field can thus be written as~\cite{serafini2017quantum}
\begin{equation} \label{app:def:electric:field}
\hat{E}(\vec{x}) = i \sum_{s,m} E_{m} \left( \hat a_{s, m}\,  - \hat a^\dag_{s,m} \right) \chi_{s,m}(\vec{x}), 
\end{equation}
where $s$ is the spin-polarization index and $m$ signifies the field-mode number, and $E_m = \sqrt{\frac{\omega_{\mathrm{m}} \hbar}{2 \epsilon_0 \, V_{\mathrm{c}}}}$ is the field amplitude with $V_{\mathrm{c}}$ being the cavity mode volume. The functions $\chi_{s,m} $ must obey the spatial solutions to the wave-equations, where the full classical solutions separate into $\vec{E}( \vec{r}, t) = \chi ( \vec{r}) \, T(t)$. 

If we assume that the polarization is given by $P(\vec{x}) = \epsilon_{\mathrm{c}} \epsilon_0  E(\vec{x})$, we obtain the simpler expression 
\begin{equation} \label{app:def:dielectric:hamiltonian:simple}
\hat{H}^i _{\mathrm{diel}} = - \frac{\epsilon_{\mathrm{c}} \epsilon_0}{2} \int_{V(\vec{r})} \, \mathrm{d} \vec{x}\, [\hat{E}(\vec{x})]^2, 
\end{equation}
where  $\epsilon_{\mathrm{c}} = 3 \frac{ \epsilon_r - 1}{\epsilon_r + 2}$, and where $\epsilon_r$ is the relative dielectric constant of the nanodiamond. 

We now assume that the electric field operators are displaced by a classical part: $\hat{a} \rightarrow \braket{\hat{a}_0} + \hat{a}$. The classical part $ \braket{\hat{a}_0}$ will form the optical trapping field, while the quantum part describes the light--matter interaction. 

Thus the classical contribution to the electrical field is given by 
\begin{equation}
\mathcal{E}(\vec{x}) = i \sqrt{\frac{\omega_{\mathrm{c}}}{2\epsilon_0 V_{\mathrm{c}}}} \left( \alpha f(\vec{x}) - \alpha^* f^*(\vec{x}) \right), 
\end{equation}
where $\alpha$ is a complex prefactor and $f(\vec{x})$ is a complex function which describe the standing waves inside the cavity. We now write our full electric field as $\hat{E}_{\mathrm{tot}}(\vec{x}) = \hat{E}(\vec{x}) + \mathcal{E}(\vec{x})$, where $\hat{E}(\vec{x})$ is the quantum contribution containing $\hat{a}$ and $\hat{a}^\dag$, and $\mathcal{E}(\vec{x})$ is the classical part.  The full Hamiltonian is now 
\begin{align}
 \hat{H}^i _{\mathrm{diel}} &= - \frac{\epsilon_{\mathrm{c}} \epsilon_0}{2} \int_{V(\vec{r})} \, \mathrm{d} \vec{x}\, [\hat{E}(\vec{x}) + \mathcal{E}(\vec{x})]^2 \nonumber \\
 &=  - \frac{\epsilon_{\mathrm{c}} \epsilon_0}{2} \int_{V(\vec{r})} \, \mathrm{d} \vec{x}\, [\hat{E}^2(\vec{x}) + \mathcal{E}^2(\vec{x}) + 2\hat{E}(\vec{x}) \mathcal{E} (\vec{x})] \, . 
\end{align}
The classical contribution, $\mathcal{E}(\vec{x})$ will yield a trapping frequency, while the operator terms $\hat{E}(\vec{x})$ will yield the light--matter interaction term for the levitated sphere. The cross-term, $\hat{E}(\vec{x}) \mathcal{E}^2 (\vec{x})$ will generate elastic scattering processes inside the cavity which converts cavity photons and tweezer photons into free modes~\cite{romero2011optically}. We shall ignore them here and focus on the generation of the trapping frequency $\omega_{\mathrm{m}}$ and the coupling $g(t)$. We begin with the trapping frequency. 

\subsection{Mechanical trapping frequency}

We now assume that the classical field has a Gaussian profile which extends in the $y$-direction for a cylindrical geometry. The cavity extends along the $z$-direction. We here follow the derivation presented in~\cite{clemente2010magnetically}. 

If we denote the radius of the cylinder by $r$, we can write down the trapping field as 
\begin{equation}
\mathcal{E}(y, r) = E_0 \frac{W_0}{W(y)} \exp[ - \frac{r^2}{W^2(y)} ] \, , 
\end{equation}
where $E_0 = \sqrt{\frac{P_t}{\epsilon_0 c \pi W_0^2} }$, $P_t$ is the trapping laser power and $W_0$ is the beam waist with the full beam as a funtion of $y $ being $W(y) = W_0 \sqrt{1 + \frac{y^2 \lambda^2}{\pi^2 W_0^4}}$. It  follows that the narrowest part of the beam $W_0$ occurs at $y = 0$, which is the minimum in the potential where the nanobead is trapped. 

We can now expand $[\mathcal{E}(y,r)]^2$ to second order in $r$ and $y$ around the origin $y_0 = r_0 = 0$. We start with the exponential, which we expand as
\begin{equation}
[\mathcal{E} (y,r)]^2 \approx E_0^2 \frac{W_0^2}{W^2(y)} \left( 1 - 2\frac{r^2}{W^2(y)} \right) \, . 
\end{equation}
Next, we expand the inverse beam width to second order in $y$:
\begin{equation}
\frac{1}{W^2(y)} \approx  \frac{1}{W_0^2} \left( 1 - \frac{y^2 \lambda^2}{2 \pi^2 W_0^4} \right) \, . 
\end{equation}
Combining the two expressions give us
\begin{align}
[\mathcal{E}(y, r)]^2 &\approx E_0^2 \left( 1  - \frac{y^2 \lambda^2}{2\pi^2 W_0^4} \right) \left( 1 - 2 \frac{r^2}{W_0^2} \left( 1 - \frac{y^2 \lambda^2}{2 \pi^2 W_0^2} \right) \right) \nonumber \\
&\approx E_0^2 - \frac{E_0^2 \, y^2 \, \lambda^2 }{\pi^2 \, W_0^2} + r^2 E_0^2 \left( \frac{4 y^2 \, \lambda^2}{\pi^2 \, W_0^4} - \frac{2}{W_0^2}\right) \, . 
\end{align}
If we now assume that $y \ll W_0$, meaning that the beam waist is much larger than the region we consider, we can approximate the above as
\begin{equation}
[\mathcal{E}(y,r)]^2 \approx E_0^2 - r^2E_0^2 \frac{2}{W_0^2} \, . 
\end{equation}
We then insert this now constant expression into the integral for the Hamiltonian and we drop all constant terms as they are just constant energy shifts. To perform this integral, we assume that the radius $R$ of the bead is much smaller than the wavelength of the light. This is often referred to as the `point--particle approximation', or the Rayleigh approximation. Essentially, this means that the field inside the bead is constant (although the field still changes in space with $x$ and $y$). Thus we can assume that wherever the sphere is located in the field, the integral just simplifies to the volume of the sphere times the field amplitude. For a derivation which includes arbitrary particle sizes, see~\cite{pflanzer2012master}. 

This gives 
\begin{align}
H_{\mathrm{trap}} \approx   \frac{\epsilon_{\mathrm{c}}}{2} \int_{V(\vec{r})} \mathrm{d} \vec{x} \, r^2 E_0^2 \frac{2}{W_0^2} \approx r^2\frac{\epsilon_{\mathrm{c}} E_0^2 }{W_0^2} V \, , 
\end{align}
where $V$ is the integration volume. The result is a harmonic trapping of the form 
\begin{equation}
\frac{1}{2} m \omega_{\mathrm{m}}^2 r^2 = \frac{\epsilon_{\mathrm{c}} E_0^2 }{W_0^2} Vr^2 \, , 
\end{equation}
where we identify the trapping frequency as 
\begin{equation}
 \omega_{\mathrm{m}}^2 = \frac{2}{m} \frac{\epsilon_{\mathrm{c}} E_0^2 }{W_0^2}V = \frac{12 I m }{\rho c \epsilon_{\mathrm{c}} W_0^2} \left( \frac{\epsilon_r - 1}{\epsilon_r + 2} \right) \, , 
\end{equation}
where $\rho = \frac{m}{V}$ is the density of the levitated object and where we have used $E_0^2 = \frac{2 I }{c \epsilon_0 }$, where $I$ is the intensity of the laser beam, and $\epsilon_{\mathrm{c}} = 3 \frac{ \epsilon_r - 1}{\epsilon_r + 2}$. 

\subsection{The light--matter interaction term}

We now come to the most important term, which is the light--matter interaction term denoted $g_0$ in this Appendix. We will continue to follow the derivation in Ref~\cite{romero2011optically} to show exactly where time--dependence could potentially be included. 

If the sphere is sufficiently small, we can choose a TEM 00 (transverse electromagnetic mode) as the cavity mode, which is aligned in the $z$-direction. In this mode, the cross-section in $x$ and $y$ is perfectly Gaussian, and it is one of the most commonly used modes in experiments. If the sphere is smaller than the laser waist and if it is placed close to the centre of the cavity, we can approximate the field at the centre of the beam by 
\begin{equation}
[E(\vec{x})]^2 \approx \frac{\omega_{\mathrm{c}}}{2 \epsilon_0 V_c} \left( 1 - \frac{2( x^2 + y^2)}{W_c^2} \right) \cos^2{(k_c z - \varphi)} \, \hat{a}^\dag \hat{a} \, . 
\end{equation}
Here, the laser waist is given by $W_c = \sqrt{\frac{\lambda L}{(2 \pi)^2}}$,  $L$ is the cavity length. $\lambda$ is the laser wavelength. We assume that the wave-vector $\vec{k}_c$ points in the $z$-direction, along the axis of the cavity, and $\varphi$ is a generic phase which determines the minimum of the potential seen by the bead. For laser-trapped nanobeads, this phase can be made time-dependent, whereas for a Paul trap, it is static. We will leave out the time-dependence for now for notational simplicity. Finally, $\hat{a}$ and $\hat{a}^\dag$ are the annihilation and creation operators of the electromagnetic field. 

To obtain the Hamiltonian term, we now integrate over the full energy within the volume of the nanobead. For a bead situated at  $\vec{r} = (x,y,z)$ leads to 
\begin{align}
\hat{H}_{\mathrm{diel}} &= - \frac{\epsilon_c \epsilon_0}{2} \int_{V(\vec{r})} \mathrm{d} \vec{x} \, [E(\vec{x})]^2 \nonumber \\
&= - \frac{\epsilon_c \epsilon_0}{2} \int_{V(\vec{r})} \mathrm{d} \vec{x} \, \frac{\omega_{\mathrm{c}}}{2 \epsilon_0 V_c} \left( 1 - \frac{2( x^2 + y^2)}{W_c^2} \right) \cos^2{(k_c z - \varphi)} \hat{a}^\dag \hat{a} \, . 
\end{align}
We now assume that the radius of the sphere $R$ is much smaller than the wavelength of the light, such that $k_c R \ll 1$. As mentioned above, this is the `point--particle approximation', or the Rayleigh approximation. 
Thus the integral simplifies to 
\begin{equation}
\hat{H}_{\mathrm{diel}}  =  - \frac{\epsilon_c \epsilon_0}{2} \int_{V(\vec{r})} \mathrm{d} \vec{x} \, \frac{\omega_{\mathrm{c}}}{2 \epsilon_0 V_c} \left( 1 - \frac{2( x^2 + y^2)}{W_c^2} \right) \cos^2{(k_c z - \varphi)} \, \hat{a}^\dag \hat{a} = \omega_c f(\vec{r}) \,  \hat{a}^\dag \hat{a} \, , 
\end{equation} 
where we have defined the function $f(\vec{r})$ as 
\begin{align} \label{def:frequency:function}
f(\vec{r}) = - \frac{V \epsilon_c}{4 V_c } \left( 1 - \frac{2 (x^2 + y^2)}{W_c^2}\right) \cos^2{(k_c z - \varphi)} \, . 
\end{align}
Now, we assume that the sphere is trapped at position $\vec{r}_0 = (x_0, y_0, z_0)^{\mathrm{T}}$, which we take to be the origin of the cavity with $x_0 = 0$, $y_0 = 0$ and $z_0 = 0$. For small perturbations to $z$, which we will later quantize, we can expand~\eqref{def:frequency:function} around $z_0 = 0$ to first order. For this to be valid, we must also expand $\varphi$ to first order. We write
\begin{align}
\cos^2{(k_c z - \varphi)} &= [\cos(k_c z) \cos(\varphi) + \sin(k_c z) \sin(\varphi)]^2\nonumber \\
&\approx \left[ \left( 1 - \frac{k_c^2 z^2}{2} \right) \left( 1 - \frac{\varphi^2}{2} \right) + k_c z \varphi \right]^2 \nonumber \\
&\approx \left[ 1 + k_c z \varphi \right]^2 \nonumber \\
&\approx 1 +2 k_c z \varphi \, . 
\end{align}
We note the linearised $z$-coordinate here, which will later become our quantum operator. We can then write down the full expression
\begin{equation}
\omega_c f(\vec{r}) \hat{a}^\dag \hat{a} = -\omega_c \frac{V \epsilon_c}{4 V_c}\left( 1 + 2k_c z \varphi  \right) \hat{a}^\dag \hat{a} \, ,. 
\end{equation}
From this term, we note that the light-interaction yields a constant reduction of the cavity resonant frequency $\omega_c$ of the form
\begin{equation}
\omega_c \rightarrow \tilde{\omega}_c = \omega_c \left( 1 - \frac{\epsilon_c V}{4 V_c}  \right) \, . 
\end{equation}
The first-order correction in $z$ can now be quantised by promoting $z$ to an operator $z \rightarrow \hat{z} = \sqrt{\frac{\hbar}{2\omega_m m}} (\hat{b}^\dag + \hat{b})$ so that we find the interaction term
\begin{equation}
\hat{H}_{int} = -\omega_c \frac{V \epsilon_c}{2 V_c}  k_c \varphi \hat{z} \, . 
\end{equation}
We now use the fact that $k_c = \frac{\omega_c}{c}$ to write
\begin{equation}
\hat{H}_{int} = - \sqrt{\frac{\hbar}{2 \omega_m m }}\frac{ \omega_c^2 V \epsilon_c \varphi }{2 V_c c} \, \hat{a} ^\dag \hat{a} \left( \hat{b}^\dag + \hat{b} \right) \, , 
\end{equation}
where we can define the final expression for the light--matter coupling:
\begin{equation}
g_0 \equiv \sqrt{\frac{\hbar}{2 \omega_m m }} \frac{ \omega_c^2 V \epsilon_c \varphi }{2 V_cc} \, . 
\end{equation}
In all traps, optical and Paul traps, the bead is trapped in a minimum of the potential. This occurs at $\varphi = \frac{\pi}{2}$. 

In optical traps, we can now modulate $\varphi \rightarrow \varphi(t)$, to change the light--matter coupling. If we let $\varphi(t) = \frac{\pi}{2} \left( 1 + \epsilon \sin{\omega_g \, t} \right)$, we obtain the scenario we investigate in Section~\ref{chap:non:Gaussianity:coupling:sec:time:dependent:coupling} in Chapter~\ref{chap:non:Gaussianity:coupling}. Finally, we note that there might be many additional ways in which the coupling can be modulated that we have not discussed in this thesis.



\chapter{Coefficients}
\label{app:coefficients}

In this Appendix, we list the $F$ and $J$-coefficients that are used throughout this thesis. They constitute the basis for any dynamical computation. 

The $F$-coefficients are obtained through solving the integrals listed in Eq. \eqref{chap:decoupling:eq:sub:algebra:decoupling:solution} in Chapter \ref{chap:decoupling}. Solving the integrals first requires determining the function $\xi(\tau)$ by solving the differential equations in Eq. \eqref{chap:decoupling:differential:equation:written:down} for $P_{11}$ and $P_{22}$, or alternatively $I_{P_{22}} = \int^\tau_0 \mathrm{d}\tau' \, P_{22}(\tau')$ in Eq. \eqref{chap:decoupling:eq:IP22}, depending on the problem at hand. The solutions for $P_{11}$ and $P_{22}$ can in turn be used to determine the $J$-coefficients through their relation to the Bogoliubov coefficients shown in  Eq. \eqref{chap:decoupling:eq:squeezing:relation}. Alternatively, the coupled differential equations in Eq. \eqref{chap:decoupling:eq:diff:equations:Js} can be solved to determine $J_b$ and $J_\pm$ directly. 

The solutions to the differential equations are in turn determined by the form of the weighting functions in the Hamiltonian in Eq. \eqref{chap:decoupling:eq:Hamiltonian}. The functions are $\mathcal{G}(t)$, which determines the form of the optomechanical coupling, $\mathcal{D}_1(t)$, which determines the form of a mechanical displacement term, and $\mathcal{D}_2(t)$, which determines the form of a single-mode mechanical squeezing term. 

In this Appendix, as in many other parts of the thesis, we present the coefficients in terms of the rescaled time $\tau = \omega_{\rm{m}} \, t$. This implies that  $\tilde{\mathcal{G}}(\tau) = \mathcal{G}(t)/\omega_{\rm{m}}$,  $\tilde{\mathcal{D}}_1(\tau) = \mathcal{D}_1(t)/\omega_{\rm{m}}$ and $\tilde{\mathcal{D}}_2( \tau) = \mathcal{D}(t)/\omega_{\rm{m}}$. 

\section{Coefficients for a constant and time-dependent nonlinear coupling}\label{app:coeff:time:dependent:c1}
Here, we list the coefficients for the case when $\tilde{\mathcal{D}}_1(\tau) = \tilde{\mathcal{D}}_2(\tau) = 0$, and for different forms of  $\tilde{\mathcal{G}}(\tau)$. 

\subsection{Constant coupling}
We begin by considering a constant coupling $\tilde{\mathcal{G}}(\tau ) \equiv \tilde{g}_0$. We find the coefficients
\begin{align} \label{app:coefficients:constant:g0}
F_{\hat N_a^2 } &=- \tilde{g}_0^2  \, \bigl[1-\textrm{sinc}(2\tau)\bigr]\,\tau \, ,  \nonumber \\
F_{\hat N_a \, \hat B_+} &= - \tilde{g}_0  \,  \sin{( \tau )} \, ,  \nonumber \\
F_{\hat N_a \, \hat B_-} &= \tilde{g}_0 \, \bigl[ \cos{(\tau)}  -1\bigr] \, .
\end{align}
The remaining coefficients are zero: $F_{\hat N_a} = F_{\hat B_+} = F_{\hat B_-} = 0$. Since $\tilde{\mathcal{D}}_2(\tau) = 0$, we find $J_b = \tau$ and $J_\pm = 0$. 

The coefficients in Eq. \eqref{app:coefficients:constant:g0} can be used to derive the time-evolution operator $\hat U(\tau)$ used in Refs \cite{bose1997preparation} and \cite{mancini1997ponderomotive}. 

\subsection{Time-dependent and resonant coupling}
We proceed to consider the dynamics when $\tilde{\mathcal{G}}(\tau) = \tilde{g}_0 ( 1 + \epsilon \,  \sin( \Omega_g \tau)  )$ and $\tilde{\mathcal{D}}_1(\tau) = \tilde{\mathcal{D}}_2(\tau) = 0$.  The coefficients are given by 
\begin{align} \label{app:coefficients:eq:time:dependent:g0}
F_{\hat{N}_a^2} &= -\tilde{g}_0^2
\bigl[ \tau -\sin (\tau) \cos (\tau) \bigr] +2 \, \epsilon \frac{\tilde{g}_0^2}{ \Omega_g }  \biggl[ \sin ^2( \tau ) \cos (\Omega_g \tau   )- 2 \sin^2 \left(\frac{\tau}{2} \right)\biggr]\nonumber \\
&- \epsilon \frac{\tilde{g}_0^2}{ \Omega_g ( 1 + \Omega_g)} \,   \sin (2\tau ) \sin ( \Omega_g \, \tau )- \epsilon \frac{4\tilde{g}_0^2}{ \Omega_g ( 1 - \Omega_g^2)} \,  \cos (\tau ) \sin^2\left( \frac{(1 - \Omega_g)\tau}{2} \right)
\nonumber \\
&+ \epsilon^2 \, \frac{\tilde{g}_0^2}{ 4 \, \Omega_g ( 1 + \Omega_g)} \, \left( 2 \, \tau-4 \sin (\tau) \cos ( \Omega_g \, \tau ) (\cos (\tau) \cos ( \Omega_g \, \tau )-2)\right) \nonumber \\
&+ \epsilon^2 \, \frac{\tilde{g}_0^2}{4 \, \Omega_g ( 1 - \Omega_g^2)} \, \biggl( 4 \, \sin (\tau) \cos ( \Omega_g \, \tau ) (\cos (\tau) \, \cos ( \Omega_g \, \tau )-2)+8 \cos (\tau) \,  \sin ( \Omega_g \, \tau )\nonumber \\
&\quad\quad\quad\quad\quad\quad\quad\quad\quad\quad+(1-2 \, \cos (2\, \tau)) \sin (2\,  \Omega_g \, \tau ) - 2 \, \tau\biggr) \nonumber \\
&+ \epsilon^2 \, \frac{\tilde{g}_0^2}{4 \, \Omega_g \, (1 - \Omega_g^2)^2} \biggl( 8 \, \Omega_g \,   \sin (\tau)\,  \cos ( \Omega_g \, \tau )-2 \, \Omega_g\,  \sin (2 \, \tau) \, \cos (2 \, \Omega_g \, \tau ) \nonumber \\
&\quad\quad\quad\quad\quad\quad\quad\quad\quad\quad-8 \cos (\tau) \sin (\Omega_g \, \tau)+2 \cos (2 \, \tau) \, \sin (2 \, \Omega_g \, \tau ) \biggr) \, ,\nonumber \\
F_{\hat N_a \, \hat{B}_+} &= -\frac{\tilde{g}_0}{1 + \Omega_g} \, \epsilon \sin(\tau) \sin(\Omega_g \, \tau) + \frac{2 \, \Omega_g \, \tilde{g}_0}{1 - \Omega_g^2} \, \epsilon \, \sin^2 \left( \frac{( 1 - \Omega_g)\tau}{2} \right) -\tilde{g}_0 \, \sin (\tau)\, ,\nonumber \\
F_{\hat N_a \, \hat{B}_- } &=  - \frac{\tilde{g}_0 }{1 - \Omega_g} \, \epsilon \,  \sin (\tau) \, \cos ( \Omega_g \, \tau) + \frac{\tilde{g}_0}{1 - \Omega_g^2} \, \epsilon \, \sin((1 + \Omega_g)\tau)  - 2 \, \tilde{g}_0 \, \sin^2 \left( \frac{\tau}{2} \right)  \, .
\end{align}
At resonance with $\Omega_g = 1$, these coefficients are given by 
\begin{align}\label{app:coefficients:eq:g0:resonance}
F_{\hat N_a^2} &=-\frac{1}{16}  \tilde{g}_0^2 \, \bigl[ 16 \, \tau-8 \sin (2\, \tau)+\epsilon \, (32-36 \cos (\tau)+4 \cos (3 \, \tau))\nonumber \\
&\quad\quad\quad\quad\quad\quad\quad +\epsilon ^2 \,   \bigl( 6 \, \tau-4 \sin (2\, \tau)+\sin (2 \, \tau)\,\cos (2 \, \tau) \bigr)\bigr]\, ,\nonumber\\
F_{\hat N_a \, \hat B_+} &= -\tilde{g}_0 \sin (\tau) \left(1+ \frac{\epsilon}{2} \sin (\tau)\right)  \, ,\nonumber \\
F_{\hat N_a \, \hat B_-} &=\frac{\tilde{g}_0}{4} \epsilon   \,  \left(  \sin (2 \, \tau)-2 \, \tau  \right) -  2  \, \tilde{g}_0 \,  \sin^2\left( \frac{\tau}{2} \right) \, .
\end{align}
Since $\tilde{\mathcal{D}}_2(\tau) = 0$, we find $J_b = \tau$ and $J_\pm = 0$. 

\section{Coefficients for a linear displacement term}
In this Section, we list the $F$ and $J$-coefficients for a constant optomechanical coupling 
\begin{equation}
\tilde{\mathcal{G}}(\tau) \equiv \tilde{g}_0 \, ,
\end{equation}
for different forms of $\tilde{\mathcal{D}}_1$. We set the mechanical squeezing to zero with $\tilde{\mathcal{D}}_2(\tau)  = 0$. 

\subsection{Constant displacement}
The coefficients for a constant mechanical displacement 
\begin{equation}
\tilde{\mathcal{D}}_1(\tau) \equiv \tilde{d}_1 \, ,
\end{equation}
are given by 
\begin{align} \label{app:coefficients:F:coefficients:constant:d1}
F_{\hat N_a} &= \, \tilde{g}_0 \, \tilde{d}_1 \left( \tau - \cos(\tau)\, \sin(\tau) \right) \, ,\nonumber \\
F_{\hat N_a^2} &= \frac{1}{2} \tilde{g}_0 ^2 \left( \sin(2 \, \tau) - 2 \, \tau \right) \, ,\nonumber \\
F_{\hat B_+} &= \tilde{d}_1 \, \sin(\tau) \, ,\nonumber \\
F_{\hat B_-} &= \tilde{d}_1 \, \left( 1- \cos(\tau)\right) \, ,\nonumber \\
F_{\hat N_a\, \hat B_+} &= - \tilde{g}_0 \, \sin(\tau) \, ,\nonumber \\
F_{\hat N_a, \hat B_-} &= - \tilde{g}_0 \, \left( 1 - \cos(\tau) \right) \, .
\end{align}
Since $\tilde{\mathcal{D}}_2(\tau) = 0$, we find $J_b = \tau$ and $J_\pm = 0$. 
\subsection{Time-dependent and resonant displacement}

We here print the $F$-coefficients for a time-dependent linear displacement term 
\begin{equation}
\tilde{\mathcal{D}}_1(\tau) = \tilde{d}_1 \, \cos(\Omega_{d_1} \, \tau) \, .
\end{equation}
For this time-dependent displacement, the $F$-coefficients are 
\begin{align} \label{app:coefficients:F:coefficients:time:dependent:d1}
F_{\hat N_a} =&-\frac{\tilde{g}_0 \, \tilde{d}_1 }{2 \,  \Omega_{d_1} \left(\Omega_{d_1}^2-1\right)}\,\biggl[ 2 \,  \Omega_{d_1}^2 \cos ^2(\tau) \,  \sin (\Omega_{d_1} \, \tau)-4 \,  \Omega_{d_1} \sin (\tau) \cos (\tau) \cos (\Omega_{d_1} \tau )  \nonumber \\
&\quad\quad\quad\quad\quad\quad\quad\quad\quad+\sin (\Omega_{d_1} \, \tau) \left(\Omega_{d_1}^2 \, \cos (2\,  \tau )-3 \, \Omega_{d_1}^2+4\right)\biggr] \, , \nonumber \\
F_{\hat N_a^2} =& \, \frac{1}{2} \tilde{g}_0^2 \left( \sin (2 \, \tau)-2 \, \tau \right)\, ,\nonumber \\
F_{\hat B_+} =&  - \tilde{d}_1 \, \frac{\Omega_{d_1} \,  \cos (\tau)  \, \sin (\Omega_{d_1} \, \tau )-\sin (\tau)  \, \cos (\Omega_{d_1} \, \tau )}{1 - \Omega_{d_1}^2} \, ,\nonumber \\
F_{\hat B_-} =&  - \tilde{d}_1 \, \frac{\Omega_{d_1} \,  \sin (\tau ) \sin (\Omega_{d_1} \, \tau)+\cos (\tau) \,  \cos (\Omega_{d_1} \, \tau )-1}{1 - \Omega_{d_1}^2}\, , \nonumber \\
F_{\hat N_a \, \hat B_+}= & -\tilde{g}_0 \, \sin (\tau) \, ,\nonumber \\
F_{\hat N_a \, \hat B_-} =& \, \tilde{g}_0 \,  (\cos (\tau)-1)  \, . 
\end{align}
At resonance with $\Omega_{d_1} = 1$, the coefficients become
\begin{align} \label{app:coefficients:d1:resonant}
F_{\hat N_a} &=-\frac{1}{4} \tilde{g}_0 \, \tilde{d}_1 \left[ \sin (3 \, \tau)-7 \sin (\tau)+4  \, \tau  \cos (\tau) \right]  \, ,\nonumber \\
F_{\hat N_a^2} &= - \frac{1}{2} \tilde g_0^2 (2\tau - \sin(2\tau)) \, ,  \nonumber\\
F_{\hat B_+} &= \frac{1}{2} \tilde{d}_1 \,  \left[ \tau +\sin (\tau ) \,  \cos (\tau) \right] \nonumber \\
F_{\hat B_-} &= \frac{1}{2} \tilde{d}_1 \,  \sin^2(\tau) \, ,\nonumber \\
F_{\hat N_a \hat B_+} &= -\tilde g_0 \sin(\tau) \, , \nonumber \\
F_{\hat N_a \hat B_-} &= \tilde g_0 (\cos(\tau) - 1) \, ,\, 
\end{align}
Since $\tilde{\mathcal{D}}_2(\tau) = 0$, we find $J_b = \tau$ and $J_\pm = 0$.

\section{Coefficients for a single-mode mechanical squeezing term}
In this Section, we consider constant and time-dependent squeezing. The pertubative solutions to the time-dependent squeezing dynamics can be found in Section \ref{app:mathieu:perturbative:solutions} in Appendix \ref{app:mathieu}, and they are only valid for $\tilde{d}_2 \ll 1$. For consistency, we will assume $\tilde{d}_2 \ll 1$ throughout this appendix, even for estimation of a constant squeezing amplitude. This assumption will also significantly simplify the expressions that follow.

\subsection{Constant squeezing}

When we consider constant squeezing, i.e. $\Omega_{d_2}=0$ with $\tilde{\mathcal{D}}_2=\tilde{d}_2$, 
we find $\xi = \cos( \sqrt{1 + 4 \tilde{d}_2} \tau) + \sin(\sqrt{1 + 4 \tilde{d}_2}\tau)/\sqrt{1 + 4 \tilde{d}_2}$. 
For $\tilde{d}_2 \ll 1$ and $\tilde{d}_2 \tau \sim 1$, this expression approximates to $\xi = e^{-i(1+2\tilde{d}_2)\tau}$. Then, the non-vanishing $F$-coefficients are
\begin{align} \label{chap:coefficients:eq:constant:d2}
F_{\hat N_a^2} &= -\tilde{g}_0^2 \,\frac{ 2(1 + 2\tilde{d}_2)\tau - \sin(2(1 + 2\tilde{d}_2)\tau) }{2(1+2\tilde{d}_2)^2}  \, ,\nonumber\\
F_{\hat N_a \hat B_+} &= -\tilde{g}_0 \, \frac{\sin((1 + 2\tilde{d}_2)\tau) }{1+2\tilde{d}_2} \, ,\nonumber \\
F_{\hat N_a \hat B_-} &= -\tilde{g}_0 \, \frac{1 - \cos((1 + 2\tilde{d}_2)\tau) }{1+2\tilde{d}_2} \, .
\end{align}
To simplify the expressions further, we discard terms proportional to $\tilde{d}_2$, while keeping only terms proportional to $\tilde{d}_2\tau$, since $\tau$ could in principle become very large. We obtain
\begin{align} \label{app:coefficients:constant:d2}
F_{\hat N_a^2} &= -\frac{1}{2}\tilde{g}_0^2 \, \left(  2(1 + 2\tilde{d}_2)\tau - \sin(2(1 + 2\tilde{d}_2)\tau) \right) \, , \nonumber\\
F_{\hat N_a \hat B_+} &= -\tilde{g}_0 \,\sin((1 + 2\tilde{d}_2)\tau) \, ,  \nonumber \\
F_{\hat N_a \hat B_-} &= -\tilde{g}_0 \, (1 - \cos((1 + 2\tilde{d}_2)\tau)) \, .
\end{align}
With the same approximations, and by using the relations in Eq.~\eqref{chap:decoupling:eq:squeezing:relation}, we obtain
\begin{align} \label{app:coefficients:eq:constant:d2:Js:coeffs}
J_+ &=  0 \,,\quad J_- =  0 \, , \quad\rm{and}\quad J_b =  (1 + 2\tilde{d}_2)\tau \, .
\end{align}

\subsection{Resonant time-dependent squeezing} 
In the next step, we will consider the resonant case. Using the approximate solution for $\Omega_{d_2}=2$, which gives the expression of $\xi(\tau)$ in Eq.~\eqref{app:mathieu:eq:RWA:solutions} and small $\tilde d_2$ given in Eq. \eqref{app:mathieu:eq:RWA:solutions} and neglecting all terms proportional to $\tilde d_2$ but keeping expressions proportional to $\tilde d_2 \tau$, we obtain for the non-vanishing  $F$-coefficients
\begin{align} \label{app:coefficients:eq:resonant:d2}
F_{\hat N_a^2} &= \frac{1}{2}\tilde{g}_0^2 \,  \left(    \cosh(2 \tilde{d}_2 \tau)\sin(2\tau) + \sinh(2 \tilde{d}_2 \tau) -2\tau \right)  \, ,\nonumber\\
F_{\hat N_a \hat B_+} &= -\tilde{g}_0 \left( \cosh(\tilde{d}_2 \tau)\sin(\tau) + \sinh(\tilde{d}_2 \tau)\cos(\tau) \right) \, , \nonumber \\
F_{\hat N_a \hat B_-} &= \tilde{g}_0 \left( \cosh(\tilde{d}_2 \tau)\cos(\tau) + \sinh(\tilde{d}_2 \tau)\sin(\tau)  -1 \right) \, .
\end{align}
Furthermore, using the relations between the Bogoliubov coefficients $\alpha$ and $\beta$ and the $J$-coefficients in Eq.  \eqref{chap:decoupling:eq:squeezing:relation}, we find under the same approximations as above, that the $J$-coefficients are given by 
\begin{align} \label{app:coefficients:eq:resonant:d2:Js}
J_+ &=  \frac{1}{2} \tilde{d}_2\tau \,,\quad J_- =  0 \, , \quad\rm{and}\quad
J_b =  \tau \, .
\end{align}
\chapter{Commutator relations and useful properties} \label{app:commutators}

In this Appendix, we list a number of useful commutator relations, expectation values, and operator identities. These identities are used throughout the thesis, especially in Chapter~\ref{chap:metrology}, where we derive the quantum Fisher information for an optomechanical system, and in Appendix~\ref{app:exp:values}, where we  compute the first and second moments of the state. 

This Appendix is structured in the following way: Our goal is for this Appendix to be self-contained, so we begin by reviewing the notation used here and in this thesis in general. We then proceed by deriving expressions for the first and second moments of a state that has evolved under the decoupled Hamiltonian in Eq.~\eqref{chap:decoupling:eq:Hamiltonian}. The next section contains a simple recipe for computing congruence relations (where an operator is multiplied on both sides by another operator), as well as an example of how the method can be used. We then derive a number of commutator relations, which will be used to compute the quantum Fisher information coefficients in Chapter~\ref{chap:metrology}.

\section{Notation}\label{app:notation}
In this Appendix and in the main text, we use the following notation:
\begin{align}\label{app:commutators:operator:Lie:algebra}
	 	\hat{N}^2_a &:= (\hat a^\dagger \hat a)^2 \nonumber \\
	\hat{N}_a &:= \hat a^\dagger \hat a &
	\hat{N}_b &:= \hat b^\dagger \hat b \nonumber\\
	\hat{B}_+ &:=  \hat b^\dagger +\hat b &
	\hat{B}_- &:= i\,(\hat b^\dagger -\hat b) &
	 & \nonumber\\
	\hat{B}^{(2)}_+ &:= \hat b^{\dagger2}+\hat b^2 &
	\hat{B}^{(2)}_- &:= i\,(\hat b^{\dagger2}-\hat b^2) \, . &
	 &  
\end{align}

\section{Basic commutator expressions}\label{app:commutator:basic:commutators}
At the heart of this investigation lies the simple fact that the following commutator relations hold:
\begin{align}
[\hat N_a , \hat a] &= \hat a \, , \nonumber \\
[\hat N_a , \hat a^\dag] &= - \hat a^\dag\, .
\end{align}
This also implies that
\begin{align} \label{app:commutators:commutator:consequence}
\hat a  \, \hat N_a = (\hat N_a + 1) \, \hat  a \, .
\end{align}

\section{Useful properties of the Weyl displacement operator}
We list some useful relations for coherent states and Weyl operators. 

\begin{itemize}
\item Weyl-representation of coherent states:
\begin{equation}
\ket{\alpha} = e^{ \alpha \, \hat a^\dag - \alpha^* \, \hat a} \ket{0} = \hat D(\alpha) \ket{0} \, .
\end{equation}
\item Overlap of two coherent states:
\begin{equation} \label{chap:commutators:coherent:states:overlap}
\braket{\beta |\alpha} = e^{- \frac{1}{2} \left( |\beta|^2 + |\alpha|^2 - 2 \beta^* \alpha \right)} \, .
\end{equation}
\item Combining two Weyl displacement operators:
\begin{equation} \label{app:commutators:weyl:operator:combinations}
\hat D(\alpha) \, \hat D(\beta) = e^{( \alpha \beta^* - \alpha^* \beta)/2}  \, \hat D( \alpha + \beta)
\end{equation}
\item The action of two displacement operators  $\hat D(\alpha)$ on the annihilation and creation operators are
\begin{align} \label{app:commutators:action:of:displacements}
\hat D(\alpha) \, \hat a\, \hat D^\dag(\alpha) &= \hat a - \alpha \, , \nonumber \\
\hat D(\alpha) \, \hat a^\dag \, \hat D^\dag(\alpha) &= \hat a^\dag -  \alpha^* \, .
\end{align}
\end{itemize}

\section{Expectation values}
To predict properties of the system, we must compute a number of properties of the evolved state. In this Section, we list a number of results that are referred to throughout this thesis. The first expectation values are used in the next Appendix~\ref{app:exp:values} to compute the first and second moments of two coherent states that evolve under the decoupled Hamiltonian in Eq.~\eqref{chap:decoupling:eq:Hamiltonian}. The second set of expectation values for a coherent state and thermal state are used in Chapter~\ref{chap:metrology} to compute the quantum Fisher information for a thermal state in the mechanics, and some remaining expressions are used in Chapter~\ref{chap:gravimetry}.

Throughout this Section, we replace the coefficients with $x$ to ensure that the expressions are presented in full generality. We do however keep the notation for the coherent states, which is $\ket{\mu_{\rm{c}}}$ and $\ket{\mu_{\rm{m}}}$, and at times where it is suitable, we replace the general coefficients with the variable names used in the main text, to enable swift computation of the expectation values.

\subsection{Expectation values for two coherent states}
In Chapters~\ref{chap:non:Gaussianity:coupling} and~\ref{chap:non:Gaussianity:squeezing}, we consider two coherent states that evolve under the Hamiltonian in Eq.~\eqref{chap:decoupling:eq:Hamiltonian}. To compute the first and second moments of these states, we must derive expressions for the following quantities. 

\subsubsection{Expectation value of $e^{- i \, x \,  \hat B_+ }  \, e^{- i \, y \,\hat  B_- }$}
We wish to calculate the expectation value of $e^{- i \, x \,  \hat B_+ }  \, e^{- i \, y \,\hat  B_- }$. 
We start by using the composition rule of two Weyl-operators in Eq.~\eqref{app:commutators:weyl:operator:combinations}
\begin{align}
\bra{\mu_{\rm{m}} } e^{- i \, x \,  \hat B_+ }  \, e^{- i \, y \,\hat  B_- } \ket{\mu_{\rm{m}}}
&= \bra{\mu_{\rm{m}}} e^{- i \, x \,  \hat B_+ - i \, y \, \hat B_-} e^{- i \, x \, y} \ket{\mu_{\rm{m}}}\nonumber \\
&=e^{- i x\, y} \bra{\mu_{\rm{m}}} e^{( - i \, x  + y ) \, \hat  b^\dagger - (- i \, x \,  +y )^*\hat  b} \ket{\mu_{\rm{m}}}  \,.
\end{align}
Now define $z = y - i \, x$ to write this as 
\begin{align}
e^{- i \, x \, y} \bra{\mu_{\rm{m}}} \hat D(z) \ket{\mu_{\rm{m}}} &= e^{ - i \, x \, y} \bra{\mu_{\rm{m}}} \hat D( z) \, \hat D(\mu_{\rm{m}}) \ket{0} \nonumber \\
 &= e^{- i \, x \, y} \,  \bra{\mu_{\rm{m}}} e^{(z  \, \mu_{\rm{m}}^* - z^*\,  \mu_{\rm{m}})/2} \ket{z + \mu_{\rm{m}}} \nonumber \\
&= e^{- i \, x \, y}e^{(z \,  \mu_{\rm{m}}^* - z^* \,  \mu_{\rm{m}})/2} e^{- \frac{1}{2} \left( |\mu_{\rm{m}}|^2 + |z + \mu_{\rm{m}}|^2 - 2 \mu_{\rm{m}}^*(z + \mu_{\rm{m}}) \right)} \, .
\end{align}
We denote this full expectation value by 
\begin{equation} \label{app:commutators:EBpBm:exp:value}
\bra{\mu_{\rm{m}} } e^{- i \, x \, \hat B_+ } \, e^{- i\,  y \, \hat B_- } \ket{\mu_{\rm{m}}}  = E_{\hat B_+ \hat B_-}\, ,
\end{equation}
where $E_{\hat B_+ \hat B_-}$ is given by 
\begin{align} \label{app:commutators:definition:of:EB+B-general}
E_{\hat B _+ \hat B_-} = \, &  \mathrm{exp} \biggl[ \frac{1}{2} \biggl( - y^2 - x^2 - 2 \,  i \,  x y - 2\, \mu_{\rm{m}} (y + i  \, x ) + 2 \,  \mu_{\rm{m}}^* ( y- i  \, x ) \biggr) \biggr] \, .
\end{align}
In the main text, and in the next section, we use the notation $x = F_{\hat N_a \, \hat B_+}$ and $y = F_{\hat N_a \, \hat B_-}$. The expression in Eq.~\eqref{app:commutators:definition:of:EB+B-general} becomes
\begin{align} \label{app:commutators:definition:of:EB+B-}
E_{\hat B _+ \hat B_-} =& \,   \mathrm{exp} \biggl[ \frac{1}{2} \biggl( - F_{\hat N _a \hat B_-}^2 - F_{\hat N_a \hat B_+}^2 - 2 \,  i \,  F_{\hat N_a \hat B_-} F_{\hat N_a \hat B_+} \nonumber\\
&\quad\quad\quad\quad- 2\, \mu_{\rm{m}} (F_{\hat N_a \hat B_-} + i F_{\hat N _a \hat B_+} ) + 2 \,  \mu_{\rm{m}}^* ( F_{\hat N_a \hat B_-} - i F_{\hat N_a \hat B_+} ) \biggr) \biggr] \, .
\end{align}
We frequently refer to this expression in the next Appendix and in the main text. 

\subsubsection{Expectation value of $e^{ -  i \, x \,  \hat B_+} \, e^{- i  \, y\, \hat  B_-} \, e^{ -  i \, x\, \hat B_+} \,e^{- i \, y \, \hat B_-}$} 
We wish to calculate the expectation value of $e^{ -  i \, x \,  \hat B_+} \, e^{- i  \, y\, \hat  B_-} \, e^{ -  i \, x\, \hat B_+} \,e^{- i \, y \, \hat B_-}$. This is one of the lengthier derivations. With the substitution, $x = F_{N_a B_+} $ and $y = F_{N_a B_-}$, and through frequent use of the Weyl operator combination rule in Eq.~\eqref{app:commutators:weyl:operator:combinations}, and overlap between two coherent states in Eq.~\eqref{chap:commutators:coherent:states:overlap}, we find
\begin{align}
&\bra{\mu_{\rm{m}}}  e^{ -  i \, x\, \hat B_+} \, e^{- i \, y \, \hat B_-} \, e^{ -  i \, x \, \hat B_+} \,e^{- i \, y \hat B_-}\,  \ket{\mu_{\rm{m}}} \nonumber \\
&=  e^{- i \, x \, y} \,  \bra{\mu_{\rm{m}}} e^{(y - i \, x) \, \hat b^\dagger - (y - i \,  x)^* \, \hat b} e^{(y - i x) \, \hat b^\dagger - (y - i \,  x)^* \, \hat b}  \, e^{- i \, x \,  y} \ket{\mu_{\rm{m}}} \nonumber \\
&=e^{- 2 \, i \, x \, y} \,   \bra{\mu_{\rm{m}}} e^{2 \, (y - i  \, x) \, \hat b^\dagger - 2 \, (y - i \, x)^* \, \hat b}  \ket{\mu_{\rm{m}}} \nonumber \\
&=  e^{- 2 \, i \, x\, y} \,  e^{(y - i \, x) \, \mu_{\rm{m}}^* - (y + i  \, x) \, \mu_{\rm{m}}}  \braket{\mu_{\rm{m}}| \mu_{\rm{m}}  + 2 \, (y- i \, x)} \nonumber \\
&= e^{- 2 \, i \, x\, y} e^{(y - i \, x) \, \mu_{\rm{m}}^* - (y + i \, x) \,  \mu_{\rm{m}}} e^{- \frac{1}{2} \left[ |\mu_{\rm{m}}|^2 +|\mu_{\rm{m}} + 2(y - i \, x)|^2 - 2\mu_{\rm{m}}^*(\mu_{\rm{m}} + 2 \, y - 2 \, i \, x) \right]} \, .
\end{align}
Denoting the expectation value by $E_{\hat B_+ \hat B_-\hat B_+ \hat B_-}$, this long expression can be simplified into:
\begin{align} \label{app:commutators:EBpBmBpBm:general}
E_{\hat B_+ \hat B_-\hat B_+ \hat B_-} &= \mathrm{exp} \left[ - 2 \,\left( x^2 + y^2 + \mu_{\rm{m}} \,  (y + i \, x)  -  \mu_{\rm{m}}^*  \, (y - i \, x) \right)\right] \, .
\end{align}
In the main text, we use the identification $x = F_{\hat N_a \, \hat B_+}$ and $y = F_{\hat N_a \, \hat B_-}$. Inserting these values, we find the expression:
\begin{align} \label{app:commutators:EBpBmBpBm}
E_{\hat B_+ \hat B_- \hat B_+ \hat B_- } &= \mathrm{exp} \biggl[ - 2 \,  \biggl( F_{\hat N_a \hat B_+}^2 + F_{\hat N_a \hat B_-}^2 + i F_{\hat N_a \hat B_+} F_{\hat N_a \hat B_-} \nonumber \\
&\quad\quad\quad\quad\quad\quad+ \mu_{\rm{m}} ( F_{\hat N_a \hat B_-} + i F_{\hat N_a \hat B_+}) - \mu_{\rm{m}}^*( F_{\hat N_a \hat B_-} - i F_{\hat N_a \hat B_+})   \biggr) \biggr] \, .
\end{align}
With the definition of $K_{\hat N_a}^2 = F_{\hat N_a \, \hat B_+}^2 + F_{\hat N_a \, \hat B_-}$, we can write the above as
\begin{equation} \label{app:commutators:EBpBmBpBm:KNa}
E_{\hat B_+ \hat B_- \hat B_+\hat B_-} = e^{-|K_{\hat N_a}|^2} \,  E_{\hat B_+ \hat B_-}^2 \, .
\end{equation}
We also make frequent reference to these expressions in the following Appendix and in the main text. 

\subsubsection{Expectation value of  $e^{- i \, x \, \hat B_+} \, e^{- i\,  y \, \hat B_- } \, \hat b^\dagger$}
We wish to calculate the expectation value of 
\begin{equation}
\bra{\mu_{\rm{m}}} e^{- i \, x\, \hat B_+} \, e^{- i\,  y \, \hat B_- } \, \hat b^\dagger \ket{\mu_{\rm{m}}} \, .
\end{equation}
We begin by combining the two displacement operators through use of Eq.~\eqref{app:commutators:weyl:operator:combinations}: 
\begin{align}
\bra{\mu_{\rm{m}}} e^{- i \, x\, \hat  B_+ } e^{- i \, y \, \hat B_- } \hat b^\dagger \ket{\mu_{\rm{m}}} 
&=e^{- i \, x \, y} \,  \bra{\mu_{\rm{m}}} e^{(y - i \, x) \, \hat  b^\dagger - (y - i \, x)^* \hat b} \hat b^\dagger \ket{\mu_{\rm{m}}}  \, .
\end{align}
We then define $z = y  - i \, x$ and the operator $\hat D(z) = e^{z \, \hat b^\dag - z^* \, \hat b}$. We then insert the identity $\hat D^\dag (z) \hat D(z) = \mathds{1} $ between $\hat b^\dag$ and the coherent state. Using the identity in Eq.~\eqref{app:commutators:action:of:displacements} to displace the operator $\hat b^\dag$, we find 
\begin{align}
e^{- i \, x \, y} \,  \bra{\mu_{\rm{m}}} \hat  D(z) \hat b^\dagger \ket{\mu_{\rm{m}}}  &=e^{-i \, x\, y} \,  \bra{\mu_{\rm{m}}} \hat D(z) \,  \hat b^\dagger \, \hat  D^\dagger(z)  \hat D(z) \ket{\mu_{\rm{m}}} \nonumber \\
&= e^{- i \, x \, y} \,  \bra{\mu_{\rm{m}}} \left( \hat b^\dag - z^* \right) \hat D (z) \ket{\mu_{\rm{m}}} \, \nonumber \\
&= e^{- i \, x \, y} \,\left(  \mu_{\rm{m}}^* - z^* \right)  \bra{\mu_{\rm{m}}}  \hat D (z) \ket{\mu_{\rm{m}}} \, .
\end{align}
However, we now note that 
\begin{equation}
e^{- i \, x\, y} \bra{\mu_{\rm{m}}} \hat D(z) \ket{\mu_{\rm{m}}} = E_{\hat B_+ \hat B_-} \, ,
\end{equation}
as defined in Eq.~\eqref{app:commutators:definition:of:EB+B-general}. 
We therefore find
\begin{equation} \label{app:commutators:exp:value:of:EBpBm:bd}
\bra{\mu_{\rm{m}}} e^{ - i \, x\, \hat B_+}  \, e^{ - i \, y \, \hat B_-} b^\dag \ket{\mu_{\rm{m}}} = \left( \mu_m^* - i \, x - y \right) \, E_{\hat B_+ \hat B_-} \, .
\end{equation}

\subsubsection{Expectation value of  $\hat N_a^2$}
We wish to calculate the expectation value of $\hat N_a^2$. 
We find that
\begin{align} \label{chap:commutators:ada2:exp:value}
\bra{\mu_{\rm{c}}}  \, \hat N_a^2 \,  \ket{\mu_{\rm{c}}} &= |\mu_{\rm{c}}|^2 \bra{\mu_{\rm{c}}} \hat a \hat a^\dag \ket{\mu_{\rm{c}}}  \nonumber \\
&= |\mu_{\rm{c}}|^2 \bra{\mu_{\rm{c}}} ( 1 + \hat N_a ) \ket{\mu_{\rm{c}}} \nonumber \\
&= |\mu_{\rm{c}}|^2 ( 1 + |\mu_{\rm{c}}|^2) \, .
\end{align}

\subsubsection{Expectation value of $e^{- i \, x\, \hat N_a}$}
Now instead acting on the optics, we find
\begin{align} \label{app:commutators:exponential:number:operator}
\bra{\mu_{\rm{c}}} e^{- i  \, x \, \hat N_a } \ket{\mu_{\rm{c}}} &= \bra{\mu_{\rm{c}}} e^{- i  \, x \,  \hat N_a } e^{- |\mu_{\rm{c}}|^2/2} \sum_n \frac{\mu_{\rm{c}}^n }{\sqrt{n!}} \ket{n} \nonumber \\
&= \bra{\mu_{\rm{c}}} e^{  - |\mu_{\rm{c}}|^2 /2} \sum_n \frac{( e^{-i \, x} \mu_{\rm{c}})^n }{\sqrt{n!}}\ket{n} \nonumber \\
&= \braket{\mu_{\rm{c}} | e^{- i  \, x} \mu_{\rm{c}}}\nonumber  \\
&= e^{ - |\mu_{\rm{c}}|^2} e^{|\mu_{\rm{c}}|^2 e^{- i \, x}}  \nonumber \\
&= e^{ |\mu_{\rm{c}}|^2 ( e^{- i \, x}  - 1)} \, .
\end{align}

\subsubsection{Expectation value of $e^{- i \, x\, \hat N_a} \hat a\, \hat N_a$}
From the relation for $\hat N_a$ and $\hat a$ in Eq.~\eqref{app:commutators:commutator:consequence}, we find that
\begin{align} \label{app:commutators:exp:value:Exp:a:Na}
\bra{\mu_{\rm{c}}}e^{- i \, x \, \hat N_a } \hat a \, \hat  N_a \ket{\mu_{\rm{c}}} &= \bra{\mu_{\rm{c}}}e^{- i \, x \, \hat N_a } (\hat N_a + 1) \hat a\ket{\mu_{\rm{c}}} \nonumber  \\
&=  \mu_{\rm{c}} \bra{\mu_{\rm{c}}} e^{- i \, x \, \hat  N_a} (\hat N_a + 1) \ket{\mu_{\rm{c}}} \nonumber \\
&= \mu_{\rm{c}} \bra{\mu_{\rm{c}}} e^{- i \, x \,  \hat  N_a } (\hat N_a + 1 ) e^{ - |\mu_{\rm{c}}|^2/2} \sum_n \frac{\mu_{\rm{c}}^n}{\sqrt{n!}} \ket{n}\nonumber \\
&= \mu_{\rm{c}} \bra{\mu_{\rm{c}}} e^{ - |\mu_{\rm{c}}|^2/2} \sum_n \frac{\mu_{\rm{c}}^n}{\sqrt{n!}} e^{- i \, x \,  n } (n  + 1 ) \ket{n}\nonumber  \\
&= \mu_{\rm{c}} \left( e^{- |\mu_{\rm{c}}|^2/2}  \sum_n \frac{( |\mu_{\rm{c}}|^2 e^{- i \, x } )^n}{n!}  \, n + \braket{\mu_{\rm{c}} | \,  e^{- i \, x \,  \mu_{\rm{c}}} }\right) \nonumber \\
&= \mu_{\rm{c}} \left(  |\mu_{\rm{c}}|^2 e^{|\mu_{\rm{c}}|^2 ( e^{- i \, x } - 1)} e^{- i \, x} + e^{ |\mu_{\rm{c}}|^2 ( e^{- i \, x} - 1 )} \right) \nonumber \\
&= \mu_{\rm{c}} \, e^{ |\mu_{\rm{c}}|^2 (e^{- i \, x} - 1)} \left( |\mu_{\rm{c}}|^2 \, e^{- i \, x} + 1\right) \, .
\end{align}

\subsection{Expectation values for coherent states}
The expectation values of the number operator with respect to coherent states are:
\begin{subequations}
\begin{align} 
\braket{\mu_{\rm{c}}|\hat N_a^4|\mu_{\rm{c}}}
&= |\mu_{\rm{c}}|^8+6|\mu_{\rm{c}}|^6+7|\mu_{\rm{c}}|^4+|\mu_{\rm{c}}|^2\;,\label{app:commutators:Na4:coherent:state} \\
\braket{\mu_{\rm{c}}|\hat N_a^3|\mu_{\rm{c}}}
&= |\mu_{\rm{c}}|^6+3|\mu_{\rm{c}}|^4+|\mu_{\rm{c}}|^2\;,\label{app:commutators:Na3:coherent:state} \\
\braket{\mu_{\rm{c}}|\hat N_a^2|\mu_{\rm{c}}}
&= |\mu_{\rm{c}}|^2(1+|\mu_{\rm{c}}|^2)\;,\label{app:commutators:Na2:coherent:state}  \\
\braket{\mu_{\rm{c}} |\hat N_a |\mu_{\rm{c}}} &=  |\mu_{\rm{c}}|^2 \; . \label{app:commutators:Na:coherent:state}
\end{align}
\end{subequations}
Furthermore, the displacement operators given
\begin{subequations}
\begin{align}
\bra{\mu_{\rm{m}}} \hat B_+ \ket{\mu_{\rm{m}}} &= \mu_{\rm{m}} + \mu_{\rm{m}}^* \, , \label{app:commutators:B+:coherent} \\
\bra{\mu_{\rm{m}}} \hat B_- \ket{\mu_{\rm{m}}} &= i \left( \mu_{\rm{m}}^* - \mu_{\rm{m}} \right)  \, \label{app:commutators:B-:coherent} .
\end{align}
\end{subequations}
The squares give
\begin{subequations}
\begin{align}
\bra{\mu_{\rm{m}}} \hat B_+^2 \ket{\mu_{\rm{m}}} &= \bra{\mu_{\rm{m}}} \left( \hat b^{\dag 2} + \hat b^\dag \hat b + \hat b \hat b^\dag + \hat b^2 \right)  \ket{\mu_{\rm{m}}}  = \mu_{\rm{m}}^2 + \mu_{\rm{m}}^{*2} + 2  \, |\mu_{\rm{m}}|^2 + 1 \, ,\label{app:commutators:B+2:coherent} \\
\bra{\mu_{\rm{m}}} \hat B_-^2 \ket{\mu_{\rm{m}}} &= \bra{\mu_{\rm{m}}} i^2 \left( \hat b^{\dag 2}  - \hat b^\dag \hat b  - \hat b \hat b^\dag + \hat b^2 \right) \ket{\mu_{\rm{m}}} = - \mu_{\rm{m}}^2 - \mu_{\rm{m}}^{*2} + 2 |\mu_{\rm{m}}|^2 +1  \,\label{app:commutators:B-2:coherent} .
\end{align}
\end{subequations}
Mixtures of $\hat B_+$ and $\hat B_-$:
\begin{subequations}
\begin{align}
\bra{\mu_{\rm{m}}} \hat B_+ \hat B_- \ket{\mu_{\rm{m}}} &=  i \left( \mu_{\rm{m}}^{*2} - \mu_{\rm{m}}^2 + 1 \right)  \, ,\label{app:commutators:B+B-:coherent} \\
\bra{\mu_{\rm{m}}} \hat B_- \hat B_+ \ket{\mu_{\rm{m}}} &= - i \left( \mu_{\rm{m}}^2 - \mu_{\rm{m}}^{*2} + 1\right)  \label{app:commutators:B-B+:coherent} \, .
\end{align}
\end{subequations}
\subsection{Expectation values for Fock states}
In order to compute the coefficients of the quantum Fisher information operator $\hat{\mathcal{H}}_\theta $ in Eq.~\eqref{chap:metrology:eq:mathcalH:with:coefficients}, we must compute the expectation value of a number of operators with respect to a Fock basis, where we define $\hat N_b \ket{n} = n \, \ket{n}$. Many of the expectation values are zero, such as $\braket{\hat B_+}$ and $\braket{\hat B_-}$. For the optics, we find
\begin{align}
\braket{n|\hat N_a|n}&= n\,,\;\label{chap:commutators:Na:Fock} \\
\braket{n|\hat N_a^2|n}&= n^2\label{chap:commutators:Na2:Fock} \, .
\end{align}
For the mechanics, we find the following expectation values:
\begin{subequations}
\begin{align}
\braket{n|\hat B_+^2|n}
&= 2n+1\;, \label{app:commutators:B+2:Fock} \\
\braket{n|\hat B_-^2|n}
&= 2n+1\;, \label{app:commutators:B-2:Fock}\\
\braket{n|(\hat B_+^{(2)})^2|n}
&= 2n^2+2n+2\;,\label{app:commutators:B+2sq:Fock} \\
\braket{n|(\hat B_-^{(2)})^2|n}
&= 2n^2+2n+2\;,\label{app:commutators:B-2sq:Fock} \\
\braket{n|\hat B_+\hat B_-|n}
&= i\;,\label{app:commutators:B+B-:Fock} \\
\braket{n|
\hat B_-\hat B_+|n}
&= -i\;,\label{app:commutators:B-B+:Fock} \\
\braket{n|\hat B_+^{(2)}\hat B_-^{(2)}|n}
&= 2i(2n+1)\;, \label{app:commutators:B+2B-2:Fock}\\
\braket{n|\hat B_-^{(2)}\hat B_+^{(2)}|n}
&= -2i(2n+1)\;,\label{app:commutators:B-2B+2:Fock}
\end{align}
\end{subequations}
We must also consider a number of expectation values with respect to mixed Fock states:
\begin{subequations}
\begin{align}
\braket{n|\hat B_+|m}
&= \sqrt{m+1}\delta_{n,m+1}+\sqrt{m}\delta_{n,m-1}\;,\label{chap:commutators:Fock:different:B+} \\
\braket{n|\hat B_-|m}
&= i\left(\sqrt{m+1}\delta_{n,m+1}-\sqrt{m}\delta_{n,m-1}\right)\;,\label{chap:commutators:Fock:different:B-} \\
\braket{n|\hat B_+^{(2)}|m}
&= \sqrt{m+1}\sqrt{m+2}\delta_{n,m+2}+\sqrt{m}\sqrt{m-1}\delta_{n,m-2}\;,\label{chap:commutators:Fock:different:B+2} \\
\braket{n|\hat B_-^{(2)}|m}
&= i \left(\sqrt{m+1}\sqrt{m+2}\delta_{n,m+2} -\sqrt{m}\sqrt{m-1}\delta_{n,m-2}\right)\;\label{chap:commutators:Fock:different:B-2}.
\end{align}
\end{subequations}

\section{A recipe for computing congruence transformations}\label{app:commutator:congruence}
In this Appendix, we detail the method used to obtain the compact expressions for multiplication by congruence, which refers to the transformation $\hat O(t) = \hat U(t)  \, \hat O \, \hat U^{-1}(t)$. Our goal is to commute the terms such that we end up with a closed-form expression for $\hat O(t)$. 

\subsection{General procedure}\label{app:commutator:general:procedure}
We start by defining 
\begin{equation}
\hat O (x) = e^{i x \hat A} \hat B e^{ - i x \hat A} \, ,
\end{equation}
where $\hat A $ and $\hat B$ are general Hermitian operators and $x$ is a real parameter. We now differentiate the expression with respect to $x$ to find
\begin{equation}
\dot{\hat{O}}(x) =  - i \,  e^{ - i  \, x \,  \hat B} [\hat B, \hat A ] \, e^{i \, x\,  \hat B} \, ,
\end{equation}
where we have defined $ \dot{\hat{O}}\equiv d \hat O(x)/dx $. We repeat the same procedure, and differentiate again to find
\begin{equation}
\ddot{\hat{O}}(x) = - e^{- i\, x \,\hat B} \left[ \hat B , \left[ \hat B, \hat A \right] \right] e^{ i \,x\, \hat B} \, .
\end{equation}
The number of differentiations required depends on the problem at hand. Due to the SU(2) structure of the operators we consider below, we find that the final commutator is now proportional to the initial operator $\hat A$. We set up the differential equation 
\begin{equation}
\ddot{\hat{O}}(x) \propto \hat O (x) \, .
\end{equation}
The form of the equation will differ depending on the commutator relations, and will also influence the trail solution we use to find the solutions. We make use of the initial conditions $\hat O(x= 0) = \hat B$ and $\dot{\hat{O}}(x = 0) =  - i \left[ \hat B , \hat A\right]$ to find the final solution. 

\subsection{A concrete example}\label{app:commutator:example}
We demonstrate the applicability of this method through the following example. We wish to derive an expression for the quantity $e^{  i \,x \,\hat B_- ^{(2)} } \,  \hat B_+^{(2)} \, e^{-i\, x\, \hat B_-^{(2)} }$. We start by defining
\begin{align}
\hat O &= e^{  i \,x \,\hat B_- ^{(2)} } \,  \hat B_+^{(2)} \, e^{-i\, x\, \hat B_-^{(2)} } \, , 
\end{align}
where $\hat B_+ = \hat b^\dag + \hat b$ and $\hat B_ -= i \,\left( \hat b^\dag - \hat b \right)$. Then, we differentiate with respect to $x$ to find
\begin{align}
\dot{\hat{O}} &= e^{i \,x \,\hat B_- ^{(2)} } \, i \,\left[ \hat B_-^{(2)} , \hat B_+^{(2)} \right]\, e^{ - i \,x \,\hat B_-^{(2)} }\, .
\end{align}
where now $\hat B_+^{(2)} = \hat b^{\dag 2} + \hat b^2$ and $\hat B^{(2)} = i\, \left( \hat b^{\dag2} - \hat b^2\right)$. It then follows that
\begin{align}
\left[  \hat b^2 , \hat b^{\dag 2} \right] =  4  \, \hat b^\dag \hat b  + 2 \, ,
\end{align}
meaning that 
\begin{align}
\left[ \hat B_-^{(2)}, \hat B_+^{(2)} \right] = - i\left( 8\,  \hat b^\dag \hat b  + 4 \right)\, .
\end{align}
Therefore, 
\begin{equation}
\dot{\hat {O}} =4 \, e^{  i x \hat B_- ^{(2)} } \,  \left( 2 \, \hat b^\dag \hat b  + 1\right)\, e^{-i x \hat B_-^{(2)} }\, .
\end{equation}
We now differentiate again. We find
\begin{equation}
\ddot{\hat{O}} =4 \, i \, e^{ i\, x \,\hat B_- ^{(2)} } \, \left[  \hat B_-^{(2)}, \left( 2 \,\hat b^\dag \hat b  + 1\right) \right]\, e^{ -i \,x\, \hat B_-^{(2)} }  = -8 \, i \, e^{ i \,x \,\hat B_- ^{(2)} } \, \left[  \hat b^\dag \hat b , \hat B_-^{(2)} \right]\, e^{- i \,x \,\hat B_-^{(2)} }\, , 
\end{equation}
and
\begin{equation}
i \,\left[ \hat b^\dag \hat b, \hat b^{\dag 2} - \hat b^2 \right] = 2 \,i \,\left( \hat b^{\dag 2} + \hat b^2 \right) = 2\, i\, \hat B_+^{(2)}\, .
\end{equation}
Therefore, we find
\begin{equation}
\ddot{\hat {O}} = 16  \, e^{ i \,x\, \hat B_- ^{(2)} } \, \hat B_+^{(2)} \, e^{-i \,x\, \hat B_-^{(2)} }  \, .
\end{equation}
The solution to this differential equation is a hyperbolic cosine and sine with two coefficients. We choose the following trial solution
\begin{equation}
\hat{\hat {O}} = \hat A \,\cosh(4 x) +\hat  B\,\sinh(4 x)\, , 
\end{equation}
where $\hat A$ and $\hat B$ are two generic operator functions. Now, considering the initial condition with $x = 0$, we find
\begin{align}
4 \,\hat B &= 4  \, \left( 2\, \hat b^\dag \hat b + 1 \right)  \, ,\nonumber \\
16\, \hat A &= 16 \,  \hat B_+^{(2)} \, .
\end{align}
The final expression reads
\begin{equation}
e^{ i\, x\, \hat B_- ^{(2)} } \,  \hat B_+^{(2)} \, e^{-i \,x \,\hat B_-^{(2)} } =   \hat B_+^{(2)} \,\cosh(4 x) +  \left( 2 \,\hat b^\dag \hat b + 1 \right)  \,\sinh(4 x) \, .
\end{equation}
The same procedure can be used to derive all the congruence relations considered below. 

\section{Congruence transformation relations}\label{app:commutator:relations}
The following commutator relations are used throughout this thesis. We list them here so that they can later be referred to in a simple manner. Since most of the relations concern the mechanical phonon mode, which we denote $\hat b$ in the text, all relations will be written in terms of $\hat b$ here. However, some of them will also be used throughout this thesis to compute properties of the optical subsystem. The operators will then be changed to $\hat a$. 

By using the method outlined in Section~\ref{app:commutator:general:procedure}, we find the following commutator relations. Since the calculations are long and repetitive, we have chosen to not include them in this thesis. 
\begin{subequations}
\begin{align} 
e^{i\, x\, \hat N_b} \, \hat b \, e^{- i \,x\, \hat N_b} = & \, e^{- i \, x} \, \hat b\label{app:commutators:eq:commutators:b}  \\
e^{i \,x\, \hat N_b} \, \hat b^\dag \, e^{- i \,x\, \hat N_b} =& \,  e^{i \, x} \, \hat b^\dag\label{app:commutators:eq:commutators:bd} \\
e^{i\, x\, \hat N_b ^2} \, \hat b \,  e^{- i \,x\,\hat N_b^2} =& \,  e^{- 2\, x \,\hat N_b} \, e^{- i \,x} \, \hat b \, ,\label{app:commutators:eq:commutators:0}\\
e^{ i\, x\,\hat B_- ^{(2)} } \,  \hat B_+^{(2)} \, e^{-i \,x\, \hat B_-^{(2)} } =& \,  \hat B_+^{(2)} \cosh(4 \,x) +  \left( 2 \, \hat N_b+ 1 \right)  \sinh(4 \,x) \, ,\label{app:commutators:eq:commutators:1}  \\
e^{i \,x \,s\hat B_+^{(2)}} \hat B_-^{(2)}  e^{- i \,x \,\hat B_+^{(2)}} =&\, \hat B_-^{(2)} \, \cosh(4 x) - \left( 2 \, \hat N_b+ 1 \right) \sinh(4\, x)  \, , \label{app:commutators:eq:commutators:2} \\
e^{i \,x\, \hat B_-^{(2)} }\, \hat N_b \, e^{- i \,x \,\hat B_-^{(2)}} =&\, \hat N_b \,  \cosh(4\,x) + \hat B_+^{(2)} \frac{1}{2} \sinh(4\,x) + \sinh^2(2\,x) \, \mathds{1} \, , \label{app:commutators:eq:commutators:3}  \\
e^{i \,x\, \hat B_+^{(2)} } \,\hat N_b \, e^{- i \,x\, \hat B_+^{(2)}} =&\, \hat N_b \,  \cosh(4\,x) -  \hat B_-^{(2)} \frac{1}{2} \sinh(4\,x) + \sinh^2(2\,x) \, \mathds{1} \, , \label{app:commutators:eq:commutators:4}  \\
e^{i\,x\,\hat B_+}\, \hat N_b \,e^{-i\,x\,\hat B_+}=& \, \hat N_b-\hat B_-\,x+x^2\,\mathds{1} \, ,\label{app:commutators:eq:commutators:5} \\
e^{i\,  x\,\hat B_- }\, \hat N_b \,e^{ -i \, x\,\hat B_-}=&\, \hat N_b+ \hat B_+\,x+x^2\,\mathds{1} \, , \label{app:commutators:eq:commutators:6} \\
e^{i\,x\,\hat B_+ }\,\hat B_+^{(2)} \,e^{-i\,x\,\hat B_+}=&\,  \hat B_+^{(2)} +2\,\hat B_-\,x-2\,x^2\,\mathds{1} \, , \label{app:commutators:eq:commutators:7} \\
e^{i \, x\, \hat  B_-}\,\hat B_+^{(2)} \,e^{  - i \, x\, \hat B_-}=&\,  \hat B_+^{(2)} +2\,\hat B_+ \,x+2\,x^2\,\mathds{1} \, ,\label{app:commutators:eq:commutators:8} \\
e^{i\,x\,\hat B_+}\,\hat B_-^{(2)} \,e^{-i\,x\,\hat B_+}=&\, \hat B_-^{(2)} -2\,\hat B_+\,x \, , \label{app:commutators:eq:commutators:9} \\
e^{ i \, x\,\hat B_-}\,\hat B_-^{(2)} \,e^{-i \, x\,\hat B_-}=&\, \hat B_-^{(2)}  + 2\,\hat B_-\,x \, ,\label{app:commutators:eq:commutators:10} \\
e^{i\,x\,\hat B_+}\,\hat B_-\,e^{-i\,x\,\hat B_+}=&\, \hat B_--2\,x\,\mathds{1} \, ,\label{app:commutators:eq:commutators:11} \\
e^{i \, x\,\hat B_-}\,\hat B_+ \,e^{-i \, x\,\hat B_- }=&\, \hat B_+ + 2\,x\,\mathds{1} \, ,\label{app:commutators:eq:commutators:12}
\end{align}
\end{subequations}
Some additional relations read:
\begin{align}
e^{- i \, x\, \hat b^\dag \hat b} \, \hat b \, e^{- i \, x\, \hat b^\dag \hat b } = e^{- 2 \, i \, x \, \hat b^\dag \hat b} \, e^{- i \, x} \, \hat b \, . \label{app:commutators:eq:commutators:13}
\end{align}
This concludes this Appendix. 
\chapter{First and second moments for optomechanical evolution}
\label{app:exp:values}

In this Appendix, we compute all first and second moment for an initially coherent state that evolves under the optomechanical Hamiltonian in Eq.~\eqref{chap:decoupling:eq:Hamiltonian}. These expectation values are used in Chapters~\ref{chap:non:Gaussianity:coupling} and~\ref{chap:non:Gaussianity:squeezing} to compute the non-Gaussianity of states evolving under this Hamiltonian. In general, knowledge of the first and second moments yield important insight into the state evolution and they are routinely measured in the laboratory. We also compute the symplectic eigenvalues of the optical and mechanical subsystems, which are used to bound the measure of non-Gaussianity in Chapter~\ref{chap:non:Gaussianity:squeezing}. This derivation was first carried out by David Edward Bruschi. 

\section{Preliminaries}
We reprint the evolution operator $\hat U(t)$ that was derived in Eq.~\eqref{chap:decoupling:eq:final:evolution:operator} here for convenience: 
\begin{align}
\hat U(t):=&e^{-i\,\omega_{l,n} \hat a^\dagger \hat a\,t}\,\hat {\tilde{U}}_{\rm{sq}}\,e^{-\frac{i}{\hbar}\,F_{\hat{N}_a}\,\hat{N}_a}\,e^{-\frac{i}{\hbar}\,F_{\hat{N}^2_a}\,\hat{N}^2_a}\,e^{-\frac{i}{\hbar}\,F_{\hat{B}_+}\,\hat{B}_+}\,e^{-\frac{i}{\hbar}\,F_{\hat{N}_a\,\hat{B}_+}\,\hat{N}_a\,\hat{B}_+}\, \nonumber \\
&\times e^{-\frac{i}{\hbar}\,F_{\hat{B}_-}\,\hat{B}_-}\,e^{-\frac{i}{\hbar}\,F_{\hat{N}_a\,\hat{B}_-}\,\hat{N}_a\,\hat{B}_-}.
\end{align}
In this Appendix, we compute the expectation values of an initially coherence state. We use that given in Eq.~\eqref{chap:introduction:eq:initial:state:coherent:coherent}, which we reprint here for convenience:
\begin{equation}
\hat \rho = \ketbra{\mu_{\rm{c}}} \otimes \ketbra{\mu_{\rm{c}}} \, .
\end{equation}
In the basis that we work in, the covariance matrix elements are given as follows. The optical subsystem is given by 
\begin{align}
\sigma_{11} &= \braket{\hat a^\dagger \hat a} + \braket{\hat a \hat a^\dagger} - 2 \braket{\hat a} \braket{\hat a^\dagger }  \, ,\nonumber \\
\sigma_{13} &= 2 \braket{\hat a^{\dagger 2} } - 2 \braket{\hat a^\dagger} \braket{\hat a^\dagger} \, ,\nonumber \\ 
\sigma_{31} &= 2 \braket{\hat a^2 } - 2 \braket{\hat a} \braket{\hat a}\, , \nonumber \\
\sigma_{33} &= \braket{\hat a\hat a^\dagger} + \braket{\hat a^\dagger \hat a } - 2 \braket{\hat a} \braket{\hat a^\dagger} \, ,
\end{align}
and the mechanical subsystem elements are given by 
\begin{align}
\sigma_{22} &= \braket{\hat b^\dagger \hat b} + \braket{\hat b \hat b^\dagger} - 2 \braket{\hat b} \braket{\hat b^\dagger } \, , \nonumber\\
\sigma_{24} &= 2 \braket{\hat b^{\dagger 2} } - 2 \braket{\hat b^\dagger} \braket{\hat b^\dagger} \, ,\nonumber\\
\sigma_{42} &= 2 \braket{\hat b^2 } - 2 \braket{\hat b} \braket{\hat b}\, ,\nonumber \\
\sigma_{44} &= \braket{\hat b\hat b^\dagger} + \braket{\hat b^\dagger \hat b} - 2 \braket{\hat b} \braket{\hat b^\dagger} \, .
\end{align}
Finally, the remaining off-diagonal elements are 
\begin{align}
\sigma_{12} &= 2\braket{\hat a^\dagger \hat b} - 2 \braket{\hat a^\dagger} \braket{\hat b}\, , \nonumber \\
\sigma_{21} &= 2\braket{\hat a\hat b^\dagger} - 2 \braket{\hat a} \braket{\hat b^\dagger}\, , \nonumber\\
\sigma_{14} &= 2 \braket{\hat a^\dagger \hat b^\dagger} - 2 \braket{\hat a^\dagger} \braket{\hat b^\dagger}\, ,\nonumber \\
\sigma_{41} &= 2 \braket{\hat a\hat b} -  2\braket{\hat a} \braket{\hat b}   \, ,
\end{align}
and
\begin{align}
\sigma_{23} &=  2\braket{\hat a^\dagger \hat b ^\dagger} - 2 \braket{\hat a^\dagger } \braket{\hat b^\dagger}\, , \nonumber\\
\sigma_{32} &= 2 \braket{\hat a\hat b} - 2 \braket{\hat a} \braket{\hat b}\, ,\nonumber \\
\sigma_{34} &=  2 \braket{\hat a\hat b^\dagger} - 2 \braket{\hat a } \braket{\hat b^\dagger}\, ,\nonumber  \\
\sigma_{43} &= 2 \braket{\hat a^\dagger \hat b } - 2 \braket{\hat a^\dagger }\braket{\hat b}   \, .
\end{align}
Given these elements, the expectation values that we have to compute are
\begin{align}
&\braket{\hat a} && \braket{\hat b} \nonumber \\
&\braket{\hat a^\dag \hat a} && \braket{\hat b^\dag \hat b} \nonumber \\
&\braket{\hat a^2}&& \braket{\hat b^2 } \nonumber \\
&\braket{\hat ab}&&  \braket{\hat a \hat b^\dagger}  \, .
\end{align}
These will then be used to compute all the covariance matrix elements. 

\section{Heisenberg evolution of the operators}
In this Section, we derive the time-evolution of $\hat a (t)$ and $\hat b(t)$. From these two operators, we can construct all first and second moments. 
In the Heisenberg picture, operators evolve with the time-evolution operator $\hat U(t)$ as
\begin{align} \label{app:exp:values:Heisenberg}
\hat{a}(t) &= \hat U(t)\, \hat a \,  \hat U^\dag(t) \, ,
\end{align}
and similarly for $\hat b$. 

We begin by deriving an expression for $\hat a(t)$. Using the relations in Eqs.~\eqref{app:commutators:eq:commutators:b},~\eqref{app:commutators:eq:commutators:bd}, and~\eqref{app:commutators:eq:commutators:0}, we find 
\begin{align}
e^{i\,\omega_{\rm{c}} \hat a^\dagger \hat a\,t} \, \hat{a} \, e^{-i\,\omega_{\rm{c}}  \, \hat a^\dagger \hat a\,t}\, &= e^{-i\,\omega_{\rm{c}}} \, \hat{a} \, , \nonumber \\
e^{i\,F_{\hat{N}_a}\,\hat{N}_a}\,
\hat{a} \, 
e^{-i\,F_{\hat{N}_a}\,\hat{N}_a}\, &= e^{- i \, F_{\hat{N}_a} } \, \hat{a}  \,, \nonumber \\ 
e^{i\,F_{\hat{N}^2_a}\,\hat{N}^2_a}\,
\hat{a} \, 
e^{-i\,F_{\hat{N}^2_a}\,\hat{N}^2_a}\, &= e^{- 2 i F_{\hat N_a^2} \hat N_a } e^{ - i F_{\hat N_a^2}} \hat{a} \, ,\nonumber  \\
e^{i\,F_{\hat{N}_a\,\hat{B}_+}\,\hat{N}_a\,\hat{B}_+}\,\hat{a}  \, e^{-i\,F_{\hat{N}_a\,\hat{B}_+}\,\hat{N}_a\,\hat{B}_+}\, &= e^{-i\,F_{\hat{N}_a\,\hat{B}_+}\,\hat{B}_+} \, \hat{a}  \, , 
\end{align}
which allows us to  compute 
\begin{align}
&e^{i\,F_{\hat{B}_-}\,\hat{B}_-}\,e^{-i\,F_{\hat{N}_a\,\hat{B}_+}\,\hat{B}_+} \, \hat{a} \, e^{-i\,F_{\hat{B}_-}\,\hat{B}_-}\, = e^{- i F_{\hat{N}_a \, \hat{B}_+} \hat{B}_+ } \, e^{- 2i \, F_{\hat{N}_a \, \hat{B}_+} \, F_{\hat{B}_-}}  \hat{a}  \, . 
\end{align}
Then, by using the properties of Weyl displacement operators in Eq.~\eqref{app:commutators:weyl:operator:combinations}, we find 
\begin{align}
e^{i F_{\hat B_-} \hat B_- } \, e^{- i F_{\hat B_+} \hat  B_+ } \, e^{- i F_{\hat B_-} \hat B_- } 
&= e^{- i F_{\hat B_+} \hat{B}_+ } e^{-2 \, i \,  F_{\hat B_+}\, F_{\hat B_-}}  \, .
\end{align}
But now again, we have a $ \hat B_+ $ displacement term. The same as the above holds:
\begin{align}
&e^{\frac{i}{\hbar}\,F_{\hat{N}_a\,\hat{B}_-}\,\hat{N}_a\,\hat{B}_-}\,e^{- i F_{\hat{N}_a \, \hat{B}_+} \hat{B}_+ } \, \hat{a} \, 
e^{-\frac{i}{\hbar}\,F_{\hat{N}_a\,\hat{B}_-}\,\hat{N}_a\,\hat{B}_-}  \nonumber  \\
&= e^{\frac{i}{\hbar}\,F_{\hat{N}_a\,\hat{B}_-}\,\hat{N}_a\,\hat{B}_-}\,e^{- i F_{\hat{N}_a \, \hat{B}_+} \hat{B}_+ } \, e^{-\frac{i}{\hbar}\,F_{\hat{N}_a\,\hat{B}_-}\,\hat{N}_a\,\hat{B}_-}\, e^{\frac{i}{\hbar}\,F_{\hat{N}_a\,\hat{B}_-}\,\hat{N}_a\,\hat{B}_-}\, \hat{a} \,
e^{-\frac{i}{\hbar}\,F_{\hat{N}_a\,\hat{B}_-}\,\hat{N}_a\,\hat{B}_-} \nonumber \\
&= e^{\frac{i}{\hbar}\,F_{\hat{N}_a\,\hat{B}_-}\,\hat{N}_a\,\hat{B}_-}\,e^{- i F_{\hat{N}_a \, \hat{B}_+} \hat{B}_+ } \, e^{-\frac{i}{\hbar}\,F_{\hat{N}_a\,\hat{B}_-}\,\hat{N}_a\,\hat{B}_-}\,  e^{- i F_{\hat{N}_a \, \hat{B}_- } \, \hat{B}_- } \hat{a} \nonumber  \\
&= e^{- i F_{\hat{N}_a \hat{B}_+ } \hat{B}_+ } \, e^{- 2 i \, F_{\hat{N}_a \, \hat{B}_- } \, F_{\hat{N}_a \, \hat{B}_+ }\, \hat{N}_a} e^{- i F_{\hat{N}_a \, \hat{B}_- } \, \hat{B}_- } \hat{a}  \, .
\end{align}
Putting everything together, the full expression for the time-evolution of $\hat a$ is
\begin{align}
\hat{a}(t) =& \, e^{-i\,\omega_{l,n}} \,
e^{- i \, F_{\hat{N}_a} } \,
e^{- 2 i F_{N_a^2} N_a } \,
e^{- 2i \, F_{\hat{N}_a \, \hat{B}_+} \, F_{\hat{B}_-}} \, 
 e^{- 2 i \, F_{\hat{N}_a \, \hat{B}_- } \, F_{\hat{N}_a \, \hat{B}_+ }\, \hat{N}_a} \,  \nonumber \\
&\times  e^{- i F_{\hat{N}_a \hat{B}_+ } \hat{B}_+ } \,
 e^{- i F_{\hat{N}_a \, \hat{B}_- } \, \hat{B}_- }
 \hat{a}  \, ,
 \end{align}
 which we rearrange into 
 \begin{align}
 \hat a(t)
 =& e^{- i \omega_{l,n} \, t} e^{- i \left( F_{\hat{N}_a} + F_{\hat{N}_a^2 } + 2 F_{\hat{n} _a \, \hat{B}_+} \, F_{\hat{B}_-} \right)} \, e^{- 2 i \left( F_{\hat{N}_a^2} + F_{\hat{N}a \, \hat{B}_+} \, F_{\hat{N}_a \, \hat{B}_- } \right)\, \hat{N}_a} \nonumber \\
 &\times e^{- i F_{\hat{N}_a \, \hat{B}_+} \, \hat{B}_+ } \, e^{- i F_{\hat{N}_a \, \hat{B}_- } \, \hat{B}_- } \, \hat{a} \, .
\end{align}
We proceed to compute  the time evolution of $\hat b$:
\begin{equation}
\hat b(t) = \hat U(t)\, \hat b \, \hat U^\dag(t) \, .
\end{equation}
The exponentials with only $\hat N_a$ and $\hat N_a^2$ commute with $\hat b$. 
We first note that the qudratic mechanical subsystem operator $\hat{\tilde{U}}_{\rm{sq}}$ acts on the operator as
\begin{equation}
\hat{\tilde{U}}_{\rm{sq}}^\dagger \, \hat b \,   \hat{\tilde{U}}_{\rm{sq}} = \alpha(t) \, \hat b + \beta(t) \, \hat b^\dagger = \left[ \alpha(t) + \beta(t) \right] \frac{\hat B_+ }{2} + i \left[ \alpha(t) - \beta(t) \right] \frac{\hat B_- }{2} \, ,
\end{equation}
where $\alpha$ and $\beta$ are the Bogoliubov coefficients defined in Eq.~\eqref{chap:decoupling:eq:bogoliubov:coeffs:expression}. Therefore we must now compute:
\begin{align}
 e^{i \left( F_{\hat N_a\, \hat B_+ } \hat N_a + \hat B_+\right) \, \hat B_+}  \, \hat B_-  \, e^{-i \left(F_{\hat N_a  \, \hat B_+ } \hat N_a  + F_{\hat B_+} \right) \, \hat B_+} &=  \hat B_- -2 \left( F_{\hat N_a \, \hat B_+} \, \hat N_a + F_{\hat B_+} \right) \, ,\nonumber \\
e^{i \left(F_{\hat N_a  \, \hat B_- } \hat N_a + \hat B_-\right) \, \hat B_-} \,  \hat B_+ \,  e^{-i \left( F_{\hat N_a \, \hat B_- } \hat N_a  + \hat F_{\hat B_-} \right)\, \hat B_-} &= \hat B_+ + 2 \, \left( F_{\hat N_a \, \hat B_-} \, \hat N_a + F_{\hat B_-} \right) \, .
\end{align}
Therefore, the final expression becomes 
\begin{align} \label{app:exp:values:time:evolved:b}
\hat{b}(t) =& \,  \alpha(t) \hat  b + \beta(t) \hat b^\dag + [\alpha(t) + \beta(t)] \left(  F_{\hat N_a \, \hat  B_-} \hat N_a + F_{\hat B_-} \right)  \nonumber \\
&- i [\alpha(t) - \beta(t)] \left( F_{\hat N_a  \, \hat B_+ }\hat  N_a + F_{\hat B_+} \right) \, .
\end{align}
With these operators, we are ready to compute the expectation values.

\section{Deriving expectation values}

Here we derive all the expectation values of the time-evolved operators for when the optics and mechanics are both in coherent states. In this Section, we use the $\braket{\bullet} $ notation to denote the expectation value for these states. As an example, a  generic operator $\hat X(t)$ has expectation value
\begin{equation}
\braket{\hat X(t)} = \bra{\mu_{\rm{c}}} \bra{\mu_{\rm{m}}}  \, \hat X(t) \, \ket{\mu_{\rm{c}}} \ket{\mu_{\rm{m}}} \, , 
\end{equation}
and the expectation values for coherent states, which we will often use, read
\begin{align}
\braket{\hat a} &= \bra{\mu_{\rm{c}}} \bra{\mu_{\rm{m}}} \, \hat a \, \ket{\mu_{\rm{c}}} \ket{\mu_{\rm{m}}}  = \mu_{\rm{c}}\, ,\nonumber \\
\braket{\hat b} &=  \bra{\mu_{\rm{c}}} \bra{\mu_{\rm{m}}} \, \hat b \, \ket{\mu_{\rm{c}}} \ket{\mu_{\rm{m}}}  = \mu_{\rm{m}} \, .
\end{align}
Furthermore, in the following Section and in the main text, we leave out the free evolution of the optical field $e^{- i \, \omega_{\rm{c}} \, t}$, as we can transform into a frame that rotates with this phase. 

\subsection{Expectation value of $\braket{\hat{a}(t)}$}

Given that
\begin{align}
\hat a(t) =& \,  e^{- i \left( F_{\hat{N}_a} + F_{\hat{N}_a^2 } + 2 F_{\hat{N} _a \, \hat{B}_+} \, F_{\hat{B}_-} \right)} \, e^{- 2 i \left( F_{\hat{N}_a^2} + F_{\hat{N}_a \, \hat{B}_+} \, F_{\hat{N}_a \, \hat{B}_- } \right)\, \hat{N}_a} \nonumber \\
&\times e^{- i F_{\hat{N}_a \, \hat{B}_+} \, \hat{B}_+ } \, e^{- i F_{\hat{N}_a \, \hat{B}_- } \, \hat{B}_- } \, \hat{a} \, .
\end{align}
We simplify this expression to 
\begin{align} \label{app:exp:values:time:evolved:a:compact}
\hat a(t) &=e^{-i \, \varphi(t)}\, e^{- i \,  \theta(t) \,  \hat{N}_a } e^{- i  \, F_{\hat{N}_a \, \hat{B}_+} \, \hat{B}_+ } \, e^{- i  \, F_{\hat{N}_a \, \hat{B}_- } \, \hat{B}_- } \, \hat{a} \, ,
\end{align}
where we have defined
\begin{align}
\varphi(t) &= \left( F_{\hat{N}_a} + F_{\hat{N}_a^2 } + 2 \,  F_{\hat{N} _a \, \hat{B}_+} \, F_{\hat{B}_-} \right) \, , \nonumber \\ 
\theta(t) &= 2  \left( F_{\hat{N}_a^2} + F_{\hat{N}_a \, \hat{B}_+} \, F_{\hat{N}_a \, \hat{B}_- } \right)\, . 
\end{align}
By using the identity in Eq.~\eqref{app:commutators:exponential:number:operator}, we find that 
\begin{align}
\bra{\mu_{\rm{c}}} e^{- i  \, \theta(t) \, \hat N_a } \ket{\mu_{\rm{c}}} &=e^{ |\mu_{\rm{c}}|^2 ( e^{- i \, \theta(t)}  - 1)} \, , 
\end{align}
and for given the expectation value for $e^{- i\, F_{\hat N_a \, \hat B_+} \, \hat B_+} e^{- i \, F_{\hat N_a \, \hat B_-} \, \hat B_-}$ in Eq.~\eqref{app:commutators:EBpBm:exp:value}, we find that the expectation value of $\hat a(t)$ is given by 
\begin{equation}
\braket{\hat a (t) } = e^{- i \,  \varphi(t)} \, e^{|\mu_{\rm{c}}|^2 (e^{- i \theta(t)}-1)} \, E_{\hat B_+ \hat B_- } \,  \mu_{\rm{c}} \, , 
\end{equation}
where the quantity $E_{\hat B_+ \hat B_-}$ is defined in Eq.~\eqref{app:commutators:definition:of:EB+B-}. 

\subsection{Expectation value of $\braket{\hat{a}^\dagger \hat{a}}$}

Given the compact expression for $\hat a(t)$ in Eq.~\eqref{app:exp:values:time:evolved:a:compact}, we write 
\begin{align}
\hat{a}^\dagger \hat{a} &= \hat a^\dag \,  e^{i  \, F_{\hat{N}_a \, \hat{B}_- } \, \hat{B}_- }  \, e^{ i  \, F_{\hat{N}_a \, \hat{B}_+} \, \hat{B}_+ } \,e^{i  \, \theta(t) \,  \hat{N}_a }  \, e^{-i\varphi(t)}\nonumber  \\
&\quad\times e^{-i \, \varphi(t)}\, e^{- i  \, \theta(t) \,  \hat{N}_a } e^{- i  \, F_{\hat{N}_a \, \hat{B}_+} \, \hat{B}_+ } \, e^{- i  \, F_{\hat{N}_a \, \hat{B}_- } \, \hat{B}_- } \, \hat{a} \, .
\end{align}
All phases cancel and we are left with $\hat a^\dag \hat a$, the expectation value of which reads
\begin{equation}
\braket{\hat{a}^\dagger \hat{a} } =  |\mu_{\rm{c}}|^2 \, .
\end{equation}
This expectation value arises from the fact that the Hamiltonian in Eq.~\eqref{chap:decoupling:eq:Hamiltonian} commutes with $\hat N_a$. 
\subsection{Expectation value of $\braket{\hat{a}^2(t)}$}

The operator is
\begin{align}
\hat{a}^2(t) =& \, e^{-i\,\varphi(t)}\, e^{- i\,\theta(t)\, \hat{N}_a } e^{- i \,F_{\hat{N}_a \, \hat{B}_+} \, \hat{B}_+ } \, e^{- i \,F_{\hat{N}_a \, \hat{B}_- } \, \hat{B}_- } \, \hat{a} \nonumber \\
&\times e^{-i\,\varphi(t)}\, e^{- i\, \theta(t)  \, \hat{N}_a } e^{- i \,F_{\hat{N}_a \, \hat{B}_+} \, \hat{B}_+ } \, e^{- i \,F_{\hat{N}_a \, \hat{B}_- } \, \hat{B}_- } \, \hat{a} \, .
\end{align}
By using the expression in Eq.~\eqref{app:commutators:eq:commutators:13}, we find 
\begin{equation}
e^{- i  \, \theta(t) \, \hat  N_a } \, \hat  a  \, e^{- i \, \theta(t) \, \hat  N_a } = e^{- 2  \, i \,  \theta(t) \, \hat  N_a }  \, e^{- i \, \theta(t)} \, \hat  a \, ,
\end{equation}
the expectation value of which reads
\begin{align}
\bra{\mu_{\rm{c}}} e^{- 2  \, i \,  \theta(t) \, \hat  a^\dagger \hat a } e^{- i \, \theta(t)} \, \hat  a\ket{\mu_{\rm{c}}} &= \mu_{\rm{c}} \bra{\mu_{\rm{c}}} e^{- 2  \, i \,  \theta(t) \, \hat a^\dagger \hat a } e^{- i  \, \theta(t)} \, \hat  a\ket{\mu_{\rm{c}}}\nonumber  \\
&=\mu_{\rm{c}}^2 e^{- i \,  \theta(t)} \bra{\mu_{\rm{c}}} e^{- 2  \, i \,  \theta(t) \, \hat  a^\dag \hat a } \ket{\mu_{\rm{c}}} \nonumber \\
&= \mu_{\rm{c}}^2 e^{- i \,  \theta(t)}  e^{|\mu_{\rm{c}}|^2 \,  (e^{- 2  \, i  \, \theta(t)} - 1)} \, .
\end{align}
Then,  we find the expectation value:
\begin{align}
\braket{\hat{a}^2(t)} &= e^{- 2 \, i \varphi(t)} \, \mu_{\rm{c}}^2 \, e^{- i \, \theta(t)}  e^{|\mu_{\rm{c}}|^2  \, (e^{- 2  \, i \,  \theta(t)} - 1)} \,  E_{\hat B_+ \hat B_-\hat B_+\hat B_- } \, , 
\end{align}
where the quantity $E_{\hat B_+\hat B_-\hat B_+\hat B_-}$ has been defined in Eq.~\eqref{app:commutators:EBpBmBpBm}. 

\subsection{Expectation value of $\braket{\hat{b}(t)}$}

Given the expression in Eq.~\eqref{app:exp:values:time:evolved:b}, for the time-evolution of $\hat b$, which we reprint here for convenience, 
\begin{align}
\hat b(t)=&\left[(\alpha(t)+\beta(t))\,(F_{\hat{B}_-}+F_{\hat{N}_a\,\hat{B}_-}\,\hat{N}_a)+i\,(\alpha(t)-\beta(t))\,(F_{\hat{B}_+}+F_{\hat{N}_a\,\hat{B}_+}\,\hat{N}_a)\right] \nonumber \\
&+ (\alpha(t)\,\hat b+\beta(t)\,\hat b^\dag) \, , 
\end{align}
we first rewrite it as 
\begin{align}
\hat b(t) = \alpha(t ) \hat b + \beta(t) \hat b^\dagger + \Gamma(t) + \Delta(t) \, \hat N_a \, ,
\end{align}
where we defined 
\begin{align} \label{app:exp:values:definition:of:Gamma:Delta}
\Gamma(t) &= (\alpha(t) + \beta(t)) F_{\hat{B}_-} - i ( \alpha(t) - \beta(t) ) \, F_{\hat{B}_+} \, , \nonumber \\
\Delta(t) &= (\alpha(t) + \beta(t)) F_{\hat{N}_a \, \hat{B}_- } - i ( \alpha(t) - \beta(t)) F_{\hat{N}_a \, \hat{B}_+ } \, .
\end{align}
The expectation value becomes
\begin{equation}
\braket{\hat b(t)} = \alpha(t) \,  \mu_{\rm{m}} + \beta(t) \, \mu_{\rm{m}}^* + \Gamma(t) + \Delta(t) \,  |\mu_{\rm{c}}|^2 \, .
\end{equation}

\subsection{Expectation value of $\braket{\hat{b}^\dag (t) \hat{b}(t)}$}
The operator is
\begin{align}
\hat b^\dagger(t)\hat  b(t) = \left( \alpha(t ) \, \hat b + \beta(t)  \, \hat b^\dagger + \Gamma(t) + \Delta(t) \, \hat  N_a\right)^\dag \left( \alpha(t ) \, \hat  b + \beta(t) \, \hat b^\dagger + \Gamma(t) + \Delta(t)  \, \hat N_a\right) \, .
\end{align}
Multiplying the terms out, and using the fact that $\hat b \hat b^\dag  = \hat b^\dag \hat b + 1$ enables us to cancel some terms to find
\begin{align}
\hat b ^\dag (t)\hat b(t) =& \,   (|\alpha(t)|^2 + |\beta(t)|^2) \hat b^\dagger \hat b + \alpha^*(t) \beta (t) \hat b^{\dagger  2} + \alpha(t) \beta^*(t) \hat \hat b^2 + |\beta(t)|^2 \nonumber \\
&+ (\alpha^*(t)\,\hat b^\dag+\beta^*(t)\,\hat b) \, \left( \Gamma(t) + \Delta(t) \hat N_a \right)  + (\alpha(t)\,\hat b+\beta(t)\,\hat b^\dag) \, \left( \Gamma(t) + \Delta(t)\hat  N_a \right)^* \nonumber \\
&+  \left| \Gamma(t) + \Delta(t) \hat N_a \right|^2 \, .
\end{align}
We now note that the following term will contain a factor of $\hat N_a^2$:
\begin{align} 
 \left| \Gamma(t) + \Delta(t) \hat N_a \right|^2 &= \left( \Gamma^*(t) + \Delta^*(t)  \, \hat N_a \right) \left( \Gamma(t) + \Delta(t) \, \hat N_a \right) \nonumber  \\
 &= |\Gamma(t)|^2 + \Gamma^*(t) \Delta(t)   \, \hat N_a + \Gamma(t) \Delta^*(t)  \, \hat N_a + |\Delta(t)|^2  \, \hat N_a^2  \, .
 \end{align}
By then using the expectation value of $\hat N_a^2$ in Eq.~\eqref{chap:commutators:ada2:exp:value}, we find that the expectation value of $\hat b(t)$ is:
\begin{align}
\braket{\hat b ^\dag(t) \hat b(t)}  =&  (|\alpha(t)|^2 + |\beta(t)|^2)  \, |\mu_{\rm{m}}|^2 + \alpha^*(t)\,  \beta (t) \, (\mu_{\rm{m}}^*)^2 + \alpha(t)\,  \beta^*(t)\,  \mu_{\rm{m}}^2 + |\beta(t)|^2 \nonumber \\
&+ (\alpha^*(t)\,\mu_{\rm{m}}^* +\beta^*(t)\,\mu_{\rm{m}}) \, \left( \Gamma(t) + \Delta(t) \, |\mu_{\rm{c}}|^2 \right)  \nonumber \\
&+ (\alpha(t)\,\mu_{\rm{m}}+\beta(t)\,\mu_{\rm{m}}^*) \, \left( \Gamma(t) + \Delta(t) \, |\mu_{\rm{c}}|^2 \right)^* \nonumber \\
&+ |\Gamma(t)|^2 + \left( \Gamma^*(t) \, \Delta(t)  + \Gamma(t)\,  \Delta^*(t)\right) \, |\mu_{\rm{c}}|^2 + |\Delta(t)|^2\,  |\mu_{\rm{c}}|^2\,  ( 1 + |\mu_{\rm{c}}|^2) \, .
\end{align}

\subsection{Expectation value of $\braket{\hat{b}^2(t)}$}

The time-evolved operator is given by 
\begin{align}
\hat{b}^2(t) =& \, \left[\alpha(t )  \, \hat b + \beta(t)\, \hat b^\dagger + \Gamma(t) + \Delta(t) \, \hat N_a\right] \nonumber \\
&\times \left[\alpha(t )\,  \hat b + \beta(t) \, \hat b^\dagger + \Gamma(t) + \Delta(t)\,  \hat N_a\right],
\end{align}
We note that the first two terms on each line will form the product
\begin{align}
(\alpha(t)\,\hat b+\beta(t)\,\hat b^\dag)(\alpha(t)\,\hat b+\beta(t)\,\hat b^\dag) 
&= \alpha^2(t) \, \hat b^2 + \alpha(t)  \, \beta(t) \,  (\hat b^\dagger \hat b + \hat b\hat b^\dagger) + \beta^2(t) \, \hat b^{\dagger 2} \nonumber \\
&= \alpha^2(t) \, \hat b^2 + \alpha(t) \,  \beta(t) \, (2\,  \hat b^\dagger \hat b + 1) + \beta^2(t)  \, \hat b^{\dagger 2}  \, , 
\end{align} 
where we have again used the fact that $\hat b \hat b^\dag = \hat b^\dag \hat b + 1$, This means that the full expression for the time-evolved operator is 
\begin{align}
\hat{b}^2(t)= & \, \alpha^2(t) \, \hat b^2 + \alpha(t) \beta(t) (2 \, \hat b^\dagger \, \hat  b + 1) + \beta^2(t)\, \hat  b^{\dagger 2}  \nonumber \\
&+ 2 \, (\alpha(t)\,\hat b+\beta(t)\,\hat b^\dag )  \, \left[\Gamma(t) + \Delta(t) \, \hat N_a \right] \nonumber \\
&+ \left[\Gamma(t) + \Delta(t) \, \hat N_a \right]^2 \, ,
\end{align}
We have that 
\begin{align}
 \left[\Gamma(t) + \Delta(t) \hat N_a \right]^2&=  \Gamma^2 (t) + 2 \Gamma(t) \Delta(t) \hat N_a + \Delta^2(t) \hat N_a^2 \, .
 \end{align}
Again using the relation in Eq.~\eqref{chap:commutators:ada2:exp:value} for the expectation value of $\hat N_a^2$, we find the expectation value
\begin{align}
 \braket{\hat b^{ 2} (t)} = & \, \alpha^2(t)  \, \mu_{\rm{m}}^2  + \alpha(t) \, \beta(t) \,(2 \, |\mu_{\rm{m}}|^2 + 1) + \beta^2(t) \, \mu_{\rm{m}}^{*2}  \nonumber \\
&+ 2\,  (\alpha(t)\,\mu_{\rm{m}}+\beta(t)\,\mu_{\rm{m}}^*) \left[\Gamma(t) + \Delta(t)\,  |\mu_{\rm{c}}|^2 \right] \nonumber \\
&+ \Gamma^2 (t) + 2 \, \Gamma(t) \, \Delta(t) \, |\mu_{\rm{c}}|^2 + \Delta^2(t) \, |\mu_{\rm{c}}|^2\, ( 1 + |\mu_{\rm{c}}|^2) \,.
\end{align}

\subsection{Expectation value of $\braket{\hat{a}(t) \hat{b}(t)}$}

The time-evolved operator is 
\begin{align} \label{app:exp:values:time:evolved:ab}
\hat a(t) \, \hat b(t) =& \, e^{-i \, \varphi(t)}\, e^{- i \,  \theta(t) \,  \hat{N}_a } e^{- i \,  F_{\hat{N}_a \, \hat{B}_+} \, \hat{B}_+ } \, e^{- i  \, F_{\hat{N}_a \, \hat{B}_- } \, \hat{B}_- } \, \hat{a} \nonumber \\
&\times \left[ (\alpha(t)\,\hat b+\beta(t)\,\hat b^\dag)+\Gamma(t) + \Delta (t)  \, \hat N_a \right] \, . 
\end{align}
We rewrite the operator in this explicit form where we can see which expectation value each term will take:
\begin{align} \label{app:exp:values:time:evolved:ab:explicit}
\hat a(t) \hat b(t) = & \, e^{- i \, \varphi(t)} \,  \biggl[ e^{-i \, \theta(t)\,\hat{N}_a}\,\hat a \bigl( \alpha(t)\,e^{-i\,F_{\hat{N}_a\,\hat{B}_+}\,\hat{B}_+}\,e^{-i\,F_{\hat{N}_a\,\hat{B}_-}\,\hat{B}_-}\hat b \nonumber \\
&\quad\quad\quad\quad\quad\quad\quad\quad\quad\quad\quad\quad\quad\quad+\beta(t)\,e^{-i\,F_{\hat{N}_a\,\hat{B}_+}\,\hat{B}_+}\,e^{-i\,F_{\hat{N}_a\,\hat{B}_-}\,\hat{B}_-}\, \hat b^\dag\bigr)\nonumber  \\
&\quad\quad\quad\quad+e^{-i  \, \theta(t)\,\hat{N}_a}\,e^{-i\,F_{\hat{N}_a\,\hat{B}_+}\,\hat{B}_+}\,e^{-i\,F_{\hat{N}_a\,\hat{B}_-}\,\hat{B}_-} \, 
\Gamma(t) \, \hat a\nonumber  \\
&\quad\quad\quad\quad+ e^{-i \,\theta(t)\,\hat{N}_a}\,e^{-i\,F_{\hat{N}_a\,\hat{B}_+}\,\hat{B}_+}\,e^{-i\,F_{\hat{N}_a\,\hat{B}_-}\,\hat{B}_-}
\Delta (t)  \, \hat a\, \hat N_a \biggr]  \, .
\end{align}
Starting from the last term in Eq.~\eqref{app:exp:values:time:evolved:ab:explicit}, we know from Eq.~\eqref{app:commutators:exp:value:Exp:a:Na}, that the expectation value of the term   $e^{- i \, \theta(t) \, \hat N_a} \, \hat a \,\hat  N_a$ gives
\begin{align}
\bra{\mu_{\rm{c}}}e^{- i \, \theta(t) \, \hat N_a } \, \hat a \,\hat  N_a \ket{\mu_{\rm{c}}} &=
 \mu_{\rm{c}} e^{ |\mu_{\rm{c}}|^2 (e^{- i \theta} - 1)} \left( |\mu_{\rm{c}}|^2 e^{- i \, \theta(t)} + 1\right) \, .
\end{align}
The second to last term in Eq.~\eqref{app:exp:values:time:evolved:ab:explicit} requires us to compute the expectation value of $e^{- i \, \theta(t) \, \hat N_a} \, \hat a$, which we know from Eq.~\eqref{app:commutators:exponential:number:operator} to be
\begin{align}
\bra{\mu_{\rm{c}}} e^{- i \, \theta(t) \, \hat N_a } \, \hat a \ket{\mu_{\rm{c}}} &= \mu_{\rm{c}} \, \bra{\mu_{\rm{c}}} e^{- i \, \theta(t) \, \hat N_a }  \ket{\mu_{\rm{c}}} \nonumber \\
&= \mu_{\rm{c}} \, e^{ |\mu_{\rm{c}}|^2 ( e^{- i \, \theta(t)}  - 1)}  \, .
\end{align}
For the second term in Eq.~\eqref{app:exp:values:time:evolved:ab:explicit}, we know that the expectation value $e^{- i\,  F_{\hat N_a \, \hat B_-} \, \hat B_- } \, \hat b^\dagger$ is given by Eq.~\eqref{app:commutators:exp:value:of:EBpBm:bd}:
\begin{align}
\bra{\mu_{\rm{m}}} e^{ - i \, F_{\hat N_a \, \hat B_+} \, \hat B_+} \, e^{- i \, F_{\hat N_a \, \hat B_-} \, \hat B_-} \, \hat b^\dag \ket{\mu_{\rm{m}}} = (\mu_{\rm{m}} ^* - i \, F_{\hat N_a \, \hat B_+} - F_{\hat N_a \, \hat B_-} ) \, E_{\hat B_+ \hat B_-} \, .
\end{align}
The first term in Eq.~\eqref{app:exp:values:time:evolved:ab:explicit} is given by the expectation value in Eq.~\eqref{app:commutators:EBpBm:exp:value}, as before, namely:
\begin{align}
\bra{\mu_{\rm{m}}} e^{- i \, F_{\hat N_a \, \hat B_+} \, \hat B_+} e^{- i \, F_{\hat N_a \, \hat B_-} \, \hat B_- } \, \hat b \ket{\mu_{\rm{m}}} = \mu_{\rm{m}}  \, E_{\hat B_+ \hat B_-}  \, ,
\end{align}
with $E_{\hat B_+ \hat B_-}$ defined in Eq.~\eqref{app:commutators:definition:of:EB+B-}. 
Putting all terms together, we write the expectation value of the time-evolved operator $\hat a(t) \hat b(t)$ as 
\begin{align}
\braket{\hat a(t) \, \hat b(t) } = &  \, e^{- i \varphi(t)} \,e^{ |\mu_{\rm{c}}|^2 ( e^{- i \, \theta(t)}  - 1)} \, \mu_{\rm{c}} \, E_{\hat B_+ \, \hat B_-}  \, \biggl[ \alpha(t)\, \mu_{\rm{m}} +\beta(t)\,(\mu_{\rm{m}}^* - i \, F_{\hat N_a \, \hat B_+} - F_{\hat N_a \, \hat B_-})  \nonumber \\
&\quad\quad\quad\quad + \Gamma(t) +\left( |\mu_{\rm{c}}|^2 \,  e^{- i\,  \theta(t)} + 1\right)\Delta (t)  \biggr]  \, .
\end{align}

\subsection{Expectation value of $\braket{\hat{a}(t) \hat{b}^\dagger(t)} $}

The expression for the time-evolved operator is
\begin{align}
\hat{a} (t)\hat{b}^\dag(t) =& \, 
 e^{- i \varphi(t)} \,e^{-i \, \theta(t)\,\hat{N}_a}\,e^{-i\,F_{\hat{N}_a\,\hat{B}_+}\,\hat{B}_+}\,e^{-i\,F_{\hat{N}_a\,\hat{B}_-}\,\hat{B}_-}
\hat a \nonumber \\
&\times \left[ \alpha^*(t)\,\hat b^\dag +\beta^*(t)\,\hat b +\Gamma^*(t) + \Delta^* (t) \, \hat N_a \right]  \, .
\end{align}
Through a similar procedure as the above, the expectation value is 
\begin{align}
\braket{\hat a (t) \, \hat b^\dag(t) }= \, & e^{- i \varphi(t)} \, e^{ |\mu_{\mathrm{c}}|^2 \, ( e^{- i \theta(t)}  - 1)} \, \mu_{\mathrm{c}} \,E_{\hat B_+ \hat B_-}\,  \biggl[  \alpha^*(t)\, (\mu_{\mathrm{m}}^* - F_{\hat N_a \, \hat B_-} - i \, F_{\hat N_a \, \hat B_+}) \nonumber \\
&\quad\quad\quad\quad+\beta^*(t)\,\mu_{\mathrm{m}} +
\Gamma^*(t)  + \left( |\mu_{\mathrm{c}}|^2 \, e^{- i \, \theta(t)} + 1\right) \, \Delta^* (t)  \biggr]  \, .
\end{align}

\subsection{Summary of expectation values for coherent states}
The expectation values are:
\begin{align} \label{app:ex:values:summary:of:exp:values}
\braket{\hat a (t) } :=&  \, e^{- i \,  \varphi (t)} \, e^{|\mu_{\rm{c}}|^2 (e^{- i \,  \theta(t)}-1)} E_{\hat B_+ \hat B_- } \mu_{\rm{c}}  \, ,\nonumber \\
\braket{\hat b(t) } :=& \,  \alpha(t) \,  \mu_{\rm{m}} + \beta(t) \,  \mu_{\rm{m}}^*  + \Gamma(t) + \Delta(t)  \, |\mu_{\rm{c}}|^2 \, , \nonumber \\ 
\braket{\hat{a}^2(t)} :=&\, e^{- 2 \, i \,  \varphi(t)} \, \mu_{\rm{c}}^2 e^{- i  \, \theta(t)} \,  e^{|\mu_{\rm{c}}|^2 \, (e^{- 2 \, i \,  \theta(t)} - 1)} \,  E_{\hat B_+ \hat B_- \hat B_+ \hat B_- } \, ,\nonumber  \\
 \braket{\hat b^{2}(t) } :=& \,  \alpha^2(t) \, \mu_{\rm{m}}^2  + \alpha(t) \,  \beta(t) \,(2 \, |\mu_{\rm{m}}|^2 + 1) + \beta^2(t) \, \mu_{\rm{m}}^{*2}  \nonumber \\
 &+ 2 \,  (\alpha(t)\,\mu_{\rm{m}}+\beta(t)\,\mu_{\rm{m}}^*) \, \left[\Gamma(t) + \Delta(t) \, |\mu_{\rm{c}}|^2 \right] \nonumber \\
 &+ \Gamma^2 (t) + 2 \,  \Gamma(t) \, \Delta(t) \,  |\mu_{\rm{c}}|^2 + \Delta^2(t) \, |\mu_{\rm{c}}|^2 \, ( 1 + |\mu_{\rm{c}}|^2) \, ,\nonumber \\
 \braket{\hat b ^\dag(t) \hat b(t)}  := \, &  (|\alpha(t)|^2 + |\beta(t)|^2)  \, |\mu_{\rm{m}}|^2 + \alpha^*(t) \,  \beta (t) (\mu_{\rm{m}}^*)^2 + \alpha(t) \, \beta^*(t)\,  \mu_{\rm{m}}^2  \nonumber \\
 &+  (\alpha^*(t)\,\mu_{\rm{m}}^* +\beta^*(t)\,\mu_{\rm{m}}) \, \left( \Gamma(t) + \Delta(t) \, |\mu_{\rm{c}}|^2 \right) \nonumber \\
& + (\alpha(t)\,\mu_{\rm{m}}+\beta(t)\,\mu_{\rm{m}}^*) \, \left( \Gamma(t) + \Delta(t) \, |\mu_{\rm{c}}|^2 \right)^* \nonumber \\
&+  (\Gamma^*(t) \, \Delta(t)  + \Gamma(t) \, \Delta^*(t))\,  |\mu_{\rm{c}}|^2 + |\Delta(t)|^2 \, |\mu_{\rm{c}}|^2\,  ( 1 + |\mu_{\rm{c}}|^2) \nonumber \\
&+ |\beta(t)|^2+|\Gamma(t)|^2  \, ,\nonumber \\
\braket{\hat a(t) \hat b(t) } := \, & e^{- i \, \varphi(t)} \,e^{ |\mu_{\rm{c}}|^2 \,  ( e^{- i \,  \theta(t)}  - 1)}  \, \mu_{\rm{c}}  \, E_{\hat B_+ \hat B_-}  \biggl[ \alpha(t)\, \mu_{\rm{m}} +\beta(t)\,(\mu_{\rm{m}}^* - i  \, F_{\hat N_a \hat B_+} - F_{\hat N_a \hat B_-})   \nonumber \\
&\quad\quad\quad\quad+ 
\Gamma(t) +\left( |\mu_{\rm{c}}|^2  \, e^{- i  \, \theta(t)} + 1\right) \, 
\Delta (t)  \biggr] \, ,  \nonumber \\
\braket{\hat a (t) \, \hat b^\dag(t) }: = \, & e^{- i \, \varphi(t)} \, e^{ |\mu_{\rm{c}}|^2  \, ( e^{- i  \, \theta(t)}  - 1)} \mu_{\rm{c}} \,E_{\hat B_+ \hat B_-} \,  \biggl[  \alpha^*(t)\, (\mu_{\rm{m}}^* - i \,  F_{\hat N_a B_+} - F_{\hat N_a B_-})  \nonumber \\
&\quad\quad\quad\quad+\beta^*(t)\,\mu_{\rm{m}}  +\Gamma^*(t)  + \left( |\mu_{\rm{c}}|^2 \,  e^{- i  \, \theta(t)} + 1\right)  \, \Delta^* (t)  \biggr]  \, .
\end{align}
We have written the expectation values in their most explicit form.  In the main text, we will make the identification $K_{\hat N_a} = F_{\hat N_a \, \hat B_-} + i \, F_{\hat N_a \, \hat B_-}$. The results can be seen in Eq.~\eqref{chap:non:Gaussianity:squeezing:expectation:values}.

\subsection{Optical and mechanical quadratures}

Using the expectation values in Eq.~\eqref{app:ex:values:summary:of:exp:values}, we now compute the optical and mechanical quadratures of the state, which represent the classical trajectory of the system in phase space. The dimensionless quadrature operators are given by $\hat x_{\rm{c}} =  \frac{1}{\sqrt{2}} \left( \hat a^\dag + \hat a \right)$, $\hat p_{\rm{c}} =  \frac{i}{\sqrt{2}} \left( \hat a^\dag - \hat a \right)$, $\hat x_{\rm{m}} =  \frac{1}{\sqrt{2}} \left( \hat b^\dag + \hat b \right)$, and $\hat p_{\rm{m}} =  \frac{i}{\sqrt{2}} \left( \hat b^\dag - \hat b \right)$. Inserting out expressions for their evolution, we find
\begin{align} \label{app:exp:values:eq:optical:quadratures}
\braket{\hat{x}_{\rm{c}}} (t)
&=\sqrt{2} \, e^{- |\mu_{\rm{c}}|^2} \,  \Re \left\{ e^{-i \, \varphi (t)} \, e^{|\mu_{\rm{c}}|^2  \, e^{-i \, \theta(t)}}  \, E_{\hat B_+ \hat B_- }\,  \mu_{\rm{c}} \right\} \, , \nonumber \\
\braket{\hat{p}_{\rm{c}}}(t) 
&= \sqrt{2} \, i \, e^{- |\mu_{\rm{c}}|^2} \, \Im \left\{   e^{i \, \varphi (t)} \, e^{|\mu_{\rm{c}}|^2  \, e^{i  \,\theta(t)}}  \, E_{\hat B_+ \hat B_- }^*\,  \mu_{\rm{c}}^* \right\} \,  , \nonumber \\
\end{align}
for the optical subsystem, and
\begin{align} \label{app:exp:values:eq:mechanical:quadratures}
\braket{\hat{x}_{\rm{m}}} &= \sqrt{2} \, |\mu_{\rm{c}}|^2\, \Re\left\{  \alpha(t) \,  \mu_{\rm{m}} + \beta(t) \,  \mu_{\rm{m}}^*  + \Gamma(t) + \Delta(t)  \, \right\} \, , \nonumber \\
\braket{\hat{p}_{\rm{m}}} &=  \sqrt{2} \, i \, |\mu_{\rm{c}}|^2 \, \Re\left\{  \alpha^*(t) \,  \mu_{\rm{m}}^* + \beta^*(t) \,  \mu_{\rm{m}}  + \Gamma^*(t) + \Delta^*(t)  \right\} \, , 
\end{align}
for the mechanical subsystem. 

We present the quadratures graphically for a selection of parameters in Figure~\ref{chap:non:Gaussianity:squeezing:fig:constant:squeezing:quadratures} and Figure~\ref{chap:gravimetry:fig:quadratures}.

\section{Deriving the covariance matrix elements} \label{app:exp:values:derivation:covariance:matrix:elements}

We can now derive the elements of the covariance matrix $\sigma$. The covariance matrix is required for computing the non-Gaussianity of the optomechanical state, which we do in Chapters~\ref{chap:non:Gaussianity:coupling} and $\ref{chap:non:Gaussianity:squeezing}$.

\subsection{Expression for $\sigma_{11}$}
Given the expectation values $\braket{\hat a^\dag(t) \hat a (t)}$ and $\braket{\hat a(t)}$ in Eq.~\eqref{app:ex:values:summary:of:exp:values}, the covariance matrix element $\sigma_{11}$ is given by 
\begin{align}
\sigma_{11} &=1 + 2 \, \braket{\hat a^\dag(t) \hat a(t)} - 2 \braket{\hat a(t)} \braket{\hat a^\dag(t)}\nonumber  \\
&= 1 +2 |\mu_{\rm{c}}|^2 - 2 e^{- i \, \varphi(t) } \, e^{|\mu_{\rm{c}}|^2 (e^{- i  \, \theta(t)}-1)} E_{\hat B_+ \hat B_- } \mu_{\rm{c}} e^{ i \,  \varphi(t) } \, e^{|\mu_{\rm{c}}|^2 (e^{ i \, \theta(t)}-1)} E_{\hat B_+ \hat B_- }^* \mu_{\rm{c}}^* \nonumber \\
&= 1 + 2 \, |\mu_{\rm{c}}|^2 - 2 \,  |\mu_{\rm{c}}|^2  \, e^{|\mu_{\rm{c}}|^2 (e^{-i \,  \theta(t)} + e^{i \, \theta(t)} - 2)} |E_{B_+ B_-}|^2 \nonumber \\
&=1 + 2 \, |\mu_{\rm{c}}|^2 \left( 1 - e^{-4 \, |\mu_{\rm{c}}|^2 \sin^2{\theta(t)/2}} |E_{B_+ B_-}|^2 \right) \, ,
\end{align}
where we have used the trigonometric identity $e^{ - i \, \theta(t)} + e^{i \, \theta(t)} - 2 =  - 4 \, \sin^2\theta(t)/2$. Furthermore, due to the symmetries of the covariance matrix, we know that $\sigma_{11} = \sigma_{33}$. 

\subsection{Expression for $\sigma_{22}$}
Given the expectation values $\braket{\hat b^\dag(t) \hat b (t)}$ and $\braket{\hat b(t)}$ in Eq.~\eqref{app:ex:values:summary:of:exp:values}, the covariance matrix element $\sigma_{22}$ is given by 
\begin{align}
\sigma_{22} =& \, 1 +  2 \,  \braket{\hat b^\dag (t)\hat b(t)} - 2  \, \braket{\hat b(t)} \braket{\hat b^\dag(t)}\nonumber  \\
=& 1 + 2  \, (|\alpha(t)|^2 + |\beta(t)|^2) |\mu_{\rm{m}}|^2 + \alpha^*(t) \beta (t) (\mu_{\rm{m}}^*)^2 + \alpha(t) \beta^*(t) \mu_{\rm{m}}^2 + |\beta(t)|^2 \nonumber \\
&+ (\alpha^*(t)\,\mu_{\rm{m}}^* +\beta^*(t)\,\mu_{\rm{m}}) \, \left( \Gamma(t) + \Delta(t) |\mu_{\rm{c}}|^2 \right) \nonumber  \\
&+2\biggl[(\alpha(t)\,\mu_{\rm{m}}+\beta(t)\,\mu_{\rm{m}}^*) \, \left( \Gamma(t) + \Delta(t) |\mu|^2 \right)^* + |\Gamma(t)|^2 \nonumber \\
&\quad\quad+ (\Gamma^*(t) \Delta(t)  + \Gamma(t) \Delta^*(t)) |\mu_{\rm{c}}|^2 + |\Delta(t)|^2 |\mu_{\rm{c}}|^2 ( 1 + |\mu_{\rm{c}}|^2) \biggr] \nonumber \\
&- 2\left(  \alpha(t) \mu_{\rm{m}} + \beta(t) \mu_{\rm{m}}^*  + \Gamma(t) + \Delta(t) |\mu|^2\right) \left(  \alpha(t) \mu_{\rm{m}} + \beta(t) \mu_{\rm{m}}^*  + \Gamma(t) + \Delta(t) |\mu_{\rm{c}}|^2\right)^* \nonumber \\
=& \,  1 + 2 \,  (|\alpha(t)|^2 + |\beta(t)|^2) |\mu_{\rm{m}}|^2 + \alpha^*(t) \beta (t) (\mu_{\rm{m}}^*)^2 + \alpha(t) \beta^*(t) \mu_{\rm{m}}^2 + |\beta(t)|^2 \nonumber \\
&+ (\alpha^*(t)\,\mu_{\rm{m}}^* +\beta^*(t)\,\mu_{\rm{m}}) \, \left( \Gamma(t) + \Delta(t) |\mu_{\rm{c}}|^2 \right) \nonumber \\
&+2\biggr[(\alpha(t)\,\mu_{\rm{m}}+\beta(t)\,\mu_{\rm{m}}^*) \, \left( \Gamma(t) + \Delta(t) |\mu_{\rm{c}}|^2 \right)^* + |\Gamma(t)|^2 \nonumber \\
&\quad\quad+ (\Gamma^*(t) \Delta(t)  + \Gamma(t) \Delta^*(t)) |\mu_{\rm{c}}|^2 + |\Delta(t)|^2 |\mu_{\rm{c}}|^2 ( 1 + |\mu_{\rm{c}}|^2) \biggl] \nonumber \\
&- 2\left(  \alpha(t) \mu_{\rm{m}} + \beta(t) \mu_{\rm{m}}^*  + \Gamma(t) + \Delta(t) |\mu_{\rm{c}}|^2\right) \nonumber \\
&\quad\quad\times \left(  \alpha(t) \mu_{\rm{m}} + \beta(t) \mu_{\rm{m}}^*  + \Gamma(t) + \Delta(t) |\mu_{\rm{c}}|^2\right)^*\nonumber  \\
&= 1 + 2|\beta(t)|^2  + 2|\Delta(t)|^2 |\mu_{\rm{c}}|^2 \, .
\end{align}
Due to the symmetries of the covariance matrix in this basis, we also know that $\sigma_{22} = \sigma_{44}$. 

\subsection{Expression for $\sigma_{31}$}
Given the expectation values $\braket{\hat a^2 (t)}$ and $\braket{\hat a(t)}$ in Eq.~\eqref{app:ex:values:summary:of:exp:values}, the covariance matrix element $\sigma_{31}$ is given by 
\begin{align}
\sigma_{31} =& \,  2 \,  \braket{\hat a^2(t)} - 2 \,  \braket{\hat a(t)}^2 \nonumber \\
=&\,  2\, \mu_{\rm{c}}^2 \, e^{- 2 \, i \, \varphi(t)}  e^{- i \, \theta(t)}  e^{|\mu_{\rm{c}}|^2 (e^{- 2 \,  i \,  \theta(t)} - 1)} \,  E_{\hat B_+ \hat B_- \hat B_+ \hat B_- } \nonumber \\
&- 2 \, e^{- i \, \varphi (t)} \, e^{|\mu_{\rm{c}}|^2 (e^{- i \, \theta(t)}-1)} E_{\hat B_+ \hat B_- } \mu_{\rm{c}} e^{- i \, \varphi(t) } \, e^{|\mu|^2 (e^{- i \,  \theta(t)}-1)} E_{\hat B_+ \hat B_- } \mu_{\rm{c}}\nonumber \\
=& \,  2 \, \mu_{\rm{c}}^2 \, e^{- 2 \, i \,  \varphi(t)} \left(  e^{- i \, \theta(t)} e^{|\mu_{\rm{c}}|^2 ( e^{-  2 \, i \, \theta(t)} - 1)} E_{\hat B_+\hat B_-\hat B_+ \hat B_-} - e^{2 \, |\mu_{\rm{c}}|^2 ( e^{- i  \, \theta(t)} - 1)} \,  E_{\hat B_+ \hat B_-}^2\right) \, .
\end{align}

\subsection{Expression for $\sigma_{42}$}
Given the expectation values $\braket{\hat b^2 (t)}$ and $\braket{\hat b(t)}$ in Eq.~\eqref{app:ex:values:summary:of:exp:values}, the covariance matrix element $\sigma_{42}$ is given by 
\begin{align}
\sigma_{42} =& \,  2 \, \braket{\hat b^2(t)}  - 2 \,  \braket{\hat b (t)}^2 \nonumber \\
=& \, 2 \,  \biggl[  \alpha^2(t) \,\mu_{\rm{m}}^2  + \alpha(t) \,\beta(t) \,(2 \, |\mu_{\rm{m}}|^2 + 1) + \beta^2(t)\, \mu_{\rm{m}}^{*2} \nonumber \\
&\quad\quad + 2\, (\alpha(t)\,\mu_{\rm{m}}+\beta(t)\,\mu_{\rm{m}}^*) \left[\Gamma(t) + \Delta(t) \,|\mu_{\rm{c}}|^2 \right] \nonumber \\
 &\quad\quad+ \Gamma^2 (t) + 2\, \Gamma(t)\, \Delta(t) \,|\mu_{\rm{c}}|^2 + \Delta^2(t) \,|\mu_{\rm{c}}|^2( 1 + |\mu_{\rm{c}}|^2)\biggr] \nonumber\\
 &- 2\, \left(  \alpha(t)\, \mu_{\rm{m}} + \beta(t)\, \mu_{\rm{m}}^*  + \Gamma(t) + \Delta(t)\, |\mu_{\rm{c}}|^2\right)^2 \nonumber\\
=& \,2 \,  \alpha(t)\, \beta(t) +2\, \Delta^2(t) \,|\mu_{\rm{c}}|^2 \, .
\end{align}

\subsection{Expression for $\sigma_{21}$}
Given the expectation values $\braket{\hat a(t) \hat b^\dag (t)}$,  $\braket{\hat b(t)}$ and $\braket{\hat a (t)}$ in Eq.~\eqref{app:ex:values:summary:of:exp:values}, the covariance matrix element $\sigma_{21}$ is given by 
\begin{align}
\sigma_{21} =& \, 2 \, \braket{\hat a(t) \hat b^\dag(t)} - 2 \,  \braket{\hat a(t)} \braket{\hat b^\dag(t)} \nonumber  \\
=& \,  2 \,  e^{- i \,  \varphi(t)} \, e^{ |\mu_{\rm{c}}|^2 ( e^{- i \,  \theta(t)}  - 1)}  \,E_{\hat B_+ \hat B_-} \, \mu_{\rm{c}}\, \biggl[  \alpha^*(t)\, (\mu_{\rm{m}}^* - i F_{\hat N_a B_+} - F_{\hat N_a B_-}) +\beta^*(t)\,\mu_{\rm{m}}  \nonumber \\
&\quad\quad\quad\quad\quad\quad\quad\quad\quad\quad\quad\quad\quad\quad\quad\quad+
\Gamma^*(t)  + \left( |\mu_{\rm{c}}|^2 e^{- i \,  \theta(t)} + 1\right) \Delta^* (t)  \biggr] \nonumber \\
&- 2 e^{- i \varphi(t) } \, e^{|\mu_{\rm{c}}|^2 (e^{- i \theta(t)}-1)} E_{\hat B_+ \hat B_- } \mu_{\rm{c}} \left( \alpha(t) \mu_{\rm{m}} + \beta(t) \mu_{\rm{m}}^*  + \Gamma(t) + \Delta(t) |\mu_{\rm{c}}|^2 \right)^* \nonumber \\
=& \,  2  \, e^{- i \,\varphi(t)} \, e^{ |\mu_{\rm{c}}|^2 ( e^{- i\, \theta(t)}  - 1)}  \,E_{\hat B_+ \hat B_-} \, \mu_{\rm{c}}\,\biggl[ \alpha^*(t)\left( -i \,F_{\hat{N}_a \hat{B}_+} - F_{\hat N_a \hat B_-} \right) \nonumber \\
&\quad\quad\quad\quad\quad\quad\quad\quad\quad\quad\quad\quad\quad\quad\quad+ |\mu_{\rm{c}}|^2 \,e^{- i \,\theta(t)} \Delta^*(t) + \Delta^*(t) - \Delta^*(t)\,|\mu_{\rm{c}}|^2 \biggr] \nonumber \\
=&\,2\, e^{- i\, \varphi(t)} \, e^{ |\mu_{\rm{c}}|^2 ( e^{- i \,\theta(t)}  - 1)}  \,E_{\hat B_+ \hat B_-} \, \mu_{\rm{c}}\, \left[ -\alpha^*(t) \,F + \Delta^*(t) \left( 1 + |\mu_{\rm{c}}|^2( e^{- i\, \theta(t)} - 1)\right)\right] \, .
\end{align}
Due to the symmetries of the covariance matrix $\sigma$, we also know that $\sigma_{21} = \sigma_{12}^*= \sigma_{43} = \sigma_{34}^*$.

\subsection{Expression for $\sigma_{41} $}
Given the expectation values $\braket{\hat a(t)  \hat b (t)}$,  $\braket{\hat b(t)}$ and $\braket{\hat a (t)}$ in Eq.~\eqref{app:ex:values:summary:of:exp:values}, the covariance matrix element $\sigma_{41}$ is given by 
\begin{align}
\sigma_{41} =& \,  2 \,  \braket{\hat a (t) \hat b(t)} - 2 \braket{\hat a(t)}\braket{\hat b(t)} \nonumber \\
=&\,  2 \, e^{- i \, \varphi(t)} \,e^{ |\mu_{\rm{c}}|^2 ( e^{- i \, \theta(t)}  - 1)}E_{\hat B_+ \hat B_-} \,  \mu_{\rm{c}} \, \biggl[ \alpha(t)\, \mu_{\rm{m}} +\beta(t)\,(\mu_{\rm{m}}^* - i F_{\hat N_a \hat B_+} - F_{\hat N_a \hat B_-})  \nonumber \\
&\quad\quad\quad\quad\quad\quad\quad\quad\quad\quad\quad\quad\quad\quad\quad\quad + \,
\Gamma(t) +\left( |\mu|^2 e^{- i \theta(t)} + 1\right)\Delta (t)  \biggr] \nonumber \\
&- 2 e^{- i \varphi(t) } \, e^{|\mu|^2 (e^{- i \theta(t)}-1)} E_{\hat B_+ \hat B_- } \mu_{\rm{c}}\left( \alpha(t) \mu_{\rm{m}} + \beta(t) \mu_{\rm{m}}^*  + \Gamma(t) + \Delta(t) |\mu_{\rm{c}}|^2 \right) \nonumber \\
=& \, 2 \, e^{- i \, \varphi(t)} \,e^{ |\mu_{\rm{c}}|^2 ( e^{- i \,  \theta(t)}  - 1)}E_{\hat B_+ \hat B_-} \,  \mu_{\rm{c}} \, \left[  \left(|\mu_{\rm{c}}|^2 e^{- i  \, \theta(t)} + 1\right)\Delta (t) -\Delta(t) |\mu_{\rm{c}}|^2  \right]  \nonumber \\
= &\, 2 \, e^{- i  \, \varphi(t)} \,e^{ |\mu_{\rm{c}}|^2 ( e^{- i \,  \theta(t)}  - 1)}E_{\hat B_+ \hat B_-} \,  \mu_{\rm{c}} \, \Delta (t) \left(|\mu_{\rm{c}}|^2 (e^{- i  \, \theta(t)} -1)-1  \right)  \, .
\end{align}

\subsection{Summary of covariance matrix elements}
In summary, the covariance matrix elements of the non-Gaussian optomechanical state are:
\begin{align} \label{app:exp:values:CM:elements}
\sigma_{11} &=1 + 2\, |\mu_{\mathrm{c}}|^2 \left( 1 - e^{-4 \, |\mu_{\mathrm{c}}|^2 \sin^2{\theta(t)/2}} \, |E_{\hat B_+ \hat B_-}|^2 \right)\, , \nonumber \\
\sigma_{31} &=  2\, \mu_{\mathrm{c}}^2 \, e^{- 2 \, i \,  \varphi(t)} \left(  e^{- i \, \theta(t)} \, e^{|\mu_{\mathrm{c}}|^2 ( e^{-  2 \, i \,  \theta(t)} - 1)} e^{-|K_{\hat N_a}|^2} \,   - e^{2 \, |\mu_{\mathrm{c}}|^2 ( e^{- i  \, \theta(t)} - 1)} \right) \, E_{\hat B_+ \hat B_-}^2\, , \nonumber \\
\sigma_{21} &=  2 \,e^{- i \,  \varphi(t)} \, e^{ |\mu_{\mathrm{c}}|^2 ( e^{- i \, \theta(t)}  - 1)}  \,E_{\hat B_+ \hat B_-} \, \mu_{\mathrm{c}}\, \left[\Delta^*(t)\left( |\mu_{\mathrm{c}}|^2 (e^{- i \,  \theta(t)} - 1) + 1  \right) - \alpha^*(t) \, K_{\hat N_a}\right]\, , \nonumber \\
\sigma_{41} &= 2\, e^{- i \, \varphi(t) } \, e^{|\mu_{\mathrm{c}}|^2 \,  (e^{- i \, \theta(t)}-1)} E_{\hat B_+ \hat B_- } \mu_{\mathrm{c}} \, \left[\Delta(t) \left( |\mu_{\mathrm{c}}|^2 (e^{- i \theta(t)} - 1) + 1\right)- \beta(t) \,  K_{\hat N_a}  \right]\, ,\nonumber\\
\sigma_{22} &= 1 + 2\, |\beta(t)|^2  + 2\, |\Delta(t)|^2 \, |\mu_{\mathrm{c}}|^2\, , \nonumber \\
\sigma_{42}  &=2\,  \alpha(t) \, \beta(t) +2\,  \Delta^2(t) \, |\mu_{\mathrm{c}}|^2 \, ,
\end{align}
where we have used the following notation:
\begin{align}
K_{\hat N_a} &= F_{\hat N_a \, \hat B_-} +  i \, F_{\hat N_a \, \hat B_+} \, , \nonumber \\
\varphi(t) =& \left( F_{\hat{N}_a} + F_{\hat{N}_a^2 } + 2 \,  F_{\hat{N} _a \, \hat{B}_+} \, F_{\hat{B}_-} \right) \nonumber \\ 
\theta(t) =& \,  2 \, \left( F_{\hat{N}_a^2} + F_{\hat{N}_a \, \hat{B}_+} \, F_{\hat{N}_a \, \hat{B}_- } \right)\,\nonumber \\
\Gamma(t) =& \,  (\alpha(t) + \beta(t)) F_{\hat{B}_-} - i ( \alpha(t) - \beta(t) ) \, F_{\hat{B}_+} \nonumber \\
\Delta(t) =& (\alpha(t) + \beta(t)) F_{\hat{N}_a \, \hat{B}_- } - i ( \alpha(t) - \beta(t)) F_{\hat{N}_a \, \hat{B}_+ } \nonumber \\
E_{\hat B _+ \hat B_-} =& \,   \mathrm{exp} \biggl[ \frac{1}{2} \biggl( - F_{\hat N _a \hat B_-}^2 - F_{\hat N_a \hat B_+}^2 - 2 \,  i \,  F_{\hat N_a \hat B_-} F_{\hat N_a \hat B_+} \nonumber\\
&\quad\quad\quad\quad- 2\, \mu_{\rm{m}} (F_{\hat N_a \hat B_-} + i F_{\hat N _a \hat B_+} ) + 2 \,  \mu_{\rm{m}}^* ( F_{\hat N_a \hat B_-} - i F_{\hat N_a \hat B_+} ) \biggr) \biggr] \, , \nonumber\\
E_{\hat B_+ \hat B_- \hat B_+ \hat B_- } &= \mathrm{exp} \biggl[ - 2 \,  \biggl( F_{\hat N_a \hat B_+}^2 + F_{\hat N_a \hat B_-}^2 + i F_{\hat N_a \hat B_+} F_{\hat N_a \hat B_-} \nonumber \\
&\quad\quad\quad\quad\quad\quad+ \mu_{\rm{m}} ( F_{\hat N_a \hat B_-} + i F_{\hat N_a \hat B_+}) - \mu_{\rm{m}}^*( F_{\hat N_a \hat B_-} - i F_{\hat N_a \hat B_+})   \biggr) \biggr] \, .
\end{align}

\section{Symplectic eigenvalues of the optical and mechanical subsystems}\label{app:exp:values:symplectic:eigenvalues}
In this Section, we use the covariance matrix elements in Eq.~\eqref{app:exp:values:CM:elements} to compute the symplectic eigenvalues of the optical and mechanical subsystems. 

\subsection{The optical symplectic eigenvalue}
We wish to compute the symplectic eigenvalue of the optical subsystem. It is given by 
\begin{equation}
\nu_{Op}^2 = \sigma_{11}^2 - |\sigma_{13}|^2  \, .
\end{equation}
From the covariance matrix elements in Eq.~\eqref{app:exp:values:CM:elements} we find
\begin{align}
\sigma_{11}^2 &= 1 + 4\, |\mu_{\mathrm{c}}|^2 \left( 1 - e^{-4|\mu_{\mathrm{c}}|^2 \sin^2{\theta(t)/2}} \,e^{- |K_{\hat N_a}|^2} \right)+ 4 \, |\mu_{\mathrm{c}}|^4 \left( 1 - e^{-4 \, |\mu_{\mathrm{c}}|^2 \sin^2{\theta(t)/2}} \,e^{- |K_{\hat N_a}|^2} \right) ^2  \, ,\nonumber \\
|\sigma_{31}|^2 &= 4\, |\mu_{\mathrm{c}}|^4 \, |E_{\hat B_+ \hat B_-}|^4 \left(  e^{i \,  \theta(t)} \, e^{|\mu_{\mathrm{c}}|^2 ( e^{2 \, i \, \theta(t)} - 1)} e^{-|K_{\hat N_a}|^2} \,   - e^{2 \, |\mu_{\mathrm{c}}|^2 ( e^{i \,  \theta(t)} - 1)} \right) \, \nonumber \\
&\quad\quad\quad\quad\quad\quad\quad\quad\times \left(  e^{- i \, \theta(t)} \, e^{|\mu_{\mathrm{c}}|^2 ( e^{-  2 \, i\, \theta(t)} - 1)} e^{-|K_{\hat N_a}|^2} \,   - e^{2 \,|\mu_{\mathrm{c}}|^2 ( e^{- i  \,\theta(t)} - 1)} \right)\nonumber \\
&= 4\, |\mu_{\mathrm{c}}|^4 \, |E_{\hat B_+ \hat B_-}|^4 \biggl( e^{-2 \, |K_{\hat N_a}|^2} \, e^{|\mu_c|^2 \left( e^{2 \, i \, \theta(t)} + e^{- 2 \, i \, \theta(t)} -2\right)}    + e^{2 \,|\mu_{\mathrm{c}}|^2 ( e^{i \, \theta(t)} + e^{- i \, \theta(t)} -  2)}  \nonumber \\
&\quad\quad\quad\quad\quad\quad\quad\quad\quad- 2 \, \Re \left\{  e^{i \,\theta(t)} \, e^{|\mu_{\mathrm{c}}|^2 ( e^{2 \, i \, \theta(t)} - 1)} e^{-|K_{\hat N_a}|^2} e^{2 \, |\mu_c|^2 (e^{- i \, \theta(t)} - 1)} \right\} \biggr)  \,.
\end{align}
By using the following trigonometric identities, 
\begin{align}
e^{2\,i \,\theta(t)} + e^{-2 \, i \, \theta(t)} - 2  &= 2\, \cos{2\,\theta(t)} - 2= - 4\, \sin^2{\theta(t)}  \, , \nonumber \\
e^{i \,\theta(t)} + e^{- i \,\theta(t) } - 2 &= 2\,( \cos\theta(t) - 1) = - 4 \sin^2{\theta(t)/2}   \, ,
\end{align}
 and from the fact that $|E_{\hat B_+ \hat B_-}|^2 = e^{- |K_{\hat N_a}|^2 }$, we obtain
\begin{align}
\sigma_{11}^2 =& \,  1 + 4\, |\mu_{\mathrm{c}}|^2 \left( 1 - e^{ -4\,|\mu_{\mathrm{c}}|^2 \sin^2{\theta(t)/2}} \,e^{- |K_{\hat N_a}|^2} \right) \nonumber \\
&+ 4 \, |\mu_{\mathrm{c}}|^4 \left( 1 - e^{ -4\,|\mu_{\mathrm{c}}|^2 \sin^2{\theta(t)/2}} \,e^{- |K_{\hat N_a}|^2} \right) ^2 \, ,\nonumber \\
|\sigma_{31}|^2= & \,  4\, |\mu_{\mathrm{c}}|^4 \, e^{-2 \,|K_{\hat N_a}|^2 } \biggl( e^{-2\,|K_{\hat N_a}|^2}\,e^{ - 4\,|\mu_c|^2  \sin^2{\theta(t)}}    + e^{-8 \,|\mu_{\mathrm{c}}|^2  \sin^2\theta(t)/2} \nonumber \\
&\quad\quad\quad\quad\quad\quad\quad\quad\quad- 2 \, \Re \left\{  e^{i\, \theta(t)} \, e^{|\mu_{\mathrm{c}}|^2 ( e^{2\,i \,\theta(t)} - 1)} e^{-|K_{\hat N_a}|^2} e^{2\, |\mu_c|^2 (e^{- i\, \theta(t)} - 1)} \right\} \biggr) \, .
\end{align}
Putting them together, we find 
\begin{align}
\nu_{Op}^2 &= 1 +4  \, |\mu_{\mathrm{c}}|^2 \left( 1 - e^{-4|\mu_{\mathrm{c}}|^2 \sin^2{\theta(t)/2}} \,e^{- |K_{\hat N_a}|^2} \right) \nonumber \\
&+ 4 \,  |\mu_c|^4 \biggl( 1 - 2 \,  e^{-4\,|\mu_{\mathrm{c}}|^2 \sin^2{\theta(t)/2}} e^{- |K_{\hat N_a}|^2 } + e^{ -8\,|\mu_{\mathrm{c}}|^2 \sin^2{\theta(t)/2}} e^{- 2\,|K_{\hat N_a}|^2 } \nonumber \\
&\quad\quad\quad\quad\quad- e^{-4 \,|K_{\hat N_a}|^2 } \,e^{- 4\, |\mu_c|^2 \,\sin^2\theta(t)} - e^{- 2 \,|K_{\hat N_a}|^2 } e^{- 8\, |\mu_{\rm{c}}|^2 \sin^2\theta(t)/2} \nonumber \\
&\quad\quad\quad\quad\quad+ 2 \, e^{- 3\,|K_{\hat N_a}|^2} \,  \Re \left\{  e^{i \,\theta(t)} \, e^{|\mu_{\mathrm{c}}|^2 ( e^{2\,i\, \theta(t)} - 1)}  e^{2\, |\mu_c|^2 (e^{- i\, \theta(t)} - 1)} \right\}  \biggr) \, ,
\end{align}
which can  be simplified into 
\begin{align}
\nu_{Op}^2 &= 1 +4  \, |\mu_{\mathrm{c}}|^2 \left( 1 - e^{-4\,|\mu_{\mathrm{c}}|^2 \sin^2{\theta(t)/2}} \,e^{- |K_{\hat N_a}|^2} \right) \nonumber \\
&+ 4 \,|\mu_c|^4 \biggl( 1 - 2\, e^{-4\,|\mu_{\mathrm{c}}|^2 \sin^2{\theta(t)/2}} e^{- |K_{\hat N_a}|^2 }  - e^{-4\, |K_{\hat N_a}|^2 } e^{- 4\, |\mu_c|^2 \sin^2\theta(t)} \nonumber \\
&\quad\quad\quad\quad\quad\quad+ 2 \, e^{- 3\,|K_{\hat N_a}|^2} \,  \Re \left[  e^{i\, \theta(t)} \, e^{|\mu_{\mathrm{c}}|^2 ( e^{2\,i\, \theta(t)} - 1)}  \, e^{2\, |\mu_c|^2 (e^{- i\, \theta(t)} - 1)} \right]  \biggr) \, .
\end{align}
Next, we compute the mechanical symplectic eigenvalue.

\subsection{The mechanical symplectic eigenvalue}
We first recall the Bogoliubov identities, which are $|\alpha(\tau)|^2 - |\beta(\tau)|^2 = 1$ and $\alpha(\tau) \beta^*(\tau) - \alpha^*(\tau) \beta(\tau) = 0$. 

The mechanical eigenvalue is given by 
\begin{equation}
\nu_{\rm{Me}}^2 = \sigma_{22}^2 - |\sigma_{42}|^2 \, .
\end{equation} 
Given the covariance matrix elements in Eq.~\eqref{app:exp:values:CM:elements}, we find 
\begin{align}
\sigma_{22}^2 &= 1 + 4 \,  |\beta|^2 + 4\, |\beta|^4+ 4\, |\Delta|^2 |\mu_{\rm{c}}|^2 + 8\, |\beta|^2 \, |\Delta|^2 |\mu_{\rm{c}}|^2   + 4 \,|\Delta|^4 |\mu_{\rm{c}}|^4 \, ,\nonumber \\
|\sigma_{42}|^2 &= 4 \,|\alpha|^2 |\beta|^2 + 4\, \alpha^* \beta^* \Delta^2 |\mu_{\rm{c}}|^2 + 4\, \alpha \beta \Delta^{*2} |\mu_{\rm{c}}|^2 + 4\, |\Delta|^4 |\mu_{\rm{c}}|^4 \, .
\end{align}
This allows us to write 
\begin{equation}
\nu_{\rm{Me}}^2 = \sigma_{22}^2 - |\sigma_{42}|^2 = 1 + 4\,\left[ \left( 1 + 2\, |\beta|^2 \right) |\Delta|^2 - 2 \, \Re \left\{ \alpha \, \beta \Delta^{*2} \right\} \right] |\mu_{\rm{c}}|^2 \, ,
\end{equation}
where we have suppressed the dependence of $\tau$ for notational clarity. 

We wish to simplify this expression by examining each term in turn and using the Bogoliubov conditions. We recall that 
\begin{equation}
\Delta = \left( \alpha + \beta \right) F_{\hat N_a \, \hat B_-} - i \, \left( \alpha - \beta \right) F_{\hat N_a \,\hat B_+} \, , 
\end{equation}
We can now use the Bogoliubov identities to show that 
\begin{equation}
|\Delta|^2 = \left( 1 +  2\,|\beta|^2 \right) \, |K_{\hat N_a}|^2 - ( \alpha \beta^* + \alpha^*  \beta ) \left( F_{\hat N_a \, \hat B_+}^2 - F_{\hat N_a \, \hat B_-}^2 \right) \, ,
\end{equation}
and
\begin{align}
\Delta^{*2} = \, & ( \alpha^*   + \beta^*)^2 F_{\hat N_a \, \hat B_-}^2  - ( \alpha^* - \beta^*)^2 F_{\hat N_a \, \hat B_+}^2 + 2 \, i \, ( \alpha^{*2} - \beta^{*2}) \, F_{\hat N_a \, \hat B_+} \, F_{\hat N_a \, \hat B_-} \nonumber \\
= \, & ( \alpha^{*2}  + \beta^{*2}   + 2\, \alpha^* \beta^*) F_{\hat N_a \, \hat B_-}^2 - ( \alpha^{*2}  + \beta^{*2}   - 2 \, \alpha^* \beta^*) \, F_{\hat N_a \, \hat B_+}^2  \nonumber \\
&+ 2 \, i \, ( \alpha^{*2} - \beta^{*2} ) F_{\hat N_a \, \hat B_+} F_{\hat N_a \, \hat B_-} \nonumber\\
= \, & \left( \alpha^{*2} + \beta^{*2}  \right) \left( F_{\hat N_a \, \hat B_-}^2 - F_{\hat N_a \, \hat B_+}^2 \right) + 2\, \alpha^* \beta^* |K_{\hat N_a}|^2 \nonumber \\ 
&+ 2\, i\, \left( \alpha^{*2} - \beta^{*2}  \right) F_{\hat N_a \, \hat B_+} \, F_{\hat N_a \, \hat B_-}  \, ,
\end{align}
where we recall that $|K_{\hat N_a}|^2 = F_{\hat N_a \, \hat B_+}^2 + F_{\hat N_a \, \hat B_-}^2 $. Furthermore, we find
\begin{align}
\alpha \beta \, \Delta^{*2} = \, & \left( |\alpha|^2\, \alpha^* \beta + \alpha\, \beta^*\, |\beta|^2 \right)\, \left( F_{\hat N_a \, \hat B_-}^2 - F_{\hat N_a \, \hat B_+} ^2\right) + 2\, |\alpha|^2\, |\beta|^2 \,|K_{\hat N_a}|^2 \nonumber \\
&+ 2\, i\, \left( |\alpha|^2 \alpha^*\, \beta - \alpha \,\beta^*\, |\beta|^2 \right) \,F_{\hat N_a \,\hat B_+} \,F_{\hat N_a \, \hat B_-} \nonumber \\
= \, & \alpha^* \,\beta \left( F_{\hat N_a \, \hat B_-}^2 - F_{\hat N_a \, \hat B_+}^2 \right) + |\beta|^2\, \left( \alpha^*\, \beta + \alpha\, \beta^* \right) \left( F_{\hat N_a \, \hat B_-}^2 - F_{\hat N_a \, \hat B_+}^2 \right)  \nonumber \\
&+ 2 \,\left( 1 + |\beta|^2 \right)\, |\beta|^2 \,|K_{\hat N_a}|^2 + 2\, i \,\alpha^*\, \beta \,F_{\hat N_a \, \hat B_+} \,F_{\hat N_a \, \hat B_-} \nonumber \\
&+ 2\, i \,|\beta|^2\, \left( \alpha^*\, \beta - \alpha \,\beta^* \right) \,F_{\hat N_a \, \hat B_+} \,F_{\hat N_a \, \hat B_-} \, ,
\end{align} 
where we used $|\alpha|^2 = |\beta|^2 + 1$ everywhere. We now note that the last term disappears because $\alpha^* \,\beta - \alpha\, \beta^* = 0$. We also note that $ \alpha^* \,\beta = \frac{1}{2} ( \alpha ^*\, \beta + \alpha\, \beta^* ) $ is real, which follows from the Bogoliubov condition $\alpha^*\, \beta = \alpha\, \beta^*$. When we take the real part, the second-to-last term disappears as well because it has an additional $i$, meaning that we are left with 
\begin{align}
\Re \left\{ \alpha\, \beta \,\Delta^{*2} \right\} =& \frac{1}{2} \left( \alpha^* \,\beta + \alpha\, \beta^* \right) \left( F_{\hat N_a \, \hat B_-}^2 - F_{\hat N_a \, \hat B_+}^2 \right)   \nonumber \\
&+ |\beta|^2 \,\left( \alpha^*\, \beta + \alpha \,\beta^* \right) \left( F_{\hat N_a \, \hat B_-}^2 - F_{\hat N_a \, \hat B_+}^2 \right) + 2\,\left( 1 + |\beta|^2 \right)\, |\beta|^2\, |K_{\hat N_a}|^2 \, .
\end{align}
We turn again to the symplectic eigenvalue, which we can now simplify as
\begin{align}
\nu_{\rm{Me}}^2 =&  1 + 4\,\left[ \left( 1 + 2 \,|\beta|^2 \right) |\Delta|^2 - 2 \, \Re \left( \alpha \, \beta \,\Delta^{*2} \right) \right] |\mu_{\rm{c}}|^2 \nonumber \\
=& 1 + 4\, \biggl[ ( 1 + 2 \,|\beta|^2 )^2\, |K_{\hat N_a}|^2  - ( 1 + 2 \,|\beta|^2) \left( \alpha\, \beta^* + \alpha^*\, \beta\right) \left( F_{\hat N_a \, \hat B_+}^2 - F_{\hat N_a \, \hat B_-}^2 \right) \nonumber \\
&\quad\quad\quad- \left( \alpha^*\, \beta + \alpha\, \beta^* \right) \left( F_{\hat N_a \, \hat B_-}^2 - F_{\hat N_a \, \hat B_+}^2 \right) \nonumber \\
&\quad\quad\quad- 2\, |\beta|^2 (\alpha^*\, \beta + \alpha \,\beta^*) \left( F_{\hat N_a \, \hat B_-} ^2 - F_{\hat N_a \, \hat B_+}^2 \right)  - 4\, (1 + |\beta|^2 ) \,|\beta|^2\, |K_{\hat N_a}|^2 \biggr]\, |\mu_{\rm{c}}|^2 \nonumber \\
=& 1 + 4\, \left[ \left( 1 + 4\, |\beta|^2 + 4\,|\beta|^4 \right)\, |K_{\hat N_a}|^2 - 4 \,\left( |\beta|^2 + |\beta|^4 \right) \,|K_{\hat N_a}|^2 \right] |\mu_{\rm{c}}|^2 \nonumber \\
=& 1 + 4 \,|K_{\hat N_a}|^2 \,|\mu_{\rm{c}}|^2 \, .
\end{align}

\subsection{Summary of symplectic eigenvalues}
The symplectic eigenvalues for the optical and mechanical subsystems are given by 
\begin{align}
\label{app:exp:values:sympelctic:eigenvalue:summary}
\nu_{\rm{Op}}^2 =&  \, 1 +4  \, |\mu_{\mathrm{c}}|^2 \left( 1 - e^{-4 \, |\mu_{\mathrm{c}}|^2 \sin^2{\theta(t)/2}} \,e^{- |K_{\hat N_a}|^2} \right) 
 \nonumber \\
 &+ 4 \, |\mu_{\mathrm{c}}|^4 \biggl( 1 - 2 \,  e^{-4 \, |\mu_{\mathrm{c}}|^2 \sin^2{\theta(t)/2}} e^{- |K_{\hat N_a}|^2 }\nonumber\\
&\quad\quad\quad\quad\quad- 2 \, e^{- 4 \, |\mu_{\rm{c}}|^2 \sin^2\theta(t)} \, e^{-4 \, |K_{\hat N_a}|^2 } \,  e^{- 3 \, |K_{\hat N_a}|^2} \, \nonumber \\
&\quad\quad\quad\quad\quad\quad\times  \Re \left\{  e^{i  \, \theta(t)} \, e^{|\mu_{\mathrm{c}}|^2 ( e^{2 \, i  \, \theta(t)} - 1)} \, e^{2 \,  |\mu_{\rm{c}}|^2 (e^{- i \,  \theta(t)} - 1)} \right\}  \biggr) \, , \nonumber \\
\nu_{\mathrm{Me}}^2=&1+4\,|K_{\hat N_a}|^2 |\mu_{\mathrm{c}}|^2 \, ,
\end{align}

\chapter{Consequences of a time-modulated squeezing term }\label{app:mathieu}

In this Appendix, we consider the consequences of the modulated single-mode mechanical squeezing term $\tilde{\mathcal{D}}_2(\tau) \left( \hat b^\dag + \hat b \right)^2$  in the Hamiltonian in Eq.~\eqref{chap:decoupling:eq:Hamiltonian}. 

We explore two specific cases in this Appendix. First, in Section~\ref{app:mathieu:time:varying:trapping:frequency},  we consider a transformation of the optomechanical Hamiltonian with a squeezing term, where we show that the squeezing can be interpreted as a modulation of the trapping frequency $\omega_{\rm{m}}$. Then, in Section~\ref{app:mathieu:perturbative:solutions}, we derive perturbative solutions to the Mathieu equation, which solve the differential equations in Eq.~\eqref{chap:decoupling:differential:equation:written:down}. 

The considerations of a time-varying mechanical trapping frequency in Section~\ref{app:mathieu:time:varying:trapping:frequency} and the second perturbative solution shown in Section~\ref{app:mathieu:perturbative:solutions} this Appendix were derived by Dennis R\"{a}tzel. 

\section{Time-varying mechanical trapping frequency} \label{app:mathieu:time:varying:trapping:frequency}

In this appendix we show how a constant squeezing can be interpreted as a shift in the mechanical oscillation frequency $\omega_{\mathrm{m}}$. The extended Hamiltonian in Eq. ~\eqref{chap:decoupling:eq:Hamiltonian} can be written as
\begin{align}
	\hat {H} &=  \hat{H}'_0 - \hbar\left(\mathcal{G}(t) \hat a^\dagger\hat a - \mathcal{D}_1(t)\right) \left(\hat b+ \hat b{}^\dagger\right),
\end{align}
where the quadratic part, which we call $\hat{H}'_0$, reads
\begin{align}
	\hat{H}'_0 &:=\hbar\,\omega_\mathrm{c} \hat a^\dagger \hat a + \hbar\,\omega_\mathrm{m} \,\hat b{}^\dagger \hat b +  \hbar \, \mathcal{D}_2(t)\left(\hat b+ \hat b{}^\dagger \right)^2. 
\end{align}
To show how the squeezing affects the mechanics, we rewrite the quadratic part as 
\begin{align}
\hat{H}_0'	&= \hbar\,\omega_\mathrm{c} \hat a^\dagger \hat a + \hbar\,\omega_\mathrm{m} \,\hat b{}^\dagger \hat b +  \hbar \, \mathcal{D}_2(t)\left(\hat b+ \hat b{}^\dagger \right)^2 \nonumber \\
&= \hbar\,\omega_\mathrm{c} \hat a^\dagger \hat a + \frac{m\omega_\mathrm{m}^2}{2} \,\left(\hat x_\mathrm{m} - \frac{i}{m\omega_\mathrm{m}}\hat p_\mathrm{m}\right)\left(\hat x_\mathrm{m} + \frac{i}{m\omega_\mathrm{m}}\hat p_\mathrm{m}\right) +  2 m\omega_\mathrm{m} \mathcal{D}_2(t) \hat x_\mathrm{m}^2 \nonumber \\
	&= \hbar\,\omega_\mathrm{c} \hat a^\dagger \hat a + \frac{1}{2m} \hat p_\mathrm{m}^2 +  \frac{m\omega_\mathrm{m}^2}{2} \,\left(1 + \frac{4\mathcal{D}_2(t)}{\omega_\mathrm{m}}\right) \hat x_\mathrm{m}^2 - \frac{\hbar \omega_\mathrm{m}}{2} \,.
\end{align}
where 
\begin{align}
&\hat{x}_{\mathrm{m}} = \sqrt{\frac{\hbar}{2\omega_{\mathrm{m}} m }} ( \hat{b}^\dag + \hat{b} ),\quad\quad \mbox{and} &&\hat{p}_{\mathrm{m}} =i  \sqrt{\frac{\hbar m \omega_{\mathrm{m}}}{2}}(\hat{b}^\dag - \hat{b}) 
\end{align}
This shows that $\mathcal{D}_2(t)$ can be understood and implemented as a possibly time-dependent modulation of the frequency $\omega_\mathrm{m}$ of the mechanical oscillator. For the case of constant squeezing $\mathcal{D}_2$, the Hamiltonian $\hat{H}'_0$ becomes time-independent and we can define $\hat b'$ and $\hat b'{}^\dagger$ with respect to $\omega_\mathrm{m}':= \omega_\mathrm{m} \sqrt{1+4\mathcal{D}_2/\omega_\mathrm{m}}$ such that
\begin{align}
	\hat{H}'_0 &:=\hbar\,\omega_\mathrm{c} \hat a^\dagger \hat a + \hbar\,\omega_\mathrm{m}' \,\hat b'{}^\dagger \hat b' + \frac{\hbar}{2}(\omega'_\mathrm{m} - \omega_\mathrm{m}) \,.
\end{align}
This transformation can be implemented by the squeezing operation $ \hat U^\dagger_{\mathrm{sq}} \hat{H}'_0 \hat U_{\mathrm{sq}}$, where
\begin{align}
\hat 	U_{\rm{sq}} &:=\exp[\frac{r}{2}( \hat b^{\dagger 2} - {\hat b}^2)]  \,,
\end{align}
which induces the mapping
\begin{align}
	\hat U^\dagger_{\rm{sq}} \, \hat b \,  \hat U_{\rm{sq}} &= \cosh(r) \, \hat b + \sinh(r) \, \hat b^{\dagger} \,.
\end{align}
When we apply this to the quadratic Hamiltonian, we obtain 
\begin{align}
\hat U^\dagger_{\rm{sq}} \hat{H}'_0 \hat U_{\rm{sq}} =& \, \hbar\,\omega_{\mathrm{c}} \hat a^\dagger \hat a + \hbar\,\omega_{\mathrm{m}} \,(\cosh(r) \hat b^\dagger + \sinh(r)\hat b) (\cosh(r)\hat b + \sinh(r)\hat b^\dagger)\\
	& +  \hbar \, \mathcal{D}_2\left(\cosh(r)\hat b + \sinh(r)\hat b^\dagger + \cosh(r)\hat b^\dagger + \sinh(r)b \right)^2  \nonumber \\
=& \,  \hbar\,\omega_{\mathrm{c}} \hat a^\dagger \hat a + \hbar\,\omega_{\mathrm{m}} \,\left((1+2\sinh^2(r)) \hat b^\dagger \hat b  +  \cosh(r)\sinh(r)\left(\hat b^{\dagger2} + \hat b^2\right) + \sinh^2(r)  \right) \nonumber \\
& +  \hbar \, \mathcal{D}_2 e^{2r} \left(2 \hat b^\dagger b + b^{\dagger^2} + b^2 + 1\right) \,.
\end{align}
To cancel the term proportional to $\hat b^{\dagger2} + \hat b^2$, we have to fix $\mathcal{D}_2 \, e^{2r} = -\omega_{\mathrm{m}} \cosh(r)\sinh(r) = -\omega_{\mathrm{m}}(e^{2r} - e^{-2r})/4$, and therefore, $e^{-2r} = \sqrt{ 1 + 4\mathcal{D}_2/\omega_{\mathrm{m}}} = \omega'_{\mathrm{m}}/\omega_{\mathrm{m}}$. With $\mathcal{D}_2 = \omega_{\mathrm{m}}((\omega'_{\mathrm{m}}/\omega_{\mathrm{m}})^2-1)/4$, we obtain
\begin{align}
	 \hat{H}''_0 :=& \hat U^\dagger_{\rm{sq}} \hat{H}'_0 \hat U_{\rm{sq}} \nonumber \\
=& \hbar\,\omega_\mathrm{c} \hat a^\dagger \hat a + \hbar\,\omega_\mathrm{m} \,\left((1+2\sinh^2(r)) \hat b^\dagger \hat b  + \sinh^2(r)  \right) +  \hbar \, \mathcal{D}_2 e^{2r} \left(2 \, \hat b^\dagger \hat b + 1\right) \nonumber \\
=& \hbar\,\omega_\mathrm{c} \hat a^\dagger \hat a + \hbar\,\omega_\mathrm{m} \,\left(\frac{1}{2}\left(\frac{\omega'_\mathrm{m}}{\omega_\mathrm{m}} + \frac{\omega_\mathrm{m}}{\omega'_\mathrm{m}}\right) \hat b^\dagger \hat b  + \frac{1}{4}\left(\frac{\omega'_\mathrm{m}}{\omega_\mathrm{m}} - 2 + \frac{\omega_\mathrm{m}}{\omega'_\mathrm{m}}\right)  \right) \nonumber \\
	& + \frac{\hbar}{4} \, \omega_\mathrm{m}\left(\frac{\omega'_\mathrm{m}}{\omega_\mathrm{m}} - \frac{\omega_\mathrm{m}}{\omega'_\mathrm{m}}\right) \left(2 \, \hat b^\dagger \hat b + 1\right) \nonumber \\
	=& \hbar\,\omega_\mathrm{c} \hat a^\dagger \hat a + \hbar\,\omega'_\mathrm{m} \, \hat b^\dagger \hat b  + \frac{\hbar}{2}(\omega'_\mathrm{m} - \omega_\mathrm{m})  \,.
\end{align}
In particular,  $\hat{H}_0 \rightarrow \hat{H}''_0 - \hbar(\omega'_\mathrm{m} - \omega_\mathrm{m})/2$ under the replacement $\omega_\mathrm{m} \rightarrow  \omega'_\mathrm{m}$. Furthermore, we find that $\hat b+ \hat b{}^\dagger$ transforms as
\begin{eqnarray}
	\hat U^\dagger_{\rm{sq}}\left(\hat b + \hat b{}^\dagger\right)\hat U_{\rm{sq}} = \sqrt{\frac{\omega_\mathrm{m}}{\omega'_\mathrm{m}}}\left(\hat b+ \hat b{}^\dagger\right)\,.
\end{eqnarray}
When applying the same transformation to the nonlinear part of the Hamiltonian, we find
\begin{align}
	\hat {H}'':= \hat U^\dagger_{\rm{sq}} \hat {H} \hat U_{\rm{sq}} &=  \hat{H}''_0 - \hbar\sqrt{\frac{\omega_\mathrm{m}}{\omega'_\mathrm{m}}}\left(\mathcal{G}(t) \hat a^\dagger\hat a - \mathcal{D}_1(t)\right) \left(\hat b+ \hat b{}^\dagger\right)\,.
\end{align}

If $\mathcal{G}(t)\propto 1/\sqrt{\omega_\mathrm{m}}$ and $\mathcal{D}_1(t)\propto 1/\sqrt{\omega_\mathrm{m}}$, which is indeed fulfilled for the interesting cases, we find that $\hat{H}_\mathrm{opt} \rightarrow \hat{H}'' - \hbar(\omega'_\mathrm{m} - \omega_\mathrm{m})/2$ under the replacement $\omega_\mathrm{m} \rightarrow  \omega'_\mathrm{m}$, where
\begin{equation}
	\hat {H}_\mathrm{opt} =  \hat{H}_0 - \hbar\left(\mathcal{G}(t) \hat a^\dagger\hat a - \mathcal{D}_1(t)\right) \left(\hat b+ \hat b{}^\dagger\right)\,.
\end{equation} 
We define $\hat {H}_\mathrm{opt}^{\omega'_\mathrm{m}}:=\hat {H}_\mathrm{opt}[\omega_\mathrm{m}\rightarrow \omega'_\mathrm{m}]$, and we obtain for the full time evolution
\begin{eqnarray}
	\hat{U}(t)&=&\overset{\leftarrow}{\mathcal{T}}\,\exp\left[-\frac{i}{\hbar}\int_0^{t} dt'\,\hat{H}(t')\right] \nonumber \\
	&=& \hat U_{\rm{sq}}\overset{\leftarrow}{\mathcal{T}}\,\exp\left[-\frac{i}{\hbar}\int_0^{t} dt'\,\hat{H}''(t')\right] \hat U^\dagger_{\rm{sq}} \nonumber \\
	&=& e^{-\frac{i}{2}(\omega'_\mathrm{m} - \omega_\mathrm{m})t} \hat U_{\rm{sq}}\overset{\leftarrow}{\mathcal{T}}\,\exp\left[-\frac{i}{\hbar}\int_0^{t} dt'\,\hat {H}_\mathrm{opt}^{\omega'_\mathrm{m}}(t')\right] \hat U^\dagger_{\rm{sq}} \nonumber  \\
	&=& e^{-\frac{i}{2}(\omega'_\mathrm{m} - \omega_\mathrm{m})t} \hat U_{\rm{sq}} \hat{U}^{\omega'_\mathrm{m}}_\mathrm{opt}(t) \hat U^\dagger_{\rm{sq}}\,.
\end{eqnarray}
For the expectation values of quadrature of a state $\hat \rho$, this leads to 
\begin{eqnarray}
	\nonumber \langle \hat{x}_\mathrm{m}(t) \rangle &=& \mathrm{Tr}(\hat{x}_\mathrm{m} \hat{U}(t) \hat \rho \hat{U}^\dagger(t)) \\
	\nonumber &=&  \mathrm{Tr}(\hat U^\dagger_{\rm{sq}} \hat{x}_\mathrm{m} \hat U_{\rm{sq}} \hat{U}^{\omega'_\mathrm{m}}_\mathrm{opt}(t) \,  \hat \rho^\mathrm{sq} \,  \hat{U}^{\omega'_\mathrm{m}\dagger}_\mathrm{opt}(t)) \nonumber \\
	&=&  \mathrm{Tr}\left(\sqrt{\frac{\hbar }{2m\omega'_\mathrm{m}}}\left(\hat b^\dagger + \hat b\right) \hat{U}^{\omega'_\mathrm{m}}_\mathrm{opt}(t)  \, \hat \rho^\mathrm{sq} \, \hat{U}^{\omega'_\mathrm{m}\dagger}_\mathrm{opt}(t)\right) \, ,\nonumber \\
	\langle \hat{p}_\mathrm{m}(t) \rangle &=& \mathrm{Tr}\left(i\sqrt{\frac{\hbar m\omega'_\mathrm{m}}{2}}\left(\hat b^\dagger - \hat b\right) \hat{U}^{\omega'_\mathrm{m}}_\mathrm{opt}(t) \,  \hat \rho^\mathrm{sq} \,  \hat{U}^{\omega'_\mathrm{m}\dagger}_\mathrm{opt}(t)\right)\, , \nonumber \\
		\langle \hat{x}_\mathrm{c}(t) \rangle &=& \mathrm{Tr}\left(\hat{x}_\mathrm{c} \hat{U}^{\omega'_\mathrm{m}}_\mathrm{opt}(t)  \, \hat \rho^\mathrm{sq}  \, \hat{U}^{\omega'_\mathrm{m}\dagger }_\mathrm{opt}(t)\right)\, , \nonumber \\
	\langle \hat{p}_\mathrm{c}(t) \rangle &=& \mathrm{Tr}\left(\hat{p}_\mathrm{c} \hat{U}^{\omega'_\mathrm{m}}_\mathrm{opt}(t)  \, \hat \rho^\mathrm{sq}  \, \hat{U}^{\omega'_\mathrm{m}\dagger}_\mathrm{opt}(t)\right) \, ,
\end{eqnarray}
where $\hat{\rho}^\mathrm{sq} := \hat U^\dagger_{\rm{sq}} \hat{\rho} \hat U_{\rm{sq}}$. For the initial separable coherent state $|\mu_\mathrm{c}\rangle|\mu_\mathrm{m}\rangle$, we find that the time evolution of the quadratures induced by the full Hamiltonian $\hat{H}$ with constant $\mathcal{D}_2$ can be obtained by calculating the corresponding time evolution of the quadratures induced by $\hat{H}$ with vanishing $\mathcal{D}_2$ by replacing $\omega_\mathrm{m}$ with $\omega'_\mathrm{m}$ and considering the squeezed coherent initial state $\hat U^\dagger_{\rm{sq}}|\mu_\mathrm{c}\rangle|\mu_\mathrm{m}\rangle =|\mu_\mathrm{c}\rangle|\mu'_\mathrm{m},r\rangle $, where $\mu'_\mathrm{m}= \mu_\mathrm{m}\cosh(r) + \mu_\mathrm{m}^*\sinh(r)$.

As a result, the techniques we have developed here can also be utilised to model all the expectation values for an optomechanical system for an initially squeezed states. 

\section{Perturbative solutions to the Mathieu equation} \label{app:mathieu:perturbative:solutions}

In Chapter~\ref{chap:non:Gaussianity:squeezing}, we found that the equations of motion in Eqs.~\eqref{chap:decoupling:differential:equation:written:down} and~\eqref{chap:decoupling:eq:IP22} become the Mathieu equation for a time-dependent squeezing term. 
The Mathieu equation is notoriously difficult to solve. It lacks general analytic solutions and is extremely difficult to solve numerically. 

In this Section, we derive perturbative solutions to the Mathieu equations using two different methods. We then compare them with the rotating-wave approximation and show that the solutions coincide for $\tau \gg1$. These approximate solutions are used in Chapter~\ref{chap:non:Gaussianity:squeezing} to compute the non-Gaussianity of an optomechanical state that evolves under the extended Hamiltonian in Eq.~\eqref{chap:decoupling:eq:Hamiltonian} where the rescaled squeezing term weighting function $\tilde{\mathcal{D}}_2(\tau) = \mathcal{D}_2(\omega_{\rm{m}} \, t)/\omega_{\rm{m}}$ is taken to be of the form
\begin{equation}
\tilde{\mathcal{D}}(\tau) = \tilde{d}_2 \, \cos(\Omega_{\tilde{d}_2} \, \tau) \, ,
\end{equation}
where $\tilde{d}_2 = d_2/\omega_{\rm{d}}$ is the squeezing amplitude, and $\Omega_{\tilde{d}_2} = \omega_{\tilde{d}_2}/\omega_{\rm{m}}$ is the rescaled oscillation frequency. 

From these assumptions, the differential equations in Eq.~\eqref{chap:decoupling:differential:equation:written:down} and Eq.~\eqref{chap:decoupling:eq:IP22} for the parameter $P_{11}$ and $I_{P_{22}}$ take on the following from:
\begin{align}
\ddot{P}_{11} &=- \left( 1 + 4 \, \tilde{d}_2 \, \cos(\Omega_{\tilde{d}_2} \, \tau) \right) \, P_{11} \, , \nonumber \\
\ddot{I}_{P_{22}} &= -\left( 1 + 4 \, \tilde{d}_2 \, \cos(\Omega_{\tilde{d}_2} \, \tau) \right) \, I_{P_{22}} \, . 
\end{align}
When $\Omega_{\mathcal{d}_2} = 2$, we identify these equations to the be Mathieu equation, which reads 
\begin{equation} \label{app:Mathieu:Mathieu:equation}
\frac{d^2 y}{dx^2} + \left[ a - 2  \, q \, \cos(2 \, x) \right] \, y= 0 \, .
\end{equation}
This equation is notoriously difficult to solve.  Numerical solutions tend to be unstable and difficult to obtain to high accuracy, since the solutions grow exponentially for certain parameter regimes. See Ref~\cite{kovacic2018mathieu} for a stability chart and further analysis of the equations. We found that although standard mathematical tools such as \textit{Mathematica} provides built-in solutions to the Mathieu equations, these could not be obtained to the degree required for accurate computations. 

\subsection{Approximate two-scale solution}
Our goal is to obtain approximate solutions to the differential equations Eq.~\eqref{chap:decoupling:differential:equation:written:down} and Eq.~\eqref{chap:decoupling:eq:IP22}. We do so by following the derivation in Ref~\cite{kovacic2018mathieu} with some modifications. 

The general form of Mathieu's equation is given by 
\begin{equation} \label{app:Mathieu:Mathieu:equation:repeat}
\frac{d^2 y}{dx^2} + \left[ a - 2  \, q \, \cos(2 \, x) \right] \, y= 0 \, .
\end{equation}
We use the general notation in this Appendix and then compare it with the notation used in the main text in Chapter~\ref{chap:non:Gaussianity:squeezing}. 

We begin by defining a slow time scale $X= q x$. We then assume that the solutions $y$ depend on both scales, such that $y(x, X)$. This means that we can treat $x$ and $X$ as independent variables and the absolute derivative $d/dx$ in Eq.~\eqref{app:Mathieu:Mathieu:equation:repeat} can be split in two:
\begin{equation}
\frac{d}{dx} = \partial_x +  q \,  \partial_X \, .
\end{equation}
Mathieu's equation Eq.~\eqref{app:Mathieu:Mathieu:equation:repeat} therefore becomes
\begin{equation}
\left( \partial_x +  q \,  \partial_X \right)^2 y(x, X) + ( a - 2 q \cos(2 x) ) \, y(x, X) = 0 \, .
\end{equation}
We then expand the solution $y(x, X)$ for small $q$ as $y(x,X) = y_0(x,X) +  q\, y_1 (x,X) + \mathcal{O}(q^2)$ and insert this into the differential equation above. Our goal is to obtain a solution for $y_0$ which incorporates a number of restrictions from the differential equation for $y_1$. 

To zeroth order, we recover the regular harmonic differential equation for $y_0$, which is the limiting case as $q \rightarrow 0$:
\begin{equation}
\partial^2_x y_0 + a \, y_0 = 0 \, ,
\end{equation}
where we know that the solutions are sinusoidal, while the coefficients must depend on $X$.  We choose the following trial solution:
\begin{equation}
y_0(x,X) = A(X) \, e^{i \sqrt{a} \, x} +A^*(X) \, e^{- i \sqrt{a} \, x} \, .
\end{equation}
Our goal is now to find explicit solutions to the complex function $A(X)$. We continue with the equation for $y_1$. We discard all terms of order $q^2$ to find
\begin{equation}
 q \, \partial^2_x y_1 + 2\, q \,  \partial_x \partial _X y_0 +   a \, q \, y_1 -2 \, q \cos(2 x) y_0 = 0
\end{equation}
We divide by $ q$ and insert our solution for $y_0$ to find 
\begin{align}
 &\partial^2_x y_1 + a \, y_1 + 2 \, i \, \sqrt{a} \, \left(\frac{\partial A(X)}{\partial X} \, e^{i \sqrt{a} x} - \frac{\partial A^*(X)}{\partial X} \, e^{ -i \sqrt{a} x}    \right)  \nonumber \\
 &\quad- 2  \, \cos(2 x)  \left( A(X) \, e^{i \sqrt{a} x}  + A^*(X) \, e^{- i \sqrt{a} x} \right) = 0
\end{align}
At this point, we specialise to $a = 1$, which corresponds to setting $\Omega_0 = 2$ in the main text (the resonance condition). We combine the exponentials to find
\begin{align}
&\partial^2_x \, y_1 + a\, y_1 + \left( 2  i \frac{\partial A(X)}{\partial X} -  \, A^*(X) \right) \, e^{i x} + \left( 2 i \frac{\partial A^*(X)}{\partial X} + \, A(X) \right) \, e^{- i x} \nonumber \\
&\quad  -  A(X) \, e^{3 i x} - A^*(X) \, e^{- 3 i x} = 0
\end{align}
In order for the solution to be stable, we require that secular terms such as resonant terms $e^{i x}$ vanish. If these do not vanish, the solution will grow exponentially~\cite{kovacic2018mathieu}.  We also neglect terms that oscillate much faster, such as $e^{3 i x}$.  This leaves us with the condition that
\begin{equation} \label{app:resonance:condition}
 \left( 2 i \frac{\partial A^*(X)}{\partial X} + \, A(X) \right)  = 0 \,  , 
\end{equation}
which can be differentiated again and solved with the trial solution $A(X) = (c_1 - i \, c_2) \, e^{X/2} + (c_3- i\,  c_4) \, e^{- X/2}$ for the  parameters $c_1,c_2,c_3$ and $c_4$. From the requirement in Eq.~\eqref{app:resonance:condition}, it is now possible to fix two of the coefficients in Eq.~\eqref{app:y0:almost:final:solution}. We differentiate $A(X)$ and use Eq.~\eqref{app:resonance:condition} to find that the conditions $c_1 =  c_2 $ and $c_3 = -c_4$ must always be fulfilled. 

We then recall that $X =  q x$ and after combining some exponentials, we obtain the full trial solution for the zeroth order term $y_0$:
\begin{align} \label{app:y0:almost:final:solution}
y_0(x) &=  A(qx) \, e^{i  \, x} +A^*(qx) \, e^{ -i  x} \, \nonumber \\
&= 2 \,  \left( c_1 \, e^{qx /2} + c_3 \, e^{- qx/2} \right) \, \cos(x) +  2 \, \left( c_1 \, e^{qx/2} - c_3 \, e^{- qx/2} \right) \sin(x) \, .
\end{align}
We now proceed to compare this solution with the parameters and initial conditions given for $P_{11}$ in Eq.~\eqref{chap:decoupling:differential:equation:written:down} and $I_{P_{22}}$ in Eq.~\eqref{chap:decoupling:eq:IP22} in the main text. 

First, we note that $q = - 2\,  \tilde{d}_2$ and that $x = \tau$. Then we consider the boundary conditions for $P_{11}$, which are $P_{11}(0) = 1$ and $\dot{P}_{11} (0) = 0$. From these conditions, we find that  $c_1 = c_3 = 1/4 $, and the the approximate solution to $P_{11}$ is given by 
\begin{equation} \label{app:mathieu:eq:P11:approx}
P_{11}(\tau) =\cos (\tau) \,  \cosh (\tilde{d}_2 \, \tau) - \sin (\tau) \,  \sinh ( \tilde{d}_2 \,  \tau) \, .
\end{equation}
The equation for $I_{P_{22}}$ has the opposite initial conditions $I_{P_{22}}(0) = 0$ and $\dot{I}_{P_{22}} = 1$. For this case, we find that $c_1 = -c_3 =  1/(4 (1 - \tilde{d}_2))$. The full solution to $I_{P_{22}}$ is therefore
\begin{equation} \label{app:mathieu:eq:IP22:approx}
I_{P_{22}}(\tau) = - \frac{1}{ 1- \tilde{d}_2}\left(\cos (\tau) \,  \sinh (\tilde{d}_2 \, \tau ) - \sin (\tau) \,  \cosh (\tilde{d}_2 \, \tau) \right)  \, , 
\end{equation}
and thus
\begin{equation} 
P_{22} = \cos (\tau) \cosh (\tilde{d}_2 \,  \tau)-\frac{\tilde{d}_2 + 1}{\tilde{d}_2-1} \sin (\tau) \sinh (\tilde{d}_2 \,  \tau) \, .
\end{equation}
Both solutions reduce to the correct expressions as $\tilde{d}_2 \rightarrow 0$. From the expression for $\xi(\tau)$ in Eq.~\eqref{chap:decoupling:eq:definition:of:xi} we then find
\begin{align}
\xi(\tau) =&\cos (\tau) \,  \cosh (\tilde{d}_2 \,  \tau) - \sin (\tau)  \, \sinh (\tilde{d}_2 \, \tau) \nonumber \\
&- \frac{i}{1- \tilde{d}_2} \,  \left( \sin (\tau) \cosh (\tilde{d}_2 \, \tau) -\cos (\tau) \sinh (\tilde{d}_2 \, \tau)\right) \, . 
\end{align} 
For very small $\tilde{d}_2 \ll 1$, which was the condition for deriving the approximate solutions in the first place, we can approximate the fraction as unity and we find the compact expression
\begin{align} \label{app:mathieu:eq:RWA:solutions}
\xi(\tau) =& \,  e^{- i \, \tau} \, \cosh (\tilde{d}_2 \, \tau) + i \, e^{i \, \tau} \,  \sinh (\tilde{d}_2 \, \tau) \, .
\end{align} 
To better understand what this approximation entails physically, we compare it with the rotating-wave approximation, which has a well-known physical interpretation. 

\subsection{Alternative solution}
There is another solution which more explicitly demonstrates how the resonance conditions helps constrain the solution. 
We write the solution to the differential equation for $P_{11}$ as
\begin{equation}
	P_{11}(\tau) = Q_{c}(\tau) \cos(\tau + \pi/4 ) + Q_{s}(\tau) \sin(\tau + \pi/4 )\,.
\end{equation}
Then, the differential equation $\ddot P_{11} + (1+f(\tau)) P_{11} = 0$ is solved approximately,
considering only terms of first order in $\tilde{d}_2$ and neglecting terms rotating with
frequency $3$ off-resonantly, by the following set of differential equations 
\begin{eqnarray}
	\dot Q_c(\tau) =  \tilde{d}_2 Q_c(\tau) \quad\mathrm{and}\quad	\dot Q_s(\tau) =  - \tilde{d}_2 Q_s(\tau)\, .
\end{eqnarray}
These can be solved as
\begin{eqnarray}
	Q_c(\tau) =  e^{\tilde{d}_2 \tau} Q_c(0) \quad\mathrm{and}\quad Q_s(\tau) = e^{- \tilde{d}_2 \tau} Q_s(0)\,.
\end{eqnarray}
As the initial conditions for $P_{11}$ are $P_{11}(0)=1$ and $\dot P_{11}(0)=0$ ,
we find 
\begin{equation}
	Q_{c}(0) + Q_{s}(0) = \sqrt{2} \, , \quad\rm{and}\quad  \left(1 - \tilde{d}_2\right) \left( Q_{c}(0) - Q_{s}(0) \right) = 0  \, ,
\end{equation}
which implies $Q_{c}(0) = Q_{s}(0) =  1/\sqrt{2}$, and
\begin{equation}
	P_{11}(\tau) = \frac{1}{\sqrt{2}}\left( e^{\tilde{d}_2\tau}  \cos(\tau + \pi/4 ) + e^{-\tilde{d}_2\tau} \sin(\tau + \pi/4 )\right)\,.
\end{equation}
The same steps as above can be applied to find an approximate solution for $I_{P_{22}}$:
\begin{equation}
	I_{P_{22}}(\tau) = \bar Q_{c}(\tau) \cos(\tau + \pi/4 ) + \bar Q_{s}(\tau) \sin(\tau + \pi/4 )\,, 
\end{equation}
with
\begin{eqnarray}
	\bar Q_c(\tau) =  e^{\tilde{d}_2\tau} \bar Q_c(0) \, , \quad\mathrm{and}\quad \bar Q_s(\tau) = e^{-\tilde{d}_2\tau} \bar Q_s(0)\,.
\end{eqnarray}
and
\begin{equation}
	\bar Q_{c}(0) + \bar Q_{s}(0) =  0 \, , \quad\rm{and}\quad  - \left(1 - \tilde d_2\right) \left( \bar Q_{c}(0) - \bar Q_{s}(0) \right) =  \sqrt{2} \, ,
\end{equation}
which leads to 
\begin{align}
	\bar Q_c(0)= & -\frac{1}{\sqrt{2}\left(1 - \tilde d_2\right)} \, ,\quad \mathrm{and} \quad \bar Q_s(0)=  \frac{1}{\sqrt{2}\left(1 - \tilde d_2\right)} \, ,
\end{align}
and
\begin{equation}
I_{P_{22}}(\tau) = -\frac{1}{\sqrt{2}\left(1 - \tilde d_2\right)}\left( e^{\tilde{d}_2\tau}  \cos(\tau + \pi/4 ) - e^{-\tilde{d}_2\tau} \sin(\tau + \pi/4 )\right)\,.
\end{equation}
We find that
\begin{align}
	\xi(\tau) =& \frac{1}{\sqrt{2}}\left( 1 + i\frac{1}{\left(1 - \tilde d_2\right)}\right) e^{\tilde{d}_2\tau}  \cos(\tau + \pi/4 ) \nonumber \\
	&+  \frac{1}{\sqrt{2}}\left( 1 - i\frac{1}{\left(1 - \tilde d_2\right)}\right) e^{-\tilde{d}_2\tau} \sin(\tau + \pi/4 ) \,  \, ,
\end{align}
which exactly coincides with the solution in Eq.~\eqref{app:mathieu:eq:RWA:solutions}.

\subsection{Comparison with the rotating-wave approximation} \label{app:mathieu:rotating:wave:approximation}
Here we compare the approximate resonance solution for $P_{11}$ in Eq.~\eqref{app:mathieu:eq:P11:approx} and $I_{P_{22}}$  in Eq.~\eqref{app:mathieu:eq:IP22:approx} with the rotating-wave approximation, which is obtained as an approximation to the Hamiltonian in Eq. ~\eqref{chap:decoupling:eq:Hamiltonian} when $\tau \gg 1$. 
In the main text, we separated the Hamiltonian Eq. ~\eqref{chap:decoupling:eq:Hamiltonian} into a part with the coupling (Eq.~\eqref{chap:decoupling:free:and:coupling:Hamiltonian}) and a squeezing term (Eq. ~\eqref{chap:decoupling:eq:Hamiltonian:squeezing:subsystem}), which for our specific choice of $\tilde{\mathcal{D}}_2(\tau) = \tilde{d}_2 \, \cos(\Omega_0 \, \tau)$ becomes
\begin{align}
\hat{{H}}_{\rm{sq}} &= \hat b^\dag \hat b + \tilde{d}_2   \cos(\Omega_0 \tau)  \left( \hat b^\dag + \hat b  \right)^2 \, .
\end{align}
We now define the free evolution Hamiltonian $\hat{H}_0 =  \hat b^\dag \hat b$ and the squeezing term $\hat{H}'_{\rm{sq}} = \tilde{d}_2  \cos(\Omega_0 \, \tau)  ( \hat b^{\dag} + \hat b  )^2$ as separate operators. 
We transform into a frame rotating with $\exp[-i \,  \tau \, \hat b^\dag \hat b ]$, which means that the squeezing term transforms into 
\begin{align} \label{eq:rotating:frame:more:exponentials}
e^{i \hat{H}_0 \, \tau} \hat{H}_{\rm{sq}}' e^{ - i \hat{H}_0 , \tau} =& \, \tilde{d}_2 \, \cos(\Omega_0 \, \tau)  \left( e^{2 \, i \, \tau} \, b^{2\dag} +  e^{- 2 \, i \, \tau} \, \hat b^2 + 2 \, \hat b^\dag \hat b + 1\right) \, .
\end{align}
For this specific case, the system becomes resonant when $\Omega_0 = \omega_0/\omega_{\mathrm{m}} = 2$. We can see this by expanding the cosine in terms of exponentials to obtain
\begin{align}
e^{i \hat{H}_0 \, \tau} \hat{H}_{\rm{sq}}' e^{ - i \hat{H}_0 , \tau} =& \frac{1}{2} \tilde{d}_2 \left[ \left( e^{i \,(2 + \Omega_0 ) \, \tau} + e^{i \, (2 - \Omega_0 )\, \tau} \right) \hat b^{2\dag} + \left( e^{  -i \, (2 + \Omega_0 ) \, \tau} + e^{- i \, ( 2 - \Omega_0 )\, \tau} \right) \hat b^2 \right] \, \nonumber \\
&+ \tilde{d}_2 \, \cos(\Omega_0 \, \tau) \left( 2 \, \hat b^\dag \hat b  + 1 \right) \, .
\end{align}
When $\Omega_0 = 2$, two of the time-dependent terms will cancel. We then perform the rotating-wave approximation, i.e.~we neglect all remaining time-dependent terms. This approximation is only valid for $\tau \gg1$. 
In the interaction frame, we find
\begin{align} \label{app:rotating:frame:approx}
\hat H_{\rm{sq}, I} = e^{i \hat H_0\, \tau} \hat{H}_{\rm{sq}}' e^{ - i \hat H_0 \tau} \approx  \frac{1}{2} \tilde{d}_2    \, (  \hat b^{\dag 2}  +  \hat b^2 )  \, .
\end{align}
In the symplectic basis $\hat \vec{r} = (\hat b, \hat b ^\dag )^{\rm{T}}$ the corresponding symplectic operator, given by $\boldsymbol{S}_{\rm{sq}} = e^{\boldsymbol{\Omega} \, \boldsymbol{H}_{\rm{sq}}}$, where $\boldsymbol{\Omega} = i \, \mathrm{diag}(-1, 1)$ and $\boldsymbol{H}_{\rm{sq}}$ is given by 
\begin{align}
\boldsymbol{H}_{\rm{sq}} =  \tilde{d}_2  \, \begin{pmatrix} 0 & 1 \\ 1  & 0 \end{pmatrix} 
\end{align}
The symplectic representation of the squeezing operator in Eq.~\eqref{main:general:expression:time:dependent:squeezing} in the lab frame  reads 
\begin{align}\label{app:lab:frame:evolution}
\boldsymbol{S}_{\mathrm{sq}}(\tau)=&\boldsymbol{S}_0(\tau) \,\begin{pmatrix}
\cosh (\tilde{d}_2 \,\tau) & -i\,\sinh(\tilde{d}_2 \,\tau) \\
i\,\sinh(\tilde{d}_2 \,\tau)  & \cosh (\tilde{d}_2 \,\tau) 
\end{pmatrix},
\end{align}
where $\boldsymbol{S}_0 = e^{-  i \, \tau}$ encodes the evolution from the Hamiltonian $\hat{H}_0$. We therefore find the Bogoliubov coefficients 
\begin{align}
\alpha(\tau)=&  \, e^{-  i \, \tau}\,\cosh (\tilde{d}_2 \,\tau) \, ,\nonumber\\
\beta(\tau)=&-i\, e^{-  i \, \tau}\,\sinh (\tilde{d}_2 \,\tau) \, ,
\end{align}
which 
evidently satisfy the Bogoliubov conditions, and obtain 
\begin{equation}
\xi = \alpha(\tau) + \beta^*(\tau) = e^{- i \, \tau} \, \cosh (\tilde{d}_2 \, \tau) + i \, e^{i \, \tau} \,  \sinh (\tilde{d}_2 \, \tau) \, .
\end{equation}
This expression exactly matches the one we derived as a perturbative solution to the Mathieu equations in Eq.~\eqref{app:mathieu:eq:RWA:solutions}. However, the requirement for the validity of the RWA is that $\tau \gg1$, while the approximate solutions are still valid for small $\tau$. We conclude that the approximate solutions only coincide with the RWA for large $\tau$, while this interpretation cannot be used when $\tau \sim1$. 

\chapter{Quantum Fisher information: Derivation and expressions for pure and mixed states}
\label{app:QFI}
\chaptermark{Quantum Fisher information: Derivation and expressions for...}

In this Appendix, derive a number of relations and expressions for the quantum Fisher information (QFI) and classical Fisher information (CFI). 

In Section \ref{app:QFI:derivation:of:QFI:mixed:states}, we derive an expression for the quantum Fisher information (QFI) for mixed states. The expression we derive are used in Chapter~\ref{chap:metrology} to compute the quantum and classical Fisher information for optomechanical systems. In Section~\ref{app:QFI:different:initial:states:expressions}, we consider three different initial states and derive a compact yet general form of the QFI for each case. Following that, in Section~\ref{app:QFI:expressions:estimation:schemes}, we consider specific estimation schemes and insert the $F$ and $J$-coefficients found in Appendix~\ref{app:coefficients} to derive often long and cumbersome expressions. We refer to these in the main text, in Chapter~\ref{chap:metrology}. Finally, in Section~\ref{app:QFI:homodyne:optimality:proof}, we prove that the CFI for a homodyne measurement is optimal when estimating a constant gravitational acceleration. 


\section{Derivation of QFI for mixed states} \label{app:QFI:derivation:of:QFI:mixed:states}
In this Section, we provide a derivation of the quantum Fisher information for mixed states in Eq.~\eqref{chap:metrology:definition:of:QFI}. We closely follow the derivation in Refs~\cite{pang2014quantum} and~\cite{jing2014quantum}. 

We reprint Eq.~\eqref{chap:introduction:eq:QFI:definition:of:QFI:mixed:states} here for reference, using the same notation as in Refs~\cite{pang2014quantum} and~\cite{jing2014quantum} to better distinguish between the eigenstates and eigenvectors:
\begin{align}\label{app:QFI:definition:of:QFI:mixed:states}
\mathcal{I}_\theta
=& \;4\sum_i p_i\left(\bra{\psi_i}\mathcal{\hat H}_\theta^2\ket{\psi_i} - \bra{\psi_i}\mathcal{\hat H}_\theta\ket{\psi_i}^2 \right)\nonumber\\
&\;-8\sum_{i\neq j}
\frac{p_i \, p_j}{p_i+p_j}
\left| \bra{\psi_i}\mathcal{\hat H}_\theta \ket{\psi_j}\right|^2\;,
\end{align}
where $\hat{\mathcal{H}}_\theta = - i \dot{\hat{U}}_\theta \hat U_\theta$.  

To derive this expression, we assume that $\ket{\psi_i}$ is the eigenbasis of the evolved state with eigenvalues $p_i$. The symmetric logarithmic derivative (SLD) is then given by 
\begin{equation}
\hat L_\theta = 2 \sum_{i,j} \frac{\bra{p_i} \p_\theta \hat \rho \ket{p_j} }{p_i + p_j} \ket{\psi_i}\bra{\psi_j}   \, .
\end{equation}
Then, the Fisher information is, in general, given by 
\begin{align}
\mathcal{I}_\theta = \mathrm{Tr} \left[ \hat L^\dag_\theta\,  \hat \rho \,  \hat L_\theta  \right] \, .
\end{align}
Follow the proof in Ref~\cite{jing2014quantum}, we first write the state in its eigenbasis
\begin{equation}
\hat \rho = \sum_i p_i \ketbra{\psi_i} \, . 
\end{equation}
Then we note that the definition of the SLD is given by 
\begin{equation}
\p_\theta \hat \rho = \frac{1}{2} \left( \hat L_\theta\hat  \rho + \hat \rho \hat L
_\theta \right) \, .
\end{equation}
Overlapping this definition with two of the eigenbasis states gives us
\begin{align}
\bra{\psi_i} \p_\theta \hat \rho \ket{\psi_j} &=  \frac{1}{2} \left( \bra{\psi_i}\hat L \hat \rho \ket{\psi_j} + \bra{\psi_i}\hat \rho \hat L \ket{\psi_j} \right) \nonumber \\
&=  \frac{1}{2} \left( \bra{\psi_i}\hat L \sum_k p_k \ketbra{\psi_k} \ket{\psi_j} + \bra{\psi_i}\sum_k p_k \ketbra{\psi_k}L \ket{\psi_j} \right) \nonumber \\
&= \frac{1}{2} \left( \bra{\psi_i}\hat L \sum_k p_k \ketbra{\psi_k} \ket{\psi_j} + \bra{\psi_i}\sum_k p_k \ketbra{\psi_k}L \ket{\psi_j} \right) \nonumber \\
&= \frac{1}{2} \left( \bra{\psi_i} \hat L p_j \ket{\psi_j} + \bra{\psi_i} p_i \hat L \ket{\psi_j} \right) \nonumber \\
&= \frac{1}{2} \left( p_i + p_j \right) L_{ij} \, ,
\end{align}
where we have defined $L_{ij} = \bra{\psi_i} \hat L \ket{\psi_j}$. We can then write the QFI as
\begin{align}
\mathcal{I}_\theta &= \mathrm{Tr} \left[ \hat L^2 \hat \rho \right] \nonumber \\
&= \sum_i \bra{\psi_i} \hat L^2 \hat \rho \ket{\psi_j} \nonumber \\
&= \sum_{ik} \bra{\psi_i} \hat L \ketbra{\psi_k} \hat L \hat \rho \ket{\psi_i} \nonumber \\
&= \sum_{ik} \bra{\psi_i} \hat L \ketbra{\psi_k} \hat L \ket{\psi_i} p_i \nonumber \\
&= \sum_{ij} L_{ij} L_{ji} p_i \, ,
\end{align}
We can now rewrite the SLD as
\begin{equation}
L_{ij} = \frac{2 \bra{\psi_i } \p_\theta \hat \rho \ket{\psi_j}}{p_i + p_j} \, .
\end{equation}
With this expression, the Fisher information can be written as
\begin{equation}
\mathcal{I}_\theta = \sum_{i= 0}^s \sum_{j= 0}^t \frac{4 p_i}{(p_i + p_j)^2} |\bra{\psi_i} \p_\theta \hat \rho \ket{\psi_j} |^2 \, .
\end{equation}
Now, we have introduced the two indices $s$ and $t$. The state only has support for $s$, so any indices $p_t = 0$ for $t \geq s$. 
This works because the operator $\p_\theta \hat \rho$ is Hermitian. Then we can show from the spectral decomposition of $\hat \rho_\theta$ that 
\begin{equation}
\bra{\psi_i} \p_\theta \hat \rho \ket{\psi_j} = \p_\theta p_i \, \delta_{ij} + ( p_j  - p_i) \, \braket{\psi_i | \p_\theta \psi_j}  \, .
\end{equation}
We can prove this by writing 
\begin{align}
\bra{\psi_i} \p_\theta \hat \rho \ket{\psi_j} &= \sum_k \bigl( \bra{\psi_i} \p_\theta p_k \ket{\psi_k} \braket{\psi_k|\psi_j} + p_k \braket{\psi_i | \p_\theta \psi_k} \braket{\psi_k | \psi_j } \nonumber \\
&\quad\quad\quad+ p_i \braket{\psi_i | \psi_k } \braket{\p_\theta \psi_k |\psi_j } \bigr) \nonumber \\
&= \p_\theta p_i \delta_{ij} + p_j \braket{\psi_i | \p_\theta \psi_k} + p_i \braket{\p_\theta \psi_i | \psi_j}  \, .
\end{align}
Now use the fact that 
\begin{equation}
\braket{\psi_i | \p_\theta \psi_j} = - \braket{\p_\theta \psi_i |\psi_j} \, , 
\end{equation} 
which can be derived by noting that the resolution of identity becomes zero when differentiated:
\begin{align}
\p_\theta \mathbb{I} &= \sum_i \bra{\psi_\theta \psi_i }\ket{\psi_i} + \ket{\psi_i }\bra{\psi_\theta \psi_i} = 0 \, .
\end{align}
Then, if we take the overlap on both sides, we can write
\begin{equation}
\sum_i \bra{\psi_j| \p_\theta \psi_i} \braket{\psi_i | \psi_k} + \braket{\psi_j | \psi_i} \braket{\p_\theta \psi_i \psi_k} = 0 \, ,
\end{equation}
which implies the identity
\begin{equation}
\braket{\p_\theta \psi_i | \psi_j} = - \braket{\psi_i | \p_\theta \psi_j} \, .
\end{equation}
Then we find that 
\begin{equation}
\bra{\psi_i} \p_\theta\hat \rho \ket{\psi_j} = \p_\theta p_i \delta_{ij} + \left( p_j - p_i \right)\braket{\psi_i | \p_\theta \psi_j}  \, .
\end{equation}
We can now insert this into the above equation to find:
\begin{align}
\mathcal{I}_\theta &=\sum_{i= 0}^s \sum_{j= 0}^t \frac{4 p_i}{(p_i + p_j)^2} |\bra{\psi_i} \p_\theta \hat \rho \ket{\psi_j} |^2 \nonumber \\
&=\sum_{i= 0}^s \sum_{j= 0}^t\frac{4 p_i}{(p_i + p_j)^2} |\p_\theta p_i \delta_{ij} + \left( p_j - p_i \right)\braket{\psi_i | \p_\theta \psi_j} |^2 \nonumber \\
&=\sum_{i= 0}^s \frac{1}{p_i } (\p_\theta p_i)^2 + \sum_{i= 0}^s \sum_{j= 0}^t \frac{4 p_i(p_i - p_j) ^2}{(p_i + p_j)^2} |\braket{\psi_i |\p_\theta \psi_j} |^2 \, .
\end{align}
But now, we note that the second sum can be divided into two sums: One that runs over the support up until $s$ and one where $j$ has support from $s+ 1 $ to $t$. 
\begin{align}
\sum_{i= 0}^s \sum_{j= 0}^t\frac{4 p_i(p_i - p_j) ^2}{(p_i + p_j)^2} |\braket{\psi_i |\p_\theta \psi_j} |^2= &\sum_{i,j= 0}^s \frac{4 p_i (p_i - p_j)^2}{(p_i + p_j)^2} |\braket{\psi_i |\p_\theta \psi_j} |^2 \nonumber \\
&+ \sum_{i = 0}^s \sum_{j = s+1}^t 4 p_i |\braket{\psi_i |\p_\theta \psi_j} |^2 \, .
 \end{align}
All $p_{s+1}$ are zero, because these probabilities stretch beyond the support of the state. 
The second sum can then be written as 
\begin{equation}
\sum_{j = s+1}^t \ketbra{\psi_j} = \mathds{I} - \sum_{j = 0}^s \ketbra{\psi_j}\, .
\end{equation}
Inserting this into the expression above, we find
\begin{align}
\sum_{i = 0}^s \sum_{j = s+1}^t 4 p_i |\braket{\psi_i |\p_\theta \psi_j} |^2 &=\sum_{i = 0}^s \sum_{j = s+1}^t 4 p_i |\braket{\psi_j |\p_\theta \psi_i} |^2 \nonumber \\
&= \sum_{i = 0}^s \sum_{j = s+1}^t 4 p_i \braket{\p_\theta \psi_i | \psi_j} \braket{\psi_j |\p_\theta \psi_i} \nonumber \\
&= \sum_{i = 0}^s  4 p_i \bra{\p_\theta \psi_i } \left(\mathbb{I} - \sum_{j = 0}^s \ketbra{\psi_j} \right)\ket{\p_\theta \psi_i} \nonumber \\
&= \sum_{i = 0}^s 4 p_i \braket{\p_\theta \psi_i | \p_\theta \psi_i} - \sum_{i,j}^s 4 p_i |\braket{\psi_j |\p_\theta \psi_i}|^2  \, .
\end{align}
Putting everything together, we find
\begin{align}
\mathcal{I}_\theta =& \sum_{i= 0}^s \frac{1}{p_i } (\p_\theta p_i)^2 + \sum_{i,j= 0}^s \frac{4 p_i (p_i - p_j)^2}{(p_i + p_j)^2} |\braket{\psi_i |\p_\theta \psi_j} |^2 \nonumber \\
&+  \sum_{i = 0}^s 4 p_i \braket{\p_\theta \psi_i | \p_\theta \psi_i} - \sum_{i,j}^s 4 p_i |\braket{\psi_j |\p_\theta \psi_i}|^2  \,.
\end{align}
To combine the second and last terms of the sums, we write
\begin{equation}
\sum_{i,j}^s \left( \frac{4 p_i ( p_i - p_j)^2}{(p_i + p_j)^2} - \frac{4 p_i (p_i + p_j)^2}{(p_i + p_j)^2} \right)= \sum_{i,j}^s16 \frac{p_i^2 p_j}{(p_i + p_j)^2} \,.
\end{equation}
Now, the following identity holds:
\begin{equation} \label{app:metrology:eq:probability:identity}
\sum_{i,k} \frac{2 p_i p_k^2}{(p_i + p_k)^2} |\braket{\psi_i | \p_\theta   \psi_k}|^2 = \sum_{i,k} \frac{p_i p_k}{p_i + p_k} |\braket{\psi_i | \p_\theta   \psi_k}|^2 \,.
\end{equation}
With this inequality, we find that the Fisher information can be written as
\begin{equation}
\mathcal{I}_\theta = \sum_{i= 0}^s \frac{1}{p_i } (\p_\theta p_i)^2 +   \sum_{i = 0}^s 4 p_i \braket{\p_\theta \psi_i | \p_\theta \psi_i} - 8 \sum_{i,j} \frac{p_i p_j}{p_i + p_j} |\bra{\psi_i | \p_\theta \psi_j} |^2 \, .
\end{equation}
We are almost there. The first part of this expression is a classical contribution to the Fisher information. We can write it as 
\begin{equation}
\sum_j^s \frac{1}{p_i} ( \p_\theta p_i)^2 = 4 \sum^2_j ( \p_\theta \sqrt{p_i})^2 \, .
\end{equation}
The two second parts are the quantum Fisher information: 
\begin{equation}
\mathcal{I}_\theta =   \sum_{i = 0}^s 4 p_i \braket{\p_\theta \psi_i | \p_\theta \psi_i} - 8 \sum_{i,j} \frac{p_i p_j}{p_i + p_j} |\bra{\psi_i | \p_\theta \psi_j} |^2 \, .
\end{equation}
We now note that the second sum has both diagonal and off-diagonal values. We write 
\begin{align}
\mathcal{I}_\theta &=   \sum_{i = 0}^s 4 p_i \braket{\p_\theta \psi_i | \p_\theta \psi_i} - 8 \left( \sum_{i} \delta_{ij} +\sum_{i \neq j} \right) \frac{p_i p_j}{p_i + p_j} |\bra{\psi_i | \p_\theta \psi_j} |^2 \nonumber \\
&=  \sum_{i = 0}^s 4 p_i \braket{\p_\theta \psi_i | \p_\theta \psi_i} - 8 \sum_i^s \frac{p_i^2}{2p_i} |\braket{\psi_i |\p_\theta \psi_j}|^2 \nonumber \\
&\quad\quad\quad- 8 \sum_{i \neq j} \frac{p_i p_j}{p_i + p_j} |\braket{\psi_i |\p_\theta \psi_j}|^2  \, .
\end{align}
This expression can be further simplified into 
\begin{align} \label{app:metrology:eq:QFI:final}
\mathcal{I}_\theta &= 4 \sum_i^s p_i \left( \braket{\p_\theta \psi_i |\p_\theta \psi_i} - |\braket{\psi_i |\p_\theta \psi_i}|^2 \right)  - 8 \sum_{i \neq j}^s \frac{p_i p_j}{p_i + p_j} |\braket{\psi_i |\p_\theta \psi_j}|^2  \, . 
\end{align}
We now note that when the states $\ket{\psi_i}$ evolve in time, we obtain $\ket{\psi_i(t)} = \hat U_\theta(t) \ket{\psi_i(0)}$, which implies $\p_\theta \ket{\psi_i(t)} = \partial_\theta \hat U_\theta(t) \ket{\psi_i(0)}$. We now define the generator $\hat{\mathcal{H}}_\theta =- i \hat U_\theta ^\dag \partial_\theta \hat U_\theta$, which will simplify the handling of this expression. By using the identity $\hat U_\theta ^\dag \hat U_\theta = \mathds{1}$, we see that
\begin{equation}
\hat U_\theta^\dag \partial \hat U_\theta = -( \partial_\theta \hat U_\theta^\dag )\hat U_\theta \, .
\end{equation}
We added the brackets to denote where the partial derivative acts. With this identity, we note that 
\begin{equation} \label{app:QFI:eq:calH2:identity}
\hat{\mathcal{H}}_\theta^2 = - \hat U_\theta ^\dag (\partial_\theta \hat U_\theta) \hat U_\theta^\dag \partial_\theta \hat U = (\p_\theta \hat U_\theta)^\dag \p_\theta \hat U_\theta \, .
\end{equation}
Thus we proceed to write
\begin{align} \label{app:QFI:final:expression}
\mathcal{I}_\theta =& \,  4 \sum_i^s \left( \braket{ \psi_i(0) \hat{\mathcal{H}}_\theta^2 |\psi_i(0)} - \bra{\psi_i(0)} \hat{\mathcal{H}}_\theta \ket{\psi_i(0)}^2  \right)  \nonumber \\
&\quad - 8 \sum_{i \neq j}^s \frac{p_i p_j}{p_i + p_j} \left| \bra{\psi_i(0)} \hat{\mathcal{H}}_\theta \ket{\psi_i(0)} \right|  \, ,
\end{align}
which is the expression we use in Chapter~\ref{chap:metrology}. Using  the notation in the main text, we identify $\psi_i(0) = \lambda_i$ and change the indices from $i $ and $j$ to $n $ and $m$, and letting the sums to go infinity with $s \rightarrow \infty$, we recover the expression in Eq.~\eqref{chap:metrology:definition:of:QFI}. 

\subsection{Relation to the pure QFI} \label{app:QFI:relation:pure:states}

In Chapter~\ref{chap:gravimetry}, we use a form of the QFI in Eq.~\eqref{chap:gravimetry:pure:QFI} that works for pure states only. We here show how the expression for the mixed-state QFI in Eq.~\eqref{app:QFI:final:expression} reduces to Eq.~\eqref{chap:gravimetry:pure:QFI} for pure states. 

We first set $\ket{\psi_0(0)} \equiv \ket{\psi_0}$ and $p_0 = 1$, while $\ket{\psi_j(0)} = 0$ and $p_j = 0$ for all other $j$. For a single vector, the last sum in Eq.~\eqref{app:QFI:final:expression} vanishes, and we are left with 
\begin{equation}
\mathcal{I}_\theta = 4 \left( \bra{\psi_0} \hat{\mathcal{H}}_\theta ^2 \ket{\psi_0}  - \bra{\psi_0} \hat{\mathcal{H}}_\theta \ket{\psi_0}^2 \right) \, .
\end{equation}
We then note that through the identity in Eq.~\eqref{app:QFI:eq:calH2:identity}, we have 
\begin{equation}
\hat{\mathcal{H}}_\theta^2 = \partial_\theta \hat U_\theta^\dag \, \partial_\theta \hat U_\theta \, .
\end{equation}
Inserting this, we have 
\begin{equation} \label{app:QFI:pure:state:QFI}
\mathcal{I}_\theta = 4 \left( \braket{\partial_\theta \Psi(t) | \partial_\theta \Psi(t)}  - |\braket{\Psi(t) |\partial_\theta \Psi(t)}|^2 \right) \,,
\end{equation}
which is the conventional expression that we use in Chapter~\ref{chap:gravimetry}

\section{QFI for different initial states} \label{app:QFI:different:initial:states:expressions}
In this Section, we compute the QFI given different initial states. The QFI for mixed states is given by the expression in Eq.~\eqref{app:QFI:final:expression}, and for pure states in Eq.~\eqref{app:QFI:pure:state:QFI}. 

The states we consider are: 
\begin{itemize}
\item[(i)] \textbf{A coherent state of the optics and mechanics}: given by 
\begin{equation} \label{app:QFI:initial:state:coherent:coherent}
\ket{\Psi} = \ket{\mu_{\rm{c}}} \otimes \ket{\mu_{\rm{m}}} \, ,
\end{equation}
\item[(ii)] \textbf{A coherent state of the optics and a thermal state of the mechanics} given by 
\begin{equation} \label{app:QFI:initial:state:coherent:thermal}
\hat \rho_{\rm{th}} = \ketbra{\mu_{\rm{c}}} \otimes \sum_n \frac{\tanh^{2n}r_T}{\cosh^2 r_T} \ketbra{n} \, .
\end{equation} 
\item[(iii)] \textbf{A Fock state of the optics and a coherent state of the mechanics} given by 
\begin{equation} \label{app:QFI:initial:state:Fock:coherent}
\ket{\Psi_{\rm{Fock}}} = \frac{1}{\sqrt{2}} \left( \ket{0} + \ket{n} \right) \otimes \ket{\mu_{\rm{m}}} \, . 
\end{equation}
\end{itemize}

From the QFI in Eq.~\eqref{app:QFI:final:expression}, we see that we can  compute a concise expression by using the operator $\hat{\mathcal{H}}_\theta$. In Chapter~\ref{chap:metrology}, we found that the operator $\hat{\mathcal{H}}_\theta$ is given by 
\begin{align} \label{app:QFI:eq:mathcalH:with:coefficients}
 \mathcal{\hat H}_\theta =& \, A\,\hat N_a^2 +B\,\hat N_a +  C_+\,\hat B_+ + C_{\hat N_a,+} \hat N_a \, \hat B_+ + C_-\,\hat B_- + C_{\hat N_a,-} \, \hat N_a \, \hat B_- + E\,\hat N_b \nonumber \\
 &+  F\,\hat B_+^{(2)} + G\,\hat B_-^{(2)} +K \, . 
\end{align}
for the dynamics generated by the Hamiltonian in Eq.~\eqref{chap:decoupling:eq:Hamiltonian}, which is the main Hamiltonian of interest in this thesis. We fond expressions for the different coefficients in Section~\ref{chap:metrology:sec:derivation:coefficients} in Chapter~\ref{chap:metrology}. It remains to find the expectation value of the operators in $\hat{\mathcal{H}}_\theta$ for each of the different initial states. 

In Chapter~\ref{chap:gravimetry}, we consider estimation of a constant linear displacement $\tilde{d}_1$. We therefore consider the case where there is no mechanical squeezing, which means that $\mathcal{D}_2 = 0$ and where we wish to estimate a parameter in the weighting function $\mathcal{D}_1$.  These considerations imply that $A = C_{\hat N_a, \pm} = E = F= G= 0$, which considerably simplifies the QFI. In what follows, we compute expressions for the coherent states in Eq.~\eqref{app:QFI:initial:state:coherent:coherent} and for the Fock and coherent state in Eq.~\eqref{app:QFI:initial:state:Fock:coherent} for the linear displacement only. For the thermal state in Eq.~\eqref{app:QFI:initial:state:coherent:coherent}, however, we compute the QFI in full generality.

\subsection{QFI for coherent states} \label{app:QFI:coherent:coherent}
We start by considering estimation of a parameter contained in the mechanical displacement weighting function $\mathcal{D}_1$ with the initially coherent state shown in Eq.~\eqref{app:QFI:initial:state:coherent:coherent}. The absence of squeezing means that the operator $\hat{\mathcal{H}}_{\mathcal{D}_1}$ for estimations of $\mathcal{D}_1$ reduces to 
\begin{equation}
\hat{\mathcal{H}}_{\mathcal{D}_1} =  B \, \hat N_a + C_+ \hat B_+ + C_- \hat B_- \, . 
\end{equation}
We now consider the expectation values of terms in $\braket{\hat{\mathcal{H}}_\theta^2}$ and $\braket{\hat{\mathcal{H}}_\theta}^2$ given the initial state. Certain cross-terms will not contribute to the QFI, and so we are left with evaluating $\braket{\hat N_a}$, $\braket{\hat N_a^2} $, $\braket{\hat B_+^2}$,  $\braket{\hat B_-^2}$, and $\braket{\hat B_+ \hat B_-}$. These expectation values are given in Eqs.~\eqref{app:commutators:Na:coherent:state},~\eqref{app:commutators:Na2:coherent:state},~\eqref{app:commutators:B+2:coherent},~\eqref{app:commutators:B-2:coherent}, and~\eqref{app:commutators:B+B-:coherent}, respectively. 

Putting everything together, we find
\begin{equation} \label{app:QFI:coherent:coherent:D1}
\mathcal{I}_{\mathcal{D}_1} = 4 \left( B^2  |\mu_{\rm{m}}| + C_+ ^2 + C_-^2 \right) \, .
\end{equation}
This expression is used in Chapter~\ref{chap:gravimetry} to compute the QFI for estimation of a constant gravitational acceleration. 

\subsection{QFI for coherent and thermal states} \label{app:QFI:coherent:thermal:state}
Next, we consider the most general estimation scenario with the initially coherent and thermal state shown in Eq.~\eqref{app:QFI:initial:state:coherent:thermal}. In other words, we derive the QFI for estimation of any of the parameters that can enter into the extended Hamiltonian in Eq.~\eqref{chap:metrology:eq:Hamiltonian}. 

We do so by taking the expectation values of $\mathcal{\hat H}_\theta$ according to Eq.~\eqref{app:QFI:final:expression}. 
In order to do so, we will need the expectation values listed in Appendix~\ref{app:commutators}. 
Noticing that the coefficients $E$ and $K$ will not contribute to the QFI, we drop them. We also note that cross-products of different operators that commute, such as the combination of $\hat N_a $ and $\hat B_+$ will cancel, since both the quantity $\braket{\hat{\mathcal{H}}_\theta^2}$ and $\braket{\hat{\mathcal{H}}_\theta}^2$ contains them. However, squares of operators, such as $\braket{\hat N_a^2}$ do not cancel, since the equivalent term $\braket{\hat N_a}^2$ takes on a different value, unless it is the eigenstate of the initial state. The shape of the QFI thus both depends on the commutativity of the operators, and the initial state. 

As a result, we use all the expectation values $\braket{\hat N_a^4}$, $\braket{\hat N_a^3}$, $\braket{\hat N_a^2}$, and $\braket{\hat N_a}$ for coherent states, listed in Eqs.~\eqref{app:commutators:Na4:coherent:state},~\eqref{app:commutators:Na3:coherent:state},~\eqref{app:commutators:Na2:coherent:state}, and~\eqref{app:commutators:Na:coherent:state}, as well as the expectation values $\braket{\hat B_+^2}$, $\braket{\hat B_-^2}$,  $\braket{(\hat B_+^{(2)})^2}$, $\braket{(\hat B_-^{(2)})^2}$, $\braket{\hat B_+ \hat B_-}$, and $\braket{\hat B_+^{(2)} \hat B_-^{(2)}}$, which are listed in Eqs.~\eqref{app:commutators:B+2:Fock},~\eqref{app:commutators:B-2:Fock},~\eqref{app:commutators:B+2sq:Fock},~\eqref{app:commutators:B-2sq:Fock},~\eqref{app:commutators:B+B-:Fock}, and~\eqref{app:commutators:B+2B-2:Fock}, respectively. With all these expressions, the first term of the QFI in Eq.~\eqref{app:QFI:final:expression} becomes:  
\begin{align}
& \braket{\lambda_n|\mathcal{\hat H_\theta}^2|\lambda_n} -\braket{\lambda_n|\mathcal{\hat H_\theta}|\lambda_n}^2 \nonumber \\
&=   A^2\left(4 \, |\mu_{\rm{c}}|^6+6 \, |\mu_{\rm{c}}|^4+|\mu_{\rm{c}}|^2\right)+2 \, AB\left(2 \, |\mu_{\rm{c}}|^4+|\mu_{\rm{c}}|^2\right)\nonumber  \\
&\quad
+B^2|\mu_{\rm{c}}|^2+(2n+1)\sum_{s\in\{+,-\}}\left(C_s+2 \, C_sC_{\hat N_a,s}|\mu_{\rm{c}}|^2+C_{\hat N_a,s}^2\left(|\mu_{\rm{c}}|^4+|\mu_{\rm{c}}|^2\right)\right) \nonumber \\
&\quad+2 \, (F^2+G^2)\left(n^2+n+1\right)\;,
\end{align}
and for evaluating $. \left| \braket{\lambda_n|\mathcal{\hat H}_\theta|\lambda_m} \right|^2$, we require the expectation values in Eqs.~\eqref{chap:commutators:Fock:different:B+},~\eqref{chap:commutators:Fock:different:B-},~\eqref{chap:commutators:Fock:different:B+2}, and~\eqref{chap:commutators:Fock:different:B-2} to find:
\begin{align}
\left. \left| \braket{\lambda_n|\mathcal{\hat H}_\theta|\lambda_m} \right|^2  \right|_{n\neq m}
=&  \left[\left(C_++C_{\hat N_a,+}|\mu_{\rm{c}}|^2\right)^2 + \left(C_-+C_{\hat N_a,-}|\mu_{\rm{c}}|^2\right)^2\right] \nonumber \\
&\quad\times\left((m+1)\delta_{n,m+1}+m\delta_{n,m-1}\right)\nonumber \\
&+ \left(F^2 +G^2\right)\left((m+1)(m+2)\delta_{n,m+2}+m(m-1)\delta_{n,m-2}\right)\;.
\end{align}
In order to evaluate the single sum, we must make sure all indices are the same. We can rewrite the last term by noting that $m = n+2$ to find:
\begin{align}
\left. \left| \braket{\lambda_n|\mathcal{\hat H}_\theta|\lambda_m} \right|^2  \right|_{n\neq m}
=& \left[\left(C_-+C_{\hat N_a,-}|\mu_{\rm{c}}|^2\right)^2 + \left(C_-+C_{\hat N_a,-}|\mu_{\rm{c}}|^2\right)^2\right]\nonumber \\
&\quad\times\left((m+1)\delta_{n,m+1}+(n+1)\delta_{m,n+1}\right) \nonumber \\
&+ \left(F^2 +G^2\right)\left((m+1)(m+2)\delta_{n,m+2}+(n+1)(n+2)\delta_{m,n+2}\right)\;.
\end{align}
We obtain that 
\begin{align}
\sum_{n\neq m}\frac{\lambda_n \lambda_m}{\lambda_n+\lambda_m}\left| \bra{\lambda_n}\mathcal{\hat H}_\theta \ket{\lambda_m}\right|^2 = 2\sum_{n}\left(\frac{\lambda_n \lambda_{n+1}}{\lambda_n+\lambda_{n+1}} C_{1,\hat{\mathcal{H}}}   + \frac{\lambda_n \lambda_{n+2}}{\lambda_n+\lambda_{n+2}}  C_{2,\hat{\mathcal{H}}} \right) \, ,
\end{align}
where
\begin{align}
C_{1,\hat{\mathcal{H}}} =& \left(
\left(C_-+C_{\hat N_a,-}|\mu_c|^2\right)^2 
+ \left(C_-+C_{\hat N_a,-}|\mu_c|^2\right)^2\right) (n+1) \, , \\
C_{2,\hat{\mathcal{H}}} =& \left(F^2  +  G^2\right)(n+1)(n+2)\;.
\end{align}
Using the fact that $\lambda_n = \frac{\tanh^{2n}(r_T)}{\cosh^2(r_T)}$ for the thermal state, we use the fact that the sums converge to:
\begin{align}
\sum_n \frac{\tanh^{2n}(r_T)}{\cosh^2(r_T)} &= 1 \, ,\nonumber \\
\sum_n n \, \frac{\tanh^{2n}(r_T)}{\cosh^2(r_T)} &= \sinh^2(r_T)  \, ,\nonumber \\
\sum_n n^2 \frac{\tanh^{2n}(r_T)}{\cosh^2(r_T)} &= \cosh(2\, r_T) \, \sinh^2(r_T) \, .
\end{align}
With these expressions, we obtain the final result: 
\begin{align} \label{app:QFI:eq:coherent:thermal:final:QFI}
\mathcal{I}_\theta =& 4  \, \biggl[ \left(4|\mu_\textrm{c}|^6+6|\mu_\textrm{c}|^4+|\mu_\textrm{c}|^2\right)A^2
+  \left(4|\mu_\textrm{c}|^4+2|\mu_\textrm{c}|^2\right)AB  +|\mu_\textrm{c}|^2 B^2  \nonumber \\
&\quad+\cosh(2 \, r_T) \sum_{s\in\{+,-\}} C^2_{\hat N_a,s}  |\mu_c|^2 +\frac{1}{\cosh(2 \, r_T)} \sum_{s\in\{+,-\}} (C_s + C_{\hat N_a,s} |\mu_c|^2)^2 \nonumber \\
&\quad  +  4\frac{\cosh ^2(2 r_T)}{\cosh ^2(2 r_T)+1} \left( F^2 + G^2 \right)   \biggr] \, , 
\end{align}
where we have condensed the expression somewhat by letting the sums run over the indices $s = +$ and  $s = -$. Given expressions for the coefficients $A$, $B$, $C_\pm$, $C_{\hat N_a, \pm}$, $F$ and $G$, the QFI can then be computed for a number of different cases. We do so in Chapter~\ref{chap:metrology}, where we consider three straight-forward applications of the expression in Eq.~\eqref{app:QFI:eq:coherent:thermal:final:QFI}.

\subsection{QFI for Fock states} \label{app:QFI:Fock:coherent:states}
Similarly to Section~\ref{app:QFI:coherent:coherent} above, we here consider measurements of linear displacements only. It follows that 
\begin{equation}
\hat{\mathcal{H}}_{\mathcal{D}_1} =  B \, \hat N_a + C_+ \hat B_+ + C_- \hat B_- \, . 
\end{equation}
We compute the expectation values of $\braket{\hat N_a}$, $\braket{\hat N_a^2} $, $\braket{\hat B_+^2}$,  $\braket{\hat B_-^2}$, and $\braket{\hat B_+ \hat B_-}$ with respect to the initial state in Eq.~\eqref{app:QFI:initial:state:coherent:thermal}.  This time, the expectation values are given in Eqs.~\eqref{chap:commutators:Na:Fock},~\eqref{chap:commutators:Na2:Fock},~\eqref{app:commutators:B+2:coherent},~\eqref{app:commutators:B-2:coherent}, and~\eqref{app:commutators:B+B-:coherent}, respectively. 

These results allow us to compute the QFI to find the general expression:
\begin{equation}
\mathcal{I}_{\mathcal{D}_1} =  n^2 \,  B^2 + 4 \left( C_+^2 + C_-^2 \right) \, .
\end{equation}

\section{Expressions for different estimation schemes with initial thermal states} \label{app:QFI:expressions:estimation:schemes}
In Section~\ref{chap:metrology:three:examples} in Chapter~\ref{chap:metrology}, we consider three example estimation schemes for parameters that are encoded in the extended optomechanical Hamiltonian in Eq.~\eqref{chap:decoupling:eq:Hamiltonian}: (i) estimation of the coupling strength $\tilde{g}_0$ for a time-modulated optomechanical coupling, (ii) estimation of the oscillation amplitude $\tilde{d}_1$ for a time-modulated mechanical displacement, and (iii) estimation of the oscillation amplitude $\tilde{d}_2$ for a constant and time-modulated single-mode mechanical squeezing. Many of the expressions that we consider are long and cumbersome, and we therefore list them in this Appendix. 

This Section exclusively contains expressions for an initial coherent state in the optics and a thermal state of the mechanic, which is the initial state considered in Eq.~\eqref{chap:metrology:eq:main:result:QFI} in Chapter~\ref{chap:metrology}. Furthermore, all the following expressions are derived from QFI in Eq.~\eqref{app:QFI:eq:coherent:thermal:final:QFI}.

\subsection{QFI for estimation of $\tilde{g}_0$}
In Section~\ref{chap:metrology:sec:example:1}, we consider estimation of the strength $\tilde{g}_0$ of an oscillating light--matter coupling. For a time-dependent nonlinear coupling $\tilde{\mathcal{G}}(\tau) = \tilde{g}_0\left( 1 + \epsilon  \, \sin (\Omega_g \, \tau ) \right) $, the $F$-coefficients are given in Eq.~\eqref{app:coefficients:eq:time:dependent:g0}, and we find $J_b = \tau$ and $J_\pm = 0$. The non-zero coefficients that enter into the QFI are given by 
\begin{align}\label{app:QFI:eq:g0:nonzero:coefficients}
A &=\frac{\tilde{g}_0}{2 \,  \Omega_g \left( 1-\Omega_g^2\right)^2} \biggl[( \Omega_g^2-1)  \biggl(4  \Omega_g^3\tau-2 \Omega_g \tau \left(\epsilon ^2+2\right)+\epsilon ^2 \sin (2 \,  \Omega_g\, \tau) \nonumber \\
&\quad\quad\quad\quad\quad\quad\quad\quad\quad\quad\quad\quad+4 \Omega_g \sin (\tau) \left(\epsilon  \sin ( \Omega_g \, \tau)-\Omega_g^2+1\right)+4 \Omega_g^2 \epsilon -8 \epsilon \biggr) \nonumber \\
&\quad\quad\quad\quad\quad\quad\quad\quad+4 \Omega_g^2 \epsilon  \cos (\tau) \left(\left(\Omega_g^2-1\right) (\cos ( \Omega_g \, \tau)-1)+\epsilon  \sin ( \Omega_g \, \tau)\right) \nonumber \\
&\quad\quad\quad\quad\quad\quad\quad\quad-4 \epsilon  \cos (\Omega_g\, \tau) \left(\Omega_g^3 \epsilon  \sin \tau)+\Omega_g^4-3 \Omega_g^2+2\right)\biggr]\;,\\\nonumber
 C_{\hat N_a,+ } &= \epsilon\frac{ \sin (\tau) \sin ( \Omega_g \, \tau)+ \Omega_g \cos (\tau) \cos ( \Omega_g \, \tau)+\Omega_g}{1-\Omega_g^2}+\sin (\tau) \;, \nonumber \\
C_{\hat N_a, -} &=1 -\epsilon \frac{  \cos (\tau) \sin ( \Omega_g \, \tau)-\Omega_g   \sin (\tau) \cos ( \Omega_g \, \tau)}{1-\Omega_g^2}-\cos (\tau) \, .
\end{align}
The full expression for the QFI is then given by 
\begin{align} \label{app:QFI:eq:g0:general:Omega}
\mathcal{I}_{\tilde{g}_0} = 
& \frac{4 \, \tilde{g}_0^2}{\Omega_g^2 \, (1 -  \Omega_g^2)^4} \left| \mu_{\rm{c}}\right| ^2 \left(4 \, \left| \mu_{\rm{c}}\right| ^4+6 \,  \left| \mu_{\rm{c}}\right| ^2+1\right) \nonumber \\
&\quad\times \biggl(2 \,  \tau \, \Omega_g^5-4 \, \tau \, \Omega_g^3+2 \,  \tau \, \Omega_g- \tau \, \Omega_g^3 \epsilon ^2+\frac{1}{2} \Omega_g^2 \, \epsilon ^2 \sin (2 \,  \Omega_g \, \tau) \nonumber \\
&\quad\quad\quad+2 \,  \Omega_g^2 \,  \epsilon ^2 \,  \cos (\tau) \,  \sin ( \Omega_g \, \tau)+\tau \, \Omega_g  \, \epsilon ^2-4 \, \Omega_g^4 \,  \epsilon \, \cos (\tau) \sin ^2(\Omega_g \, \tau/2) \nonumber \\
&\quad\quad\quad-2 \, \left(\Omega_g^2-1\right) \Omega_g \sin (\tau) \left(\Omega_g^2-\epsilon \,  \sin ( \Omega_g \, \tau)-1\right)+4 \, \Omega_g^2 \, \epsilon   \, \cos (\tau) \, \sin ^2(\Omega_g \, \tau/2)\nonumber \\
&\quad\quad\quad-\epsilon \, \cos ( \Omega_g \, \tau) \left(2 \, \Omega_g^3 \, \epsilon \,  \sin (\tau)+\epsilon   \, \sin ( \Omega_g \, \tau)+2\,  \Omega_g^4-6 \,  \Omega_g^2+4\right) \nonumber \\
&\quad\quad\quad+2\, \Omega_g^4 \, \epsilon -6 \,  \Omega_g^2\, \epsilon +4\,  \epsilon \biggr)^2 \, \nonumber \\
&+  \, 4 \,  |\mu_{\rm{c}}|^2 \, \cosh(2 \,  r_T) \left( 1 + \frac{|\mu_{\rm{c}}|^2 }{\cosh^2 ( 2 \,  r_T)} \right) \nonumber \\
&\quad\times \biggl[ \left( 1 - \cos(\tau) - \epsilon \frac{\Omega_g \cos(\Omega_g \tau) \sin (\tau) - \cos(\tau) \sin(\Omega_g \, \tau) }{\Omega_g^2 - 1} \right)^2 \nonumber \\
&\quad\quad\quad+ \left( \sin( \tau) + \epsilon \frac{\Omega_g (1 - \cos(\tau)  \cos(\Omega_g \, \tau)) - \sin (\tau) \sin(\Omega_g \,  \tau) }{\Omega_g^2 - 1} \right)^2 \biggr]  \,  .
\end{align}
At resonance with $\Omega_{g} = 1$, the coefficients simplify to those shown in Eq.~\eqref{app:coefficients:eq:g0:resonance} and the QFI reduces to 
\begin{align} \label{app:QFI:eq:g0:general:Omega:resonance}
\mathcal{I}_{\tilde{g}_0}^{(\rm{res})} =& \frac{\left| \mu_{\rm{c}}\right| ^2}{16}  \biggl[\tilde{g}_0^2 \left(4 \left|  \mu_{\rm{c}}\right| ^4+6 \left|  \mu_{\rm{c}}\right| ^2+1\right) \nonumber \\
&\quad\quad\quad\quad\times \biggl(4 \tau \epsilon ^2-3 \epsilon ^2 \sin (2 \, \tau )-8 \, \tau \,  \epsilon  \sin (\tau)-32 \,  \epsilon  \cos (\tau) \nonumber \\
&\quad\quad\quad\quad\quad\quad+  2 \,  \epsilon  (\tau \, \epsilon +2) \cos (2 \, \tau)+16 \, \tau -16 \sin (\tau)+28 \epsilon \biggr)^2 \nonumber \\
&\quad\quad+16 \,  \cosh (2 r_T) \left(\left| \mu_{\rm{c}}\right| ^2 \frac{1}{\cosh^2(2 r_T)} +1\right) \bigl(\sin ^2(\tau) (\epsilon  \sin (\tau)+2)^2 \nonumber \\
&\quad\quad\quad\quad\quad\quad\quad\quad\quad\quad\quad\quad+(\tau \epsilon -\cos (\tau) (\epsilon  \sin (\tau)+2)+2)^2\bigr)\biggr] \, .
\end{align}
\subsection{QFI for estimation of $\tilde{d}_1$}
In Section~\ref{chap:metrology:sec:example:2}, we consider estimation the strength $\tilde{d}_1$ of an oscillating mechanical displacement of the form $\tilde{\mathcal{D}}_1 = \tilde{d}_1 \, \cos( \Omega_{d_1} \, \tau)$. Given the $F$-coefficients in Eq.~\eqref{app:coefficients:F:coefficients:time:dependent:d1},  and $J_b = \tau$ and $J_\pm = 0$, we find  the following non-zero coefficients:
\begin{align}\label{app:QFI:eq:d1:nonzero:coefficients}
 B&=2  \, \tilde{g}_0 \,  \frac{ \sin (\Omega_{d_1} \, \tau )+\Omega_{d_1} (\Omega_{d_1} (\cos (\tau)-1) \sin (\Omega_{d_1} \, \tau )-\sin (\tau) \cos (\Omega_{d_1} \, \tau))}{\Omega_{d_1} \left(\Omega_{d_1}^2-1\right)}
 \;,\nonumber \\
C_+ &=-\frac{\sin (\tau) \cos (\Omega_{d_1}\, \tau )-\Omega_{d_1} \cos (\tau) \sin (\Omega_{d_1} \, t )}{1-\Omega_{d_1}^2}\;, \nonumber \\ 
C_- &= \frac{\Omega_{d_1} \sin (\tau) \sin (\Omega_{d_1} \, \tau)+\cos (\tau) \cos (\Omega_{d_1} \,\tau)-1}{1-\Omega_{d_1}^2} \, .
\end{align}
The QFI becomes
\begin{align}\label{app:QFI:eq:formula:oscillating:d1}
\mathcal{I}_{\tilde{d}_1} 
=& \,  \frac{4 }{\Omega_{d_1}^2\left(1-\Omega_{d_1}^2\right)^2} \biggl[4 \, \tilde{g}_0^2  \, |\mu_{\rm{c}}|^2 \,  \biggl(\sin (\Omega_{d_1} \, \tau ) \,  \left(\Omega_{d_1}^2 (1-\cos (\tau )) - 1\right) \nonumber \\
&\quad\quad\quad\quad\quad\quad\quad\quad\quad\quad\quad\quad\quad\quad\quad\quad\quad\quad\quad+\Omega_{d_1} \,  \sin (\tau ) \,  \cos (\Omega_{d_1} \, \tau)\biggr)^2 \nonumber \\
&+ \frac{\Omega_{d_1}^2}{2 \, \cosh(2  \, r_T)} \biggl(3 -\left(\Omega_{d_1}^2-1\right) \cos (2 \Omega_{d_1} \, \tau)-4 \,  \Omega_{d_1} \sin (\tau) \sin ( \Omega_{d_1} \, \tau) \nonumber \\
&\quad\quad\quad\quad\quad\quad\quad\quad\quad\quad\quad\quad\quad\quad\quad-4 \, \cos (\tau) \cos ( \Omega_{d_1} \, \tau)+\Omega_{d_1}^2\biggr)\biggr] \, .
\end{align}
For a constant displacement with $\Omega_{d_1}=0$, the $F$-coefficients are given in Eq.~\eqref{chap:gravimetry:F:coefficients} and we find
\begin{align}
\mathcal{I}_{\tilde{d}_1}^{\rm{(con)}} =& 16 \,\biggl(\tilde{g}_0^2 \left| \mu_c \right| ^2 (\tau-\sin (\tau))^2 +\frac{\sin ^2\left(\tau/2\right)}{\cosh(2 \, r_T)} \biggr) \, .
\end{align}
At resonance with $\Omega_{d_1} = 1$, the $F$-coefficients are given in Eq.~\eqref{app:coefficients:d1:resonant} and the QFI becomes
\begin{align} \label{app:QFI:d1:res}
\mathcal{I}_{\tilde{d}_1}^{(\rm{res})} =&   4 \, \tilde{g}_0^2 \, |\mu_{\rm{c}}|^2 \left(\tau+\sin(\tau)\left(\cos(\tau)-2\right)\right)^2   \nonumber \\
&+\frac{1}{\cosh{(2 r_T)}} \left(\tau^2+2\tau\sin(\tau)\cos(\tau)+\sin^2(\tau)\right) \, .
\end{align}

\subsection{QFI for estimation of $\tilde{d}_2$}

For estimation of a constant mechanical single-mode squeezing strength $\tilde{d}_2$, many of the terms in the QFI coefficients in Eq.~\eqref{chap:metrology:eq:QFI:coefficients} are zero, namely $A = B = C_{\hat N_a, - } = G = F = 0$.  The $F$-coefficients are given in Eq.~\eqref{app:coefficients:constant:d2}, and the $J$-coefficients are given in Eq.~\eqref{app:coefficients:eq:constant:d2:Js:coeffs}. Subsequently, the only non-zero coefficient is $C_{\hat N_a, m} = 2 \,  \tilde{g}_0 \,  \tau$. We then find the QFI
\begin{align} \label{app:QFI:d2:con:app}
\mathcal{I}_{\tilde{d}_2}^{(\rm{const,app})} =   8 \,  \tilde{g}_0^2 \,  \tau^2  \, |\mu_{\rm{c}}|^2 \frac{1}{\cosh (2  \, r_T)} \left(1+ 2 |\mu_{\rm{c}}|^2+\cosh (4 \, r_T)\right) \, .
\end{align}

When $\tilde{\mathcal{D}}_2(\tau) = \tilde{d}_2 \, \cos( \Omega_{d_2} \, \tau)$ is modulated at resonance with $\Omega_{d_2} = 2$, the only non-zero coefficients are $A$, $C_{\hat N_a, +}$, and $F$. The $F$-coefficients are given in Eq.~\eqref{app:coefficients:eq:resonant:d2}, and the $J$-coefficients are given in Eq.~\eqref{app:coefficients:eq:resonant:d2:Js}. With these expressions, we find 
\begin{align}
A &= - \tilde{g}_0^2 \, \tau \, ,\nonumber \\
C_{\hat N_a, +} &= \tilde{g}_0 \, \tau \, , \nonumber \\
F &= - \tau/2 \, .
\end{align}
The QFI becomes
\begin{align} \label{app:QFI:eq:resonant:d2:QFI}
\mathcal{I}_{\tilde{d}_2}^{(\rm{res,app})} =&   \,4 \tau^2\biggl( \tilde{g}_0^4 (4|\mu_c|^6 + 6|\mu_c|^4 + |\mu_c|^2)  \nonumber \\
&\quad\quad\quad\quad+ \tilde{g}_0^2 |\mu_c|^2 \, \frac{|\mu_c|^2 + \cosh(2r_T)^2}{\cosh(2r_T)} + \frac{\cosh^2(2r_T)}{\cosh^2(2r_T)+1} \biggr) \, .
\end{align}
This concludes our derivations of the different expressions for the QFI. 

\section{Proof of the optimality of homodyne detection for estimating a constant gravitational acceleration} \label{app:QFI:homodyne:optimality:proof}
\sectionmark{Proof of the optimality of homodyne detection for estimating...}
In this Section, we prove that the classical Fisher information (CFI) for a homodyne measurement saturates the QFI for estimation of a constant gravitational acceleration with an optomechanical system . We assume that the initial state is a coherent state of the optics and mechanics, as that given in Eq.~\eqref{app:QFI:initial:state:coherent:coherent}. 

We show in Section~\ref{chap:gravimetry:CFI:homodyne:detection} in Chapter~\ref{chap:gravimetry} that the CFI for homodyne detection of the dimensionless parameter $\tilde{d}_1$ is given by 
\begin{align} \label{app:QFI:eq:homodyne:CFI}
I_{\tilde{d}_1} =&- 4 \, \tilde{g}_0^2  \left( \tau - \sin \tau \right)^2 \, e^{-|\mu_{\rm{c}}|^2} \int \mathrm{d}x_\lambda \frac{\left[ \sum_{n,n^\prime} (n-n') \, c_{n,n^\prime} \, d_{n,n^\prime}(x_\lambda) \, f_{n,n^\prime}\right]^2}{\sum_{n,n^\prime} c_{n,n^\prime} \, d_{n,n^\prime}(x_\lambda) \, f_{n,n^\prime} },
\end{align}
where
\begin{subequations}
\begin{align}
c_{n,n^\prime} &= \frac{(\mu_{\rm{c}}^*)^{n^\prime} \mu_{\rm{c}}^n }{\sqrt{n!n^\prime!}} \, e^{i \left[\tilde{g}_0^2 (n^2 - n^{\prime2} ) - 2 \, \tilde{g}_0 \, \tilde{d}_1 (n-n^\prime) \right] (\tau - \sin \tau)} \, , \label{app:QFI:eq:definition:of:c} \\
d_{n,n^\prime}(x_\lambda) &= \frac{e^{- x_\lambda^2} }{\pi^{1/2}}  \frac{H_n(x_\lambda )\,  H_{n^\prime}(x_\lambda ) \, e^{- i \lambda  (n-n^\prime)}}{2^{(n+n^\prime)/2} \, \sqrt{n!n^\prime!}} \, , \label{app:QFI:eq:definition:of:d} \\
f_{n,n^\prime} &=   e^{\tilde{g}_0 \, (n-n^\prime) \left( \eta \, \mu_{\rm{m}} - \eta^*  \, \mu_{\rm{m}}^* \right)/2 } \, e^{- |\phi_{n^\prime}|^2/2 - |\phi_n|^2/2 + \phi_{n^\prime}^* \phi_n } \, \label{app:QFI:eq:definition:of:f}.
\end{align} 
\end{subequations}
Our goal is to prove that the CFI in Eq.~\eqref{app:QFI:eq:homodyne:CFI} coincides with the QFI in Eq.~\eqref{chap:gravimetry:eq:QFI:coherent:coherent:2pi} in Chapter~\ref{chap:gravimetry} at $\tau = 2\pi$. 
Before proceeding, we list some useful properties of the Hermite polynomials. The generating function for the Hermite polynomial is given by 
\begin{equation} \label{app:QFI:eq:Hermite:generating}
\sum_{n = 0}^\infty  y^n \frac{H_n(x)}{n!} = e^{2 \, x\, y - y^2}  \, .
\end{equation}
Taking the derivative of both sides results in 
\begin{equation} \label{app:QFI:eq:Hermite:generating:derivative}
 \sum_{n= 0}^\infty y^{n-1} \frac{H_n(x)}{ (n-1)!} = 2 \, (  x -  y)e^{2 \, x \, y - y^2}  \, .
\end{equation}
We are now ready to continue with the proof. We assume that $\tilde{d}_1$ and $\tilde{g}_0$ are integers. This means that the phases in Eq.~\eqref{app:QFI:eq:definition:of:c} are all unity, and at $\tau = 2\pi$, we are left with 
\begin{subequations}
\begin{align}
c_{n,n^\prime} &= \frac{(\mu_{\rm{c}}^*)^{n^\prime} \mu_{\rm{c}}^n }{\sqrt{n!n^\prime!}}  \, , \label{app:QFI:eq:definition:of:c:2pi} \\
d_{n,n^\prime}(x_\lambda) &= \frac{e^{- x_\lambda^2} }{\pi^{1/2}}  \frac{H_n(x_\lambda )\,  H_{n^\prime}(x_\lambda ) \, e^{- i \lambda  (n-n^\prime)}}{2^{(n+n^\prime)/2} \, \sqrt{n!n^\prime!}} \, , \label{app:QFI:eq:definition:of:d:2pi} \\
f_{n,n^\prime} &=  1 \, \label{app:QFI:eq:definition:of:f:2pi}.
\end{align} 
\end{subequations}
We begin by examining the derivative $\partial_{\tilde{d}_1} p(x_\lambda|\tilde{d}_1)$, which we printed in Eq.~\eqref{chap:gravimetry:eq:derivative:of:p} in Chapter~\ref{chap:gravimetry}. At $\tau = 2\pi$, and with these assumptions, it is given by 
\begin{align}
\partial_{\tilde{d}_1} p(x_\lambda | \tilde{d}_1) = \sum_{n,n^\prime} (n-n^\prime) \frac{(\mu_{\rm{c}}^*)^{n^\prime} \mu_{\rm{c}}^n}{n!n^\prime!} \frac{e^{- x^2_\lambda}}{\pi^{1/2}} \frac{H_n(x_\lambda) H_{n^\prime}(x_\lambda) e^{-i \lambda(n-n^\prime)} }{2^{(n+n^\prime)/2}} \, .
\end{align}
The above can be divided into two terms due to the $(n- n^\prime)$-bracket which we call
\begin{align} \label{app:QFI:CFI:optimality:terms}
A_1 &= \sum_{n,n^\prime} n  \frac{(\mu_{\rm{c}}^*)^{n^\prime} \mu_{\rm{c}}^n}{n!n^\prime!} \frac{e^{- x^2_\lambda}}{\pi^{1/2}} \frac{H_n(x_\lambda) H_{n^\prime}(x_\lambda) e^{-i \lambda(n-n^\prime)} }{2^{(n+n^\prime)/2}} \,,  \nonumber \\
A_2 &=- \sum_{n,n^\prime} n^\prime \frac{(\mu_{\rm{c}}^*)^{n^\prime} \mu_{\rm{c}}^n}{n!n^\prime!} \frac{e^{- x^2_\lambda}}{\pi^{1/2}} \frac{H_n(x_\lambda) H_{n^\prime}(x_\lambda) e^{-i \lambda(n-n^\prime)} }{2^{(n+n^\prime)/2}} \, .
\end{align}
We start with $A_1$. First, we separate the two sums such that they run over one Hermite polynomial each:
\begin{align}
A_1 &= \frac{e^{- x^2_\lambda}}{\sqrt{\pi}} \sum_l \frac{(\mu_{\rm{c}}^*)^{n^\prime}}{n^\prime!} \frac{H_{n^\prime} e^{i \lambda n^\prime}}{2^{n^\prime/2}} \sum_{n = 0}^\infty \frac{\mu_{\rm{c}}^n}{(n-1)!} \frac{H_n e^{- i \lambda n }}{ 2^{n/2}} \, .
\end{align}
We then use the generating function in Eq.~\eqref{app:QFI:eq:Hermite:generating} to find
\begin{align}
A_1 &=  \frac{e^{- x^2_\lambda}}{\pi^{1/2}} e^{\sqrt{2} \, x_\lambda(\mu_{\rm{c}}^*  \, e^{i \lambda} ) - (\mu_{\rm{c}}^*)^2 \,  2^{-1} \, e^{2 i \lambda}} \sum_{n = 0}^\infty \frac{\mu_{\rm{c}}^n}{(n-1)!} \frac{H_n (x_\lambda)e^{- i \lambda n }}{ 2^{n/2}} \, .
\end{align}
Then, we similarly use the differentiated generating function in Eq.~\eqref{app:QFI:eq:Hermite:generating:derivative} to write 
\begin{align}
A_1=   & \, \frac{e^{- x^2_\lambda}}{\pi^{1/2}} e^{ \sqrt{2} \, x_\lambda \, \mu_{\rm{c}}^* \,  e^{i  \, \lambda}  - \mu_{\rm{c}}^{*2} \,  e^{2 i \,  \lambda} 2^{-1}} \nonumber \\
&\times \mu_{\rm{c}} e^{- i \, \lambda} \left(  \sqrt{2} \, x_\lambda - \mu_{\rm{c}} e^{- i \lambda} \right) e^{  \sqrt{2} \, x_\lambda \mu_{\rm{c}} e^{- i \, \lambda} -  \mu_{\rm{c}}^2 e^{- 2 \, i \, \lambda} \, 2^{-1 }} \, .
\end{align}
The exponentials can be combined into the following expression:
\begin{align}
A_1=   & \, \frac{e^{- x^2_\lambda}}{\pi^{1/2}}  \,  e^{ \Re \{ \,  2\sqrt{2}  \, x_\lambda \,  \mu_{\rm{c}} e^{- i \, \lambda} -  \mu_{\rm{c}}^2 e^{-  2\, i \, \lambda} \}} \, \mu_{\rm{c}}  \, e^{- i \, \lambda} \left(  \sqrt{2} \, x_\lambda - \mu_{\rm{c}} e^{- i \lambda} \right)  \, .
\end{align}
We then consider the other term $A_2$ in Eq.~\eqref{app:QFI:CFI:optimality:terms}. We again write the term as two separate sums and use the generating function in Eq.~\eqref{app:QFI:eq:Hermite:generating} to find:
\begin{align}
A_2 =& -\frac{e^{- x^2_\lambda}}{\pi^{1/2}} \sum_{n^\prime = 0} \frac{(\mu_{\rm{c}}^*)^{n^\prime} }{(n^\prime-1)!}  \frac{H_{n^\prime}(x_\lambda) }{2^{n^\prime/2}} \sum_{n = 0}^\infty \frac{\mu_{\rm{c}}^n}{n!} \frac{H_n(x_\lambda) e^{-i \lambda(n-n^\prime)} }{2^{n/2}} \nonumber \\
&=  -\frac{e^{- x^2_\lambda}}{\pi^{1/2}} \sum_{n^\prime = 0} \frac{(\mu_{\rm{c}}^*)^{n^\prime} }{(n^\prime-1)!}  \frac{H_{n^\prime}(x_\lambda) }{2^{n^\prime/2}} e^{2 x_\lambda \mu_{\rm{c}} 2^{-1/2} e^{- i \, \lambda}  - \mu_{\rm{c}}^2 2^{-1} e^{- 2 i \, \lambda}} \, .
\end{align}
Then, using the derivative of the generating function in Eq.~\eqref{app:QFI:eq:Hermite:generating:derivative}, we find 
\begin{align}
A_2 = & -\frac{e^{- x^2_\lambda}}{\pi^{1/2}} e^{ \Re \{ \,  2\sqrt{2}  \, x_\lambda \,  \mu_{\rm{c}} e^{- i \, \lambda} -  \mu_{\rm{c}}^2 e^{-  2\, i \, \lambda} \}} \,   \mu_{\rm{c}}^* e^{i \, \lambda} \left( \sqrt{2} \, x_\lambda - \mu_{\rm{c}}^* \,  e^{i \, \lambda} \right) \, .
\end{align}
Putting the two terms together, we find 
\begin{align} \label{app:QFI:eq:nominator}
\partial_{\tilde{d}_1} p(x_\lambda |\tilde{d}_1 ) =& \,  A_1 + A_2 \nonumber \\
=& \,   e^{- x_\lambda^2} \, e^{ \Re \{ \,  2\sqrt{2}  \, x_\lambda \,  \mu_{\rm{c}} e^{- i \, \lambda} -  \mu_{\rm{c}}^2 e^{-  2\, i \, \lambda} \}} \nonumber \\
&\times \left[ \mu_{\rm{c}}  e^{- i \, \lambda} \left(  \sqrt{2} \, x_\lambda  - \mu_{\rm{c}} e^{- i \, \lambda } \right) - \mu_{\rm{c}}^* \, e^{i \, \lambda} \left( \sqrt{2} \, x_\lambda - \mu_{\rm{c}}^*\, e^{i \, \lambda} \right)  \right]\, . 
\end{align}
We now examine the denominator of Eq.~\eqref{app:QFI:eq:homodyne:CFI}. At $\tau = 2 \pi$, and with $\tilde{g}_0$ and $\tilde{d}_1$ being integer values, we can simplify the probability function $p(x_\lambda |\tilde{d}_1)$ from Eq.~\eqref{chap:gravimetry:eq:probability:distribution} as follows:
\begin{align} \label{app:QFI:eq:probability:distribution}
p(x_\lambda| \tilde{d}_1) &=  e^{- |\mu_{\rm{c}}|^2} \sum_{n,n'}  \frac{\mu_{\rm{c}}^n(\mu_{\rm{c}}^*)^{n^\prime} }{n!n^\prime! } \frac{e^{- x_\lambda^2} }{\pi^{1/2}}  \frac{H_n(x_\lambda ) \,H_{n^\prime}(x_\lambda ) \, e^{- i \lambda  (n-n^\prime)}}{2^{(n+n^\prime)/2}} \nonumber \\
&=  \frac{e^{- x_\lambda^2}}{\pi^{1/2}} \, e^{ \Re \{ \,  2\sqrt{2}  \, x_\lambda \,  \mu_{\rm{c}} e^{- i \, \lambda} -  \mu_{\rm{c}}^2 e^{-  2\, i \, \lambda} \}} \, ,
\end{align}
through the same methods as we used above. This term is cancelled when the nominator Eq.~\eqref{app:QFI:eq:nominator} is squared. Therefore, the CFI in Eq.~\eqref{app:QFI:eq:homodyne:CFI} becomes:
\begin{align}
I_{\tilde{d}_1} &= - 16 \pi^2 \tilde{g}_0^2 e^{- |\mu_{\rm{c}}|^2 } \frac{1}{\sqrt{\pi}}\int \mathrm{d} x_\lambda \, e^{- x_\lambda^2} \, e^{ \Re \{ \,  2\sqrt{2}  \, x_\lambda \,  \mu_{\rm{c}} e^{- i \, \lambda} -  \mu_{\rm{c}}^2 e^{-  2\, i \, \lambda} \}} \nonumber \\
&\quad\quad\quad\quad\quad\quad\quad\quad\times \left[ \mu_{\rm{c}}  e^{- i \, \lambda} \left(  \sqrt{2} \, x_\lambda  - \mu_{\rm{c}} e^{- i \, \lambda } \right) - \mu_{\rm{c}}^* \, e^{i \, \lambda} \left( \sqrt{2} \, x_\lambda - \mu_{\rm{c}}^*\, e^{i \, \lambda} \right)  \right]^2  \, .
\end{align}
Through use of standard integrals, we find that this expression simplifies into 
\begin{equation} 
I_{\tilde{d}_1} =\frac{8 \,  \pi^2 \tilde{g}_0^2 m}{\hbar \omega_{\rm{m}}^3}  (i \, e^{-i \,  \lambda } \mu_{\rm{c}} - i \, e^{i \, \lambda} \mu_{\rm{c}}^*)^2 \, , 
\end{equation}
which can be compared with the expression for the QFI in Eq.~\eqref{chap:gravimetry:eq:QFI:coherent:coherent:2pi}. 
	
This concludes our proof of the optimality of the homodyne measurement for estimating a constant gravitational acceleration.

\end{appendix}


\addcontentsline{toc}{chapter}{Bibliography}


\setstretch{1} 
\bibliographystyle{nicebib} 
\bibliography{thesis_bibliography}

\begin{thebibliography}{100}
\providecommand{\url}[1]{\texttt{#1}}
\providecommand{\urlprefix}{}
\providecommand{\arxivId}[2][]{\url{#2}}

\bibitem{qvarfort2018gravimetry}
S.~Qvarfort, A.~Serafini, P.~F. Barker \emph{et~al.}
\newblock \href{https://doi.org/10.1038/s41467-018-06037-z}{Gravimetry through
  non-linear optomechanics}.
\newblock \emph{Nature Communications}, \textbf{9}, 3690 (2018).

\bibitem{qvarfort2019enhanced}
S.~Qvarfort, A.~Serafini, A.~Xuereb \emph{et~al.}
\newblock \href{https://doi.org/10.1088/1367-2630/ab1b9e}{Enhanced continuous
  generation of non-{G}aussianity through optomechanical modulation}.
\newblock \emph{New Journal of Physics} (2019).

\bibitem{qvarfort2019time}
S.~Qvarfort, A.~Serafini, A.~Xuereb \emph{et~al.}
\newblock \href{https://arxiv.org/abs/1908.00790}{Time-evolution of nonlinear
  optomechanical systems: Interplay of arbitrary mechanical squeezing and
  non-{G}aussianity}.
\newblock \emph{arXiv preprint arXiv:1908.00790} (2019).

\bibitem{schneiter2019optimal}
F.~Schneiter, S.~Qvarfort, A.~Serafini \emph{et~al.}
\newblock \href{https://doi.org/10.1103/PhysRevA.101.033834}{Optimal estimation
  with quantum optomechanical systems in the nonlinear regime}.
\newblock \emph{Physical Review A}, \textbf{101}, 033834 (2020).

\bibitem{qvarfort2018mesoscopic}
S.~Qvarfort, S.~Bose, and A.~Serafini.
\newblock \href{https://arxiv.org/abs/1812.09776}{Mesoscopic entanglement from
  central potential interactions}.
\newblock \emph{arXiv preprint arXiv:1812.09776} (2018).

\bibitem{aspelmeyer2014cavity}
M.~Aspelmeyer, T.~J. Kippenberg, and F.~Marquardt.
\newblock \href{https://doi.org/10.1103/RevModPhys.86.1391}{Cavity
  optomechanics}.
\newblock \emph{Reviews of Modern Physics}, \textbf{86}, 1391 (2014).

\bibitem{landau2013quantum}
L.~D. Landau and E.~M. Lifshitz.
\newblock \emph{Quantum mechanics: non-relativistic theory}, volume~3.
\newblock Elsevier (2013).

\bibitem{bassi2013models}
A.~Bassi, K.~Lochan, S.~Satin \emph{et~al.}
\newblock \href{https://doi.org/10.1103/RevModPhys.85.471}{Models of
  wave-function collapse, underlying theories, and experimental tests}.
\newblock \emph{Reviews of Modern Physics}, \textbf{85}, 471 (2013).

\bibitem{bahrami2014proposal}
M.~Bahrami, M.~Paternostro, A.~Bassi \emph{et~al.}
\newblock \href{https://doi.org/10.1103/PhysRevLett.112.210404}{Proposal for a
  noninterferometric test of collapse models in optomechanical systems}.
\newblock \emph{Physical Review Letters}, \textbf{112}, 210404 (2014).

\bibitem{goldwater2016testing}
D.~Goldwater, M.~Paternostro, and P.~Barker.
\newblock \href{https://doi.org/10.1103/PhysRevA.94.010104}{Testing
  wave-function-collapse models using parametric heating of a trapped
  nanosphere}.
\newblock \emph{Physical Review A}, \textbf{94}, 010104 (2016).

\bibitem{carlesso2018non}
M.~Carlesso, M.~Paternostro, H.~Ulbricht \emph{et~al.}
\newblock \href{https://doi.org/10.1088/1367-2630/aad863}{Non-interferometric
  test of the continuous spontaneous localization model based on rotational
  optomechanics}.
\newblock \emph{New Journal of Physics}, \textbf{20}, 083022 (2018).

\bibitem{pearle1989combining}
P.~Pearle.
\newblock \href{https://doi.org/10.1103/PhysRevA.39.2277}{Combining stochastic
  dynamical state-vector reduction with spontaneous localization}.
\newblock \emph{Physical Review A}, \textbf{39}, 2277 (1989).

\bibitem{ghirardi1990markov}
G.~C. Ghirardi, P.~Pearle, and A.~Rimini.
\newblock \href{https://doi.org/10.1103/PhysRevA.42.78}{Markov processes in
  hilbert space and continuous spontaneous localization of systems of identical
  particles}.
\newblock \emph{Physical Review A}, \textbf{42}, 78 (1990).

\bibitem{diosi1989models}
L.~Diosi.
\newblock
  \href{http://www.rmki.kfki.hu/~diosi/prints/1989pra40p1165.pdf}{Models for
  universal reduction of macroscopic quantum fluctuations}.
\newblock \emph{Physical Review A}, \textbf{40}, 1165 (1989).

\bibitem{penrose1996gravity}
R.~Penrose.
\newblock \href{https://doi.org/10.1007/BF02105068}{On gravity's role in
  quantum state reduction}.
\newblock \emph{General relativity and gravitation}, \textbf{28}, 581 (1996).

\bibitem{vinante2016upper}
A.~Vinante, M.~Bahrami, A.~Bassi \emph{et~al.}
\newblock \href{https://doi.org/10.1103/PhysRevLett.116.090402}{Upper bounds on
  spontaneous wave-function collapse models using millikelvin-cooled
  nanocantilevers}.
\newblock \emph{Physical Review Letters}, \textbf{116}, 090402 (2016).

\bibitem{helou2017lisa}
B.~Helou, B.~Slagmolen, D.~E. McClelland \emph{et~al.}
\newblock \href{https://doi.org/10.1103/PhysRevD.95.084054}{Lisa pathfinder
  appreciably constrains collapse models}.
\newblock \emph{Physical Review D}, \textbf{95}, 084054 (2017).

\bibitem{kaltenbaek2016macroscopic}
R.~Kaltenbaek, M.~Aspelmeyer, P.~F. Barker \emph{et~al.}
\newblock \href{https://doi.org/10.1140/epjqt/s40507-016-0043-7}{Macroscopic
  quantum resonators (maqro): 2015 update}.
\newblock \emph{EPJ Quantum Technology}, \textbf{3}, 5 (2016).

\bibitem{bose2017spin}
S.~Bose, A.~Mazumdar, G.~W. Morley \emph{et~al.}
\newblock
  \href{https://journals.aps.org/prl/abstract/10.1103/PhysRevLett.119.240401}{Spin
  entanglement witness for quantum gravity}.
\newblock \emph{Physical Review Letters}, \textbf{119}, 240401 (2017).

\bibitem{marletto2017gravitationally}
C.~Marletto and V.~Vedral.
\newblock
  \href{https://journals.aps.org/prl/abstract/10.1103/PhysRevLett.119.240402}{Gravitationally
  induced entanglement between two massive particles is sufficient evidence of
  quantum effects in gravity}.
\newblock \emph{Physical Review Letters}, \textbf{119}, 240402 (2017).

\bibitem{nguyen2019entanglement}
H.~C. Nguyen and F.~Bernards.
\newblock \href{https://arxiv.org/abs/1812.09776}{Entanglement dynamics of two
  mesoscopic objects with gravitational interaction}.
\newblock \emph{arXiv preprint arXiv:1906.11184} (2019).

\bibitem{krisnanda2019observable}
T.~Krisnanda, G.~Y. Tham, M.~Paternostro \emph{et~al.}
\newblock \href{https://arxiv.org/abs/1906.08808}{Observable quantum
  entanglement due to gravity}.
\newblock \emph{arXiv preprint arXiv:1906.08808} (2019).

\bibitem{dienes1997string}
K.~R. Dienes.
\newblock \href{https://doi.org/10.1016/S0370-1573(97)00009-4}{String theory
  and the path to unification: a review of recent developments}.
\newblock \emph{Physics Reports}, \textbf{287}, 447 (1997).

\bibitem{rovelli2008loop}
C.~Rovelli.
\newblock \href{https://doi.org/10.12942/lrr-2008-5}{Loop quantum gravity}.
\newblock \emph{Living reviews in relativity}, \textbf{11}, 5 (2008).

\bibitem{carney2018massive}
D.~Carney, P.~C.~E. Stamp, and J.~M. Taylor.
\newblock \href{https://doi.org/10.1088%2F1361-6382%2Faaf9ca}{Tabletop
  experiments for quantum gravity: a user's manual}.
\newblock \emph{Classical and Quantum Gravity}, \textbf{36}, 034001 (2019).

\bibitem{mason2019continuous}
D.~Mason, J.~Chen, M.~Rossi \emph{et~al.}
\newblock \href{https://doi.org/10.1038/s41567-019-0533-5}{Continuous force and
  displacement measurement below the standard quantum limit}.
\newblock \emph{Nature Physics}, page~1 (2019).

\bibitem{Cervantes2014}
F.~Guzm{\'a}n~Cervantes, L.~Kumanchik, J.~Pratt \emph{et~al.}
\newblock \href{https://arxiv.org/abs/1303.1188}{High sensitivity
  optomechanical reference accelerometer over 10 khz}.
\newblock \emph{Applied Physics Letters}, \textbf{104}, 221111 (2014).

\bibitem{kippenberg2008cavity}
T.~J. Kippenberg and K.~J. Vahala.
\newblock \href{https://doi.org/10.1126/science.1156032}{Cavity optomechanics:
  back-action at the mesoscale}.
\newblock \emph{science}, \textbf{321}, 1172 (2008).

\bibitem{yin2013optomechanics}
Z.-Q. Yin, A.~A. Geraci, and T.~Li.
\newblock \href{https://doi.org/10.1142/S0217979213300181}{Optomechanics of
  levitated dielectric particles}.
\newblock \emph{International Journal of Modern Physics B}, \textbf{27},
  1330018 (2013).

\bibitem{millen2019optomechanics}
J.~Millen, T.~S. Monteiro, R.~Pettit \emph{et~al.}
\newblock \href{https://arxiv.org/abs/1907.08198}{Optomechanics with levitated
  particles}.
\newblock \emph{arXiv preprint arXiv:1907.08198} (2019).

\bibitem{bowen2015quantum}
W.~P. Bowen and G.~J. Milburn.
\newblock \emph{Quantum {O}ptomechanics}.
\newblock CRC Press (2015).

\bibitem{law1995interaction}
C.~Law.
\newblock \href{https://doi.org/10.1103/PhysRevA.51.2537}{Interaction between a
  moving mirror and radiation pressure: A {H}amiltonian formulation}.
\newblock \emph{Physical Review A}, \textbf{51}, 2537 (1995).

\bibitem{romero2011optically}
O.~Romero-Isart, A.~C. Pflanzer, M.~L. Juan \emph{et~al.}
\newblock \href{https://doi.org/10.1103/PhysRevA.83.013803}{Optically
  levitating dielectrics in the quantum regime: Theory and protocols}.
\newblock \emph{Physical Review A}, \textbf{83}, 013803 (2011).

\bibitem{Millen2015iontrap}
J.~Millen, P.~Fonseca, T.~Mavrogordatos \emph{et~al.}
\newblock \href{https://doi.org/10.1103/PhysRevLett.114.123602}{Cavity cooling
  a single charged levitated nanosphere}.
\newblock \emph{Physical Review Letters}, \textbf{114}, 123602 (2015).

\bibitem{vanner2011pulsed}
M.~R. Vanner, I.~Pikovski, G.~D. Cole \emph{et~al.}
\newblock \href{https://doi.org/10.1073/pnas.1105098108}{Pulsed quantum
  optomechanics}.
\newblock \emph{Proceedings of the National Academy of Sciences}, \textbf{108},
  16182 (2011).

\bibitem{bose1997preparation}
S.~Bose, K.~Jacobs, and P.~Knight.
\newblock \href{https://doi.org/10.1103/PhysRevA.56.4175}{Preparation of
  nonclassical states in cavities with a moving mirror}.
\newblock \emph{Physical Review A}, \textbf{56}, 4175 (1997).

\bibitem{mancini1997ponderomotive}
S.~Mancini, V.~Man'{k}o, and P.~Tombesi.
\newblock \href{https://doi.org/10.1103/PhysRevA.55.3042}{Ponderomotive control
  of quantum macroscopic coherence}.
\newblock \emph{Physical Review A}, \textbf{55}, 3042 (1997).

\bibitem{favero2009optomechanics}
I.~Favero and K.~Karrai.
\newblock \href{https://doi.org/10.1038/nphoton.2009.42}{Optomechanics of
  deformable optical cavities}.
\newblock \emph{Nature Photonics}, \textbf{3}, 201 (2009).

\bibitem{arcizet2006radiation}
O.~Arcizet, P.-F. Cohadon, T.~Briant \emph{et~al.}
\newblock \href{https://doi.org/10.1038/nature05244}{Radiation-pressure cooling
  and optomechanical instability of a micromirror}.
\newblock \emph{Nature}, \textbf{444}, 71 (2006).

\bibitem{barker2010cavity}
P.~Barker and M.~Shneider.
\newblock \href{https://doi.org/10.1103/PhysRevA.81.023826}{Cavity cooling of
  an optically trapped nanoparticle}.
\newblock \emph{Physical Review A}, \textbf{81}, 023826 (2010).

\bibitem{yin2013large}
Z.-q. Yin, T.~Li, X.~Zhang \emph{et~al.}
\newblock \href{https://doi.org/10.1103/PhysRevA.88.033614}{Large quantum
  superpositions of a levitated nanodiamond through spin-optomechanical
  coupling}.
\newblock \emph{Physical Review A}, \textbf{88}, 033614 (2013).

\bibitem{neukirch2015multi}
L.~P. Neukirch, E.~Von~Haartman, J.~M. Rosenholm \emph{et~al.}
\newblock \href{https://doi.org/10.1038/nphoton.2015.162}{Multi-dimensional
  single-spin nano-optomechanics with a levitated nanodiamond}.
\newblock \emph{Nature Photonics}, \textbf{9}, 653 (2015).

\bibitem{schliesser2010cavity}
A.~Schliesser and T.~J. Kippenberg.
\newblock \href{https://doi.org/10.1016/S1049-250X(10)05810-6}{Cavity
  optomechanics with whispering-gallery mode optical micro-resonators}.
\newblock \href{https://doi.org/10.1016/S1049-250X(10)05810-6}{In
  \emph{Advances In Atomic, Molecular, and Optical Physics}}, volume~58, pages
  207--323. Elsevier (2010).

\bibitem{barker2010doppler}
P.~Barker.
\newblock \href{https://doi.org/10.1103/PhysRevLett.105.073002}{Doppler cooling
  a microsphere}.
\newblock \emph{Physical Review Letters}, \textbf{105}, 073002 (2010).

\bibitem{tsaturyan2017ultracoherent}
Y.~Tsaturyan, A.~Barg, E.~S. Polzik \emph{et~al.}
\newblock \href{https://doi.org/10.1038/nnano.2017.101}{Ultracoherent
  nanomechanical resonators via soft clamping and dissipation dilution}.
\newblock \emph{Nature nanotechnology}, \textbf{12}, 776 (2017).

\bibitem{jayich2008dispersive}
A.~Jayich, J.~Sankey, B.~Zwickl \emph{et~al.}
\newblock \href{https://doi.org/10.1088/1367-2630/10/9/095008}{Dispersive
  optomechanics: a membrane inside a cavity}.
\newblock \emph{New Journal of Physics}, \textbf{10}, 095008 (2008).

\bibitem{eichenfield2009picogram}
M.~Eichenfield, R.~Camacho, J.~Chan \emph{et~al.}
\newblock \href{https://doi.org/10.1038/nature08061}{A picogram-and
  nanometre-scale photonic-crystal optomechanical cavity}.
\newblock \emph{Nature}, \textbf{459}, 550 (2009).

\bibitem{safavi2014two}
A.~H. Safavi-Naeini, J.~T. Hill, S.~Meenehan \emph{et~al.}
\newblock \href{https://doi.org/10.1103/PhysRevLett.112.153603}{Two-dimensional
  phononic-photonic band gap optomechanical crystal cavity}.
\newblock \emph{Physical Review Letters}, \textbf{112}, 153603 (2014).

\bibitem{van2016unifying}
R.~Van~Laer, R.~Baets, and D.~Van~Thourhout.
\newblock \href{https://doi.org/10.1103/PhysRevA.93.053828}{Unifying brillouin
  scattering and cavity optomechanics}.
\newblock \emph{Physical Review A}, \textbf{93}, 053828 (2016).

\bibitem{purdy2010tunable}
T.~P. Purdy, D.~Brooks, T.~Botter \emph{et~al.}
\newblock \href{https://doi.org/10.1103/PhysRevLett.105.133602}{Tunable cavity
  optomechanics with ultracold atoms}.
\newblock \emph{Physical Review Letters}, \textbf{105}, 133602 (2010).

\bibitem{brennecke2008cavity}
F.~Brennecke, S.~Ritter, T.~Donner \emph{et~al.}
\newblock \href{https://doi.org/10.1126/science.1163218}{Cavity optomechanics
  with a {B}ose-{E}instein condensate}.
\newblock \emph{Science}, \textbf{322}, 235 (2008).

\bibitem{groblacher2009observation}
S.~Gr{\"o}blacher, K.~Hammerer, M.~R. Vanner \emph{et~al.}
\newblock \href{https://doi.org/10.1038/nature08171}{Observation of strong
  coupling between a micromechanical resonator and an optical cavity field}.
\newblock \emph{Nature}, \textbf{460}, 724 (2009).

\bibitem{chan2011laser}
J.~Chan, T.~M. Alegre, A.~H. Safavi-Naeini \emph{et~al.}
\newblock \href{https://doi.org/10.1038/nature10461f}{Laser cooling of a
  nanomechanical oscillator into its quantum ground state}.
\newblock \emph{Nature}, \textbf{478}, 89 (2011).

\bibitem{chang2010cavity}
D.~E. Chang, C.~Regal, S.~Papp \emph{et~al.}
\newblock \href{http://www.pnas.org/content/107/3/1005.full}{Cavity
  opto-mechanics using an optically levitated nanosphere}.
\newblock \emph{Proceedings of the National Academy of Sciences}, \textbf{107},
  1005 (2010).

\bibitem{pontin2019ultra}
A.~Pontin, N.~Bullier, M.~Toro{\v{s}} \emph{et~al.}
\newblock \href{https://arxiv.org/abs/1907.06046}{An ultra-narrow line width
  levitated nano-oscillator for testing dissipative wavefunction collapse}.
\newblock \emph{arXiv preprint arXiv:1907.06046} (2019).

\bibitem{bullier2019characterisation}
N.~Bullier, A.~Pontin, and P.~Barker.
\newblock \href{https://arxiv.org/abs/1906.09580}{Characterisation of a charged
  particle levitated nano-oscillator}.
\newblock \emph{arXiv preprint arXiv:1906.09580} (2019).

\bibitem{pino2018chip}
H.~Pino, J.~Prat-Camps, K.~Sinha \emph{et~al.}
\newblock \href{https://doi.org/10.1088/2058-9565/aa9d15}{On-chip quantum
  interference of a superconducting microsphere}.
\newblock \emph{Quantum Science and Technology}, \textbf{3}, 025001 (2018).

\bibitem{munstermann1999dynamics}
P.~M{\"u}nstermann, T.~Fischer, P.~Maunz \emph{et~al.}
\newblock
  \href{https://journals.aps.org/prl/abstract/10.1103/PhysRevLett.82.3791}{Dynamics
  of single-atom motion observed in a high-finesse cavity}.
\newblock \emph{Physical Review Letters}, \textbf{82}, 3791 (1999).

\bibitem{pearle2012simple}
P.~Pearle.
\newblock \href{https://doi.org/10.1088/0143-0807/33/4/805}{Simple derivation
  of the lindblad equation}.
\newblock \emph{European Journal of Physics}, \textbf{33}, 805 (2012).

\bibitem{gardiner2004quantum}
C.~Gardiner, P.~Zoller, and P.~Zoller.
\newblock \emph{Quantum noise: a handbook of Markovian and non-Markovian
  quantum stochastic methods with applications to quantum optics}, volume~56.
\newblock Springer Science \& Business Media (2004).

\bibitem{johansson2013qutip}
J.~R. Johansson, P.~D. Nation, and F.~Nori.
\newblock \href{https://doi.org/10.1016/j.cpc.2012.11.019}{Qutip 2: A python
  framework for the dynamics of open quantum systems}.
\newblock \emph{Computer Physics Communications}, \textbf{184}, 1234 (2013).

\bibitem{zurek1981pointer}
W.~H. Zurek.
\newblock
  \href{https://journals.aps.org/prd/abstract/10.1103/PhysRevD.24.1516}{Pointer
  basis of quantum apparatus: Into what mixture does the wave packet collapse?}
\newblock \emph{Physical Review D}, \textbf{24}, 1516 (1981).

\bibitem{paz1993reduction}
J.~P. Paz, S.~Habib, and W.~H. Zurek.
\newblock
  \href{https://journals.aps.org/prd/abstract/10.1103/PhysRevD.47.488}{Reduction
  of the wave packet: Preferred observable and decoherence time scale}.
\newblock \emph{Physical Review D}, \textbf{47}, 488 (1993).

\bibitem{anglin1996decoherence}
J.~Anglin and W.~Zurek.
\newblock \href{https://arxiv.org/abs/quant-ph/9510021}{Decoherence of quantum
  fields: Pointer states and predictability}.
\newblock \emph{Physical Review D}, \textbf{53}, 7327 (1996).

\bibitem{zurek1993coherent}
W.~H. Zurek, S.~Habib, and J.~P. Paz.
\newblock
  \href{https://journals.aps.org/prl/abstract/10.1103/PhysRevLett.70.1187}{Coherent
  states via decoherence}.
\newblock \emph{Physical Review Letters}, \textbf{70}, 1187 (1993).

\bibitem{lemons1997paul}
D.~S. Lemons and A.~Gythiel.
\newblock \href{https://doi.org/10.1119/1.18725}{Paul langevin’s 1908 paper
  “on the theory of brownian motion”[“sur la th{\'e}orie du mouvement
  brownien,” cr acad. sci.(paris) 146, 530--533 (1908)]}.
\newblock \emph{American Journal of Physics}, \textbf{65}, 1079 (1997).

\bibitem{gardiner1985input}
C.~Gardiner and M.~Collett.
\newblock \href{https://doi.org/10.1103/PhysRevA.31.3761}{Input and output in
  damped quantum systems: Quantum stochastic differential equations and the
  master equation}.
\newblock \emph{Physical Review A}, \textbf{31}, 3761 (1985).

\bibitem{walls2007quantum}
D.~F. Walls and G.~J. Milburn.
\newblock \emph{Quantum {O}ptics}.
\newblock Springer Science \& Business Media (2007).

\bibitem{serafini2017quantum}
A.~Serafini.
\newblock \emph{Quantum {C}ontinuous {V}ariables: {A Primer of Theoretical
  Methods}}.
\newblock CRC Press (2017).

\bibitem{wigner1997quantum}
E.~P. Wigner.
\newblock \href{https://doi.org/10.1007/978-3-642-59033-7_9}{On the quantum
  correction for thermodynamic equilibrium}.
\newblock \href{https://doi.org/10.1007/978-3-642-59033-7_9}{In \emph{Part I:
  Physical Chemistry. Part II: Solid State Physics}}, pages 110--120. Springer
  (1997).

\bibitem{kenfack2004negativity}
A.~Kenfack and K.~{\.Z}yczkowski.
\newblock \href{https://doi.org/10.1088/1464-4266/6/10/003}{Negativity of the
  wigner function as an indicator of non-classicality}.
\newblock \emph{Journal of Optics B: Quantum and Semiclassical Optics},
  \textbf{6}, 396 (2004).

\bibitem{dell2010teleportation}
F.~Dell'Anno, S.~De~Siena, G.~Adesso \emph{et~al.}
\newblock \href{https://doi.org/10.1103/PhysRevA.82.062329}{Teleportation of
  squeezing: Optimization using non-{G}aussian resources}.
\newblock \emph{Physical Review A}, \textbf{82}, 062329 (2010).

\bibitem{lloyd1999quantum}
S.~Lloyd and S.~L. Braunstein.
\newblock Quantum computation over continuous variables.
\newblock In \emph{Quantum Information with Continuous Variables}, pages 9--17.
  Springer (1999).

\bibitem{menicucci2006universal}
N.~C. Menicucci, P.~van Loock, M.~Gu \emph{et~al.}
\newblock \href{https://doi.org/10.1103/PhysRevLett.97.110501}{Universal
  quantum computation with continuous-variable cluster states}.
\newblock \emph{Physical Review Letters}, \textbf{97}, 110501 (2006).

\bibitem{niset2009no}
J.~Niset, J.~Fiur{\'a}{\v{s}}ek, and N.~J. Cerf.
\newblock \href{https://doi.org/10.1103/PhysRevLett.102.120501}{No-go theorem
  for {G}aussian quantum error correction}.
\newblock \emph{Physical Review Letters}, \textbf{102}, 120501 (2009).

\bibitem{eisert2002conditions}
J.~Eisert and M.~B. Plenio.
\newblock \href{https://doi.org/10.1103/PhysRevLett.89.097901}{Conditions for
  the local manipulation of {G}aussian states}.
\newblock \emph{Physical Review Letters}, \textbf{89}, 097901 (2002).

\bibitem{fiuravsek2002gaussian}
J.~Fiur{\'a}{\v{s}}ek.
\newblock \href{https://doi.org/10.1103/PhysRevLett.89.137904}{{G}aussian
  transformations and distillation of entangled {G}aussian states}.
\newblock \emph{Physical Review Letters}, \textbf{89}, 137904 (2002).

\bibitem{giedke2002characterization}
G.~Giedke and J.~I. Cirac.
\newblock \href{https://doi.org/10.1103/PhysRevA.66.032316}{Characterization of
  {G}aussian operations and distillation of {G}aussian states}.
\newblock \emph{Physical Review A}, \textbf{66}, 032316 (2002).

\bibitem{zhuang2018resource}
Q.~Zhuang, P.~W. Shor, and J.~H. Shapiro.
\newblock \href{https://doi.org/10.1103/PhysRevA.97.052317}{Resource theory of
  non-{G}aussian operations}.
\newblock \emph{Physical Review A}, \textbf{97}, 052317 (2018).

\bibitem{takagi2018convex}
R.~Takagi and Q.~Zhuang.
\newblock \href{http://dx.doi.org/10.1103/PhysRevA.97.062337}{Convex resource
  theory of non-{G}aussianity}.
\newblock \emph{Physical Review A}, \textbf{97}, 062337 (2018).

\bibitem{albarelli2018resource}
F.~Albarelli, M.~G. Genoni, M.~G. Paris \emph{et~al.}
\newblock \href{https://doi.org/10.1103/PhysRevA.98.052350}{Resource theory of
  quantum non-{G}aussianity and {W}igner negativity}.
\newblock \emph{Physical Review A}, \textbf{98}, 052350 (2018).

\bibitem{sabapathy2011robustness}
K.~K. Sabapathy, J.~S. Ivan, and R.~Simon.
\newblock \href{https://doi.org/10.1103/PhysRevLett.107.130501}{Robustness of
  non-{G}aussian entanglement against noisy amplifier and attenuator
  environments}.
\newblock \emph{Physical Review Letters}, \textbf{107}, 130501 (2011).

\bibitem{nha2010linear}
H.~Nha, G.~Milburn, and H.~Carmichael.
\newblock \href{https://doi.org/10.1088/1367-2630/12/10/103010}{Linear
  amplification and quantum cloning for non-{G}aussian continuous variables}.
\newblock \emph{New Journal of Physics}, \textbf{12}, 103010 (2010).

\bibitem{silberhorn2001generation}
C.~Silberhorn, P.~K. Lam, O.~Weiss \emph{et~al.}
\newblock \href{https://doi.org/10.1103/PhysRevLett.86.4267}{Generation of
  continuous variable {E}instein--{P}odolsky--{R}osen entanglement via the
  {K}err nonlinearity in an optical fiber}.
\newblock \emph{Physical Review Letters}, \textbf{86}, 4267 (2001).

\bibitem{lvovsky2001quantum}
A.~I. Lvovsky, H.~Hansen, T.~Aichele \emph{et~al.}
\newblock \href{https://doi.org/10.1103/PhysRevLett.87.050402}{Quantum state
  reconstruction of the single-photon fock state}.
\newblock \emph{Physical Review Letters}, \textbf{87}, 050402 (2001).

\bibitem{heersink2003polarization}
J.~Heersink, T.~Gaber, S.~Lorenz \emph{et~al.}
\newblock \href{https://doi.org/10.1103/PhysRevA.68.013815}{Polarization
  squeezing of intense pulses with a fiber-optic {S}agnac interferometer}.
\newblock \emph{Physical Review A}, \textbf{68}, 013815 (2003).

\bibitem{ourjoumtsev2006quantum}
A.~Ourjoumtsev, R.~Tualle-Brouri, and P.~Grangier.
\newblock \href{https://doi.org/10.1103/PhysRevLett.96.213601}{Quantum homodyne
  tomography of a two-photon fock state}.
\newblock \emph{Physical Review Letters}, \textbf{96}, 213601 (2006).

\bibitem{ourjoumtsev2007generation}
A.~Ourjoumtsev, H.~Jeong, R.~Tualle-Brouri \emph{et~al.}
\newblock \href{https://doi.org/10.1038/nature06054}{Generation of optical
  ‘{S}chr{\"o}dinger cats’ from photon number states}.
\newblock \emph{Nature}, \textbf{448}, 784 (2007).

\bibitem{parigi2007probing}
V.~Parigi, A.~Zavatta, M.~Kim \emph{et~al.}
\newblock \href{https://doi.org/10.1126/science.1146204}{Probing quantum
  commutation rules by addition and subtraction of single photons to/from a
  light field}.
\newblock \emph{Science}, \textbf{317}, 1890 (2007).

\bibitem{mari2012positive}
A.~Mari and J.~Eisert.
\newblock \href{https://doi.org/10.1103/PhysRevLett.109.230503}{Positive
  {W}igner functions render classical simulation of quantum computation
  efficient}.
\newblock \emph{Physical Review Letters}, \textbf{109}, 230503 (2012).

\bibitem{zurek2001sub}
W.~H. Zurek.
\newblock \href{https://doi.org/10.1038/35089017}{Sub-{P}lanck structure in
  phase space and its relevance for quantum decoherence}.
\newblock \emph{Nature}, \textbf{412}, 712 (2001).

\bibitem{toscano2006sub}
F.~Toscano, D.~A. Dalvit, L.~Davidovich \emph{et~al.}
\newblock \href{https://doi.org/10.1103/PhysRevA.73.023803}{Sub-{P}lanck
  phase-space structures and {H}eisenberg-limited measurements}.
\newblock \emph{Physical Review A}, \textbf{73}, 023803 (2006).

\bibitem{howard2018hypercube}
L.~Howard, T.~Weinhold, J.~Combes \emph{et~al.}
\newblock \href{https://arxiv.org/abs/1811.03011}{Hypercube states for
  sub-{P}lanck sensing}.
\newblock \emph{arXiv preprint arXiv:1811.03011} (2018).

\bibitem{ludwig2008optomechanical}
M.~Ludwig, B.~Kubala, and F.~Marquardt.
\newblock \href{https://doi.org/10.1088/1367-2630/10/9/095013}{The
  optomechanical instability in the quantum regime}.
\newblock \emph{New Journal of Physics}, \textbf{10}, 095013 (2008).

\bibitem{palomaki2013coherent}
T.~Palomaki, J.~Harlow, J.~Teufel \emph{et~al.}
\newblock \href{https://doi.org/10.1038/nature11915}{Coherent state transfer
  between itinerant microwave fields and a mechanical oscillator}.
\newblock \emph{Nature}, \textbf{495}, 210 (2013).

\bibitem{sankey2010strong}
J.~C. Sankey, C.~Yang, B.~M. Zwickl \emph{et~al.}
\newblock \href{https://doi.org/10.1038/nphys1707}{Strong and tunable nonlinear
  optomechanical coupling in a low-loss system}.
\newblock \emph{Nature Physics}, \textbf{6}, 707 (2010).

\bibitem{doolin2014nonlinear}
C.~Doolin, B.~Hauer, P.~Kim \emph{et~al.}
\newblock \href{https://doi.org/10.1103/PhysRevA.89.053838}{Nonlinear
  optomechanics in the stationary regime}.
\newblock \emph{Physical Review A}, \textbf{89}, 053838 (2014).

\bibitem{brawley2016nonlinear}
G.~Brawley, M.~Vanner, P.~E. Larsen \emph{et~al.}
\newblock \href{https://doi.org/10.1038/ncomms10988}{Nonlinear optomechanical
  measurement of mechanical motion}.
\newblock \emph{Nature Communications}, \textbf{7}, 10988 (2016).

\bibitem{leijssen2017nonlinear}
R.~Leijssen, G.~R. La~Gala, L.~Freisem \emph{et~al.}
\newblock \href{https://doi.org/10.1038/ncomms16024}{Nonlinear cavity
  optomechanics with nanomechanical thermal fluctuations}.
\newblock \emph{Nature Communications}, \textbf{8}, ncomms16024 (2017).

\bibitem{kippenberg2007cavity}
T.~J. Kippenberg and K.~J. Vahala.
\newblock \href{https://doi.org/10.1364/OE.15.017172}{Cavity opto-mechanics}.
\newblock \emph{Optics Express}, \textbf{15}, 17172 (2007).

\bibitem{williamson1936algebraic}
J.~Williamson.
\newblock \href{https://doi.org/10.2307/2371062}{On the algebraic problem
  concerning the normal forms of linear dynamical systems}.
\newblock \emph{American Journal of Mathematics}, \textbf{58}, 141 (1936).

\bibitem{adesso2014continuous}
G.~Adesso, S.~Ragy, and A.~R. Lee.
\newblock \href{https://doi.org/10.1142/S1230161214400010}{Continuous variable
  quantum information: {G}aussian states and beyond}.
\newblock \emph{Open Systems \& Information Dynamics}, \textbf{21}, 1440001
  (2014).

\bibitem{marquardt2009optomechanics}
F.~Marquardt and S.~M. Girvin.
\newblock \href{https://arxiv.org/abs/0905.0566}{Optomechanics (a brief
  review)}.
\newblock \emph{arXiv preprint arXiv:0905.0566} (2009).

\bibitem{meystre2013short}
P.~Meystre.
\newblock \href{https://doi.org/10.1002/andp.201200226}{A short walk through
  quantum optomechanics}.
\newblock \emph{Annalen der Physik}, \textbf{525}, 215 (2013).

\bibitem{genoni2008quantifying}
M.~G. Genoni, M.~G. Paris, and K.~Banaszek.
\newblock \href{https://doi.org/10.1103/PhysRevA.78.060303}{Quantifying the
  non-{G}aussian character of a quantum state by quantum relative entropy}.
\newblock \emph{Physical Review A}, \textbf{78}, 060303 (2008).

\bibitem{park2019faithful}
J.~Park, J.~Lee, K.~Baek \emph{et~al.}
\newblock \href{https://doi.org/10.1103/PhysRevA.100.012333}{Faithful measure
  of quantum non-{G}aussianity via quantum relative entropy}.
\newblock \emph{Physical Review A}, \textbf{100}, 012333 (2019).

\bibitem{marian2013relative}
P.~Marian and T.~A. Marian.
\newblock \href{https://doi.org/10.1103/PhysRevA.88.012322}{Relative entropy is
  an exact measure of non-gaussianity}.
\newblock \emph{Physical Review A}, \textbf{88}, 012322 (2013).

\bibitem{Adesso:Ragy:2014}
G.~Adesso, S.~Ragy, and A.~R. Lee.
\newblock \href{https://doi.org/10.1142/S1230161214400010}{Continuous variable
  quantum information: {G}aussian states and beyond}.
\newblock \emph{Open Systems \& Information Dynamics}, \textbf{21}, 1440001
  (2014).

\bibitem{ahmadi2014quantum}
M.~Ahmadi, D.~E. Bruschi, and I.~Fuentes.
\newblock \href{https://doi.org/10.1103/PhysRevD.89.065028}{Quantum metrology
  for relativistic quantum fields}.
\newblock \emph{Physical Review D}, \textbf{89}, 065028 (2014).

\bibitem{datta2016notes}
A.~Datta.
\newblock Quantum metrology foundations (2016).
\newblock Lecture notes for a course on Quantum Sensing taught at University
  College London.

\bibitem{gobel2015quantum}
E.~O. G{\"o}bel and U.~Siegner.
\newblock \emph{Quantum {M}etrology: {F}oundation of {U}nits and
  {M}easurements}.
\newblock John Wiley \& Sons (2015).

\bibitem{simon2017quantum}
D.~S. Simon, G.~Jaeger, and A.~V. Sergienko.
\newblock \emph{Quantum Metrology, Imaging, and Communication}.
\newblock Springer (2017).

\bibitem{aad2012observation}
G.~Aad, T.~Abajyan, B.~Abbott \emph{et~al.}
\newblock \href{https://doi.org/10.1016/j.physletb.2012.08.020}{Observation of
  a new particle in the search for the standard model {H}iggs boson with the
  {ATLAS} detector at the {LHC}}.
\newblock \emph{Physics Letters B}, \textbf{716}, 1 (2012).

\bibitem{giovannetti2004quantum}
V.~Giovannetti, S.~Lloyd, and L.~Maccone.
\newblock \href{10.1126/science.1104149}{Quantum-enhanced measurements: beating
  the standard quantum limit}.
\newblock \emph{Science}, \textbf{306}, 1330 (2004).

\bibitem{braginskiui1975quantum}
V.~B. Braginski{\u\i} and Y.~I. Vorontsov.
\newblock
  \href{https://doi.org/10.1070/PU1975v017n05ABEH004362}{Quantum-mechanical
  limitations in macroscopic experiments and modern experimental technique}.
\newblock \emph{Soviet Physics Uspekhi}, \textbf{17}, 644 (1975).

\bibitem{caves1981quantum}
C.~M. Caves.
\newblock \href{https://doi.org/10.1103/PhysRevD.23.1693}{Quantum-mechanical
  noise in an interferometer}.
\newblock \emph{Physical Review D}, \textbf{23}, 1693 (1981).

\bibitem{barnett2003ultimate}
S.~M. Barnett, C.~Fabre, and A.~Ma{\i}tre.
\newblock \href{https://doi.org/10.1140/epjd/e2003-00003-3}{Ultimate quantum
  limits for resolution of beam displacements}.
\newblock \emph{The European Physical Journal D-Atomic, Molecular, Optical and
  Plasma Physics}, \textbf{22}, 513 (2003).

\bibitem{yurke19862}
B.~Yurke, S.~L. McCall, and J.~R. Klauder.
\newblock \href{https://doi.org/10.1103/PhysRevA.33.4033}{Su (2) and su (1, 1)
  interferometers}.
\newblock \emph{Physical Review A}, \textbf{33}, 4033 (1986).

\bibitem{dowling1998correlated}
J.~P. Dowling.
\newblock \href{https://doi.org/10.1103/PhysRevA.57.4736}{Correlated
  input-port, matter-wave interferometer: Quantum-noise limits to the
  atom-laser gyroscope}.
\newblock \emph{Physical Review A}, \textbf{57}, 4736 (1998).

\bibitem{bollinger1996optimal}
J.~J. Bollinger, W.~M. Itano, D.~J. Wineland \emph{et~al.}
\newblock \href{https://doi.org/10.1103/PhysRevA.54.R4649}{Optimal frequency
  measurements with maximally correlated states}.
\newblock \emph{Physical Review A}, \textbf{54}, R4649 (1996).

\bibitem{ou1997fundamental}
Z.~Ou.
\newblock \href{https://doi.org/10.1103/PhysRevA.55.2598}{Fundamental quantum
  limit in precision phase measurement}.
\newblock \emph{Physical Review A}, \textbf{55}, 2598 (1997).

\bibitem{boixo2007generalized}
S.~Boixo, S.~T. Flammia, C.~M. Caves \emph{et~al.}
\newblock \href{https://doi.org/10.1103/PhysRevLett.98.090401}{Generalized
  limits for single-parameter quantum estimation}.
\newblock \emph{Physical Review Letters}, \textbf{98}, 090401 (2007).

\bibitem{roy2008exponentially}
S.~Roy and S.~L. Braunstein.
\newblock \href{https://doi.org/10.1103/PhysRevLett.100.220501}{Exponentially
  enhanced quantum metrology}.
\newblock \emph{Physical Review Letters}, \textbf{100}, 220501 (2008).

\bibitem{zwierz2010general}
M.~Zwierz, C.~A. P{\'e}rez-Delgado, and P.~Kok.
\newblock \href{https://doi.org/10.1103/PhysRevLett.105.180402}{General
  optimality of the {Heisenberg} limit for quantum metrology}.
\newblock \emph{Physical Review Letters}, \textbf{105}, 180402 (2010).

\bibitem{barnett2009quantum}
S.~M. Barnett and S.~Croke.
\newblock \href{https://doi.org/10.1364/AOP.1.000238}{Quantum state
  discrimination}.
\newblock \emph{Advances in Optics and Photonics}, \textbf{1}, 238 (2009).

\bibitem{nielsen2002quantum}
M.~A. Nielsen and I.~Chuang.
\newblock Quantum computation and quantum information (2002).

\bibitem{shannon1948mathematical}
C.~E. Shannon.
\newblock \href{https://doi.org/10.1002/j.1538-7305.1948.tb01338.x}{A
  mathematical theory of communication}.
\newblock \emph{Bell system technical journal}, \textbf{27}, 379 (1948).

\bibitem{zegers2015fisher}
P.~Zegers.
\newblock \href{https://doi.org/10.3390/e17074918}{Fisher information
  properties}.
\newblock \emph{Entropy}, \textbf{17}, 4918 (2015).

\bibitem{paris2009quantum}
M.~G. Paris.
\newblock \href{https://doi.org/10.1142/S0219749909004839}{Quantum estimation
  for quantum technology}.
\newblock \emph{International Journal of Quantum Information}, \textbf{7}, 125
  (2009).

\bibitem{alipour2014quantum}
S.~Alipour, M.~Mehboudi, and A.~Rezakhani.
\newblock \href{https://doi.org/10.1103/PhysRevLett.112.120405}{Quantum
  metrology in open systems: dissipative cram{\'e}r-rao bound}.
\newblock \emph{Physical Review Letters}, \textbf{112}, 120405 (2014).

\bibitem{beau2017nonlinear}
M.~Beau and A.~del Campo.
\newblock \href{https://doi.org/10.1103/PhysRevLett.119.010403}{Nonlinear
  quantum metrology of many-body open systems}.
\newblock \emph{Physical Review Letters}, \textbf{119}, 010403 (2017).

\bibitem{liu2016quantum}
J.~Liu, J.~Chen, X.-X. Jing \emph{et~al.}
\newblock \href{https://doi.org/10.1088/1751-8113/49/27/275302}{Quantum fisher
  information and symmetric logarithmic derivative via anti-commutators}.
\newblock \emph{Journal of Physics A: Mathematical and Theoretical},
  \textbf{49}, 275302 (2016).

\bibitem{pang2014quantum}
S.~Pang and T.~A. Brun.
\newblock \href{https://doi.org/10.1103/PhysRevA.90.022117}{Quantum metrology
  for a general {H}amiltonian parameter}.
\newblock \emph{Physical Review A}, \textbf{90}, 022117 (2014).

\bibitem{jing2014quantum}
L.~Jing, J.~Xiao-Xing, Z.~Wei \emph{et~al.}
\newblock \href{https://doi.org/10.1088/0253-6102/61/1/08}{Quantum {F}isher
  information for density matrices with arbitrary ranks}.
\newblock \emph{Communications in Theoretical Physics}, \textbf{61}, 45 (2014).

\bibitem{cramer1946contribution}
H.~Cram{\'e}r.
\newblock
  \href{http://www.tandfonline.com/doi/pdf/10.1080/03461238.1946.10419631}{A
  contribution to the theory of statistical estimation}.
\newblock \emph{Scandinavian Actuarial Journal}, \textbf{1946}, 85 (1946).

\bibitem{wolfowitz1947efficiency}
J.~Wolfowitz.
\newblock \href{http://doi.org/10.1214/aoms/1177730439}{The efficiency of
  sequential estimates and wald's equation for sequential processes}.
\newblock \emph{The Annals of Mathematical Statistics}, pages 215--230 (1947).

\bibitem{rao1992information}
C.~R. Rao.
\newblock
  \href{https://link.springer.com/chapter/10.1007/978-1-4612-0919-5_16}{Information
  and the accuracy attainable in the estimation of statistical parameters}.
\newblock
  \href{https://link.springer.com/chapter/10.1007/978-1-4612-0919-5_16}{In
  \emph{Breakthroughs in statistics}}, pages 235--247. Springer (1992).

\bibitem{wei1963lie}
J.~Wei and E.~Norman.
\newblock \href{https://doi.org/10.1063/1.1703993}{Lie algebraic solution of
  linear differential equations}.
\newblock \emph{Journal of Mathematical Physics}, \textbf{4}, 575 (1963).

\bibitem{wilcox1967exponential}
R.~Wilcox.
\newblock \href{https://doi.org/10.1063/1.1705306}{Exponential operators and
  parameter differentiation in quantum physics}.
\newblock \emph{Journal of Mathematical Physics}, \textbf{8}, 962 (1967).

\bibitem{fulton2013representation}
W.~Fulton and J.~Harris.
\newblock \emph{Representation theory: {A} first course}, volume 129.
\newblock Springer Science \& Business Media (2013).

\bibitem{gutowski2007symmetry}
J.~B. Gutowski.
\newblock Symmetry and particle physics.
\newblock \emph{DAMTP, Centre for Mathematical Sciences, University of
  Cambridge} (2007).

\bibitem{bose2003vacuum}
S.~Bose, A.~Carollo, I.~Fuentes-Guridi \emph{et~al.}
\newblock \href{https://doi.org/10.1080/09500340308234561}{Vacuum induced
  {Berry} phase: {Theory} and experimental proposal}.
\newblock \emph{Journal of Modern Optics}, \textbf{50}, 1175 (2003).

\bibitem{Bruschi2013Time}
D.~E. Bruschi, A.~R. Lee, and I.~Fuentes.
\newblock \href{https://doi.org/10.1088/1751-8113/46/16/165303}{Time evolution
  techniques for detectors in relativistic quantum information}.
\newblock \emph{Journal of Physics A: Mathematical and Theoretical},
  \textbf{46}, 165303 (2013).

\bibitem{bruschi2018mechano}
D.~E. Bruschi and A.~Xuereb.
\newblock \href{https://doi.org/10.1088/1367-2630/aaca27}{"{M}echano-optics":
  an optomechanical quantum simulator}.
\newblock \emph{New Journal of Physics}, \textbf{20}, 065004 (2018).

\bibitem{fuentes2007family}
I.~Fuentes-Schuller and P.~Barberis-Blostein.
\newblock \href{https://doi.org/10.1088/1751-8113/40/27/F04}{A family of
  many-body models which are exactly solvable analytically}.
\newblock \emph{Journal of Physics A: Mathematical and Theoretical},
  \textbf{40}, F601 (2007).

\bibitem{barberis2008mode}
P.~Barberis-Blostein and I.~Fuentes-Schuller.
\newblock \href{https://doi.org/10.1103/PhysRevA.78.013641}{Mode-exchange
  collisions in an exactly solvable two-mode bose-einstein condensate}.
\newblock \emph{Physical Review A}, \textbf{78}, 013641 (2008).

\bibitem{wollman2015quantum}
E.~E. Wollman, C.~Lei, A.~Weinstein \emph{et~al.}
\newblock \href{https://doi.org/10.1126/science.aac5138}{Quantum squeezing of
  motion in a mechanical resonator}.
\newblock \emph{Science}, \textbf{349}, 952 (2015).

\bibitem{rashid2016experimental}
M.~Rashid, T.~Tufarelli, J.~Bateman \emph{et~al.}
\newblock \href{https://doi.org/10.1103/PhysRevLett.117.273601}{Experimental
  realization of a thermal squeezed state of levitated optomechanics}.
\newblock \emph{Physical Review Letters}, \textbf{117}, 273601 (2016).

\bibitem{bruschi2019time}
D.~E. Bruschi.
\newblock \href{https://doi.org/10.1063/1.5106409}{Time evolution of coupled
  multimode and multiresonator optomechanical systems}.
\newblock \emph{Journal of Mathematical Physics}, \textbf{60}, 062105 (2019).

\bibitem{lemonde2016enhanced}
M.-A. Lemonde, N.~Didier, and A.~A. Clerk.
\newblock \href{https://doi.org/10.1038/ncomms11338}{Enhanced nonlinear
  interactions in quantum optomechanics via mechanical amplification}.
\newblock \emph{Nature Communications}, \textbf{7}, 11338 (2016).

\bibitem{yin2017nonlinear}
T.-S. Yin, X.-Y. L{\"u}, L.-L. Zheng \emph{et~al.}
\newblock \href{https://doi.org/10.1103/PhysRevA.95.053861}{Nonlinear effects
  in modulated quantum optomechanics}.
\newblock \emph{Physical Review A}, \textbf{95}, 053861 (2017).

\bibitem{latmiral2016probing}
L.~Latmiral, F.~Armata, M.~G. Genoni \emph{et~al.}
\newblock \href{https://doi.org/10.1103/PhysRevA.93.052306}{Probing
  anharmonicity of a quantum oscillator in an optomechanical cavity}.
\newblock \emph{Physical Review A}, \textbf{93}, 052306 (2016).

\bibitem{liao2014modulated}
J.-Q. Liao, K.~Jacobs, F.~Nori \emph{et~al.}
\newblock \href{https://doi.org/10.1088/1367-2630/16/7/072001}{Modulated
  electromechanics: large enhancements of nonlinearities}.
\newblock \emph{New Journal of Physics}, \textbf{16}, 072001 (2014).

\bibitem{barbieri2010non}
M.~Barbieri, N.~Spagnolo, M.~G. Genoni \emph{et~al.}
\newblock \href{https://doi.org/10.1103/PhysRevA.82.063833}{Non-{G}aussianity
  of quantum states: an experimental test on single-photon-added coherent
  states}.
\newblock \emph{Physical Review A}, \textbf{82}, 063833 (2010).

\bibitem{millen2015cavity}
J.~Millen, P.~Z.~G. Fonseca, T.~Mavrogordatos \emph{et~al.}
\newblock \href{https://doi.org/10.1103/PhysRevLett.114.123602}{Cavity cooling
  a single charged levitated nanosphere}.
\newblock \emph{Physical Review Letters}, \textbf{114}, 123602 (2015).

\bibitem{hensinger2006t}
W.~Hensinger, S.~Olmschenk, D.~Stick \emph{et~al.}
\newblock \href{https://doi.org/10.1063/1.2164910}{T-junction ion trap array
  for two-dimensional ion shuttling, storage, and manipulation}.
\newblock \emph{Applied Physics Letters}, \textbf{88}, 034101 (2006).

\bibitem{walther2012controlling}
A.~Walther, F.~Ziesel, T.~Ruster \emph{et~al.}
\newblock \href{https://doi.org/10.1103/PhysRevLett.109.080501}{Controlling
  fast transport of cold trapped ions}.
\newblock \emph{Physical Review Letters}, \textbf{109}, 080501 (2012).

\bibitem{tatham2012nonclassical}
R.~Tatham, L.~Mi{\v{s}}ta~Jr, G.~Adesso \emph{et~al.}
\newblock \href{https://doi.org/10.1103/PhysRevA.85.022326}{Nonclassical
  correlations in continuous-variable non-{G}aussian werner states}.
\newblock \emph{Physical Review A}, \textbf{85}, 022326 (2012).

\bibitem{genoni2007measure}
M.~G. Genoni, M.~G. Paris, and K.~Banaszek.
\newblock \href{https://doi.org/10.1103/PhysRevA.76.042327}{Measure of the
  non-{G}aussian character of a quantum state}.
\newblock \emph{Physical Review A}, \textbf{76}, 042327 (2007).

\bibitem{park2017quantifying}
J.~Park, J.~Lee, S.-W. Ji \emph{et~al.}
\newblock \href{https://doi.org/10.1103/PhysRevA.96.052324}{Quantifying
  non-{G}aussianity of quantum-state correlation}.
\newblock \emph{Physical Review A}, \textbf{96}, 052324 (2017).

\bibitem{genoni2013detecting}
M.~G. Genoni, M.~L. Palma, T.~Tufarelli \emph{et~al.}
\newblock \href{https://doi.org/10.1103/PhysRevA.87.062104}{Detecting quantum
  non-{G}aussianity via the wigner function}.
\newblock \emph{Physical Review A}, \textbf{87}, 062104 (2013).

\bibitem{yadin2018operational}
B.~Yadin, F.~C. Binder, J.~Thompson \emph{et~al.}
\newblock \href{http://dx.doi.org/10.1103/PhysRevX.8.041038}{Operational
  resource theory of continuous-variable nonclassicality}.
\newblock \emph{Physical Review X}, \textbf{8}, 041038 (2018).

\bibitem{nunnenkamp2011single}
A.~Nunnenkamp, K.~B{\o}rkje, and S.~M. Girvin.
\newblock \href{https://doi.org/10.1103/PhysRevLett.107.063602}{Single-photon
  optomechanics}.
\newblock \emph{Physical Review Letters}, \textbf{107}, 063602 (2011).

\bibitem{fonseca2016nonlinear}
P.~Fonseca, E.~Aranas, J.~Millen \emph{et~al.}
\newblock \href{https://doi.org/10.1103/PhysRevLett.117.173602}{Nonlinear
  dynamics and strong cavity cooling of levitated nanoparticles}.
\newblock \emph{Physical Review Letters}, \textbf{117}, 173602 (2016).

\bibitem{aranas2016split}
E.~Aranas, P.~Fonseca, P.~Barker \emph{et~al.}
\newblock \href{https://doi.org/10.1088/1367-2630/18/11/113021}{Split-sideband
  spectroscopy in slowly modulated optomechanics}.
\newblock \emph{New Journal of Physics}, \textbf{18}, 113021 (2016).

\bibitem{weis2010optomechanically}
S.~Weis, R.~Rivi{\`e}re, S.~Del{\'e}glise \emph{et~al.}
\newblock \href{https://doi.org/10.1126/science.1195596}{Optomechanically
  induced transparency}.
\newblock \emph{Science}, \textbf{330}, 1520 (2010).

\bibitem{karuza2013optomechanically}
M.~Karuza, C.~Biancofiore, M.~Bawaj \emph{et~al.}
\newblock \href{https://doi.org/10.1103/PhysRevA.88.013804}{Optomechanically
  induced transparency in a membrane-in-the-middle setup at room temperature}.
\newblock \emph{Physical Review A}, \textbf{88}, 013804 (2013).

\bibitem{hughes2014quantum}
C.~Hughes, M.~G. Genoni, T.~Tufarelli \emph{et~al.}
\newblock \href{https://doi.org/10.1103/PhysRevA.90.013810}{Quantum
  non-{G}aussianity witnesses in phase space}.
\newblock \emph{Physical Review A}, \textbf{90}, 013810 (2014).

\bibitem{araki2002entropy}
H.~Araki and E.~H. Lieb.
\newblock Entropy inequalities.
\newblock In \emph{Inequalities}, pages 47--57. Springer (2002).

\bibitem{fogliano2019cavity}
F.~Fogliano, B.~Besga, A.~Reigue \emph{et~al.}
\newblock \href{https://arxiv.org/abs/1904.01140}{Cavity nano-optomechanics in
  the ultrastrong coupling regime with ultrasensitive force sensors}.
\newblock \emph{arXiv preprint} (2019).

\bibitem{PhysRevA.98.052350}
F.~Albarelli, M.~G. Genoni, M.~G.~A. Paris \emph{et~al.}
\newblock \href{https://doi.org/10.1103/PhysRevA.98.052350}{Resource theory of
  quantum non-{G}aussianity and wigner negativity}.
\newblock \emph{Physical Review A}, \textbf{98}, 052350 (2018).

\bibitem{aasi2013enhanced}
J.~Aasi, J.~Abadie, B.~Abbott \emph{et~al.}
\newblock \href{https://doi.org/10.1038/nphoton.2013.177}{Enhanced sensitivity
  of the ligo gravitational wave detector by using squeezed states of light}.
\newblock \emph{Nature Photonics}, \textbf{7}, 613 (2013).

\bibitem{clerk2010introduction}
A.~A. Clerk, M.~H. Devoret, S.~M. Girvin \emph{et~al.}
\newblock \href{https://doi.org/10.1103/RevModPhys.82.1155}{Introduction to
  quantum noise, measurement, and amplification}.
\newblock \emph{Reviews of Modern Physics}, \textbf{82}, 1155 (2010).

\bibitem{moore2016tuneable}
C.~Moore and D.~E. Bruschi.
\newblock
  \href{https://www.researchgate.net/profile/David_Bruschi/publication/290028661_Tuneable_interacting_bosons_for_relativistic_and_quantum_information_processing/links/56b85fcc08ae44bb330ca325/Tuneable-interacting-bosons-for-relativistic-and-quantum-information-processing.pdf}{Tuneable
  interacting bosons for relativistic and quantum information processing}.
\newblock \emph{arXiv preprint arXiv:1601.01919} (2016).

\bibitem{emary2003chaos}
C.~Emary and T.~Brandes.
\newblock \href{https://doi.org/10.1103/PhysRevE.67.066203}{Chaos and the
  quantum phase transition in the dicke model}.
\newblock \emph{Physical Review E}, \textbf{67}, 066203 (2003).

\bibitem{van2010bose}
T.~van Zoest, N.~Gaaloul, Y.~Singh \emph{et~al.}
\newblock \href{https://doi.org/10.1126/science.1189164}{{B}ose-{E}instein
  condensation in microgravity}.
\newblock \emph{Science}, \textbf{328}, 1540 (2010).

\bibitem{lenef1997rotation}
A.~Lenef, T.~D. Hammond, E.~T. Smith \emph{et~al.}
\newblock \href{https://doi.org/10.1103/PhysRevLett.78.760}{Rotation sensing
  with an atom interferometer}.
\newblock \emph{Physical Review Letters}, \textbf{78}, 760 (1997).

\bibitem{bruschi2014testing}
D.~E. Bruschi, C.~Sab{\'\i}n, A.~White \emph{et~al.}
\newblock \href{https://doi.org/10.1088/1367-2630/16/5/053041}{Testing the
  effects of gravity and motion on quantum entanglement in space-based
  experiments}.
\newblock \emph{New Journal of Physics}, \textbf{16}, 053041 (2014).

\bibitem{howl2019exploring}
R.~Howl, R.~Penrose, and I.~Fuentes.
\newblock \href{https://doi.org/10.1088/1367-2630/ab104a}{Exploring the
  unification of quantum theory and general relativity with a bose--einstein
  condensate}.
\newblock \emph{New Journal of Physics}, \textbf{21}, 043047 (2019).

\bibitem{ratzel2019testing}
D.~R{\"a}tzel and I.~Fuentes.
\newblock \href{https://doi.org/10.1088/2399-6528/aaff1f}{Testing small scale
  gravitational wave detectors with dynamical mass distributions}.
\newblock \emph{Journal of Physics Communications}, \textbf{3}, 025009 (2019).

\bibitem{wiseman2009quantum}
H.~M. Wiseman and G.~J. Milburn.
\newblock \emph{Quantum measurement and control}.
\newblock Cambridge University Press (2009).

\bibitem{giovannetti2011advances}
V.~Giovannetti, S.~Lloyd, and L.~Maccone.
\newblock \href{https://doi.org/10.1038/nphoton.2011.35}{Advances in quantum
  metrology}.
\newblock \emph{Nature {P}hotonics}, \textbf{5}, 222 (2011).

\bibitem{liu2019nanoscale}
Y.-X. Liu, A.~Ajoy, P.~Cappellaro \emph{et~al.}
\newblock \href{https://arxiv.org/abs/1908.00790}{Nanoscale vector dc
  magnetometry via ancilla-assisted frequency up-conversion}.
\newblock \emph{Physical Review Letters}, \textbf{122}, 100501 (2019).

\bibitem{fiderer2018quantum}
L.~J. Fiderer and D.~Braun.
\newblock \href{https://doi.org/10.1038/s41467-018-03623-z}{Quantum metrology
  with quantum-chaotic sensors}.
\newblock \emph{Nature Communications}, \textbf{9}, 1351 (2018).

\bibitem{arcizet2006high}
O.~Arcizet, P.~F. Cohadon, T.~Briant \emph{et~al.}
\newblock \href{https://doi.org/10.1103/PhysRevLett.97.133601}{High-sensitivity
  optical monitoring of a micromechanical resonator with a quantum-limited
  optomechanical sensor}.
\newblock \emph{Physical Review Letters}, \textbf{97}, 133601 (2006).

\bibitem{ranjit2016zeptonewton}
G.~Ranjit, M.~Cunningham, K.~Casey \emph{et~al.}
\newblock \href{https://doi.org/10.1103/PhysRevA.93.053801}{Zeptonewton force
  sensing with nanospheres in an optical lattice}.
\newblock \emph{Physical Review A}, \textbf{93}, 053801 (2016).

\bibitem{hempston2017force}
D.~Hempston, J.~Vovrosh, M.~Toro{\v{s}} \emph{et~al.}
\newblock \href{https://doi.org/10.1063/1.4993555}{Force sensing with an
  optically levitated charged nanoparticle}.
\newblock \emph{Applied Physics Letters}, \textbf{111}, 133111 (2017).

\bibitem{marshman2018mesoscopic}
R.~J. Marshman, A.~Mazumdar, G.~W. Morley \emph{et~al.}
\newblock \href{https://arxiv.org/abs/1807.10830}{Mesoscopic interference for
  metric and curvature (mimac) \& gravitational waves}.
\newblock \emph{arXiv:1807.10830} (2018).

\bibitem{bernad2018optimal}
J.~Z. Bern{\'a}d, C.~Sanavio, and A.~Xuereb.
\newblock \href{https://doi.org/10.1103/PhysRevA.97.063821}{Optimal estimation
  of the optomechanical coupling strength}.
\newblock \emph{Physical Review A}, \textbf{97}, 063821 (2018).

\bibitem{rivas2010precision}
{\'A}.~Rivas and A.~Luis.
\newblock \href{https://doi.org/10.1103/PhysRevLett.105.010403}{Precision
  quantum metrology and nonclassicality in linear and nonlinear detection
  schemes}.
\newblock \emph{Physical Review Letters}, \textbf{105}, 010403 (2010).

\bibitem{genoni2009enhancement}
M.~G. Genoni, C.~Invernizzi, and M.~G. Paris.
\newblock \href{https://doi.org/10.1103/PhysRevA.80.033842}{Enhancement of
  parameter estimation by {K}err interaction}.
\newblock \emph{Physical Review A}, \textbf{80}, 033842 (2009).

\bibitem{rossi2016enhanced}
M.~A. Rossi, F.~Albarelli, and M.~G. Paris.
\newblock \href{https://doi.org/10.1103/PhysRevA.93.053805}{Enhanced estimation
  of loss in the presence of {K}err nonlinearity}.
\newblock \emph{Physical Review A}, \textbf{93}, 053805 (2016).

\bibitem{farace2012enhancing}
A.~Farace and V.~Giovannetti.
\newblock \href{https://doi.org/10.1103/PhysRevA.86.013820}{Enhancing quantum
  effects via periodic modulations in optomechanical systems}.
\newblock \emph{Physical Review A}, \textbf{86}, 013820 (2012).

\bibitem{helstrom1976quantum}
C.~W. Helstrom.
\newblock \emph{Quantum detection and estimation theory}.
\newblock Academic press (1976).

\bibitem{holevo2011probabilistic}
A.~S. Holevo.
\newblock \emph{Probabilistic and statistical aspects of quantum theory},
  volume~1.
\newblock Springer Science \& Business Media (2011).

\bibitem{braunstein1994statistical}
S.~L. Braunstein and C.~M. Caves.
\newblock \href{https://doi.org/10.1103/PhysRevLett.72.3439}{Statistical
  distance and the geometry of quantum states}.
\newblock \emph{Physical Review Letters}, \textbf{72}, 3439 (1994).

\bibitem{braunstein1996generalized}
S.~L. Braunstein, C.~M. Caves, and G.~J. Milburn.
\newblock \href{https://doi.org/10.1006/aphy.1996.0040}{Generalized uncertainty
  relations: theory, examples, and {L}orentz invariance}.
\newblock \emph{Annals of Physics}, \textbf{247}, 135 (1996).

\bibitem{peres2006quantum}
A.~Peres.
\newblock \emph{Quantum theory: Concepts and methods}, volume~57.
\newblock Springer Science \& Business Media (2006).

\bibitem{braun2018quantum}
D.~Braun, G.~Adesso, F.~Benatti \emph{et~al.}
\newblock Quantum-enhanced measurements without entanglement.
\newblock \emph{Reviews of Modern Physics}, \textbf{90}, 035006 (2018).

\bibitem{giovannetti2006quantum}
V.~Giovannetti, S.~Lloyd, and L.~Maccone.
\newblock \href{https://doi.org/10.1103/PhysRevLett.96.010401}{Quantum
  metrology}.
\newblock \emph{Physical Review Letters}, \textbf{96}, 010401 (2006).

\bibitem{luis2004nonlinear}
A.~Luis.
\newblock \href{https://doi.org/10.1016/j.physleta.2004.06.080}{Nonlinear
  transformations and the {Heisenberg} limit}.
\newblock \emph{Physics Letters A}, \textbf{329}, 8 (2004).

\bibitem{braun2011heisenberg}
D.~Braun and J.~Martin.
\newblock \href{https://doi.org/10.1038/ncomms1220}{Heisenberg-limited
  sensitivity with decoherence-enhanced measurements}.
\newblock \emph{Nature Communications}, \textbf{2}, 223 (2011).

\bibitem{fraisse2015coherent}
J.~M.~E. Fra{\"\i}sse and D.~Braun.
\newblock \href{https://doi.org/10.1002/andp.201500169}{Coherent averaging}.
\newblock \emph{Annalen der Physik}, \textbf{527}, 701 (2015).

\bibitem{o2010quantum}
A.~D. O’Connell, M.~Hofheinz, M.~Ansmann \emph{et~al.}
\newblock \href{https://doi.org/10.1038/nature08967}{Quantum ground state and
  single-phonon control of a mechanical resonator}.
\newblock \emph{Nature}, \textbf{464}, 697 (2010).

\bibitem{ray2006tide}
R.~Ray and S.~Luthcke.
\newblock \href{https://doi.org/10.1111/j.1365-246X.2006.03229.x}{Tide model
  errors and {GRACE} gravimetry: towards a more realistic assessment}.
\newblock \emph{Geophysical Journal International}, \textbf{167}, 1055 (2006).

\bibitem{crowley2006land}
J.~W. Crowley, J.~X. Mitrovica, R.~C. Bailey \emph{et~al.}
\newblock \href{https://doi.org/10.1029/2006GL027070}{Land water storage within
  the {C}ongo {B}asin inferred from {GRACE} satellite gravity data}.
\newblock \emph{Geophysical Research Letters}, \textbf{33} (2006).

\bibitem{chen2006satellite}
J.~Chen, C.~Wilson, and B.~Tapley.
\newblock \href{https://doi.org/10.1126/science.1129007}{Satellite gravity
  measurements confirm accelerated melting of greenland ice sheet}.
\newblock \emph{science}, \textbf{313}, 1958 (2006).

\bibitem{iess2018measurement}
L.~Iess, W.~Folkner, D.~Durante \emph{et~al.}
\newblock \href{https://doi.org/10.1038/nature25776}{Measurement of
  {J}upiter’s asymmetric gravity field}.
\newblock \emph{Nature}, \textbf{555}, 220 (2018).

\bibitem{biswas2012towards}
T.~Biswas, E.~Gerwick, T.~Koivisto \emph{et~al.}
\newblock \href{https://doi.org/10.1103/PhysRevLett.108.031101}{Towards
  singularity-and ghost-free theories of gravity}.
\newblock \emph{Physical Review Letters}, \textbf{108}, 031101 (2012).

\bibitem{peters2001high}
A.~Peters, K.~Y. Chung, and S.~Chu.
\newblock
  \href{http://iopscience.iop.org/article/10.1088/0026-1394/38/1/4/meta}{High-precision
  gravity measurements using atom interferometry}.
\newblock \emph{Metrologia}, \textbf{38}, 25 (2001).

\bibitem{mcguirk2002sensitive}
J.~McGuirk, G.~Foster, J.~Fixler \emph{et~al.}
\newblock \href{https://arxiv.org/abs/physics/0105088}{Sensitive
  absolute-gravity gradiometry using atom interferometry}.
\newblock \emph{Physical Review A}, \textbf{65}, 033608 (2002).

\bibitem{bidel2013compact}
Y.~Bidel, O.~Carraz, R.~Charriere \emph{et~al.}
\newblock \href{https://doi.org/10.1063/1.4801756}{Compact cold atom gravimeter
  for field applications}.
\newblock \emph{Applied Physics Letters}, \textbf{102}, 144107 (2013).

\bibitem{hu2013demonstration}
Z.-K. Hu, B.-L. Sun, X.-C. Duan \emph{et~al.}
\newblock
  \href{https://journals.aps.org/pra/abstract/10.1103/PhysRevA.88.043610}{Demonstration
  of an ultrahigh-sensitivity atom-interferometry absolute gravimeter}.
\newblock \emph{Physical Review A}, \textbf{88}, 043610 (2013).

\bibitem{abend2016atom}
S.~Abend, M.~Gebbe, M.~Gersemann \emph{et~al.}
\newblock
  \href{https://journals.aps.org/prl/abstract/10.1103/PhysRevLett.117.203003}{Atom-chip
  fountain gravimeter}.
\newblock \emph{Physical Review Letters}, \textbf{117}, 203003 (2016).

\bibitem{johnsson2016macroscopic}
M.~T. Johnsson, G.~K. Brennen, and J.~Twamley.
\newblock \href{https://arxiv.org/abs/1412.6864}{Macroscopic superpositions and
  gravimetry with quantum magnetomechanics}.
\newblock \emph{Scientific Reports}, \textbf{6} (2016).

\bibitem{LaCoste2016}
I.~{Micro-g LaCoste}.
\newblock \href{http://www.microglacoste.com/fg5x.php}{{FG5-X Absolute
  Gravimeter}} (2014).
\newblock \urlprefix\url{http://www.microglacoste.com/fg5x.php}.

\bibitem{dimopoulos2008general}
S.~Dimopoulos, P.~W. Graham, J.~M. Hogan \emph{et~al.}
\newblock \href{https://doi.org/10.1103/PhysRevD.78.042003}{General
  relativistic effects in atom interferometry}.
\newblock \emph{Physical Review D}, \textbf{78}, 042003 (2008).

\bibitem{bruschi2014quantum}
D.~E. Bruschi, A.~Datta, R.~Ursin \emph{et~al.}
\newblock \href{https://doi.org/10.1103/PhysRevD.90.124001}{{Quantum estimation
  of the Schwarzschild spacetime parameters of the Earth}}.
\newblock \emph{Physical Review D}, \textbf{90}, 124001 (2014).

\bibitem{howl2016gravity}
R.~Howl, L.~Hackerm{\"u}ller, D.~E. Bruschi \emph{et~al.}
\newblock \href{https://doi.org/10.1103/PhysRevD.84.044013}{Gravity in the
  quantum lab}.
\newblock \emph{Advances in Physics: X}, \textbf{3}, 1383184 (2018).

\bibitem{seveso2016can}
L.~Seveso and M.~G. Paris.
\newblock \href{https://arxiv.org/abs/1612.07331}{Can quantum probes satisfy
  the weak equivalence principle?}
\newblock \emph{arXiv preprint} (2016).

\bibitem{seveso2017quantum}
L.~Seveso, V.~Peri, and M.~G. Paris.
\newblock
  \href{http://iopscience.iop.org/article/10.1088/1751-8121/aa6cc5/meta}{Quantum
  limits to mass sensing in a gravitational field}.
\newblock \emph{Journal of Physics A: Mathematical and Theoretical},
  \textbf{50}, 235301 (2017).

\bibitem{asenbaum2017phase}
P.~Asenbaum, C.~Overstreet, T.~Kovachy \emph{et~al.}
\newblock
  \href{https://journals.aps.org/prl/abstract/10.1103/PhysRevLett.118.183602}{Phase
  shift in an atom interferometer due to spacetime curvature across its wave
  function}.
\newblock \emph{Physical Review Letters}, \textbf{118}, 183602 (2017).

\bibitem{tan2017relativistic}
Y.-J. Tan, C.-G. Shao, and Z.-K. Hu.
\newblock
  \href{https://journals.aps.org/prd/abstract/10.1103/PhysRevD.95.024002}{Relativistic
  effects in atom gravimeters}.
\newblock \emph{Physical Review D}, \textbf{95}, 024002 (2017).

\bibitem{joshi2017space}
S.~K. Joshi, J.~Pienaar, T.~C. Ralph \emph{et~al.}
\newblock \href{https://arxiv.org/abs/1703.08036}{Space quest mission proposal:
  Experimentally testing decoherence due to gravity}.
\newblock \emph{arXiv preprint} (2017).

\bibitem{Arvanitaki2013}
A.~Arvanitaki and A.~A. Geraci.
\newblock \href{https://arxiv.org/abs/1207.5320}{Detecting high-frequency
  gravitational waves with optically levitated sensors}.
\newblock \emph{Physical Review Letters}, \textbf{110}, 071105 (2013).

\bibitem{jacobs2016quantum}
K.~Jacobs, R.~Balu, and J.~D. Teufel.
\newblock \href{https://arxiv.org/abs/1612.07246}{Quantum-enhanced
  accelerometry with a non-linear electromechanical circuit}.
\newblock \emph{arXiv preprint} (2016).

\bibitem{Scala2013a}
M.~Scala, M.~S. Kim, G.~W. Morley \emph{et~al.}
\newblock \href{https://doi.org/10.1103/PhysRevLett.111.180403}{{Matter-wave
  interferometry of a levitated thermal nano-oscillator induced and probed by a
  spin}}.
\newblock \emph{Physical Review Letters}, \textbf{111}, 1 (2013).

\bibitem{montenegro2015entanglement}
V.~Montenegro, A.~Ferraro, and S.~Bose.
\newblock \href{https://arxiv.org/abs/1503.04462}{Entanglement distillation in
  optomechanics via unsharp measurements}.
\newblock \emph{arXiv preprint} (2015).

\bibitem{barnett2002methods}
S.~M. Barnett and P.~M. Radmore.
\newblock \emph{Methods in theoretical quantum optics}, volume~15.
\newblock Oxford University Press (2002).

\bibitem{cahill1969ordered}
K.~E. Cahill and R.~J. Glauber.
\newblock \href{https://doi.org/10.1103/PhysRev.177.1857}{Ordered expansions in
  boson amplitude operators}.
\newblock \emph{Physical Review}, \textbf{177}, 1857 (1969).

\bibitem{fehlberg1969low}
E.~Fehlberg.
\newblock
  \href{https://ntrs.nasa.gov/archive/nasa/casi.ntrs.nasa.gov/19690021375.pdf}{Low-order
  classical runge-kutta formulas with stepsize control and their application to
  some heat transfer problems} (1969).

\bibitem{press2007numerical}
W.~H. Press, S.~A. Teukolsky, W.~T. Vetterling \emph{et~al.}
\newblock \emph{Numerical recipes 3rd edition: {T}he art of scientific
  computing}.
\newblock Cambridge University Press (2007).

\bibitem{chiow2011102}
S.-w. Chiow, T.~Kovachy, H.-C. Chien \emph{et~al.}
\newblock
  \href{https://journals.aps.org/prl/abstract/10.1103/PhysRevLett.107.130403}{102$\hbar$k
  large area atom interferometers}.
\newblock \emph{Physical Review Letters}, \textbf{107}, 130403 (2011).

\bibitem{kritsotakis2018optimal}
M.~Kritsotakis, S.~S. Szigeti, J.~A. Dunningham \emph{et~al.}
\newblock \href{https://doi.org/10.1103/PhysRevA.98.023629}{Optimal matter-wave
  gravimetry}.
\newblock \emph{Physical Review A}, \textbf{98}, 023629 (2018).

\bibitem{zhao2009vibration}
Y.~Zhao, J.~Zhang, A.~Stejskal \emph{et~al.}
\newblock \href{https://doi.org/10.1364/OE.17.008970}{A vibration-insensitive
  optical cavity and absolute determination of its ultrahigh stability}.
\newblock \emph{Optics Express}, \textbf{17}, 8970 (2009).

\bibitem{pontin2018levitated}
A.~Pontin, L.~S. Mourounas, A.~Geraci \emph{et~al.}
\newblock \href{https://doi.org/10.1088/1367-2630/aaa71c}{Levitated
  optomechanics with a fiber fabry-perot interferometer}.
\newblock \emph{New Journal of Physics} (2018).

\bibitem{hu2014gravitational}
B.~Hu.
\newblock \href{https://doi.org/10.1088/1742-6596/504/1/012021}{Gravitational
  decoherence, alternative quantum theories and semiclassical gravity}.
\newblock \textbf{504}, 012021 (2014).

\bibitem{pfister2016universal}
C.~Pfister, J.~Kaniewski, M.~Tomamichel \emph{et~al.}
\newblock \href{https://www.nature.com/articles/ncomms13022}{A universal test
  for gravitational decoherence}.
\newblock \emph{Nature Communications}, \textbf{7} (2016).

\bibitem{shalashilin2009quantum}
D.~V. Shalashilin.
\newblock \href{https://doi.org/10.1063/1.3153302}{Quantum mechanics with the
  basis set guided by {E}hrenfest trajectories: {T}heory and application to
  spin-boson model}.
\newblock \emph{The Journal of Chemical Physics}, \textbf{130}, 244101 (2009).

\bibitem{ye2012modeling}
S.-Y. Ye, D.~Shalashilin, and A.~Serafini.
\newblock \href{https://doi.org/10.1103/PhysRevA.86.032312}{Modeling of
  quantum-information processing with {E}hrenfest guided trajectories: {A} case
  study with spin-boson-like couplings}.
\newblock \emph{Physical Review A}, \textbf{86}, 032312 (2012).

\bibitem{shahidani2014steady}
S.~Shahidani, M.~Naderi, M.~Soltanolkotabi \emph{et~al.}
\newblock \href{https://doi.org/10.1364/JOSAB.31.001087}{Steady-state
  entanglement, cooling, and tristability in a nonlinear optomechanical
  cavity}.
\newblock \emph{JOSA B}, \textbf{31}, 1087 (2014).

\bibitem{plunien1986casimir}
G.~Plunien, B.~M{\"u}ller, and W.~Greiner.
\newblock \href{https://doi.org/10.1016/0370-1573(86)90020-7}{The {C}asimir
  effect}.
\newblock \emph{Physics Reports}, \textbf{134}, 87 (1986).

\bibitem{clemente2010magnetically}
L.~Clemente.
\newblock
  \href{https://clemente.io/Bachelor_Lucas_Clemente.pdf}{\emph{Magnetically
  levitated micro-objects in the quantum regime}}.
\newblock Ph.D. thesis, Bachelor’s thesis (Ludwig-Maximilians-Universit{\"a}t
  M{\"u}nchen) (2010).

\bibitem{pflanzer2012master}
A.~C. Pflanzer, O.~Romero-Isart, and J.~I. Cirac.
\newblock \href{https://doi.org/10.1103/PhysRevA.86.013802}{Master-equation
  approach to optomechanics with arbitrary dielectrics}.
\newblock \emph{Physical Review A}, \textbf{86}, 013802 (2012).

\bibitem{kovacic2018mathieu}
I.~Kovacic, R.~Rand, and S.~M. Sah.
\newblock \href{https://doi.org/10.1115/1.4039144}{Mathieu's equation and its
  generalizations: Overview of stability charts and their features}.
\newblock \emph{Applied Mechanics Reviews}, \textbf{70}, 020802 (2018).

\end{thebibliography}





\end{document}